\documentclass[12pt]{article}
\pdfoutput=1
\usepackage{amssymb}
\usepackage{amsmath}
\usepackage{latexsym}
\usepackage[usenames]{color}
\usepackage{fancybox}
\usepackage{simplewick}
\usepackage{comment}
\usepackage{cite}
\usepackage{framed}
\definecolor{shadecolor}{rgb}{0.9,0.9,0.95}
\usepackage{setspace}
\usepackage{pdflscape}
\usepackage{multirow}
\usepackage{tabularx,booktabs}
\usepackage{bm}
\definecolor{darkgreen}{rgb}{0,0.5,0}
\definecolor{darkblue}{cmyk}{0.9,0.9,0,0}
\definecolor{darkred}{rgb}{0.6,0,0.3}

\usepackage{graphicx}
\usepackage[setpagesize=false,pagebackref=false, linktocpage, bookmarksopen=true, colorlinks=true, linkcolor=darkblue,citecolor=darkblue,urlcolor=darkblue]{hyperref}
\usepackage{hyperref}

\newcommand{\tr}{{\rm tr}}

\renewcommand{\thefootnote}{\arabic{footnote}}

\def\del{\partial}

\def\fn#1{\footnote{#1}}
\def\nn{\nonumber}
\def\eqref#1{(\ref{#1})}
\def\comma{\,,}
\def\period{\,.}

\def\I{{\bf 1}}
\def\II{{\bf 2}}
\def\III{{\bf 3}}
\def\zero{{\bf 0}}
\def\ket#1{|#1\rangle}

\def\beq{\begin{equation}}
\def\eeq{\end{equation}}
\def\pmatrix#1#2{\left(
\begin{array}{#1}
#2\end{array}
\right)}
\def\red#1{\textcolor[rgb]{1, 0, 0}{#1}}
\def\blue#1{\textcolor[rgb]{0,0,1}{#1}}
\numberwithin{equation}{section}

\begin{document}
\thispagestyle{empty}

\renewcommand{\thefootnote}{\fnsymbol{footnote}}
\setcounter{page}{1}
\setcounter{footnote}{0}
\setcounter{figure}{0}
\begin{flushright}
{\tt CERN-TH-2019-093
\\
\tt Imperial-TP-EV-2019-01
}
\end{flushright}
\vspace{0.7cm}
\begin{center}
{\large\textbf{\mathversion{bold}
Structure Constants  in $\mathcal{N}=4$ SYM at Finite Coupling as Worldsheet $g$-Function
}\par}

\vspace{1.3cm}

\textrm{Yunfeng Jiang$^{\textcolor[rgb]{0.8,0,0.1}{\blacktriangleright}}$, Shota Komatsu$^{\textcolor[rgb]{0,0,0.8}{\blacktriangledown}}$, Edoardo Vescovi$^{\textcolor[rgb]{0.2,0.8,0}{\blacklozenge}}$}
\\ \vspace{1cm}
\footnotesize{\textit{
$^{\textcolor[rgb]{0.8,0,0.1}{\blacktriangleright}}$Theoretical Physics Department, CERN,  1211 Geneva 23, Switzerland\\
$^{\textcolor[rgb]{0,0,0.8}{\blacktriangledown}}$School of Natural Sciences, Institute for Advanced Study, Princeton, New Jersey 08540, USA\\
$^{\textcolor[rgb]{0.2,0.8,0}{\blacklozenge}}$The Blackett Laboratory, Imperial College, London SW7 2AZ, United Kingdom\\
}
\vspace{1cm}
}

{\tt Yunfeng.Jiang AT cern.ch, shota.komadze AT gmail.com, e.vescovi AT imperial.ac.uk}

\par\vspace{1.5cm}

\textbf{Abstract}\vspace{2mm}
\end{center}
\noindent
We develop a novel nonperturbative approach to a class of three-point functions in planar $\mathcal{N}=4$ SYM based on Thermodynamic Bethe Ansatz (TBA). More specifically, we study three-point functions of a non-BPS single-trace operator and two determinant operators dual to maximal Giant Graviton D-branes in AdS$_5\times$S$^{5}$. They correspond to disk one-point functions on the worldsheet and admit a simpler and more powerful integrability description than the standard single-trace three-point functions. We first introduce two new methods to efficiently compute such correlators at weak coupling; one based on large $N$ collective fields, which provides an example of open-closed-open duality discussed by Gopakumar, and the other based on combinatorics. The results so obtained exhibit a simple determinant structure and indicate that the correlator can be interpreted as a generalization of $g$-functions in 2d QFT; an overlap between an integrable boundary state and a state corresponding to the single-trace operator. We then determine the boundary state at finite coupling using the symmetry, the crossing equation and the boundary Yang-Baxter equation. With the resulting boundary state, we derive the ground-state $g$-function based on TBA and conjecture its generalization to other states. This is the first fully nonperturbative proposal for the structure constants of operators of finite length. The results are tested extensively at weak and strong couplings. Finally, we point out that determinant operators can provide better probes of sub-AdS locality than single-trace operators and discuss possible applications.

\setcounter{page}{1}
\renewcommand{\thefootnote}{\arabic{footnote}}
\setcounter{footnote}{0}
\setcounter{tocdepth}{3}
\newpage

\parskip 5pt plus 1pt   \jot = 1.5ex
\tableofcontents

\newpage
\section{Introduction\label{sec:intro}}
Understanding the low-energy physics of Quantum Chromodynamics (QCD) is an important theoretical challenge with diverse physical implications. This however is a difficult question since QCD is strongly coupled at low energy and lacks controllable perturbative expansions. One way of making progress is to draw lessons from the analysis of simpler theories such as supersymmetric gauge theories or theories in 't Hooft's large $N$ limit \cite{tHooft:1973alw}. In particular, the large $N$ limit provides a useful expansion parameter $1/N$ for otherwise strongly-coupled field theories and makes manifest some of the salient features of the low-energy dynamics. For instance, the theory in the large $N$ limit can be thought of as a theory of almost stable mesons and glueballs whose interactions are weak and  suppressed by $1/N$. This resembles, at least qualitatively, what we expect in the low-energy QCD. We should however note that there is yet another important class of excitations in the real-world QCD, namely baryons, which are color singlets made up of $N$ quarks,
\beq
(\text{Baryon})\sim \epsilon^{i_1,\cdots ,i_N}q_{i_1}\cdots q_{i_N}\period
\eeq
 Unlike mesons and glueballs, the description of baryons in the large $N$ limit is less straightforward: Being composed of a large number of quarks, they are heavy excitations and are described either as a classical solitonic configuration of the pion effective Lagrangian or as a collection of $N$ quarks moving in a mean-field potential \cite{Witten:1979kh,Witten:1983tx}\fn{See also a recent interesting proposal on the description of high-spin baryons as quantum-hall droplets \cite{Komargodski:2018odf}.}. Although both approaches capture important qualitative features, it is not necessarily easy to make the argument more rigorous\fn{An important exception is the large $N$ QCD in two dimensions discussed in \cite{Witten:1979kh}, in which one can resum the 't Hooft expansions and formulate a relativistic Hartree equation which becomes exact in the large $N$ limit.} and perform a systematic computation based on such physical pictures.

The purpose of this paper is to analyze analogues of baryons in $\mathcal{N}=4$ supersymmetric Yang-Mills theory ($\mathcal{N}=4$ SYM) in four dimensions, and develop nonperturbative methods to study such observables with the help of powerful integrability machineries. Being a conformal field theory, natural physical quantities in $\mathcal{N}=4$ SYM are local operators rather than particle excitations. In particular, the most basic operators in the large $N$ limit are {\it single-trace operators}, which are of the form
\beq
\mathcal{O}\sim {\rm tr}\left(\Phi \cdots \Phi\right)\period
\eeq
These operators play similar roles to  mesons and glueballs in the large $N$ QCD, and much like mesons and glueballs, their three-point couplings (or more precisely the OPE structure constants) are suppressed by $1/N$. On the other hand, a counterpart of baryons in $\mathcal{N}=4$ SYM, which only contains matters in the adjoint representation, is a {\it determinant operator} defined by
\beq
\det \Phi \sim \epsilon^{i_1,\cdots ,i_N}\epsilon^{j_1,\cdots,j_N} \Phi_{i_1,j_1}\cdots \Phi_{i_N,j_N}\period
\eeq
As we explain shortly, the determinant operators also play an important role in planar $\mathcal{N}=4$ SYM and in its relation to holography.

In the past fifteen years, significant progress has been made in computing the spectrum and the correlation functions of single-trace operators. This was achieved largely by the successful application of integrability techniques \cite{Beisert:2010jr}, which were originally developed in the study of two-dimensional quantum field theories. By contrast, the analysis of determinant operators is much less developed. The spectrum of small deformations of determinant operators was analyzed using various approaches \cite{Balasubramanian:2002sa,Berenstein:2003ah,Berenstein:2004kk,Balasubramanian:2004nb,Berenstein:2005vf,deMelloKoch:2007rqf, deMelloKoch:2007nbd,Bekker:2007ea,Koch:2010gp,DeComarmond:2010ie,Carlson:2011hy, deMelloKoch:2011ci,deMelloKoch:2012ck,deCarvalho:2018xwx} including integrability \cite{Hofman:2007xp,Mann:2006rh,Berenstein:2005fa,Agarwal:2006gc,Okamura:2006zr,Berenstein:2006qk,Chen:2007ec,Nepomechie:2009zi,Correa:2009dm,Galleas:2009ye,Bajnok:2012xc,Bajnok:2013wsa,Bajnok:2013sza,Zhang:2015fea,Bajnok:2015kfz}. However, not much work has been done on correlation functions\fn{See also \cite{Bak:2011yy} for the study of correlation functions of open strings attached to determinant operators and a single-trace operator. More recently an attempt was made to study such correlators using the hexagon formalism \cite{Kim:2019gcq}.} except in the protected BPS sector \cite{Bissi:2011dc,Caputa:2012yj,deMelloKoch:2019dda} and in the free field limit, where various interesting structures were found using algebraic \cite{Kimura:2007wy}, combinatorial \cite{Corley:2001zk,deMelloKoch:2004crq,Kimura:2016bzo} and coherent state \cite{Berenstein:2013md} approaches.  In this paper, we first develop two new methods, one based on the large $N$ collective fields and the other based on direct Wick contractions, to study the perturbative correlation functions of determinant operators efficiently. We then introduce a nonperturbative method based on integrability, in particular on Thermodynamic Bethe Ansatz (TBA), to analyze the three-point function of two BPS determinant operators and a general non-BPS single-trace operator $\mathcal{O}$. A typical representative of such three-point functions is
\beq
\langle \det Z (x_1)\,  \det \bar{Z} (x_2) \, \mathcal{O}(x_3)\rangle \comma
\eeq
with $Z$ and $\bar{Z}$ being a complex scalar field and its conjugate.
\subsection{Determinants are interesting...\label{subsec:motivation}}
Before discussing the contents of this paper, let us give a couple of more physical motivations behind our analysis of determinant operators.

\begin{figure}[t]
\centering
\includegraphics[clip,height=4cm]{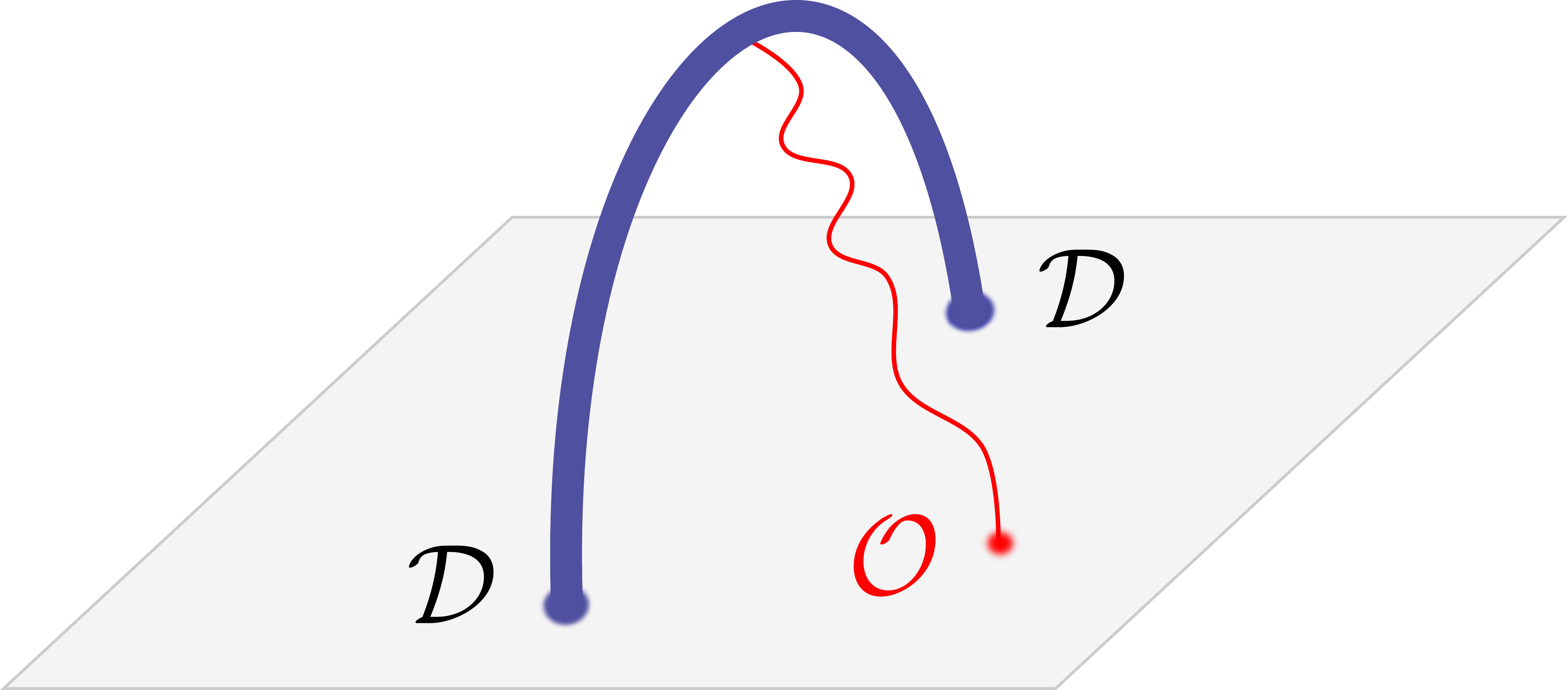}
\caption{The AdS description of the three-point function of two determinant operators and one single-trace operator. The determinant operators create a D-brane in the bulk which travels along the geodesics. The red wavy line represents a closed string worldsheet which gets attached to the D-brane.}
\label{fig:fig1}
\end{figure}

The first motivation comes from, as we already mentioned, the similarity to baryons in the large $N$ QCD: Developing useful techniques for determinant operators has potential bearings on the analysis of baryons and more general complicated composite operators in large $N$ quantum field theories. In addition, the integrability of $\mathcal{N}=4$ SYM makes it possible to perform the analysis without relying on the perturbative expansion, providing a possibility to gain insights into nonperturbative properties of baryons and determinants from explicit computations.

The second motivation is due to the fact that the determinant operator is a localized classical probe in AdS which allows us to extract sub-AdS physics in the AdS/CFT correspondence: The scaling dimension of the determinant operators is $O(N)$, which is much larger than $1$ in the large $N$ limit. According to the standard AdS/CFT dictionary, this translates to the fact that the Compton wave length of the particle dual to the determinant operator is much smaller than the AdS radius, and can be treated as a local classical particle travelling along the geodesics. At the same time, the determinant operator does not backreact or deform the AdS geometry since the dimension of the operator is much smaller than the inverse Newton constant $N^2\sim 1/G_{\rm Newton}$. For the specific determinant operators that we discuss in this paper, one can make these claims more quantitative since the precise AdS dual is known \cite{Balasubramanian:2001nh}: They are dual to the so-called Giant Graviton D3-brane in AdS$_5\times $S$^5$ \cite{McGreevy:2000cw}, which is point-like in the AdS subspace being in line with what we said above (see also figure \ref{fig:fig1}).
 These properties make the determinant operator an ideal probe of the local bulk physics.  A similar point of view was taken in a series of interesting works by Ferrari \cite{Ferrari:2012nw, Ferrari:2013hg, Ferrari:2013wla, Ferrari:2013aba, Ferrari:2013waa} in which he analyzed various gauge-theory observables dual to D-branes and succeeded in reading off the bulk metric from gauge theories in some simple setups. In addition the determinant operators in the matrix models were previously utilized to explore the non-perturbative structure of the bulk spacetime in $(p,q)$ minimal string theory \cite{Maldacena:2004sn}. Although we will not directly address such a question in this paper, it is likely that the correlators of the determinant operators and the techniques developed in this paper would provide much cleaner and simpler setups than the correlators of single-trace operators for extracting the local bulk physics. See the conclusion section for further discussions on this point.

\begin{figure}[t]
\centering
\includegraphics[clip,height=1.8cm]{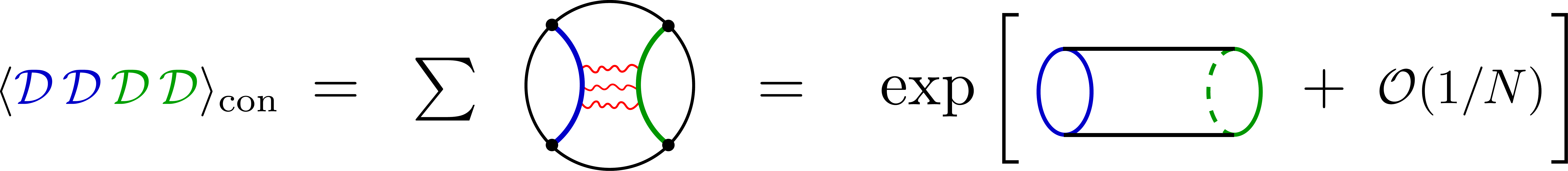}
\caption{The connected four-point function of determinant operators. On the AdS side, it is given by a sum of exchanges of closed string states. The leading contribution at large $N$ can be computed by exponentiating the cylinder partition function.}
\label{fig:fig2}
\end{figure}

We should also mention that operators with analogous properties are often discussed in the AdS$_3$/CFT$_2$ correspondence. The counterparts in that context are the operators with $\Delta \sim \sqrt{c}$ with $c$ being the central charge of CFT$_2$. Using the Virasoro symmetry, it was shown that the four-point function of such operators in the large $c$ limit can be computed by exponentiating the graviton exchange diagram \cite{Fitzpatrick:2014vua}. It was later discussed in \cite{Maxfield:2017rkn} through the worldline analysis in AdS that a similar exponentiation should hold also in holographic\fn{Here we are following the standard terminology in which ``holographic CFTs'' means the large $N$ CFT with a large gap in the spectrum.} CFTs in higher dimensions. As we explain in this paper, our semi-classical approach allows us to show such an exponentiation already at weak coupling, thereby providing additional supports to such claims. See figure \ref{fig:fig2} for further explanations.

\begin{figure}[t]
\centering
\includegraphics[clip,height=3cm]{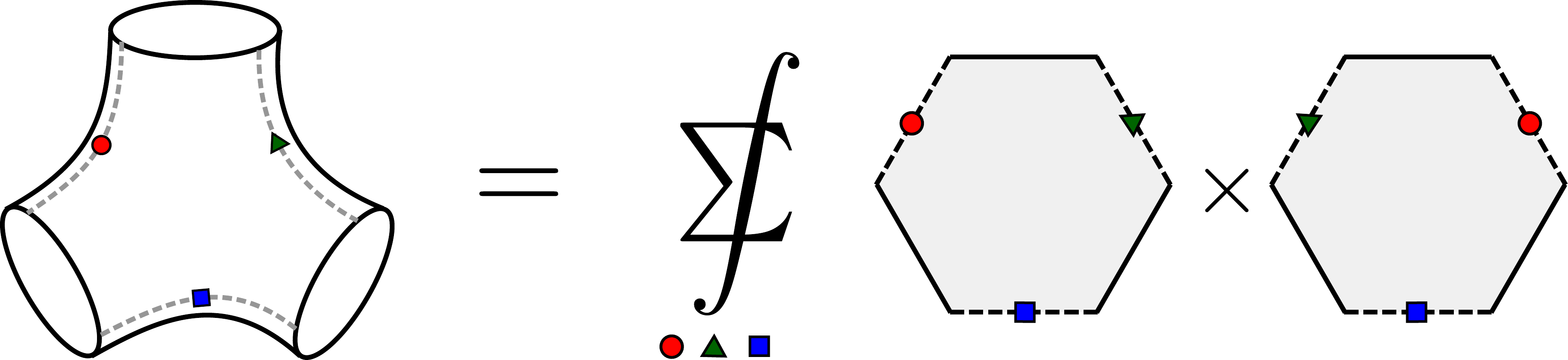}
\caption{The hexagon approach to the three-point function. In this approach one cuts the worldsheet into hexagonal patches and glue then together by summing over intermediate states (the states along the dashed lines in the figure).}
\label{fig:fig3}
\end{figure}

The third motivation comes from the relation to integrability: Somewhat surprisingly, the three-point function of two determinants and one single-trace operator admits a much simpler integrability description than the three-point functions of single-trace operators commonly studied in the literature. To understand this seemingly counterintuitive fact, let us briefly review the current status of the integrability-based approach to the correlation functions.
For the correlation functions of single-trace operators, there exists an integrability-based framework called the hexagonalization formalism\cite{Basso:2015zoa,Fleury:2016ykk,Eden:2016xvg}, which does not rely on the weak-coupling expansion. In this formalism, one constructs a complicated string worldsheets describing planar (and non-planar \cite{Bargheer:2017nne,Eden:2017ozn,Bargheer:2018jvq}) correlation functions by gluing together hexagonal patches (see figure \ref{fig:fig3})\fn{See also \cite{Bajnok:2015hla,Bajnok:2017mdf}, in which similar gluing procedures were discussed in the context of the lightcone string vertex.}. The gluing procedure involves a summation over the intermediate states and the result is therefore given by an infinite sum of such states \cite{Basso:2015zoa,Fleury:2016ykk}. As the contribution from each state can be computed at finite coupling, the formalism enables us to study the regimes which cannot be accessed with the usual perturbative expansions, most notably the strong-coupling limit \cite{Jiang:2016ulr}. However, it is still unsatisfactory as a method to compute the finite-coupling correlation functions since the summation over the intermediate states is intractable except in certain limiting situations \cite{Coronado:2018ypq,Bargheer:2019kxb,Kostov:2019auq}.

\begin{figure}[t]
\centering
\includegraphics[clip,height=1.7cm]{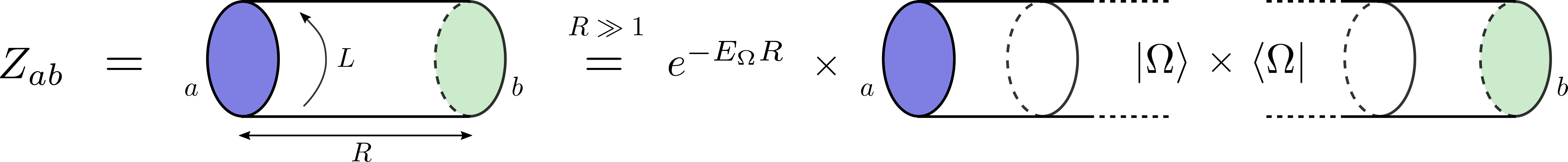}
\caption{The $g$-function and the cylinder partition function in 2d QFT. When the length of the cylinder is large $R\gg 1$, the partition function can be approximated by the overlaps between the boundary states and the ground state. This allows us to compute the $g$-functions in integrable QFTs by taking the limit of the cylinder partition function, which in turn can be computed by the Thermodynamic Bethe ansatz.}
\label{fig:figextra}
\end{figure}

By contrast, as we will show in this paper, the worldsheet topology for the three-point function of determinants and a single-trace operator is a disk with one puncture, which is topologically equivalent to a semi-infinite cylinder (see figure \ref{fig:figextra}). This is also the topology relevant for the standard analysis of the spectrum in integrable QFTs based on Thermodynamic Bethe Ansatz (TBA). This suggests that one may be able to study them in a more conventional TBA formalism without resorting to the somewhat exotic hexagonalization procedure. This expectation turns out to be correct: As we discuss in sections \ref{sec:integrability} and \ref{sec:TBA}, the three-point function studied in this paper can be interpreted as (a generalization of) the so-called {\it $g$-function} in the two-dimensional QFT \cite{affleck1991universal}, which is defined as an overlap between a boundary state and the ground state on a cylinder and is known to quantify the degrees of freedom associated with a boundary \cite{Friedan:2003yc,Casini:2016fgb}. The methods to compute the $g$-function in integrable QFTs \cite{LeClair:1995uf,Dorey:2004xk,Pozsgay:2010tv,Woynarovich:2010wt,Kostov:2018dmi} were partially developed in the literature based on TBA \cite{Chatterjee:1995be,Dorey:1997yg,Dorey:1999cj,Woynarovich:2004gc,Dorey:2009vg,Dorey:2010ub,Kostov:2019fvw} and we will demonstrate that they can be applied also to our problem after appropriate generalizations. The advantage of applying the TBA formalism is that it automatically resums the intermediate states, unlike the hexagon formalism in which one needs to compute each term separately.

\subsection{...but they are hard\label{subsec:hard}}
Although it is important to study the determinant operators, we should note that computing the correlators of determinants at large $N$ is a rather nontrivial task even at weak coupling.
 This is a well-known difficulty in the large $N$ expansion of baryonic observables \cite{Witten:1979kh}, but it would be useful for us to give a brief review and emphasize several important aspects.

To see the difficulty in a simple set up, let us compare the general structure\fn{Precisely speaking, the determinant operators that we study in this paper are BPS and therefore there are no perturbative corrections to the two-point functions. However, the general  discussion presented here also applies to the three-point function of two determinants and one single-trace operator, which does have perturbative corrections.} of the perturbative two-point functions of single-trace operators and determinant operators in the planar limit.
At tree-level, the computation of the two-point function of single-trace operators is straightforward: One simply needs to perform {\it planar} Wick contractions; namely we contract fields inside the traces so that the propagators do not cross each other. On the other hand, the two-point function of determinants is already complicated at tree level since there is no notion of ``ordering'' of fields inside determinants, and one has to consider various contraction patterns and see which ones survive in the large $N$ limit.  This however is purely a technical problem which can be solved by carefully working out the combinatorics. It is at loop levels where intrinsic difficulty shows up.

\begin{figure}[t]
\centering
\includegraphics[clip,height=5cm]{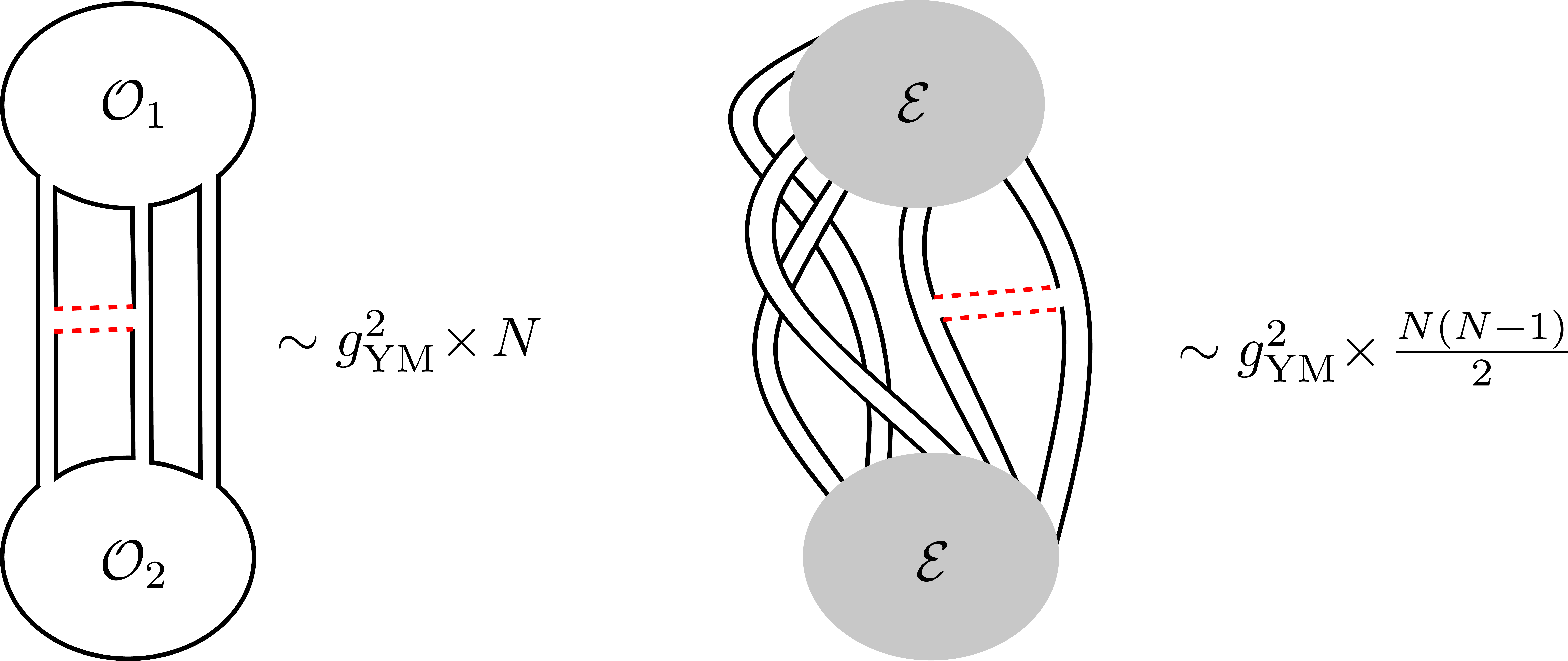}
\caption{The comparison of large $N$ two-point functions of single-trace operators (left panel) and determinant operators (right panel). Unlike the single-trace two-point function, Wick contractions of determinant operators are complicated owing to the epsilon tensors. In addition, the one-loop corrections from the gluon exchange are enhanced by a combinatorial factor $N(N-1)/2$.}
\label{fig:fig4}
\end{figure}

At one loop, the tree-level diagrams get dressed by the interaction vertices. For simplicity, let us focus on the corrections coming from exchanging a single gluon between two propagators. In the case of single-trace operators, such an exchange graph can only be drawn between neighboring fields in the traces owing to the planarity of the diagram. As a result, the diagram acquires two interaction vertices and one extra face, leading to the correction
\beq
\left.(\text{2pt of ${\rm tr}$'s})\right|_{\text{1-loop}}\propto g_{\rm YM}^2 N =\lambda\comma
\eeq
which is proportional to the 't Hooft coupling $\lambda$ and is of $O(1)$ in the planar limit. On the other hand, for determinant operators, one can in principle draw an exchange graph between any pair of the propagators since fields inside the determinants are permutation invariant up to sign. Therefore, the result is now multiplied by a combinatorial factor $N(N-1)/2$:
\beq
\left.(\text{2pt of ${\rm det}$'s})\right|_{\text{1-loop}}\propto g_{\rm YM}^2 \frac{N(N-1)}{2} \sim N \lambda \period
\eeq
See also figure \ref{fig:fig4}.
 We can now see the problem: For the two-point function of determinant operators, the one-loop correction is multiplied by an extra factor of $N$ and is divergent in the planar limit. In fact, the problem only gets worse as we increase the loop order since the expansion in general takes the following form,
 \beq
 (\text{2pt of ${\rm det}$'s}) =(\text{Tree-level})\left[1+ \lambda (c_{1}^{(0)} N+c_{1}^{(1)}) + \lambda^2(c_{2}^{(0)}N^2+c_{2}^{(1)}N +c_{2}^{(2)})+\cdots\right]\comma
 \eeq
 with $c_{a}^{(b)}$ being $O(1)$ numbers.

 At first sight, this may seem to suggest that the perturbative expansion is never valid for the determinant operators. However, it is a bit too hasty to draw such a conclusion: In quantum field theories, it is often the case that even if the naive expansion behaves badly at each order in perturbation theory the resummed quantity is well-defined and admits a simple perturbative expansion. One such example is the computation of anomalous dimensions of local operators: If we naively perform the perturbative expansion of the two-point functions, we obtain $(\log |x_{12}|)^{k}$ terms at $k$-th loop order, which are badly divergent at long distance. However, these logarithms can be resummed and get exponentiated, leading to the anomalous dimension of the operator $\gamma (\lambda)$:
 \beq
 \begin{aligned}
 &1+ \lambda \left[c_{1}^{(0)} \log \Lambda|x_{12}|+c_{1}^{(1)}\right] + \lambda^2\left[c_{2}^{(0)}(\log \Lambda|x_{12}|)^2+c_{2}^{(1)}\log \Lambda|x_{12}| +c_{2}^{(2)}\right]+\cdots\\
 &\hspace{10cm}\propto \frac{1}{|x_{12}|^{2\gamma(\lambda)}}\period
\end{aligned}
 \eeq
As is well-known,  it is the anomalous dimension $\gamma (\lambda)$, not the bare two-point function itself, which has a well-defined perturbative expansion.
Such an exponentiation\fn{For baryons in the large $N$ QCD, the exponentiation was already discussed in \cite{Witten:1979kh} based on general physical arguments.} is indeed expected for determinants in $\mathcal{N}=4$ SYM as was discussed in \cite{Aharony:2002nd} using the combinatorial argument: Namely the two-point function of the determinants takes the following form,
\beq\label{eq:expecteddettwo}
(\text{2pt of ${\rm det}$'s}) =\exp\left[ N f_0(\lambda)+f_1 (\lambda)+\frac{f_2 (\lambda)}{N}+\cdots\right]\comma
\eeq
with $f_{k}(\lambda)$ being $O(1)$ and expandable in powers of $\lambda$.

Let us take a close look at the expected structure of the two-point function \eqref{eq:expecteddettwo}. One notable feature is that the leading large $N$ answer gives a large exponent proportional to $N$. Such a large exponent is indicative of the semi-classical expansions e.g.~the WKB approximation, and suggests the existence of some alternative description of this correlator which becomes classical in the large $N$ limit. We should however note that it  cannot be a classical solution of the original Yang-Mills theory since the classical solution of the Yang-Mills theory gives a contribution $e^{N^2}$, not $e^{N}$. This poses a question:

{\it What becomes classical in the large $N$ limit of correlators of determinants?}

\noindent The answer to this question is known on the AdS side: As we mentioned above, the determinants (and baryons) are expected to be dual to D-branes, which are heavy and classical in the large $N$ limit. We also know that the baryons can be described by a classical solitonic configuration of the pion effective Lagrangian in the large $N$ QCD. However, it is not a priori clear how to obtain these effective descriptions starting from the original Yang-Mills theory.

One of the main purposes of this paper is to derive such a classical description starting from the original $\mathcal{N}=4$ SYM. Besides manifesting the classical nature of the problem, the classical description allows us to bypass complicated combinatorial arguments and perform the weak-coupling computation more straightforwardly as we see in section \ref{sec:effective}.
\subsection{Outline and brief summary\label{subsec:outline}}
Let us now explain the outline and preview some of the main results derived in this paper.

First in section \ref{sec:general}, we explain the setup of our problem and summarize general properties of three-point functions studied in this paper, such as the spacetime and R-symmetry dependences, symmetries, and the AdS dual.

In the subsequent two sections, we develop two new methods to efficiently compute the correlator of determinant operators at weak coupling.
In section \ref{sec:effective}, we introduce a semi-classical method for determinant operators. In this method, we first represent the determinant operators as Gaussian integrals of zero-dimensional fermions, and perform a sequence of integrating-in-and-out procedures. The resulting action is given in terms of auxiliary bilocal bosonic fields and turns out to become classical in the large $N$ limit. Physically, it can be interpreted as a version of open string field theory (OSFT) on Giant Graviton D-branes: For the correlation function of $k$ determinant operators, the bosonic fields are valued in $k\times k$ matrices, reproducing the expected Chan-Paton structure of OSFT. We also discuss other physical implications of the method including the relation to the graph duality, tricks known in the matrix model literature \cite{brezin2000characteristic,Brezin:2007aa,Brezin:2007wv} and the ``open-closed-open'' duality proposed by Gopakumar \cite{Gopakumar}.

One interesting outcome of this approach is that the correlator of $k$ determinant operators and one single-trace operator can be computed by applying the following substitution rule to the single-trace operator:
\beq\label{eq:introsubstitute}
{\rm tr}\left(\Phi^{I_1}\Phi^{I_2}\cdots \Phi^{I_L}\right) \quad \mapsto \quad -{\rm Tr}_{k}\left[M^{I_1}M^{I_2}\cdots M^{I_{L}}\right]\period
\eeq
Here $M_I$'s are $c$-number matrices of size $k$ which are dynamically determined by the positions of determinant operators. As will be discussed in section \ref{subsec:analogy}, the right hand side of \eqref{eq:introsubstitute} can be interpreted as a gauge-invariant observable on Giant Graviton D-branes, not of the original $\mathcal{N}=4$ SYM, and provides an analogue of the Ellwood invariant in OSFT \cite{Hashimoto:2001sm,Gaiotto:2001ji,Ellwood:2008jh}. It is also reminiscent of {\it large $N$ master fields} discussed previously for simple large $N$ theories \cite{Gopakumar:1994iq}.  The same correlator can be written alternatively  as an overlap between a state describing the single-trace operator and the following {\it matrix product state},
\beq
| M\rangle =-\sum_{I_1,\ldots ,I_L}{\rm Tr}_{k}\left[M^{I_1}\cdots M^{I_L}\right]|\Phi^{I_1}\cdots\Phi^{I_L}\rangle\period
\eeq
This is similar to one-point functions in the presence of a domain-wall defect \cite{deLeeuw:2015hxa,Buhl-Mortensen:2015gfd,Buhl-Mortensen:2016pxs,deLeeuw:2016umh,Buhl-Mortensen:2016jqo,deLeeuw:2016ofj,Buhl-Mortensen:2017ind,deLeeuw:2018mkd,Grau:2018keb}, but one important difference is that here the matrices $M$'s are purely {\it emergent} while, in the defect one-point function, they simply come from the expectation value of the scalar fields in $\mathcal{N}=4$ SYM.
For more details, see section \ref{subsec:derivation}.

In section \ref{sec:PCGG}, we develop an alternative approach based on direct Wick contractions, building upon earlier works on determinant operators in the literature. The key idea is to first perform Wick contractions between a pair of determinant operators. The resulting object, which we call the {\it partially-contracted Giant Graviton} (PCGG), is a sum of multi-trace operators and is amenable to the standard 't Hooft expansions. Although the approach is less physically intuitive, it has an advantage that it reduces the problem to the computation of correlators of single- and multi-trace operators and allows us to recycle the existing results in the literature.

Then in section \ref{sec:weak}, we apply these methods to the three-point function of two determinant operators and one single-trace operator at tree level. Interestingly, the two methods naturally lead to two different descriptions of the same quantity: As briefly described above, the computation in the semi-classical approach boils down to evaluating the single-trace operator on an emergent classical background, or equivalently an overlap between a matrix product state and a single-trace state. On the other hand, the PCGG approach leads to an overlap between a generalization of the N\'{e}el states and a single-trace state. The actual computation is performed in the SU(2), SL(2) and SO(6) sectors. For the SU(2) sector, the generalized N\'{e}el state that we obtain is the same as the one studied in \cite{deLeeuw:2015hxa,Foda:2015nfk} while the states for SL(2) and SO(6) sectors are new and provide natural generalizations of the N\'{e}el state for each sector: For instance, the generalized N\'{e}el state in the SL(2) sector is given by
\beq
| \text{N\'{e}el}_{\rm SL(2)} \rangle=\sum_{\substack{\text{all states with}\\|n_i-n_j|:\text{even}}}  \left| \cdots \circ\, \underset{n_1}{\bullet}\,\circ\cdots \circ\,\underset{n_2}{\bullet}\,\circ\cdots\circ\,\underset{n_3}{\bullet}\,\circ\cdots\right>\comma
\eeq
where $\bullet$ and $\circ$ denote the sites with and without magnons respectively.

The result for the structure constants exhibits two important features: First, it vanishes unless the Bethe roots (of the single-trace operator) are parity symmetric. Second, it is given by a ratio of two determinants, each of which resembles the Gaudin formula for the norm of the Bethe state. More concretely, the results in all the three sectors can be uniformly expressed in terms of (middle-node) Bethe roots $u_s$'s as follows:
\beq\label{eq:dotree}
\left.\mathfrak{D}_{\mathcal{O}_{\bf u}}\right|_{\text{tree level}}=-\frac{i^{J}+(-i)^{J}}{2^{M}\sqrt{L}}\sqrt{\left(\prod_{1\leq s\leq \frac{M}{2}}\frac{u_s^2+\frac{1}{4}}{u_s^2}\right)\frac{\det G_{+}}{\det G_{-}}}\period
\eeq
Here $L$ is the length of the operator and $M$ is the number of magnons while $J$ is the U(1) R-charge of the operator. The only sector-dependent part is the ratio of determinants $G_{\pm}$ whose explicit forms are given in \eqref{eq:Gpmsu2}, \eqref{eq:GpmSL2} and \eqref{eq:GpmSO6}. In Appendix \ref{ap:Qfunctions}, we also provide an alternative expression for the SU(2) sector in terms of integrals of so-called $Q$-functions.

In section \ref{sec:integrability}, we explain how these two features can be understood naturally from the viewpoint of integrable quantum field theories: Namely we discuss that the three-point functions studied in this paper can be interpreted as overlaps between an integrable boundary state and a state corresponding to the single-trace operators, which are generalizations of $g$-functions in integrable QFT. We then explain the TBA formalism for $g$-functions and provide a qualitative explanation on how the aforementioned two features can be reproduced from such a physical picture. Our derivation of TBA and $g$-function is slightly unconventional and simplifies several aspects of the derivations found in the literature. In particular, it leads to a new form of the effective action in which the $Y$-function plays the role of a fundamental field variable and from which TBA and the $g$-function can be computed in a more transparent manner. We also provide a more rigorous derivation of $g$-functions in Appendix \ref{ap:gfunctionderivation}.

After that, in section \ref{sec:bootstrap} we determine the details of the relevant integrable boundary state. We first impose the symmetry and the so-called Watson's equation to constrain the reflection matrix. Since the symmetry group is identical to the one of the single-trace three-point function, we find a solution identical to the hexagon form factor; a quantity relevant for the single-trace three-point function. However, we also find a one-parameter family of other solutions: This is because magnon rapidities are always parity symmetric in our setup, and the constraints from the aforementioned conditions are somewhat weaker. We then impose the boundary Yang-Baxter equation and show that it is actually one of the latter solutions which we need to pick. This solution admits a simple ``unfolding'' interpretation analogous to the reflection matrix for the half-BPS Wilson loop \cite{Drukker:2012de,Correa:2012hh}.
Finally we determine the overall phase, called the boundary dressing phase, by solving the crossing equation.

In section \ref{sec:asymptotic}, based on the reflection matrix that we determined, we present a concrete proposal for the nonperturbative structure constant in the asymptotic limit, in which the length of the single-trace operator is large. The resulting formula is a simple generalization of \eqref{eq:dotree} and reads
\beq\label{eq:introasympt}
\left.\mathfrak{D}_{\mathcal{O}_{{\bf u}}}\right|_{\text{asymptotic}}=-\frac{i^{J}+(-i)^{J}}{\sqrt{L}}\sqrt{\left(\prod_{1\leq s\leq \frac{M}{2}}\frac{u_s^2+\frac{1}{4}}{u_s^2}\sigma_B^2(u_s)\right)\frac{\det G_{+}}{\det G_{-}}}\period
\eeq
See \eqref{eq:fullbdress} and \eqref{eq:defGpmAsympt} for definitions of $\sigma_B(u)$ and $G_{\pm}$. We also show that one can eliminate the dependence on the Bethe roots at the nested levels and express the final result purely in terms of the middle-node Bethe roots.

In section \ref{sec:TBA}, we apply the general framework of the $g$-function in section \ref{sec:integrability} to our problem and derive a nonperturbative expression for the structure constant which is valid for operators of finite length. We first formulate a (mirror) boundary TBA and derive a nonperturbative expression for the $g$-function, which is given by a ratio of Fredholm determinants. As is the case with the usual TBA arguments, the $g$-function is derived first for the ground state. We then generalize the results to the excited states in the SL(2) sector using the analytic continuation arguments \cite{Dorey:1996re,Dorey:1997rb}. The result is a natural generalization of \eqref{eq:introasympt} and is given by
\beq\label{eq:introfinite}
\begin{aligned}
\left.\mathfrak{D}_{\mathcal{O}_{\bf u}}\right|_{\text{SL(2), finite size}}=&-\frac{i^{J}+(-i)^{J}}{\sqrt{J}}\exp\left[\sum_{a=1}^{\infty}\int_0^{\infty}\frac{du}{2\pi}\Theta_a (u) \log (1+Y_{a,0}(u))\right]\\
&\times\sqrt{\left(\prod_{1\leq s\leq \frac{M}{2}}\frac{u_s^2+\frac{1}{4}}{u_s^2}\sigma_B^2(u_s)\right)\frac{\det \left[1-\hat{G}^{\bullet}_{-}\right]}{\det \left[1-\hat{G}^{\bullet}_{+}\right]}}\comma
\end{aligned}
\eeq
where $\det \left[ 1-\hat{G}_{\pm}^{\bullet}\right]$ are Fredholm determinants and $Y_{a,0}$'s are the middle-node $Y$-functions. See section \ref{subsec:SL2finite} for details. We then derive the asymptotic formula from the excited state $g$-functions by rewriting it in terms of the {\it exact Gaudin norm}. We also provide a related discussion in Appendix \ref{ap:Eliminating} and show that one can express the result purely in terms of $Y$-functions for the middle node.

In section \ref{sec:check}, we perform extensive tests of our proposal. This includes the comparisons with
\begin{enumerate}
\item direct one-loop computation of structure constants in the SU(2) sector,
\item the superconformal block expansion of one-loop four-point functions of two determinants and two single-trace operators of arbitrary lengths,
\item the OPE expansion of the two-loop four-point function of two determinants and two ${\bf 20}^{\prime}$ operators.
\end{enumerate}
In all the cases, we found perfect agreement. In particular, the second test allows us to perform extensive tests in  higher-rank sectors while the third test serves as a check of the boundary dressing phase of the reflection matrix, which starts contributing at two loops. Since the relevant perturbative data are not available in the literature, we performed all these computations by ourselves generalizing the existing techniques such as the Lagrangian insertion formalism and the lightcone OPE analysis. The details of the computation are provided in the appendices. We also performed more direct tests of the reflection matrix and the boundary dressing phase by computing the reflection matrix in SO(6) and SO(4,2) sectors at tree level and the phase shift at strong coupling.

Along the way, we found (up to two loops) that the structure constant of two determinants and a twist $2$ operator exhibits a simple large spin behavior,
\beq
\log \left[\frac{\mathfrak{D}_{\mathcal{O}_{2,S}}}{\left.\mathfrak{D}_{\mathcal{O}_{2,S}}\right|_{\rm tree}}\right]\sim f(\lambda) \log S +\tilde{f}(\lambda) +O(1/S)\comma
\eeq
which is significantly simpler than the behavior of the structure constant of single-trace operators \cite{Alday:2013cwa}.
See section \ref{subsec:2loop4pt} for more details.

Finally in section \ref{sec:conclusion}, we summarize the results and chart a path forward. In particular, we point out a possibility of applying our framework to the four-point function of determinant operators and discuss its connection to various interesting physics; to name a few, the phase transition of geodesic Witten diagrams, the Hagedorn instability on the worldsheet, the Regge physics and the BFKL limit, the light-ray operators, the bulk-point limit and its possible relation to the Loschmidt echo.

Several appendices are included to explain technical details. In particular, in Appendix \ref{ap:gfunctionNested} we derive the $g$-function for theories described by the nested Bethe ansatz. We also present an integral representation for the overlap between the N\'{e}el state and the Bethe state in terms of $Q$-functions in Appendix \ref{ap:Qfunctions}.
\section{Generalities\label{sec:general}}
In this section, we explain the setup of our problem and give a brief summary of its basic properties.
\subsection{Setup, kinematics and symmetry\label{subsec:setup}}
The determinant operators that we study in this paper are of the form,
\beq\label{eq:defofD}
\mathcal{D}_i (x_i,Y_i)=\det \left(Y_i\cdot \Phi (x_i)\right)\comma
\eeq
where each $Y_i$ is a six-component complex vector satisfying
\beq
Y_i\cdot Y_i\equiv \sum_{I=1}^{6}Y_i^{I}Y_i^{I}=0\comma
\eeq
and parameterizes the orientation in the R-symmetry space. The dot product $(Y_i\cdot \Phi)$ stands for the standard inner product $\sum_{I=1}^{6} Y_i^{I} \Phi^{I}$ where $\Phi^{I}$'s are the six real scalars in $\mathcal{N}=4$ SYM.

With $Y_i$ being a null vector, the operator \eqref{eq:defofD} is half-BPS and has a protected conformal dimension $\Delta_{\mathcal{D}_i}=N$, where $N$ is the rank of the gauge group U($N$)\fn{Note that, unlike baryonic operators constructed out of fundamental quarks, the determinant operator can be defined even in theories with U($N$) gauge groups. The difference between SU($N$) and U($N$) is not important in this paper since we only discuss the leading large $N$ answers.}. Thus the two-point function of these operators is given by
\beq
\langle \mathcal{D}_1 (x_1,Y_1)\mathcal{D}_2(x_2,Y_2)\rangle =\mathfrak{n}_{\mathcal{D}}\times (d_{12})^{N} \comma
\eeq
where $d_{ij}$ is the free-field Wick contraction,
\beq
d_{ij}\equiv \frac{Y_i\cdot Y_j}{x_{ij}^2}\comma
\eeq
while $\mathfrak{n}_{\mathcal{D}}$ is the normalization constant, which can be set to unity by rescaling the operators.

\paragraph{BPS kinematics}As already mentioned, the main subject of this paper is the correlation function of two such determinant operators and one single-trace operator $\mathcal{O}$. In particular, we are interested in cases where the single-trace operator is non-BPS. In the integrability-based approach, we describe non-BPS operators by adding ``impurities'' in the $1/2$-BPS operator. It is therefore useful to first discuss $1/2$-BPS single-trace operators,
 \beq
 \mathcal{O}_{\circ}(x_3,Y_3)= {\rm tr}\left((Y_3\cdot \Phi)^{L}(x_3)\right)\period
 \eeq
 Here we put a subscript $\circ$ in order to emphasize that the operator is BPS.
As is well-known, the three-point functions of half-BPS operators are tree-level exact and take the following form:
\beq\label{eq:structureBPS}
\langle \mathcal{D}_1(x_1,Y_1) \mathcal{D}_2(x_2,Y_2) \mathcal{O}_{\circ}(x_3,Y_3)\rangle =\underbrace{\mathfrak{n}_{\mathcal{D}}(\mathfrak{n}_{\mathcal{O}_{\circ}})^{\frac{1}{2}}}_{\text{Normalization}}\times \underbrace{(d_{12})^{N}\left(\frac{d_{23}d_{31}}{d_{12}}\right)^{\frac{L}{2}}}_{\text{kinematical factor}}\times  \underbrace{\mathfrak{D}_{\mathcal{O}_{\circ}}}_{\text{Structure constant}}\period
\eeq
As indicated, it consists of three different terms: The first term $\mathfrak{n}_{\mathcal{D}}(\mathfrak{n}_{\mathcal{O}_{\circ}})^{\frac{1}{2}}$ represents the normalizations\fn{$\mathfrak{n}_{\mathcal{O}_{\circ}}$ is defined in terms of the two-point function
\beq
\langle {\rm tr}\left((Y_1\cdot \Phi)^{L}(x_1)\right){\rm tr}\left((Y_2\cdot \Phi)^{L}(x_2)\right)\rangle =\mathfrak{n}_{\mathcal{O}_{\circ}}\times (d_{12})^{L}\period
\eeq
} of the operators, which are unphysical and can be set to be $1$ by rescaling. The second term is the standard space-time and R-symmetry dependence of the three-point function,
\beq
(d_{12})^{N}\left(\frac{d_{23}d_{31}}{d_{12}}\right)^{\frac{L}{2}}=\frac{(Y_1\cdot Y_2)^{N-\frac{L}{2}}(Y_2\cdot Y_3)^{\frac{L}{2}}(Y_3\cdot Y_1)^{\frac{L}{2}}}{x_{12}^{2N-L}x_{23}^{L}x_{31}^{L}}\comma
\eeq
and the last term $\mathfrak{D}_{\mathcal{O}_{\circ}}$ is the structure constant, which is a physical CFT data. Its explicit expression at large $N$ will be given in section \ref{subsubsec:2det1BPS}.

\paragraph{Twisted translation} Of course, what we explained above---the factorization of the three-point function into a kinematical part and a structure constant---is a well-known consequence of the conformal Ward identities which applies also to non-BPS operators. However, for non-BPS operators, an explicit expression for the kinematical factor depends on details of the operator such as spin and R-charges, making it difficult to write down a simple universal expression like \eqref{eq:structureBPS}. One way to circumvent this problem is to consider some canonical configuration in which some of the consequences of the Ward identities take a simple form. A configuration that is useful for us is the so-called {\it twisted-translated frame}, which was introduced in \cite{Drukker:2009sf} and utilized in the integrability context in \cite{Basso:2015zoa}. For BPS three-point functions, one can go to that frame by putting three operators on a single line and aligning $Y_i$'s along a particular U$(1)$ direction inside SO$(6)$ by making use of the conformal and R-symmetry transformations.  The resulting configuration is given by
\beq
\mathcal{D}_1 =\det \mathfrak{Z} (a_1) \comma\quad \mathcal{D}_2 =\det \mathfrak{Z} (a_2) \comma\quad \mathcal{O}_{\circ}={\rm tr}\left(\mathfrak{Z}^{L}(a_3)\right)\period
\eeq
Here $\mathfrak{Z}$ is the twisted-translated scalar defined by
\beq
\begin{aligned}
\mathfrak{Z}(a)&\equiv \frac{(1+\kappa^2 a^2)\Phi^1+i (1-\kappa^2 a^2)\Phi^2+2 i \kappa a \Phi^4}{\sqrt{2}}(0,a,0,0)\\
&=(Z+\kappa^2 a^2 \bar{Z}+\kappa a (Y-\bar{Y}))(0,a,0,0)\comma
\end{aligned}
\eeq
where $X$, $Y$, $Z$ and their conjugates are complex scalar fields,
\beq\label{eq:defcomplexscalars}
\begin{aligned}
&Z=\frac{\Phi^1+i\Phi^2}{\sqrt{2}}\comma\quad \bar{Z}=\frac{\Phi^1 -i\Phi^2}{\sqrt{2}}\comma\quad Y=\frac{\Phi^3+i\Phi^4}{\sqrt{2}}\comma\quad \bar{Y}=\frac{\Phi^3-i\Phi^4}{\sqrt{2}}\comma\\
&X=\frac{\Phi^5+i\Phi^6}{\sqrt{2}}\comma\quad \bar{X}=\frac{\Phi^5-i\Phi^6}{\sqrt{2}}\period
\end{aligned}
\eeq
Here $\kappa$ is a tunable parameter with mass dimension $1$ which can be set to unity by performing a further dilatation. The configuration can also be described in terms of the {\it twisted translation} generator $\mathcal{T}$,
\beq
\mathcal{T}\equiv -i\epsilon_{\alpha\dot{\alpha}}P^{\dot{\alpha}\alpha}+\kappa \epsilon_{\dot{a}a}R^{a\dot{a}}\comma
\eeq
as
\beq
\mathfrak{Z} (a)=e^{\mathcal{T} a}\cdot Z (0) \cdot e^{-\mathcal{T} a}\period
\eeq
 In the twisted-translated frame, the normalized BPS three-point function takes a simple form,
\beq
\frac{\langle \mathcal{D}_1(a_1) \mathcal{D}_2(a_2) \mathcal{O}_{\circ}(a_3)\rangle}{(\mathfrak{n}_{\mathcal{D}}^2\mathfrak{n}_{\mathcal{O}_{\circ}})^{\frac{1}{2}}} = \kappa^{2N+L}  \mathfrak{D}_{\mathcal{O}_{\circ}}\period
\eeq

For non-BPS operators, we define the twisted-translated frame by first constructing a non-BPS operator by adding ``impurities'' (denoted in red below) to ${\rm tr}(Z^{L})$ at the origin and then performing the twisted translation:
\beq
\begin{aligned}
&\mathcal{O}(0)=\left.{\rm tr}\left(ZZ\red{X}Z\red{D_{+}}Z\cdots \right)+\cdots\right|_{x^{\mu}=0}\\
& \mapsto\quad \mathcal{O} (a)=e^{\mathcal{T} a}\,\mathcal{O}(0)\,e^{-\mathcal{T}a}\period
\end{aligned}
\eeq
The two-point function of such operators are constrained by the Ward identity to be
\beq
\langle \mathcal{O}(a_1)\mathcal{O}(a_2)\rangle=\frac{\kappa^{2J}\mathfrak{n}_{\mathcal{O}}}{(a_1-a_2)^{2 (\Delta-J)}}\comma
\eeq
where $\mathfrak{n}_{\mathcal{O}}$ is the normalization constant and $\Delta$ and $J$ represent two quantum numbers of $\mathcal{O}(0)$:$\Delta$ is the conformal dimension and $J$ is the charge of the U$(1)$ rotation which rotates $Z$ and $\bar{Z}$. Similarly, one can show that the three-point function of $\mathcal{O}$ and two determinants has the following structure\fn{Note that we included all possible tensor structures into $\mathfrak{D}_{\mathcal{O}}$, which appears when the operator $\mathcal{O}$ carries Lorentz and/or R-symmetry indices. }:
\beq\label{eq:NaiveNonBPS}
\frac{\langle \mathcal{D}_1(a_1) \mathcal{D}_2(a_2) \mathcal{O}(a_3)\rangle}{(\mathfrak{n}_{\mathcal{D}}^2\mathfrak{n}_{\mathcal{O}})^{\frac{1}{2}}} = \underbrace{\kappa^{2N+J}\left[\frac{(a_1-a_2)^2}{(a_2-a_3)^2(a_3-a_1)^2}\right]^{(\Delta-J)/2}}_{\text{Kinematical factor}}\,\, \underbrace{\mathfrak{D}_{\mathcal{O}}}_{\text{Structure constant}}\period
\eeq
One unsatisfactory feature of the expression \eqref{eq:NaiveNonBPS} is that it depends explicitly on the rank of the gauge group $N$ which becomes infinite in the planar limit. This dependence can be eliminated by taking a ratio with the two-point function of twisted-translated determinant operators, $\langle\mathcal{D}_1(a_1)\mathcal{D}_2(a_2) \rangle=\mathfrak{n}_{\mathcal{D}}\kappa^{2N}$:
\beq\label{eq:ratio3ptand2pt}
\frac{\langle \mathcal{D}_1 \mathcal{D}_2 \mathcal{O}\rangle}{\langle\mathcal{D}_1\mathcal{D}_2 \rangle\sqrt{\mathfrak{n}_{\mathcal{O}}}} = \kappa^{J}\left[\frac{(a_1-a_2)^2}{(a_2-a_3)^2(a_3-a_1)^2}\right]^{(\Delta-J)/2} \mathfrak{D}_{\mathcal{O}}\period
\eeq
The result is now $O(1)$ at large $N$ limit and is expressed purely in terms of quantum numbers of $\mathcal{O}$. We will later see that it is this ratio which admits nonperturbative integrability description.

\paragraph{Symmetry} Another advantage of the twisted-translated frame is that it makes it easier to see the underlying symmetry of the problem. This was in fact the original motivation for introducing the twisted translation in \cite{Basso:2015zoa}.

\begin{figure}[t]
\centering
\includegraphics[clip,height=2.5cm]{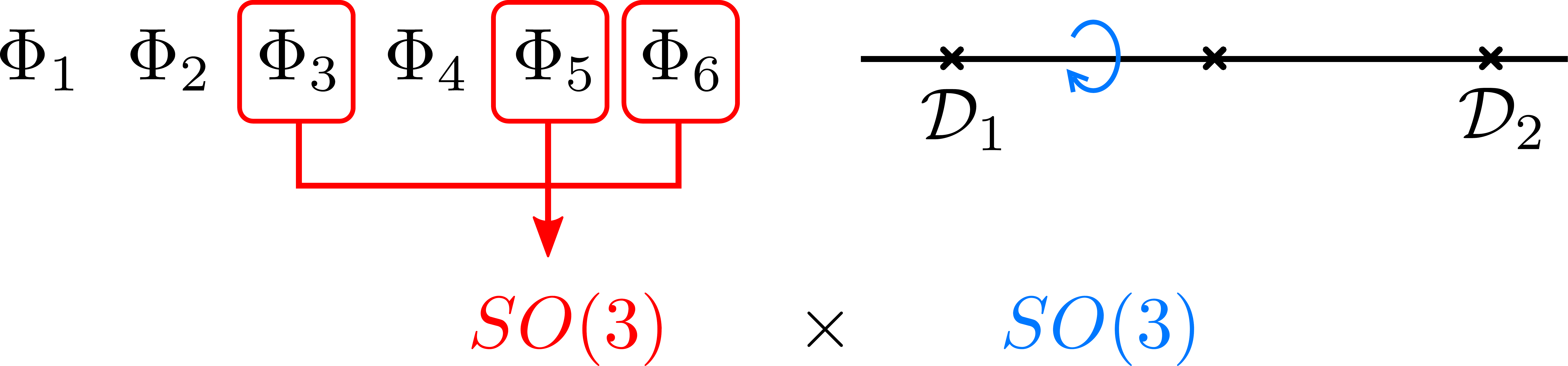}
\caption{The bosonic parts of the diagonal PSU$(2|2)$ symmetry.}
\label{fig:fig5}
\end{figure}

The relevant symmetries of our problem are the symmetries that leave the BPS three-point function invariant.
The bosonic part of the symmetries can be understood intuitively (see also figure \ref{fig:fig5}): Since we placed the three operators on a single line, there is a rotation symmetry SO$(3)$ around this line. In addition, since we are only using three scalars $\Phi_{1,2,4}$ to define the twisted translated frame, there is still a residual symmetry SO$(3)$ which rotates the remaining three. As explained in \cite{Basso:2015zoa}, these symmetry groups, when combined with super(conformal) symmetries, form a diagonal subgroup PSU$(2|2)_{D}$ of the PSU$(2|2)^{2}$ symmetry, which is the symmetry group governing the spectral problem of the single-trace operators. Using the notation of \cite{Basso:2015zoa}, they can be expressed explicitly as
\beq
\begin{aligned}
&\mathcal{L}^{\alpha}{}_{\beta}\equiv L^{\alpha}{}_{\beta}+\dot{L}^{\dot{\alpha}}{}_{\dot{\beta}}\comma\quad &\mathcal{Q}^{\alpha}{}_{a}\equiv Q^{\alpha}{}_{a}+i\kappa \epsilon^{\alpha\dot{\beta}}\epsilon_{a\dot{b}}\dot{S}^{\dot{b}}{}_{\dot{\beta}}\comma\\
&\mathcal{R}^{a}{}_{b}\equiv R^{a}{}_{b}+\dot{R}^{\dot{a}}{}_{\dot{b}}\comma\quad &\mathcal{S}^{a}{}_{\alpha}\equiv S^{a}{}_{\alpha}+\frac{i}{\kappa}\epsilon^{a\dot{b}}\epsilon_{\alpha\dot{\beta}}\dot{Q}^{\dot{\beta}}{}_{\dot{b}}\period
\end{aligned}
\eeq
Here the generators that appear on the right hand sides are the PSU$(2|2)^2$ generators which we review in Appendix \ref{ap:Smatrix}. Throughout this paper, we use the standard convention for the epsilon tensor,
\beq
\epsilon^{12}=\epsilon^{\dot{1}\dot{2}}=-\epsilon_{12}=-\epsilon_{\dot{1}\dot{2}}=1\period
\eeq

Although non-BPS three-point functions are not invariant under such symmetries, one can constrain their structures by using representation theory of the symmetry group as we show in section \ref{sec:bootstrap}.
\subsection{AdS description}
Let us also give a brief review of the AdS description of our problem. It was proposed in \cite{Balasubramanian:2001nh}, based on topological arguments in orbifold theories and free-field computations, that the determinant operators in $\mathcal{N}=4$ SYM correspond to the {\it maximal Giant Graviton} in AdS$_5\times$S$^5$, which is a D3-brane extended in a $S^3$ subspace in $S^5$ and particle-like in AdS. The proposal was later tested by various weak and strong coupling computations \cite{Balasubramanian:2002sa,Berenstein:2003ah,Berenstein:2005vf,deMelloKoch:2007rqf, deMelloKoch:2007nbd,Bekker:2007ea,Koch:2010gp,DeComarmond:2010ie,Carlson:2011hy, deMelloKoch:2011ci,deMelloKoch:2012ck,deCarvalho:2018xwx,Bissi:2011dc,Caputa:2012yj,deMelloKoch:2019dda}.

\begin{figure}[t]
\centering
\includegraphics[clip,height=4cm]{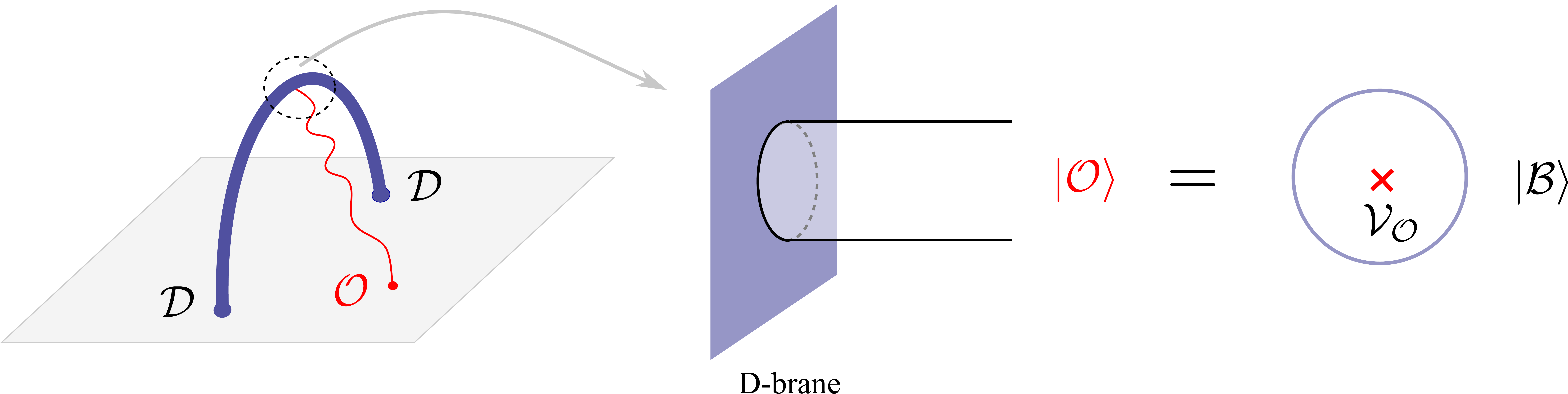}
\caption{The AdS description of the three-point function. The determinant operators $\mathcal{D}$ correspond to a geodesics of the Giant Graviton (denoted by a thick blue curve) while the single-trace operator corresponds to a closed string (denoted by a red wavy line). The worldsheet of the closed string has a boundary at which it gets attached to the Giant Graviton D-brane. It can be viewed either as a disk with one puncture (the right figure) or as a semi-infinite cylinder (the middle figure). In the latter description, the three-point function corresponds to an overlap between the boundary state $\langle\mathcal{B}|$ and a closed string state $|\mathcal{O}\rangle$. }
\label{fig:fig6}
\end{figure}

This leads to the following natural guess for the dual description of the three-point function of two determinants and one single-trace operator: We expect that the two determinants on the boundary create a classical configuration of the Giant Graviton which is a geodesics in AdS connecting the two insertion points. On the other hand, the single-trace operator would create a closed string which gets attached to the geodesics of the Giant Graviton (see figure \ref{fig:fig6}). Based on such a description, the three-point function of two determinant operators and a BPS single-trace operator was analyzed in \cite{Bissi:2011dc,Caputa:2012yj} leading to a match\fn{Initially, the results from the field theory and the AdS computation did not match \cite{Bissi:2011dc} for the extremal three-point functions. However, the discrepancy was later resolved in \cite{Kristjansen:2015gpa} by using a careful limiting procedure first discussed in \cite{Lin:2012ey}. We thank Charlotte Kristjansen for explaining this point.} between the field theory result and the AdS result \cite{Caputa:2012yj,Kristjansen:2015gpa}. The relevant classical configuration of the Giant Graviton can be obtained by Wick-rotating the solutions as given in \cite{Bissi:2011dc}.

The computations in \cite{Bissi:2011dc} were performed from the D-brane point of view, namely using the DBI action of the Giant Graviton. This was possible because they only considered the BPS single-trace operators dual to graviton states in AdS whose effects are already included in the DBI action. However, if one wants to compute the correlator of a non-BPS operator dual to a stringy state in AdS, the DBI action is not complete and one has to use the worldsheet description. From the worldsheet point of view, the interaction between the string and the D-brane is described by a disk one-point function, where the vertex operator is inserted at the center of the disk and describes the string state dual to the single-trace operator. Alternatively one can think of it as an overlap between the boundary state describing the Giant Graviton and the closed string state. See figure \ref{fig:fig6}.

Before ending this section, let us stress that the relevance of the disk worldsheet and the boundary state, which is quite natural on the AdS side, is not at all obvious on the field theory side. One of the main goals of the subsequent two sections is to develop field-theoretical techniques which manifest such features.
\section{Large $N$ Effective Theory for Giant Gravitons\label{sec:effective}}
We now introduce the semi-classical approach to the correlation functions of determinant operators. In this section, we only perform the computation at tree level, but the method can be readily generalized to include loop corrections as we comment at the end.

\subsection{Derivation\label{subsec:derivation}}
To illustrate the general structure of the resulting effective action, we consider the correlation function of $m$ determinant operators and one single-trace operator,
\beq
G_m \equiv \langle \mathcal{D}_1 (x_1,Y_1)\cdots \mathcal{D}_m (x_m,Y_m)\, \mathcal{O}(y)\rangle\comma
\eeq
where $\mathcal{D}_i = \det (Y_i\cdot \Phi)(x_i)$ and $\mathcal{O}(y)$ is a general single-trace operator made up of scalars and derivatives\fn{Note that we can replace covariant derivatives in $\mathcal{O}$ with usual derivatives since we only analyze the tree-level correlators. Also, one can in principle consider operators with fermions but the correlator of such a operator vanishes at tree level.}.
The derivation of the effective action consists of four steps of integrating in and out, each of which is quite elementary. As we comment later, the trick itself is not new and well-known in the context of matrix models but, as far as we know, it was never used effectively in higher-dimensional field theories. The physical meaning of these integrating in-and-out procedures, such as the interpretation from string theory and the relation to the graph duality, will be explained in later subsections.
\paragraph{Step 1: Integrating in fermion}
As the first step, we express it as a path integral,
\beq
G_m = \frac{1}{Z_{\Phi}}\int D\Phi^{I} \left(\prod_k \mathcal{D}_k \right)\mathcal{O} \,\,\exp\left[-\frac{1}{g_{\rm YM}^2}\int d^4 x\,\,{\rm tr}\left( \del_{\mu}\Phi^{I}\del^{\mu}\Phi_I\right)\right]\comma
\eeq
with $Z_{\Phi}$ being the partition function,
\beq
Z_{\Phi}=\int D\Phi^{I} \,\,\exp\left[-\frac{1}{g_{\rm YM}^2}\int d^4 x\,\, {\rm tr}\left(\del_{\mu}\Phi^{I}\del^{\mu}\Phi_I\right)\right]\period
\eeq
To proceed, we express the determinant operators as the integrals of zero-dimensional fermion $\chi_k$,
\beq
\mathcal{D}_k =\det (Y_k\cdot \Phi)=\int d\bar{\chi}_kd\chi_k \exp \left[\bar{\chi}_k (Y_k\cdot \Phi)\chi_k\right]\comma
\eeq
where $\chi_k$'s ($\bar{\chi}_k$'s) are in the (anti)fundamental representation of U$(N)$ and the product $\bar{\chi}_k (Y_k\cdot \Phi)\chi_k$ signifies
\beq
\bar{\chi}_{k,a}\, (Y_k\cdot \Phi)^{a}{}_{b}\, \chi_k^{b}\comma
\eeq
with $a$ and $b$ being the U$(N)$ indices. After the rewriting, we obtain
\beq\label{eq:fermionexponentiated}
\begin{aligned}
G_m=&\frac{1}{Z_{\Phi}}\int D\Phi^{I} d\bar{\chi}_kd\chi_k \,\,\mathcal{O}\\
& \exp \left[-\frac{1}{g_{\rm YM}^2}\int d^{4} x\left({\rm tr}\left(\del_{\mu}\Phi^{I}\del^{\mu}\Phi_I\right) -g_{\rm YM}^2 \sum_{k}\delta^{4}(x-x_k)\bar{\chi}_k(Y_k\cdot \Phi)\chi_k\right)\right]\period
\end{aligned}
\eeq
\paragraph{Step 2: Integrating out $\mathcal{N}=4$ SYM fields} The next step is to integrate out fields in $\mathcal{N}=4$ SYM, namely $\Phi^{I}$'s in \eqref{eq:fermionexponentiated}. This can be done most efficiently by completing the square in the following way,
\beq
\begin{aligned}
-&\frac{1}{g_{\rm YM}^2}\int d^{4} x \left({\rm tr}\left(\del_{\mu}\Phi^{I}\del^{\mu}\Phi_I\right)-g_{\rm YM}^2\sum_{k}\delta^{4}(x-x_k)\bar{\chi}_k(Y_k\cdot \Phi)\chi_k\right)\\
&=\frac{1}{g_{\rm YM}^2}\int d^{4} x\,\, {\rm tr}\left((\Phi^{I}-S^{I})\Box (\Phi_{I}-S_{I})-S^{I}\Box S_{I}\right)\comma
\end{aligned}
\eeq
with
\beq\label{eq:defofSfield}
S^{I}(x)=\frac{g_{\rm YM}^2}{2}\frac{\sum_k Y_k^{I}\chi_k\bar{\chi}_k\delta^{4}(x-x_k)}{\Box}=\frac{g_{\rm YM}^2}{8\pi^2}\sum_k\frac{Y_k^{I}\chi_k \bar{\chi_k}}{|x-x_k|^2}\period
\eeq
Here we used the identity
\beq\label{eq:dividentity}
\Box\frac{1}{|x|^2}=-4\pi^2 \delta^4 (x)\period
\eeq
Note that the U($N$) indices of $\chi$ and $\bar{\chi}$ in \eqref{eq:defofSfield} are not contracted, $\chi_k \bar{\chi}_k \equiv\chi_k^{a}\bar{\chi}_{k,b}$.

We can then perform the integration of the shifted field $\Phi^{I}-S^{I}$. This produces two effects. First it gives a one-loop partition function of $\Phi^{I}-S^{I}$, cancelling $1/Z_{\Phi}$ factor in \eqref{eq:fermionexponentiated}. Second, it replaces the $\Phi$ fields inside $\mathcal{O}(y)$ with $S$ fields, for instance,
\beq
\mathcal{O}(y)={\rm tr}\left(\Phi^{I_1}\del_{\mu}\Phi^{I_2}\cdots \right)(y) \quad \mapsto \quad\mathcal{O}^{S}(y)={\rm tr}\left(S^{I_1}\del_{\mu}S^{I_2}\cdots \right)(y)\period
\eeq
This second point perhaps deserves a further explanation. Normally, when we try to perform the integral of $\Phi-S$, we first express $\Phi$'s in $\mathcal{O}(y)$ as a sum of $(\Phi-S)+S$, and then perform the Wick contraction of $\Phi-S$. However, in this case there  is no Wick contraction to be performed since the operators in $\mathcal{N}=4$ SYM are normal-ordered and therefore we are not allowed to perform the self-contractions. Note that this is only true when there is  only one single-trace operator. If there are multiple single-trace operators, we do need to perform the contractions among those operators when performing the integral of $\Phi-S$.

As a result, we obtain the following integrals of fermions,
\beq
G_m=\int d\bar{\chi}_kd\chi_k\, \mathcal{O}^{S} \exp \left[-\frac{1}{g_{\rm YM}^2}\int d^{4} x\,{\rm tr}\left(S^{I}\Box S_{I}\right)\right]\comma
\eeq
which can be expressed more explicitly using \eqref{eq:defofSfield} and \eqref{eq:dividentity} as follows:
\beq
G_m=\int d\bar{\chi}_kd\chi_k\, \mathcal{O}^{S} \exp \left[-\frac{g^{2}}{N}\sum_{i\neq j}\underbrace{\frac{Y_i\cdot Y_j}{x_{ij}^2}}_{=\,\,d_{ij}} (\bar{\chi}_i\chi_j)(\bar{\chi}_j\chi_i)\right]\period
\eeq
Here we introduced the notation for the 't Hooft coupling commonly used in the integrability literature,
\beq
g^2\equiv\frac{\lambda}{16\pi^2}=\frac{g_{\rm YM}^2 N}{16\pi^2}\period
\eeq
\paragraph{Step 3: Integrating in bosonic bilocal fields}
As the third step, we perform the standard Hubbard-Stratonovich transformation. We first integrate in the auxiliary bosonic fields $\rho_{ij}$ with $i,j=1,\ldots , m$ and $i\neq j$ as
\beq\label{eq:hubbard1}
G_{m}=\frac{1}{Z_{\rho}}\int d\rho d\bar{\chi}_kd\chi_k\,\mathcal{O}^{S}\,\exp \left[\frac{N}{g^2}\sum_{i\neq j}\rho_{ij}\rho_{ji}-\frac{g^2}{N}\sum_{i\neq j}d_{ij}(\bar{\chi}_i\chi_j)(\bar{\chi}_j\chi_i)\right]\comma
\eeq
with
\beq\label{eq:freepartitionrho}
Z_{\rho}=\int d\rho\,\exp \left[\frac{N}{g^2}\sum_{i\neq j}\rho_{ij}\rho_{ji}\right]=\left(\frac{g^2\pi}{2N}\right)^{m(m-1)/2}\period
\eeq
We then shift $\rho$ in \eqref{eq:hubbard1} as
\beq
\rho_{ij}\to \rho_{ij}-\frac{g^2}{N}(d_{ij})^{\frac{1}{2}}\bar{\chi}_i\chi_j\comma
\eeq
and rewrite the action as
\beq\label{eq:hubbard2}
G_m =\frac{1}{Z_{\rho}}\int d\rho d\bar{\chi}_kd\chi_k\,\mathcal{O}^{S}\,\exp \left[\frac{N}{g^2}{\rm Tr}_{m}\left[\rho^2\right]+2{\rm Tr}_{m}\left[\hat{\rho}(\bar{\chi}\chi)\right]\right]\comma
\eeq
with\fn{Note that the indices in \eqref{eq:defhatrho} are not summed over.}
\beq\label{eq:defhatrho}
\hat{\rho}_{ij}\equiv (d_{ij})^{\frac{1}{2}}\rho_{ij}\period
\eeq
Here we are regarding $\rho_{ij}$ and $(\bar{\chi}_i\chi_j)$ as $m\times m$ matrices without diagonal entries and the trace is over such matrix indices, namely ${\rm Tr}_{m}[\rho^2]\equiv \sum_{i\neq j} \rho_{ij}\rho_{ji}$ and ${\rm Tr}_{m}\left[\hat{\rho}(\bar{\chi}\chi)\right]\equiv \sum_{i\neq j}\hat{\rho}_{ij}(\bar{\chi}_{j}\chi_i)$.

Note that the quadratic term of $\rho$ comes with a factor of $N$, indicating that $\rho$ becomes classical in the large $N$ limit. We will see this more explicitly in the next step.
\paragraph{Step 4: Integrating out fermions}
Since the new action \eqref{eq:hubbard2} is quadratic in fermions $\chi_k$, one can easily integrate them out. As a result, we obtain
\beq
G_m=\frac{1}{Z_{\rho}}\int d\rho\,\, \langle \mathcal{O}^{S}\rangle_{\chi}\, \exp\left(NS_{\rm eff}[\rho]\right)\comma
\eeq
where the large $N$ effective action $S_{\rm eff}[\rho]$ consists of the quadratic term of $\rho$ and the term coming from the one-loop fermion determinants,
\beq
S_{\rm eff}=\frac{1}{g^2} {\rm Tr}_m \left(\rho^2\right) +{\rm Tr}_{m}\log(- 2\hat{\rho})\period
\eeq
The expectation value $\langle \mathcal{O}^{S}\rangle_{\chi}$ is defined by
\beq
\langle \mathcal{O}^{S}\rangle_{\chi} \equiv \frac{\int d\chi_{k}d\chi_k \mathcal{O}^{S}\exp\left[2{\rm Tr}_m [\hat{\rho}(\bar{\chi}\chi)]\right]}{\int d\chi_{k}d\chi_k \exp\left[2{\rm Tr}_m [\hat{\rho}(\bar{\chi}\chi)]\right]}\comma
\eeq
and can be computed by performing the Wick contraction of $\mathcal{O}^{S}$ using
\beq\label{eq:wickchi}
\bcontraction{}{\bar{\chi}}{_{k,a}\quad}{\chi}
\bar{\chi}_{k,a}\quad \chi_{l,b}=-\frac{\delta_{ab}}{2} \left(\hat{\rho}^{-1}\right)_{k,l}\comma
\eeq
where $a$ and $b$ are U$(N)$ indices and $\hat{\rho}^{-1}$ is a matrix inverse of $\hat{\rho}$, which is a $m\times m$ matrix.

In general, one needs to sum up various different Wick contractions of $\mathcal{O}^{S}$. There is however a drastic simplification in the large $N$ limit. To understand this, let us analyze how the computation works for the following operator $\mathcal{O}$:
\beq
\begin{aligned}
\mathcal{O}=&{\rm tr}\left(\cdots \textcolor[rgb]{1,0,0}{\Phi^{I_1}}\textcolor[rgb]{0,0,1}{\Phi^{I_2}}\Phi^{I_3}\cdots\right)=\,\,\cdots (\textcolor[rgb]{1,0,0}{\Phi^{I_1}})^{a}{}_{b}(\textcolor[rgb]{0,0,1}{\Phi^{I_2}})^{b}{}_{c}(\Phi^{I_3})^{c}{}_{d}\cdots\period
\end{aligned}
\eeq
In the second equality, we wrote down explicitly the U($N$) indices of each field. We also colored each field to make it easier to track contributions from each field. Now, after integrating out $\mathcal{N}=4$ SYM fields, $\Phi$'s are replaced with $S$'s given in \eqref{eq:defofSfield}, leading to the following sequence of fermions:
\beq
\mathcal{O}^{S}= \sum_{k_s}\left(\prod_{s}\frac{g_{\rm YM}^2}{8\pi^2}\frac{Y_{k_s}^{I_s}}{|y-x_s|^2}\right)\times  (\cdots \textcolor[rgb]{1,0,0}{\chi_{k_1}^{\textcolor[rgb]{0,0,0}{a}}\bar{\chi}_{k_1\textcolor[rgb]{0,0,0}{,b}}}\,\,\textcolor[rgb]{0,0,1}{\chi_{k_2}^{\textcolor[rgb]{0,0,0}{b}}\bar{\chi}_{k_2\textcolor[rgb]{0,0,0}{,c}}}\,\,\textcolor[rgb]{0,0,0}{\chi_{k_3}^{\textcolor[rgb]{0,0,0}{c}}\bar{\chi}_{k_3\textcolor[rgb]{0,0,0}{,d}}}\cdots) \period
\eeq
Note that each $k_s$ runs from $1$ to $m$ with $m$ being the number of determinant operators in the original correlator.
We then compute the expectation value using the Wick contraction \eqref{eq:wickchi}. It turns out that the leading contribution at large $N$ comes from pair-wise contractions of neighboring fields with the same indices. Namely we just need to consider the following contraction:
\beq
\bcontraction{\cdots }{\red{\bar{\chi}}}{_{\red{k_1},b}}{\blue{\chi}}
\bcontraction{\cdots \red{\bar{\chi}}_{\red{k_1},b}\blue{\chi}_{\blue{k_2}}^{b}}{\blue{\bar{\chi}}}{_{\blue{k_2},c}}{\chi}
\cdots \red{\bar{\chi}}_{\red{k_1},b}\blue{\chi}_{\blue{k_2}}^{b}\blue{\bar{\chi}}_{\blue{k_2},c}\chi^{c}_{k_3}\cdots\quad  \propto \quad \delta^{b}_{b}\delta^{c}_{c}=N^2\period
\eeq
Other contractions pair up more than two fields, and leads to a smaller power of $N$, for instance
\beq
\acontraction{\cdots \red{\bar{\chi}}_{\red{k_1},b}}{\blue{\chi}}{_{\blue{k_2}}^{b}}{\blue{\bar{\chi}}}
\bcontraction{\cdots }{\red{\bar{\chi}}}{_{\red{k_1},b}\blue{\chi}_{\blue{k_2}}^{b}\blue{\bar{\chi}}_{\blue{k_2},c}}{\chi}
\cdots \red{\bar{\chi}}_{\red{k_1},b}\blue{\chi}_{\blue{k_2}}^{b}\blue{\bar{\chi}}_{\blue{k_2},c}\chi^{c}_{k_3}\cdots\quad  \propto \quad \delta^{b}_{c}\delta^{c}_{b}=N\period
\eeq
Now, after performing such large $N$ contractions, we obtain the following sequence of $m\times m$ matrices $\hat{\rho}^{-1}$,
\beq
\langle \mathcal{O}^{S}\rangle_{\chi}=-\sum_{k_s}\left(\prod_{s}\frac{g_{\rm YM}^2 N}{16\pi^2}\frac{Y_{k_s}^{I_s}}{|y-x_s|^2}\right)\times\left(\cdots \left(-\hat{\rho}^{-1}\right)_{\red{k_1},\blue{k_2}}\left(-\hat{\rho}^{-1}\right)_{\blue{k_2},k_3}\cdots \right)\period
\eeq
(Note that an overall extra minus sign comes from reshuffling the order of fermions.) This can be expressed simply as the following matrix trace,
\beq\label{eq:matrixtraceformula}
\langle \mathcal{O}^{S}\rangle_{\chi}=-{\rm Tr}_{m}\left[\cdots M^{I_1}M^{I_2}M^{I_3}\cdots\right]\comma
\eeq
with\fn{Here ${\rm diag}$ denotes a diagonal matrix and the dot product means a matrix multiplication.}
\beq\label{eq:conversionMtorhohat}
M^{I}\equiv {\rm diag}\left(\frac{g^2Y_1^{I}}{|y-x_1|^2},\frac{g^2Y_2^{I}}{|y-x_2|^2},\ldots,\frac{g^2Y_m^{I}}{|y-x_m|^2}\right) \,\cdot \,\left(-\hat{\rho}^{-1}\right)\period
\eeq
This matrix-trace representation \eqref{eq:matrixtraceformula} resembles the one that appeared in the computation of one-point functions in the presence of a domain wall in \cite{deLeeuw:2015hxa}. There is however one important qualitative difference: In the case of one-point functions, the domain wall creates a classical profile of fields in $\mathcal{N}=4$ SYM and the computation naturally boils down to evaluating the single-trace operator on that profile. Because of this, the resulting matrix trace is taken over a subgroup of the original U$(N)$ gauge group. On the other hand, in our setup nothing in the original $\mathcal{N}=4$ SYM Lagrangian is classical and the classical background only emerges after a series of rewriting that we explained above. Correspondingly, the matrix space over which the trace is taken is also emergent, having no relation to the U$(N)$ gauge group of $\mathcal{N}=4$ SYM. As we discuss later, the $\rho$ fields can be interpreted as open string fields on the Giant Graviton D-branes, and from this perspective, we can interpret the matrix trace ${\rm Tr}_{m}$ as defining a gauge-invariant observable in that open string field theory, not in the original $\mathcal{N}=4$ SYM.

Having computed the expectation value $\langle \mathcal{O}^{S}\rangle$, one can now perform the $\rho$-integral by using the saddle-point approximation. Since $\langle \mathcal{O}^{S}\rangle$ does not produce a large exponent, we can simply look for the saddle point of the effective action $S_{\rm eff}$,
\beq\label{eq:saddlepoint}
\frac{\del S_{\rm eff}}{\del \rho_{ji}}=0\qquad \iff \qquad \frac{\rho_{ij}}{g^2}=-\left(\frac{1}{2\hat{\rho}}\right)_{ij}d_{ij}\period
\eeq
Then, the final answer in the large $N$ limit is given by
\beq
G_m \overset{N\to \infty}{=}\left.\langle \mathcal{O}^{S}\rangle_{\chi}\right|_{\rho=\rho^{\ast}} e^{NS_{\rm eff}[\rho^{\ast}]}\comma
\eeq
with $\rho^{\ast}$ being the solution to the saddle-point equation \eqref{eq:saddlepoint}. In general, there can be multiple saddle points and one needs to find the one which gives the dominant contribution. However, for $m=2,3$, we will see below that there is only one saddle point and it gives the correct result for the corresponding correlators.

Before ending this subsection, let us make a few remarks: First, although we mostly discussed single-trace operators which only contain scalar fields, the computation can be readily generalized to operators with derivatives. We will see this in explicit examples in section \ref{subsec:sl2tree}.
Second, as with the analysis in \cite{deLeeuw:2015hxa}, one can alternatively represent $\langle \mathcal{O}^{S}\rangle$ as the following overlap between a matrix product state and a state corresponding to the operator $\mathcal{O}$,
 \beq\label{eq:overlapmatrixproduct}
 \langle \mathcal{O}^{S}\rangle_{\chi}= \langle M(\rho)|\mathcal{O}\rangle \comma
\eeq
 with
 \beq
 | M(\rho)\rangle\equiv \sum_{I_1,\ldots,I_m} -{\rm Tr}_{m}\left[ M^{I_1}M^{I_2}M^{I_3}\cdots\right] | \Phi^{I_1}\cdots \Phi^{I_m}\rangle \comma\qquad |\mathcal{O}\rangle\equiv |\Phi^{I_1}\cdots \Phi^{I_m}\rangle\period
 \eeq
It is amusing to note that the way matrix product states show up in our context has some similarity with the way they are used in condensed matter physics: The matrix product state is often used as a variational ansatz for the spin-chain ground state. In that context, the matrix $M^{I}$ is a function of several unfixed parameters and one determines them by minimizing the expectation value of the energy. Also here, the matrix $M^{I}$ is a function of the ``parameter'' $\rho$, and one determines it by extremizing the effective action $S_{\rm eff}[\rho]$.
\subsection{String theory interpretation\label{subsec:interpretation}}
The sequence of rewritings that we explained above might seem like technical computational tricks. However, we now point out that there is a natural interpretation from string theory which suggests that each step of rewritings has a physical meaning.

\begin{figure}[t]
\centering
\includegraphics[clip,height=5.7cm]{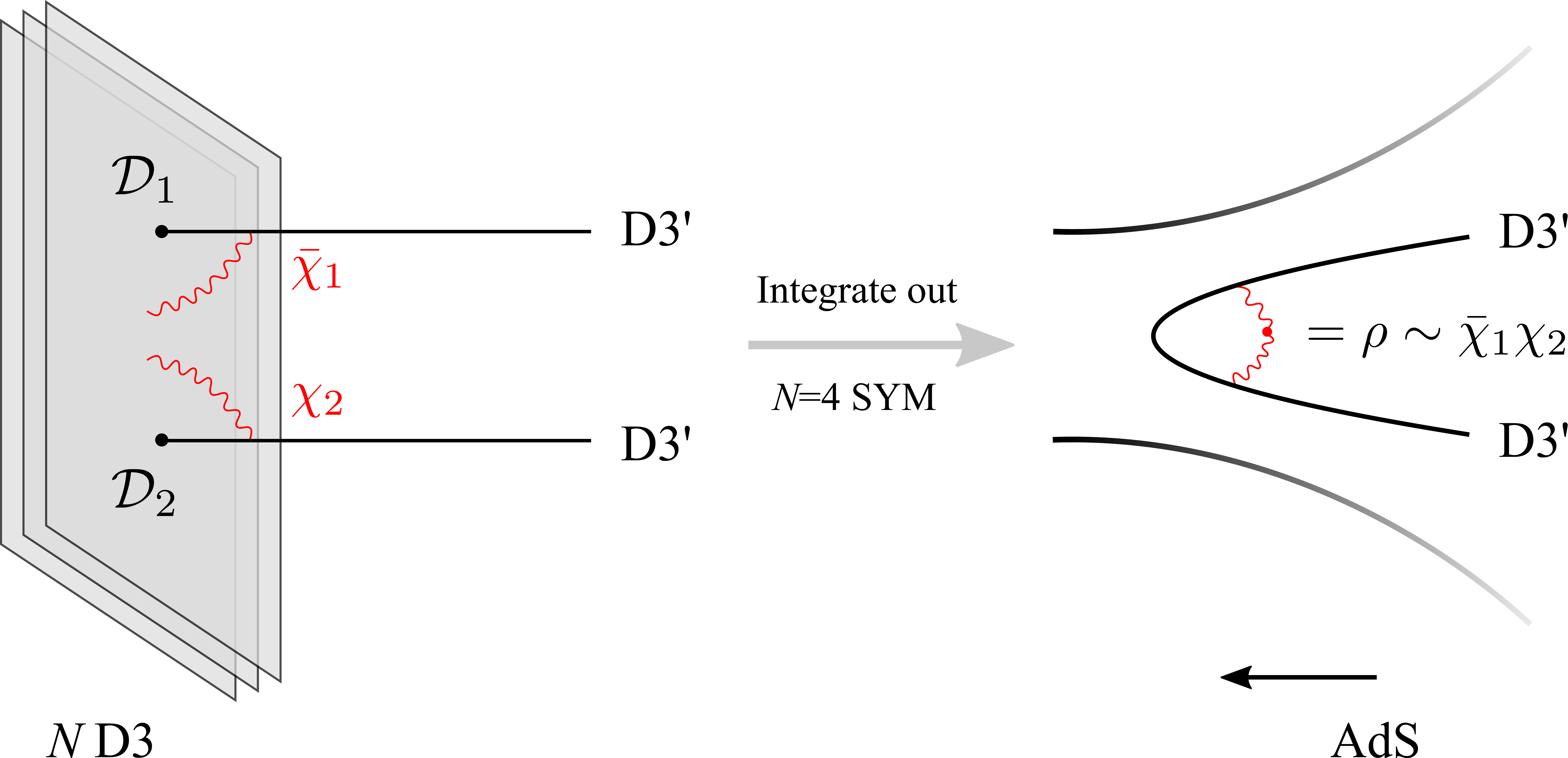}
\caption{The string theory interpretation of our rewritings. The determinant operators would correspond to two extra D3 branes intersecting perpendicularly with a stack of D3 branes. After integrating out $\mathcal{N}=4$ SYM fields, we are left with a single connected D3 brane in AdS and the $\rho$-fields on top of it, which are mesons made out of  ``quarks'' $\chi_{1,2}$ and $\bar{\chi}_{1,2}$. The mechanism is similar to the chiral symmetry breaking in holographic QCD, such as the Sakai-Sugimoto model \cite{Sakai:2004cn}.}
\label{fig:fig7}
\end{figure}

The basic idea is to interpret the determinant operators as point-like intersections between a stack of $N$ D3 branes, which describe $\mathcal{N}=4$ SYM, and probe D3 branes (to be denoted by D3$^{\prime}$ branes). In this configuration, in addition to open strings connecting $N$ D3 branes which become $\mathcal{N}=4$ SYM fields in the near horizon limit, we also have open strings connecting D3 branes and the D3$^{\prime}$ branes. Such open strings give rise to localized degrees of freedom at the intersections which are in the fundamental representation of the U$(N)$ gauge group. The fermionic fields $\chi$ that we introduced above have precisely such a property and can therefore be understood as such open strings.

Next in step 2, we integrated out $\mathcal{N}=4$ SYM fields. In the D-brane picture, this amounts to integrating out the degrees of freedom on $N$ D3 branes and replacing it with a classical curved background. After this manipulation, we expect that the probe D3$^{\prime}$ branes get reconnected\fn{In terms of $\rho$-fields, this reconnection corresponds to the off-diagonal $\rho$-fields getting nonzero expectation values.} as shown in figure \ref{fig:fig7}.
We are then left with open string excitations on the reconnected D3$^{\prime}$ branes, which are ``bound states'' of two fermions. Such bound states are nothing but the $\rho$-fields that we introduced in step 3 through the Hubbard-Stratonovich transformation. This suggests that one can interpret the effective action for $\rho$-fields as the action of open string field theory on the reconnected D3$^{\prime}$ branes, which become Giant Graviton D3 branes in AdS$_5\times$S$^{5}$ after taking the near horizon limit. This interpretation is further supported by the fact that the resulting $\rho$-fields are valued in $k\times k$ matrices when there are $k$ determinant operators. From the bulk point of view, this is just a reflection of the fact that open strings on $k$ D-branes carry Chan-Paton factors\fn{The emergence of the gauge symmetry for the Giant Gravitons was discussed also in \cite{Balasubramanian:2004nb,Berenstein:2013md} from different approaches. It would be interesting to understand the relation to our method.} and can be thought of as $k\times k$ matrices.

Of course, what we said here is at best qualitative. It would be better to make these arguments more precise by finding the relevant D3-brane configurations in flat space and analyzing the open string spectrum. In addition, it would be interesting to know how closely related operators such as subdeterminant operators and Schur polynomial operators\fn{See also section \ref{subsubsec:other} for discussions of such operators in our formalism.} are realized in a similar D-brane picture. Such analysis was indeed carried out for the BPS Wilson loops in \cite{Gomis:2006sb} and we expect that it can be generalized to determinant operators.
\subsection{Matrix model, graph duality and ``open-closed-open'' duality\label{subsec:matrixmodel}}
As we already mentioned above, our trick of rewritings appeared previously in the literature on matrix models. For instance, the trick was utilized in a series of works by Brezin and Hikami on the computation of intersection numbers of the moduli space of curves \cite{brezin2000characteristic,Brezin:2007aa,Brezin:2007wv}. It also played a key role in the duality between $(2,2k+1)$ minimal string theory and matrix models \cite{Maldacena:2004sn}: The minimal string theory is a noncritical string theory whose worldsheet theory consists of Liouville theory and a minimal model. When the minimal model is a $(2,2k+1)$ minimal model, the theory is conjectured to have two different dual descriptions: The first description is a double-scaling limit of the standard one-matrix model, which can be viewed as a theory on ZZ branes in the minimal string theory. The second description is the so-called Kontsevich matrix model, which corresponds to a theory on FZZT branes. The two descriptions appear to be quite different; in particular, while the first description involves the large $N$ limit, the size of the matrix  in the second description is not necessarily large. This raises a question of whether one can go from one description to the other. This question was addressed and solved in \cite{Maldacena:2004sn} using precisely the same rewritings as the one described above: The strategy was to realize the FZZT branes in the first description by inserting determinant operators, and then perform a sequence of rewritings to obtain a dual description\fn{In our language this is nothing but the integral of $\rho$ fields.}, which coincides with the Kontsevich model after taking the double-scaling limit.

This duality between matrix models was revisited recently in an unpublished work by Gopakumar \cite{Gopakumar} in which he pointed out that the duality can be understood pictorially as a graph duality. Since his argument applies also to our problem and gives another useful perspective, we will explain it below. We want to emphasize that the explanation in the following paragraph is just a review of his argument and does not contain original materials.

\begin{figure}[t]
\centering
\includegraphics[clip,height=3.5cm]{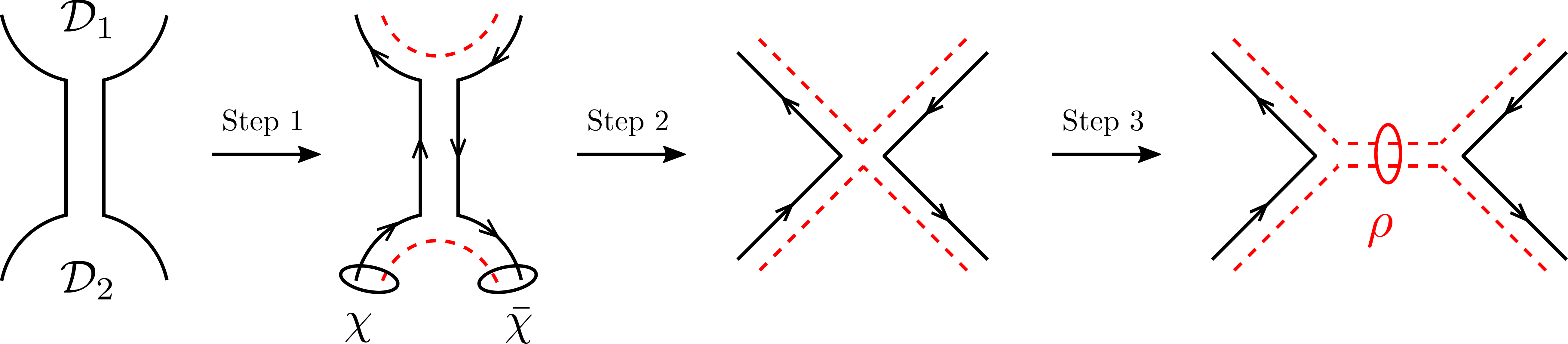}
\caption{Graphical interpretation of the rewriting procedure. The fermion fields $\chi$ and $\bar{\chi}$ correspond to a double line made up of a dashed line and a black line. As a result of integrating in and out, we resolve the quartic interaction and introduce a double line made up of two dashed lines. This corresponds to the $\rho$ field that we obtained in the end.}
\label{fig:fig8}
\end{figure}

\begin{figure}[t]
\centering
\includegraphics[clip,height=3.5cm]{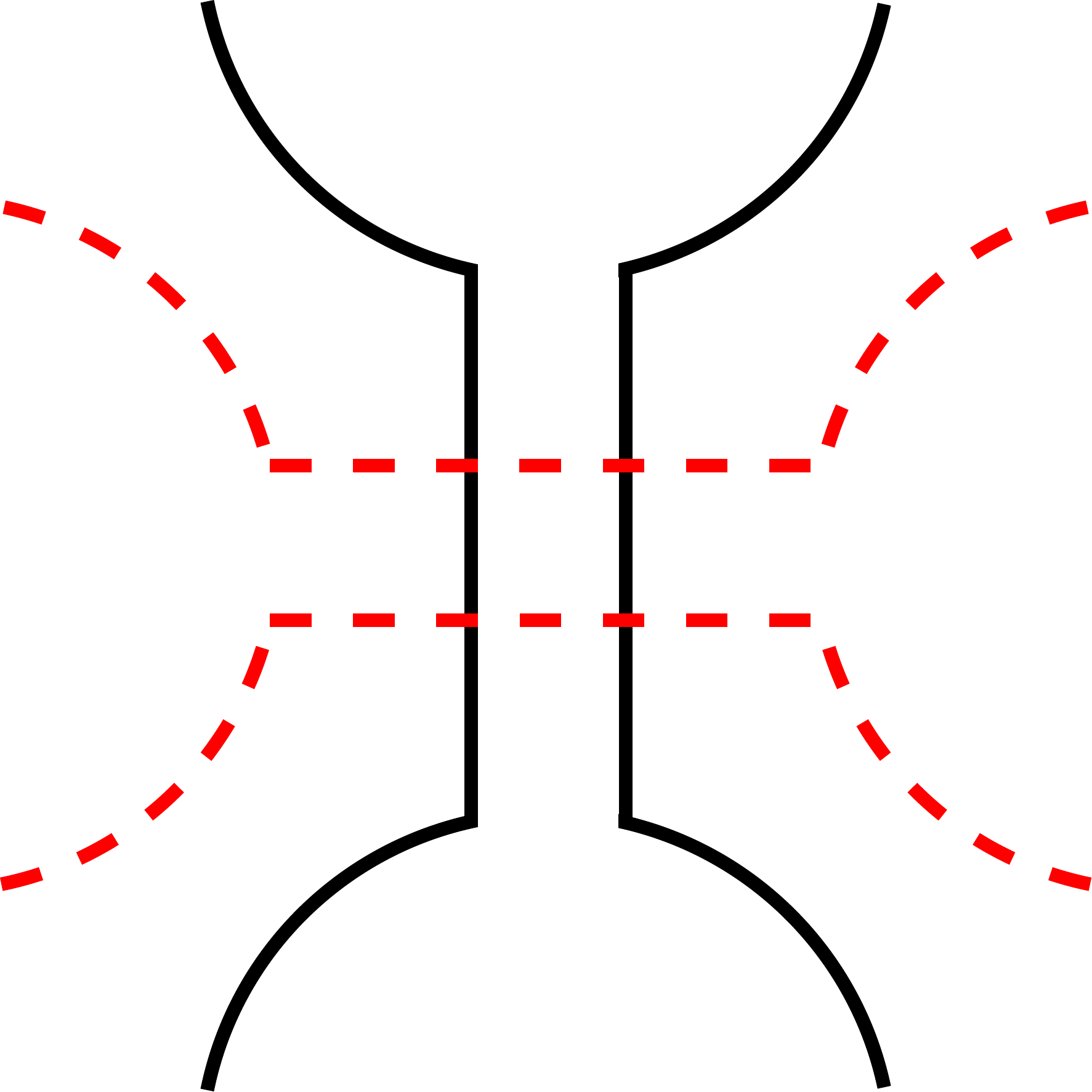}
\caption{Rewriting as a graph duality. The diagram of $\rho$ fields is a dual graph of the original Feynman diagram; the faces of the original graph become the vertices and vice versa.}
\label{fig:fig10}
\end{figure}

The Wick contractions among determinant operators are complicated collections of propagators. In order to simplify them, we introduced the fermions $\chi$'s and replaced each determinant with a Gaussian integral of $\chi$'s. As shown in figure \ref{fig:fig8}, graphically this amounts to adding new lines (denoted by red dashed lines) to the endpoints of the original propagators (denoted by black lines). After doing so, we have two kinds of double lines; the original double lines which describe the $\mathcal{N}=4$ SYM fields and the double lines made up of a black line and a red dashed line, which carry only one U$(N)$ index and correspond to $\chi$. Next in step 2, we integrated out the original $\mathcal{N}=4$ SYM fields. This corresponds to eliminating the black double lines and creating four-fermi interactions of $\chi$. Such four-fermi interactions can be resolved by inserting new double lines made up of two red dashed lines. This is precisely our step 3, in which we integrated in $\rho$ fields. After doing so, we can integrate out $\chi$ fields and this creates non-linear interactions of $\rho$'s as shown in figure \ref{fig:fig8}. After this sequence of rewritings, the original graph is converted to its dual graph, for which the roles of vertices and faces are swapped (see figure \ref{fig:fig10}).

\begin{figure}[t]
\centering
\includegraphics[clip,height=3.5cm]{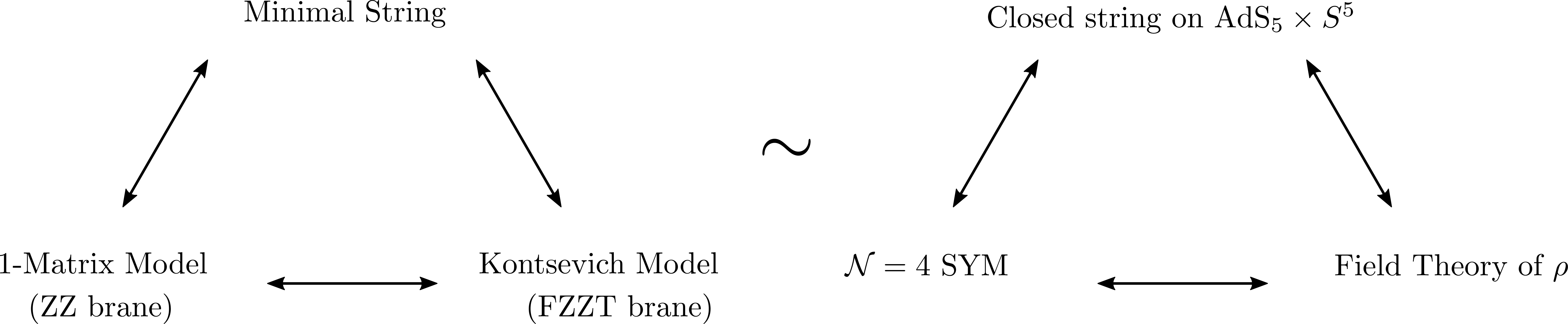}
\caption{Open-closed-open duality. The field theory of $\rho$ can be regarded as an analogue of the Kontsevich model in the minimal string duality.}
\label{fig:fig9}
\end{figure}

The duality---or more precisely the triality---between the two matrix models and the minimal string was called ``open-closed-open'' duality in \cite{Gopakumar}. He further suggested that an analogous relation might exist also for the AdS$_5$/CFT$_4$ correspondence and mentioned that it would be interesting to find the missing ``open'' description. In a sense, what we did in this section is one realization\fn{In fact it was even mentioned in \cite{Gopakumar} that the Giant Graviton might provide the missing ``open'' description.} of such an idea in a simple toy example (see figure \ref{fig:fig9}). It would be interesting to further explore it. See the conclusion section (section \ref{sec:conclusion}) for further discussions.
\subsection{Analogies with open string field theory\label{subsec:analogy}}
In section \ref{subsec:interpretation}, we explained that our large $N$ effective action can be interpreted as open string field theory on Giant Gravitons, based on the D-brane picture and the Chan-Paton structure. Here we discuss that the analogy goes further by pointing out similarities with works done in open string field theory.

Open string field theory (OSFT) is a second-quantized formulation of open strings \cite{Witten:1985cc}. For bosonic OSFT, the action and the equation of motion take simple forms and they read
\beq\label{eq:OSFTeom}
\begin{aligned}
&S=\frac{1}{2}\langle \Psi , Q_B\Psi\rangle +\frac{1}{3}\langle \Psi ,\Psi \ast\Psi\rangle\comma\qquad \qquad Q_B \Psi +\Psi \ast \Psi=0\comma
\end{aligned}
\eeq
where $\Psi$ is a string field, $Q_B$ is a  BRST operator of a chosen open string background, and $\ast$ is the star-product which maps a product of two string fields to a single string field. The symbol $\langle A,B\rangle$ denotes a BPZ inner product of the worldsheet CFT.

An interesting feature of OSFT is the way it describes different string backgrounds: In the first-quantized formalism, different open string backgrounds correspond to different conformal boundary conditions on the worldsheet, or equivalently boundary CFTs (BCFTs), and one needs to find them by imposing a set of consistency conditions such as the Cardy condition. On the other hand, in OSFT one can find different backgrounds by simply\fn{Of course it is hard in practice to solve the equation of motion, but here we wanted to emphasize it is conceptually much simpler than finding all consistent conformal boundary conditions.} solving the equation of motion \eqref{eq:OSFTeom}; the trivial solution $\Psi=0$ corresponds to the original background we started with while other solutions describe non-perturbatively distinct backgrounds.

This feature allows us to systematically search different string backgrounds with the help of numerical/analytical techniques. At the same time it also raises a puzzle: How and why are these two descriptions equivalent? One way to compare the two approaches is to compute the BRST cohomology of a new background using a shifted BRST operator,
\beq
Q^{\rm new}_{B}=Q_B +\{\Psi_0,\bullet\}\comma
\eeq
with $\Psi_0$ being a solution to \eqref{eq:OSFTeom}, and compare it with the spectrum of a candidate BCFT. This does provide a way to identify solutions in OSFT with BCFTs, but it is not necessarily the simplest way. Another possible way is to compute some physical observables on both sides and compare. This was indeed an approach discussed by Ellwood: In \cite{Ellwood:2008jh}, he considered a class of gauge-invariant observables in open string field theory which were first introduced in \cite{Hashimoto:2001sm,Gaiotto:2001ji}. They are often called Ellwood invariants, whose formal expression reads
\beq\label{eq:ellwoodinv}
W(\Psi_0,\mathcal{V}_{\rm cl})=\langle \mathcal{I}|\mathcal{V}_{\rm cl}(i)|\Psi_0\rangle\comma
\eeq
with $\mathcal{I}$ being the identity open string field and $\mathcal{V}_{\rm cl}$ being an on-shell closed-string vertex operator inserted at the midpoint of open string. The Ellwood invariant is a functional of the solution $\Psi_0$ and roughly speaking it can be interpreted as a closed string vertex operator evaluated on the classical background sourced by the open string field $\Psi_0$. The conjecture by Ellwood states that these observables are the same as disk one-point functions in a BCFT, or equivalently, an overlap between the closed string state $\mathcal{V}_{\rm cl}$ and a boundary state corresponding to the solution $\Psi_0$:
\beq
W(\Psi_0,\mathcal{V}_{\rm cl})\quad \sim\quad \langle B_{\Psi_0}|\mathcal{V}_{\rm cl}\rangle \period
\eeq
Although the conjecture is not proven yet, this gives an efficient way to read off the boundary state from a given classical solution $\Psi_0$ as was demonstrated in \cite{Kudrna:2012re}. A more explicit connection between OSFT and BCFTs was made in a beautiful paper \cite{Kiermaier:2008qu} in which they put forward a prescription to construct a relevant boundary state starting from a solution to OSFT. In their construction, the boundary state for the solution $\Psi_0$ is given in the form,
\beq\label{eq:KOZformula}
|B_{\Psi_0}\rangle \sim \oint{\rm P}\exp \left(-\int dt \left[\mathcal{L}_R (t)+\{\mathcal{B}_{R},\Psi_0\}\right]\right)\comma
\eeq
where $\mathcal{L}_R$ and $\mathcal{B}_R$ are certain line integrals of a stress energy tensor and a b-ghost while $\oint$ stands for identification of half-string boundaries, which roughly can be thought of as a trace over the half-string Hilbert space. The detailed explanation of this expression is certainly beyond the scope of this paper and we simply refer the readers to the original article. What is important for us is that the expression provides a generalization of Wilson loops in gauge theories to open string field theory, and realizes the idea of expressing closed strings using the Wilson loop \cite{Polyakov:1980ca} in a concrete form.

We can now see some analogy with our story. In section \ref{subsec:derivation}, we showed that the computation of correlators of determinants and a single-trace operator reduces to evaluating a certain matrix trace \eqref{eq:matrixtraceformula}, which is a  function of $\rho$ and can be interpreted as a gauge-invariant observable on Giant Graviton D-branes. Upon identification of $\rho$ with the string field $\Psi_0$, one can see that this is analogous to the Ellwood invariant given in \eqref{eq:ellwoodinv}. In our case, it is rather straightforward to rewrite such a matrix trace as an overlap between a matrix product state and a closed string state as shown in \eqref{eq:overlapmatrixproduct}. The resulting matrix product state can be interpreted as a boundary state describing the Giant Graviton D-branes. Interestingly, the structure of the matrix product state
\beq
| M(\rho)\rangle\equiv \sum_{I_1,\ldots,I_L} -{\rm Tr}_{m}\left[M^{I_1}(\rho)\,M^{I_2}(\rho)\,M^{I_3}(\rho)\cdots\right] | \Phi^{I_1}\cdots \Phi^{I_L}\rangle
\eeq
 resembles the boundary state constructed in \cite{Kiermaier:2008qu}. Namely, the matrices $M^{I}(\rho)$'s can be viewed as discretization of the path-ordered exponential in \eqref{eq:KOZformula}. In other words, we propose that the physical interpretation of the matrix product state is the ``Wilson loop'' on the Giant Graviton D-branes.

 Of course, we admit that the analogies are rather formal and qualitative. It would be interesting if one could make the arguments more quantitative. Even more interesting would be to reconstruct full-fledged open string field theory on Giant Gravitons in AdS purely from $\mathcal{N}=4$ SYM. We hope that the analogies presented here may provide some useful guidance towards such an ambitious goal.
\subsection{Examples\label{subsec:example1}}
We now see how the method works in several simple examples.
\subsubsection{Two- and three-point functions of determinants}
Let us first discuss the two- and the three-point functions of determinants without any single-trace insertions. In the absence of single-trace operators, the computation boils down to solving the saddle-point equation \eqref{eq:saddlepoint} and evaluating the effective action (and the fluctuation around it).

For the two-point functions, there are only two $\rho$ variables $\rho_{12}$ and $\rho_{21}$, and the saddle-point equation reads
\beq\label{eq:saddlepntrho21}
\begin{aligned}
\rho_{12}\rho_{21}=-\frac{g^2}{2}\period
\end{aligned}
\eeq
Plugging this into the effective action, we get
\beq
S_{\rm eff}[\rho^{\ast}]=-1+\log (2g^2d_{12})\qquad \iff \qquad \left.\langle\mathcal{D}_{1}\mathcal{D}_{2} \rangle\right|_{N\to \infty}=e^{N S_{\rm eff}[\rho^{\ast}]}=(2g^2d_{12})^{N}e^{-N}\period
\eeq
Note that the result correctly reproduces the expected space-time dependence of the two-point function of determinant operators.

We can also compute the one-loop fluctuation around this saddle-point. For this purpose, we express $\rho_{12}$ and $\rho_{21}$ as
\beq
\rho_{12}=i (a+ib)\comma\quad \rho_{21}=i(a-ib)\comma
\eeq
and perform the integral of $a$ and $b$ around the saddle point. Note that the extra factors of $i$ are necessary in order to make the integral convergent and it is related to the fact that the appropriate contour for $\rho$ is not along the real axis\fn{This can be seen explicitly in the free partition function \eqref{eq:freepartitionrho}.}. In the new variables $a$ and $b$, the saddle point is given by
\beq
a^2+b^2=\frac{g^2}{2}\period
\eeq
As this equation shows, there is actually a flat direction around the saddle point and it is therefore convenient to introduce the polar coordinates $a+ib =r e^{i\theta}$. Then the one-loop integral gives
\beq
(\text{\tt one-loop})=\underbrace{2\pi}_{\text{$\theta$ integral}} \underbrace{\frac{g}{\sqrt{2}}}_{\text{saddle-point value of $r$}}\int d\delta r \exp \left[-\frac{4N}{g^2}(\delta r)^2\right]=\frac{\pi^{3/2}g^2}{\sqrt{2N}}\period
\eeq
Combining this with $1/Z_{\rho}$ factor, we obtain
\beq\label{eq:twopointsemiclassicalresult}
\left.\langle\mathcal{D}_{1}\mathcal{D}_{2} \rangle\right|_{O(1/N)}= (2g^2 d_{12})^N\frac{\sqrt{2\pi N}e^{-N}}{2}\period
\eeq
Later in section \ref{subsec:examplePCGG}, we will see that this result agrees with the result from the direct Wick contractions.

Similarly one can compute the three-point function of determinant operators. For the three-point functions, there are $6$ relevant $\rho$ variables $\rho_{12}$, $\rho_{23}$, $\rho_{31}$, $\rho_{21}$, $\rho_{32}$ and $\rho_{13}$. The saddle-point equations for the first three read
\beq
\begin{aligned}
\frac{\rho_{12}}{g^2}&=-\frac{\rho_{13}\rho_{32}}{2(\rho_{12}\rho_{23}\rho_{31}+\rho_{13}\rho_{32}\rho_{21})}\comma\\
\frac{\rho_{23}}{g^2}&=-\frac{\rho_{21}\rho_{13}}{2(\rho_{12}\rho_{23}\rho_{31}+\rho_{13}\rho_{32}\rho_{21})}\comma\\
\frac{\rho_{31}}{g^2}&=-\frac{\rho_{32}\rho_{21}}{2(\rho_{12}\rho_{23}\rho_{31}+\rho_{13}\rho_{32}\rho_{21})}\period
\end{aligned}
\eeq
The other three equations can be obtained by permuting the indices of these equations. The solution to those six equations is given by
\beq
\rho_{12}\rho_{21}=\rho_{23}\rho_{32}=\rho_{31}\rho_{13}=-\frac{g^2}{4}\period
\eeq
Evaluating the action at the saddle point, we obtain
\beq
S_{\rm eff}[\rho^{\ast}]=-\frac{3}{2}+\log (2(d_{12}d_{23}d_{31})^{1/2} g^2)\comma
\eeq
which leads to the following expression for the correlator:
\beq
\left.\langle \mathcal{D}_1 \mathcal{D}_2 \mathcal{D}_3\rangle\right|_{N\to \infty}=e^{N S_{\rm eff}[\rho^{\ast}]}=(2g^2 (d_{12}d_{23}d_{31})^{1/2})^{N}e^{-\frac{3}{2}N}\period
\eeq
Again the expected space-time dependence is correctly reproduced from this saddle-point analysis.
Dividing this by the normalization of two-point functions, we can also read off the structure constant,
\beq
\left.\frac{\langle \mathcal{D}_1 \mathcal{D}_2 \mathcal{D}_3\rangle}{\mathfrak{n}_{\mathcal{D}}^{3/2}}\right|_{N\to \infty}=(d_{12}d_{23}d_{31})^{N/2}\times 2^{-N/2}\period
\eeq
The fact that the structure constant $2^{-N/2}$ is exponentially small is consistent with the general expectation for the three-point function of heavy operators. It would be interesting if one can reproduce this number at strong coupling, namely by finding a classical solution describing the three-point function of Giant Graviton D-branes and evaluating its action.
\subsubsection{2 determinants and 1 BPS single-trace\label{subsubsec:2det1BPS}}
Let us now consider the three-point function of two determinants and one single-trace operator. The detailed analysis for general non-BPS operators will be performed in section \ref{sec:weak}, and here we focus on the BPS single-trace operators,
\beq
\mathcal{O}_{\circ}(x_3)={\rm tr}\left((Y_3\cdot \Phi)^{L}\right)(x_3)\comma
\eeq
just to illustrate how the method works. In this case, the matrix trace \eqref{eq:matrixtraceformula} can be simplifies as
\beq\label{eq:matrixtraceforBPS}
\langle \mathcal{O}_{\circ}^{S}\rangle_{\chi}=-{\rm Tr}_{2}\left[(Y_3\cdot M)^{L}\right]\comma
\eeq
with
\beq
Y_3\cdot M=g^{2}\pmatrix{cc}{d_{31}&0\\0&d_{23}}\pmatrix{cc}{0&-(\sqrt{d_{12}}\rho_{21})^{-1}\\-(\sqrt{d_{12}}\rho_{12})^{-1}&0}\period
\eeq
To evaluate \eqref{eq:matrixtraceforBPS}, we diagonalize the matrix $(Y_3\cdot M)$. We then obtain eigenvalues,
\beq
e_{\pm}(\rho)=\pm\sqrt{\frac{d_{23}d_{31}}{d_{12}\rho_{12}\rho_{21}}}\comma
\eeq
which can be evaluated on the saddle point \eqref{eq:saddlepntrho21} as
\beq
e_{\pm}(\rho^{\ast})=\pm i \sqrt{2}g\sqrt{\frac{d_{23}d_{31}}{d_{12}}}\period
\eeq
As a result, we obtain the following expression for the three-point function in the large $N$ limit:
\beq
\left.\langle \mathcal{D}_1 \mathcal{D}_2 \mathcal{O}_{\circ}\rangle\right|_{N\to \infty} =-2^{\frac{L}{2}}g^{L}\left(i^{L}+(-i)^{L}\right)\left(\frac{d_{31}d_{23}}{d_{12}}\right)^{L/2} \underbrace{(2g^2 d_{12})^{N}e^{-N}}_{=\,\,e^{NS_{\rm eff}[\rho^{\ast}]}}\period
\eeq
Also in this case, one can see that the expected spacetime dependence \eqref{eq:structureBPS} is correctly reproduced from this computation. The last term on the right hand side comes from $e^{NS_{\rm eff}[\rho^{\ast}]}$, and can be eliminated by taking a ratio with $\langle \mathcal{D}_1 \mathcal{D}_2 \rangle$.

To read off the structure constant, it is convenient to normalize $\mathcal{O}_{\circ}$ by dividing the result by the normalization constant $\mathfrak{n}_{\mathcal{O}_{\circ}}$ of the two-point function
\beq\label{eq:twopntnormal}
\langle{\rm tr}\left((Y_1\cdot \Phi)^{L}\right)(x_1)\,{\rm tr}\left((Y_2\cdot \Phi)^{L}\right)(x_2) \rangle=\underbrace{L(2g^2)^{L}}_{=\,\,\mathfrak{n}_{\mathcal{O}_{\circ}}}\times d_{12}^{L}\period
\eeq
We then get
\beq\label{eq:finalbpsexercise}
\left.\frac{\langle \mathcal{D}_1 \mathcal{D}_2 \mathcal{O}_{\circ}\rangle}{\langle \mathcal{D}_1\mathcal{D}_2\rangle\sqrt{\mathfrak{n}_{\mathcal{O}_{\circ}}}}\right|_{N\to \infty} =\mathfrak{D}_{\mathcal{O}_{\circ}}\left(\frac{d_{31}d_{23}}{d_{12}}\right)^{L/2}\comma\qquad \mathfrak{D}_{\mathcal{O}_{\circ}}=-\frac{\left(i^{L}+(-i)^{L}\right)}{\sqrt{L}}\comma
\eeq
with $\mathfrak{D}_{\mathcal{O}_{\circ}}$ being the structure constant.

A few comments are in order: First the factor $1/\sqrt{L}$ simply comes from the fact that there are $L$ different contractions for the two-point function \eqref{eq:twopntnormal}, which are related to each other by a cyclic permutation of fields inside a trace. This structure is universal and persists also for the non-BPS three-point functions as we will see in section \ref{sec:weak}.
Second, owing to the factor $(i^{L}+(-i)^{L})$ which comes from a sum of two eigenvalues of $(Y_3\cdot M)$, the structure constant vanishes unless the length of the single-trace operator $L$ is even. This selection rule can also be reproduced from more direct Wick contractions as we see in the next section. Third, when we introduce a non-perturbative integrability description in section \ref{sec:bootstrap}, we will see that different eigenvalues correspond to different boundary states and the sum $(i^{L}+(-i)^{L})$ can be interpreted as coming from a sum of two different boundary states. It is interesting that the semi-classical approach allows us to see such an intricate structure already at weak coupling.
\subsubsection{Four-point functions of determinants}
Let us also discuss qualitative features of the four-point functions of determinants (without any single-trace insertions). In this case, there are $12$ independent $\rho$ fields and the saddle-point equation
\beq
\frac{\rho_{ij}}{g^2}=-\left(\frac{1}{2\hat{\rho}}\right)_{ij}d_{ij}
\eeq
takes a complicated form. Furthermore, the equation admits a multiple of physically distinct solutions and one has to look for the one which gives the most dominant contribution.

\begin{figure}[t]
\centering
\includegraphics[clip,height=4cm]{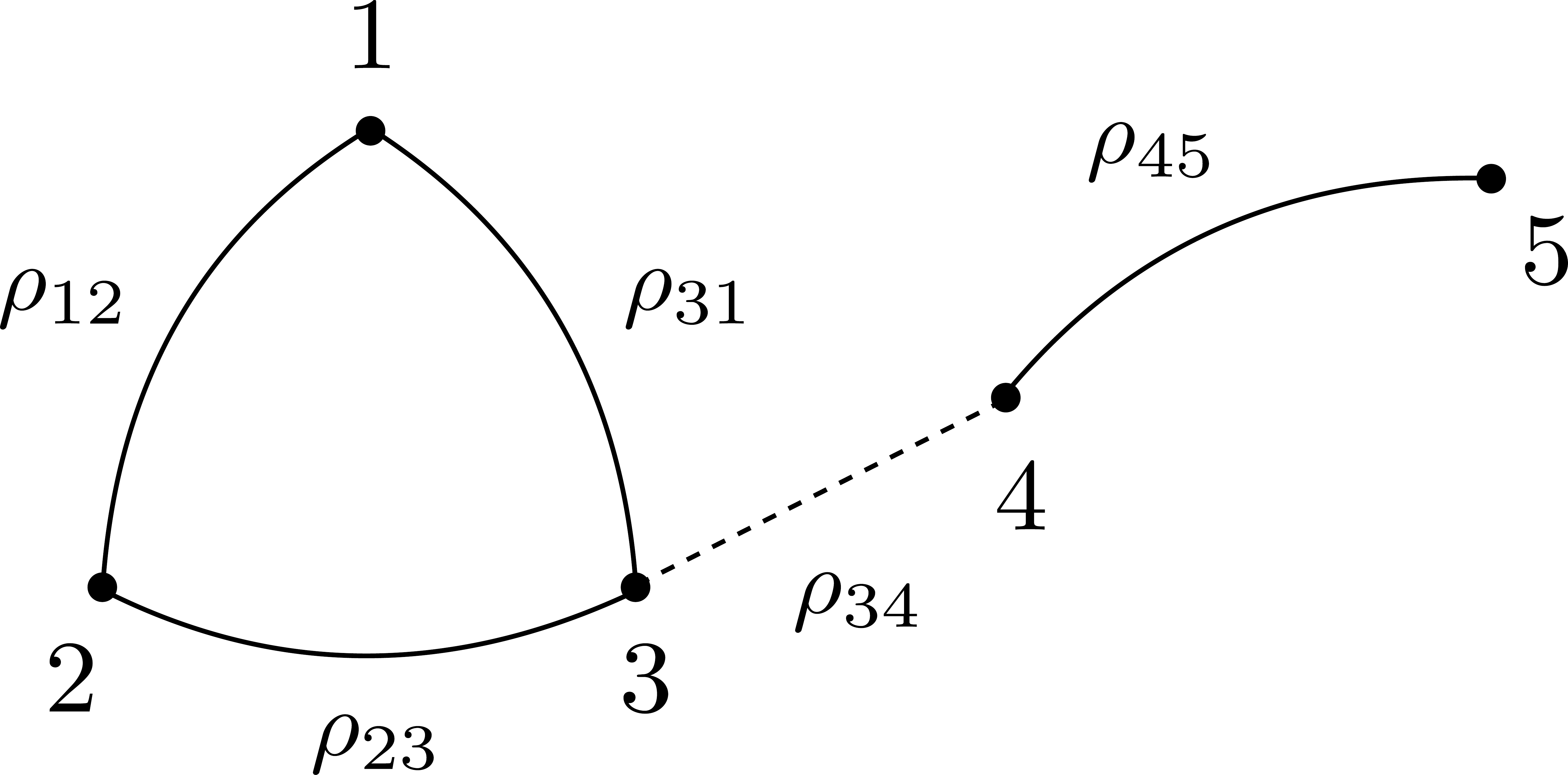}
\caption{The structure of a solution to the saddle-point equation for the five-point function, which corresponds to a disconnected diagram. The black solid lines denote $\rho$ fields which are nonzero while the dashed line and the lines which are not drawn correspond to $\rho$ fields which vanish. The one shown in figure corresponds to a product of a three-point function and a two-point function.}
\label{fig:fig11}
\end{figure}

Nevertheless, there are three simple solutions which have a clear physical interpretation:
\beq\label{eq:disconnectedsolutions}
\begin{aligned}
\text{Solution 1:}\quad \rho_{12}\rho_{21}=\rho_{34}\rho_{43}=-g^2/2\comma\qquad \text{others}=0\comma\\
\text{Solution 2:}\quad \rho_{13}\rho_{31}=\rho_{24}\rho_{42}=-g^2/2\comma\qquad \text{others}=0\comma\\
\text{Solution 3:}\quad \rho_{14}\rho_{41}=\rho_{23}\rho_{32}=-g^2/2\comma\qquad \text{others}=0\period
\end{aligned}
\eeq
As can be seen from the structure of the solutions, they have a ``block-diagonal'' structure and can be obtained by superposing the solutions for the two-point functions \eqref{eq:saddlepoint}. Correspondingly, the saddle-point actions are given by a product of two-point functions, for instance
\beq
\text{Solution 1:}\quad e^{N S_{\rm eff}[\rho^{\ast}]}=(2g^{2}d_{12})^{N}(2g^{2}d_{34})^{N}e^{-2N}\period
\eeq
Physically, they represent disconnected diagrams in which the determinant operators are pair-wise contracted.
More generally one can show that the saddle-point equation for higher-point functions always admits solutions which are just superpositions of solutions for lower-point functions (see figure \ref{fig:fig11} for an example). This reflects the fact on the field theory that the disconnected diagrams for higher-point functions factorize into a product of lower-point functions.

\begin{figure}[t]
\centering
\includegraphics[clip,height=3.8cm]{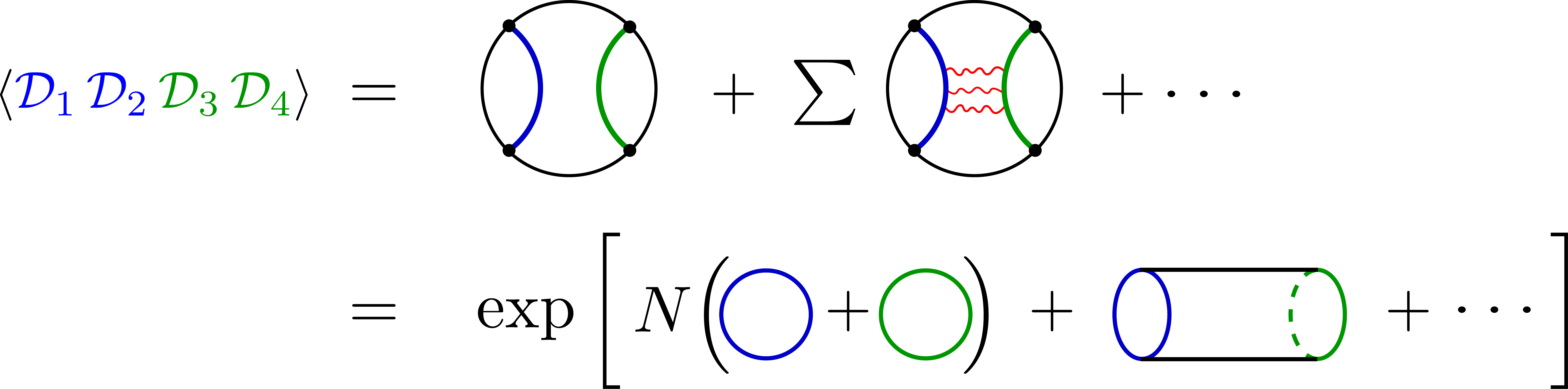}
\caption{The four-point function of determinant operators. The leading contribution is given by a disconnected diagram, which corresponds to two disconnected disks from the worldsheet point of view. The next leading contribution is given by a cylinder, which potentially encodes various interesting physics.}
\label{fig:fig12}
\end{figure}

Although such disconnected diagrams are not quite interesting by themselves, we expect that the fluctuation around such a saddle point contains rich physics: For instance, the leading $1/N$ fluctuations around the saddle point \eqref{eq:disconnectedsolutions} describes an exchange of single-trace states between two pairs of determinants (see figure \ref{fig:fig12}). On the AdS side, the corresponding process is an exchange of a single string state between two geodesic Witten diagrams created by the Giant Graviton D-branes. Such a process is expected to exhibit various interesting phenomena such as the tachyonic instability of open string modes and the phase transitions between different geodesic Witten diagrams. We will discuss them in more detail in the conclusion section (section \ref{sec:conclusion}).

We also want to point out that the analysis of the four-point function reveals an interesting interplay among three different concepts; the correlation of operators, the condensation of $\rho$ fields, and the reconnection of D-branes. Whenever different determinants are correlated, the corresponding off-diagonal $\rho$-field gets a nonzero expectation value. On the AdS side, this corresponds to a reconnection of the corresponding D-branes. The relation between the condensation of open strings and the reconnection of D-branes was discussed before in a different context (see for instance \cite{Hashimoto:1997gm}), but the connection to correlation of operators in the dual CFT has never been discussed. In a sense, what we are observing here is a toy-model version of the relation between geometry and quantum correlation, which is often discussed in the context of entanglement entropy in recent years \cite{VanRaamsdonk:2010pw}.

\subsubsection{Other applications\label{subsubsec:other}}
We now briefly mention other applications of our approach. A more detailed analysis will be presented in an upcoming paper \cite{JKV2}.
\paragraph{Higher loop}
In this section we only discussed the tree-level correlators. To include the loop corrections, one simply needs to bring down the interaction Lagrangian from the action and consider it as a correlator with multiple single-trace insertions. Alternatively, one could try to use a more sophisticated Lagrangian insertion approach. It would be interesting to work it out and develop perturbative techniques for the correlator of determinant operators.
\paragraph{Generating function and Schur polynomial}
In addition to determinant operators, there are closely related operators called subdeterminant operators\fn{Using the identities \eqref{eq:antisymmetricdelta} and \eqref{eq:antisymmetricdelta2} that we discuss in section \ref{subsec:derivationPCGG}, we can express it alternatively as
\beq
\det{}_{M} (Z)\equiv \frac{1}{M! (N-M)!} \epsilon_{a_1,\ldots, a_{M}, c_{M+1},\ldots, c_{N}}\epsilon^{b_1,\ldots, b_{M}, c_{M+1},\ldots, c_{N}} Z^{a_1}{}_{b_1}\cdots Z^{a_M}{}_{b_M}
\period
\eeq
This is in fact the original expression introduced in \cite{Balasubramanian:2001nh}.
}
\beq\label{eq:defsubdet}
\det{}_{M}(Z) \equiv \frac{1}{M!}\delta^{[b_1\cdots b_M]}_{[a_1\cdots a_M]}Z^{a_1}_{b_1}\cdots Z^{a_{M}}_{b_{M}}
\qquad \quad \left(\delta^{[b_1 \cdots b_M]}_{[a_1\cdots a_M]}\equiv  \sum_{\sigma\in S_{M}}(-1)^{|\sigma|}\delta^{b_1}_{a_{\sigma_1}}\cdots \delta^{b_{M}}_{a_{\sigma_{M}}}\right)
\comma
\eeq
which are also dual to (non-maximal) Giant Gravitons in AdS. A convenient way to deal with such operators is to consider a ``generating function''
\beq\label{eq:singlegenerating}
\mathcal{D}[z]\equiv \det (z-Z)=z^{N}\sum_{M=0}^{N}(-z)^{-M}\det{}_{M} (Z)\comma
\eeq
where $z$ is a c-number parameter. Such an operator is commonly used in the matrix model literature. The generating function $\mathcal{D}[z]$ can be studied straightforwardly in our formalism: The only modification is that the fermion action now gets a mass term $z(\bar{\chi}{\chi})$. This makes the resulting saddle-point equation slightly more complicated but does not add any conceptual difficulty.

There are also a whole class of operators, often called Schur polynomial operators, which are labelled by the Young tableau. The determinant and subdeterminant operators are a particular subclass of such operators which correspond to the Young tableau with a single column. Although general Schur polynomial operators are not contained in a single generating function \eqref{eq:singlegenerating}, one can extract them from a product of generating functions\fn{After the submission of this paper to arXiv, an interesting paper \cite{Chen:2019gsb} appeared in arXiv in which the idea explained here was worked out and elaborated. It was also shown in \cite{Chen:2019gsb} that subdeterminant operators and other $1/2$-BPS Schur polynomial operators do not correspond to integrable boundary states.},
\beq
\det (z_1 -Z)\det (z_2 -Z)\cdots \det (z_m-Z)\period
\eeq
A particularly interesting class of operators are the ones with $O(N^2)$ since they are dual to the deformation of the geometry called $1/2$-BPS bubbling geometries. It would be interesting to see if our approach can be used to efficiently compute the correlators of such operators.
\paragraph{From determinant to single trace} What we explained so far is a trick to recast the determinant operators into some emergent classical backgrounds. One could then wonder if a similar trick works also for single-trace operators, which are more commonly studied in planar $\mathcal{N}=4$ SYM. If the operators are BPS, there is indeed a way to apply our approach.

For this purpose, we first consider the {\it resolvent operator}, which is a generating function of BPS single-trace operators,
\beq
R[z]\equiv {\rm tr}\left(\frac{1}{z-Z}\right)=\frac{1}{z}\sum_{M=0}^{\infty}z^{-M}{\rm tr}\left(Z^{M}\right)\period
\eeq
In the matrix-model literature, there are two known approaches to relate the resolvent operator with the generating function \eqref{eq:singlegenerating}. The first approach is the {\it replica method}, which is based on the following identity:
\beq
{\rm tr}\left(\frac{1}{z-Z}\right)=\lim_{n\to 0}\frac{1}{n}\frac{\del}{\del z}\left(\det (z-Z)\right)^{n} \period
\eeq
The idea is to first study the right hand side for positive integer $n$ and then analytically continue the final result to obtain the result. In our context, the size of the matrix of $\rho$-variables is determined by the number of determinant operators. Therefore, the limit $n\to 0$ corresponds to a limit in which the size of the $\rho$-matrix goes to zero. Although this might sound like an unusual limit, it was actually used effectively in the study of matrix models \cite{Brezin:2007aa,Brezin:2007wv}. Also, a similar limit was discussed recently in \cite{Bargheer:2019kxb} in the study of correlators of large-charge single-trace operators. The matrix-model duality utilized in \cite{Bargheer:2019kxb} is closely related to the integrating-in-and-out tricks that we explained above, and from that point of view, it is not too surprising that the same limit shows up in their computation.

One drawback of the replica method is that it is not always easy to compute the final result as a function of the replica number $n$. The other approach, which avoids such an analytic continuation in $n$, is the {\it supersymmetry method}, which was also used in the recent study of doubly-nonperturbative effects in Jackiw-Teitelboim gravity \cite{Saad:2019lba}. In this approach, one expresses the resolvent as.
\beq
{\rm tr}\left(\frac{1}{z-Z}\right)=\left.\frac{\del}{\del z}\frac{\det (z-Z)}{\det (z^{\prime}-Z)}\right|_{z^{\prime}\to z}\period
\eeq
We can then rewrite $\det (z-Z)$ as a fermion one-loop determinant while expressing $1/(\det (z^{\prime}-Z))$ as a boson one-loop determinant. The tricks that we explained above are applicable also in this case, and the upshot is that the resulting $\rho$-matrix becomes a supermatrix. This is actually consistent with the string-theory interpretation: The inverse of a determinant is expected to describe a {\it ghost D-brane} and the theory living on a combined system of D-branes and ghost D-branes is known to be a supergroup gauge theory \cite{Vafa:2001qf,Okuda:2006fb}.
More detailed analysis and further explanation will be presented in \cite{JKV2}.

Note that the two approaches described here are standard techniques used in the analysis of disordered systems. In that context, there is yet another commonly-used approach, called {\it dynamic} or {\it stochastic} method (see for instance \cite{cond-mat/0209399}). It might be interesting to apply that method to our problem.
\section{Alternative Approach: Partially-Contracted Giant Gravitons\label{sec:PCGG}}
In this section, we explain an alternative approach to compute the correlation function of two determinants and several single-trace operators building on the analysis in \cite{Berenstein:2003ah}. As compared to the semi-classical method presented in the previous section, it is less intuitive and does not straightforwardly generalize to higher-point correlators of determinants. However, it also has the advantage that the computation boils down to evaluating correlators of single- and multi-trace operators and makes it easier to recycle the results in the literature. Because of this, one-loop computations presented later in this paper were performed using this approach. It also serves as a cross check of the results obtained in the semi-classical approach.
\subsection{Partial contraction of determinant operators\label{subsec:derivationPCGG}}
Let us discuss correlators of two determinant operators and (a multiple of) single-trace operators
\beq\label{eq:toanalyzePCGG}
G=\langle \mathcal{D}_1 (x_1) \mathcal{D}_2 (x_2) \prod_{k}\mathcal{O}_k\rangle\comma
\eeq
with
\beq
\mathcal{D}_1 (x_1)=\det (Y_1\cdot \Phi)(x_1) \comma\qquad \mathcal{D}_2(x_2)=\det (Y_2 \cdot \Phi)(x_2)\period
\eeq
The key idea of the method is to perform the computation in two steps: We first perform (partial) free Wick contractions between two determinant operators, and then take into account other contractions and loop corrections. To perform the computation, we make use of various identities worked out previously in the literature on Giant Gravitons; in particular in \cite{Berenstein:2003ah}.

As the first step, we rewrite determinant operators using the epsilon symbols as
\beq
\begin{aligned}
\mathcal{D}_1 (x_1) &=\frac{1}{N!}\epsilon_{a_1\cdots a_N}\epsilon^{b_1\cdots b_N}\Phi^{a_1}{}_{b_1}\cdots \Phi^{a_N}{}_{b_N} \comma\\
\mathcal{D}_2 (x_2) &=\frac{1}{N!}\epsilon_{\bar{a}_1\cdots \bar{a}_N}\epsilon^{\bar{b}_1\cdots \bar{b}_N}\bar{\Phi}^{\bar{a}_1}{}_{\bar{b}_1}\cdots \bar{\Phi}^{\bar{a}_N}{}_{\bar{b}_N} \period
\end{aligned}
\eeq
Here we used the shorthand notation $\Phi \equiv Y_{1} \cdot \Phi (x_{1})$ and $\bar{\Phi} \equiv Y_{2} \cdot \Phi (x_{2})$. Now, to express the result of the computation we define a quantity $\mathcal{G}_{\ell}$ called the {\it partially contracted Giant Graviton} (PCGG) of length $2\ell$, which is defined by contracting $N-\ell$ pairs of fields in determinants while leaving $\ell$ pairs free. The fields that are left free will later be contracted with other operators or interaction vertices. The result of the partial contractions of two determinant operators can be expressed as
\beq\label{eq:PCGGexpansion}
\left.\mathcal{D}_1 (x_1)\mathcal{D}_2 (x_2)\right|_{\text{partial contractions}}=\sum_{\ell=0}^{N}\left(\frac{g_{\rm YM}^2d_{12}}{8\pi^2}\right)^{N-\ell}\mathcal{G}_{\ell}(x_1,x_2)\comma
\eeq
with\fn{We colored the indices in \eqref{eq:defofPCGG} and \eqref{eq:simplifyPCGG} to clarify the patterns of index contractions.}
\beq\label{eq:defofPCGG}
\begin{aligned}
&\mathcal{G}_{\ell}(x_1,x_2) \equiv \frac{1}{(N!)^2}\left(\begin{array}{c}N\\N-\ell\end{array}\right)^{2}\epsilon_{\red{a_1\cdots a_{N-\ell}}c_1\cdots c_{\ell}}\epsilon^{\red{b_1\cdots b_{N-\ell}}d_1\ldots d_{\ell}}\epsilon_{\red{\bar{a}_1\cdots \bar{a}_{N-\ell}}\bar{c}_1\cdots \bar{c}_{\ell}}\epsilon^{\red{\bar{b}_1\cdots \bar{b}_{N-\ell}}\bar{d}_1\ldots \bar{d}_{\ell}}\\
&\qquad \times\frac{\langle \Phi^{\red{a_1}}{}_{\red{b_1}}\cdots \Phi^{\red{a_{N-\ell}}}{}_{\red{b_{N-\ell}}}\bar{\Phi}^{\red{\bar{a}_1}}{}_{\red{\bar{b}_1}}\cdots \bar{\Phi}^{\red{\bar{a}_{N-\ell}}}{}_{\red{\bar{b}_{N-\ell}}}\rangle_{0}}{\left(g_{\rm YM}^2d_{12}/(8\pi^2)\right)^{N-\ell}}\,\,\Phi^{c_1}{}_{d_1}\cdots \Phi^{c_{\ell}}{}_{d_{\ell}}\bar{\Phi}^{\bar{c}_1}{}_{\bar{d}_1}\cdots \bar{\Phi}^{\bar{c}_{\ell}}{}_{\bar{d}_{\ell}}\period
\end{aligned}
\eeq
Here $\langle \qquad \rangle_{0}$ stands for the tree-level Wick contraction and the prefactor comes from the free propagators
\beq
\langle \Phi^{a}{}_{b}\bar{\Phi}^{c}{}_{d}\rangle_0=\frac{g_{\rm YM}^2d_{12}}{8\pi^2}\delta^{a}_{d}\delta^{c}_{b}\period
\eeq

The combinatorial factor $\left(\begin{array}{c}N\\N-\ell\end{array}\right)^{2}$ in \eqref{eq:defofPCGG}  is due to the fact that one can in principle contract any $N-\ell$ fields inside each determinant while in the formula the first $N-\ell$ fields are contracted\fn{Different choices of fields lead to the same answer owing to the anti-symmetry of epsilon tensors.}.
Using the identities of epsilon tensors discussed in \cite{Berenstein:2003ah}, one can evaluate the right hand side of \eqref{eq:defofPCGG} explicitly. Relegating the derivation to Appendix \ref{ap:PCGG}, here we present the final formula:
\beq\label{eq:PCGGfinal}
\mathcal{G}_{\ell}(x_1,x_2)=(N-\ell)! (-1)^{\ell}\sum_{\substack{k_1,\ldots,k_{\ell}\\\sum_{s}s k_s=\ell}}\prod_{m=1}^{\ell}\frac{\left(-{\rm tr}\left[(\Phi\bar{\Phi})^m\right]\right)^{k_m}}{m^{k_m}k_m!}\period
\eeq
This is the main result of this section.

\begin{figure}[t]
\centering
\includegraphics[clip,height=3cm]{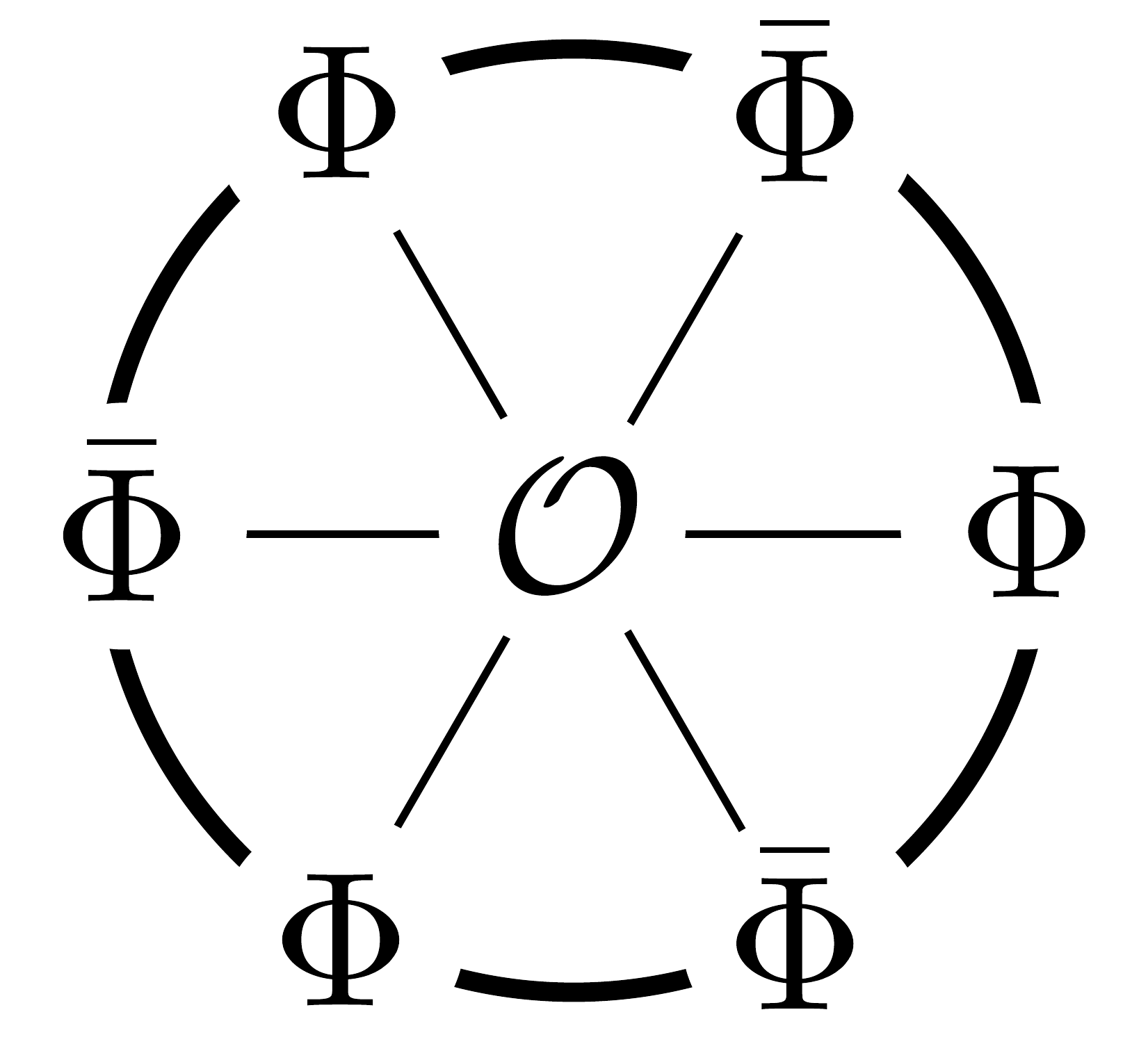}
\caption{The contractions between a single-trace operator and a PCGG. The alternating sequence of $\Phi$ and $\bar{\Phi}$ creates a boundary of the 't Hooft surface and the contraction has topology of a disk with one puncture.}
\label{fig:figextra2}
\end{figure}

The expression \eqref{eq:PCGGfinal} allows us to express the partially contracted Giant Graviton  as a sum of non-local multi-trace operators. There are two important features worth emphasizing: First the multi-trace terms in \eqref{eq:PCGGfinal} {\it do not} come with extra powers of $N$ as compared to the single-trace term. This allows us to use the standard planar expansion to compute correlators with single-trace operators \eqref{eq:toanalyzePCGG}: For instance, for the tree-level three-point function of two determinants and one single-trace operator at large $N$, we can focus on the single-trace term in \eqref{eq:PCGGfinal} since the correlator between a multi-trace operator and a single-trace operator is suppressed by powers of $1/N$. We should contrast this with the naive perturbation mentioned in the introduction, in which non-planar diagrams cannot be neglected because of the enhancement due to combinatorial factors.
Second the final result is given by an alternating sequence of fields $\Phi$ and $\bar{\Phi}$, see figure \ref{fig:figextra2}. The relevance of alternating structures in the study of determinants is not a priori obvious, but it also showed up in the previous studies \cite{Berenstein:2003ah} in a slightly different manner. We also want to point out that such an alternating sequence appeared already in an interesting pioneering work\fn{To our knowledge, this is the first paper in which the relation between baryonic operators and holes in the 't Hooft surface was pointed out. However the paper appears to be largely unknown.} \cite{Kazakov:1982pq} on the relation between baryons and the worldsheet with boundaries.

Before ending this subsection, let us also make one remark on the loop corrections. As mentioned earlier, the loop corrections can be taken into account by contracting the PCGG given in \eqref{eq:defofPCGG} with interaction vertices. However, for the computation of four-point functions which we explain in Appendix \ref{apsubsec:4ptPCGG1loop}, it turned out ot be more convenient to include the interaction vertices already when we perform the partial contractions \eqref{eq:PCGGexpansion}. If we perform the computation in this way, the PCGG itself would be expanded in powers of $g_{\textrm{YM}}^2$ as
\begin{gather}
\label{eq:PCGGtree1}
\mathcal{G}_{\ell}\left(x_1,x_2\right)
=
\mathcal{G}_{\ell}^{(0)}\left(x_1,x_2\right)
+
\mathcal{G}_{\ell}^{(1)}\left(x_1,x_2\right)
+
\dots
\end{gather}
where $\mathcal{G}_{\ell}^{(0)}$ and $\mathcal{G}_{\ell}^{(1)}$ are produced by contractions at tree-level and one-loop order respectively. $\mathcal{G}_{l}^{(0)}$ coincides with \eqref{eq:defofPCGG} while the expression for $\mathcal{G}_{\ell}^{(1)}$ will be given in Appendix \ref{ap:PCGG}.
\subsection{Examples\label{subsec:examplePCGG}}
Let us now see the use of the method in several examples.
\subsubsection{Two-point functions of determinants}
As the simplest example, let us evaluate the tree-level two-point functions of determinant operators. This amounts to taking $\ell=0$ term in the PCGG expansion \eqref{eq:PCGGexpansion}. Setting $\ell=0$ in \eqref{eq:PCGGfinal}, we obtain
\beq\label{eq:PCGGtree2pt}
\begin{aligned}
\langle \mathcal{D}_1(x_1)\mathcal{D}_{2}(x_2)\rangle_0&=\left(\frac{g_{\rm YM}^2d_{12}}{8\pi^2}\right)^{N}N!=(2g^2d_{12})^{N}\sqrt{2\pi N}e^{-N}\left[1+O(1/N)\right]\period
\end{aligned}
\eeq
In the last equality, we used the Stirling formula $N! \sim \sqrt{2\pi N} N^{N}e^{-N}$. The result \eqref{eq:PCGGtree2pt} is in perfect agreement with the result that we obtained in the semiclassical method \eqref{eq:twopointsemiclassicalresult}.
\subsubsection{2 determinants and 1 BPS single-trace}
Let us next study the correlator of two determinant operators and one single-trace BPS operator $\langle \mathcal{D}_1 (x_1)\mathcal{D}_2 (x_2) \mathcal{O}_{\circ}(x_3)\rangle$ at tree level with $\mathcal{O}_{\circ}(x_3)\equiv {\rm tr}\left[(Y_3\cdot \Phi)^{L}(x_3)\right]$. The first thing to notice is that the PCGG expansion only produces the operators of even lengths. Therefore unless $L$ is even, the result vanishes being in agreement with the result that we got from the semiclassical approach.

As mentioned above, at large $N$ we can focus on the single-trace term in the PCGG expansion. Thus the result at large $N$ is given by
\beq
\begin{aligned}
&\langle \mathcal{D}_1 (x_1)\mathcal{D}_2 (x_2) \mathcal{O}_{\circ}(x_3)\rangle|_{N\to\infty}= \\
&-\left(\frac{g_{\rm YM}^2d_{12}}{8\pi^2}\right)^{N-\ell}\frac{(N-\ell)!(-1)^{\ell}}{\ell}\left<{\rm tr}\left[\left((Y_1\cdot \Phi)(Y_2\cdot \Phi)\right)^{\ell}\right] \,\, {\rm tr}\left[(Y_3\cdot \Phi)^{L}(x_3)\right]\right>_0\comma
\end{aligned}
\eeq
with $L=2\ell$. Evaluating the right hand side using the planar contraction, we get\fn{Here we used the leading order Stirling approximation and the identity
\beq
2 (-1)^{L/2}=i^{L}+(-i)^{L}\comma
\eeq
which holds for even integers $L$.
}
\beq
\langle \mathcal{D}_1 (x_1)\mathcal{D}_2 (x_2) \mathcal{O}_{\circ}(x_3)\rangle|_{N\to\infty}=-2^{\frac{L}{2}}g^{L}\left(i^{L}+(-i)^{L}\right)\left(\frac{d_{31}d_{23}}{d_{12}}\right)^{L/2} (2g^2 d_{12})^{N}e^{-N}\comma
\eeq
which is in agreement with the result of the semi-classical method.

It is rather interesting that the PCGG approach leads directly to a two-point function of single-trace operators, which can be immediately recast as an overlap of two spin-chain states, while the semi-classical approach gives a matrix-trace representation \eqref{eq:matrixtraceformula}, which can be interpreted as an expectation value of the operator in the presence of the emergent classical background. As mentioned in section \ref{subsec:analogy}, the equivalence between the two is reminiscent of the Ellwood conjecture in open string field theory.
\section{Weak Coupling Analysis\label{sec:weak}}
We now apply the techniques developed in the previous two sections to compute the three-point functions of two determinants and one non-BPS single-trace operator at the {\it leading order} at weak coupling: More precisely, we perform the computation at {\it tree level}, but the single-trace operators that we use are the {\it one-loop} eigenstates of the dilatation operator. This is simply because the tree-level dilatation operator has huge degeneracies and one first needs to lift them by going to one loop in order to perform a systematic weak-coupling expansion \cite{Escobedo:2010xs}. Note that this is a standard recipe for degenerate perturbation theory in quantum mechanics.

Two interesting outcomes of the computation is that
\begin{itemize}
\item The results are nonzero only for parity-symmetric states.
\item The results are given by a ratio of simple determinants (see \eqref{eq:finaldOsu2tree}, \eqref{eq:finaldOsl2tree} and \eqref{eq:finaldOso6tree}).
\end{itemize}

To perform the computation, we use the twisted-translated frame introduced in section \ref{subsec:setup}. We also specialize the position of the single-trace operator $\mathcal{O}$ to be at the origin: Namely we consider
\beq
\langle\mathcal{D}_1 (a_1) \mathcal{D}_2 (a_2)\mathcal{O}(0) \rangle\qquad \qquad \left(\mathcal{D}_{1,2}(a_{1,2}) =\det \mathfrak{Z}(a_{1,2})\right)\comma
\eeq
with\fn{See also section \ref{subsec:setup} for more detailed explanation of the setup.} $\mathfrak{Z}(a)\equiv Z+\kappa^2 a^2\bar{Z}+\kappa a (Y-\bar{Y})$. In this setup, the operator $\mathcal{O}$ can be described as an excited state on the standard $Z$-vacuum ${\rm tr}\left(Z^{L}\right)$, making it easier to make connection with the commonly-used spin-chain description of single-trace operators in the literature.
\subsection{SU(2) sector\label{subsec:su2tree}}
Let us first discuss operators made up only of two complex scalars $Z$ and $Y$:
\beq
\mathcal{O}={\rm tr}(ZYZZY\cdots)+\cdots\period
\eeq
\subsubsection{Bethe states} To compute three-point functions, we need to use operators which are eigenstates of the dilatation operator. At the leading order at weak coupling, such states are known to be in one-to-one correspondence with eigenstates of the Heisenberg XXX spin chain under the identification
\beq
\begin{aligned}
Z \quad \leftrightarrow \quad \uparrow\comma\qquad Y \quad \leftrightarrow \quad \downarrow\comma\qquad {\rm tr}\left(Z\red{Y}ZZ\red{YY}Z\cdots\right) \quad \leftrightarrow \quad |\!\uparrow\red{\downarrow}\uparrow\uparrow\red{\downarrow\downarrow}\uparrow\cdots \rangle\period
\end{aligned}
\eeq
Moreover, since the XXX spin chain is integrable, one can systematically construct eigenstates using the so-called Bethe ansatz\fn{For a concise review of the Bethe ansatz in the context of AdS/CFT integrability, see section 2 of \cite{Escobedo:2010xs}.}. The eigenstates given by the Bethe ansatz (to be called {\it Bethe states}) are of a plane-wave form. Namely we regard the down spins as quasi-particles excitations on the up-spin ground state and write down an ansatz for the wave function as a sum of plane waves,
\beq\label{eq:betheansatzsu2}
\begin{aligned}
|\Psi\rangle&= \sum_{n_1<\cdots <n_M} \psi(n_1,\ldots, n_M)|\!\uparrow\cdots \underset{n_1}{\downarrow}\underset{\cdots\vspace{10pt}}{\cdots}\underset{n_M}{\downarrow}\cdots \uparrow\rangle\,,
\end{aligned}
\eeq
with $M$ being the number of down spins and
\beq\label{eq:psisu2}
\begin{aligned}
\psi (n_1,\ldots, n_M)=\sum_{\sigma \in S_M} \prod_{\substack{j<k\\\sigma_k<\sigma_j}}
S_{\text{SU(2)}}(u_{\sigma_k},u_{\sigma_j})\prod_{j=1}^{M}e^{ip(u_{\sigma_j})n_j}\,.
\end{aligned}
\eeq
Here the momentum $p$ and the S-matrix $S_{\text{SU(2)}}$ are given in terms of auxiliary variables called {\it rapidities} or {\it Bethe roots}:
\beq\label{eq:smatrixsu2}
S_{\rm SU(2)}(u_1,u_2)=\frac{u_1-u_2-i}{u_1-u_2+i}\comma\qquad e^{ip (u)}= \frac{u+i/2}{u-i/2}\period
\eeq
In order for $|\Psi\rangle$ to be an eigenstate of the Hamiltonian, the rapidities must obey the so-called {\it Bethe equation}\fn{More precisely the ansatz \eqref{eq:psisu2} with $S_{\rm SU(2)}$ and $e^{ip}$ given by \eqref{eq:smatrixsu2} automatically provides the eigenstate of an infinitely long spin chain. The Bethe equation is an additional requirement coming from the periodicity of the wave function on a finite chain of length $L$.},
\beq\label{eq:su2betheeq}
e^{ip(u_{j}) L}\prod_{k\neq j}S_{\rm SU(2)}(u_j,u_k)=1\comma
\eeq
where $L$ is the length of the spin chain (or equivalently the number of fields inside the single-trace operator).
Imposing the Bethe equation \eqref{eq:su2betheeq} is sufficient to construct the eigenstate of the Hamiltonian. It turns out, however, that we need to impose an additional condition, called the {\it level matching condition}, in order to have a well-defined single-trace operator:
\beq\label{eq:zeromomentum}
\prod_{j=1}^{M}e^{ip(u_j)}=1\period
\eeq
The condition comes from the cyclicity of the trace and requires that the state is invariant under shifting the overall state by one spin-chain site\fn{If this condition is not met, the state does not make sense as a single-trace operator; in other words, we simply get zero if we try to construct an operator from a state.}.

As is clear from the construction above, the Bethe state and the corresponding single-trace operator are function(al)s of the set of rapidity variables ${\bf u}\equiv \{u_1,\ldots, u_M\}$. Therefore, in the following discussion we denote them as
\beq
|\Psi\rangle\,\,\, \to \,\,\, |{\bf u}\rangle\comma\qquad \qquad \mathcal{O}\,\,\, \to \,\,\, \mathcal{O}_{\bf u}\period
\eeq
\subsubsection{Matrix trace and matrix product state} Having explained the structure of the single-trace operator $\mathcal{O}$, we now evaluate its three-point function with two determinant operators. Let us first perform the computation using the semi-classical method in section \ref{sec:effective}.

In the semi-classical method, the computation boils down to evaluating the matrix trace given in \eqref{eq:matrixtraceformula}, which can be obtained by replacing the fields inside the trace with appropriate $2\times 2$ matrices\fn{Here and below, we always evaluate $\langle \mathcal{O}\rangle_{\chi}$ at the saddle-point solution $\rho^{\ast}$ and omit writing the symbol $|_{\rho^{\ast}}$.} (see also \eqref{eq:matrixtraceforBPS}):
\beq\label{eq:su2replacement}
{\rm tr}\left(Z\red{Y}ZZ\red{YY}Z\cdots \right)\quad \mapsto \quad -{\rm Tr}_{m=2}\left[M_{Z}\red{M_{Y}}M_{Z}M_{Z}\red{M_{Y}M_{Y}}M_{Z}\cdots \right]\period
\eeq
Using the Wick contractions
\beq\label{eq:contractionrulewicksu2}
\bcontraction{}{\mathfrak{Z}}{(a)\quad }{Z}
\bcontraction{\mathfrak{Z}(a)\quad Z=\kappa^2\comma\qquad}{\mathfrak{Z}}{(a)\quad }{Y}
\mathfrak{Z}(a)\quad Z=\kappa^2\comma\qquad \mathfrak{Z}(a)\quad Y=-\kappa /a \comma
\eeq
and the general formula \eqref{eq:conversionMtorhohat}, we can compute $M_{Z,Y}$ as
\beq
\begin{aligned}
M_Z=\pmatrix{cc}{\kappa^2&0\\0&\kappa^2}\cdot m_{\rho^{\ast}}\comma\qquad
M_Y=\pmatrix{cc}{-\kappa /a_1&0\\0&-\kappa / a_2}\cdot m_{\rho^{\ast}}\comma
\end{aligned}
\eeq
with
\beq\label{eq:matrixmrhostar}
m_{\rho^{\ast}}=-\frac{g^2}{\kappa}\pmatrix{cc}{0&1/\rho_{21}^{\ast}\\1/ \rho_{12}^{\ast}&0}\period
\eeq
Here $\rho^{\ast}$ are the saddle-point values of the $\rho$-field satisfying \eqref{eq:saddlepntrho21}, $\rho^{\ast}_{12}\rho^{\ast}_{21}=-g^2/2$.
 To evaluate this matrix trace, it is convenient to use the basis in which $M_Z$ is diagonal\fn{Note that the matrix trace is invariant under such a change of basis.}:
 \beq\label{eq:diagonalizedMzMy}
\begin{aligned}
\bar{M}_Z=&i\sqrt{2}g\kappa\pmatrix{cc}{1&0\\0&-1}\comma\\
\bar{M}_Y=&\frac{ig}{\sqrt{2}}\pmatrix{cc}{-\left(a_1^{-1}+a_2^{-1}\right)&-i\left(a_1^{-1}-a_2^{-1}\right)\\-i\left(a_1^{-1}-a_2^{-1}\right)&\left(a_1^{-1}+a_2^{-1}\right)}\comma
\end{aligned}
 \eeq

 As explained in section \ref{subsec:derivation}, one can also recast this matrix trace as an overlap between a single-trace state and a matrix product state, which in the spin chain notation reads
 \beq
 \langle\mathcal{O}_{\bf u}\rangle_{\chi}=\langle \bar{M}|{\bf u}\rangle\comma
 \eeq
with
 \beq\label{eq:generalMbarsu2}
 \langle\bar{M}|=-{\rm Tr}_{m=2}\left[\prod_{s=1}^{L}\Big(\langle \uparrow\!\!|_{s}\otimes \bar{M}_{Z}+\langle \downarrow\!\!|_{s}\otimes \bar{M}_{Y}\Big)\right]\period
 \eeq
 Here $\langle \uparrow\!\!|_s$ and $\langle \downarrow\!\!|_s$ denote states on the $s$-th site of the spin chain. In this form, we can clearly see the resemblance with the matrix product state discussed in the context of one-point functions in the presence of a domain-wall defect \cite{deLeeuw:2015hxa}. The relation becomes even more transparent if we further specialize the positions of the determinant operators to be
 \beq\label{eq:definitionofsymmetricspecial}
 a_1 =- a_2 \period
 \eeq
In what follows we call it the {\it symmetric configuration} and abbreviate it by ${\tt sym}$.
In this configuration, $\bar{M}_Y$ becomes purely off-diagonal and leads to
\beq\label{eq:specialMPS}
\left.\langle \bar{M}|\right|_{{\tt sym}}=-(2\sqrt{2}ig\kappa)^{L}{\rm Tr}_{m=2}\left[\prod_{s=1}^{L}\Big(\langle \uparrow\!\!|_{s}\otimes t_1-i z\langle \downarrow\!\!|_{s}\otimes t_2\Big)\right]\comma
\eeq
where $t_{1,2}$ and $z$ are\fn{$\sigma_{1,3}$ are the Pauli matrices.}
\beq
\begin{aligned}
t_1 \equiv \sigma_3 /2 \comma\qquad t_2 \equiv \sigma_1/2\comma\qquad z\equiv\frac{1}{\kappa a_1}\period
\end{aligned}
\eeq
Remarkably, the right hand side of \eqref{eq:specialMPS} coincides (up to an overall factor) with the generalized matrix product state defined in \cite{deLeeuw:2015hxa}.

This coincidence allows us to immediately write down an overlap with the Bethe state $|{\bf u}\rangle$ using the results in \cite{deLeeuw:2015hxa,brockmann2014neel,brockmann2014gaudin,pozsgay2014overlaps,pozsgay2018overlaps,Piroli:2018ksf,Foda:2015nfk}. The results obey two important selection rules:
\begin{enumerate}
\item It is nonzero only when the length of the spin chain $L$ and the number of magnons $M$ are both even.
\item It is nonzero only when the rapidities are parity symmetric, namely \\${\bf u}=\{u_1,-u_1,u_2,-u_2,\ldots,u_{M/2},-u_{M/2}\}$.
\end{enumerate}
When both of these conditions are met, the result is given by the following determinant expression\fn{Written explicitly, $\del_u p(u)=-1/(u^2+1/4)$.},
\beq
\begin{aligned}
\left.\langle \bar{M}|{\bf u}\rangle\right|_{\tt sym}=&-\frac{(\sqrt{2}g)^{L}\kappa^{L-M}}{a_1^{M}}\left[i^{L-M}+(-i)^{L-M}\right]\left(\prod_{s=1}^{M/2}\frac{u_s-\frac{i}{2}}{ u_s\,\del_u p(u_s)} \right)
\det G_{+}^{\rm SU(2)}\comma
\end{aligned}
\eeq
where we introduced $\frac{M}{2}\times \frac{M}{2}$ matrices $G_{\pm }^{\rm SU(2)}$ by\fn{Note that we have different normalizations for $G_{\pm}^{\rm SU(2)}$ as compared to $G_{\pm}$ in \cite{deLeeuw:2015hxa}. The ones in \cite{deLeeuw:2015hxa} are normalized in the momentum basis whereas ours are normalized in the rapidity basis. The difference of the normalizations drop out if we take the ratio $\det G_{+}/\det G_{-}$.}
\beq\label{eq:Gpmsu2}
\left(G_{\pm}^{\rm SU(2)}\right)_{ij}=\left[L\del_{u}p(u_i)+\sum_{k=1}^{\frac{M}{2}}\mathcal{K}_{+}^{\rm SU(2)}(u_i,u_k)\right]\delta_{ij}-\mathcal{K}_{\pm}^{\rm SU(2)}(u_i,u_j)\comma
\eeq
with
\beq
\mathcal{K}_{\pm}^{\rm SU(2)}(u,v)\equiv\frac{1}{i}\left( \del_{u} \log S_{\rm SU(2)}(u,v)\pm \del_u\log S_{\rm SU(2)}(u,-v)\right)\period
\eeq

To read off the structure constant, we need to divide the answer by the normalization of the two-point function, which reads\fn{The overall factor $L$ comes from cyclic permutations of fields inside the trace and $(2g^2)^{L}$ comes from the normalization of propagators. See also \eqref{eq:finalbpsexercise}.}
\beq
\mathfrak{n}_{\mathcal{O}_{\bf u}}=L(2g^{2})^{L} \langle {\bf u}|{\bf u}\rangle\comma
\eeq
where $\langle {\bf u}|{\bf u}\rangle$ is the so-called Gaudin norm of the spin-chain state\cite{gaudin1976diagonalisation,Korepin:1982gg}, which takes a factorized form $\det G_{+}^{\rm SU(2)}\det G_{-}^{\rm SU(2)}$ for parity-symmetric states (see Appendix \ref{ap:norm}).
As a result we get the following expression for the ratio of correlators,
\beq\label{eq:beforefinalsu2tree}
\left.\frac{\langle \mathcal{D}_1\mathcal{D}_2\mathcal{O}_{\bf u}\rangle}{\langle \mathcal{D}_1\mathcal{D}_2\rangle \sqrt{\mathfrak{n}_{\mathcal{O}_{\bf u}}}}\right|_{\tt sym}=-\frac{\kappa^{L-M}}{a_1^{M}}\frac{\left[i^{L-M}+(-i)^{L-M}\right]}{\sqrt{L}}\sqrt{\left(\prod_{1\leq s\leq \frac{M}{2}}\frac{u_s^2+\frac{1}{4}}{u_s^2}\right)\frac{\det G_{+}^{\rm SU(2)}}{\det G_{-}^{\rm SU(2)}}}\period
\eeq
Comparing this with the general form of the ratio of correlators \eqref{eq:ratio3ptand2pt} using $\Delta=L$ and $J=L-M$, we conclude that the structure constant is given by
\beq\label{eq:finaldOsu2tree}
\mathfrak{D}_{\mathcal{O}_{\bf u}}=-\frac{i^{J}+(-i)^{J}}{2^{M}\sqrt{L}}\sqrt{\left(\prod_{1\leq s\leq \frac{M}{2}}\frac{u_s^2+\frac{1}{4}}{u_s^2}\right)\frac{\det G_{+}^{\rm SU(2)}}{\det G_{-}^{\rm SU(2)}}}\period
\eeq
The factor $i^{J}+(-i)^{J}=i^{L-M}+(-i)^{L-M}$ is a natural generalization of the factor that we encountered for the BPS three-point function \eqref{eq:finalbpsexercise}.
We will later see that this can be interpreted as a sum of contributions from two boundary states.
 Note also that the factor $\sqrt{L}$ in the denominator comes from the normalization of the two-point function as is the case with the BPS three-point function.

So far, we have been working in the symmetric configuration \eqref{eq:definitionofsymmetricspecial} in order to make use of the results in \cite{deLeeuw:2015hxa}. However, once we get the result \eqref{eq:finaldOsu2tree}, we can reverse-engineer it using the general structure of the correlator \eqref{eq:ratio3ptand2pt} to predict the overlap between a Bethe state and a general matrix product state \eqref{eq:generalMbarsu2}. This leads to a conclusion that the overlap between a Bethe state $|{\bf u}\rangle$ and a one-parameter family of matrix product states
\beq
\langle {\rm MPS}(x)|\equiv {\rm Tr}_{m=2}\left[\prod_{s=1}^{L}\Big(\langle \uparrow\!\!|_{s}\otimes t_1+\langle \downarrow\!\!|_{s}\otimes (t_2+x\,t_1)\Big)\right]\comma
\eeq
does not depend on the parameter $x$. This is actually easy to prove since $\langle {\rm MPS}(x)|$ can be expressed as
\beq
\langle {\rm MPS}(x)|=\langle {\rm MPS}(0)| +\langle \cdots|S^{+} \comma
\eeq
and the Bethe state $|{\bf u} \rangle$ satisfies the highest weight condition $S^{+}|{\bf u}\rangle=0$.
\subsubsection{PCGG and generalized N\'{e}el state}
Let us now see how the computation goes in the PCGG approach. As discussed in section \ref{sec:PCGG}, at large $N$ we can focus on the single-trace term of the PCGG. This reduces the computation of the three-point function in the twisted translated frame to the following correlator:
\beq\label{eq:tocomputesecgneel}
\begin{aligned}
&\langle \mathcal{D}_1 \mathcal{D}_2\mathcal{O}_{\bf u}\rangle=-\left(2g^2 N\kappa^2\right)^{N-\frac{L}{2}}\frac{2(N-\tfrac{L}{2})!(-1)^{L/2}}{L}\left<{\rm tr}\left[\left(\mathfrak{Z}(a_1)\mathfrak{Z}(a_2)\right)^{L/2}\right] \,\, \mathcal{O}_{\bf u}(0)\right>\period
\end{aligned}
\eeq
Using the contraction rules \eqref{eq:contractionrulewicksu2}, one can translate the correlator into the spin-chain language,
\beq\label{eq:fromtwopntntoneel}
\left<{\rm tr}\left[\left(\mathfrak{Z}(a_1)\mathfrak{Z}(a_2)\right)^{L/2}\right] \,\, \mathcal{O}_{\bf u}(0)\right>=\frac{L(2g^2\kappa^2)^{L}}{2(-\kappa a_1)^{M}}\times \langle \text{N\'{e}el}_{a_1/a_2}|{\bf u}\rangle\comma
\eeq
where the {\it weighted N\'{e}el state} $\langle \text{N\'{e}el}_x|$ is defined by
\beq\label{eq:weightedneel}
\langle \text{N\'{e}el}_x | \equiv \sum_{\psi:\text{ all states}}\left(x^{N^{\psi}_{\rm even}}+x^{N^{\psi}_{\rm odd}}\right)\langle \psi|\period
\eeq
Here $N^{\psi}_{\text{even, odd}}$ denote the numbers of down spins at even/odd sites in the state $\langle \psi |$.

Upon setting $x=0$, the weighted N\'{e}el state can be expressed as a sum of two terms,
\beq
\langle \text{N\'{e}el}_0 | = \sum_{N^{\psi}_{\rm even}=0}\langle \psi|+\sum_{N^{\psi^{\prime}}_{\rm odd}=0}\langle \psi^{\prime}|\comma
\eeq
where each term is a sum of states with down spins only at even/odd sites.
This coincides with (a sum of) generalized N\'{e}el states defined in \cite{deLeeuw:2015hxa}:
\beq
\langle \text{N\'{e}el}_0 |=\sum_{M} \underbrace{\sum_{\substack{\substack{n_1<\cdots<n_M\\|n_i-n_j|:\text{even}}}} \left< \uparrow \cdots \underset{n_1}{\downarrow}\cdots \underset{n_M}{\downarrow}\cdots \uparrow\right|}_{\text{generalized N\'{e}el state in \cite{deLeeuw:2015hxa}}}
\eeq
For other values of $x$, \eqref{eq:weightedneel} provides a one-parameter generalization of generalized N\'{e}el state.

By performing the computation for various states, we found that the overlap between the weighted N\'{e}el state and a Bethe state is given by the following simple expression\fn{In principle, one should be able to prove this equality by expressing $\langle \text{N\'{e}el}_{x}| $ in terms of $\langle \text{N\'{e}el}_{0}|$ and the action of global symmetry generators $S_3$ and $S_{+}$. However, we will not attempt it in this paper.},
\beq\label{eq:weightedneelnicerelation}
\langle \text{N\'{e}el}_{x}|{\bf u}\rangle=(1-x)^{M}\langle \text{N\'{e}el}_{0}|{\bf u}\rangle\comma
\eeq
where $\langle \text{N\'{e}el}_{0}|{\bf u}\rangle$ is given by \cite{deLeeuw:2015hxa}
\beq
\langle \text{N\'{e}el}_{0}|{\bf u}\rangle=2\left(\frac{i}{2}\right)^{M}\left(\prod_{s=1}^{M/2}\frac{u_s-\frac{i}{2}}{ u_s\,\del_u p(u_s)}\right)\det G^{\rm SU(2)}_{+}\period
\eeq
Plugging this expression into \eqref{eq:tocomputesecgneel} and \eqref{eq:fromtwopntntoneel} and normalizing the correlator\fn{Note that in the twisted translated frame, the two-point function of determinants \eqref{eq:PCGGtree2pt} reads
\beq
\langle\mathcal{D}_1\mathcal{D}_2\rangle=\left(2g^2\kappa^2\right)^{N}N!\period
\eeq}, we obtain
\beq
\frac{\langle \mathcal{D}_1\mathcal{D}_2\mathcal{O}_{\bf u}\rangle}{\langle \mathcal{D}_1\mathcal{D}_2\rangle \sqrt{\mathfrak{n}_{\mathcal{O}_{\bf u}}}}=-\kappa^{L-M}\left(\frac{a_1-a_2}{2a_1 a_2}\right)^{M}\frac{\left[i^{L-M}+(-i)^{L-M}\right]}{\sqrt{L}}\sqrt{\left(\prod_{1\leq s\leq \frac{M}{2}}\frac{u_s^2+\frac{1}{4}}{u_s^2}\right)\frac{\det G_{+}^{\rm SU(2)}}{\det G_{-}^{\rm SU(2)}}}\comma\nn
\eeq
which agrees with the result obtained from the semi-classical approach \eqref{eq:beforefinalsu2tree} upon setting $a_1=-a_2$.

An interesting outcome of our computation is that the semi-classical approach naturally leads to the matrix product states introduced in \cite{deLeeuw:2015hxa} while the PCGG approach gives generalized N\'{e}el states. In order for them to reproduce the same answers, it is important that the two states are related\fn{For an explicit relation between the two states, see (5.12) in \cite{deLeeuw:2015hxa}.} up to an overall factor $2^{L-M}i^{M}$ and non-highest weight contributions as was discovered in \cite{deLeeuw:2015hxa}. In particular, the constant of proportionality between the two states, $2^{L-M}i^{M}$, is precisely what is needed to reproduce the correct spacetime dependence. This shows that the relation between the two states is not just a technical coincidence, but it is actually a consequence of the spacetime Ward identity of $\mathcal{N}=4$ SYM.

\subsection{SL(2) sector\label{subsec:sl2tree}}
We now perform a similar analysis for operators in the SL(2) sector,
\beq
\mathcal{O}={\rm tr}\left(Z D_{+}Z Z (D_{+})^3Z\cdots\right)+\cdots\comma
\eeq
with $D_{+}$ being a covariant derivative defined by $D_{+}\equiv (D_{1}-iD_{2})/2$. Since we only analyze the tree-level correlators, we can also replace the covariant derivatives with ordinary derivatives $\del_{+}$.
\subsubsection{Bethe state}
Operators in the SL(2) sector can be mapped to the so-called SL(2) spin chain. One notable difference from the SU(2) spin chain is that each site in the SL(2) spin chain can host an arbitrary number of magnon excitations. This reflects the fact that one can act an arbitrary number of derivatives to a given $Z$-field. A more precise map between the field and the spin is
\beq\label{eq:mapoffieldsSL2}
\frac{(D_{+})^{n}Z}{n!}\quad \leftrightarrow \quad|n\rangle\comma\qquad {\rm tr}\left(Z\red{D_{+}Z}Z\red{\frac{(D_{+})^3Z}{3!}}Z\cdots\right)\quad \leftrightarrow\quad |0\,\red{1}\,0\,\red{3}\,0\cdots\rangle\comma
\eeq
where $|n\rangle$ is an $n$ magnon state of a single-site SL(2) spin chain. The factorial factor $1/n!$ in the relation \eqref{eq:mapoffieldsSL2} comes from the normalization of states: In the spin-chain language, the derivative $D_{+}$ corresponds to one of the generators in SL$(2,R)$, $S_{+}$, and the field $(D_{+})^{n}Z$ is mapped to a state $(S_{+})^{n}|0\rangle$. Computing the norm of this spin-chain state using the SL(2,R) algebra, we get $(n!)^2$.
On the other hand, the $n$-th excited state is unit-normalized, $\langle n|n\rangle=1$. This difference of the normalizations accounts for the factorial in \eqref{eq:mapoffieldsSL2}.

With slight modifications, the Bethe ansatz can be applied also to the SL(2) spin chain. To write down the Bethe state, we again express the state in terms of positions of magnons as
\beq
|\Psi\rangle =\sum_{n_1\leq \cdots \leq n_{S}}\psi(n_1,\cdots,n_{S})|0\cdots 0\,\underset{n_1}{1}\,0\cdots0\,\underset{n_S}{1}\,0\cdots0\rangle\period
\eeq
Note that, in the SL(2) spin chain several magnons can live at the same site, and therefore the summation includes $n_i=n_j$\fn{For instance, for $n_1=n_2$, the summand is given by
\beq
\psi(n_1,n_1,n_3\cdots,n_S)|0\cdots 0\,\underset{n_1}{\red{2}}\,0\cdots0\,\underset{n_S}{1}\,0\cdots0\rangle\period
\eeq }.
The Bethe wave function $\psi$ is again a sum of plane waves,
\beq
\begin{aligned}
\psi (n_1,\ldots, n_M)=\sum_{\sigma \in S_M} \prod_{\substack{j<k\\\sigma_k<\sigma_j}}
S_{\text{SL(2)}}(u_{\sigma_k},u_{\sigma_j})\prod_{j=1}^{M}e^{ip(u_{\sigma_j})n_j}\,,
\end{aligned}
\eeq
where the SL(2) S-matrix $S_{\text{SL(2)}}$ is given by the following expression\fn{The momentum $p$ is given by the same expression as before \eqref{eq:smatrixsu2}.}:
\beq
S_{\rm SL(2)}(u_1,u_2)=\frac{u_1-u_2+i}{u_1-u_2-i}\period
\eeq
With the replacement of the S-matrix, the Bethe equation takes the same form,
\beq
e^{ip(u_j)}\prod_{k\neq j}S_{\rm SL(2)}(u_j,u_k)=1\comma
\eeq
and we again need to impose the zero-momentum condition \eqref{eq:zeromomentum} in order to have a well-defined single-trace operator.
\subsubsection{Matrix trace and matrix product state}
Let us now analyze the three-point function using the semi-classical approach. As with the SU(2) sector, the computation boils down to evaluating a certain matrix trace. The only thing we need to figure out is how each field, which now comes with derivatives, gets mapped to a $2\times2$ matrix. This can be deduced straightforwardly by repeating the analysis in section \ref{sec:effective} using the following Wick contraction:
\beq\label{eq:wickSL(2)}
\bcontraction[2ex]{}{\mathfrak{Z}}{(a)\qquad \frac{(D_{+})^{n}}{n!}}{Z}
\mathfrak{Z}(a)\qquad \frac{(D_{+})^{n}}{n!}Z(0)=\kappa^2\left(ia\right)^{-n}\period
\eeq
This results in the following substitution rule,
\beq
\mathcal{O}={\rm tr}\left(Z\red{D_{+}Z}Z\red{\frac{(D_{+})^3Z}{3!}}Z\cdots \right)\quad \mapsto \quad \langle\mathcal{O}\rangle_{\chi}=-{\rm Tr}_{m=2}\left[M_{0}\red{M_{1}}M_{0}\red{M_{3}}M_{0}\cdots \right]\period
\eeq
where $M_{n}$ is given by
\beq
\begin{aligned}
M_{n}=&\pmatrix{cc}{\kappa^2(ia_1)^{-n}&0\\0&\kappa^2(ia_2)^{-n}}\cdot m_{\rho^{\ast}}\comma
\end{aligned}
\eeq
with $m_{\rho^{\ast}}$ being given in \eqref{eq:matrixmrhostar}.

As is the case with the SU(2) sector, it is convenient to use the basis in which $M_0$ is diagonal. The result reads
\beq
\begin{aligned}
\bar{M}_n=&\frac{ig\kappa }{\sqrt{2}}\pmatrix{cc}{(ia_1)^{-n}+(ia_2)^{-n}&i\left[(ia_1)^{-n}-(ia_2)^{-n}\right]\\i\left[(ia_1)^{-n}-(ia_2)^{-n}\right]&-(ia_1)^{-n}-(ia_2)^{-n}}\period
\end{aligned}
\eeq
Using these matrices, one can write down a matrix product state of the spin chain whose overlap with a Bethe state gives $\langle \mathcal{O}_{\bf u}\rangle_{\chi}=\langle \bar{M}|{\bf u}\rangle$:
\beq
\begin{aligned}
\langle \bar{M}|\equiv -{\rm Tr}_{m=2}\left[\prod_{s=1}^{L}\left(\sum_{n=0}^{\infty}\langle n|_s\otimes \bar{M}_{n}\right)\right]\period
\end{aligned}
\eeq
Here $\langle n|_s$ is the $n$-th excited state on the $s$-th site of the spin chain.

Interestingly, also in the SL(2) sector the matrix product state simplifies in the symmetric configuration $a_1=-a_2$: In this configuration, $\bar{M}_{\rm even}$ is purely diagonal while $\bar{M}_{\rm odd}$ is purely off-diagonal. As a result, we get
\beq
\left.\langle \bar{M}|\right|_{\tt sym}=-(2\sqrt{2}i g\kappa)^{L}{\rm Tr}_{m=2}\left[\prod_{s=1}^{L}\left(\sum_{n:\text{ even}}\frac{\langle n|_s}{(ia_1)^{n}}\otimes t_1+i\sum_{n:\text{ odd}}\frac{\langle n |_s}{(ia_1)^{n}}\otimes t_2\right)\right]\comma
\eeq
with $t_{1}=\sigma_3/2$ and $t_{2}=\sigma_1/2$.
The result turns out to take an even simpler form if we use $\langle n| =\langle 0|(S_{-})^{n}/n!$. We then have an expression
\beq\label{eq:MsymSL2nice}
\left.\langle \bar{M}|\right|_{\tt sym}=-(2\sqrt{2}i g\kappa)^{L}{\rm Tr}_{m=2}\left[\prod_{s=1}^{L}\left(\langle 0|_s\cos \left(\frac{S_{-}^{(s)}}{a_1}\right)\otimes t_1+\langle 0|_s \sin \left(\frac{S_{-}^{(s)}}{a_1}\right)\otimes t_2\right)\right]\comma
\eeq
where $S_{-}^{(s)}$ is a spin-lowering operator acting on the $s$-th site.

Having obtained an explicit representation for the matrix product state, we can now experimentally compute overlaps with various Bethe states and see if the result admits a simple expression as is the case with the SU(2) sector. The answer turned out to be positive: We again found the selection rule which forces both $L$ and $S$ to be even and the set of rapidities to be parity-symmetric. We also numerically observed that the overlap is given by the following formula:
\beq\label{eq:symoverlapSL2}
\begin{aligned}
\left.\langle \bar{M}|{\bf u}\rangle\right|_{\tt sym}=&-\frac{(\sqrt{2}g\kappa)^{L}}{a_1^{S}}\left[i^{L}+(-i)^{L}\right]\left(\prod_{s=1}^{S/2}\frac{u_s+\frac{i}{2}}{ u_s\,\del_u p(u_s)} \right)
\det G_{+}^{\rm SL(2)}\comma
\end{aligned}
\eeq
where $G_{\pm}^{\rm SL(2)}$ are $\frac{S}{2}\times\frac{S}{2}$ matrices given by
\beq\label{eq:GpmSL2}
\left(G_{\pm}^{\rm SL(2)}\right)_{ij}=\left[L\del_{u}p(u_i)+\sum_{k=1}^{\frac{M}{2}}\mathcal{K}_{+}^{\rm SL(2)}(u_i,u_k)\right]\delta_{ij}-\mathcal{K}_{\pm}^{\rm SL(2)}(u_i,u_j)\comma
\eeq
with
\beq
\mathcal{K}_{\pm}^{\rm SL(2)}(u,v)\equiv \frac{1}{i}\left(\del_{u} \log S_{\rm SL(2)}(u,v)\pm \del_u\log S_{\rm SL(2)}(u,-v)\right)\period
\eeq
Although the result is similar to the one for the SU(2) sector, some of the factors\fn{For instance we had $i^{L-M}+(-i)^{L-M}$ in the SU(2) sector while we now have $i^{L}+(-i)^{L}$.} which were given previously by $L-M$ in the SU(2) sector now get replaced with $L$, not with $L-S$. This is because those factors are associated with the R-charge $J$ of the operator, not the length $L$, as will become clear when we discuss the nonperturbative approach in section \ref{sec:bootstrap}.

Dividing the result \eqref{eq:symoverlapSL2} by the normalization of the two-point function,
\beq
\mathfrak{n}_{\mathcal{O}_{\bf u}}=L(2g^2)^{L}\langle {\bf u}|{\bf u}\rangle\comma
\eeq
with $\langle {\bf u}|{\bf u}\rangle$ being the Gaudin norm (see Appendix \ref{ap:norm} for an explicit expression), we obtain
\beq\label{eq:beforefinalSL2tree}
\left.\frac{\langle \mathcal{D}_1\mathcal{D}_2\mathcal{O}_{\bf u}\rangle}{\langle \mathcal{D}_1\mathcal{D}_2\rangle \sqrt{\mathfrak{n}_{\mathcal{O}_{\bf u}}}}\right|_{\tt sym}=-\frac{\kappa^{L}}{a_1^{S}}\frac{\left[i^{L}+(-i)^{L}\right]}{\sqrt{L}}\sqrt{\left(\prod_{1\leq s\leq \frac{S}{2}}\frac{u_s^2+\frac{1}{4}}{u_s^2}\right)\frac{\det G_{+}^{\rm SL(2)}}{\det G_{-}^{\rm SL(2)}}}\period
\eeq
From this, we can read off the structure constant as follows:
\beq\label{eq:finaldOsl2tree}
\mathfrak{D}_{\mathcal{O}_{\bf u}}=-\frac{i^{J}+(-i)^{J}}{2^{S}\sqrt{L}}\sqrt{\left(\prod_{1\leq s\leq \frac{S}{2}}\frac{u_s^2+\frac{1}{4}}{u_s^2}\right)\frac{\det G_{+}^{\rm SL(2)}}{\det G_{-}^{\rm SL(2)}}}\period
\eeq

To our knowledge, the matrix product state \eqref{eq:MsymSL2nice} and the determinant formula for its overlap \eqref{eq:symoverlapSL2} never appeared  before in the literature. Here we checked the determinant formula numerically, but it would be desirable to derive it analytically\fn{Another interesting question is whether the matrix product state \eqref{eq:MsymSL2nice} is annihilated by odd higher conserved charges as is the case with the SU(2) matrix product state \cite{deLeeuw:2015hxa}. We expect the answer to be yes. It would be nice to work that out explicitly.}. It would also be interesting to consider the application of our formula to condensed matter or statistical physics. Recently it was pointed out that the SL(2) spin chain is related to certain integrable stochastic processes \cite{Frassek:2019vjt,Frassek:2019isa}. It might be possible to use our matrix product state (or its appropriate generalization) as an initial probability distribution and study its relaxation to a nonequilibrium steady state using integrability.
\subsubsection{PCGG and generalized N\'{e}el state}
We now apply the PCGG method to the SL(2) sector. The advantage of the PCGG is that the computation in any sector boils down to the (non-local) two-point function $\left<{\rm tr}\left[\left(\mathfrak{Z}(a_1)\mathfrak{Z}(a_2)\right)^{L/2}\right] \,\, \mathcal{O}_{\bf u}(0)\right>$. Using the Wick contraction rules \eqref{eq:wickSL(2)}, one can express this two-point function in terms of an overlap in the SL(2) spin chain,
\beq
\left<{\rm tr}\left[\left(\mathfrak{Z}(a_1)\mathfrak{Z}(a_2)\right)^{L/2}\right] \,\, \mathcal{O}_{\bf u}(0)\right>=\frac{L(2g^2\kappa^2)^{L}}{2(i a_1)^{S}}\times \langle \text{N\'{e}el}_{a_1/a_2}|{\bf u}\rangle\period
\eeq
Here $\langle \text{N\'{e}el}_{x}|$ is the SL(2) version of the weighted N\'{e}el state defined by
\beq
\langle \text{N\'{e}el}_{x}|\equiv \sum_{\psi:\text{ all states}}\left(x^{N^{\psi}_{\rm even}}+x^{N^{\psi}_{\rm odd}}\right)\langle \psi|\comma
\eeq
with $N^{\psi}_{\text{even, odd}}$ being the numbers of magnons at even/odd sites in the state $\langle \psi |$. What is interesting is that the structure of the weighted SL(2) N\'{e}el state is exactly the same as the one for the SU(2) sector. Furthermore, it turns out that the relation \eqref{eq:weightedneelnicerelation} holds also for the SL(2) N\'{e}el state,
\beq\label{eq:neelSL2relation}
\langle \text{N\'{e}el}_{x}|{\bf u}\rangle=(1-x)^{M}\langle \text{N\'{e}el}_{0}|{\bf u}\rangle\period
\eeq
The state $\langle \text{N\'{e}el}_{0}|$ turned out to be a direct analogue of the generalized N\'{e}el state in the SU(2) sector and admits the following representation,
\beq
\langle \text{N\'{e}el}_0 |=\sum_{\substack{\text{all states with}\\|n_i-n_j|:\text{even}}}  \left< \circ \cdots \circ\, \underset{n_1}{\bullet}\,\circ\cdots \circ\,\underset{n_2}{\bullet}\,\circ\cdots\right|\comma
\eeq
where $\bullet$ denotes a site occupied by magnons while $\circ$ denotes an empty site.
As a result of numerical computations, we found that the overlap $\langle \text{N\'{e}el}_{0}|{\bf u}\rangle$ is given by
\beq\label{eq:neelSL2overlap}
\langle \text{N\'{e}el}_{0}|{\bf u}\rangle=2\left(\frac{i}{2}\right)^{S}\left(\prod_{s=1}^{M/2}\frac{u_s+\frac{i}{2}}{ u_s\,\del_u p(u_s)}\right)\det G^{\rm SL(2)}_{+}\comma
\eeq
which again resembles the result for the SU(2) sector. Let us emphasize that the results \eqref{eq:neelSL2relation} and \eqref{eq:neelSL2overlap} were obtained by numerical computations and it would be interesting to prove them analytically.

Putting together these ingredients, we obtain the following ratio for the correlator,
\beq
\frac{\langle \mathcal{D}_1\mathcal{D}_2\mathcal{O}_{\bf u}\rangle}{\langle \mathcal{D}_1\mathcal{D}_2\rangle \sqrt{\mathfrak{n}_{\mathcal{O}_{\bf u}}}}=-\kappa^{L}\left(\frac{a_1-a_2}{2a_1 a_2}\right)^{S}\frac{\left[i^{L}+(-i)^{L}\right]}{\sqrt{L}}\sqrt{\left(\prod_{1\leq s\leq \frac{S}{2}}\frac{u_s^2+\frac{1}{4}}{u_s^2}\right)\frac{\det G_{+}^{\rm SL(2)}}{\det G_{-}^{\rm SL(2)}}}\comma\nn
\eeq
which is in agreement with the result of the semi-classical approach \eqref{eq:beforefinalSL2tree}.
\subsection{SO(6) sector\label{subsec:so6tree}}
We now extend the analysis to operators in the SO(6) sector by allowing arbitrary scalars to appear inside the trace:
\beq
\mathcal{O}={\rm tr}\left(Z\red{YX\bar{X}}\bar{Z}\red{\bar{Y}}\cdots\right)+\cdots\period
\eeq
\subsubsection{Bethe state}
\begin{figure}[t]
\centering
\includegraphics[clip,height=3cm]{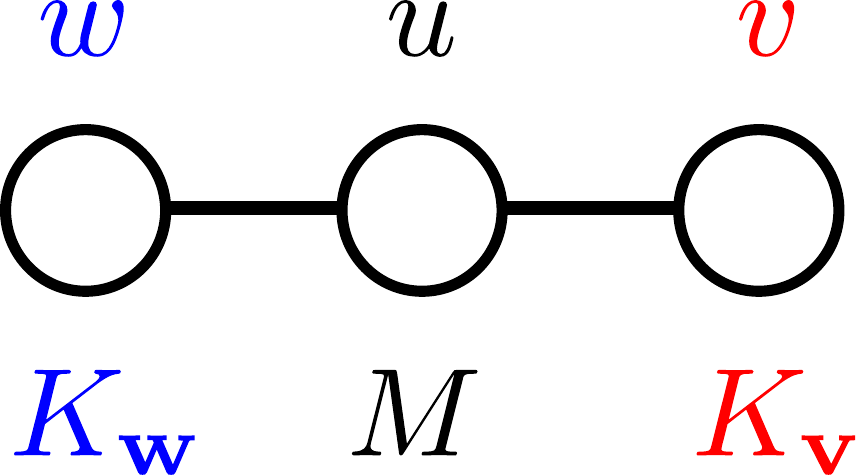}
\caption{Dynkin diagram of SO$(6)$ and rapidities. The rapidities associated with the middle node describe the actual propagation of magnons on a spin chain while the rapidities associated with the wings correspond to the spin-wave excitations which describe the index structure of magnon wave functions.}
\label{fig:fig14}
\end{figure}

The generalization to the SO(6) sector is not just a technical complication: It allows us to explore qualitatively different features of our problem. Unlike the SU(2) and SL(2) sectors studied above, magnons in the SO(6) sector carry indices and the states are described by the {\it nested Bethe ansatz}. Correspondingly, there are {\it three} sets of rapidities, each of which is associated with a node of the SO(6) Dynkin diagram (see figure \ref{fig:fig14}). The one associated with the middle node (to be denoted by ${\bf u}$) corresponds to the momenta of excitations in the spin chain while the other two (denoted by $\red{\bf v}$ and $\blue{\bf w}$ respectively) correspond to the momenta of ``spin waves'', which move on top of magnon excitations and describe the index structure.

More concretely the relation between scalar fields and patterns of rapidities is given as follows:
\beq
\begin{aligned}
 X\,\, \leftrightarrow\,\, u\comma\qquad \bar{X}\,\, \leftrightarrow\,\, \overset{\red{v},\blue{w}}{u}\comma\qquad Y\,\, \leftrightarrow\,\, \overset{\red{v}}{u}\comma\qquad \bar{Y}\,\, \leftrightarrow\,\, \overset{\blue{w}}{u}\comma\qquad \bar{Z}\,\, \leftrightarrow\,\,\overset{\red{v},\blue{w}}{u_1, u_2}\period
\end{aligned}
\eeq
This relation should be understood as follows: A magnon which only carries one middle-node rapidity $u$ corresponds to the $X$-field while, if it carries $\red{v}$ and $\blue{w}$ in addition to $u$, it corresponds to the $\bar{X}$ field. Two different magnons with the middle-node rapidities $u_1$ and $u_2$ can be at the same site if and only if $\red{v}$ and $\blue{w}$ are also at the same site. In case this happens, it describes the $\bar{Z}$ field. Of course, if there are no magnons at all, that would describe the vacuum, namely the $Z$ field.

Rather surprisingly, the coordinate Bethe ansatz for the SO(6) spin chain relevant for $\mathcal{N}=4$ SYM was not developed until recently  \cite{Basso:2017khq}\fn{In \cite{Basso:2017khq}, the algebraic Bethe ansatz and the relation to the vertex model were also worked out.}. Since the explicit structure of the Bethe state is rather complicated, we will not display it here, referring the interested readers to Appendix E of \cite{Basso:2017khq}. Instead, here we show the Bethe equation which now consists of three sets of equations:
\beq\label{eq:so6betheeq}
\begin{aligned}
1&=e^{i\phi_{u_j}}\equiv\left(\frac{u_j+i/2}{u_j-i/2}\right)^{L}\prod_{k\neq j}^{M}\frac{u_j-u_k-i}{u_j-u_k+i}\prod_{k=1}^{\red{K_{\bf v}}}\frac{u_j-\red{v_k}+i/2}{u_j-\red{v_k}-i/2}\prod_{k=1}^{\blue{K_{\bf w}}}\frac{u_j-\blue{w_k}+i/2}{u_j-\blue{w_k}-i/2}\comma\\
1&=e^{i\phi_{\red{v_j}}}\equiv\prod_{k=1}^{M}\frac{\red{v_j}-u_k+i/2}{\red{v_j}-u_k-i/2}\prod_{k\neq j}^{\red{K_{\bf v}}}\frac{\red{v_j}-\red{v_k}-i}{\red{v_j}-\red{v_k}+i}\comma\\
1&=e^{i\phi_{\blue{w_j}}}\equiv\prod_{k=1}^{M}\frac{\blue{w_j}-u_k+i/2}{\blue{w_j}-u_k-i/2}\prod_{k\neq j}^{\blue{K_{\bf w}}}\frac{\blue{w_j}-\blue{w_k}-i}{\blue{w_j}-\blue{w_k}+i}\comma
\end{aligned}
\eeq
Here $M$, $\red{K_{\bf v}}$ and $\blue{K_{\bf w}}$ are the numbers of ${\bf u}$, $\red{\bf v}$ and $\blue{\bf w}$ respectively, and we introduced the symbols for the phase factors $\phi_{u,\red{v},\blue{w}}$ for later convenience.

In what follows, we denote the Bethe state by $|{\bf u},\red{\bf v}, \blue{\bf w}\rangle$ and the corresponding operator by  $\mathcal{O}_{{\bf u},\red{\bf v}, \blue{\bf w}}$ in order to manifest the dependence on the rapidities.
\subsubsection{Matrix trace, matrix product state and generalized N\'{e}el state}
We now apply the approaches in sections \ref{sec:effective} and \ref{sec:PCGG} to the SO(6) sector. Here we focus on the derivation of the matrix product state and the generalized N\'{e}el state, postponing writing the final results for overlaps and the structure constant.

Let us first write down the matrix-trace representation. To derive it, we use the following Wick contractions in addition to the ones in \eqref{eq:contractionrulewicksu2}
\beq\label{eq:so6contractionrule}
\bcontraction{}{\mathfrak{Z}}{(a)\quad }{\bar{Z}}
\bcontraction{\mathfrak{Z}(a)\quad \bar{Z}=1/a^2\comma\qquad }{\mathfrak{Z}}{(a)\quad }{\bar{Y}}
\mathfrak{Z}(a)\quad \bar{Z}=1/a^2\comma\qquad \mathfrak{Z}(a)\quad \bar{Y}=\kappa /a \qquad \text{others}=0\period
\eeq
As a result we get the substitution rule
\beq\label{eq:substitutionruleSO(6)}
\begin{aligned}
&Z \mapsto M_{Z}\comma\quad X\mapsto 0\comma\quad \bar{X} \mapsto 0\comma\quad Y\mapsto M_{Y}\comma\quad \bar{Y}\mapsto M_{\bar{Y}}  \quad \bar{Z}\mapsto M_{\bar{Z}}\comma \\
&\mathcal{O}={\rm tr}\left(ZY\bar{Y}\bar{Z}\cdots\right) \quad \mapsto \quad \langle\mathcal{O} \rangle_{\chi}=-{\rm Tr}_{m=2}\left[M_{Z}M_{Y}M_{\bar{Y}}M_{\bar{Z}}\cdots\right]\comma
\end{aligned}
\eeq
with
\beq
\begin{aligned}
&M_{Z}=\kappa^2  m_{\rho^{\ast}}\comma\quad &&M_{\bar{Z}}={\rm diag}\left(\frac{1}{(\kappa a_1)^2},\frac{1}{(\kappa a_2)^{2}}\right)\cdot M_{Z}\comma\\
& M_{Y}=-{\rm diag}\left(\frac{1}{\kappa a_1},\frac{1}{\kappa a_2}\right)\cdot M_{Z}\comma\quad &&M_{\bar{Y}}={\rm diag}\left(\frac{1}{\kappa a_1},\frac{1}{\kappa a_2}\right)\cdot M_{Z}\period
\end{aligned}
\eeq
The substitution rule \eqref{eq:substitutionruleSO(6)} immediately implies that the result for operators with $X$ or $\bar{X}$ is always zero.

Now, to write down a simple matrix product state, it is again convenient to go to a basis in which $M_Z$ is diagonal. In that basis, $\bar{M}_{Z}$ and $\bar{M}_{Y}$ are given by \eqref{eq:diagonalizedMzMy} while the other two are
\beq
\bar{M}_{\bar{Y}}=-\bar{M}_{Y}\comma\qquad \bar{M}_{\bar{Z}}=\frac{ig}{\sqrt{2}\kappa}\pmatrix{cc}{a_1^{-2}+a_2^{-2}&i\left(a_1^{-2}-a_2^{-2}\right)\\i\left(a_1^{-2}-a_2^{-2}\right)&-\left(a_1^{-2}+a_2^{-2}\right)}\period
\eeq
Using these, we get the following matrix product state\fn{Here we are using the convention in which overlaps read $\langle \bar{Z} |Z\rangle=\langle \bar{X}|X\rangle=\langle \bar{Y}|Y\rangle=1$ and $\langle Z|Z\rangle=0$.},
\beq
\begin{aligned}
\langle\bar{M}|=-{\rm Tr}_{m=2}\left[\prod_{s=1}^{L}\Big(\langle \bar{Z}|_{s}\otimes \bar{M}_{Z}+\langle Z|_{s}\otimes \bar{M}_{\bar{Z}}+\langle \bar{Y}|_{s}\otimes \bar{M}_{Y}+\langle Y|_{s}\otimes \bar{M}_{\bar{Y}}\Big)\right]\comma
 \end{aligned}
\eeq
whose overlap with a Bethe state $\langle \bar{M}|{\bf u},\red{\bf v},\blue{\bf w}\rangle$ gives $\langle \mathcal{O}_{{\bf u},\red{\bf v},\blue{\bf w}}\rangle_{\chi}$.
This further simplifies in the symmetric configuration $a_1=-a_2$ as follows:
\beq
\left.\langle \bar{M}|\right|_{{\tt sym}}=-(2\sqrt{2}ig\kappa)^{L}\,{\rm Tr}_{m=2}\left[\prod_{s=1}^{L}\Big(\left(\langle \bar{Z}|_{s}+z^2\langle Z|\right)\otimes t_1-i z\left(\langle \bar{Y}|_{s}-\langle Y|_{s}\right)\otimes t_2\Big)\right]\comma
\eeq
with $t_{1,2}=\sigma_{3,1}/2$ and $z=1/(\kappa a_1)$. Note that this matrix product state is different\fn{A general framework to analyze the integrable matrix product states in spin chains was proposed recently in \cite{Pozsgay:IntB}, and it would be interesting to see the relation to the matrix product state obtained here.} from the one that appeared in \cite{deLeeuw:2018mkd} in the analysis of the SO(6) sector.

Alternatively, one can use the PCGG approach to compute the three-point function. The PCGG approach always leads to the same non-local two-point function \eqref{eq:tocomputesecgneel}. Using the Wick contraction rules \eqref{eq:contractionrulewicksu2} and \eqref{eq:so6contractionrule}, one can recast this two-point function as the following overlap of the spin chain
\beq\label{eq:SO6Neeloverlap}
\langle{\rm tr}\left[\left(\mathfrak{Z}(a_1)\mathfrak{Z}(a_2)\right)^{L/2}\right] \,\, \mathcal{O}_{{\bf u},\red{\bf v},\blue{\bf w}}(0)\rangle=\frac{L(2g^2\kappa^2)^{L}}{2(\kappa a_1)^{M}}\times\langle \text{N\'{e}el}_{a_1/a_2}|{\bf u},\red{\bf v},\blue{\bf w}\rangle\comma
\eeq
with
\beq
\langle \text{N\'{e}el}_x|=\langle \bar{\mathfrak{Z}}_{1}\bar{\mathfrak{Z}}_{x}\bar{\mathfrak{Z}}_{1}\bar{\mathfrak{Z}}_{x}\cdots| +\langle \bar{\mathfrak{Z}}_{x}\bar{\mathfrak{Z}}_{1}\bar{\mathfrak{Z}}_{x}\bar{\mathfrak{Z}}_{1}\cdots|\comma
\qquad\qquad
 \bar{\mathfrak{Z}}_{x}\equiv x^2 Z+x(Y-\bar{Y})+\bar{Z}\period
\eeq
As with the SU(2) and SL(2) sectors, this overlap turns out to satisfy the identity\fn{As with the SL(2) sector, this identity was checked only through the numerical computation, and it would be better to prove it analytically.}
\beq
\langle\text{N\'{e}el}_{x}|{\bf u},\red{\bf v},\blue{\bf w}\rangle=(1-x)^{M}\langle\text{N\'{e}el}_{0}|{\bf u},\red{\bf v},\blue{\bf w}\rangle\period
\eeq
The state $\langle\text{N\'{e}el}_{0}|$ is a sum of two states each of which contains magnons only at even or odd sites. In this sense, this is a natural generalization of the N\'{e}el state to the SO(6) sector, and it would be interesting to study its property in more details.
\subsubsection{Result for overlaps and structure constant}
As a result of numerical computations, we found that the relevant overlaps again obey selection rules:
\begin{enumerate}
\item Both $L$ and $M$ must be even.
\item The rapidities of the middle node must be parity-symmetric, ${\bf u}=\{u_1,-u_1,u_2,-u_2,\cdots\}$.
\item The rapidities of the left node are $(-1)$ times the rapidities of the right node: $\blue{\bf w}=-\red{\bf v}$. This also implies $\red{K_{\bf v}}=\blue{K_{\bf w}}$.
\end{enumerate}
A point worth noting is that this selection rules are {\it similar but different from} the ones found for the defect one-point functions in \cite{deLeeuw:2018mkd}, where the overlaps in the SO(6) sector are found to be nonzero only when all the rapidity sets are separately parity-symmetric. The difference comes from the difference of the underlying symmetry: In our setup, the SU(2)$\times$SU(2) symmetry\fn{This is a subgroup of SO(6) which is associated with the left and the right Dynkin nodes.} governing the index structure of magnons gets broken to the diagonal SU$(2)$ while in the defect one-point function it is broken down to U$(1)$. In section \ref{sec:asymptotic}, we will see that the symmetry structure and its compatibility indeed imply the selection rule $\blue{\bf w}=-\red{\bf v}$. It would be desirable to develop a similar understanding for the ones found in \cite{deLeeuw:2018mkd}.

To write down the results, it is convenient to introduce the generalization of $G_{\pm}$ matrices which are $(\red{K_{\bf v}}+\frac{M}{2})\times(\red{K_{\bf v}}+\frac{M}{2})$ matrices\fn{Note that we can further simplify the expression since the rapidities on the two different wings are not coupled, namely $\del_{\red{v}}\phi_{\blue{w}}=\del_{\blue{v}}\phi_{\red{w}}=0$.},
\beq\label{eq:GpmSO6}
G_{\pm}^{\rm SO(6)}\equiv\left. \pmatrix{cc}{\del_{\red{\hat{v}_i}}\left(\phi_{\red{\hat{v}_j}}\pm\phi_{\blue{\hat{w}_j}}\right)&\del_{\red{\hat{v}_i}}\left(\phi_{\hat{u}_{2j-1}}\pm\phi_{\hat{u}_{2j}}\right)\\\del_{\hat{u}_{2i-1}}\left(\phi_{\red{\hat{v}_j}}\pm\phi_{\blue{\hat{w}_j}}\right)&\del_{\hat{u}_{2i-1}}\left(\phi_{\hat{u}_{2j-1}}\pm \phi_{\hat{u}_{2j}}\right)}\right|_{\substack{\hat{u}_{2j-1}=-\hat{u}_{2j}=u_j\\\red{\hat{v}_j}=-\blue{\hat{w}_j}=\red{v_j}}}\comma
\eeq
with $\phi_{u,\red{v},\blue{w}}$ being the phase factors associated with the Bethe equations \eqref{eq:so6betheeq}. As indicated, in this formula one has to compute derivatives assuming all the rapidities to be independent. Only after taking derivatives do we impose the selection rule, $\hat{u}_{2j-1}=-\hat{u}_{2j}=u_j$ etc.

We wish to emphasize that the structures of these matrices are \emph{different} from the ones in the defect one-point functions \cite{deLeeuw:2018mkd}. This makes clear that the ratio of determinants---which is sometimes regarded as universal and insensitive to the details of the boundary states---{\it does depend} on the underlying symmetry of the problem.

As a result of numerical computations, we found that the overlaps are given by
\beq
\begin{aligned}
\left.\langle \bar{M}|{\bf u},\red{\bf v},\blue{\bf w}\rangle\right|_{\tt sym}&=-\frac{(\sqrt{2}g)^{L}\kappa^{L-M}}{a_1^{M}}\left[i^{L-M}+(-i)^{L-M}\right]\left(\prod_{s=1}^{M/2}\frac{u_s-\frac{i}{2}}{ u_s} \right)
\det G_{+}^{\rm SO(6)}\comma\\
\langle \text{N\'{e}el}_{0}|{\bf u},\red{\bf v},\blue{\bf w}\rangle&=2\left(\frac{i}{2}\right)^{M}\left(\prod_{s=1}^{M/2}\frac{u_s-\frac{i}{2}}{ u_s}\right)\det G^{\rm SO(6)}_{+}\period
\end{aligned}
\eeq
Multiplying the necessary factors and dividing by the Gaudin norm\fn{More precisely, we divide by the normalization of the two-point function which is
\beq
\mathfrak{n}_{\mathcal{O}_{{\bf u},\red{\bf v},\blue{\bf w}}}=L (2g^2)^{L}\langle {\bf u},\red{\bf v},\blue{\bf w}|{\bf u},\red{\bf v},\blue{\bf w}\rangle
\eeq} (see Appendix \ref{ap:norm}), we obtain the following expression for the ratio of correlators
\beq
\frac{\langle \mathcal{D}_1\mathcal{D}_2\mathcal{O}_{{\bf u},\red{\bf v},\blue{\bf w}}\rangle}{\langle \mathcal{D}_1\mathcal{D}_2\rangle \sqrt{\mathfrak{n}_{\mathcal{O}_{{\bf u},\red{\bf v},\blue{\bf w}}}}}=-\kappa^{L-M}\left(\frac{a_1-a_2}{2a_1 a_2}\right)^{M}\frac{\left[i^{L-M}+(-i)^{L-M}\right]}{\sqrt{L}}\sqrt{\left(\prod_{1\leq s\leq \frac{M}{2}}\frac{u_s^2+\frac{1}{4}}{u_s^2}\right)\frac{\det G_{+}^{\rm SO(6)}}{\det G_{-}^{\rm SO(6)}}}\comma\nn
\eeq
which reproduces the correct spacetime dependence \eqref{eq:ratio3ptand2pt}. From this, the structure constant $\mathfrak{D}_{\mathcal{O}_{{\bf u},\red{\bf v},\blue{\bf w}}}$ can be read off as follows:
\beq\label{eq:finaldOso6tree}
\mathfrak{D}_{\mathcal{O}_{{\bf u},\red{\bf v},\blue{\bf w}}}=-\frac{i^{J}+(-i)^{J}}{2^{M}\sqrt{L}}\sqrt{\left(\prod_{1\leq s\leq \frac{M}{2}}\frac{u_s^2+\frac{1}{4}}{u_s^2}\right)\frac{\det G_{+}^{\rm SO(6)}}{\det G_{-}^{\rm SO(6)}}}\period
\eeq
Remarkably, the result takes exactly the same form as the results in the SU(2) and SL(2) sectors! This is in marked contrast to the results for the defect one-point functions \cite{deLeeuw:2018mkd} and we will explain the origin of this simplicity in section \ref{sec:TBA}.

In section \ref{sec:asymptotic}, we will propose a generalization of this result to finite coupling and to all sectors, and perform extensive tests of the proposal in section \ref{sec:check}.
\section{Relation to Integrable Boundary\label{sec:integrability}}
We now explain that the two important features observed in the previous section---the selection rule of the rapidities and the determinant structure of the final result---can be understood naturally from a point of view of integrable boundaries of 2d integrable field theories.
\subsection{From overlap to boundary scattering\label{subsec:overlaptoreflection}}
\begin{figure}[t]
\centering
\includegraphics[clip,height=5cm]{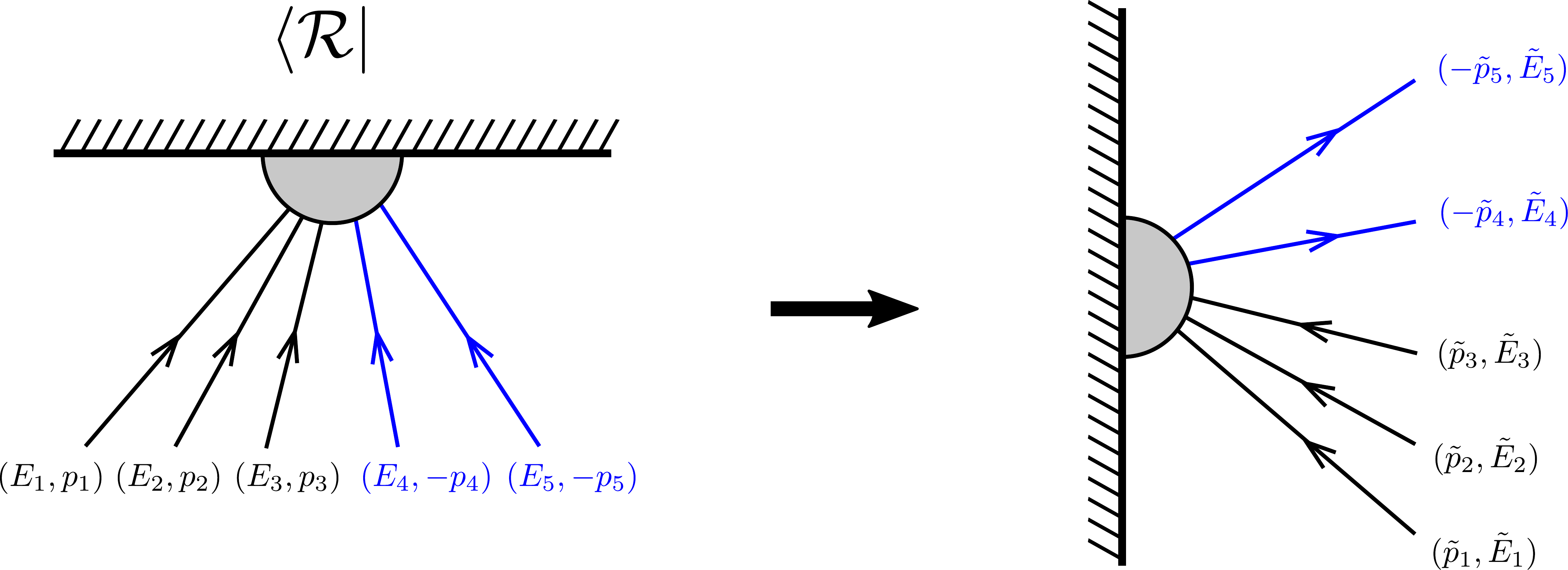}
\caption{The relation between the form factor and the reflection matrix. In order to relate the two, we perform the double-Wick rotation, and appropriate crossing transformations which map particles with negative energy to the out state. As a result, we obtain a process in which particles get scattered off a boundary.}
\label{fig:fig15}
\end{figure}

To understand the connection between the three-point function and the integrable boundaries, it is useful to start our discussion with something slightly more general; an overlap between an asymptotic state and some translationally invariant state $|\mathcal{R}\rangle$ in a generic integrable field theories in the infinite volume,
\beq\label{eq:formfactorR}
F_{\mathcal{R}}(p_1,p_2,\cdots)=\langle \mathcal{R}|(E_1,p_1),(E_2,p_2),\cdots \rangle \period
\eeq
Here the set $(E_k,p_k)$ denotes the energy and the momentum of the $k$-th particle.
This can be expressed alternatively as a {\it form factor}\fn{This point of view will be expanded and utilized in section \ref{subsec:formVSreflection}.} of a  state-creating operator $\mathbb{R}$, $\langle \mathcal{R}|=\langle \Omega |\mathbb{R}$, as
\beq
F_{\mathcal{R}}(p_1,p_2,\cdots)=\langle \Omega |\mathbb{R}|(E_1,p_1),\cdots\rangle\period
\eeq
Since the state is translationally invariant, the asymptotic state must have a zero total momentum in order for the overlap \eqref{eq:formfactorR} to be nonzero. Then the particles can be grouped into the ones with positive momenta and the ones with negative momenta as follows\fn{For simplicity we do not consider particles with zero momentum, but this does not really affect our discussion.}:
\begin{align}
&|\underbrace{(E_1,p_1),\ldots,(E_m,p_m)}_{\rm positive},\underbrace{(E_{m+1},-p_{m+1}),\cdots,(E_{m+n},-p_{m+n})}_{\rm negative}\rangle\comma\\
&\sum_{k=1}^{m}p_k=\sum_{k=m+1}^{m+n}p_k\period\label{eq:momentumconserv}
\end{align}
Note that, although momenta can take both signs, the energies of the particles are all positive, namely $E_k>0$ for all $k$. For a pictorial explanation, see figure \ref{fig:fig15}.

Now, to relate this to the boundary scattering, we rotate the whole picture by $90$ degrees; namely we perform the Wick rotation and swap the roles of space and time. After doing so, the energy and the momenta of the original theory are mapped to the energy $\tilde{E}$ and the momenta $\tilde{p}$ of the rotated theory in the following way\fn{Here we are following the standard but slightly misleading convention in which the mirror theory is defined by a {\it combination} of the double Wick rotation and the parity transformation. If we just perform the double-Wick rotation, we will get $E\to -i\tilde{p}$ and $p\to i\tilde{E}$. In relativistic field theories, these two choices correspond to two different ways to perform the mirror transformation, $\theta \to \theta +i\pi/2$ and $\theta\to i\pi/2-\theta$. The former choice is more convenient but it is the latter choice which corresponds to the honest Wick rotation. For more details, see for instance \cite{vanTongeren:2016hhc}.}:
\beq\label{eq:doublewickrotation}
E\to i\tilde{p} \comma\qquad p\to i\tilde{E}\comma
\eeq
In what follows, we call the rotated theory the {\it mirror theory} while we call the original theory the {\it physical theory}, following the convention in AdS/CFT integrability.

The Wick rotated overlap has the structure,
\beq
\tilde{F}_{\mathcal{R}}(\tilde{p}_1,\tilde{p}_2,\cdots)=\langle \Omega |\mathbb{R}|\underbrace{(\tilde{p}_1,\tilde{E}_1),\cdots,(\tilde{p}_m,\tilde{E}_m)}_{\rm positive}, \underbrace{(\tilde{p}_{m+1},-\tilde{E}_{m+1}),\cdots, (\tilde{p}_{m+n},-\tilde{E}_{m+n})}_{\rm negative}\rangle\comma
\eeq
with $\tilde{p_k}\equiv -i E_k$ and $\tilde{E}_k\equiv -ip_k$. We now see that the energies of the particles can take both signs\fn{Precisely speaking, here both the energies and momenta are pure imaginary since we are interpreting physical excitations from the point of view of the mirror theory. Nevertheless it still makes sense to talk about whether all the particles come with the same sign or not.} while the momenta of the particles have the same sign.
To recast it into a more physically-looking process, we perform the crossing transformation to the ``negative-energy'' particles and bring them to the bra:
\beq
\tilde{F}_{\mathcal{R}}(\tilde{p}_1,\tilde{p}_2,\cdots)=\langle (-\tilde{p}_{m+n},\tilde{E}_{m+n}),\cdots, (-\tilde{p}_{m+1},\tilde{E}_{m+1}) |\mathbb{R}|(\tilde{p}_1,\tilde{E}_1),\cdots,(\tilde{p}_m,\tilde{E}_m)\rangle\period
\eeq
The resulting expression can be interpreted as a transition process in which the signs of the momenta get all flipped, whereas the energy is conserved $\sum_{k=1}^{m}\tilde{E}_k=\sum_{k=m+1}^{m+n}\tilde{E}_k$ owing to \eqref{eq:momentumconserv}. This is precisely what we expect for a scattering off a boundary and it is also evident from figure \ref{fig:fig15}. This shows that the overlap (or the form factor) \eqref{eq:formfactorR} in the original theory can be understood as a reflection amplitude in the mirror theory.

The discussion so far applies to general overlaps which do not necessarily exhibit selection rules.  In the presence of the selection rules, the asymptotic states need to be parity-symmetric; namely, for each particle, there is a corresponding particle with the opposite momentum. Starting from such a special kinematics and applying the argument above, we arrive at the following reflection process:
\beq
\tilde{F}_{\mathcal{R}}(\tilde{p}_1,\tilde{p}_2,\cdots)=\langle (-\tilde{p}_{1},\tilde{E}_{1}),\cdots, (-\tilde{p}_{m},\tilde{E}_{m}) |\mathbb{R}|(\tilde{p}_1,\tilde{E}_1),\cdots,(\tilde{p}_m,\tilde{E}_m)\rangle\period
\eeq
Notable features of this process are that no particle is created or annihilated in the process, and the momenta of individual particles get flipped separately. Such features are reminiscent of the (bulk) S-matrix in integrable quantum field theories, and much like in that case, they imply the existence of infinitely many conservation laws. More specifically, they imply the existence of infinitely many ``even'' charges\fn{This is basically the mirror-theory version of the statement that the state $|\mathcal{R}\rangle$ is annihilated by infinitely many ``odd'' charges, which was proven at weak coupling for the defect one-point functions.}:
\beq\label{eq:conservationlaws}
\sum_{k:\text{ incoming}}(\tilde{p}_k)^{n} =\sum_{k:\text{ outcoming}}(\tilde{p}_k)^{n} \qquad n:\text{ even}\period
\eeq
The boundary conditions which admit such conservation laws are known as {\it integrable boundary conditions} and discussed in detail in \cite{Ghoshal:1993tm} (see also \cite{Piroli:2017sei,Pozsgay:IntB} for integrable boundaries in lattice models).
We therefore conclude that

\noindent {\it The existence of the selection rules\fn{The argument presented in this subsection does not quite explain the selection rules for the Bethe roots at the nested level observed in the SO(6) sector. This comes from detailed symmetry structures of the boundary state which we discuss in section \ref{sec:bootstrap}.} in the overlap is a strong indication that the problem is related to an integrable boundary condition in the mirror theory (or equivalently to an overlap with a integrable boundary state\fn{The relation between integrable initial states for quantum quenches and integrable boundaries in lattice models was discussed in \cite{Piroli:2017sei}.}).}

\noindent In the subsequent sections, we will assume that this is the case and solve our problem at finite coupling using integrability.

\begin{figure}[t]
\centering
\includegraphics[clip,height=9cm]{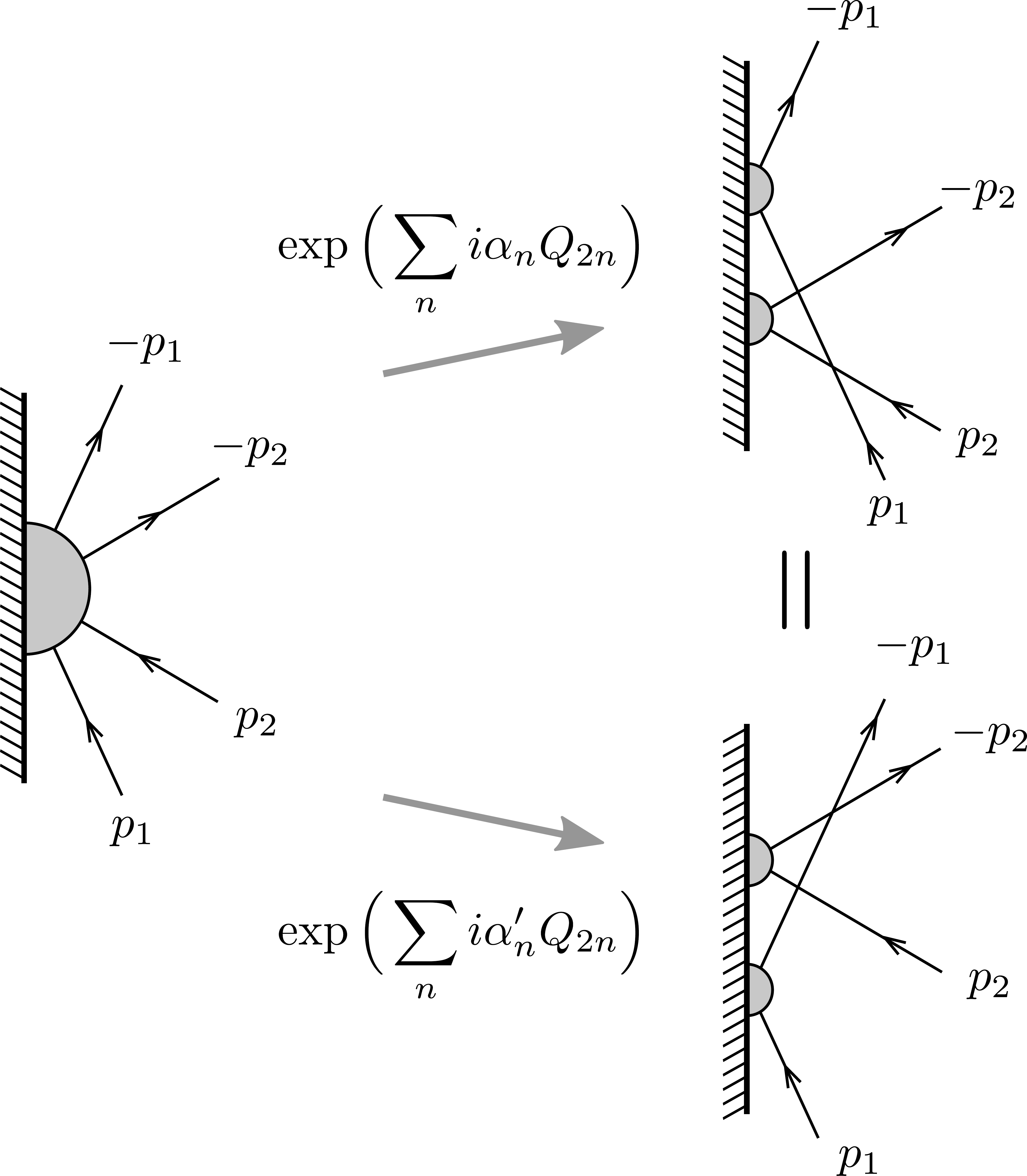}
\caption{Conserved charges and the factorized reflection. By acting the higher conserved charges to the reflection process, one can factorize it into one-particle reflection processes. There are two different ways of performing the factorization and the equality between the two is the boundary Yang-Baxter equation.}
\label{fig:fig16}
\end{figure}

Before concluding this subsection, let us explain one important dynamical implication of the conservation laws \eqref{eq:conservationlaws}. This can be seen most easily by generalizing the argument by Shankar and Witten on the bulk scattering \cite{Shankar:1977cm}. Consider a $2\to 2$ reflection process in which two localized wave packets
\beq
\langle x|\psi_i\rangle=\int dp\,\, e^{-a^2 (p-p_i)^2}e^{ip (x-x_i)} \qquad (i=1,2)\comma
\eeq
 scatter off an integrable boundary and evolve into two other localized wave packets labelled by $i=3,4$. If we act a higher conserved charge with spin $2n$ to such a process, each plane wave gets multiplied by a factor $e^{i \alpha p^{2n}}$, and as a result the center of the wave packet gets shifted as
\beq
\delta x_i =\left.\alpha \del_p p^{2n}\right|_{p=p_i}=2\alpha n p_i^{2n-1}\period
\eeq
Since the shift depends on the momenta of individual particles, this will move particles relative to one another. However, a crucial difference from the bulk scattering is that the shift is always given by an odd power of the momentum. This means that the shifts for the particles with momentum $p$ and $-p$ are always correlated, and one cannot move them far apart. Taking this into account and applying appropriate linear combinations of higher spin conserved charges to the reflection process, one can factorize the $2\to 2$ reflection process into a product of two successive $1\to 1$ reflection processes. Importantly, there are two different ways of factorization as shown in figure \ref{fig:fig16}. Clearly their spacetime interpretations are different, but nevertheless they should give the same reflection amplitude since the two processes are related by the action of conserved charges. This gives a nontrivial equality for the reflection amplitudes and is called the {\it boundary Yang-Baxter equation}.
The boundary Yang-Baxter equation will be used in section \ref{subsec:matrixpart} to ``bootstrap'' the boundary reflection amplitude of our problem.

\subsection{Thermodynamic Bethe ansatz and $g$-function\label{subsec:TBAgeneral}}
So far we have seen that the selection rules can be naturally interpreted as an integrable boundary condition for the mirror theory. We now show how such a physical picture also explains the determinant structure of the overlap. The discussion here is mostly qualitative and focuses on a simple toy example. A more rigorous discussion can be found in Appendix \ref{ap:gfunctionderivation}, and the application to our problem is explained in section \ref{sec:TBA}.
\subsubsection{Cylinder partition function and $g$-function}
\begin{figure}[t]
\centering
\includegraphics[clip,height=3.5cm]{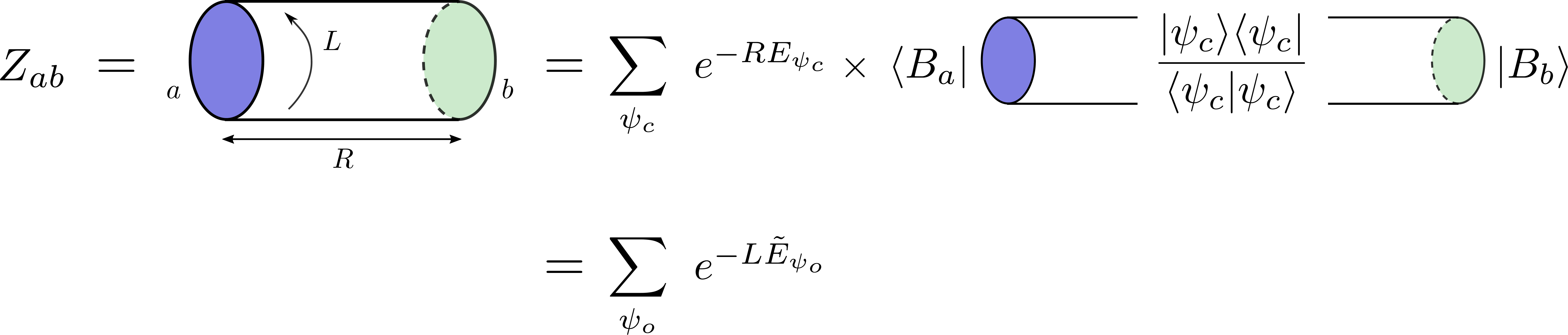}
\caption{The cylinder partition function and the $g$-function. The cylinder partition function can be expanded in two different channels: In the closed string channel, it is given in terms of overlaps between the boundary states and the closed string states and the propagation factor $e^{-R E_{\psi_c}}$. In the open string channel, it is a standard thermal partition function.}
\label{fig:fig17}
\end{figure}

For this purpose, we consider the partition function of a cylinder with a circumference $L$ and a length $R$, where the two ends of the cylinder are contracted with integrable boundary states $\langle B_a|$ and $|B_b\rangle$. See figure \ref{fig:fig17} for pictorial explanation. This partition function, to be denoted by $Z_{ab}$, admits two different expansions. The first one is the expansion in the closed-string channel, which reads
\beq
Z_{ab}=\sum_{\psi_c:\text{closed strings}} e^{-RE_{\psi_c}(L)}\frac{\langle B_a|\psi_c\rangle\langle \psi_c|B_b\rangle}{\langle \psi_c|\psi_c\rangle}\period
\eeq
In the limit $R\gg1$, the expansion is dominated by the contribution from the ground state $|\Omega\rangle$, and we get
\beq
Z_{ab}\overset{R\to \infty}{\sim}e^{-RE_{\Omega}}g_a g_b^{\ast}\comma
\eeq
with
\beq
g_a \equiv \frac{\langle B_a |\Omega\rangle}{\sqrt{\langle\Omega|\Omega\rangle}}\comma \qquad g_b \equiv \frac{\langle B_b |\Omega\rangle}{\sqrt{\langle\Omega|\Omega\rangle}}\period
\eeq
This shows that the overlaps between the boundary states and the ground state can be read off from the $R\to \infty$ limit of the cylinder partition function. In what follows, we focus on the computation of the ground-state overlap first and discuss the generalization to other states afterwards.

The quantities $g_{a,b}$ are called {\it $g$-functions}: The $g$-function, also known as the ground state degeneracy or boundary entropy, was first introduced by Affleck and Ludwig in the study of Kondo problem \cite{affleck1991universal}. It is known to be a useful measure of the boundary degrees of freedom and plays the role similar to the central charge $c$ of theories without boundaries. In relativistic field theories, it was proven in \cite{Friedan:2003yc,Casini:2016fgb} that the $g$-function decreases along the RG flow, much like the famous $c$-theorem by Zamolodchikov \cite{Zamolodchikov:1986gt}.

 Now, to compute the $g$-function using integrability, one needs to use the expansion in the other channel, namely the open-string channel or equivalently the mirror channel. In this channel, the partition function can be interpreted as a thermal partition function of a theory with boundaries at temperature $T=1/L$,
 \beq
 Z_{ab}=\sum_{\psi_o:\text{open strings}}e^{-L \tilde{E}_{\psi_o}(R)}\period
 \eeq
 Equating the two expansions and taking the limit, we obtain the following expression for the $g$-functions:
 \beq\label{eq:basistba}
 e^{-RE_{\Omega}}g_a g_b^{\ast} =\underbrace{\lim_{R\to\infty} \sum_{\psi_o:\text{open strings}}e^{-L \tilde{E}_{\psi_o}(R)}}_{\text{thermal partition function in the infinite volume}}\period
 \eeq
 Now, the crucial point is that the $R\to \infty$ limit corresponds to the infinite volume limit in the open-string channel and therefore can be studied using the S-matrix description of integrable quantum field theories, or more precisely the {\it thermodynamic Bethe ansatz} (TBA).

Before explaining the TBA approach to the $g$-function, let us make three clarifying remarks: First, in simple nonrelativistic systems such as the spin chains with the nearest-neighbor interaction, the standard Bethe ansatz can be directly applied to a system in a finite volume. However this is not the case for relativistic field theories or long-range spin chains, which are relevant for $\mathcal{N}=4$ SYM at finite 't Hooft coupling: In such theories, there are genuine finite-volume corrections known as the {\it wrapping corrections}, which come from virtual particles circulating the cylinder. These corrections give an infinite series of corrections to the answer obtained by the Bethe ansatz. Thus, for such theories, it is actually crucial to go to the mirror channel in which the spatial volume is infinite, in order to perform any reliable computations using integrability.

Second, the trick of going to the mirror channel and analyzing the infinite-volume thermal system is also used in the computation of the finite-size {\it spectrum} in integrable QFTs. In such cases, one simply needs to compute the leading large $R$ behavior of the thermal sum---namely, the right hand side of \eqref{eq:basistba}---and read off the exponent $e^{-RE_0}$. This is usually achieved by expressing the thermal sum as a path integral of the density of excitations $\rho$ and compute its saddle point $\rho^{\ast}$,
\beq
\lim_{R\to\infty} \sum_{\psi_o:\text{open strings}}e^{-L \tilde{E}_{\psi_o}(R)} \sim \lim_{R\to\infty}\int \mathcal{D}\rho\,\, e^{-RS[\rho]}\sim e^{-R S[\rho^{\ast}]}
\eeq
 In our case however, we are interested in the coefficients multiplying the exponential, and therefore we have to compute the next leading correction in the large $R$ expansion. As we explain below, this amounts to computing the fluctuation around the saddle point and also taking into account $O(1)$ prefactors which are normally neglected in the standard TBA analysis.

Third, we should emphasize that the use of TBA in our problem is different from the so-called {\it boundary Thermodynamic Bethe ansatz} (BTBA), which was introduced by LeClair {\it et al} in \cite{LeClair:1995uf} and applied to the cusped Wilson loop in $\mathcal{N}=4$ SYM \cite{Drukker:2012de,Correa:2012hh} and the spectrum on D-branes \cite{Bajnok:2012xc,Bajnok:2013wsa}. In our case, we take the open-string length $R$ to be infinite and apply the TBA in the open-string channel in order to read off the overlaps in the closed-string channel. By contrast, in the BTBA, the closed-string length $L$ is taken to be infinite and the TBA was used to read off the finite-volume spectrum of the open string. Our use of TBA never appeared before in the context of $\mathcal{N}=4$ SYM.
\subsubsection{Thermodynamic Bethe ansatz for $g$-function}
Let us now explain how to compute the $g$-functions using integrability.

The computation of the $g$-function in integrable QFTs was first attempted in \cite{LeClair:1995uf} based on Thermodynamic Bethe ansatz. The result however turned out to be incomplete as was pointed out in \cite{Woynarovich:2004gc}. A more systematic analysis was carried out in \cite{Dorey:2004xk} in which they performed an explicit computation of the thermal sum for states with a few excitations and conjectured a correct form of the $g$-function, which includes further corrections to the result in \cite{Woynarovich:2004gc}. Later, the same result was reproduced in \cite{Pozsgay:2010tv} by a careful analysis of Thermodynamic Bethe ansatz, giving support for the conjecture. Recently, a more rigorous derivation of the $g$-function was put forward in \cite{Kostov:2018dmi}, in which they used the combinatorial techniques such as the cluster expansion  and the matrix tree theorem\fn{See also \cite{Vu:2018iwv} for a beautiful application of the cluster expansion and the matrix tree theorem to the generalized hydrodynamics in integrable systems. It would be interesting to see if such a generalized hydrodynamics can capture an interesting kinematical limit of the correlation functions in $\mathcal{N}=4$ SYM.} \cite{Kostov:2018ckg} and computed a complete thermal sum of states.

In Appendix \ref{ap:gfunctionderivation}, we present a derivation which is in a sense hybrid of the methods in \cite{Pozsgay:2010tv}, \cite{Woynarovich:2010wt} and \cite{Kostov:2018dmi}. It has an advantage that the discussion can be made rigorous throughout the analysis but nevertheless avoids the use of sophisticated combinatorial techniques. However, since the derivation still involves several steps, here we present a more heuristic argument based on the standard derivation of TBA. Although less rigorous\fn{The derivation in this section uses the standard concepts in the TBA analysis, such as the density of energy levels and the density of holes, which are physically well-motivated but might be hard to digest in the beginning. For this reason, the readers with less familiarity with the standard TBA argument might actually find the derivation in Appendix \ref{ap:gfunctionderivation} simpler since it does not rely on such concepts.}, the derivation in this section shares two important benefits with the derivation in Appendix \ref{ap:gfunctionderivation}: First it leads to a simple TBA action for which one can straightforwardly compute the fluctuation around the saddle point. Second the so-called $Y$-function naturally shows up in the derivation as a fundamental field variable.
To elucidate the basic idea in a simple setup, we discuss a theory with a single species of particles without any bound states. However, the result can be readily generalized to theories with multiple species and bound states as we see in section \ref{sec:TBA}.

\begin{figure}[t]
\centering
\includegraphics[clip,height=5cm]{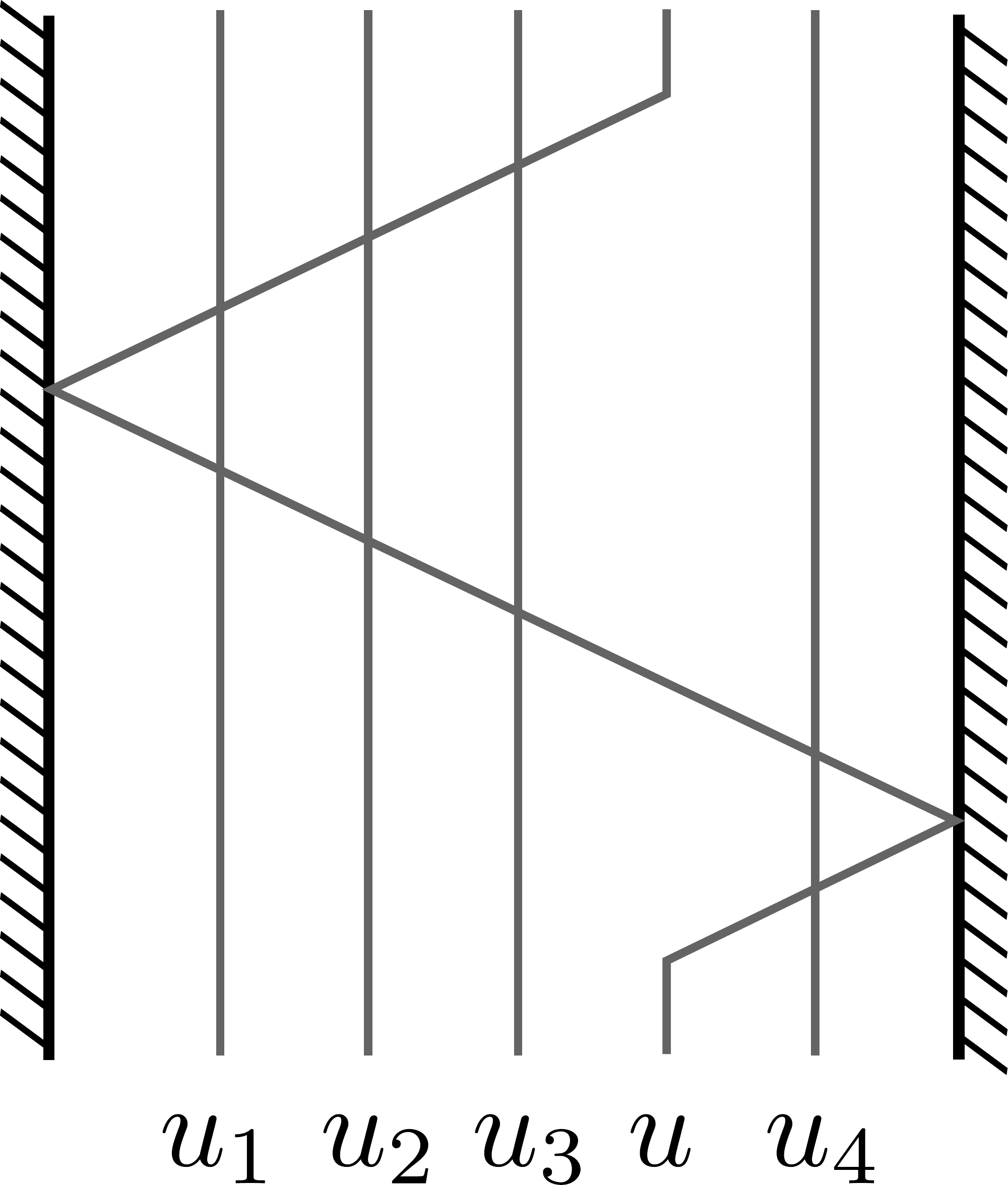}
\caption{A pictorial explanation of the boundary Bethe equation. When a magnon moves along the chain and comes back to its original position, it scatters with other magnons twice (in the opposite directions) and also gets reflected at the boundaries. The phase shift in this process is given by the right hand side of \eqref{eq:boundarybetheeqtoy}.}
\label{fig:fig18}
\end{figure}

In the large $R$ limit, each state in the open-string channel can be described as a collection of excitations on the vacuum with a set of rapidities ${\bf u}=\{u_1,\ldots ,u_M\}$. This in particular means that the energy of each state is given by a sum of the energies of individual excitations,
\beq
\tilde{E}=\sum_{k=1}^{M}\tilde{E}(u_k)\period
\eeq
We should however keep in mind that the rapidities cannot take arbitrary values; they need to satisfy the quantization condition called the boundary Bethe equation (see figure \ref{fig:fig18} for a pictorial explanation)
\beq\label{eq:boundarybetheeqtoy}
\begin{aligned}
1=&\tilde{R}^{(L)}_{a} (-u_j)\tilde{R}^{(R)}_{b}(u_j)e^{2i\tilde{p}(u_j)R}\prod_{k\neq j}\tilde{S}(u_j,u_k)\tilde{S}(u_j,-u_k)=\mathcal{R}(u_j)\prod_{k}\mathcal{S}(u_j,u_k)\period
\end{aligned}
\eeq
Here $\tilde{R}^{(L,R)}_{a,b}$ and $\tilde{S}(u,v)$ are the reflection amplitude and the S-matrix in the mirror channel, and in the second equality we introduced a compact notation:
\beq
\mathcal{S}(u,v)\equiv \tilde{S}(u,v)\tilde{S}(u,-v)\comma\qquad \mathcal{R}(u)\equiv \tilde{R}^{(L)}_{a}(-u)\tilde{R}^{(R)}_{b}(u)e^{2i\tilde{p}(u)R}/\mathcal{S}(u,u)\period
\eeq
 The superscripts $(L,R)$ specify whether the reflection happen at the left or the right boundary.  Owing to the parity symmetry and the unitarity, the reflection matrices (both $\tilde{R}_a$ and $\tilde{R}_b$) satisfy
\beq\label{eq:propertiesR}
\tilde{R}^{(R)}(u)=\tilde{R}^{(L)}(-u)=1/\tilde{R}^{(R)}(-u)=1/\tilde{R}^{(L)}(u)\period
\eeq
Here the first equality comes from the parity symmetry while the latter two equalities are due to the unitarity. Using this relation, one can always express the left reflection matrix in terms of the right reflection matrix and vice versa. For instance, $\mathcal{R}(u)$ can be expressed purely in terms of the left reflection matrices as
\beq
\mathcal{R}(u)=\left(\tilde{R}^{(L)}_a (u)\tilde{R}^{(L)}_b(u)\mathcal{S}(u,u)\,e^{-2i\tilde{p}(u)R}\right)^{-1}\period
\eeq
The parity and the unitarity also give constraints on the bulk S-matrix:
\beq\label{eq:propertiesS1}
\begin{aligned}
\text{Unitarity: }& \tilde{S}(u,v)\tilde{S}(v,u)=1\comma\qquad
\text{Parity: }\tilde{S}(u,-v)=\tilde{S}(v,-u)\period
\end{aligned}
\eeq
In addition, it satisfies the following important property,
\beq\label{eq:propertiesexclusion}
 \text{Exclusion: }\tilde{S}(u,u)=-1\period
\eeq

In the limit $R\to \infty$, we expect that the thermal partition function receives main contributions from finite particle-density states, namely states with $M\sim R \gg 1$. To describe such a state, it is convenient to introduce a rapidity density defined by
\beq
\rho (u)=\frac{1}{R}\sum_{k}\delta (u-u_k)\period
\eeq
Using the density $\rho$, one can express the energy of the state as $\tilde{E}=R\int du\, \tilde{E}(u)\rho (u)$. One might then think that the thermal partition function in the $R\to\infty$ limit is simply given by
\beq
Z_{ab}=\sum_{\psi_o}e^{-L\tilde{E}_{\psi_o}(R)}\quad \overset{\displaystyle \text{?}}{\sim}\quad \int \mathcal{D}\rho \,\, e^{-R\int du\,L \tilde{E}(u)\rho (u)}\period
\eeq
Unfortunately, this turns out to be incorrect for the following reason: In the limit $R\to\infty$, the typical distance between rapidities scale as $1/R$. Therefore, for any small but finite interval $\Delta u$, there will be a large number of rapidities of order $R/\Delta u$. Because of this, a large number of microscopically different rapidity configurations lead to the same macroscopic density $\rho$ and one needs to take into account such degeneracy.

To do so, we first write the logarithm of the Bethe equation in terms of the density as
\beq\label{eq:logBethe}
2\pi n_j =\frac{1}{i}\log \mathcal{R}(u_j) +\frac{R}{i}\int_0^{\infty} du^{\prime} \log \mathcal{S}(u_j,u^{\prime}) \rho (u^{\prime}) \comma\qquad (j=1,\ldots, M)\period
\eeq
with $n_j$ being integer. Note that the range of integration is $(0,\infty)$ since particles with rapidities $u$ and $-u$ are identified\fn{This is because particles flip the signs of rapidities when reflected by the boundaries.} in the presence of boundaries. Physically, $n_j$ can be interpreted as the mode numbers of excitations: For any given state, the set ${\bf n}=\{n_1,\ldots ,n_M\}$ provides a subset of all possible positive\fn{Again we restrict ourselves to positive integers since particles with the mode number $\pm n$ are identified in the boundary scattering problem.} integers and we interpret them as ``occupied modes''. On the other hand, the (positive) integers that do not belong to {\bf n} are interpreted as ``unoccupied modes'' or ``holes''. For each occupied mode, there is a corresponding rapidity $u_j$ determined by the equation \eqref{eq:logBethe}. We now extend this relation also to holes by defining a rapidity for any (positive) mode number $n$ as
\beq\label{eq:generalizedrapidity}
2\pi n = \frac{1}{i}\log \mathcal{R}(u^{(n)}) +\frac{R}{i}\int_0^{\infty} du^{\prime} \log \mathcal{S}(u^{(n)},u^{\prime}) \rho (u^{\prime}) \qquad n \in \mathbb{Z}_{>0}\period
\eeq
Although this is a standard argument in TBA, this idea of defining rapidities for both occupied and unoccupied modes might appear a bit artificial at first sight: In particular it might be a priori unclear whether the rapidities for holes have any real physical meaning. To understand this point, let us consider a situation in which we already have a large ($M\sim R\gg 1$) number of excitations and want to add one more excitation. Since there are already a large number of excitations, one can regard the added excitation as a small perturbation and approximate the Bethe equation by its linearized form. It then turns out that the linearized equation precisely coincides with the equation \eqref{eq:generalizedrapidity}. This shows that the rapidities for holes can be interpreted as the positions at which one can add extra excitations and the equation \eqref{eq:generalizedrapidity} governs small deformations of a given solution to the Bethe equation.

With this remark, we can now follow the standard arguments used in TBA: In a small but finite interval of the rapidity $[u,u+\Delta u]$, there are $R\rho(u) \Delta u$ number of occupied modes. On the other hand, the number of available modes in this interval is given by $R\rho_{\rm tot} (u) \Delta u$ where
\beq
\begin{aligned}
&\rho_{\rm tot}\equiv \frac{1}{R}\frac{dn}{du}=\sigma (u)+\mathcal{K}_{+}\ast \rho (u)\comma\\
\text{with} \quad&\sigma (u)\equiv \frac{1}{2\pi i R}\del_{u}\log \mathcal{R}(u)-\frac{\delta (u)}{2R}\comma\qquad \mathcal{K}_{+}(u,v)\equiv \frac{1}{i }\del_{u}\log \mathcal{S}(u,v)\comma
\end{aligned}
\eeq
and $A\ast B$ stands for a convolution integral $\int_{0}^{\infty}\frac{dv}{2\pi}A(u,v) B(v)$. The offset $-\delta(u)/2R$ in $\sigma (u)$ comes from the fact that, although the Bethe equation admits $u=0$ as a solution\fn{More precisely this corresponds to excluding solutions with the mode number $n=0$. See Appendix \ref{ap:gfunctionderivation} for a more rigorous treatment.}, it does not correspond to a physical state since the corresponding wave function vanishes.
Using this, we can compute the microscopic degeneracy for a given macroscopic $\rho$ as
\beq
\omega [u, u+\Delta u] =\pmatrix{c}{R\rho_{\rm tot}(u)\Delta u\\R\rho (u)\Delta u}\period
\eeq
Normally we then approximate this binomial factor using the Stirling approximation assuming that the number of modes in the interval is huge. This however is not very efficient for the computation of $g$-functions, in which one has to keep the subleading term in the Stirling approximation. Instead, here we express the binomial factor as the following integral
\beq
\omega [u, u+\Delta u] =\pmatrix{c}{R\rho_{\rm tot}(u)\Delta u\\R\rho (u)\Delta u} =\oint \frac{d\eta}{2\pi} e^{-iR \rho (u)\Delta u\eta} \left(1+e^{i\eta}\right)^{R\rho_{\rm tot}\Delta u}\comma
\eeq
where $\eta$ is integrated from $[0,2\pi]$. As a result, we obtain the following path-integral representation for the degeneracy:
\beq
\begin{aligned}
\Omega [\rho]&\equiv \lim_{\Delta u\to 0}\prod_{u}\omega [u,u+\Delta u]\\
&=\int \mathcal{D}\eta (u) \,\,\exp \left[R\int_0^{\infty} du \left\{ (\sigma (u)+\mathcal{K}_{+}\ast \rho (u))\log (1+e^{i\eta (u)})-i\eta (u)\rho(u)\right\}\right]\period
\end{aligned}
\eeq

We now combine the degeneracy factor with the energy factor and write down the path-integral representation of the thermal sum,
\beq
Z_{ab}=\sum_{\psi_{o}}e^{-LE_{\psi_o}(R)} \overset{R\to\infty}{=}\int \mathcal{D}\rho\mathcal{D}\eta\,\,\mathcal{N}\times  e^{-R S[\rho,\eta]}\comma
\eeq
with
\beq\label{eq:newTBAaction}
S[\rho,\eta]=\int_0^{\infty} du\Big[L\tilde{E} (u)\rho(u)+i\eta (u)\rho(u)-(\sigma (u)+\mathcal{K}_{+}\ast \rho (u))\log (1+e^{i\eta (u)}) \Big]\period
\eeq
Here $\mathcal{N}$ is a subtle normalization factor that we discuss later. In the leading large $R$ limit, one can approximate the path integral by its saddle point. An important feature of our action \eqref{eq:newTBAaction} is that it is linear in $\rho$. This makes the computations much simpler than the standard formulation. For instance, if we take a variation with respect to $\rho$, we immediately get the equation
\beq\label{eq:oursimpleTBA}
\frac{\delta S}{\delta \rho}=0\qquad \iff\qquad0 =L\tilde{E}(u)+\log Y(u) -\log \left(1+Y\right)\ast \mathcal{K}_{+}(u)\comma
\eeq
with $Y\equiv e^{i\eta}$, which coincides with the TBA equation once we identify $Y(u)$ with the so-called $Y$-function. In other words,  the $Y$-function, which is of prime importance in TBA and its generalization, is a fundamental field variable in our formulation. If, on the other hand, we take a variation with respect to $\eta$, we get
\beq
\frac{\delta S}{\delta \eta}=0\qquad \iff \qquad \rho (u)=(\sigma (u)+\mathcal{K}_{+}\ast\rho(u))\frac{Y(u)}{1+Y(u)}\comma
\eeq
 which is nothing but the relation between the $Y$-function and the density of rapidities $\rho(u)$.
To compute the saddle-point value of the action, we  just need to drop all the terms that contain $\rho$ since the action is linear in $\rho$. We then get
\beq
\left.S[\rho,\eta]\right|_{\rm saddle}=-\int_0^{\infty} du\,\, \sigma (u)\log (1+Y(u))\comma
\eeq
which matches with the result in the literature.

The beauty of our formulation is that it makes manifest the simplicity of the answers obtained by the TBA formalism, which are somewhat obscured in the standard formulation. We emphasize that it is mainly due to the auxiliary field $\eta$. The inclusion of the auxiliary field in the action is reminiscent of the Yang-Yang functional for the TBA system discussed in the study of supersymmetric gauge theories \cite{Nekrasov:2011bc} and the null polygonal Wilson loops in $\mathcal{N}=4$ SYM at strong coupling \cite{Alday:2010ku}. The details are however slightly different and it would be interesting to clarify the relation\fn{As we explain below, our action is useful also for the computation of quadratic fluctuations around the saddle point. It is therefore interesting to see if our formulation of the action can capture quantum corrections to the results in \cite{Nekrasov:2011bc} and \cite{Alday:2010ku}.} between the two.

Let us now compute the fluctuation around the saddle point. Since the action is linear in $\rho$, the Hessian takes the following simple form,
\beq
{\tt Hessian}=\pmatrix{ll}{{\displaystyle \frac{\delta^2 S}{\delta \eta \delta \eta}}&{\displaystyle \frac{\delta^2 S}{\delta \eta \delta \rho}}\vspace{10pt}\\{\displaystyle\frac{\delta^2 S}{\delta \rho \delta \eta}}&{\displaystyle \frac{\delta^2 S}{\delta \rho \delta \rho}=0}}\period
\eeq
From this, we can compute the fluctuation around the saddle point as\fn{The factor $1/i$ arises from diagonalizing the Hessian in the basis $\rho\pm\eta /\sqrt{2}$ and computing the Gaussian integrals.}
\beq
{\tt fluctuation}=\left(\det \frac{1}{i}\frac{\delta^2 S}{\delta \eta\delta\rho}\right)^{-1}\period
\eeq
This can be computed by taking a functional variation of the TBA equation \eqref{eq:oursimpleTBA}, $\delta S/\delta \rho$, with respect to $\eta$. As a result, we obtain the Fredholm determinant
\beq
{\tt fluctuation}=\left(\det \left[1-\hat{G}_{+}\right]\right)^{-1}\comma
\eeq
with\fn{Precisely speaking, we need to perform the ``transposition'' to the integral kernel after taking a functional variation in order to bring it into a form \eqref{eq:actionGplusexplicit}. This however does not affect the Fredholm determinant.}
\beq\label{eq:actionGplusexplicit}
\hat{G}_{+}\cdot f(u)\equiv \int_{0}^{\infty}\frac{dv}{2\pi} \frac{\mathcal{K}_{+}(u,v)}{1+1/Y(v)}f(v)\period
\eeq
It is interesting to note that the way we derived the Fredholm determinant is reminiscent of the Gaudin norms of the spin chain, which can be computed by taking a variation of the Bethe equation. We will later see that this point of view is useful when writing down the $g$-function for more complicated theories.

To obtain an expression for the $g$-function, we also need to include the normalization factor $\mathcal{N}$. Physically, it comes from the difference of the normalizations between the rapidity basis and the mode-number basis of the states in the mirror channel, and it is given by a continuum limit of the Gaudin norm. The necessity of this factor was first discussed in \cite{Pozsgay:2010tv} and a rigorous derivation was given in \cite{Woynarovich:2010wt}. It is however not easy to derive this factor within the heuristic argument presented in this section. So in this section we just present the result relegating the derivation to Appendix \ref{ap:gfunctionderivation}. The result reads
\beq
\mathcal{N}=\det \left[1-\hat{G}_{-}\right]\comma
\eeq
with
\beq\label{eq:actionGminusexplicit}
\hat{G}_{-}\cdot f(u)\equiv \int_{0}^{\infty} \frac{dv}{2\pi}\frac{\mathcal{K}_{-}(u,v)}{1+1/Y(v)}f(v) \comma\qquad \mathcal{K}_{-}(u,v) \equiv \frac{1}{i}\del_{u}\log \frac{\tilde{S} (u,v)}{\tilde{S}(u,-v)}\period
\eeq

Combining everything, we finally get the expression for the thermal partition function in the large $R$ limit,
\beq
e^{-R E_{\Omega}}g_ag_b^{\ast}= e^{-R \left.S\right|_{\rm saddle}} \times \frac{\det \left[1-\hat{G}_{-}\right]}{\det \left[1-\hat{G}_{+}\right]}\period
\eeq
Separating $\left.RS\right|_{\rm saddle}$ into $O(R)$ and $O(1)$ pieces, we obtain the following expression for the ground-state energy and the $g$-function\fn{The expression for the TBA energy can be expressed in a more standard way by rewriting \eqref{eq:tbaebergyinnew} as
\beq
E_{\Omega}=-\int_{-\infty}^{\infty} \frac{du}{2\pi} \del_u \tilde{p}(u) \log (1+Y(u))\comma
\eeq
by extending the definition of the $Y$-function as $Y(-u)\equiv Y(u)$.
},
\begin{align}
E_{\Omega}&=-\int_{0}^{\infty} \frac{du}{\pi} \del_u \tilde{p}(u) \log (1+Y(u))\comma\label{eq:tbaebergyinnew}\\
g_a&=\exp \left[\int^{\infty}_{0}\frac{du}{2\pi}\Theta_a (u)\log (1+Y(u))\right] \times \sqrt{ \frac{\det \left[1-\hat{G}_{-}\right]}{\det \left[1-\hat{G}_{+}\right]}}\comma\label{eq:gfunctionrank1}
\end{align}
with
\beq\label{eq:defofthetasimple}
\Theta_a (u)= -\frac{1}{ i}\del_u\log \tilde{R}^{(L)}_a (u)-\pi\delta (u)-\left.\frac{1}{ i}\del_u \log \tilde{S} (u,v)\right|_{v=-u}\period
\eeq
Interestingly, the expression \eqref{eq:gfunctionrank1} resembles the results obtained at weak coupling in section \ref{sec:weak}: Both are given by a ratio of determinants multiplied by some factor. The resemblance becomes more apparent if we write the kernel $\mathcal{K}_{\pm}$ explicitly in terms of the S-matrix:
\beq
\mathcal{K}_{\pm}(u,v)=\frac{1}{i}\left[\del_u \log \tilde{S}(u,v)\pm \del_u\log \tilde{S}(u,-v)\right]\period
\eeq
The details are however somewhat different: Here we have Fredholm determinants while in section \ref{sec:weak} we had standard determinants. In addition, the ratio we have here is of the form $\det [1-\hat{G}_{-}]/\det [1-\hat{G}_{+}]$ but the ratio we had in section \ref{sec:weak} is more like an inverse; $\det G_{+}/\det G_{-}$.

In the next subsection, we explain that these differences can be naturally explained once we extend \eqref{eq:tbaebergyinnew} and \eqref{eq:gfunctionrank1} to excited states, and the two formula are indeed related.
\subsubsection{Useful rewriting}
Before proceeding to the discussions on the excited states, let us mention that there exists another expression for the ratio of the Fredholm determinants, which will turn out to be useful later in section \ref{subsec:N=4gfunction}.

The idea is simple: We rewrite the ratio of the Fredholm determinants as
\beq
\frac{\det \left[1-\hat{G}_{-}\right]}{\det \left[1-\hat{G}_{+}\right]}=\frac{\det \left[1-\hat{G}_{+}\right]\det \left[1-\hat{G}_{-}\right]}{\left(\det \left[1-\hat{G}_{+}\right]\right)^2}\period
\eeq
Of course, this is just a trivial rewriting, but the key observation is that one can replace a product of the Fredholm determinants in the numerator with
\beq\label{eq:combiningtwodets}
\det \left[1-\hat{G}_{+}\right]\det \left[1-\hat{G}_{-}\right]=\det \left[1-\hat{G}\right]\comma
\eeq
where $\hat{G}$ is an integral kernel acting on the \emph{full real axis} whose action is given by
\beq
\hat{G}\cdot f(u)=\int^{\infty}_{-\infty}\frac{dv}{2\pi}\frac{K(u,v)}{1+1/Y(v)}f(v)\comma
\eeq
with $Y(v)\equiv Y(-v)$ $(v<0)$ and
\beq
K(u,v)\equiv \frac{1}{2}\left[\mathcal{K}_{+}(u,v)+\mathcal{K}_{-}(u,v)\right]=\frac{1}{i}\del_u \tilde{S}(u,v)\period
\eeq
Postponing the derivation of \eqref{eq:combiningtwodets}, let us first discuss the implication of this rewriting. As already mentioned, the Fredholm determinant $\det\left[ 1-G_{+}\right]$ can be obtained by taking a functional variation of the TBA equation in the presence of boundaries \eqref{eq:oursimpleTBA}. As it turns out, $\det \left[1-\hat{G}\right]$ can also be interpreted as a functional variation of the TBA equation but now for standard periodic boundary conditions,
\beq
0=L\tilde{E}(u)+\log Y(u)-\log (1+Y)\star K(u)\comma
\eeq
 with $\star$ being the convolution along the full real axis. Therefore, after rewriting both the numerator and the denominator of $\det [1-\hat{G}]/ (\det [1-\hat{G}_{+}])^2$ are associated with the variations of the TBA. This feature makes it simpler to generalize the $g$-functions to more complicated systems such as $\mathcal{N}=4$  SYM as we will see in section \ref{subsec:N=4gfunction}.

\paragraph{Derivation of \eqref{eq:combiningtwodets}}To show the relation \eqref{eq:combiningtwodets}, we first decompose the action of $\hat{G}$ depending on whether the arguments of the functions are positive or negative:
\beq
\begin{aligned}
\hat{G}\cdot f_{+}(u)&=\int^{\infty}_{0}\frac{dv}{2\pi}\frac{K(u,v)}{1+1/Y(v)}f_{+}(v)+\int^{\infty}_{0}\frac{dv}{2\pi}\frac{K(u,-v)}{1+1/Y(v)}f_{-}(v)\comma\\
\hat{G}\cdot f_{-}(u)&=\int^{\infty}_{0}\frac{dv}{2\pi}\frac{K(-u,v)}{1+1/Y(v)}f_{+}(v)+\int^{\infty}_{0}\frac{dv}{2\pi}\frac{K(-u,-v)}{1+1/Y(v)}f_{-}(v)\period
\end{aligned}
\eeq
Here both $u$ and $v$ are positive real and $f_{\pm }(v)\equiv f (\pm v)$ with $v>0$. This can be combined into a $2\times 2$ matrix structure
\beq
\hat{G}\cdot \pmatrix{c}{f_{+}(u)\\f_{-}(u)}=\int_0^{\infty} \frac{dv}{2\pi}\frac{1}{1+1/Y(v)} \pmatrix{cc}{K(u,v)&K(u,-v)\\K(u,-v)&K(u,v)}\pmatrix{c}{f_{+}(v)\\f_{-}(v)}\comma
\eeq
where we have used the  equalities between kernels
\beq
K(-u,v)=K(u,-v)\comma\quad K(-u,-v)=K(u,v)\comma
\eeq
which can be shown by using the parity invariance\fn{For instance, the first equality can be shown as follows:
\beq
\begin{aligned}
K(-u,v)&\equiv \frac{1}{i}\del_u^{\prime}\log \tilde{S}(u^{\prime},v)|_{u^{\prime}=-u}= -\frac{1}{i}\del_u\log\tilde{S}(-u,v)\\
&\overset{\rm parity}{=}-\frac{1}{i}\del_u\log\tilde{S}(v,-u)\overset{\rm unitarity}{=}\frac{1}{i}\del_u\log \tilde{S}(u,-v)=K(u,-v)\period\nn
\end{aligned}
\eeq
} of the S-matrix. Now, using this $2\times 2$ matrix representation, one can compute $\log \det \left[1-\hat{G}\right]$ as
\beq\label{eq:logdetproductfredrel}
\begin{aligned}
\log \det \left[1-\hat{G}\right]=-\sum_{n=1}^{\infty}\frac{1}{n}{\rm Tr}\left[(\hat{G})^{n}\right]=-\sum_{n=1}^{\infty}\frac{1}{n}\left({\rm Tr}\left[(\hat{G}_{+})^{n}\right]+{\rm Tr}\left[(\hat{G}_{-})^{n}\right]\right)\comma
\end{aligned}
\eeq
where, in the second equality, we diagonalized the matrix of $K$'s to compute the trace ${\rm Tr}\left[(\hat{G})^{n}\right]$:
\beq
\pmatrix{cc}{K(u,v)&K(u,-v)\\K(u,-v)&K(u,v)}\quad \overset{\rm diagonalization}{\mapsto} \quad \pmatrix{cc}{\mathcal{K}_{+}(u,v)&0\\0&\mathcal{K}_{-}(u,v)}\period
\eeq
Clearly, the right hand side of \eqref{eq:logdetproductfredrel} is the expansion of $\log \left(\det \left[1-\hat{G}_{+}\right]\det \left[1-\hat{G}_{-}\right]\right)$ thereby proving the statement \eqref{eq:combiningtwodets}.
\subsection{Analytic continuation and excited state $g$-function\label{subsec:analyticcont}}
\subsubsection{Analytic continuation of TBA}
For the finite-volume spectrum, the generalization of TBA to excited states was discussed first by Dorey and Tateo \cite{Dorey:1996re,Dorey:1997rb}. The key idea in their approach is to analytically continue some parameter, such as the mass or the coupling constant, to complex values: Suppose that we start with the ground state and adiabatically continue the parameter to complex values. In many physical situations, there are branch cuts in the complex parameter plane, and by crossing them and coming back to the same position on a different Riemann sheet, we can transform the original ground state to an excited state \cite{Dorey:1996re}.
Although it is often difficult to specify which analytic continuation we need to perform in order to obtain a desired state, the net effect of the analytic continuation is simple and well-understood: In the process of analytic continuation, some poles in the integrand, both in the TBA equation \eqref{eq:oursimpleTBA} and in the energy formula \eqref{eq:tbaebergyinnew}, cross the integration contour and produce extra contributions.

\begin{figure}[t]
\begin{minipage}{0.4\hsize}
\centering
\includegraphics[clip,height=3cm]{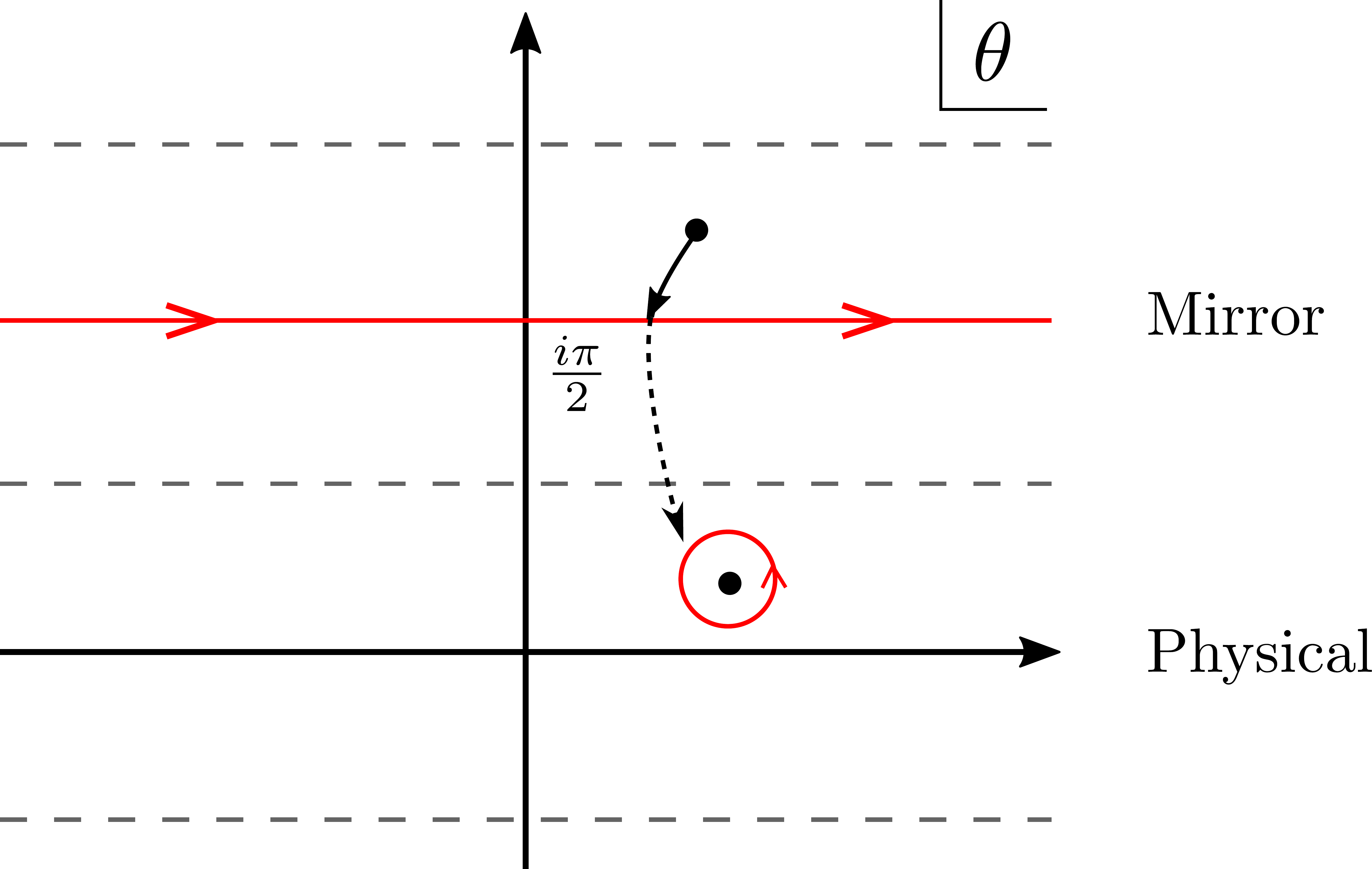}\\\vspace{10pt}
$(a)$ Relativistic
\end{minipage}
\begin{minipage}{0.6\hsize}
\centering
\includegraphics[clip,height=3.3cm]{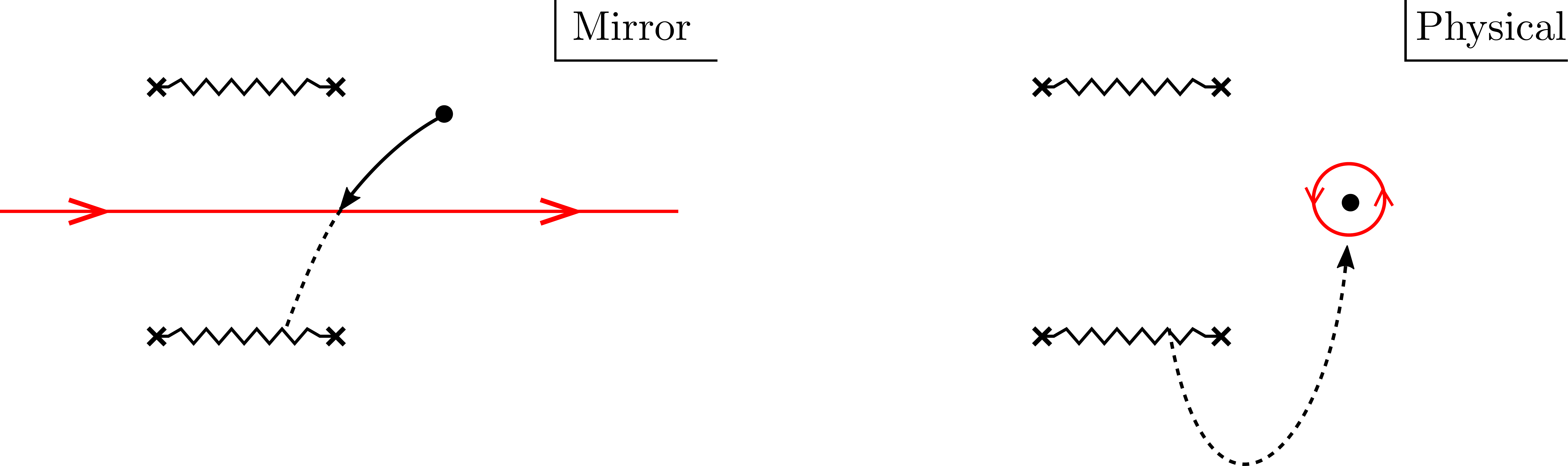}\\
$(b)$ $\mathcal{N}=4$ SYM
\end{minipage}
\caption{The analytic continuation for the relativistic theories $(a)$ and $\mathcal{N}=4$ SYM (b). In both cases, the red lines are the integration contours in the mirror channel. The poles cross the contour from above and end up in the physical kinematics.}
\label{fig:fig19}
\end{figure}

To see this more explicitly, it is useful to rewrite \eqref{eq:tbaebergyinnew} using integration by parts\fn{We used $p(0)=0$ to kill boundary contributions which arise from integration by parts.}
\beq\label{eq:toanalyticallycontinue}
E_{\Omega}=\int_{0}^{\infty}\frac{du}{\pi} \tilde{p}(u)\frac{\del_u Y(u)}{1+Y(u)}\period
\eeq
We can now see that the integrand has poles with a unit residue at $1+Y(u)=0$. Without analytic continuation, such poles sit somewhere away from the integration contour but they can move and cross the contour once we analytically continue. When this happens, the equation \eqref{eq:toanalyticallycontinue} gets transformed into
\beq
E=2i\sum_{k}\tilde{p}(\tilde{u}_k)+\int_{0}^{\infty}\frac{du}{\pi} \tilde{p}(u)\frac{\del_u Y(u)}{1+Y(u)}\comma
\eeq
where $\tilde{u}_k$ are the positions of the poles and we assumed that they crossed the contour from above. This is indeed what we expect physically as explained in figure \ref{fig:fig19}. Using the relation between the energies and momenta in the two channels \eqref{eq:doublewickrotation}, one can rewrite this as
\beq\label{eq:analyticallycontinuedenergy}
E=2\sum_k E(u_k)-\int_0^{\infty}\frac{du}{\pi}\del_u \tilde{p}(u)\log (1+Y(u))\comma
\eeq
where the rapidities in the physical channel $u_k$ were defined by $i\tilde{p}(\tilde{u}_k)=E(u_k)$.
This gives the expression for the energy of excited states.

The analytic continuation also affects the TBA equation \eqref{eq:oursimpleTBA}, which determines the $Y$-function. This is again easy to see once we rewrite the convolution using integration by parts,
\beq
-\log (1+Y(u))\ast \mathcal{K}_{+}=\int \frac{dv}{2\pi i} \frac{\del_v Y(v)}{1+Y(v)}\left(\log \tilde{S}(v,u)+\log \tilde{S}(v,-u)\right)\period
\eeq
We can then see the poles explicitly and follow the same steps as above. As a result, we arrive at the following TBA equation for excited states:
\beq\label{eq:toyexcitedTBA}
0=\log Y(u) + L\tilde{E}(u)+\sum_k \log \Big(\tilde{S}(\tilde{u}_k,u)\tilde{S}(\tilde{u}_k,-u)\Big)-\log (1+Y)\ast \mathcal{K}_{+}(u)\period
\eeq

Let us now check that the excited-state TBA correctly reproduces the standard Bethe ansatz equation in the physical channel upon taking the infinite volume limit $L\to \infty$. Owing to the TBA equation \eqref{eq:toyexcitedTBA}, the $Y$-function behaves as $Y\sim e^{-L m}$ with $m$ being the mass of excitations and therefore is exponentially small in the infinite volume limit. This allows us to drop the last term in the TBA equation \eqref{eq:toyexcitedTBA}. We then get the infinite-volume expression for the $Y$-function,
\beq
Y(u)=\frac{e^{-L\tilde{E}(u)}}{\prod_{k}\tilde{S}(\tilde{u}_k,u)\tilde{S}(\tilde{u}_k,-u)}\period
\eeq
Evaluating this equation at $u=\tilde{u}_j$, we obtain
\beq\label{eq:analyticallycontinuedbethe}
1+e^{L\tilde{E}(\tilde{u}_j)}\prod_{k}\tilde{S}(\tilde{u}_k,\tilde{u}_j)\tilde{S}(\tilde{u}_k,-\tilde{u}_j)=0\period
\eeq
To bring it into a standard form of the Bethe equation, we first replace $\tilde{E}(\tilde{u}_j)$ by $-ip(u_j)$ using the relation \eqref{eq:doublewickrotation}. In addition, we also need to re-express the S-matrix $\tilde{S}(\tilde{u}_k,\pm \tilde{u}_j)$ in terms of the rapidities in the physical channel $u_k$'s. To be concrete, let us for now consider relativistic field theories in which the energy and the momentum are parameterized by the rapidity $\theta$ as
\beq
E=m\cosh\theta\comma\qquad p=m\sinh \theta\period
\eeq
In such theories, the physical excitations carry rapidities $\theta -i\pi/2$ with $\theta \in \mathbb{R}$ when described in the mirror channel. Note that this is simply the inverse of the {\it mirror transformation} $\theta\to \theta +i\pi/2$ in the literature, which describes the mirror excitations in the physical channel. Using this more standard convention, we can re-express the S-matrices as
\beq
\tilde{S}(\tilde{u}_k,\tilde{u}_j)\mapsto \tilde{S}\left(\theta_k -i\frac{\pi}{2},\theta_j -i\frac{\pi}{2}\right)\comma\qquad \tilde{S}\left(\tilde{u}_k,-\tilde{u}_j\right)\mapsto \tilde{S}\left(\theta_k -i\frac{\pi}{2},-\theta_j \red{+}i\frac{\pi}{2}\right)\period
\eeq
Here we highlighted in color that the second argument in the second S-matrix comes with a positive imaginary part.
Now, using the unitarity, we can rewrite the first S-matrix as
\beq
\tilde{S}\left(\theta_k -i\frac{\pi}{2},\theta_j -i\frac{\pi}{2}\right)=S\left(\theta_k ,\theta_j \right)=\left(S(\theta_j,\theta_k)\right)^{-1}\period
\eeq
Here $S(\theta,\theta^{\prime})$ is the S-matrix in the physical channel\fn{Precisely speaking, in relativistic invariant theories, the physical and mirror S-matrices are equivalent, $\tilde{S}=S$. Here we distinguished the two in order to make clear that the derivation works also for more general theories.}
On the other hand, for the second S-matrix, we need to use the crossing relation\fn{For relativistic invariant theories with a single species of particles, the crossing transformation reads $S(u,v\pm i\pi)=S(v,u)$.} and the parity \eqref{eq:propertiesS1} to eliminate the mismatch of imaginary parts:
\beq\label{eq:rewritingcrossingparitycomplicated}
\begin{aligned}
\tilde{S}\left(\theta_k -i\frac{\pi}{2},-\theta_j +i\frac{\pi}{2}\right)&\overset{\rm crossing}{=}
\tilde{S}\left(-\theta_j -i\frac{\pi}{2},\theta_k -i\frac{\pi}{2}\right)=\left(S\left(\theta_k ,-\theta_j\right)\right)^{-1}\\
&\overset{\rm parity}{=}\left(S\left(\theta_j,-\theta_k\right)\right)^{-1}\period
\end{aligned}
\eeq

Putting things together and using $S(\theta,\theta)=-1$, we find that the equation \eqref{eq:analyticallycontinuedbethe} can be rewritten as
\beq\label{eq:bethefromanalytic}
e^{ip (\theta_j)L}S(\theta_j,-\theta_j)\prod_{k\neq j} S(\theta_j,\theta_k)S(\theta_j,-\theta_k)=1\period
\eeq
This can be interpreted as the Bethe equation in the physical channel for a parity symmetric state, ${\bf \theta}=\{\theta_1,-\theta_1,\ldots\}$. The appearance of a parity-symmetric state is consistent with the discussion in section \ref{subsec:formVSreflection}. It is also consistent with the factor of $2$ in \eqref{eq:analyticallycontinuedenergy}; it simply means that, for every particle with a positive rapidity $u_k$, there is a corresponding particle with a rapidity $-u_k$.
The argument above works as long as the following relations are satisfied:
\beq\label{eq:universalrelation}
\tilde{S}(\tilde{u}_k,\tilde{u}_j)=\left(S(u_j,u_k)\right)^{-1}\comma\qquad \tilde{S}(\tilde{u}_k,-\tilde{u}_j)=\left(S(u_j,-u_k)\right)^{-1}\period
\eeq
Note that the second equality requires the crossing symmetry of the S-matrix as shown in \eqref{eq:rewritingcrossingparitycomplicated}.

Let us also point out that one can write a finite-volume version of the Bethe equation \eqref{eq:bethefromanalytic} by reinstating the last factor in \eqref{eq:toyexcitedTBA}. Switching back to the general notation, the result (in the logarithmic form) reads
\beq\label{eq:finitevolumephase}
2\pi n_j=\phi (u_j)\equiv p(u_j)L+\frac{1}{i}\sum_{k\neq j}\left[\log S(u_j,u_k)+\log S(u_j,-u_k)\right]+\frac{1}{i}\log (1+Y)\ast \mathcal{K}_{+}(u_j)\comma
\eeq
with $n_j$ being a (positive) integer. In the infinite volume limit, we simply drop the last term which describes the interaction between physical and mirror particles and the result reduces to the standard Bethe equation.
\subsubsection{Analytic continuation of $g$-function and determinant formula}
Having understood the analytic continuation of TBA, it is now rather straightforward to perform the analytic continuation of the $g$-function \eqref{eq:gfunctionrank1}.

Let us first discuss the prefactor containing $\Theta_a$. Since the structure of this factor is the same as that for the energy, one can simply follow the arguments above. Namely we first perform integration by parts to $\Theta_a$ to get
\beq\label{eq:logprefactor}
\begin{aligned}
\log ({\tt prefactor})&\equiv \int^{\infty}_{0}\frac{du}{2\pi}\Theta_a (u)\log (1+Y(u))\\
&=-\frac{\log (1+Y(0))}{2} +\int_0^{\infty}\frac{du}{2\pi i}\log \left(\tilde{R}^{(L)}_a (u)\sqrt{\tilde{S}(u,-u)}\right) \frac{\del_u Y(u)}{1+Y(u)}\period
\end{aligned}
\eeq
Perhaps the square root of $\tilde{S}(u,u)$ deserves explanation:  When we perform integration by parts, we need to rewrite $\left.i \del_u \log \tilde{S}(u,v)\right|_{v=-u}$ in \eqref{eq:defofthetasimple} using the parity symmetry,
\beq
\left.i \del_u \log \tilde{S}(u,v)\right|_{v=-u}=\frac{i}{2}\del_u \log \tilde{S}(u,-u)\comma
\eeq
where, on the right hand side, the derivative acts on both arguments. This is the origin of the square root in \eqref{eq:logprefactor}. Then, after the analytic continuation, we get
\beq
\log ({\tt prefactor})= \sum_{k}\log\left(\tilde{R}_a^{(L)}(\tilde{u}_k) \sqrt{\tilde{S} (\tilde{u}_k,-\tilde{u}_k)}\right)+\int^{\infty}_{0}\frac{du}{2\pi} \Theta_a (u)\log (1+\log Y(u))\period
\eeq
To recast it in a standard form, we express the first term using rapidities in the physical channel. This can be achieved by using the relation \eqref{eq:universalrelation}, and we get
\beq
({\tt prefactor})=\left(\prod_{k} \frac{F_a (u_k)}{\sqrt{S(u_k,-u_k)}}\right)\exp\left[\int^{\infty}_{0}\frac{du}{2\pi}\Theta_a (u)\log (1+Y(u))\right]\comma
\eeq
where we introduced a ``form factor'' in the physical channel by\fn{Normally $F_a (u)$ is denoted by ``$K$'' in the literature. However, we chose to use this notation in order to avoid the clash of notations.}
\beq
F_a(u_k)\equiv \tilde{R}^{(L)}_a(\tilde{u}_k)\period
\eeq
As will be discussed in section \ref{sec:bootstrap}, this form factor satisfies the so-called Watson relation, $S(u_k,-u_k)=F_a (u_k)/F_a (-u_k)$. Using this relation, we arrive at the final expression for the prefactor
\beq
({\tt prefactor})=\sqrt{\prod_{k} F_a (u_k)F_a (-u_k)}\exp\left[\int^{\infty}_{0}\frac{du}{2\pi}\Theta_a (u)\log (1+Y(u))\right]\period
\eeq

Let us next discuss the ratio of Fredholm determinants. The idea is simple; the Fredholm determinants are defined through integral kernels whose actions are given in \eqref{eq:actionGplusexplicit} and \eqref{eq:actionGminusexplicit}. If we perform the analytic continuation, poles of the kernel cross the contour and deform the action on functions. To be more explicit, let us recall the action of the integral kernel $\hat{G}_{\pm}$:
\beq
\hat{G}_{\pm}\cdot f (u)=\int_0^{\infty}\frac{dv}{2\pi}\frac{\mathcal{K}_{\pm}(u,v)}{1+1/Y(v)} f(v)\comma
\eeq
As can be seen from this, the kernel can have poles at $1+Y(\tilde{u})=0$. If such poles cross the contour, its action gets deformed as
\beq
\hat{G}_{\pm}^{\bullet}\cdot f (u)=\sum_{k}\frac{i \mathcal{K}_{\pm}(u,\tilde{u}_k)}{\del_u \log Y(\tilde{u}_k)}f(\tilde{u}_k)+\int_0^{\infty}\frac{dv}{2\pi}\frac{\mathcal{K}_{\pm}(u,v)}{1+1/Y(v)} f(v)\comma
\eeq
where the superscript $\bullet$ signifies that the kernel was analytically continued. This can also be expressed in terms of the finite-volume phase factor $\phi$ in \eqref{eq:finitevolumephase} as\fn{Note that a similar kernel played a key role in the recent study on the diagonal finite volume form factor in the sinh-Gordon model \cite{Bajnok:2019yik}, which is based on the hidden Grassmann structure \cite{Jimbo:2010jv} and generalizes earlier results for the ground state \cite{Negro:2013wga}. One can bring our results to a form closer to theirs by first performing the integral convolution and reducing it to a finite-dimensional determinant \cite{Jiang:2019zig}.}
\beq\label{eq:actionofGbullet}
\hat{G}_{\pm}^{\bullet}\cdot f (u)=\sum_{k}\frac{\mathcal{K}_{\pm}(u,\tilde{u}_k)}{\del_{u}\phi(u_j)}f(\tilde{u}_k)+\int_0^{\infty}\frac{dv}{2\pi}\frac{\mathcal{K}_{\pm}(u,v)}{1+1/Y(v)} f(v)\comma
\eeq

Combining the two factors, we conjecture that the excited-state $g$-function is given by the following simple formula:
\beq\label{eq:excitedfinitevolumeconj}
g_a =\exp\left[\int^{\infty}_{0}\frac{du}{2\pi}\Theta_a (u)\log (1+Y(u))\right]\sqrt{\prod_{k=1}^{M/2} F_a (u_k)F_a (-u_k)\frac{\det \left[1-\hat{G}_{-}^{\bullet}\right]}{\det \left[1-\hat{G}_{+}^{\bullet}\right]}}\period
\eeq
A conjecture for the excited state $g$-function was put forward previously in \cite{Kostov:2018dmi}. However, their expression does not match with what we wrote here: First, their formula does not include $F_a$ factors. Second, the action of their kernel does not contain the first term on the right hand side of \eqref{eq:actionofGbullet}. As we see below, these two features are essential in order to reproduce the structures of the results in section \ref{sec:weak}. Because of this, we think the results in \cite{Kostov:2018dmi} are incomplete.

 Let us finally explain how \eqref{eq:excitedfinitevolumeconj} reproduces the structure that we observed at weak coupling of $\mathcal{N}=4$ SYM in section \ref{sec:weak}. To see the connection, we again take the infinite volume limit $L\to \infty$. As mentioned already, in the infinite volume limit, the $Y$-function is exponentially small. We can therefore simply drop the first term in \eqref{eq:excitedfinitevolumeconj}. To see what happens for the ratio of determinants, it is useful to rewrite it as
 \beq\label{eq:ratioasymptoticlimit}
 \frac{\det \left[1-\hat{G}_{-}^{\bullet}\right]}{\det \left[1-\hat{G}_{+}^{\bullet}\right]} =\exp\left[\sum_{n}\frac{1}{n}\left({\rm Tr}\left[(G_{+}^{\bullet})^{n}\right]-{\rm Tr}\left[(G_{-}^{\bullet})^{n}\right]\right)\right]\period
 \eeq
 We can then compute each term on the right hand side as iterated integrals and sums. In the infinite volume limit, the terms that involve integrals become exponentially small due to the suppression coming from the $Y$-function. Therefore the result is given purely by iterated sums,
 \beq\label{eq:iteratedsums}
 {\rm Tr}\left[(G_{\pm}^{\bullet})^{n}\right]=\sum_{k_1,\ldots,k_n}\frac{\mathcal{K}_{\pm}(\tilde{u}_{k_1},\tilde{u}_{k_2})}{\del_u \phi (u_{k_1})}\frac{\mathcal{K}_{\pm}(\tilde{u}_{k_2},\tilde{u}_{k_3})}{\del_u \phi (u_{k_2})}\cdots \frac{\mathcal{K}_{\pm}(\tilde{u}_{k_n},\tilde{u}_{k_1})}{\del_u \phi (u_{k_n})}\period
 \eeq
 Now, the first important observation is that one can re-express the kernels $\mathcal{K}_{\pm}$ in terms of physical rapidities $u_k$'s as follows:
 \beq
 \begin{aligned}
 \mathcal{K}_{\pm}(\tilde{u}_k,\tilde{u}_j)=&\frac{1}{i}\left[\del_{u_k} \log \tilde{S}(\tilde{u}_k,\tilde{u}_j)\pm \del_{u_k}\log \tilde{S}(\tilde{u}_k,-\tilde{u}_j)\right]\\
 \overset{\text{\eqref{eq:universalrelation}}}{=}&-\frac{1}{i}\left[\del_{u_k} \log S(u_j,u_k)\pm \del_{u_k}\log S(u_j,-u_k)\right]\\
 =&\frac{1}{i}\left[\del_{u_k} \log S(u_k,u_j)\mp \del_{u_k}\log S(u_k,-u_j)\right]\equiv \mathcal{K}_{\mp} (u_k,u_j)\period
 \end{aligned}
 \eeq
 Here in the second equality we used \eqref{eq:universalrelation} while in the third equality we used the unitarity $S(u,v)=1/S(v,u)$ for the first term and the parity $S(u,-v)=S(v,-u)$ for the second term. Surprisingly, this rewriting swaps $\mathcal{K}_{+}$ and $\mathcal{K}_{-}$! This is precisely what is needed to relate the $g$-function formula and the results at weak coupling in $\mathcal{N}=4$ SYM.

 The second important observation is that the iterated sum \eqref{eq:iteratedsums} can be regarded as a trace of a product of finite dimensional matrices. From this, it immediately follows that the ratio \eqref{eq:ratioasymptoticlimit} reduces in the limit to a ratio of finite-dimensional matrix determinants. Using the asymptotic form of $\phi(u)$ (see the discussion below \eqref{eq:finitevolumephase}) and multiplying the common factors $\del_u \phi (u_k)$ to the numerator and the denominator, we finally obtain the following expression,
 \beq
 \frac{\det \left[1-\hat{G}_{-}^{\bullet}\right]}{\det \left[1-\hat{G}_{+}^{\bullet}\right]}\overset{L\to\infty}{=}\frac{\det G_{+}}{\det G_{-}}\comma
 \eeq
 with
 \beq
 \begin{aligned}
 \left(G_{\pm}\right)_{ij}\equiv \underbrace{\left[L\del_u p(u_i)+\sum_{k}\mathcal{K}_{+}(u_i,u_k)\right]}_{\del_u\phi (u_i)}\delta_{ij}-\mathcal{K}_{\pm}(u_i,u_j)\period
 \end{aligned}
 \eeq

 Combining the two factors, we finally obtain the excited-state $g$-function in the infinite volume limit,
 \beq\label{eq:toyasymptotic}
 \frac{\langle B|{\bf u}\rangle}{\sqrt{\langle {\bf u}|{\bf u}\rangle}}\overset{L\to\infty}{=}\sqrt{\prod_{k=1}^{M/2} F_a (u_k)F_a (-u_k)\frac{\det G_{+}}{\det G_{-}}}\period
 \eeq
 This beautifully reproduces the structure that we observed in section \ref{sec:weak}!

 So far, we have been discussing a toy example with a single species of particles and no bound states. In addition, the match with the weak-coupling answer is only qualitative since we did not specify the form factor $F_a$. In the subsequent sections, we apply the general formalism of TBA and $g$-function to our problem, by determining the form factor $F_a$ explicitly and generalizing the results to theories with multiple particles and bound states.
\section{Bootstrapping the Boundary State\label{sec:bootstrap}}
We now set out to apply the framework discussed in section \ref{sec:integrability} to our problem; the three-point function of two determinant operators and one non-BPS single-trace operator. As the first step, we determine the reflection matrix at finite 't Hooft coupling using the symmetry and integrability.
\subsection{Kinematics, form factor and reflection matrix\label{subsec:formVSreflection}}
Before discussing the details on the reflection matrix, it is useful to elaborate on the relation between the form factor and the reflection matrix, specializing to the case of the spin chain for $\mathcal{N}=4$ SYM.
\subsubsection{Kinematics and dispersion relation}
In order to explain our convention, let us give a brief review of the kinematics and the dispersion relation of the $\mathcal{N}=4$ SYM spin chain.

At finite coupling, the momentum and the energy of magnons are parametrized by the rapidity $u$ in the following way\fn{Note that here we are using the convention $E=(\Delta-J)/2$, which is natural on the gauge-fixed string worldsheet.},
\beq
e^{ip}=\frac{x^{+}}{x^{-}}\comma\qquad \qquad E=\frac{1}{2}\frac{1+\frac{1}{x^{+}x^{-}}}{1-\frac{1}{x^{+}x^{-}}}\comma
\eeq
where $x(u)$ is the Zhukowsky variable defined by
\beq\label{eq:defZhukovsky}
u=g \left(x(u)+\frac{1}{x(u)}\right)\quad \iff\quad x(u)=\frac{u+\sqrt{u^2-4g^2}}{2g}\comma
\eeq
and the notation $f^{\pm}$ means $f(u\pm i/2)$.
The coupling constant $g$ is defined by
\beq
g^2\equiv \frac{\lambda}{16\pi^2}\period
\eeq
As can be seen from \eqref{eq:defZhukovsky}, the momentum and the energy contain two branch cuts in the rapidity plane, one for $x^{+}$ and the other for $x^{-}$. If we cross those branch cuts, the corresponding Zhukovsky variable gets inverted as $x(u)\to 1/x(u)$. This property allows us to define the crossing and the mirror transformation: Consider the analytic continuation in which one crosses both cuts once. In this process (to be denoted by $2\gamma$), the Zhukovsky variables get transformed as
\beq
\text{\bf Crossing}:\quad x^{+}(u^{2\gamma})=1/x^{+}(u)\comma\qquad x^{-}(u^{2\gamma})=1/x^{-}(u)\period
\eeq
As shown above, this can be interpreted as the crossing transformation, which maps a particle to an antiparticle (or equivalently an incoming particle to an outgoing particle). We can see this explicitly from the transformations of the momentum and the energy:
\beq
e^{ip(u^{2\gamma})}=\frac{x^{-}}{x^{+}}=e^{-ip(u)}\comma\qquad E(u^{2\gamma})=\frac{1}{2}\frac{1+x^{+}x^{-}}{1-x^{+}x^{-}}=-E(u)\period
\eeq

\begin{figure}[t]
\centering
\includegraphics[clip,height=5cm]{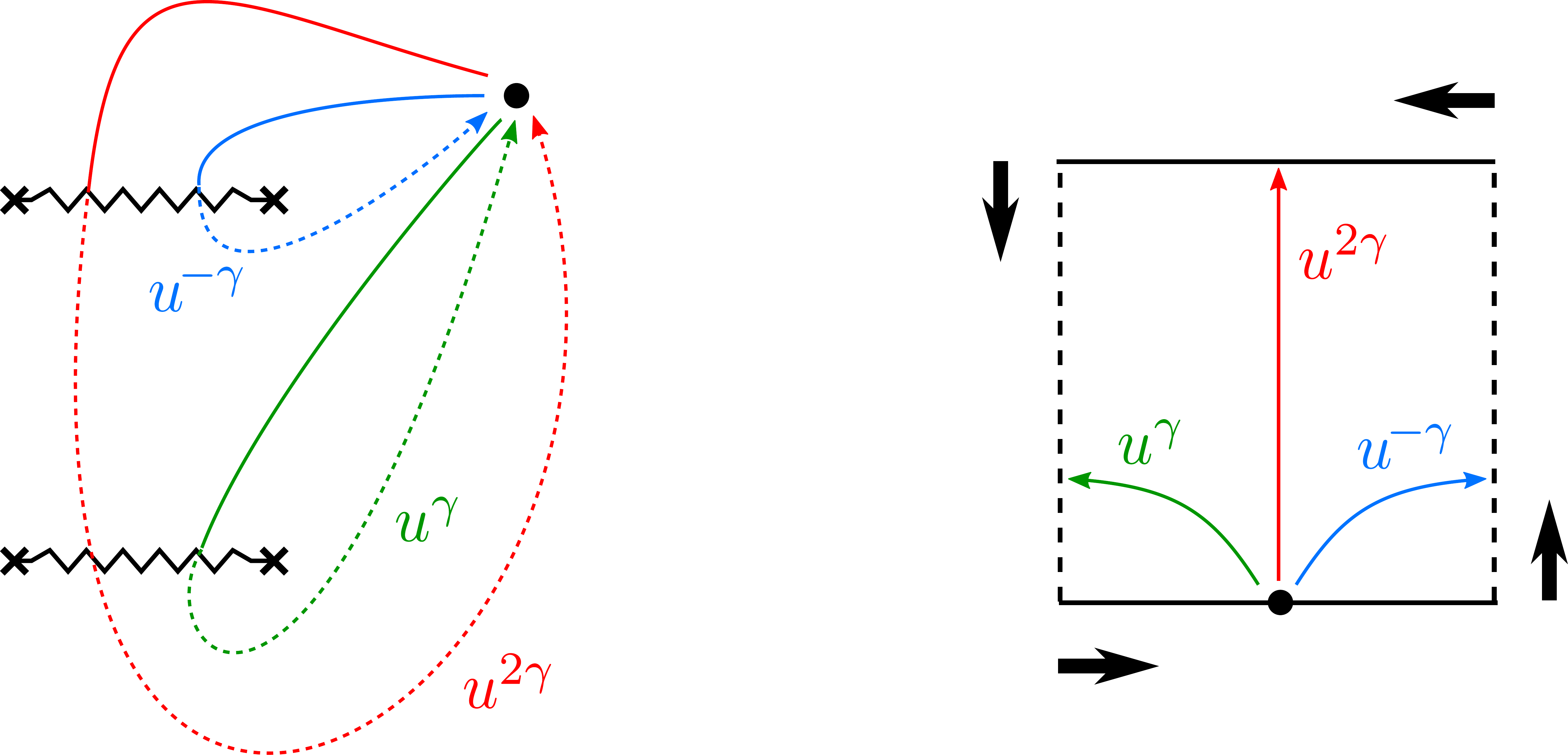}
\caption{The crossing and the mirror transformations. The left figure shows the analytic continuation in the rapidity plane while the right figure shows the spacetime interpretation of each manipulation. In the left figure, the upper branch cut comes from $x^{-}$ while the lower branch cut comes from $x^{+}$. In the right figure, we also indicated the direction of space coordinates in each channel by thick black arrows.}
\label{fig:fig21}
\end{figure}

 One can also perform a ``half'' of this transformation in which one crosses only one of the two branch cuts. Such processes correspond to the mirror transformations in which one maps a magnon in the physical channel to a magnon in the mirror channel:
\beq
\begin{aligned}
\text{\bf Mirror}_{+}:&\quad x^{+}(u^{\gamma})=1/x^{+}(u)\comma\qquad x^{-}(u^{\gamma})=x^{-}(u)\comma\\
\text{\bf Mirror}_{-}:&\quad x^{+}(u^{-\gamma})=x^{+}(u)\comma\qquad x^{-}(u^{-\gamma})=1/x^{-}(u)\period
\end{aligned}
\eeq
For a pictorial explanation, see figure \ref{fig:fig21}. The energy and momentum of the mirror theory are defined through the mirror transformations as\fn{See also \eqref{eq:doublewickrotation}.}
\beq
\tilde{E}(u)\equiv- ip(u^{\gamma})=\log \left( x^{+}(u)x^{-}(u)\right)\comma\qquad \qquad \tilde{p}(u)\equiv -i E(u^{\gamma})= \frac{1}{2i} \frac{1+\frac{x^{+}(u)}{x^{-}(u)}}{1-\frac{x^{+}(u)}{x^{-}(u)}}\period
\eeq

In order to discuss the boundary scattering, we also need the parity transformation. We denote it by $u\to \bar{u}$ and define it as follows:
\beq
\text{\bf Parity}: \quad x^{+}(\overline{u})=-x^{-}(u)\comma\qquad x^{-}(\overline{u})=-x^{+}(u)\period
\eeq
For a magnon in the physical channel, the parity transformation is simply given by $\overline{u}=-u$, but we choose to use this notation since it allows us to distinguish different orders of manipulations such as $\overline{u^{\gamma}}$ and $\overline{u}^{\gamma}$ (see also the discussion in the next paragraph). One can check explicitly that it gives the correct momentum and energy for the parity-transformed state both in the physical and the mirror channels:
\beq
\begin{aligned}
&e^{ip (\overline{u})}=\frac{x^{-}}{x^{+}}=e^{-ip(u)}\comma\qquad &&E(\overline{u})=\frac{1}{2}\frac{1+\frac{1}{x^{+}x^{-}}}{1-\frac{1}{x^{+}x^{-}}}=E(u)\comma\\
&\tilde{p}(\overline{u})=-iE(\overline{u}^{\gamma})=-\tilde{p}(u)\comma\qquad &&\tilde{E}(\overline{u})=-ip(\overline{u}^{\gamma})=\tilde{E}(u)\period
\end{aligned}
\eeq

\begin{figure}[t]
\centering
\includegraphics[clip,height=6cm]{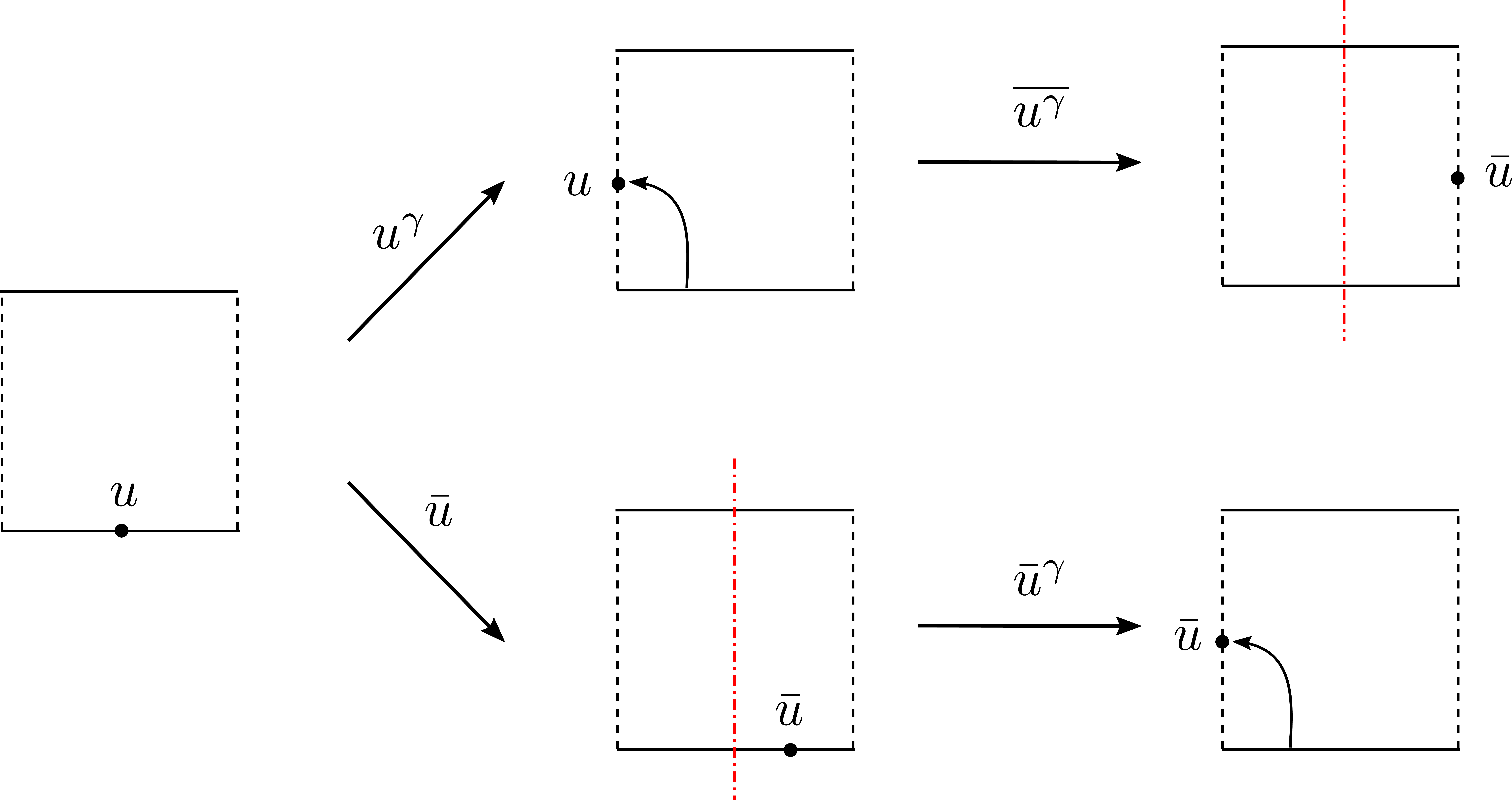}
\caption{Noncommutativity of the mirror transformation and the parity transformation. The red dashed line denotes the reflection axis for the parity transformation. If we first perform the mirror transformation and then perform the parity transformation, we follow the upper path and the particle ends up on the right edge. On the other hand, if we perform the transformations in the opposite order, the particle ends up on the left edge.}
\label{fig:fig22}
\end{figure}

It is worth emphasizing that {\it the parity transformation and the mirror transformation do not commute}. One can see this explicitly by comparing the Zhukovsky variables $x^{\pm}(\overline{u^{\gamma}})$ and $x^{\pm}(\overline{u}^{\gamma})$\fn{In the former case, we perform first the mirror transformation and then the parity transformation while in the latter case, we apply the transformations in the opposite order.
}:
\beq
\begin{aligned}
&x^{+}(\overline{u^{\gamma}})=-x^{-}(u^{\gamma})=-x^{-}(u)\comma\qquad &&x^{-}(\overline{u^{\gamma}})=-x^{+}(u^{\gamma})=-1/x^{+}(u)\comma\\
&x^{+}(\overline{u}^{\gamma})=1/x^{+}(\overline{u})=-1/x^{-}(u)\comma\qquad &&x^{-}(\overline{u}^{\gamma})=x^{-}(\overline{u})=-x^{+}(u)\period
\end{aligned}
\eeq
By inspecting the transformations of the Zhukovsky variables, one can in fact show that
\beq
\overline{u^{\gamma}}=\overline{u}^{-\gamma}\comma\qquad \overline{u^{-\gamma}}=\overline{u}^{\gamma}\period
\eeq
The relation can also be understood pictorially; see figure \ref{fig:fig22}.

In relativistic field theories, the counterparts of the manipulations that we described here are given by
\beq
\begin{aligned}
\text{\bf Crossing}: \quad \theta^{2\gamma} = \theta +\pi i\comma\qquad \text{\bf Mirror}_{\pm}:\quad \theta^{\pm \gamma} = \theta\pm\frac{i\pi}{2}\comma\qquad \text{\bf Parity}:\quad \overline{\theta} = -\theta\period
\end{aligned}\nn
\eeq
Also here we can see that the crossing and the mirror transformations do not commute and the relation $\overline{\theta^{\gamma}}=\overline{\theta}^{-\gamma}$ is satisfied.
\subsubsection{Form factor and reflection matrix for $\mathcal{N}=4$ SYM spin chain}
In section \ref{subsec:overlaptoreflection}, we gave a qualitative explanation for the relation between the form factor and the boundary reflection. We now specialize the discussion to $\mathcal{N}=4$ SYM taking into account the index structure and the charge conjugation.
\paragraph{Magnons for $\mathcal{N}=4$ SYM}
Elementary magnons in the $\mathcal{N}=4$ SYM spin chain transform in bifundamental representations under the PSU$(2|2)^2$ symmetry of the BPS two-point function \cite{Beisert:2005tm,Beisert:2006qh}. To manifest the structure of the group and the representation, it is customary to represent magnons as
\beq
\mathcal{X}_{\bf A}=\chi_A \dot{\chi}_{\dot{A}}\comma
\eeq
where ${\bf A}\equiv A\dot{A}$ is a collective notation of the indices. Here $\chi$ and $\dot{\chi}$ are the fundamental representations of the left and the right PSU$(2|2)$ and consist of two bosonic ($\varphi$) and two fermionic ($\psi$) components:
\beq
\chi_{A}=(\varphi^1 ,\varphi^2,\psi^1,\psi^2)\comma\qquad \dot{\chi}_{\dot{A}}=(\dot{\varphi}^1 ,\dot{\varphi}^2,\dot{\psi}^1,\dot{\psi}^2)\period
\eeq
They are related to the fields in $\mathcal{N}=4$ SYM as
\beq
\begin{aligned}
&\varphi^{1}\dot{\varphi}^{1}=X\comma\quad \varphi^{1}\dot{\varphi}^{2}=Y\comma\quad \varphi^{2}\dot{\varphi}^{1}=\bar{Y}\comma\quad\varphi^{2}\dot{\varphi}^{2}=-\bar{X}\comma\\
&\psi^{\alpha}\dot{\psi}^{\dot{\alpha}}=D^{\alpha\dot{\alpha}}Z\qquad \qquad \psi^{\alpha}\dot{\varphi}^{\dot{a}}\comma \,\dot{\varphi}^{a}\psi^{\dot{\alpha}}=\text{fermion}\period
\end{aligned}
\eeq
Here $D$ is a covariant derivative and $X$, $\bar{X}$, $Y$, $\bar{Y}$ and $Z$ are complex scalars defined in \eqref{eq:defcomplexscalars}. We also denote scalars as $\Phi^{a\dot{a}}(\varphi^{a}\dot{\varphi}^{\dot{a}})$ in the rest of this paper.

\begin{figure}[t]
\centering
\includegraphics[clip,height=3.5cm]{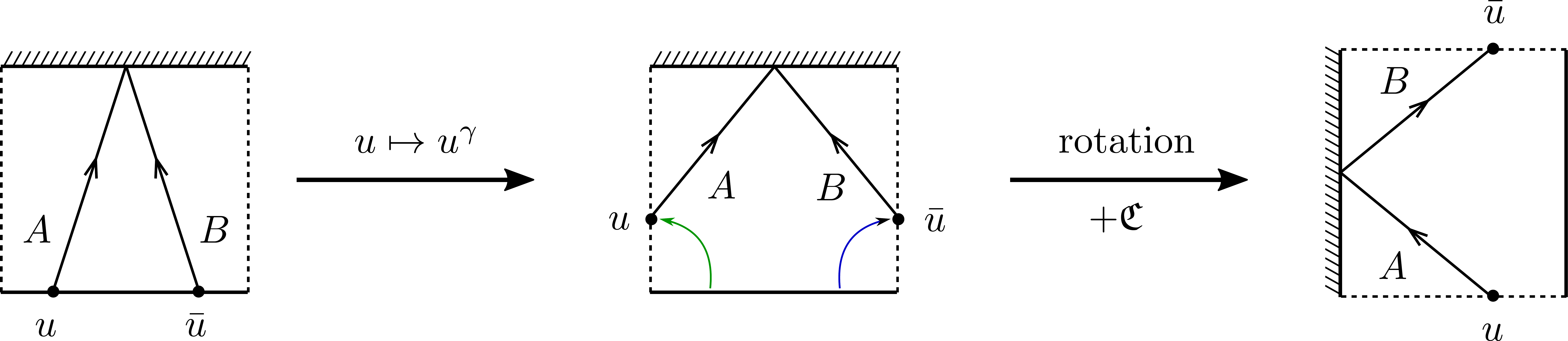}
\caption{The relation between the form factor and the reflection matrix. After the mirror transformation, the particle with rapidity $u$ moves to the left edge while the particle with rapidity $\bar{u}$ moves to the right edge. If we further perform the charge conjugation (and rotate the whole picture by 90 degrees), we end up with a standard reflection process.}
\label{fig:fig23}
\end{figure}

\paragraph{Reflection matrix from form factor}To see the relation between the form factor and the reflection matrix, let us start with the following form factor depicted in figure \ref{fig:fig23},
\beq
F_{{\bf A}{\bf B}}(u)=\langle \mathcal{G}|\mathcal{X}_{\bf A}(u)\mathcal{X}_{\bf B}(\bar{u})\rangle\comma
\eeq
where $\langle \mathcal{G}|$ is a boundary state describing the Giant Gravitons. We then perform the mirror transformation $u\to u^{\gamma}$ to get
\beq
F_{{\bf A}{\bf B}}(u^{\gamma})=\langle \mathcal{G}|\mathcal{X}_{\bf A}(u^{\gamma})\mathcal{X}_{\bf B}(\bar{u}^{-\gamma})\rangle\comma
\eeq
As shown in figure \ref{fig:fig23}, after the transformation, the first particle can be regarded as a particle on the left edge with rapidity $u$ while the second particle can be regarded as a particle on the right edge with rapidity $\bar{u}$. Rotating the whole picture by 90 degrees, we can then interpret this process as a reflection process in which the particle  with rapidity $u$ scatters off the left boundary. To read off the reflection matrix, there is one more step to go: We need to flip the orientation of the arrow for the outgoing particle. This amounts to performing the charge conjugation to the index of the outgoing particle ${\bf B}$. As a result, we arrive at the relation
\beq
\left[R_L(u)\right]_{{\bf A}}^{{\bf B}}=F_{{\bf A}{\bf C}}(u^{\gamma})\mathfrak{C}^{{\bf C}{\bf B}}\comma
\eeq
with $\mathfrak{C}^{{\bf A}{\bf B}}$ being the charge conjugation matrix. Switching back to the bi-fundamental notation, we get
\beq\label{eq:relationformreflection}
\left[R_{L}(u)\right]_{A\dot{A}}^{B\dot{B}}=F_{A\dot{A},C\dot{C}}(u^{\gamma})\mathcal{C}^{CB}\mathcal{C}^{\dot{C}\dot{B}}\period
\eeq
where $\mathcal{C}^{AB}$ is the charge conjugation matrix of a single PSU$(2|2)$ which reads
\beq
\mathcal{C}^{AB}=\pmatrix{cc}{i\epsilon_{ab}&0\\0&\epsilon_{\alpha\beta}}\period
\eeq

The reflection matrix $R_{R}(u)$ at the right boundary can be obtained from $R_{L}(u)$ by using the parity transformation
\beq
R_{R}(u)=R_{L}(\bar{u})=\left(R_{L}(u)\right)^{-1}\period
\eeq
We thus focus on the left reflection matrix in what follows. We should however keep in mind that
the left reflection matrix always enters the Bethe equation in a form $R_{L}(\bar{u})=\left(R_{L}(u)\right)^{-1}$. See for instance \eqref{eq:boundarybetheeqtoy}.
\subsubsection{Form factor axioms}
In \cite{Ghoshal:1993tm}, several conditions satisfied by the integrable reflection matrix were written down. In what follows, we translate them into the form-factor language since the latter formulation turns out to be easier to deal with in our setup.

\begin{figure}[t]
\centering
\includegraphics[clip,height=5cm]{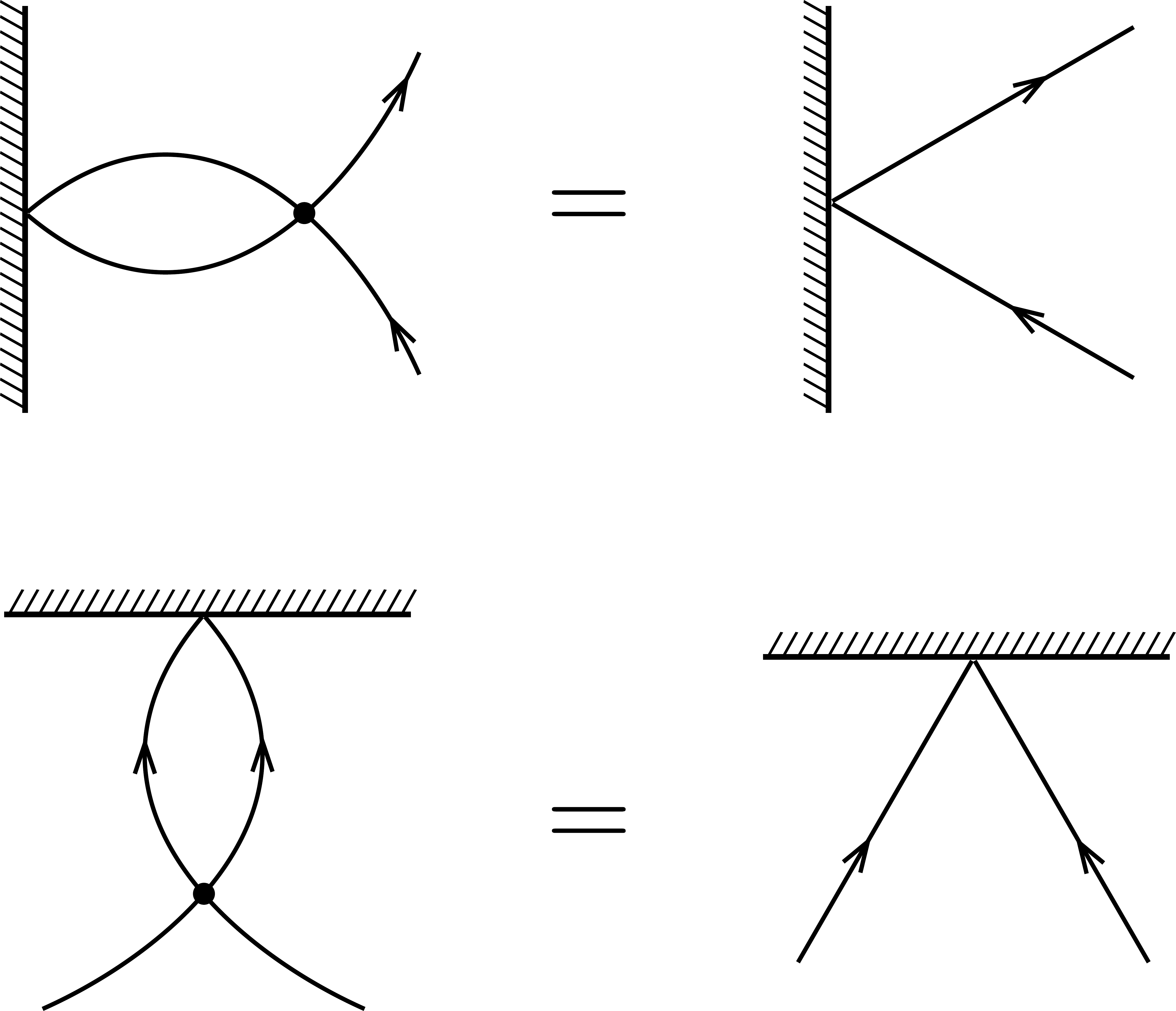}
\caption{The boundary crossing relation and the Watson equation. The boundary crossing relation in the reflection picture (the upper figure) is equivalent to the Watson equation in the form factor picture (the lower figure).}
\label{fig:fig24}
\end{figure}

\begin{figure}[t]
\centering
\includegraphics[clip,height=6cm]{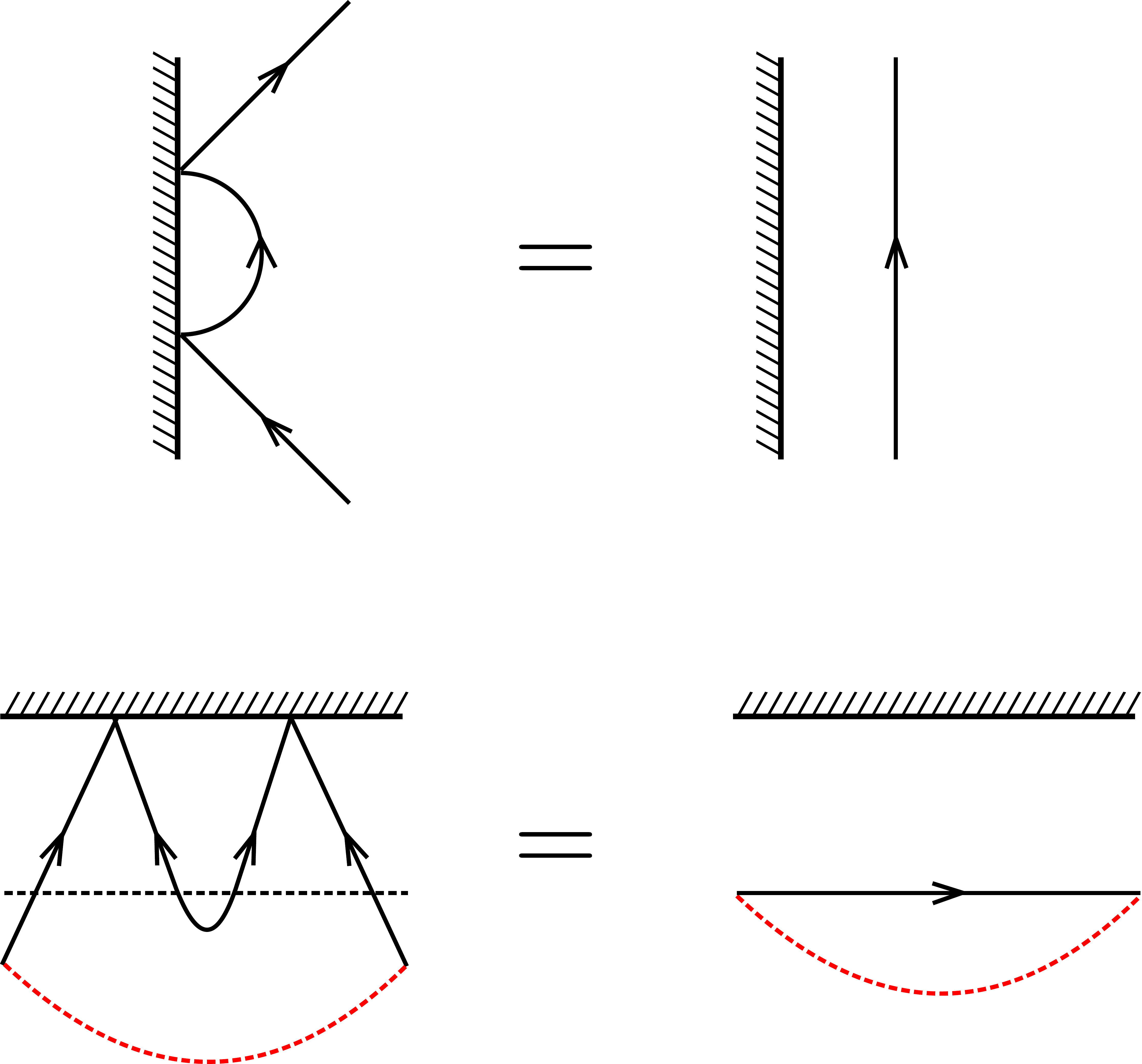}
\caption{The boundary unitarity and the decoupling condition. By cutting the lower figure along the black dashed line, we can see that the boundary unitarity in the reflection picture (the upper figure) is equivalent to the decoupling condition for a {\it pair} of particle-antiparticle pairs in the form factor picture (the lower figure). The extra red dashed curve denotes the contraction with $\delta_{\bf C}^{\bf A}$.}
\label{fig:fig25}
\end{figure}

\paragraph{Boundary crossing $=$ Watson}
One of the conditions is called the boundary crossing relation, which is given by figure \ref{fig:fig24}. As was already noted in \cite{Ghoshal:1993tm}, this condition takes a simpler form in the form factor picture. Although the condition was called ``boundary cross-unitarity'' in \cite{Ghoshal:1993tm}, this is nothing but the well-known Watson equation, which states that the permutation of particles generates the bulk S-matrix. Written explicitly, the relation reads
\beq
\begin{aligned}
&\langle \mathcal{G}|\mathcal{X}_{\bf A}(u)\mathcal{X}_{\bf B}(\bar{u})\rangle =\langle \mathcal{G}|\mathbb{S}|\mathcal{X}_{\bf A} (u)\mathcal{X}_{\bf  B} (\bar{u})\rangle\quad \iff\quad F_{{\bf A}{\bf B}}(u)=   \mathbb{S}_{{\bf A}{\bf B}}^{{\bf C}{\bf D}}(u,\bar{u})F_{{\bf C}{\bf D}}(\bar{u})\period
\end{aligned}
\eeq
where $\mathbb{S}$ is the two-particle (bulk) S-matrix. See figure \ref{fig:fig24}.
\paragraph{Boundary unitarity $=$ Decoupling}
The second important condition is the boundary unitarity condition, which states that
\beq
[R_L (u)]^{{\bf A}}_{{\bf B}} [R_L(\bar{u})]^{\bf B}_{\bf C}=\delta^{\bf A}_{\bf C}\period
\eeq
Using the relation between the form factor and the reflection matrix (see figure \ref{fig:fig25}), one can rewrite this in terms of $F$ in the following way:
\beq\label{eq:decouplingN=4SYM}
F_{{\bf A}{\bf B}}(u^{\gamma})
\mathfrak{C}^{{\bf B}{\bf B}^{\prime}}F_{{\bf B}^{\prime}{\bf C}^{\prime}}(\bar{u}^{\gamma})\mathfrak{C}^{{\bf C}^{\prime}{\bf C}}=\delta_{\bf A}^{\bf C}\quad \iff \quad F_{{\bf A}{\bf B}}(u)
\mathfrak{C}^{{\bf B}{\bf B}^{\prime}}F_{{\bf B}^{\prime}{\bf C}^{\prime}}(\bar{u}^{2\gamma})\mathfrak{C}^{{\bf C}^{\prime}{\bf C}}=\delta_{\bf A}^{\bf C}\period
\eeq
To understand its physical meaning, it is useful to contract both sides with $\delta^{{\bf A}}_{{\bf C}}$, and express it as a pairwise contracted four-particle form factor:
\beq
\begin{aligned}
F_{{\bf A}{\bf B}}(u)
\mathfrak{C}^{{\bf B}{\bf B}^{\prime}}F_{{\bf B}^{\prime}{\bf C}^{\prime}}(\bar{u}^{2\gamma})\mathfrak{C}^{{\bf C}^{\prime}{\bf A}}=&\mathfrak{C}^{{\bf B}{\bf B}^{\prime}}\mathfrak{C}^{{\bf C}^{\prime}{\bf A}}\langle\mathcal{G}|\mathcal{X}_{{\bf A}}(u)\mathcal{X}_{\bf B}(\bar{u})\mathcal{X}_{{\bf B}^{\prime}}(\bar{u}^{2\gamma})\mathcal{X}_{{\bf C}^{\prime}}(u^{-2\gamma})\rangle\\
=&\delta_{\bf A}^{\bf C}\delta_{\bf C}^{\bf A}\period
\end{aligned}
\eeq
Now the crucial observation is that the index contractions and the rapidities of the particles are arranged such that these pairs individually form {\it singlets} which carry zero net charge under any symmetry. Therefore, the boundary unitarity simply states that adding {\it a pair} of singlets decouple from the rest and do nothing to the form factor. This is physically reasonable since the singlets are always produced by the vacuum fluctuation and should not have any physical consequence. This property of the singlet was previously used to formulate the crossing equation for the S-matrix \cite{Janik:2006dc,Beisert:2006qh} and for the hexagon form factor \cite{Basso:2015zoa}.

\paragraph{Remark on the decoupling condition}
In the form factor of a local operator or in the hexagon form factor, the ``decoupling'' condition involves taking residue since the form factor diverges when a singlet is formed. This divergence is an IR divergence: Initially, particles are attached to a point in space (in the case of local operators) or to a half space (in the case of hexagon form factors). When a particle and an anti-particle form a singlet, they decouple from the operator and start moving in the full space. This produces a divergence which is proportional to the volume of space (which is infinite).

\begin{figure}[t]
\centering
\includegraphics[clip,height=2.5cm]{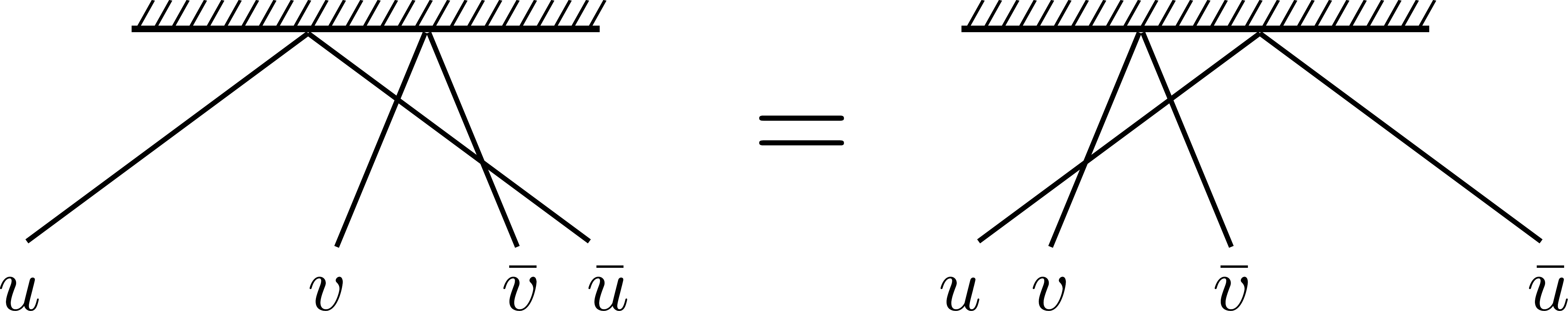}
\caption{The boundary Yang-Baxter equation in the form factor picture.}
\label{fig:fig26}
\end{figure}

On the other hand, the decoupling equation we derived does not involve taking residue. In fact, our form factor is completely finite even if we have a singlet. This is due to the fact that the boundary state is fully non-local and translationally invariant: Because of this, even when there are no singlets, particles can explore the full space. Thus, if we defined the form factor of a boundary state in the same way as the form factor of a local operator, there would always be a divergence proportional to the volume. To define a finite quantity, we need to divide it by the volume factor. What we are calling ``the form factor of the boundary state'' is in fact the quantity after division. This explains the reason why we do not encounter any extra divergence in the decoupling condition.
\paragraph{Boundary Yang-Baxter equation}
In a similar vein, one can translate the boundary Yang-Baxter equation to the form factor language. It simply becomes the relation between two different ways of computing the four-particle form factors:
\beq
\langle \mathcal{G}|\mathbb{S}_{24}\mathbb{S}_{34}|\mathcal{X}_{1}(u)\mathcal{X}_{2}(v)\mathcal{X}_{3}(\bar{v})\mathcal{X}_{4}(\bar{u})\rangle=\langle \mathcal{G}|\mathbb{S}_{13}\mathbb{S}_{12}|\mathcal{X}_{1}(u)\mathcal{X}_{2}(v)\mathcal{X}_{3}(\bar{v})\mathcal{X}_{4}(\bar{u})\rangle\period
\eeq
Here $\mathbb{S}_{ij}$ is a two-particle S-matrix between $\mathcal{X}_{i}$ and $\mathcal{X}_j$. See also a pictorial explanation in figure \ref{fig:fig26}.

\subsection{Bootstrapping the matrix structure\label{subsec:matrixpart}}
We now determine the matrix structure of the form factor by requiring the symmetry, Watson equation and the boundary Yang-Baxter relation.
\subsubsection{Symmetry constraints}
The matrix structure of the form factor is constrained by the underlying symmetry. As explained in section \ref{subsec:setup}, the symmetry of three-point functions is the PSU$(2|2)_{D}$ symmetry, which is the diagonal subgroup of PSU$(2|2)^2$ symmetry governing the spin chain of $\mathcal{N}=4$ SYM. A crucial insight by Beisert is that this PSU$(2|2)^2$ symmetry gets centrally extended once we consider an infinitely long spin chain and cut open the trace. Physically, the central charges are the analogues of ``asymptotic symmetry''; namely they are field-dependent gauge transformations which do not die off at spatial infinity of the spin chain\fn{See for example \cite{Komatsu:2017buu} for explanation on this point.}.

It turns out that the PSU$(2|2)_{D}$ symmetry of our problem is also centrally extended. This can be seen by computing the anticommutators of the fermionic charges
\beq
\mathcal{Q}^{\alpha}{}_{a}=Q^{\alpha}{}_{a}+i\kappa \epsilon^{\alpha\dot{\beta}}\epsilon_{a\dot{b}}\dot{S}^{\dot{b}}{}_{\dot{\beta}}\comma\qquad \mathcal{S}^{a}{}_{\alpha}=S^{a}{}_{\alpha}+\frac{i}{\kappa}\epsilon^{a\dot{b}}\epsilon_{\alpha\dot{\beta}}\dot{Q}^{\dot{\beta}}{}_{\dot{b}}\comma
\eeq
which gives
\beq
\begin{aligned}
&\{\mathcal{Q}^{\alpha}{}_{a}\comma \mathcal{Q}^{\beta}{}_{b}\}=\epsilon^{\alpha\beta}\epsilon_{ab}\mathcal{P}\comma\qquad
\{\mathcal{S}^{a}{}_{\alpha}\comma \mathcal{S}^{b}{}_{\beta}\}=\frac{1}{\kappa^2}\epsilon_{\alpha\beta}\epsilon^{ab}\mathcal{P}\period
\end{aligned}
\eeq
Here $\mathcal{P}$ is a linear combination of the central charges for PSU$(2|2)^2$,
\beq
\mathcal{P}=P-\kappa^2 K\period
\eeq

We can then impose the invariance of this centrally extended PSU$(2|2)_D$ symmetry to relate different matrix elements. In practice this amounts to imposing
\beq
\langle \mathcal{G}|t|\mathcal{X}_{\bf A}(u)\mathcal{X}_{\bf B}(\bar{u})\rangle=0\comma \qquad t\in \mathfrak{psu}(2|2)_{\rm D}+\mathcal{P}\comma
\eeq
with $\mathfrak{psu}(2|2)_{D}$ being the Lie algebra of PSU$(2|2)_{D}$.

Since the underlying symmetry is identical, the analysis is quite similar to the one for the hexagon form factor. However, there is one important difference: In the case of the hexagon form factor, the invariance under $\mathcal{P}$ provided a nontrivial constraint which allows us to read off the rules to pull out the so-called $\mathcal{Z}$-markers. On the other hand, in our problem, the action of $\mathcal{P}$ reads
\beq
\begin{aligned}
&\langle \mathcal{G}|\mathcal{P}|\mathcal{X}(u)\mathcal{X}(\bar{u})\rangle=\\
&g\underbrace{\left(1-\frac{x^{+}(u)x^{+}(\bar{u})}{x^{-}(u)x^{-}(\bar{u})}\right)}_{=0}\langle \mathcal{G}|Z\mathcal{X}(u)\mathcal{X}(\bar{u})\rangle+g\kappa^2\underbrace{\left(1-\frac{x^{-}(u)x^{-}(\bar{u})}{x^{+}(u)x^{+}(\bar{u})}\right)}_{=0}\langle \mathcal{G}|Z^{-}\mathcal{X}(u)\mathcal{X}(\bar{u})\rangle\comma
\end{aligned}
\eeq
which identically vanishes. Because of this, one needs to impose a rule to pull out $\mathcal{Z}$-markers separately, which we write in the following way:
\beq\label{eq:pullingoutZ}
\langle \mathcal{G}|\mathcal{Z}\mathcal{X}(u)\mathcal{X}(\bar{u})\rangle=i\kappa z
\langle \mathcal{G}|\mathcal{X}(u)\mathcal{X}(\bar{u})\rangle\comma
\eeq
where $z$ is an unfixed constant and $z=\pm 1$ corresponds to the solution for the hexagon form factor. With this extra rule, one can proceed in the same way as the hexagon form factor and determine all the matrix elements as a function of $z$ up to an overall scalar factor.

The result of the analysis has the following structure:
\beq\label{eq:ansatzfullwithg}
\langle \mathcal{G}|\mathcal{X}_{A\dot{A}}(u)\mathcal{X}_{B\dot{B}}(\bar{u})\rangle=g_0(u) \times (-1)^{|\dot{A}||B|}\times M[\mathcal{X}_{A\dot{A}}\mathcal{X}_{B\dot{B}}](u)\period
\eeq
Here $g_0(u)$ is an unfixed overall scalar factor and $|\dot{A}|$ and $|B|$ are the fermion numbers of each index. The matrix part $M[\mathcal{X}_{A\dot{A}}\mathcal{X}_{B\dot{B}}]$ is given by
\beq\label{eq:matrixpartwithz}
\begin{aligned}
M[\Phi^{a\dot{a}}\Phi^{b\dot{b}}]&=\frac{1}{2}\left(A_{\mathcal{G}}+B_{\mathcal{G}}\right)\epsilon^{a\dot{b}}\epsilon^{b\dot{a}}+\frac{1}{2}\left(A_{\mathcal{G}}-B_{\mathcal{G}}\right)\epsilon^{a\dot{a}}\epsilon^{b\dot{b}}\comma\\
M[D^{\alpha\dot{\alpha}}D^{\beta\dot{\beta}}]&=\frac{1}{2}\left(D_{\mathcal{G}}+E_{\mathcal{G}}\right)\epsilon^{\alpha\dot{\beta}}\epsilon^{\beta\dot{\alpha}}+\frac{1}{2}\left(D_{\mathcal{G}}-E_{\mathcal{G}}\right)\epsilon^{\alpha\dot{\alpha}}\epsilon^{\beta\dot{\beta}}\comma\\
M[\Phi^{a\dot{a}}D^{\beta\dot{\beta}}]&=G_{\mathcal{G}}\,\,\epsilon^{a\dot{a}}\epsilon^{\beta\dot{\beta}}\comma\\
M[D^{\alpha\dot{\alpha}}\Phi^{b\dot{b}}]&=L_{\mathcal{G}}\,\,\epsilon^{\alpha\dot{\alpha}}\epsilon^{b\dot{b}}\comma\\
M[\Psi^{a\dot{\alpha}}\Psi^{b\dot{\beta}}]&=C_{\mathcal{G}}\,\,\epsilon^{ab}\epsilon^{\dot{\alpha}\dot{\beta}}\comma\\
M[\Psi^{a\dot{\alpha}}\bar{\Psi}^{\beta\dot{b}}]&=H_{\mathcal{G}}\,\,\epsilon^{a\dot{b}}\epsilon^{\beta\dot{\alpha}}\comma\\
M[\bar{\Psi}^{\alpha\dot{a}}\Psi^{b\dot{\beta}}]&=K_{\mathcal{G}}\,\,\epsilon^{b\dot{a}}\epsilon^{\alpha\dot{\beta}}\comma\\
M[\bar{\Psi}^{\alpha\dot{a}}\bar{\Psi}^{\beta\dot{b}}]&=F_{\mathcal{G}}\,\,\epsilon^{\dot{a}\dot{b}}\epsilon^{\alpha\beta}\comma
\end{aligned}
\eeq
with
\beq
\begin{aligned}\label{eq:finalwithz}
&A_{\mathcal{G}}=1\comma\quad
B_{\mathcal{G}}=-\frac{x^{-}+z^4(x^{+})^3}{ z^2x^{+}(1+x^{+}x^{-})} \comma\quad E_{\mathcal{G}}=\frac{z^4x^{+}+(x^{-})^3}{z^2 x^{-}(1+x^{+}x^{-})}\\
 &C_{\mathcal{G}}=F_{\mathcal{G}}=-\frac{i\left[z^4(x^{+})^2-(x^{-})^2\right]}{2z^2\sqrt{x^{+}x^{-}}(1+x^{+}x^{-})}\comma\quad D_{\mathcal{G}}=-1\comma\\
&G_{\mathcal{G}}=L_{\mathcal{G}}=\frac{x^{-}+z^2x^{+}}{2 z\sqrt{x^{+}x^{-}}}\comma\quad H_{\mathcal{G}}=K_{\mathcal{G}}=\frac{x^{-}-z^2x^{+}}{2 z\sqrt{x^{+}x^{-}}}\period
\end{aligned}
\eeq
As expected, the matrix part reduces to the one for the hexagon form factor \cite{Basso:2015zoa} upon setting $z=\pm 1$. However, at this point, the value of $z$ is completely arbitrary\fn{It can even depend on the rapidity $u$.}.
\subsubsection{Watson equation and boundary Yang-Baxter equation}
Another constraint comes from the Watson equation
\beq
\langle \mathcal{G}|\mathcal{X}_{\bf A}(u)\mathcal{X}_{\bf B}(\bar{u})\rangle =\langle \mathcal{G}|\mathbb{S}|\mathcal{X}_{\bf A} (u)\mathcal{X}_{\bf  B} (\bar{u})\rangle\period
\eeq
If we plug our ansatz \eqref{eq:matrixpartwithz} to this equation, we get separate constraints for the matrix part $M$ and the scalar factor $g_0$. The constraint on $g_0$ is rather simple and reads
\beq\label{eq:watsonN=4g}
\frac{g_0 (u)}{g_0(\bar{u})}=S_0 (u,\bar{u})\comma
\eeq
where $S_0$ is the scalar factor for the bulk S-matrix (see Appendix \ref{ap:Smatrix} for an explicit expression). The constraints of the matrix part are rather complicated and we will not write them down, but it turns out that all of them are satisfied regardless of the value of $z$.

The analysis so far does not assume that the boundary state is integrable. However, we know, from the results at weak coupling in section \ref{sec:weak} and the argument presented in section \ref{sec:integrability}, that the relevant boundary state is likely to be an integrable boundary state. Thus, below we assume that it is true and impose the most significant consequence of the boundary integrability, namely the boundary Yang-Baxter equation. The analysis is straightforward but somewhat laborious. We therefore simply state the outcome:

\noindent {\it The boundary Yang-Baxter equation is satisfied only when $z=+i$ or $-i$.}

\noindent This in particular excludes the solutions corresponding to the hexagon form factor, $z=\pm 1$.

To summarize, the matrix part for our problem is given by \eqref{eq:finalwithz} with $z=\pm i$. In section \ref{subsec:unfolding}, we provide a more intuitive understanding of why this solution satisfies the boundary Yang-Baxter equation.
\subsubsection{Physical interpretation}
Having determined the matrix part, let us now pause and try to understand the physical meaning of the solution we obtained. As shown in \eqref{eq:pullingoutZ}, the parameter $z$ originates from the rule for pulling out $\mathcal{Z}$ markers. Although not completely justified, we could imagine repeatedly applying this rule to a chain of $J$ $\mathcal{Z}$ markers and write\fn{Precisely speaking, there is also an extra $i^{J}$ factor in front. This is due to the fact that our $\mathcal{Z}$ marker does not correspond to an insertion of a $Z$-field. It rather corresponds to an insertion of $i Z$. One should be able to see this by carefully comparing the transformation laws of the supercharges, but we will not attempt to do so in this paper.}
\beq
\langle \mathcal{G}|\underbrace{\mathcal{Z}\cdots \mathcal{Z}}_{J}\rangle\sim (z\kappa)^{J}\langle \mathcal{G}|\varnothing \rangle\comma
\eeq
where $\langle \mathcal{G}|\varnothing \rangle$ is the three-point function of Giant Gravitons and a ``length-$0$'' operator, which is basically the two-point function of Giant Gravitons and normalized to be $1$. Upon setting $z=\pm i$, this yields a factor $(i\kappa)^{J}$ and $(-i\kappa)^{J}$.

We already encountered something similar: In the computation in the SU(2) sector at weak coupling in section \ref{subsec:su2tree}, the final result was given by a sum of two terms, each of which is proportional to $(i\kappa)^{J}$ and $(-i\kappa)^{J}$. There, the two contributions arose from the $2\times 2$ structure of the matrix trace, which upon diagonalization gave two eigenvalues proportional to $i\kappa$ and $-i\kappa$. This strongly suggests that what we are seeing here is a finite-coupling version of the same phenomenon. Namely, we interpret the two choices $z=\pm i$ as representing the two eigenvalues of the matrix trace. As discussed in section \ref{subsec:interpretation}, the matrix trace can be interpreted as a gauge-invariant observable of open string field theory on Giant Gravitons.
To guarantee the gauge invariance, we always need to take a trace, namely sum the two eigenvalue contributions.
This also implies that the three-point function corresponds to a sum of two boundary states,
\beq
\langle \mathcal{G}|_{\rm true}=\langle \mathcal{G}|_{z=+i}+\langle \mathcal{G}|_{z=-i}\period
\eeq
In what follows, we assume that this is the case and sum the two contributions when we write down the final result.

Given this observation, it would be interesting to ask if the existence of the two solutions for the hexagon form factor, $z=\pm 1$, has any physical consequence. In particular, it is worth exploring if extra signs observed for the single-trace three-point functions involving fermions in \cite{Caetano:2016keh} could be explained by this. Note also that a similar $\mathbb{Z}_2$ structure showed up in the computation of four-point functions \cite{Fleury:2016ykk}.
\subsection{Unfolding picture\label{subsec:unfolding}}
The solution for the matrix part turns out to be related to the Beisert's S-matrix in a simple way. In particular it admits an ``unfolding'' interpretation as is the case with the reflection matrix for the cusped Wilson loop.

To see this, it is useful to go from the form factor picture  to the boundary reflection picture using \eqref{eq:relationformreflection}. By inspecting the action of the reflection matrix spelled out in Appendix \ref{ap:reflection}, one finds that the following identity holds\fn{Here we wrote the result for $z=i$. For $z=-i$, one needs to flip the roles of the dotted indices and undotted indices. We also omitted writing prefactors which arise from the redefinition of the basis and reordering of the particles. See Appendix \ref{ap:reflection} for all these technicalities.}
\beq\label{eq:unfoldingformula}
[R_{L}(u)]^{B\dot{B}}_{A\dot{A}}\left(=[R_{R}(\bar{u})]^{B\dot{B}}_{A\dot{A}}\right)=r_0(\bar{u}) \times \mathcal{S}_{A\dot{A}}^{\dot{B}B}(\bar{u}^{\gamma},u^{\gamma})\comma
\eeq
where the scalar factor $r_0(\bar{u})$ is given by
\beq\label{eq:defroref}
r_0 (\bar{u})\equiv g_0(u^{\gamma})\times \frac{(x^{-}+1/x^{-})(1+x^{+}x^{-})}{2(x^{+}+x^{-})}\period
\eeq
while $\mathcal{S}$ is the matrix part of Beisert's PSU(2$|$2) S-matrix (for a definition, see Appendix \ref{ap:Smatrix}). Note also that we defined $r_0$ so that the scalar factor for the {\it right} reflection matrix is $r_0(u)$.

\begin{figure}[t]
\centering
\includegraphics[clip,height=3.5cm]{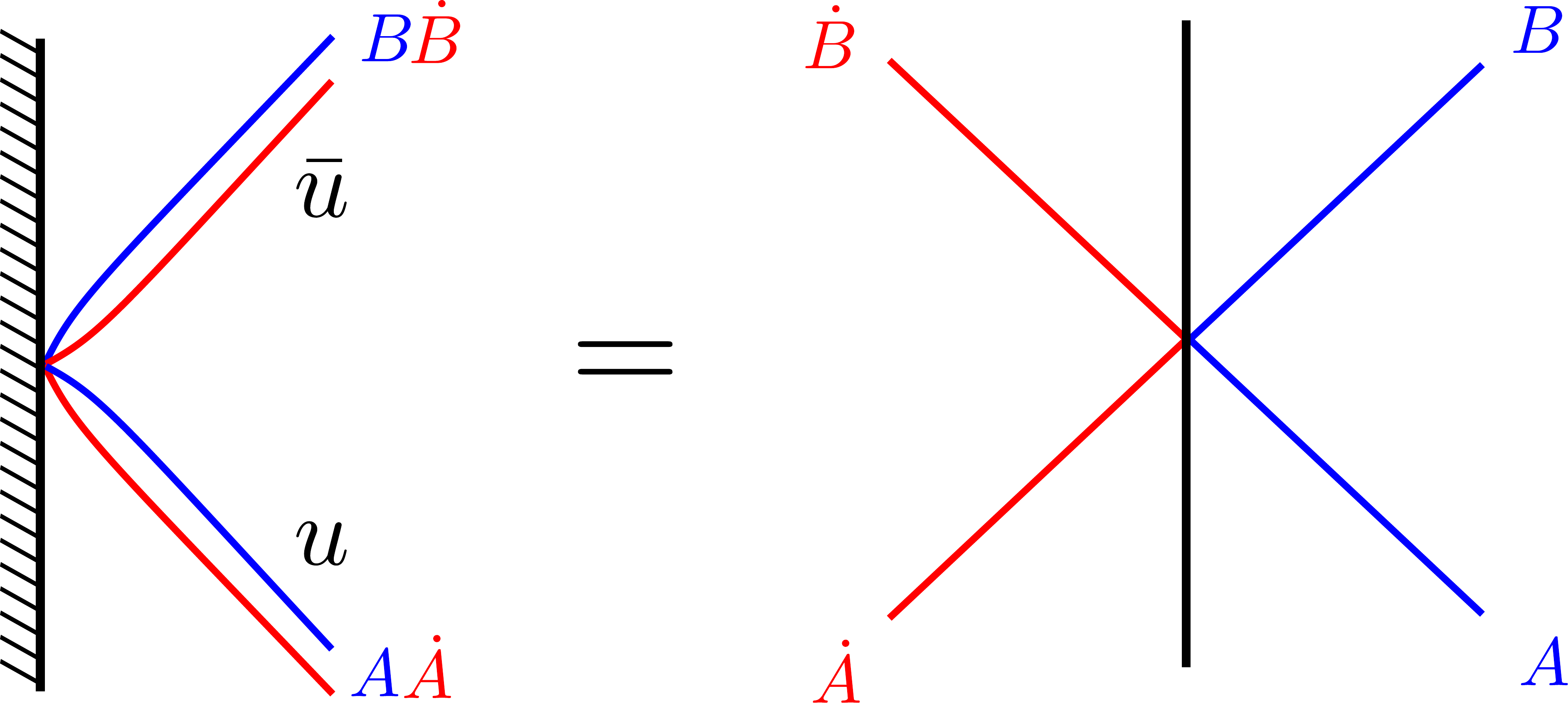}
\caption{The unfolding interpretation of the reflection matrix. One can regard the reflection process of a single magnon with PSU$(2|2)^2$ indices as the scattering process of two magnons, each with a single PSU$(2|2)$ index.}
\label{fig:fig27}
\end{figure}

Remarkably, the right hand side of \eqref{eq:unfoldingformula} is precisely the ``square-root'' of the bulk S-matrix in the mirror channel. This property allows us to perform the ``unfolding trick'', as was done in \cite{Drukker:2012de,Correa:2012hh} for the cusped Wilson loop. Namely, we can view the reflection process as a scattering process on a spin chain with a single PSU$(2|2)_D$ symmetry, in which a magnon with rapidity $u$ scatters with a magnon with rapidity $\bar{u}$. See figure \ref{fig:fig27}.

This also makes it trivial to see why the boundary Yang-Baxter equation is satisfied. After unfolding, it is simply a consequence of the standard Yang-Baxter equation for Beisert's S-matrix. As we see in section \ref{sec:TBA}, this unfolding property also makes it easier to write down the asymptotic Bethe ansatz in the mirror channel and determine the TBA equations.

\begin{figure}[t]
\centering
\includegraphics[clip,height=3.3cm]{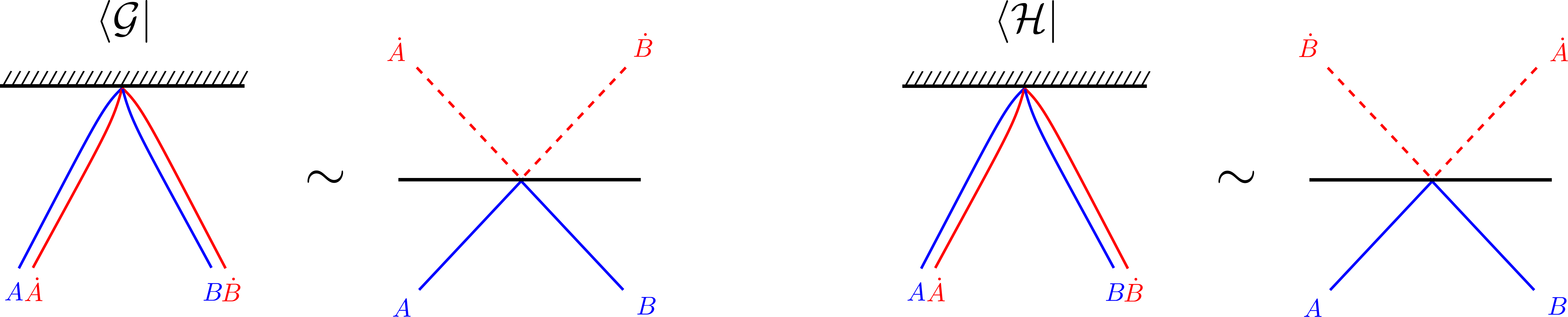}
\caption{The comparison of the matrix structures for the Giant Graviton form factor (left) and the hexagon form factor (right). The two are related by a swap of the dotted indices as shown in the figure. Because of this difference, the matrix structure for the hexagon form factor does not admit a simple interpretation as an unfolded reflection process. }
\label{fig:fig28}
\end{figure}

Before concluding our discussion on the matrix part, let us make one clarifying remark. At first sight, the fact that the reflection matrix is proportional to Beisert's S-matrix might seem like a trivial consequence of the symmetry; the symmetry of our problem is the centrally-extended PSU$(2|2)$, which we know to be strong enough to uniquely fix the matrix structure. However, it is a bit too hasty to draw this conclusion. As we saw above, the constrained kinematics of the boundary scattering allows for extra solutions to the symmetry constraints and we need to pick a right solution to see the unfolding property. It is also amusing to note that the solution we did not pick---the one corresponding to the hexagon form factor---has a similar unfolding property in the {\it physical channel} while the one we picked exhibits a nice property in the {\it mirror channel}. See figure \ref{fig:fig28} for the comparison.

\subsection{Bootstrapping boundary dressing phase\label{subsec:bootstrapphase}}
We now sketch the derivation of the overall scalar factor $g_0 (u)$ in \eqref{eq:ansatzfullwithg}. Details of the computation can be found in Appendix \ref{ap:crossing}.
\paragraph{Ansatz for Watson equation}
The first constraint on $g_0(u)$ comes from the Watson equation \eqref{eq:watsonN=4g}, which reads
\beq
\frac{g_0(u)}{g_0 (\bar{u})}=S_0 (u,\bar{u})=\left(\frac{x^{+}}{x^{-}}\right)^2\frac{x^{-}+\frac{1}{x^{-}}}{x^{+}+\frac{1}{x^{+}}}\frac{1}{\left(\sigma (u,\bar{u})\right)^2}\comma
\eeq
where $\sigma (u,v)$ is the dressing phase for the bulk S-matrix.
This can be solved explicitly by the following simple ansatz:
\beq
g_0 (u)=\frac{x^{+}}{x^{-}}\frac{u-\frac{i}{2}}{u}\frac{\sigma_B(u)}{\sigma (u,\bar{u})}\comma
\eeq
Here $\sigma_B(u)$ is an unfixed function satisfying
\beq\label{eq:Watsonsigma}
\sigma_B(u)=\sigma_B(\bar{u})\period
\eeq
In what follows, we call $\sigma_B(u)$ {\it the boundary dressing phase}.
\paragraph{Decoupling condition}
To determine $\sigma_B(u)$, we impose the decoupling condition \eqref{eq:decouplingN=4SYM}. Since the equation involves the summation over indices, in general one needs to consider a sum of different states. However, for special choices of external indices, we do not need to perform a sum. One such choice is the following four-particle form factor
\beq
\langle \mathcal{G}| \psi^{1}\dot{\psi}^{1}(u)\quad \psi^2 \dot{\psi}^2(\bar{u})\quad \psi^1\dot{\psi}^1(\bar{u}^{2\gamma})\quad \psi^2 \dot{\psi}^2(u^{-2\gamma})\rangle\period
\eeq
The decoupling condition equates this form factor to $1$. Plugging in our ansatz and simplifying the answer (see Appendix \ref{ap:crossing} for details), we get the {\it crossing equation} for $\sigma_B(u)$,
\beq
\sigma_B(u)\sigma_B (u^{2\gamma})=\frac{u}{u+i/2}\frac{u}{u-i/2}\period
\eeq
\paragraph{Minimal solution}
We now solve the crossing equation using the techniques developed in related contexts \cite{Volin:2009uv}. Our strategy is to first assume the following ansatz,
\beq
\sigma_B^{\rm min} (u)=\frac{G(x^{+})}{G(x^{-})}\comma
\eeq
derive a functional equation for $G(x)$ and solve it. The details of the computation can be found in Appendix \ref{ap:crossing}. As a result we obtain the following {\it minimal solution} to the crossing equation:
\beq\label{eq:Gxlargex}
\begin{aligned}
&\sigma^{\rm min}_B(u) =\frac{G(x^{+})}{G(x^{-})}\comma\\
&G(x)=\exp \left[\oint \frac{dz}{2\pi i}\frac{1}{x-z}\log \mathfrak{G}(g(z+z^{-1}))\right]\comma\qquad \mathfrak{G}(u)\equiv\frac{\Gamma \left(\frac{1}{2}-iu\right)\Gamma \left(1+iu\right)}{\Gamma \left(\frac{1}{2}+iu\right)\Gamma \left(1-iu\right)}\period
\end{aligned}
\eeq
This expression for $G(x)$ is valid only when $|x|>1$. For $|x|<1$, the result is given by
\beq\label{eq:Gxsmallx}
G(x)=\frac{\Gamma \left(\frac{1}{2}-iu\right)\Gamma \left(1+iu\right)}{\Gamma \left(\frac{1}{2}+iu\right)\Gamma (1-iu)}\exp \left[\oint \frac{dz}{2\pi i}\frac{1}{x-z}\log \mathfrak{G}(g(z+z^{-1}))\right] \qquad |x|<1\period
\eeq
One can explicitly check that the minimal solution $\sigma^{\rm min}(u)$ satisfies both the crossing equation and the Watson equation \eqref{eq:Watsonsigma}.
\paragraph{CDD ambiguity} The minimal solution $\sigma^{\rm min}(u)$ gives one particular solution to the crossing equation. This however is not the unique solution: One can construct an infinite family of solutions by multiplying to $\sigma_{B}^{\rm min}(u)$ a factor $\sigma_{\rm CDD}(u)$ which satisfies
\beq
\sigma_{\rm CDD}(u)=\sigma_{\rm CDD}(\bar{u})\comma\qquad \sigma_{\rm CDD}(u)\sigma_{\rm CDD}(u^{2\gamma})=1\period
\eeq
This is the boundary analogue of the famous Castillejo-Dalitz-Dyson (CDD) ambiguity \cite{Castillejo:1955ed} for the bulk S-matrix. A particularly simple class of CDD factors are given by
\beq
\sigma_{\rm CDD}(u)=e^{f(u)}\comma
\eeq
where $f(u)$ is any odd function of the magnon energy
\beq
E(u)=\frac{1}{2}\frac{1+\frac{1}{x^{+}x^{-}}}{1-\frac{1}{x^{+}x^{-}}}\period
\eeq
It turns out that the consistency with the weak-coupling answers forces us to pick one such a factor
\beq\label{eq:ourCDD}
\sigma_{\rm CDD}(u)=2^{-4E(u)}=\exp\left(-2\log 2\times \frac{1+\frac{1}{x^{+}x^{-}}}{1-\frac{1}{x^{+}x^{-}}}\right)\period
\eeq
\paragraph{Full solution and expansions}
Putting things together, we arrive at the following expression for the dressing phase,
\beq\label{eq:fullbdress}
\sigma_B(u)=2^{-4E(u)}\times \frac{G(x^{+})}{G(x^{-})}\comma
\eeq
with $G(x)$ given by \eqref{eq:Gxlargex} and \eqref{eq:Gxsmallx}. At weak and strong couplings, this dressing phase can be expanded as
\begin{align}\label{eq:ourdressexpand}
\sigma_B(u)&=\frac{1}{4}\left[1-g^{4}\frac{12}{u^2+\frac{1}{4}}\zeta_3+g^6\left(\frac{4(1-12u^3)}{(u^2+\frac{1}{4})^3}\zeta_3+\frac{120}{u^2+\frac{1}{4}}\zeta_5\right)+\cdots\right]\comma\\
\log \sigma_B(u)&=O(g^{-1})\period
\end{align}
Here and below $\zeta_n$ denotes a zeta function $\zeta (n)$.
In section \ref{sec:check}, we test the result both at weak and strong couplings, confirming the validity of our solution.
\paragraph{Scalar factor for the reflection matrix} Using the result for the scalar factor for the form factor, one can compute the scalar factor for the reflection matrix defined in \eqref{eq:defroref}:
\beq\label{eq:finalr0ubar}
\begin{aligned}
r_0 (\bar{u})=&\frac{(x^{-}+1/x^{-})(1+1/x^{+}x^{-})}{2(x^{+}+x^{-})}\frac{u-\frac{i}{2}}{u}\frac{\sigma_B(u^{\gamma})}{\sigma (u^{\gamma},\bar{u}^{-\gamma})}\period
\end{aligned}
\eeq
This can be expressed in a remarkably simple form by rewriting the bulk dressing phase as
\beq
\sigma (u^{\gamma},\bar{u}^{-\gamma})=\sigma (u^{\gamma},\bar{u}^{-\gamma})\sigma (u^{\gamma},\bar{u}^{\gamma})\sigma (\bar{u}^{\gamma},u^{\gamma})=\frac{\sigma (\bar{u}^{\gamma},u^{\gamma})}{\sigma (\bar{u}^{-\gamma},u^{\gamma})\sigma (\bar{u}^{\gamma},u^{\gamma})}\comma
\eeq
and using the crossing equation of the bulk S-matrix,
\beq
\sigma (u_1,u_2)\sigma (u_1^{2\gamma},u_2)=\frac{(1-1/x^{+}_1x^{+}_2)(1-x_1^{-}/x_2^{+})}{(1-1/x^{+}_1x^{-}_2)(1-x_1^{-}/x_2^{-})}\period
\eeq
As a result we get
\beq\label{eq:finalr0u}
r_0 (\bar{u})=\frac{u-\frac{i}{2}}{u}\frac{\sigma_B(u^{\gamma})}{\sigma (\bar{u}^{\gamma},u^{\gamma})}\qquad \iff\qquad r_0 (u)=\frac{u+\frac{i}{2}}{u}\frac{\sigma_B(\bar{u}^{\gamma})}{\sigma (u^{\gamma},\bar{u}^{\gamma})}\period
\eeq
\subsection{Spacetime dependence from CDD factor\label{subsec:CDD}}
At first sight, the CDD factor $2^{-4E(u)}$ might seem like an ad hoc way to match the result with perturbative data. However, as we see below, it turns out to be deeply rooted in the spacetime physics of $\mathcal{N}=4$ SYM.

To see this, let us consider a more general solution to the crossing equation which can be obtained by multiplying an extra CDD factor,
\beq
\sigma_B(u) \mapsto \sigma^{(\alpha)}_B(u)= \alpha^{4E(u)}\sigma_B(u)\period
\eeq
We saw in the toy example in section \ref{sec:integrability} that the two-particle form factor appears multiplicatively in the final answer for the overlap \eqref{eq:toyasymptotic}. As will be discussed in section \ref{sec:asymptotic}, this is also true for $\mathcal{N}=4$ SYM at finite coupling. We therefore expect that the modification of the dressing phase changes the final answer purely by a multiplicative factor,
\beq
\mathfrak{D}_{\mathcal{O}}\mapsto \alpha^{2\sum_{k=1}^{M}E(u_k)}\mathfrak{D}_{\mathcal{O}}=\alpha^{\Delta-J}\mathfrak{D}_{\mathcal{O}}\period
\eeq
As indicated, the exponent of the multiplicative factor is $\Delta-J$ of the single-trace operator $\mathcal{O}$.

\begin{figure}[t]
\centering
\includegraphics[clip,height=4cm]{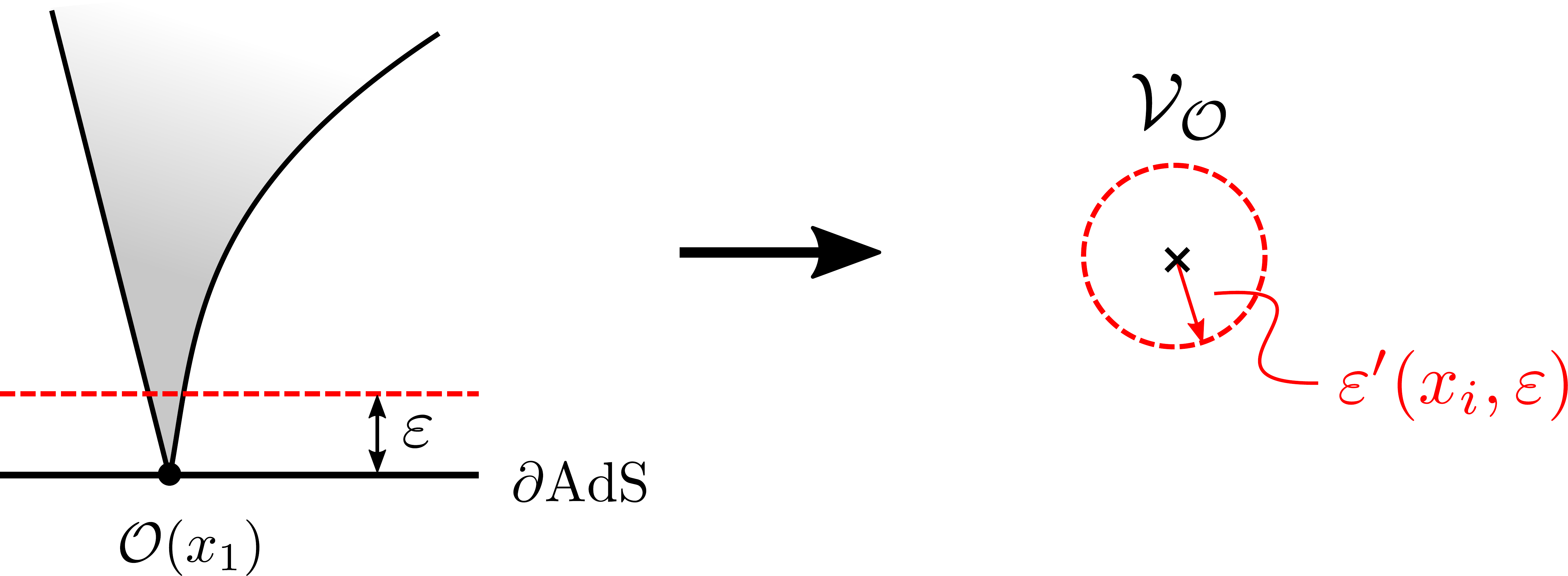}
\caption{The cut-off in AdS provides a cut-off on the worldsheet. However the relation between the two depends on the positions of operator insertions.}
\label{fig:fig29}
\end{figure}

  Now the crucial observation is that this multiplicative factor is of the same form as the factor that appears in the ratio between the three-point function and the two-point function of determinant operators given in \eqref{eq:ratio3ptand2pt},
  \beq
  \frac{\langle \mathcal{D}_1\mathcal{D}_2\mathcal{O}(0)\rangle}{\langle \mathcal{D}_1\mathcal{D}_2\rangle}\propto \left(\frac{a_2-a_1}{a_1 a_2}\right)^{\Delta-J}\mathfrak{D}_{\mathcal{O}}\period
  \eeq
  Here $a_1$ and $a_2$ are the positions of two determinant operators and we placed the single-trace operator at the origin. This means that, by multiplying an appropriate CDD factor to the boundary dressing phase, we can reproduce not just the structure constant, but also the spacetime dependence!
  A particularly interesting choice is $\alpha=2$. With this choice, the extra CDD factor {\it cancels} the CDD factor \eqref{eq:ourCDD} and the boundary dressing phase reduces to the minimal solution,
  \beq\label{eq:symmetricismin}
  \sigma_B^{(\alpha=2)}=\sigma_B^{\rm min}\period
  \eeq
  On the other hand, this choice corresponds to a ``symmetric'' configuration of the three-point function in which the determinant operators are at $a_1=1$ and $a_2=-1$.

There is a simple way to understand this relation between the CDD ambiguity and the spacetime dependence. The extra CDD factor that we introduced for the form factor translates to the following phase factor for the reflection matrix in the mirror channel:
\beq
\alpha^{4E(u)}\mapsto e^{4i \tilde{p}(u)\log \alpha}\period
\eeq
Being of the form $\exp (i\tilde{p}\,\,\ast)$, this factor can be simply absorbed by the change of the length in the mirror channel. Thus, what all these are suggesting is that the spacetime dependence is related to the change of the length in the mirror channel. This is actually what was observed in the computation of correlation functions at strong coupling in \cite{Janik:2011bd,Kazama:2011cp,Kazama:2012is}: To compute the worldsheet action, one needs to introduce a cut-off near the boundary. This {\it spacetime} cutoff translates into cutoffs on the {\it worldsheet}, but the actual values of the worldsheet cutoffs depend on the positions of the operators (see figure \ref{fig:fig29}). This effectively changes the length in the mirror channel providing a direct connection between the spacetime dependence and the length in the mirror channel. One can also explain the relation purely from field theory: See the first exercise problem in \cite{Komatsu:2017buu}.

\begin{figure}[t]
\centering
\includegraphics[clip,height=4cm]{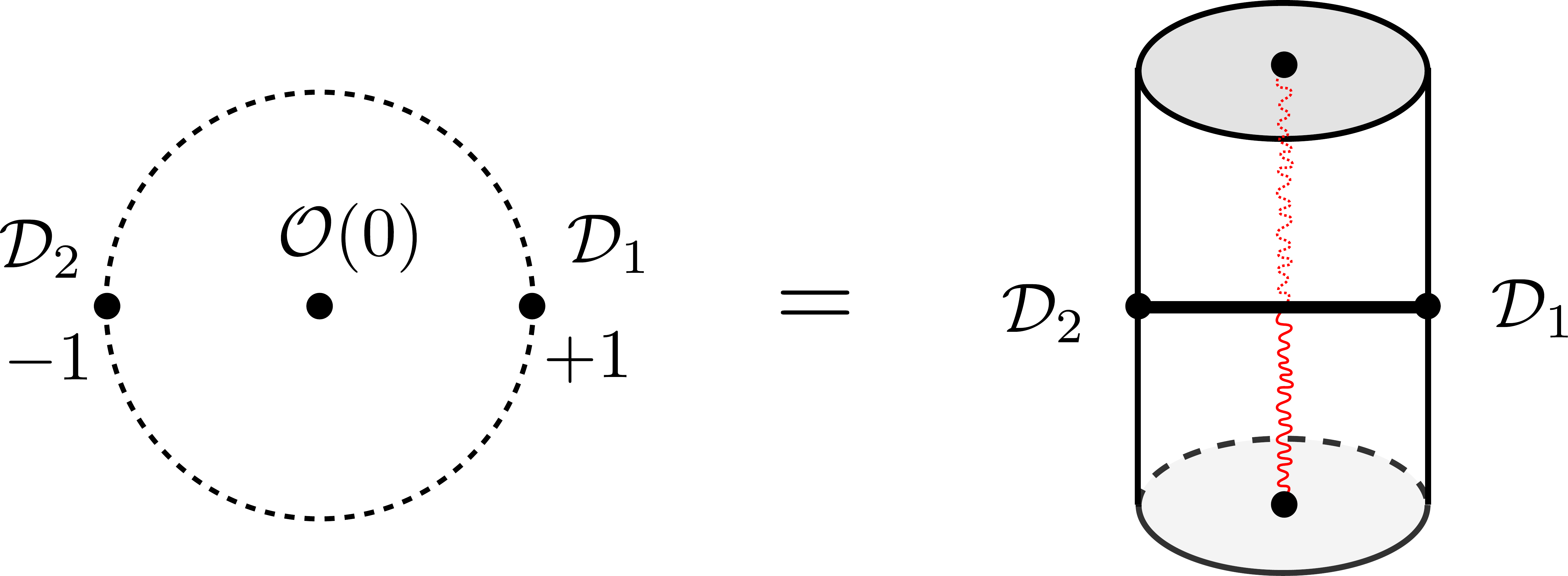}
\caption{The symmetric configuration and the AdS picture. For the symmetric configuration, the string worldsheet (drawn in the global AdS) is cut into halves by the geodesics of the Giant Graviton.}
\label{fig:fig30}
\end{figure}

This point of view also explains why the symmetric configuration $a_1=-a_2=1$ is special: In the symmetric configuration, the string worldsheet in AdS is precisely cut in the middle as shown in figure \ref{fig:fig30}. We therefore expect that the answer should be the simplest, and indeed that is what we observed in \eqref{eq:symmetricismin}.

It would be interesting to see if other observables in $\mathcal{N}=4$ SYM can have CDD factors with a natural spacetime interpretation. This is particularly so in view of recent interest in the $T\bar{T}$ deformation of the two-dimensinal QFTs \cite{Zamolodchikov:2004ce,Smirnov:2016lqw,Cavaglia:2016oda} (see \cite{Jiang:2019hxb} for a review).

\section{Asymptotic Structure Constants\label{sec:asymptotic}}
Having determined the reflection amplitude, we are now in a position to present one of the main results of this paper:
We conjecture the {\it asymptotic formula} for the structure constant at finite coupling, which is valid when the length of the operator is large. In this section, after briefly reviewing the necessary backgrounds, we simply present the conjecture postponing justification and tests of the formula to subsequent sections (sections \ref{sec:TBA} and \ref{sec:check} respectively).
\subsection{Asymptotic Bethe equation\label{subsec:ABAphysical}}
Before writing down our results, let us briefly recall the asymptotic Bethe equation at finite coupling, also known as the Beisert-Staudacher equation \cite{Beisert:2005fw}.

\begin{figure}[t]
\centering
\includegraphics[clip,height=2.5cm]{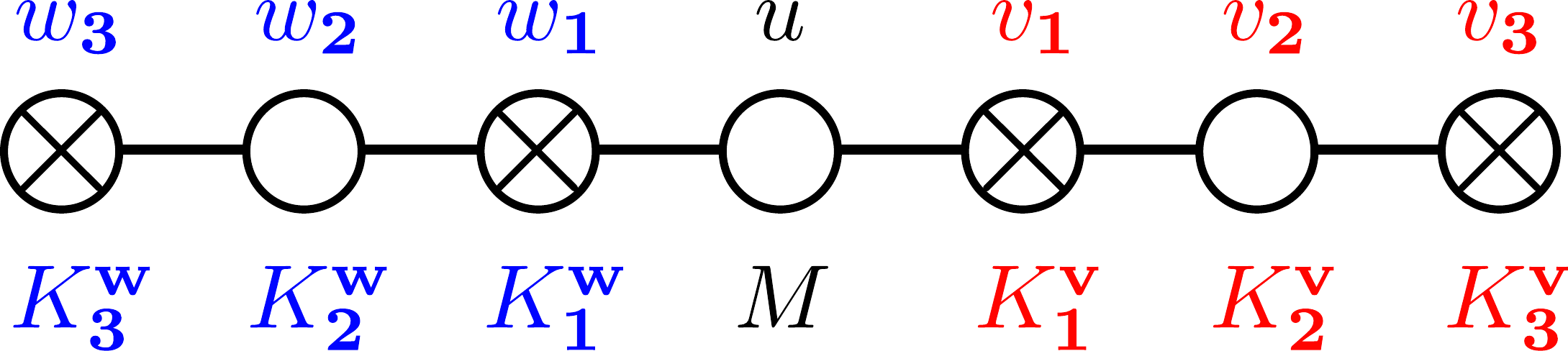}
\caption{Dynkin diagram for PSU$(2,2|4)$ and rapidities.}
\label{fig:fig31}
\end{figure}

When the length of the single-trace operator is sufficiently large, the spectrum of the operator is governed by the Bethe equation even at finite coupling. The equation at finite coupling was derived in \cite{Beisert:2005fw}, and it consists of seven sets of rapidities, $\{\blue{{\bf w}_{\III,\II,\I}},{\bf u},\red{{\bf v}_{\I,\II,\III}}\}$, each of which is associated with the Dynkin nodes of the superconformal algebra PSU$(2,2|4)$, see figure \ref{fig:fig31}. The ``middle-node'' rapidities ${\bf u}$ describe the physical momenta of magnons on the chain while the other rapidities describe the ``spin waves'', or in other words the index structures of magnons.

In the case of the Lie superalgebra, there are several different ways to draw a Dynkin diagram and we use the so-called SL(2) grading\fn{This however is not the grading used for the superconformal blocks. We therefore need to shift the quantum numbers appropriately when we compare the integrability prediction with the superconformal block decomposition of the four-point functions. See section \ref{subsec:1loop4pt} for more details.} throughout this paper, as shown in the figure.  In this grading, the quantum number of the single-trace operator are given in terms of the number of the rapidities $M$ and $K_{{\bf v},{\bf w}}^{\I,\II,\III}$ as
\beq\label{eq:quantumNum}
\begin{aligned}
\text{R-symmetry $[q_1,p,q_2]$}:\quad &q_1=K_{{\bf w}}^{\III}-2K_{{\bf w}}^{\II}+K_{{\bf w}}^{\I}\comma\\
&p=L+K_{\bf w}^{\II}-K_{\bf w}^{\I}+K_{\bf v}^{\II}-K_{\bf v}^{\I}\comma\\
&q_2=K_{{\bf v}}^{\III}-2K_{{\bf v}}^{\II}+K_{{\bf v}}^{\I}\comma\\
\text{Conformal $[s_1,r,s_2]$}:\quad &s_1=M-K_{{\bf w}}^{\III}-K_{{\bf w}}^{\I}\comma\\
&r=-L-2M+K_{\bf w}^{\I}+K_{\bf v}^{\I}-\delta \Delta\comma\\
&s_2=M-K_{{\bf v}}^{\III}-K_{{\bf v}}^{\I}\comma
\end{aligned}
\eeq
where $\delta \Delta$ is the anomalous dimension. For the traceless symmetric representation of the Lorentz group, what we normally call spin corresponds to $s=s_1=s_2$. For later convenience, let us also give an expression for the R-charge $J$, which rotates $Z$ and $\bar{Z}$:
\beq
J=L-\frac{1}{2}\left(K_{{\bf w}}^{\I}-K_{{\bf w}}^{\III}+K_{{\bf v}}^{\I}-K_{{\bf v}}^{\III}\right)\period
\eeq

Written explicitly, the Bethe equation (in the so-called spin-chain frame) reads
\begingroup \allowdisplaybreaks
\begin{flalign}\label{eq:BSequation}
1=e^{i\phi_{\red{v_{\III, j}}}}\equiv&\prod_{k=1}^{M}S^{\III, \zero}(\red{v_{\III,j}},u_k)\prod_{k=1}^{K_{\II}^{\bf v}}S^{\III, \II}(\red{v_{\III,j}},\red{v_{\II,k}})\comma\nonumber\\
1=e^{i\phi_{\red{v_{\II, j}}}}\equiv&\prod_{k\neq j}^{K_{\II}^{\bf v}}S^{\II, \II}(\red{v_{\II,j}},\red{v_{\II,k}})\prod_{k=1}^{K_{\I}^{{\bf v}}}S^{\II, \I}(\red{v_{\II,j}},\red{v_{\I,k}})\prod_{k=1}^{K_{\III}^{{\bf v}}}S^{\II, \III}(\red{v_{\II,j}},\red{v_{\III,k}})\comma\nonumber\\
1=e^{i\phi_{\red{v_{\I, j}}}}\equiv&\prod_{k=1}^{M}S^{\I, \zero}(\red{v_{\I,j}},u_k)\prod_{k=1}^{K_{\II}^{\bf v}}S^{\I, \II}(\red{v_{\I,j}},\red{v_{\II,k}})\comma\nonumber\\
1=e^{i \phi_{u_j}}\,\,\,\equiv&\left(\frac{x^{+}(u_j)}{x^{-}(u_j)}\right)^{L}\prod_{k\neq j}^{M}S_0 (u_j,u_k)\\
&\times\prod_{k=1}^{K_{\I}^{\bf v}} S^{\zero, \I}(u_j,\red{v_{\I,k}})\prod_{k=1}^{K_{\III}^{\bf v}} S^{\zero, \III}(u_j,\red{v_{\III,k}})\prod_{k=1}^{K_{\I}^{\bf w}} S^{\zero, \I}(u_j,\blue{w_{\I,k}})\prod_{k=1}^{K_{\III}^{\bf w}} S^{\zero, \III}(u_j,\blue{w_{\III,k}})\comma\nonumber\\
1=e^{i\phi_{\blue{w_{\I, j}}}}\equiv&\prod_{k=1}^{M}S^{\I, \zero}(\blue{w_{\I,j}},u_k)\prod_{k=1}^{K_{\II}^{\bf w}}S^{\I, \II}(\blue{w_{\I,j}},\blue{w_{\II,k}})\comma\nonumber\\
1=e^{i\phi_{\blue{w_{\II, j}}}}\equiv&\prod_{k\neq j}^{K_{\II}^{\bf w}}S^{\II, \II}(\blue{w_{\II,j}},\blue{w_{\II,k}})\prod_{k=1}^{K_{\I}^{{\bf w}}}S^{\II, \I}(\blue{w_{\II,j}},\blue{w_{\I,k}})\prod_{k=1}^{K_{\III}^{{\bf w}}}S^{\II, \III}(\blue{w_{\II,j}},\blue{w_{\III,k}})\comma\nonumber\\
1=e^{i\phi_{\blue{w_{\III, j}}}}\equiv&\prod_{k=1}^{M}S^{\III, \zero}(\blue{w_{\III,j}},u_k)\prod_{k=1}^{K_{\II}^{\bf w}}S^{\III, \II}(\blue{w_{\III,j}},\blue{w_{\II,k}})\period\nonumber
\end{flalign}
\endgroup
with $S_0$ being the overall scalar phase given in Appendix \ref{ap:Smatrix} and
\beq\label{eq:BSequationSmatrix}
\begin{aligned}
&S^{\zero,\I}(u,v)=\left(S^{\I,\zero}(v,u)\right)^{-1}=\frac{x^{-}(u)-x(v)}{x^{+}(u)-x(v)}\comma\\
&S^{\zero,\III}(u,v)=\left(S^{\III,\zero}(v,u)\right)^{-1}=\frac{1-1/x^{-}(u)x(v)}{1-1/x^{+}(u)x(v)}\comma\\
&S^{\II,\II}(u,v)=\frac{u-v-i}{u-v+i}\comma\\
& S^{\II,\I}(u,v)=S^{\II,\III}(u,v)=\left(S^{\I,\II}(v,u)\right)^{-1}=\left(S^{\III,\II}(v,u)\right)^{-1}=\frac{u-v+i/2}{u-v-i/2}\period
\end{aligned}
\eeq

In addition to these equations, one also need to impose the ``level-matching condition'',
\beq\label{eq:levelmatching}
\text{\bf Level matching}: \quad \prod_{j=1}^{M}\frac{x_j^{+}}{x_j^{-}}=1\period
\eeq

\paragraph{Dynamical transformation}
The equations \eqref{eq:BSequation} are invariant under the following transformations, which take the Bethe roots in the third nested levels $\red{{\bf v}_{\III}}$ and $\blue{{\bf w}_{\III}}$ and bring them to the first levels $\red{{\bf v}_{\I}}$ and $\blue{{\bf w}_{\I}}$,
\beq
\begin{aligned}
&\begin{cases}\red{\bf v_{\I}}=\{v_{\I,1},\ldots, v_{\I,K_{\I}}\}\\\red{\bf v_{\III}}=\{v_{\III,1},\ldots, v_{\III,K_{\III}}\}\\ L\end{cases}\quad \mapsto \quad \begin{cases}\red{\bf v_{\I}}=\{v_{\I,1},\ldots, v_{\I,K_{\I}},\tilde{v}_{\III,K_{\III}}\}\\\red{\bf v_{\III}}=\{v_{\III,1},\ldots, v_{\III,K_{\III}-1}\}\\ L+1\end{cases}\comma\\
&\begin{cases}\blue{\bf w_{\I}}=\{w_{\I,1},\ldots, w_{\I,K_{\I}}\}\\\blue{\bf w_{\III}}=\{w_{\III,1},\ldots, w_{\III,K_{\III}}\}\\ L\end{cases}\quad \mapsto \quad \begin{cases}\blue{\bf w_{\I}}=\{w_{\I,1},\ldots, w_{\I,K_{\I}},\tilde{w}_{\III,K_{\III}}\}\\\blue{\bf w_{\III}}=\{w_{\III,1},\ldots, w_{\III,K_{\III}-1}\}\\ L+1\end{cases}\comma
\end{aligned}
\eeq
where the rapidities with tilde are defined by
\beq
x(\tilde{v}_{\III,K_{\III}})=\frac{1}{x(v_{\III,K_{\III}})}\comma\qquad\qquad  x(\tilde{w}_{\III,K_{\III}})=\frac{1}{x(w_{\III,K_{\III}})}\period
\eeq
These transformations are called the dynamical transformation, and they physically represent the fact that the length of the spin chain can fluctuate at finite coupling and there will be mixing between operators of different lengths.

By repeated applications of the dynamical transformation, one can completely eliminate the Bethe roots at the third nested levels and trade them with the roots in the first nested levels. Such transformations are important when we state the selection rule as we see below.
\subsection{Main result\label{subsec:asymptoticmain}}
We now state our conjecture for the structure constant, which is expected to be valid when the operator is sufficiently long and the wrapping corrections are neglected.
\paragraph{Selection rule}
Let us first discuss the selection rule. In the analysis of the SO(6) sector at tree level, we found that the structure constant is nonzero only when the middle node rapidities are parity symmetric and the Bethe roots of the left and the right wings are parity-conjugate to each other.
A natural generalization to the full PSU(2,2$|$4) sector at finite coupling is to impose that ${\bf u}$ is parity symmetric and $\red{\bf v}$ and $\blue{\bf w}$ are related by $\blue{{\bf w}_{\I,\II,\III}}=\red{\bar{\bf v}_{\I,\II,\III}}$. As we see in section \ref{subsec:ABA}, this is also motivated by the unfolding interpretation of the Bethe equation in the mirror channel since the unfolding naturally identifies the left wing with the parity-transformed right wing. We however need to be careful when we state this rule since the Bethe roots at the first and the third levels are not invariant under the dynamical transformation. The correct and unambiguous way to state the selection rule is as follows:

\noindent The structure constant is nonzero only when
\begin{enumerate}
\item The middle node Bethe roots are parity symmetric; ${\bf u}=\{u_1,\bar{u}_1,\cdots, u_{\frac{M}{2}},\bar{u}_{\frac{M}{2}}\}$.
\item The Bethe roots for the left and the right wings are parity-conjugate to each other ($\blue{\bf w}=\bar{\red{\bf v}}$) {\it after performing the dynamical transformations and eliminating the roots at the third levels $\red{{\bf v}_{\III}}$ and $\blue{{\bf w}_{\III}}$}.
\end{enumerate}

In section \ref{subsec:1loop4pt}, we compare our integrability prediction with the OPE expansion of the four-point function and find that the OPE data are indeed zero when there are no Bethe roots satisfying the selection rule.
\paragraph{Asymptotic formula}
Without further delay, we present the asymptotic formula:
\beq\label{eq:asymptoticformula}
\left.\mathfrak{D}_{\mathcal{O}}\right|_{\text{asymptotic}}=-\frac{i^{J}+(-i)^{J}}{\sqrt{L}}\sqrt{\left(\prod_{1\leq s\leq \frac{M}{2}}\frac{u_s^2+\frac{1}{4}}{u_s^2}\sigma_B^2(u_s)\right)\frac{\det G_{+}}{\det G_{-}}}\period
\eeq
Here $J$ and $L$ are the $R$-charge and the length of the operator $\mathcal{O}$, and $\sigma_B$ is the boundary dressing phase \eqref{eq:fullbdress}. The matrices $G_{\pm}$ are defined by\fn{Note that $\del_{\red{v}}\phi_{\blue{w}}=\del_{\blue{v}}\phi_{\red{w}}=0$ since the rapidities on the two different wings are not coupled. However, here we wrote such terms in the formula in order to express the matrices in a more symmetric form.}
\beq\label{eq:defGpmAsympt}
\begin{aligned}
&G_{\pm}\equiv\\
& \pmatrix{cccc}{\del_{\red{\hat{v}_{\III, k}}}\left(\phi_{\red{\hat{v}_{\III,j}}}\pm\phi_{\blue{\hat{w}_{\III,j}}}\right)&\del_{\red{\hat{v}_{\III, k}}}\left(\phi_{\red{\hat{v}_{\II,j}}}\pm\phi_{\blue{\hat{w}_{\II,j}}}\right)&0&\del_{\red{\hat{v}_{\III,k}}}\left(\phi_{\hat{u}_{2j-1}}\pm\phi_{\hat{u}_{2j}}\right)\\
\del_{\red{\hat{v}_{\II, k}}}\left(\phi_{\red{\hat{v}_{\III,j}}}\pm\phi_{\blue{\hat{w}_{\III,j}}}\right)&\del_{\red{\hat{v}_{\II, k}}}\left(\phi_{\red{\hat{v}_{\II,j}}}\pm\phi_{\blue{\hat{w}_{\II,j}}}\right)&\del_{\red{\hat{v}_{\II, k}}}\left(\phi_{\red{\hat{v}_{\I,j}}}\pm\phi_{\blue{\hat{w}_{\I,j}}}\right)&0\\
 0&\del_{\red{\hat{v}_{\I, k}}}\left(\phi_{\red{\hat{v}_{\II,j}}}\pm\phi_{\blue{\hat{w}_{\II,j}}}\right)&\del_{\red{\hat{v}_{\I, k}}}\left(\phi_{\red{\hat{v}_{\I,j}}}\pm\phi_{\blue{\hat{w}_{\I,j}}}\right)&\del_{\red{\hat{v}_{\I,k}}}\left(\phi_{\hat{u}_{2j-1}}\pm\phi_{\hat{u}_{2j}}\right)\\
\del_{\hat{u}_{2k-1}}\left(\phi_{\red{\hat{v}_{\III,j}}}\pm\phi_{\blue{\hat{w}_{\III,j}}}\right) &0&\del_{\hat{u}_{2k-1}}\left(\phi_{\red{\hat{v}_{\I,j}}}\pm\phi_{\blue{\hat{w}_{\I,j}}}\right)&\del_{\hat{u}_{2k-1}}\left(\phi_{\hat{u}_{2j-1}}\pm \phi_{\hat{u}_{2j}}\right)}\comma
\end{aligned}
\eeq
where the parity conditions
\beq
\hat{u}_{2j-1}=\overline{\hat{u}}_{2j}=u_{2j-1}\comma\qquad \red{\hat{v}_{\I,j}}=\blue{\overline{\hat{w}}_{\I,j}}=\red{v_{\I,j}}\quad \text{etc.}
\eeq
are imposed after computing the derivatives.

Let us now make several remarks on different factors in the formula and explain how the formula was conjectured: First, the prefactor $i^{J}+(-i)^{J}$ comes from the fact that the relevant boundary state is a sum of two different boundary states $\langle \mathcal{G}|_{z=+i}+\langle \mathcal{G}|_{z=-i}$ as discussed in section \ref{subsec:matrixpart}. Second, the extra factor $1/\sqrt{L}$ is a standard factor which comes from the normalization of the two-point function, which in turn comes from the number of Wick contractions related to each other by the cyclic permutations. In next section, we also show that this factor is in fact necessary\fn{This factor is also needed to realize the invariance under the diagonal SU($2|2$) symmetry (in particular that changes the length of the operator). See Appendix B in \cite{Basso:2017khq} for a related discussion.} in order to rewrite the result in the spin-chain frame into the one in the string frame. For the comparison with the perturbative data, it is often convenient to strip off this factor and define the {\it length-stripped structure constant} ${\sf d}_{\mathcal{O}}$ defined by
\beq\label{eq:lengthstrippedddef}
{\sf d}_{\mathcal{O}}\equiv \sqrt{L}\mathfrak{D}_{\mathcal{O}}\period
\eeq

Third, the factor inside the square root $\prod_s \frac{u_s^2+\frac{1}{4}}{u_s^2}\sigma_B^2(u_s)$ is a product of the boundary scalar factor
\beq
\frac{u_s^2+\frac{1}{4}}{u_s^2}\sigma_B^2(u_s) =g_0 (u_s)g_0 (\bar{u}_s)\period
\eeq
Note that we only included the overall scalar factor because, in the diagonalized reflection matrix, the reflection amplitudes at the nested level are trivial. See section \ref{subsec:ABA} for more detailed explanation. Fourth, the ratio of the determinant is a natural generalization of a similar ratio which showed up in the SO(6) sector. It is further supported by the fact that, for the Bethe roots satisfying the aforementioned selection rule, the Gaudin norm can be factorized into
\beq
\left.\langle{\bf u},\red{\bf v},\blue{\bf w} |{\bf u},\red{\bf v},\blue{\bf w}\rangle\right|_{{\bf u}=\overline{\bf u},\blue{\bf w}=\overline{\red{\bf v}}}=\det G_{+}\det G_{-}\period
\eeq
See Appendix \ref{ap:norm} for a derivation. Finally, there is an overall minus sign. This sign comes from the ambiguity of taking the square root and we simply chose it in order to match it with the weak-coupling counterparts in section \ref{sec:weak}.
\subsection{From spin-chain frame to string frame\label{subsec:framechange}}
The results above are written in the spin-chain frame, in which the length of the spin chain $L$ is given by the number of fields. Although the spin-chain frame makes it easier to perform the comparison with the weak-coupling result, it has one drawback that the ``number of fields'' is not a well-defined quantum number and therefore can be ambiguous\fn{We should nevertheless emphasize that, in the asymptotic regime where the length of the operator is large, the final results computed from the spin-chain frame are well-defined and unambiguous.} at finite coupling. In particular, to compare the results with the nonperturbative $g$-function discussed in section \ref{sec:TBA}, it is more convenient to use the string frame, in which the length of the chain is given by the R-charge $J$.

At the level of the Bethe equation, the transition from the spin-chain frame to the string frame is well-understood: We simply need to replace the length $L$ with the R-charge $J$ and redefine the S-matrices as
\beq\label{eq:fromspintostringS}
\begin{aligned}
S_{\rm string}^{\zero,\I}(u,v)&= S_{\rm spin}^{\zero,\I}(u,v)\sqrt{\frac{x^{-}(u)}{x^{+}(u)}}\comma\qquad S_{\rm string}^{\I,\zero}(v,u)= S_{\rm spin}^{\I,\zero}(v,u)\sqrt{\frac{x^{+}(u)}{x^{-}(u)}}\comma\\
S_{\rm string}^{\zero,\III}(u,v)&= S_{\rm spin}^{\zero,\III}(u,v)\sqrt{\frac{x^{+}(u)}{x^{-}(u)}}\comma\qquad S_{\rm string}^{\III,\zero}(v,u)= S_{\rm spin}^{\III,\zero}(v,u)\sqrt{\frac{x^{-}(u)}{x^{+}(u)}}\comma
\end{aligned}
\eeq
where $S_{\rm spin}$ are the S-matrices in the spin-chain frame given in \eqref{eq:BSequationSmatrix}.
One can explicitly check that the Bethe equation is {\it invariant} under such transformations provided that the middle-node Bethe roots satisfy the level matching condition \eqref{eq:levelmatching}.

Since the Bethe equation is invariant. one might think that this implies that the determinants $\det G_{\pm}$ are also invariant under the change of the frames. This however is not correct: When we compute the determinants, we need to compute the derivatives of the phase factors $\phi_{{\bf u},\red{\bf v},\blue{\bf w}}$ {\it without  imposing the Bethe equations or the level-matching conditions}. This gives rise to a small difference in $\det G_{+}$, but not in $\det G_{-}$.

Let us see this more explicitly: The phase factors on the left and the right wings are modified under the transformations \eqref{eq:fromspintostringS} in the following way,
\beq\label{eq:phaseshiftspinstring}
\begin{aligned}
&\phi_{\red{v_{\I,j}}}\mapsto  \phi_{\red{v_{\I,j}}}+\frac{P_{\rm tot}}{2}\comma\quad \phi_{\red{v_{\III,j}}}\mapsto  \phi_{\red{v_{\III,j}}}-\frac{P_{\rm tot}}{2}\comma\\
&\phi_{\blue{w_{\I,j}}}\mapsto  \phi_{\blue{w_{\I,j}}}+\frac{P_{\rm tot}}{2}\comma\quad\phi_{\blue{w_{\III,j}}}\mapsto  \phi_{\blue{w_{\III,j}}}-\frac{P_{\rm tot}}{2}\comma
\end{aligned}
\eeq
with
\beq
e^{iP_{\rm tot}}=\prod_{k=1}^{M}\frac{x^{+}(u_k)}{x^{-}(u_k)}\period
\eeq
while the ones for the middle node are invariant. From this, one can immediately conclude that the determinant $\det G_{-}$ is invariant since the shifts of $\phi_{\red{v}}$ and $\phi_{\blue{w}}$ cancel out:
\beq
\left.\det G_{-}\right|_{\text{spin-chain}}=\left.\det G_{-}\right|_{\text{string}}\period
\eeq
On the other hand, the shifts add up in $\det G_{+}$ and therefore it gets modified when changing the frames. To compute how it changes, it is useful to rewrite $\det G_{+}$ in both frames by replacing the phase factor $\phi_{\hat{u}_{M-1}}+\phi_{\hat{u}_{M}}$ in the last column of $G_{+}$ in the following way:
\beq\label{eq:lastcolumn}
\begin{aligned}
\phi_{\hat{u}_{M-1}}+\phi_{\hat{u}_{M}}\quad \mapsto \,\,&\sum_{j=1}^{\frac{M}{2}}(\phi_{\hat{u}_{2j-1}}+\phi_{\hat{u}_{2j}})+\sum_{j=1}^{K^{{\bf v}}_{\I}}(\phi_{\red{v_{\I,j}}}+\phi_{\blue{w_{\I,j}}})+\sum_{j=1}^{K^{{\bf v}}_{\III}}(\phi_{\red{v_{\III,j}}}+\phi_{\blue{w_{\III,j}}})\\
&=\begin{cases}LP_{\rm tot}\qquad &\text{(spin chain)}\\JP_{\rm tot}\qquad &\text{(string)}\end{cases}
\end{aligned}
\eeq
Note that this does not modify the value of the determinant since it simply amounts to adding different columns to the last column. After this manipulation, we can add or subtract this last column from any of other columns without modifying the final result. By doing so, we can eliminate the differences of the phases \eqref{eq:phaseshiftspinstring} between the spin-chain frame and the string frame. We thus conclude that, after the rewriting, the only difference of $\det G_{+}$ in the two frames comes from the last column \eqref{eq:lastcolumn}. By comparing them, we conclude that
\beq
\left.\det G_{+}\right|_{\text{spin-chain}}=\frac{L}{J}\left.\det G_{+}\right|_{\text{string}}\period
\eeq
The extra factor $L/J$ nicely converts the spin-chain length $L$ appearing in the asymptotic formula \eqref{eq:asymptoticformula} into the R-charge $J$:
\beq
\left.\mathfrak{D}_{\mathcal{O}}\right|_{\text{asymptotic}}=-\frac{i^{J}+(-i)^{J}}{\sqrt{J}}\sqrt{\left(\prod_{1\leq s\leq \frac{M}{2}}\frac{u_s^2+\frac{1}{4}}{u_s^2}\sigma_B^2(u_s)\right)\left.\frac{\det G_{+}}{\det G_{-}}\right|_{\rm string}}\period
\eeq
In this form, the result depends only on the quantum numbers of the single-trace operator and is unambiguous at finite coupling.
\subsection{Eliminating the nested levels\label{subsec:eliminating}}
The determinants $\det G_{\pm}$ in \eqref{eq:asymptoticformula} depend explicitly on the rapidities at the nested levels $\red{\bf v}$ and $\blue{\bf w}$, which in turn depend on the choice of the grading structure of the super Dynkin diagrams\fn{Or equivalently they depend on the choice of the definition of the ``highest weights''.}. It is therefore more convenient to express the result purely in terms of the rapidities at the middle node. In what follows, we work in the spin chain frame, but the argument straightforwardly applies also to the expressions in the string frame.

For this purpose, we first rewrite the ratio of determinants as
\beq
\frac{\det G_{+}}{\det G_{-}}=\frac{\det G}{(\det G_{-})^2}\comma
\eeq
where $\det G=\det G_{+}\det G_{-}$ is the full PSU(2,2$|$4) Gaudin norm. In the next step, we use a trick discussed in \cite{Basso:2017khq} and factorize $\det G$ into an {\it induced Gaudin determinant} and a Gaudin determinant for the nested Bethe roots. For this, we first note that the full Gaudin determinant is a Jacobian between the phase factors $\phi_{{\bf u},\red{\bf v},\blue{\bf w}}$ to the rapidities,
\beq\label{eq:jacobiangaudin}
(\det G )\times d{\bf u}\wedge d\red{\bf v}\wedge d\blue{\bf w}= d\phi_{\bf u}\wedge d\phi_{\red{\bf v}}\wedge d\phi_{\blue{\bf w}}\comma
\eeq
where $d{\bf u}$'s and $d\phi_{{\bf u}}$'s denote a wedge product
\beq
d{\bf u}\equiv du_1\wedge \cdots \wedge du_M\comma\qquad d\phi_{{\bf u}}\equiv d\phi_{u_1}\wedge\cdots \wedge d\phi_{u_{M}}\period
\eeq
Now, to see the factorization, we first rewrite the left hand side of \eqref{eq:jacobiangaudin} as follows,
\beq
(\det G )\times d{\bf u}\wedge d\red{\bf v}\wedge d\blue{\bf w}=\frac{\det G}{J_{\red{\bf v}|{\bf u}}J_{\blue{\bf w}|{\bf u}}}d{\bf u}\wedge d\phi_{\red{\bf v}}\wedge d\phi_{\blue{\bf w}}\period
\eeq
Since we only performed the change of variables to $\red{\bf v}$ and $\blue{\bf w}$, the Jacobians $J_{\red{\bf v}|{\bf u}}$ and $J_{\blue{\bf w}|{\bf u}}$ are given by determinants
\beq
J_{\red{\bf v}|{\bf u}}\equiv \det \frac{\del \phi_{\red{v_j}}}{\del \red{v_k}}\left(=\langle \red{\bf v}|\red{\bf v}\rangle\right)\comma\qquad J_{\blue{\bf w}|{\bf u}}\equiv \det \frac{\del \phi_{\blue{w_j}}}{\del \blue{w_k}}\left(=\langle \blue{\bf w}|\blue{\bf w}\rangle\right)\comma
\eeq
where the derivatives are computed keeping the middle-node rapidities ${\bf u}$ fixed. As indicated, these can be identified with the Gaudin norms for the nested Bethe roots $\langle \red{\bf v}|\red{\bf v}\rangle$ and $\langle \blue{\bf w}|\blue{\bf w}\rangle$. We then trade ${\bf u}$ with $ \phi_{\bf u}$,
\beq\label{eq:gaudinrewritecompare}
(\det G )\times d{\bf u}\wedge d\red{\bf v}\wedge d\blue{\bf w}= \frac{\det G}{\langle \red{\bf v}|\red{\bf v}\rangle\langle \blue{\bf w}|\blue{\bf w}\rangle}\times \frac{1}{J_{{\bf u}|\phi_{\red{\bf v}\cup \blue{\bf w}}}}d\phi_{\bf u}\wedge d\phi_{\red{\bf v}}\wedge d\phi_{\blue{\bf w}}\period
\eeq
Since we already eliminated $\red{\bf v}$ and $\blue{\bf w}$, the Jacobian $J_{{\bf u}|\phi_{\red{\bf v}\cup \blue{\bf w}}}$ is given by a determinant of derivatives with $\phi_{\red{\bf v},\blue{\bf w}}$ fixed rather than fixing rapidities $\red{\bf v}$ and $\blue{\bf w}$:
\beq
J_{{\bf u}|\phi_{\red{\bf v}\cup \blue{\bf w}}} \equiv \det \tilde{G}_{\bf u}=\left.\det \frac{\del \phi_{u_j}}{\del u_k}\right|_{\phi_{\red{\bf v},\blue{\bf w}}}\period
\eeq
In other words, the determinant $J_{{\bf u}|\phi_{\red{\bf v}\cup \blue{\bf w}}} $ can be computed in the following procedures: We first solve the Bethe equations at the nested levels and express $\red{\bf v}$ and $\blue{\bf w}$ as functions of ${\bf u}$. This results in the effective Bethe equation which only involves ${\bf u}$. The Gaudin norm of this effective Bethe equation is $J_{{\bf u}|\phi_{\red{\bf v}\cup \blue{\bf w}}}$ and was called the {\it induced Gaudin norm} in \cite{Basso:2017khq}. Comparing \eqref{eq:jacobiangaudin} and \eqref{eq:gaudinrewritecompare}, we have
\beq
\det G=\langle \red{\bf v}|\red{\bf v}\rangle\langle \blue{\bf w}|\blue{\bf w}\rangle\det \tilde{G}_{\bf u}\period
\eeq

Next we perform a similar rewriting for $\det G_{-}$. To do so, let us recall the structure of $G_{-}$:
\beq
\begin{aligned}
&G_{-}\equiv\\
& \pmatrix{cccc}{\del_{\red{\hat{v}_{\III, k}}}\phi_{\red{\hat{v}_{\III,j}}}&\del_{\red{\hat{v}_{\III, k}}}\phi_{\red{\hat{v}_{\II,j}}}&0&\del_{\red{\hat{v}_{\III,k}}}\left(\phi_{\hat{u}_{2j-1}}-\phi_{\hat{u}_{2j}}\right)\\
\del_{\red{\hat{v}_{\II, k}}}\phi_{\red{\hat{v}_{\III,j}}}&\del_{\red{\hat{v}_{\II, k}}}\phi_{\red{\hat{v}_{\II,j}}}&\del_{\red{\hat{v}_{\II, k}}}\phi_{\red{\hat{v}_{\I,j}}}&0\\
 0&\del_{\red{\hat{v}_{\I, k}}}\phi_{\red{\hat{v}_{\II,j}}}&\del_{\red{\hat{v}_{\I, k}}}\phi_{\red{\hat{v}_{\I,j}}}&\del_{\red{\hat{v}_{\I,k}}}\left(\phi_{\hat{u}_{2j-1}}-\phi_{\hat{u}_{2j}}\right)\\
\del_{\hat{u}_{2k-1}}\left(\phi_{\red{\hat{v}_{\III,j}}}-\phi_{\blue{\hat{w}_{\III,j}}}\right) &0&\del_{\hat{u}_{2k-1}}\left(\phi_{\red{\hat{v}_{\I,j}}}-\phi_{\blue{\hat{w}_{\I,j}}}\right)&\del_{\hat{u}_{2k-1}}\left(\phi_{\hat{u}_{2j-1}}- \phi_{\hat{u}_{2j}}\right)}\comma
\end{aligned}
\eeq
with the parity condition imposed {\it after computing the derivatives}. The crucial observation is that this determinant coincides with the Gaudin determinant computed by {\it first imposing the parity condition and then taking the derivatives}. Written more explicitly, we have the following relation:
\beq\label{eq:newGminus}
G_{-}=\pmatrix{cccc}{\del_{\red{v_{\III, k}}}\phi^{p}_{\red{v_{\III,j}}}&\del_{\red{v_{\III, k}}}\phi^{p}_{\red{v_{\II,j}}}&0&\del_{\red{v_{\III,k}}}\phi^{p}_{\hat{u}_{j}}\\
\del_{\red{v_{\II, k}}}\phi^{p}_{\red{v_{\III,j}}}&\del_{\red{v_{\II, k}}}\phi^{p}_{\red{v_{\II,j}}}&\del_{\red{v_{\II, k}}}\phi^{p}_{\red{v_{\I,j}}}&0\\
 0&\del_{\red{v_{\I, k}}}\phi^{p}_{\red{v_{\II,j}}}&\del_{\red{v_{\I, k}}}\phi^{p}_{\red{v_{\I,j}}}&\del_{\red{v_{\I,k}}}\phi^{p}_{u_{j}}\\
\del_{u_{k}}\phi^{p}_{\red{v_{\III,j}}} &0&\del_{u_{k}}\phi^{-}_{\red{\hat{v}_{\I,j}}}&\del_{u_{k}}\phi^{p}_{u_{j}}}\comma
\eeq
with
\beq\label{eq:BSequation2}
\begin{aligned}
1=e^{i\phi^{p}_{\red{v_{\III, j}}}}\equiv&\prod_{k=1}^{M/2}S^{\III, \zero}(\red{v_{\III,j}},u_k)S^{\III, \zero}(\red{v_{\III,j}},\bar{u}_k)\prod_{k=1}^{K_{\II}^{\bf v}}S^{\III, \II}(\red{v_{\III,j}},\red{v_{\II,k}})\comma\\
1=e^{i\phi^{p}_{\red{v_{\II, j}}}}\equiv&\prod_{k\neq j}^{K_{\II}^{\bf v}}S^{\II, \II}(\red{v_{\II,j}},\red{v_{\II,k}})\prod_{k=1}^{K_{\I}^{{\bf v}}}S^{\II, \I}(\red{v_{\II,j}},\red{v_{\I,k}})\prod_{k=1}^{K_{\III}^{{\bf v}}}S^{\II, \III}(\red{v_{\II,j}},\red{v_{\III,k}})\comma\\
1=e^{i\phi^{p}_{\red{v_{\I, j}}}}\equiv&\prod_{k=1}^{M/2}S^{\I, \zero}(\red{v_{\I,j}},u_k)S^{\I, \zero}(\red{v_{\III,j}},\bar{u}_k)\prod_{k=1}^{K_{\II}^{\bf v}}S^{\I, \II}(\red{v_{\I,j}},\red{v_{\II,k}})\comma\\
1=e^{i \phi^{p}_{u_j}}\,\,\,\equiv&\left(\frac{x^{+}(u_j)}{x^{-}(u_j)}\right)^{L}S_0(u_j,\bar{u}_j)\prod_{k\neq j}^{M/2}S_0 (u_j,u_k)S_0 (u_j,\bar{u}_k)\times\\
&\prod_{k=1}^{K_{\I}^{\bf v}} S^{\zero, \I}(u_j,\red{v_{\I,k}})S^{\zero, \I}(u_j,\red{\bar{v}_{\I,k}})\prod_{k=1}^{K_{\III}^{\bf v}} S^{\zero, \III}(u_j,\red{v_{\III,k}})S^{\zero, \III}(u_j,\red{\bar{v}_{\III,k}})\period
\end{aligned}
\eeq
Note that in \eqref{eq:newGminus} the derivative $\del_{u_k}$ acts\fn{Similarly $\del_{\red{v_k}}$ acts both on $\red{v_k}$ and $\red{\bar{v}_k}$.} both on $u_k$ and $\bar{u}_k=-u_k$.

Having identified $\det G_{-}$ with a kind of the Gaudin determinant, we can simply apply the same logic and show the factorization of $\det G_{-}$ into two parts
\beq
\det G_{-}=\langle \red{\bf v}|\red{\bf v}\rangle\times \det \tilde{G}_{-}\comma
\eeq
where $\langle \red{\bf v}|\red{\bf v}\rangle$ is the Gaudin norm for the nested Bethe roots on the right wing while the induced Gaudin determinant $\det \tilde{G}_{-}$ is given by
\beq
\det \tilde{G}_{-}=\left.\det \frac{\del \phi^{p}_{u_j}}{\del u_k}\right|_{\phi^{p}_{\red{\bf v}}}\comma
\eeq
which can be computed by first solving for $\red{\bf v}$ and then evaluating the derivatives.

As a result, we find that the Gaudin norms for the nested Bethe roots cancel in the ratio,
\beq
\frac{\det G_{+}}{\det G_{-}}=\frac{\det G}{(\det G_{-})^2}=\frac{\det \tilde{G}_{\bf u}}{(\det \tilde{G}_{-})^2}\comma
\eeq
where we used $\langle \blue{\bf w}|\blue{\bf w}\rangle=\langle \red{\bf v}|\red{\bf v}\rangle$, which holds for the parity-symmetric states. This allows us to eliminate the dependence on the nested Bethe roots from the asymptotic formula,
 \beq
 \left.\mathfrak{D}_{\mathcal{O}}\right|_{\text{asymptotic}}=-\frac{i^{J}+(-i)^{J}}{\sqrt{L}}\sqrt{\left(\prod_{1\leq s\leq \frac{M}{2}}\frac{u_s^2+\frac{1}{4}}{u_s^2}\sigma_B^2(u_s)\right)\frac{\det \tilde{G}_{{\bf u}}}{(\det \tilde{G}_{-})^2}}\comma
 \eeq
 thereby making the result manifestly invariant under the change of gradings. In the next section, we will see that a similar rewriting is possible also for the non-perturbative $g$-functions.
\section{Exact $g$-Function for Giant Gravitons\label{sec:TBA}}
We now apply the general framework of $g$-functions outlined in section \ref{sec:integrability} to our problem and write down the nonperturbative expression for the three-point function which applies also to operators of finite size. The strategy is as follows: In section \ref{subsec:ABA} we write down the asymptotic Bethe equation in the mirror channel (mirror ABA). Then in section \ref{subsec:derivationTBA}, we derive the TBA based on the mirror Bethe equation. After doing so, we compute the $g$-function for the ground state in section \ref{subsec:N=4gfunction}. We then present the main result in this paper; we conjecture a generalization to the SL(2) excited states at finite coupling in section \ref{subsec:SL2finite} based on the analytic continuation trick \cite{Dorey:1996re,Dorey:1997rb}, mimicking what was done for the spectrum \cite{Gromov:2009zb}.  Finally in section \ref{subsec:reltoAsympt}, we provide brief discussions on the relation between this conjecture and the asymptotic formula in the previous section.
\subsection{Mirror asymptotic Bethe ansatz\label{subsec:ABA}}
To write down the Bethe equation in the mirror channel, we need to diagonalize both the S-matrix and the reflection matrix. As was pointed out in the study of cusped Wilson loops, this task greatly simplifies when the reflection matrix admits the unfolding interpretation. The argument is basically the same as the one in \cite{Drukker:2012de,Correa:2012hh}, the only difference being that we work in the mirror channel while \cite{Drukker:2012de,Correa:2012hh} analyze the physical channel. Thus we only sketch the derivation, referring interested readers to the references \cite{Drukker:2012de,Correa:2012hh}.

\subsubsection{Fundamental magnons} Let us first consider a state with $M$ fundamental magnons with rapidities ${\bf u}=\{u_1,\ldots, u_M\}$ in the mirror channel (or equivalently the open-string channel) in order to illustrate some of the key features of the mirror ABA. As in section \ref{sec:integrability}, we denote the length of the cylinder by $R$ and the circumference by $L$. In order for this description to be valid, we assume that the length is large, namely $R\gg 1$. The key insight coming from the unfolding picture is that one can view this open string alternatively as a closed string with length $2R$ and $2M$ magnons with rapidities $\{u_1,\ldots, u_M,\bar{u}_M,\ldots, \bar{u}_1\}$:
\beq
|\mathcal{X}_{A_1\dot{A}_1}(u_1)\cdots \mathcal{X}_{A_M\dot{A}_M}(u_M)\rangle_{\rm open} \mapsto |\mathcal{X}_{A_1}(u_1)\cdots \mathcal{X}_{A_M}(u_M)\mathcal{X}_{\dot{A}_M}(\bar{u}_M)\cdots \mathcal{X}_{\dot{A}_1}(\bar{u}_1)\rangle_{\rm closed}\period
\eeq
As indicated, after unfolding, each magnon only carries one SU$(2|2)$ index.

This observation makes the diagonalization of matrix structures trivial; we simply need to apply the nested Bethe ansatz to a periodic spin chain with a single SU$(2|2)$ symmetry. Normally, this can be done by introducing as many rapidities as the number of Dynkin nodes. However, for the centrally extended SU$(2|2)$ spin chain of our interest, the rapidities for the first node and the last node can be combined into one set: In the physical channel, we already saw this in section \ref{subsec:ABAphysical} as the invariance of the Bethe equations under the dynamical transformation.

\begin{figure}[t]
\centering
\includegraphics[clip,height=4.5cm]{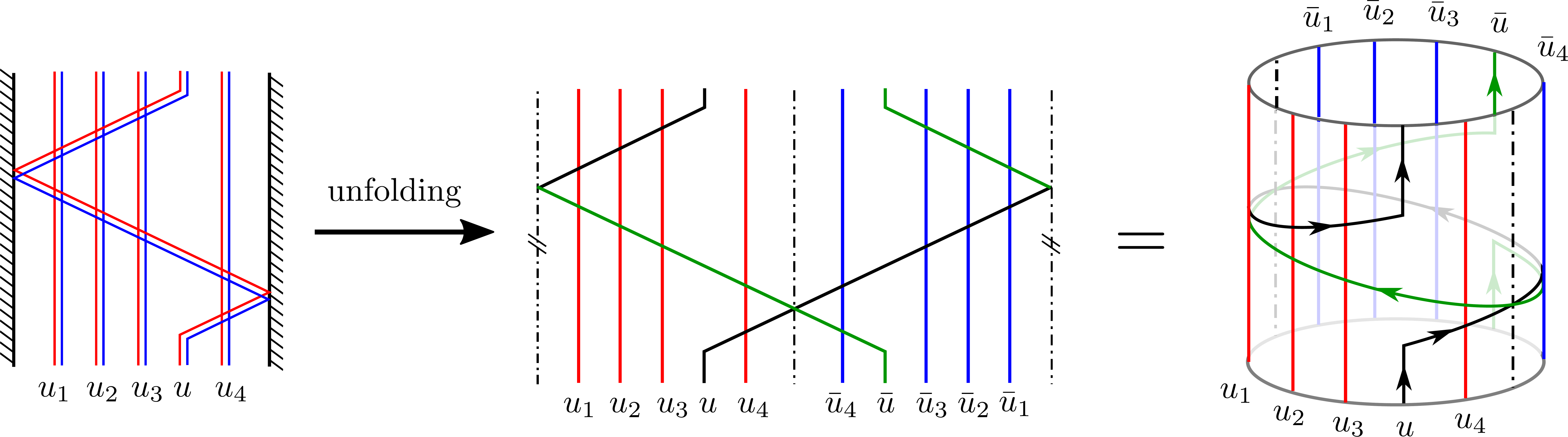}
\caption{The unfolding interpretation of the boundary Bethe equation in the mirror channel. After unfolding, the open chain with length $R$ becomes a closed chain with length $2R$. The boundary Bethe equation can be derived by imposing the periodicity condition for a pair of excitations $u$ and $\bar{u}$.}
\label{fig:fig32}
\end{figure}

Thus here we introduce two, rather than three, sets of rapidities at the nested level, which we denote by ${\bf v}_{\I}=\{v_{\I, 1},\ldots, v_{\I,K_{\I}}\}$ and ${\bf v}_{\II}=\{v_{\II,1},\ldots, v_{\II,K_{\II}}\}$. The relevant S-matrices can be computed simply by performing the mirror transformation of the middle-node rapidities, $x^{+}(u)\to 1/x^{+}(u)$, to the corresponding S-matrices in the physical channel \eqref{eq:BSequationSmatrix}:
\beq\label{eq:Smatrixmirrordiag}
\begin{aligned}
&\tilde{S}_0(u,v)=S_0(u^{\gamma},v^{\gamma})\comma\\
&\tilde{S}^{\zero,\I}(u,v)=\left(\tilde{S}^{\I,\zero}(v,u)\right)^{-1}=\sqrt{x^{+}(u)x^{-}(u)}\frac{x^{-}(u)-x(v)}{\frac{1}{x^{+}(u)}-x(v)}\comma\\
&\tilde{S}^{\II,\II}(u,v)=\frac{u-v-i}{u-v+i}\comma\qquad  \tilde{S}^{\II,\I}(u,v)=\left(\tilde{S}^{\I,\II}(v,u)\right)^{-1}=\frac{u-v+i/2}{u-v-i/2}\period
\end{aligned}
\eeq
(Here and below we work in the string frame.)
The only  difference between the standard periodic chain and the unfolded open chain is the overall scalar factor $S_0$ and $r_0$. This leads to an effective rule that, when the magnon scattering occurs at the position of the original boundary, we multiply $r_0$ while if it happens at other positions we multiply $\sqrt{S_0}$, see figure \ref{fig:fig32} for further explanation.

With these ingredients, we can now derive the Bethe equation by imposing the periodicity in the original spin chain. In the unfolded description, this amounts to imposing the periodicity condition for a {\it pair of magnons}, which are moving in the opposite directions, see figure \ref{fig:fig32}. As a result we obtain the following set of equations\fn{They are basically the mirror version of (56)-(58) in \cite{Correa:2012hh}.}:
\beq
\begin{aligned}
1=&\,e^{2i\tilde{p}(u_j)R} r_0 (u_j)^2\prod_{k\neq j}^{M}\tilde{S}_0 (u_j,u_k)\tilde{S}_0(u_j,\bar{u}_k)\prod_{k=1}^{K_{\I}}\frac{\tilde{S}^{\zero,\I}(u_j,v_{\I,k})}{\tilde{S}^{\zero,\I}(\bar{u}_j,v_{\I,k})}\comma\\
1=&\prod_{k=1}^{M}\tilde{S}^{\I,\zero}(v_{\I,j},u_k)\tilde{S}^{\I,\zero}(v_{\I,j},\bar{u}_k)\prod_{k=1}^{K_{\II}}\tilde{S}^{\I,\II}(v_{\I,j},v_{\II,k})\comma\\
1=&\prod_{k=1}^{K_{\I}}\tilde{S}^{\II,\I}(v_{\II,j},v_{\I,k})\prod_{k\neq j}^{K_{\II}}\tilde{S}^{\II,\II}(v_{\II,j},v_{\II,k})\period
\end{aligned}
\eeq
A point worth emphasizing is that only the middle node rapidities come in parity symmetric pairs $\{u,\bar{u}\}$ in this equation. This is in stark contrast to the situation discussed in \cite{Kostov:2018dmi} where the boundaries preserve the full bulk symmetry and the Bethe roots at all levels come in pair. Because of this, the Fredholm determinants that we derive later in this section are structurally different from the ones derived in \cite{Kostov:2018dmi}.

For the derivation of TBA, it is useful to rewrite the first equation using the parity invariance of the S-matrix as
\beq\label{eq:paritysimpler}
1=\,e^{2i\tilde{p}(u_j)R} r_0 (u_j)^2\prod_{k\neq j}^{M}\tilde{S}_0 (u_j,u_k)\tilde{S}_0(u_j,\bar{u}_k)\prod_{k=1}^{K_{\I}}\tilde{S}^{\zero,\I}(u_j,v_{\I,k})\tilde{S}^{\zero,\I}(u_j,\bar{v}_{\I,k})\comma
\eeq
where the parity transformation for the nested root is defined by $\bar{v}\equiv -v$ and $x(\bar{v})\equiv -x(v)$.

The unfolding structure shown in figure \ref{fig:fig32} suggests that the left wing of the Dynkin diagram of the bulk magnon is identified with the parity-transformed right wing. This is one of the motivations for the selection rule that we conjectured in section \ref{subsec:asymptoticmain}.
\subsubsection{General mirror ABA}
For the purpose of deriving the TBA, we also need to include the bound states in the mirror ABA. In what follows, in order to clarify the relations to the results in the literature, we use the notations in the review \cite{Bajnok:2010ke}.
\paragraph{Review of mirror ABA for periodic chain} Before discussing the general ABA for our problem, let us briefly review the bound-state spectrum and their S-matrices for the periodic chain in the mirror channel.
\begin{enumerate}
\item Bound states and fermions: In the mirror channel, there are four types of particle states, including three infinite types of bound states $(\bullet_Q,\triangleright_M,\circ_N)$ and also fermionic excitations $y_{\delta}$ $\delta\in\{\pm\}$. The bound states $\bullet_{Q}$ are the momentum-carrying bound states while others correspond to the bound states at the nested level (which correspond to the ``spin-wave'' excitations). Note also that there are two sets of bound states at the nested level, each of which is associated with the left or the right wing of the Dynkin diagram.
The relation to the notations for the rapidities of fundamental magnons we have been using so far is given by
\beq
\begin{aligned}
\bullet_1 &\mapsto u_k \comma\quad y_{+}\mapsto \{ \red{v_{\I,k}},\blue{w_{\I,k}}\}\comma\quad y_{-}\mapsto \{ \red{v_{\III,k}},\blue{w_{\III,k}}\}\comma\quad\circ_1 \mapsto \{\red{v_{\II,k}},\blue{w_{\II,k}}\}\period
\end{aligned}
\eeq
\item S-matrices: The S-matrices of these bound states can be determined by fusing the S-matrices of the fundamental excitations. Since their explicit forms are not necessary in this paper, we will not display them here and refer the interested readers to the review \cite{Bajnok:2010ke}. As shown below, some of the S-matrices turn out to be trivial:
\beq
\begin{array}{c|cccc}
&\bullet_{Q_2}&\triangleright_{M_2}&\circ_{N_2}&y_{\delta_2}\\\hline
\bullet_{Q_1}&S_{Q_1Q_2}^{\bullet\bullet}&S_{Q_1M_2^{\prime}}^{\bullet\triangleright}&1&S_{Q_1\delta_2}^{\bullet y}\\
\triangleright_{M_1}&S_{M_1Q_2}^{\triangleright\bullet}&S_{M_1M_2}^{\triangleright\triangleright}&1&S_{M_1\delta_2}^{\triangleright y}\\
\circ_{N_1}&1&1&S_{N_1N_2}^{\circ\circ}&S_{N_1\delta_2}^{\circ y}\\
y_{\delta_1}&S_{\delta_1Q_2}^{y\bullet}&S_{\delta_1M_2}^{y\triangleright}&S_{\delta_1N_2}^{y\circ}&1
\end{array}
\eeq
\item Asymptotic Bethe ansatz: Using the bound states and the S-matrices, we can write down the full ABA in the mirror channel which describes the scattering of a given type of particle (fundamental or bound state) with \emph{all} other types of particles. For instance, the ABA of a $\bullet_{Q}$ particle is given by
\beq
\begin{aligned}
\label{eq:ABAperiod}
1=&\,e^{i\tilde{p}_{Q_j}(u_{j})R}\prod_{k\neq j} S^{\bullet\bullet}_{Q_jQ_k}(u_{j},u_{k})\\
&\times\prod_{l} S^{\bullet y}_{Q_j\delta_l}(u_{j},\red{v_l^{y}})\prod_{m}S^{\bullet\triangleright}_{Q_j M_m}(u_{j},\red{v_{m}^{\triangleright}})\prod_n S^{\bullet\circ}_{Q_j, N_n}(u_{Q_j},\red{v_{n}^{\circ}})\\
&\times\prod_{l} S^{\bullet y}_{Q_j\delta_l}(u_{j},\blue{w_l^{y}})\prod_{m}S^{\bullet\triangleright}_{Q_j M_m}(u_{j},\blue{w_{m}^{\triangleright}})\prod_n S^{\bullet\circ}_{Q_j, N_n}(u_{j},\blue{w_{n}^{\circ}})\period
\end{aligned}
\eeq
Here $\tilde{p}_{Q}(u)$ is the fused mirror momentum given by
\beq
\tilde{p}_Q(u)\equiv -i\left[\frac{Q}{2}+\frac{g}{i}\left(\frac{1}{x^{[-Q]}}-x^{[+Q]}\right)\right]\comma
\eeq
with $f^{[\pm Q]}(u)\equiv f(u\pm iQ /2)$.
 The factors on the second line come from the bound states in the right PSU(2$|$2) while the factors on the third line come from those in the left PSU(2$|$2).
\end{enumerate}

\paragraph{Mirror ABA with boundaries}
We now introduce the bound states for the mirror ABA of our problem. The differences from the periodic chain are
\begin{enumerate}
\item Bound state: The types of bound states are basically the same as in the periodic case, namely $\star=(\bullet_Q,\triangleright_M,\circ_N,y_{\delta})$. The main difference is that now we have only one set of bound states at the nested levels. Another small difference is that for the middle-node bound states $\bullet_Q$, we can restrict their rapidity to the range $u_{Q}\ge 0$.
\item S-matrices: They are exactly the same as the periodic case.
\item Asymptotic Bethe ansatz: Even in the presence of bound states, we can use the unfolding picture to write down the asymptotic Bethe ansatz. As a result, we obtain a set of equations in which only the momentum-carrying bound states $\bullet_{Q}$ come in parity symmetric pairs.
\end{enumerate}
We then obtain the following equation for the middle-node bound states\fn{Here we already used the parity invariance of the S-matrix to bring it into a form of \eqref{eq:paritysimpler}.}:
\beq
\begin{aligned}
1=&\,e^{2i\tilde{p}_{Q_j}(u_{j})R}\times r_{Q_j}(u_j)^2\prod_{k\ne j}S^{\bullet\bullet}_{Q_jQ_k}(u_{j},u_{k})S^{\bullet\bullet}_{Q_jQ_k}(u_{j},\bar{u}_{k})\\
&\times \prod_lS_{Q_j\delta_l}^{\bullet y}(u_{j},v^{y}_l)S_{Q_j\delta_l}^{\bullet y}(u_{j},\bar{v}^{y}_l)
\prod_m S_{Q_j M_m}^{\bullet\triangleright}(u_{j},v^{\triangleright}_{m})S_{Q_j M_m}^{\bullet\triangleright}(u_{j},\bar{v}^{\triangleright}_{m})\period
\end{aligned}
\eeq
The reflection amplitude for the bound state $r_Q(u)$ can be obtained by the standard fusing procedure and reads
\beq
r_Q(u)=\left(\prod_{k=-\frac{Q-1}{2}}^{\frac{Q-1}{2}}\frac{u+ik+\frac{i}{2}}{u+ik}\right)\frac{\sigma_{B,Q}(\bar{u}^{\gamma})}{\sigma_{QQ}(u^{\gamma},\bar{u}^{\gamma})}\comma
\eeq
where $\sigma_{ab}(u,v)=e^{i\chi (u^{[a]},v^{[b]})+i\chi (u^{[-a]},v^{[-b]})-i\chi (u^{[-a]},v^{[b]})-i\chi (u^{[a]},v^{[-b]})}$ is the bulk dressing phase for bound states\fn{See for instance \cite{Vieira:2010kb} for definitions of the $\chi$ function.} while $\sigma_{B,Q}(u)$ is defined by
\beq
\sigma_{B,Q}(u)\equiv 2^{-E_Q(u)}\frac{G(x^{[+Q]})}{G(x^{[-Q]})}\comma
\eeq
with
\beq
E_Q(u)=\frac{Q}{2}+\frac{g}{i}\left(\frac{1}{x^{[-Q]}}-\frac{1}{x^{[+Q]}}\right)\period
\eeq
The other Bethe equations are given by
\begin{align}\label{eq:ABAfullmirrorN=4}
1=&\,\prod_{k}S^{\triangleright\bullet}_{M_jQ_k}(v_{j}^{\triangleright},u_{k})S^{\triangleright\bullet}_{M_jQ_k}(v_{j}^{\triangleright},\bar{u}_{k})
\prod_{l}S_{M_j\delta_l}^{\triangleright y}(v_{j}^{\triangleright},v_l^{y})\prod_{m\ne j}S_{M_j M_m}^{\triangleright\triangleright}(v_{j}^{\triangleright},v^{\triangleright}_{m})\prod_n S_{M_j N_n}^{\triangleright\circ}(v_{j}^{\triangleright},v_{n}^{\circ}),\nn\\
1=&\,\prod_k S^{y\bullet}_{\delta_j Q_k}(v_j^{y},u_{k})S^{y\bullet}_{\delta_j Q_k}(v_j^{y},\bar{u}_{k})
\prod_m S_{\delta_j M_m}^{y\triangleright}(v_j^{y},v_{m}^{\triangleright})\prod_{n}S_{\delta_j N_n}^{y\circ}(v_j,w_{N_n}),\\
1=&\,\prod_l S^{\circ y}_{N_j\delta_l}(v_{j}^{\circ},v^{y}_l)\prod_m S^{\circ\triangleright}_{N_jM_m}(v_{j}^{\circ},v_{m}^{\triangleright})
\prod_{n\ne j}S^{\circ\circ}_{N_j N_n}(v_{j}^{\circ},v_{n}^{\circ})\period\nn
\end{align}
\subsection{Thermodynamic Bethe ansatz\label{subsec:derivationTBA}}
As with the toy example in section \ref{sec:integrability}, the next step would be to take the thermodynamic limit of the Bethe equation \eqref{eq:ABAfullmirrorN=4} by introducing densities of roots, write down a path integral for the densities using the auxiliary variables $\eta$, and compute the saddle point and fluctuations. These are straightforward yet laborious tasks whose derivations can be lengthy. We therefore relegate the discussion along this line to Appendix \ref{ap:gfunctionNested}. Instead, we now explain a shortcut to obtain the final answer, which is based on the following observations:
\begin{enumerate}
\item The resulting TBA equation must describe the spectral problem in the closed string sector, which is already solved. The only difference is that our TBA equation would only describe the spectrum of the parity-symmetric states. In practice, this amounts to performing some $Z_2$ identification of $Y$-functions in the standard TBA.
\item The Fredholm determinant coming from the fluctuations can be computed by taking functional variations of the TBA equations.
\end{enumerate}
Both of these features can be seen explicitly in the toy example in section \ref{sec:integrability}, in which there is only a single species of particles. In addition, we show in Appendix \ref{ap:gfunctionNested} that these features survive even if there are multiple species of particles.

Let us see quickly how the $Z_2$ identification works for the toy example in section \ref{sec:integrability}. The standard TBA in that case is given by
\beq\label{eq:usualTBAintoy}
0 =L\tilde{E}(u)+\log Y(u) -\log \left(1+Y\right)\star K(u)\comma
\eeq
where $K(u,v)=\frac{1}{i}\del_{u}\log S(u,v)$ and $\star$ denotes the convolution along the {\it full real axis},
\beq
A\star K(u)\equiv \int_{-\infty}^{\infty}\frac{dv}{2\pi i}A(v)K(v,u)\period
\eeq
On the other hand, the TBA obtained in section \ref{sec:integrability} reads
\beq\label{eq:paritysymmetricTBAtoy}
0 =L\tilde{E}(u)+\log Y(u) -\log \left(1+Y\right)\ast \mathcal{K}_{+}(u)\period
\eeq
Recall that $\ast$ denotes the convolution from $0$ to $\infty$ and $\mathcal{K}_{+}(u,v)=K(u,v)+K(u,-v)$. To obtain \eqref{eq:paritysymmetricTBAtoy} from \eqref{eq:usualTBAintoy}, one simply needs to assume the parity-invariance of the $Y$-function, $Y(u)=Y(-u)$ and rewrite the convolution term as follows:
\beq
\begin{aligned}
\log (1+Y)\star K(u)&=\int_{0}^{\infty}\frac{dv}{2\pi i}\log Y(v)\,\,K(v,u)+\int_{-\infty}^{0}\frac{dv}{2\pi i}\log Y(v)\,\,K(v,u)\\
&=\int_{0}^{\infty}\frac{dv}{2\pi i}\log Y(v)\,\,K(v,u)+\int_{0}^{\infty}\frac{dv}{2\pi i}\log Y(v) \,\,\underbrace{K(-v,u)}_{=K(v,-u)}\\
&=\log (1+Y)\ast \mathcal{K}_{+}(u)
\end{aligned}
\eeq
In passing to the last line, we used the parity invariance of the S-matrix $S(-v,u)=\left[S(v,-u)\right]^{-1}$ to rewrite the kernel as follows:
\beq
K(-v,u)= -\frac{1}{i}\del_v \log S(-v,u)=+\frac{1}{i}\del_v\log S(v,-u)=K(v,-u)\period
\eeq

\begin{figure}[t]
\centering
\includegraphics[clip,height=5cm]{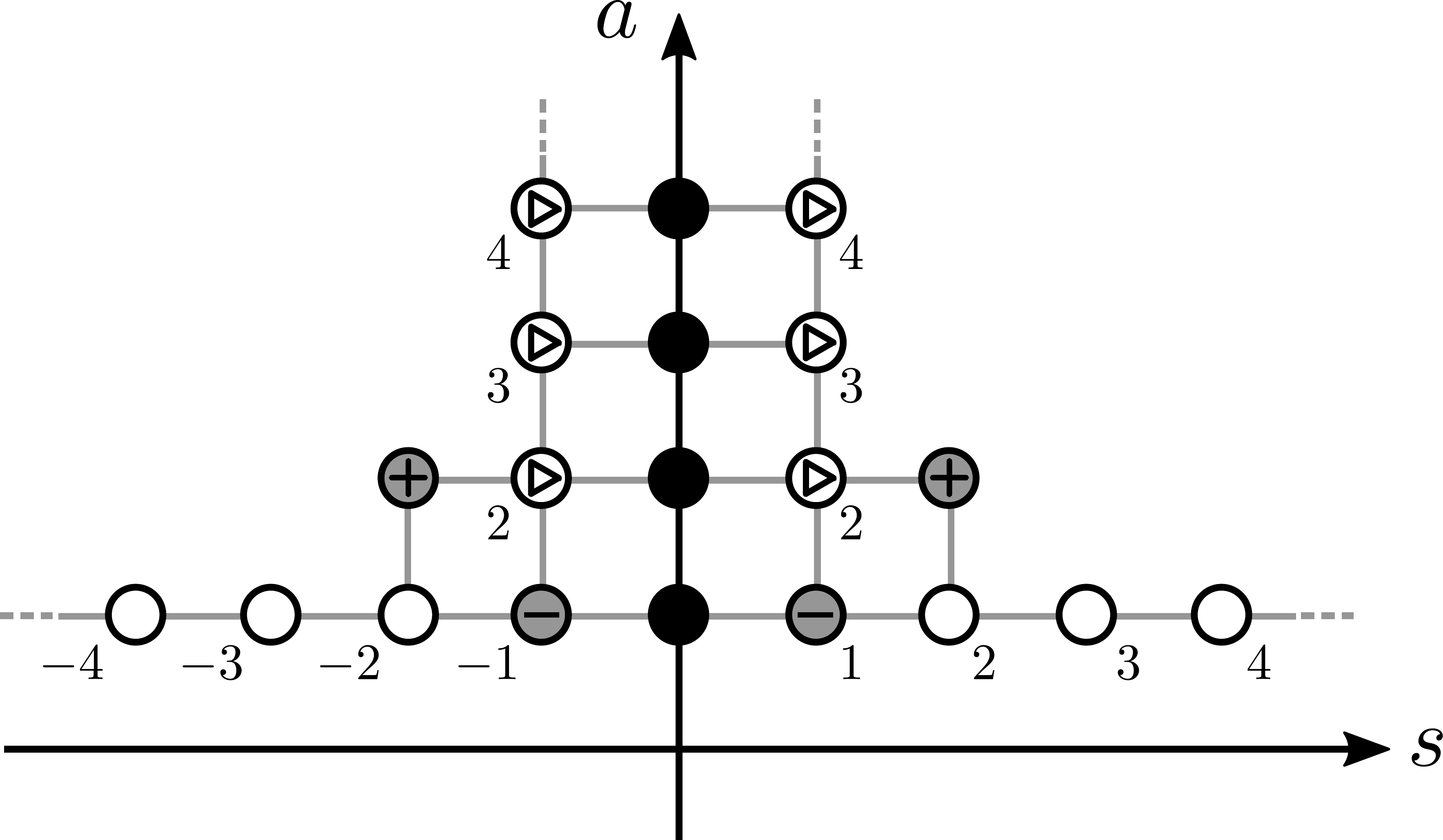}
\caption{Bound state $Y$-functions and $Y$-functions in the T-hook. $Y$-functions in $\mathcal{N}=4$ SYM are defined for each node of the T-hook shown above. The identification with the bound-state label can be read off from the symbols at each node.}
\label{fig:thook}
\end{figure}

This exercise illustrates that the simple $Z_2$ identification allows us to convert the usual TBA into the TBA relevant for the $g$-function. In our case, the $Z_2$ identification also needs to reflect the unfolding structure of the mirror ABA. By embedding the bound-state $Y$-functions into the $Y$-functions in the T-hook (see figure \ref{fig:thook})\fn{The two signs correspond to the left and the right wings.},
\beq
\begin{aligned}
Y_Q^{\bullet}\mapsto Y_{Q,0}\comma\quad &Y_M^{\triangleright}\mapsto Y_{M+1,\pm 1}\comma\quad Y_N^{\circ}\mapsto Y_{1,\pm (N+1)}\comma \quad Y_+^{y}\mapsto Y_{2,\pm 2}\comma\quad Y_-^{y}\mapsto Y_{1,\pm 1}\comma
\end{aligned}
\eeq
we can express the relevant $Z_2$ identification for our problem as follows:
\beq
Y_{a,s}(u)=Y_{a,-s}(\bar{u})\period
\eeq
With this $Z_2$ identification, we can ``fold'' the standard TBA \cite{Gromov:2009tv,Gromov:2009bc,Gromov:2009zb,Bombardelli:2009ns,Arutyunov:2009ur} to get the TBA of our problem. The result reads
\beq
\begin{aligned}
\log Y_{a,0}=\varphi^{p}_{a,0}\equiv &\,-L\tilde{E}_a+\log(1+Y_{b,0})\ast (\blue{\mathcal{K}_+})^{\bullet\bullet}_{b,a}
+\log(1+Y_{m,1})\star(\blue{\mathcal{K}_+})^{\triangleright\bullet}_{m-1,a}\\
&\,+\log(1+1/Y_{2,2})\,\,\widehat{\star}\,\,(\blue{\mathcal{K}_+})^{y\bullet }_{+a}+\log(1+Y_{1,1})\,\,\widehat{\star}\,\,(\blue{\mathcal{K}_+})^{y\bullet }_{-a},\\
\log Y_{a,1}=\varphi^{p}_{a,1}\equiv &\,\log(1+Y_{b,0})\ast(\textcolor{blue}{\mathcal{K}_+})^{\bullet\triangleright}_{b,a-1}
+\log(1+Y_{b,1})\star K^{\triangleright\triangleright}_{b-1,a-1}\\
&\,+\log\frac{1+Y_{1,1}}{1+1/Y_{2,2}}\,\widehat{\star}\, K^{y\triangleright }_{-,a-1},\\
\log Y_{1,s}=\varphi^{p}_{1,s}\equiv &\,\log(1+1/Y_{1,t})\star K^{\circ\circ}_{t-1,s-1}+\log\frac{1+Y_{1,1}}{1+1/Y_{2,2}}\,\widehat{\star}\, K^{y\circ }_{-,s-1},\\
\log Y_{2,2}=\varphi^{p}_{2,2}\equiv&\,-\log(1+Y_{a,0})\ast (\textcolor{blue}{\mathcal{K}_+})^{\bullet y}_{a,+}-\log\frac{1+1/Y_{1,m}}{1+Y_{m,1}}\star K^{\circ y}_{m-1,+}+\pi i,\\
\log Y_{1,1}=\varphi^{p}_{1,1}\equiv &\,\log(1+Y_{a,0})\ast (\textcolor{blue}{\mathcal{K}_+})^{\bullet y}_{a,-}+\log\frac{1+1/Y_{1,m}}{1+Y_{m,1}}\star K^{\circ y}_{m-1,-}+\pi i\comma
\end{aligned}
\eeq
where the kernels $K$'s are the logarithmic derivatives of the S-matrices defined, for instance, by
\beq
K_{a,b}^{\triangleright\triangleright}(u,v) \equiv \frac{1}{i}\del_u \log S^{\triangleright\triangleright}_{a,b} (u,v)\qquad \text{etc.}\comma
\eeq
and the indices runs over positive integers and the summation over the repeated indices are assumed. The symbols $\star$, $\ast$, and $\widehat{\star}$ denote the convolutions along $[-\infty,\infty]$, $[0,\infty]$ and $[-2g,2g]$ respectively while the kernel $\blue{\mathcal{K}_+}$ is defined by
\begin{align}
\mathcal{K}_+(u,v)=K(u,v)+K(u,-v)\period
\end{align}
We also introduced the TBA analogues of phase factors $\varphi^{p}$'s by the right hand sides of the TBA equations.
We also checked these equations by going through the standard derivation.

\subsection{Exact $g$-function\label{subsec:N=4gfunction}}
Having derived the TBA equations, we are now in a position to write down the exact $g$-function for the ground state. In the toy example in section \ref{sec:integrability}, the result for the $g$-function consists of two parts,
\beq
g_{\rm toy} =\underbrace{\exp\left[\int_0^{\infty}\frac{du}{2\pi}\Theta (u)\log (1+Y(u))\right]}_{=\,{\tt prefactor}}\times \underbrace{\sqrt{\frac{\det \left[1-\hat{G}_{-}\right]}{\det \left[1-\hat{G}_{+}\right]}}}_{=\,{\tt Fredholm}}\comma
\eeq
where $\Theta$ is the log derivative of the reflection amplitude while $\hat{G}_{\pm}$ give the Fredholm determinants. Among the two contributions, the prefactor part is rather easy since it simply comes from the $O(1)$ piece of the saddle-point action. In the present case, the reflection amplitudes are nontrivial only for the middle-node bound states and we thus have
\beq
{\tt prefactor}=\exp\left[\sum_{a}\int_0^{\infty}\frac{du}{2\pi}\Theta_a (u) \log (1+Y_{a,0}(u))\right]\comma
\eeq
with\fn{Note that since $r_a (u)$ satisfies $r_a (u)r_{a}(\bar{u})=1$, we can alternatively express the first term as $+\frac{1}{i}\del_u r_a (u)$. Note also that, in section \ref{sec:integrability}, the last term in \eqref{eq:defofthetaGG} was written in a form $-\frac{1}{i}\del_u \log \left.S_{aa}^{\bullet\bullet}(u,v)\right|_{v=-u}$. One can show the equivalence of the two expressions by using the parity invariance of the S-matrix. }
\beq\label{eq:defofthetaGG}
\Theta_a (u)=-\frac{1}{i}\del_u \log r_a (\bar{u})-\pi \delta (u)-\frac{1}{2i}\del_u \log S_{aa}^{\bullet\bullet} (u,\bar{u})\period
\eeq
See Appendix \ref{ap:gfunctionNested} for further explanations.

As discussed in section \ref{subsec:TBAgeneral}, the Fredholm determinant for the toy example can be re-expressed as
\beq
\left.{\tt Fredholm}\right|_{\rm toy}=\frac{\sqrt{\det \left[1-\hat{G}\right]}}{\det \left[1-\hat{G}_{+}\right]}\comma
\eeq
and the numerator and the denominator can be computed simply by taking functional variations of the periodic TBA and the TBA with boundaries respectively. As we discuss in Appendix \ref{ap:gfunctionNested}, this feature continues to hold also for the nested Bethe ansatz systems. We thus obtain the following representations for the Fredholm determinant part
\beq
{\tt Fredholm}=\frac{\sqrt{\det \left[1-\hat{G}\right]}}{\det \left[1-\hat{G}_{+}\right]}\comma
\eeq
where the kernel $\hat{G}_{+}$ is given by the functional variation
\beq
\left[\hat{G}_{+}\right]_{(a,s),(b,t)}(u,v) =\left[\frac{\delta \varphi^{p}_{a,s}(u)}{\delta(\log Y_{b,t}(v))}\right]^{T}\period
\eeq
The (left) action of the operator $\hat{G}_{+}$ involves an integral over $v$ and a summation over the indices $(b,t)$.
Here, to bring the final result into a form more standard in the literature, we performed the ``transposition''\fn{This is a functional analogue of the transposition of a matrix and does not change the value of the determinant.} $T$, which technically amounts to swapping $u$ and $v$ and the corresponding indices in the bilocal kernel.
 For instance, for $(a,0)$ and $(b,0)$, we have
\beq
\left[\frac{\delta \varphi^{p}_{a,0}(u)}{\delta(\log Y_{b,0}(v))}\right]^{T}=\left[\frac{\left(\mathcal{K}_{+}\right)^{\bullet\bullet}_{b,a}(v,u)}{1+1/Y_{b,0}(v)}\right]^{T} =\frac{\left(\mathcal{K}_{+}\right)^{\bullet\bullet}_{a,b}(u,v)}{1+1/Y_{b,0}(v)}\period
\eeq
Similarly, $\hat{G}$ can be computed by taking a functional variation of the phase factors for the periodic TBA $\varphi_{a,s}$ which we list in Appendix \ref{ap:TBAfull}:
\beq
\left[\hat{G}\right]_{(a,s),(b,t)}(u,v)=\left[\frac{\delta \varphi_{a,s}(u)}{\delta(\log Y_{b,t}(v))}\right]^{T}\period
\eeq

\paragraph{Final result}Combining the two factors, we obtain the following expression for the ground state $g$-function (or equivalently the structure constant for the BPS operator),
\beq\label{eq:BPSnonperturbativeg}
\mathfrak{D}_{\mathcal{O}_{\circ}}=-\frac{i^{J}+(-i)^{J}}{\sqrt{J}}\times g\comma
\eeq
with
\beq\label{eq:gfiniteGandGplus}
g=\exp\left[\sum_{a=1}^{\infty}\int_0^{\infty}\frac{du}{2\pi}\Theta_a (u) \log (1+Y_{a,0}(u))\right] \times \frac{\sqrt{\det\left[1-\hat{G}\right]}}{\det \left[1-\hat{G}_{+}\right]}\period
\eeq
As is the case with the asymptotic formula, we included the universal prefactor $-(i^{J}+(-i)^{J})/\sqrt{J}$, which comes from the cyclicity of the single-trace operator and the fact that the relevant boundary state is a sum of two boundary states.

From the field-theory analysis, we know that the structure constant of the BPS operator is protected; namely $g$ in \eqref{eq:BPSnonperturbativeg} must be unity. To check this, we use the fact that the middle-node $Y$-functions all vanish for the ground states, $Y_{a,0}=0$. This immediately sets the term involving $\Theta_a$ to be unity. We can also see that the ratio of determinants becomes unity in the following way: Setting $Y_{a,0}=0$ kills all the kernels involving $Y_{a,0}$. As a result, the Fredholm determinant $\det [1-\hat{G}]$ gets factorized into contributions from the left and the right wings which are identical and we get
\beq
\det \left[1-\hat{G}\right]\quad \overset{Y_{a,0}=0}{=}\quad\left(\det \left[1-\hat{G}_{+}\right]\right)^2\period
\eeq

We could also rewrite the final result into a more standard form by factorizing the $1-\hat{G}$ into $(1-\hat{G}_{+})\times (1-\hat{G}_{-})$ where the action of $\hat{G}_{-}$ is defined by replacing the kernel $\mathcal{K}_{+}$ in $\hat{G}_{+}$ with
\beq
\mathcal{K}_{+}\mapsto\mathcal{K}_{-}(u,v)\equiv K (u,v)-K(u,-v)\period
\eeq
We then get
\beq\label{eq:gfiniteGplusminus}
g=\exp\left[\sum_{a}\int_0^{\infty}\frac{du}{2\pi}\Theta_a (u) \log (1+Y_{a,0}(u))\right] \times \sqrt{\frac{\det\left[1-\hat{G}_{-}\right]}{\det \left[1-\hat{G}_{+}\right]}}\period
\eeq

\paragraph{Clarifying remark} Let us also address the subtlety in the derivation of the $g$-function in the nested Bethe ansatz system pointed out recently in \cite{Kostov:2019fvw}. In the derivation of the $g$-function, we used the string-hypothesis for the bound states at the nested level. In order to justify the use of the string hypothesis, one needs to have a large number of momentum-carrying magnons. This is true as long as the temperature of the system (or equivalently the circumference of the cylinder $R$) is O(1), but it can potentially break down in the extreme low temperature limit (which corresponds to the limit $R\to \infty$). This basically implies that the result computed in this method can be contaminated by some ``offset'' which changes the value at $R\to \infty$:
\beq
g_{\rm computed}(R)=g_{\text{offset}} \times g_{\rm true}(R)\period
\eeq
This was indeed the case in the SU(2)$_k$ WZNW model studied in \cite{Kostov:2019fvw}, in which they found $g_{\text{offset}}$ to be infinite and regularized it by twisting the periodicity.

By contrast, in the case at hand, such an offset seems absent since, as we saw above, the result correctly reproduces the structure constants of BPS operators of any length. In a sense, our setup is much nicer than what they considered since the supersymmetry plays the role of a regulator and cancels various potential divergences. We should also note that the detailed structure of our $g$-function is rather different from what they considered in \cite{Kostov:2019fvw}. Most notably, the $\Theta$-part of our result only contains the momentum-carrying $Y$-functions making it obviously convergent in the large $R$ limit.
\subsection{Excited state $g$-function in SL(2) sector\label{subsec:SL2finite}}
Let us now try to generalize the result to the excited states by employing the idea of the analytic continuation trick.
Although the basic idea is simple---we analytically continue the TBA and pick up whatever poles which cross the integration contour---the details of the procedure are not clearly understood for general operators. Therefore, here we restrict ourselves to the operators in the SL(2) sector, for which the analytic continuation is well-understood \cite{Gromov:2009zb}\fn{This is up to potential subtleties discussed in \cite{Arutyunov:2009ax}, which come from new poles and zeros crossing the integration contour above critical couplings. When this happens, our expression needs to be modified by the additional source terms.}. In what follows, we use the expression \eqref{eq:gfiniteGplusminus}, which involves $\hat{G}_{\pm}$, but it is also straightforward to apply the analytic continuation to the original expression \eqref{eq:gfiniteGandGplus}.

The analytic continuation employed in \cite{Gromov:2009zb} simply amounts to adding the contributions from poles to $\log (1+Y_{1,0})$ in the TBA equations. Following exactly the same logic as in section \ref{subsec:analyticcont}, one can show that this modifies the action of all the kernels involving $(1+1/Y_{1,0})^{-1}$ in the following way,
\beq\label{eq:excitedkernelsl2}
\begin{aligned}
& \int_0^{\infty} \frac{dv}{2\pi}\frac{\left(\mathcal{K}_{\pm}\right)_{x,1}^{X,\bullet}(u,v)}{1+1/Y_{1,0}(v)}f(v)\quad \mapsto
\\&\qquad \sum_{k=1}^{M/2}\frac{i\left(\mathcal{K}_{\pm}\right)_{x,1}^{X,\bullet}(u,u_k)}{\del_u \log Y_{1,0}(u_k)}f(u_k)+\int_0^{\infty} \frac{dv}{2\pi}\frac{\left(\mathcal{K}_{\pm}\right)_{x,1}^{X,\bullet}(u,v)}{1+1/Y_{1,0}(v)}f(v)\comma
\end{aligned}
\eeq
where $x$ and $X$ can take various indices and symbols representing different bound states. Note also that $u_k$ and $\bar{u}_k$ are living on the physical sheet and the relevant quantities are defined through the analytic continuation. Following the convention in section \ref{subsec:analyticcont}, we denote the kernels after these modifications by $\hat{G}_{\pm}^{\bullet}$.

For the $\Theta$-part, again following the logic in section \ref{subsec:analyticcont}, we find that the contributions from the poles give extra factors\fn{Here we used $\tilde{S}_0(u,v)\equiv S_0 (u^{\gamma},v^{\gamma})$ (see \eqref{eq:Smatrixmirrordiag}), and the crossing equation for the bulk dressing phase together with the zero momentum condition.}
\beq
\begin{aligned}
\prod_{k=1}^{M/2}\left(r_0(\overline{u_k^{-\gamma}})\sqrt{\tilde{S}_0(u_k^{-\gamma},\overline{u_k^{-\gamma}})}\right)&=\prod_{k=1}^{M/2}\frac{(x_k^{-}+1/x_k^{-})(1+x_k^{+}/x_k^{-})}{2(x_k^{-}+1/x_k^{+})}g_0 (u)\sqrt{S_0 (u_k,\bar{u}_k^{2\gamma})}\\
&=\prod_{k=1}^{M/2}\frac{u_k^2+1/4}{u_k^2}\sigma_B^2(u_k)\period
\end{aligned}
\eeq

Combining everything, we arrive at the following formula for the structure constant in the SL(2) sector at finite coupling:
\beq\label{eq:finalSL2finite}
\begin{aligned}
\mathfrak{D}_{\mathcal{O}_{\bf u}}=&-\frac{i^{J}+(-i)^{J}}{\sqrt{J}}\exp\left[\sum_{a=1}^{\infty}\int_0^{\infty}\frac{du}{2\pi}\Theta_a (u) \log (1+Y_{a,0}(u))\right]\\
&\times\sqrt{\left(\prod_{1\leq s\leq \frac{M}{2}}\frac{u_s^2+\frac{1}{4}}{u_s^2}\sigma_B^2(u_s)\right)\frac{\det \left[1-\hat{G}^{\bullet}_{-}\right]}{\det \left[1-\hat{G}^{\bullet}_{+}\right]}}\period
\end{aligned}
\eeq
This is the main result in this paper. We should however note that, as is well-known, the procedures of the analytic continuation contain some ambiguities and therefore the result here should be regarded as a conjecture. It would be an important future problem to put this formula to extensive tests by analyzing various limits and also by performing numerical computations.
\subsection{From exact result to the asymptotic formula\label{subsec:reltoAsympt}} Let us now discuss the relation to the asymptotic formula derived in section \ref{sec:asymptotic}. In the toy examples analyzed in section \ref{sec:integrability}, one could recover the asymptotic formula from the excited-state $g$-function by dropping the integral convolutions in the Fredholm determinants and recasting the remaining contribution into finite-dimensional determinants. In the case at hand, such an argument still applies to the prefactor since the integrals involving $\Theta_a$ in \eqref{eq:finalSL2finite} all vanish in the asymptotic limit. However it is rather nontrivial to see that the Fredholm determinants reduce to the correct finite-dimensional determinants because of the following reasons:
\begin{enumerate}
\item Some of the convolution kernels contain only the $Y$-functions at the nested levels, which are not suppressed in the asymptotic limit. Therefore, we cannot simply drop those integrals in the Fredholm determinants.
\item In the toy example in section \ref{sec:integrability}, we used the crossing symmetry of the S-matrix to convert the kernels $\mathcal{K}_{\pm}$ in the mirror channel to $\mathcal{K}_{\mp}$ in the physical channel. The same argument does not apply in the present case since the scalar factor of the S-matrix $S_0$ does not satisfy the crossing symmetry by itself; it only satisfies the crossing symmetry after including the matrix part.
\end{enumerate}
These difficulties can be overcome by re-expressing the Fredholm determinants in terms of the {\it exact Gaudin determinant}. We will illustrate the main idea by working with the simply toy model introduced in section~\ref{subsec:analyticcont}. Generalization to the cases with nested levels is straightforward and will be discussed in appendix~\ref{ap:asymptotic}.\par

To start with, we split the action of the kernel $\hat{G}^{\bullet}_{\pm}$ into a \emph{discrete sum} and an \emph{continuous integral}
\begin{align}
\hat{G}_{\pm}^{\bullet}\cdot f(u)=\mathsf{S}_{\pm}\cdot f(u)+\mathsf{I}_{\pm}\cdot f(u)
\end{align}
where from (\ref{eq:actionofGbullet}), we have
\begin{align}
\label{eq:kernelsIP}
\mathsf{S}_{\pm}\cdot f(u)\equiv&\, \sum_{k=1}^{M/2}\frac{i\mathcal{K}_{\pm}(u,\tilde{u}_k)}{\partial_u\log Y(\tilde{u}_k)}f(\tilde{u}_k)
=\sum_{k=1}^{M/2}\frac{\mathcal{K}_{\pm}(u,\tilde{u}_k)}{\partial_u\phi(u_j)}f(\tilde{u}_k),\\\nonumber
\mathsf{I}_{\pm}\cdot f(u)\equiv&\,\int_0^{\infty}\frac{dv}{2\pi}\frac{\mathcal{K}_{\pm}(u,v)}{1+1/Y(v)}f(v).
\end{align}
We can then factorize the Fredholm determinant as
\begin{align}
\det(1-\hat{G}_{\pm}^{\bullet})=\det(1-\mathsf{I}_{\pm})\det(1-\widehat{\mathsf{S}}_{\pm})\comma
\end{align}
where $\widehat{\mathsf{S}}_{\pm}$ are the \emph{dressed} summation kernel defined by
\begin{align}
\widehat{\mathsf{S}}_{\pm}\equiv \frac{\mathsf{S}_{\pm}}{1-\mathsf{I}_{\pm}}=\mathsf{S}_{\pm}\ast\left(1+\mathsf{I}_{\pm}+\mathsf{I}_{\pm}^2+\cdots \right).
\end{align}
Therefore, the ratio of Fredholm determinants can be written as
\begin{align}
\label{eq:decomposeToy}
\frac{\det(1-\hat{G}_-^{\bullet})}{\det(1-\hat{G}_+^{\bullet})}
=\frac{\det(1-\mathsf{I}_-)}{\det(1-\mathsf{I}_+)}\times
\frac{\det(1-\widehat{\mathsf{S}}_-)}{\det(1-\widehat{\mathsf{S}}_+)}
\end{align}
The first ratio on the right hand side of (\ref{eq:decomposeToy}) becomes trivial in the asymptotic limit, as was already discussed in section~\ref{subsec:analyticcont}. The second ratio can be rewritten as
\begin{align}
\label{eq:ratiodet2}
\frac{\det\big(1-\widehat{\mathsf{S}}_-\big)}{\det\big(1-\widehat{\mathsf{S}}_+\big)}=
\frac{\det\big(\partial_u\phi(u_j)\delta_{jk}-\widehat{\mathcal{K}}_-(\tilde{u}_j,\tilde{u}_k)\big)}
{\det\big(\partial_u\phi(u_j)\delta_{jk}-\widehat{\mathcal{K}}_+(\tilde{u}_j,\tilde{u}_k)\big)}
\end{align}
where
\begin{align}
\widehat{\mathcal{K}}_{\pm}=\frac{\mathcal{K}_{\pm}}{1-\mathsf{I}_{\pm}}=\mathcal{K}_{\pm}\ast\big(1+\mathsf{I}_{\pm}+\mathsf{I}^2_{\pm}+\cdots\big)
\end{align}
The crucial observation is that the determinants in (\ref{eq:ratiodet2}) can be identified with the Gaudin determinants associated to the \emph{exact quantization condition}.
\paragraph{Exact Gaudin determinant} In finite volume, the exact quantization condition for the rapidities of a \emph{parity symmetric} state is given by (\ref{eq:finitevolumephase}):
\begin{align}
\label{eq:exactquant1}
\phi(u_j)=2\pi n_j,\qquad n_j\in\mathbb{Z}
\end{align}
where
\begin{align}
\label{eq:exactquant2}
\phi(u)=p(u)L+\frac{1}{i}\sum_{k=1}^{M/2}\log\left[S(u,u_k)S(u,-u_k)\right]+\frac{1}{i}\log(1+Y)\ast \mathcal{K}_+(u).
\end{align}
Using this quantization condition, the exact Gaudin determinant is given by
\begin{align}
\det_{j,k}\left(\frac{\partial\phi(u_j)}{\partial u_k}\right).
\end{align}
Each matrix element is given by
\begin{align}
\label{eq:partialphiful}
\frac{\partial\phi(u_j)}{\partial u_k}=\partial_u\phi(u_j)\delta_{jk}-\mathcal{K}_-(u_k,u_j)+\frac{1}{i}\int_0^{\infty}\frac{\partial\epsilon(v)}{\partial u_k}\frac{\mathcal{K}_+(v,\tilde{u}_j)}{1+1/Y(v)}\frac{dv}{2\pi}.
\end{align}
Notice that the integral over $v$ is in the mirror channel and we have introduced the pseudo-energy $\epsilon(u)=\log Y(u)$ for simplicity. The excited state pseudo energy $\epsilon(u)$ depends on the rapidities $u_k$. To proceed, we need to calculate the `back reaction' $\partial\epsilon(u)/\partial u_k$, which can be done by taking the derivative with respect to the TBA equation (\ref{eq:toyexcitedTBA})
\begin{align}
\frac{\partial\epsilon(u)}{\partial u_k}+i\mathcal{K}_+(\tilde{u}_k,u)-\int_0^{\infty}\frac{dv}{2\pi}\frac{\mathcal{K}_+(v,u)}{1+1/Y(v)}\frac{\partial\epsilon(v)}{\partial u_k}=0.
\end{align}
Using (\ref{eq:kernelsIP}), it can be rewritten as
\begin{align}
\frac{\partial\epsilon(u)}{\partial u_k}=-i\mathcal{K}_+(\tilde{u}_k,u)+\frac{\partial\epsilon}{\partial u_k}\ast\mathsf{I}_+(u).
\end{align}
A formal solution can be obtained by iteration
\begin{align}
\frac{\partial\epsilon(u)}{\partial u_k}=-i\mathcal{K}_+\ast\big(1+\mathsf{I}_+ +\mathsf{I}_+^2+\cdots \big).
\end{align}
Plugging this into (\ref{eq:partialphiful}) we find
\begin{align}
\frac{\partial\phi(u_j)}{\partial u_k}=&\,\partial_u\phi(u_j)\delta_{jk}-\mathcal{K}_-(u_k,u_j)-\mathcal{K}_+\ast\big(\mathsf{I}_+ +\mathsf{I}_+^2+\cdots \big)(\tilde{u}_k,\tilde{u}_j)\\\nonumber
=&\,\partial_u\phi(u_j)\delta_{jk}-\widehat{\mathcal{K}}_-(u_k,u_j)
\end{align}
where in the second line we have used the relation
\begin{align}
\mathcal{K}_{\pm}(\tilde{u}_j,\tilde{u}_k)=\mathcal{K}_{\mp}(u_j,u_k).
\end{align}
Therefore we have proved that
\begin{align}
\label{eq:coreRes}
\left(\frac{\partial\phi(u_j)}{\partial u_k}\right)^T=\partial_u\phi(u_j)\delta_{jk}-\widehat{\mathcal{K}}_-(u_j,u_k)
\end{align}

\paragraph{Final rewriting}
We have shown that the \emph{denominator} of (\ref{eq:ratiodet2}) can be written as exact Gaudin determinant corresponding to the exact quantization conditions (\ref{eq:exactquant1}) and (\ref{eq:exactquant2}). The \emph{numerator} cannot be written directly as a Gaudin determinant for a certain quantization condition. To proceed, we use the rewriting in section \ref{subsec:analyticcont}
\begin{align}
\label{eq:rewrite}
\frac{\det\big[1-\hat{G}_-^{\bullet}\big]}{\det\big[1-\hat{G}_+^{\bullet}\big]}
=\frac{\det\big[1-\hat{G}_-^{\bullet}\big]\det\big[1-\hat{G}_+^{\bullet}\big]}{\big(\det\big[1-\hat{G}_+^{\bullet}\big]\big)^2}
=\frac{\det\big[1-\hat{G}^{\bullet}\big]}{\big(\det\big[1-\hat{G}_+^{\bullet}\big]\big)^2}
\end{align}
where the kernel $\hat{G}^{\bullet}$ is defined as
\begin{align}
G^{\bullet}\cdot f(u)=\sum_{k=1}^M\frac{i\mathcal{K}(u,\tilde{u}_k)}{\partial_u\log Y(\tilde{u}_k)}f(\tilde{u}_k)+\int_{-\infty}^{\infty}\frac{dv}{2\pi}\frac{\mathcal{K}(u,v)}{1+1/Y(v)}f(v)
\end{align}
Notice that the sum is over the full set of $M$ physical rapidities $\{u_1,-u_1,\cdots,u_{M/2},-u_{M/2}\}$ and the integral is over the whole real axis, which are different from (\ref{eq:kernelsIP}). The kernel $\mathcal{K}$ is given by
\begin{align}
\mathcal{K}=\frac{1}{2}(\mathcal{K}_+ + \mathcal{K}_-)
\end{align}
One can prove
\begin{align}
\det\big[1-\hat{G}_-^{\bullet}\big]\det\big[1-\hat{G}_+^{\bullet}\big]=\det\big[1-\hat{G}^{\bullet}\big]
\end{align}
using a similar approach to the proof of (\ref{eq:combiningtwodets}).
The right hand side of (\ref{eq:rewrite}) can be written as
\begin{align}
\frac{\det\big[1-G^{\bullet}\big]}{\left(\det\big[1-G_+^{\bullet}\big]\right)^2}
=\frac{\det\big[1-\mathsf{I}\big]}{\left(\det\big[1-\mathsf{I}_+\big]\right)^2}\times
\frac{\det\big[1-\widehat{\mathsf{S}}\big]}{\big(\det\big[1-\widehat{\mathsf{S}}_+\big]\big)^2}
\end{align}
Repeating the derivation of (\ref{eq:coreRes}), we find
\begin{align}
\label{eq:detSS}
\frac{\det\big[1-\widehat{\mathsf{S}}\big]}{\big(\det\big[1-\widehat{\mathsf{S}}_+\big]\big)^2}
=\frac{\det\big[\partial_{u_k}\tilde{\phi}(u_j)\big]}{\big(\det\big[\partial_{u_k}\phi(u_j)\big]\big)^2}
\end{align}
where $\tilde{\phi}(u)$ is given by
\begin{align}
\tilde{\phi}(u)=p(u)L+\frac{1}{i}\sum_{k=1}^M\log S(u,u_k)+\frac{1}{i}\log(1+Y)\ast\mathcal{K}(u).
\end{align}
Therefore we see that the determinants in (\ref{eq:detSS}) are given by the \emph{exact Gaudin determinants}, which correspond to the exact quantization conditions of the rapidities in finite volume. The key point is that, in the asymptotic limit $L\gg 1$, the exact quantization condition becomes the asymptotic Bethe equations. As a result, we have
\begin{align}
\label{eq:summationPart}
\frac{\det\big[\partial_{u_k}\tilde{\phi}(u_j)\big]}{\big(\det\big[\partial_{u_k}\phi(u_j)\big]\big)^2}\overset{L\to\infty}{\rightarrow}
\frac{\det\big[\partial_{u_k}\tilde{\phi}^{\text{asym}}(u_j)\big]}{\big(\det\big[\partial_{u_k}\phi^{\text{asym}}(u_j)\big]\big)^2}=\frac{\det G_+}{\det G_-}
\end{align}
At the same time, as mentioned before, the integral parts are exponentially suppressed and become trivial in the asymptotic limit
\begin{align}
\label{eq:integralPart}
\frac{\det\big[1-\mathsf{I}\big]}{\left(\det\big[1-\mathsf{I}_+\big]\right)^2}\overset{L\to\infty}{\rightarrow} 1
\end{align}
Combining (\ref{eq:summationPart}) and (\ref{eq:integralPart}), we see that in the asymptotic limit we have
\begin{align}
\lim_{L\to \infty}\frac{\det\big[1-\hat{G}_-^{\bullet}\big]}{\det\big[1-\hat{G}_+^{\bullet}\big]}=\frac{\det G_+}{\det G_-}
\end{align}

\section{Further Checks\label{sec:check}}
In this section, we perform extensive tests of our conjecture at finite coupling. More precisely, we test the asymptotic formula and also the reflection matrix at weak and strong couplings and find perfect agreements. Unfortunately, there is currently no available data that we can use to test the finite-size effects. Nevertheless, since the asymptotic formula and the nonperturbative $g$-function are closely related---the asymptotic formula was conjectured by inspecting the structure of the exact $g$-function, and it is likely, although not proven yet, that the two are related by the analytic continuation---the validity of the asymptotic formula is a strong indication of the correctness of the whole formalism.
\subsection{SU(2) sector at one loop\label{subsec:su21loop}}
The first test is the one-loop correction to the result in the SU(2) sector. As with the single-trace three-point functions \cite{Okuyama:2004bd,Alday:2005nd,Gromov:2012uv}, there are two sources of corrections at one loop: The first one is a correction to the Bethe state itself since we now need an eigenstate of the two-loop dilatation operator. The second one is a correction from the one-loop Feynman diagrams. We performed the computation explicitly in Appendix \ref{ap:1loop3pt} for the operators in the SO(6) sector, and found that this second correction produces an insertion of the one-loop Hamiltonian. Written explicitly, we found that all we need to do is to perform the following substitution:
\beq
\begin{aligned}
\frac{\langle \text{N\'{e}el}_{0} |{\bf u}\rangle}{\sqrt{\langle {\bf u}|{\bf u}\rangle}}\quad  \mapsto &\quad \frac{\Big(\langle \text{N\'{e}el}_0|-\frac{1}{2}\langle \text{N\'{e}el}_0|\mathbb{H}\Big)|{\bf u};g^2\rangle}{\sqrt{\langle{\bf u};g^2|{\bf u};g^2\rangle}} \\
&=\frac{\langle \text{N\'{e}el}_{0} |{\bf u};g^2\rangle}{\sqrt{\langle {\bf u};g^2|{\bf u};g^2\rangle}} -\frac{1}{2}\frac{\langle \text{N\'{e}el}_{0}|\mathbb{H} |{\bf u}\rangle}{\sqrt{\langle {\bf u}|{\bf u}\rangle}}+O(g^4)\period
\end{aligned}
\eeq
Here $|{\bf u};g^2\rangle$ is a loop-corrected Bethe state and $\mathbb{H}$ is the one-loop Hamiltonian
\beq
\mathbb{H}|{\bf u}\rangle =\gamma_{\bf u}|{\bf u}\rangle \comma\qquad \gamma_{\bf u} =\sum_{k}\frac{2g^2}{u_k^2+\frac{1}{4}}\period
\eeq
We should emphasize that we have the {\it full Hamiltonian} of the spin chain unlike the case of the single-trace three-point function, in which we insert the {\it Hamiltonian densities} at the splitting points. This is mainly because every neighboring sites in the N\'{e}el state come from different operators and therefore can be regarded as an analogue of the splitting points.

Among the two contributions, the overlap between the N\'{e}el state and the loop-corrected Bethe state was computed\fn{We also performed extensive tests of the formula by ourselves.} in \cite{Buhl-Mortensen:2017ind} in the context of the defect one-point function. The result reads
\beq
\frac{\langle \text{N\'{e}el}_{0} |{\bf u};g^2\rangle}{\sqrt{\langle {\bf u};g^2|{\bf u};g^2\rangle}}=2 \left(\frac{i}{2}\right)^{M}\left(1+\frac{\gamma_{\bf u}}{2}\right)\sqrt{\left(\prod_{1\leq s\leq \frac{M}{2}}\frac{u_s^2+\frac{1}{4}}{u_s^2}\right)\left.\frac{\det G_{+}^{\rm SU(2)}}{\det G_{-}^{\rm SU(2)}}\right|_{O(g^2)}}\period
\eeq
Here $G_{\pm}|_{O(g^2)}$ are the one-loop counterparts of the tree-level $G_{\pm}$, which are defined by the loop corrected momentum and the S-matrix and evaluated at the two-loop Bethe roots.

The other contribution is even easier; since the Bethe state is an eigenstate of the Hamiltonian, we can simply replace $\mathbb{H}$ with $\gamma_{\bf u}$. Combining the two contributions, we find that the terms proportional to $\gamma_{\bf u}$ cancel. As a result, we obtain the following formula for the one-loop structure constant\fn{Note that ${\sf d}_{\mathcal{O}}$ is the {\it length-stripped} structure constant defined in \eqref{eq:lengthstrippedddef}.}:
\beq
\mathfrak{D}_{\mathcal{O}_{\bf u}}=-\frac{i^{J}+(-i)^{J}}{2^{M}\sqrt{L}}\sqrt{\left(\prod_{1\leq s\leq \frac{M}{2}}\frac{u_s^2+\frac{1}{4}}{u_s^2}\right)\left.\frac{\det G_{+}^{\rm SU(2)}}{\det G_{-}^{\rm SU(2)}}\right|_{O(g^2)}}\period
\eeq
This is in precise agreement with our asymptotic formula at one loop.
\subsection{Direct test of reflection matrix at weak coupling\label{subsec:reflectionweak}}
We can also check directly the matrix part of the reflection matrix that we determined in section \ref{sec:bootstrap}. The strategy is to first construct the two-particle states in an {\it infinitely long} spin chain and contract it with the N\'{e}el state,
\beq\label{eq:neelsimplesectioncheck}
{\rm tr}\left(\mathfrak{Z}(+1)\mathfrak{Z}(-1)\mathfrak{Z}(+1)\mathfrak{Z}(-1)\cdots\right)\period
\eeq
The computations are performed in the SO(6) and the SO(4,2) sectors at tree level. In this subsection, we do not keep track of the overall normalization, and instead focus on relative coefficients between different matrix structures.

\paragraph{SO(6) sector}
The two-particle eigenstate in the one-loop SO(6) sector, which is computed in Appendix \ref{apsubsec:2particleSO6}, takes the following form,
\beq
| \Psi_{c\dot{c}|d\dot{d}} \rangle =\sum_{n}\chi_{c\dot{c}|d\dot{d}} (n)|\ldots , \underset{n}{\bar{Z}},\ldots \rangle +\sum_{n<m}\psi_{c\dot{c}|d\dot{d}}^{a\dot{a}|b\dot{b}}(n,m)|\cdots, \underset{n}{\Phi}^{a\dot{a}},\cdots,\underset{m}{\Phi}^{b\dot{b}},\cdots\rangle\comma
\eeq
with
\beq
\begin{aligned}
\psi^{a\dot{a}|b\dot{b}}_{c\dot{c}|d\dot{d}}(n,m) &= \delta^{a}_{c}\delta^{b}_{d}\delta^{\dot{a}}_{\dot{c}}\delta^{\dot{b}}_{\dot{d}} e^{ip n +iqm}+ S^{ab}_{cd}\dot{S}^{\dot{a}\dot{b}}_{\dot{c}\dot{d}} e^{ipm+iqn}\comma \\
 \chi_{c\dot{c}|d\dot{d}}(n)&=\tilde{\chi}_{c\dot{c}|d\dot{d}}e^{i (p+q)n}\period
 \end{aligned}
\eeq
For explicit expressions for $\tilde{\chi}$ and $S$, see \eqref{eq:chiSso6exp}.

Contracting this state with $q=-p$ against the N\'{e}el state \eqref{eq:neelsimplesectioncheck}, we find that the overlap is given by
\beq
{\tt overlap}\propto \sum_{n}\chi_{c\dot{c}|d\dot{d}} (n) +\sum_{n<m}(-1)^{n+m}\epsilon_{a\dot{a}}\epsilon_{b\dot{b}}\psi_{c\dot{c}|d\dot{d}}^{a\dot{a}|b\dot{b}}(n,m)\period
\eeq
The form factor can be determined by reading off a term which is proportional to the length\fn{As mentioned in the discussion about the decoupling condition in section \ref{subsec:formVSreflection}, the form factor of the boundary state is defined by removing the volume factor from the overlap.} $L\gg 1$ of the chain. The result reads
\beq\label{eq:frombrute}
{\tt overlap}\propto\frac{i}{2u}\epsilon^{c\dot{c}}\epsilon^{d\dot{d}}-\epsilon^{c\dot{d}}\epsilon^{d\dot{c}}\comma
\eeq
where $u$ is defined by $e^{ip}=(u+i/2)/(u-i/2)$.

On the other hand, the tree-level expression for the matrix part \eqref{eq:finalwithz} reads
\beq
A_{\mathcal{G}}=1\comma\qquad B_{\mathcal{G}}=\frac{u+i/2}{u-i/2}\period
\eeq
This leads to an expression
\beq\label{eq:frommatrix}
\langle \mathcal{G}|\Phi^{c\dot{c}} (u)\Phi^{d\dot{d}}(\bar{u})\rangle\propto \frac{-u}{u-i/2}\left(\frac{i}{2u}\epsilon^{c\dot{c}}\epsilon^{d\dot{d}}-\epsilon^{c\dot{d}}\epsilon^{d\dot{c}}\right)\period
\eeq
Comparing \eqref{eq:frombrute} and \eqref{eq:frommatrix}, we see that the relative coefficients between matrix structures perfectly agree.
\paragraph{SO(4,2) sector}
One can also perform the same analysis in the SO(4,2) sector. The two-particle state is given by
\beq
\begin{aligned}
\ket{\Psi_{\gamma\dot{\gamma}|\delta\dot{\delta}}}=& \sum_n \chi^{\alpha\dot{\alpha}|\beta\dot{\beta}}_{\gamma\dot{\gamma}|\delta\dot{\delta}}(n) \ket{\ldots, \underbrace{D^{\alpha\dot{\alpha}}D^{\beta\dot{\beta}}Z}_{n},\ldots}\\&+\sum_{n<m}\psi_{\gamma\dot{\gamma}| \delta\dot{\delta}}^{\alpha\dot{\alpha}|\beta\dot{\beta}}(n,m)\ket{\ldots,\underbrace{D^{\alpha\dot{\alpha}}Z}_{n},\ldots,\underbrace{D^{\beta\dot{\beta}}Z}_{m},\ldots}\comma
 \end{aligned}
\eeq
with
\beq
\begin{aligned}
\psi^{\alpha\dot{\alpha}|\beta\dot{\beta}}_{\gamma\dot{\gamma}|\delta\dot{\delta}}(n,m) &= \delta^{\alpha}_{\gamma}\delta^{\beta}_{\delta}\delta^{\dot{\alpha}}_{\dot{\gamma}}\delta^{\dot{\beta}}_{\dot{\delta}} e^{ip n +iqm}+ S^{\alpha\beta}_{\gamma\delta}\dot{S}^{\dot{\alpha}\dot{\beta}}_{\dot{\gamma}\dot{\delta}} e^{ipm+iqn}\comma \\
 \chi^{\alpha\dot{\alpha}|\beta\dot{\beta}}_{\gamma\dot{\gamma}|\delta\dot{\delta}}(n)&=\tilde{\chi}^{\alpha\dot{\alpha}|\beta\dot{\beta}}_{\gamma\dot{\gamma}|\delta\dot{\delta}}e^{i (p+q)n}\period
 \end{aligned}
\eeq
For details, see Appendix \ref{apsubsec:2particlSO42}.

Now, to compute the contraction with the N\'{e}el state, we need to know how the derivative acts on the propagators. This is mostly easily figured out by first going to the vector notation of the derivatives\fn{The prefactor $1/2$ is just a convenient convention.},
\beq
D^{\alpha\dot{\alpha}}\equiv \frac{1}{2}\pmatrix{cc}{-\del_0+\del_3&\del_1-i\del_2\\\del_1+i\del_2&-\del_0-\del_3}\comma
\eeq
 and then act on the propagator. The action on the propagator is given by
 \beq
 \begin{aligned}
&\del_{i\neq 2} \mapsto 0\comma\qquad \del_{2}\mapsto \begin{cases}-2\quad & \text{for }\mathfrak{Z}(+1)\\2\quad& \text{for }\mathfrak{Z}(-1)\end{cases}\\
&(\del_{i\neq 2})^2\mapsto -2\comma\quad \del_{2}^2 \mapsto 6\period
\end{aligned}
 \eeq
 This translates to the following conversion rule:
 \beq
 |\ldots,\underbrace{D^{\alpha\dot{\alpha}}Z}_{n},\ldots\rangle\mapsto  i \epsilon^{\alpha\dot{\alpha}}(-1)^{n}\comma\qquad |\ldots, D^{\alpha\dot{\alpha}}D^{\beta\dot{\beta}}Z,\ldots\rangle\mapsto -\left(\epsilon^{\alpha\dot{\alpha}}\epsilon^{\beta\dot{\beta}}+\epsilon^{\alpha\dot{\beta}}\epsilon^{\beta\dot{\alpha}}\right)\period
 \eeq

 Using the conversion rule and setting $q=-p$, we find that the overlap is proportional to the following answer\fn{Here again we focused on terms proportional to the spin-chain length $L\gg1$.}:
 \beq
 {\tt overlap}\propto \frac{i}{2u}\epsilon^{\alpha\dot{\alpha}}\epsilon^{\beta\dot{\beta}}+\epsilon^{\alpha\dot{\beta}}\epsilon^{\beta\dot{\alpha}}\period
 \eeq
 On the other hand, the weak-coupling expansion of the matrix part gives
 \beq
 \langle \mathcal{G}|D^{\alpha\dot{\alpha}}(u)D^{\beta\dot{\beta}}(\bar{u})\rangle\propto -\frac{2}{u+i/2}\left(\frac{i}{2u}\epsilon^{\alpha\dot{\alpha}}\epsilon^{\beta\dot{\beta}}+\epsilon^{\alpha\dot{\beta}}\epsilon^{\beta\dot{\alpha}}\right)\period
 \eeq
 Again the relative coefficients are in complete agreement.

 Before ending this subsection, let us make a clarifying remark: Since we only discussed the relative coefficients, the checks performed here might seem rather trivial. However, we emphasize that, if we took other solutions to the symmetry constraints such as the one for the hexagon form factor, the relative coefficients would be different. Therefore, the agreement of the relative coefficients does provide a nontrivial check of our choice of the matrix structure.
\subsection{Four-point functions at tree level and one loop\label{subsec:1loop4pt}}
We now want to test our conjecture for the asymptotic formula in the higher rank sectors by comparing the results with the superconformal block expansion of the four-point functions at tree level and one loop.

\paragraph{Set-up}More specifically, we study the four-point functions of two determinant operators and two BPS single-trace operators of equal lengths,
\beq
\mathcal{O}_{\circ}^{(p)}(x,Y)\equiv {\rm tr}\left((Y\cdot \Phi)^{p}\right)(x)\period
\eeq
In the large $N$ limit, the four-point function consists of a disconnected part, which is a product of two-point functions, and a connected part $G_{\{p,p\}}$ whose leading contribution is proportional\fn{Precisely speaking, there is a $O(1)$ connected term when $p$ is even. See Appendix \ref{ap:4pttreeloop}. However such contributions only affect the OPE data with the twist $\tau\geq 2p$ and therefore does not affect the analysis in this section.} to $1/N$:
\beq
\begin{aligned}
&\langle \mathcal{D}_1 (x_1,Y_1)\mathcal{D}_2 (x_2,Y_2)\mathcal{O}_{\circ}^{(p)}(x_3,Y_3)\mathcal{O}_{\circ}^{(p)}(x_4,Y_4)\rangle={\tt disconnected}+\red{\frac{p}{N}}G_{\{p,p\}}\\
&\Big({\tt disconnected}\equiv\langle \mathcal{D}_1 (x_1,Y_1)\mathcal{D}_2 (x_2,Y_2)\rangle\langle\mathcal{O}_{\circ}^{(p)}(x_3,Y_3)\mathcal{O}_{\circ}^{(p)}(x_4,Y_4) \rangle\Big)\period
\end{aligned}
\eeq
As indicated in red, we defined the connected part by factorizing out the factor $p/N$ in order to make it easier to compare the result with integrability. The factor $1/N$ accounts for the large $N$ counting and $p$ is a combinatorial factor of Wick contractions which originates from the cyclicity of the trace\fn{See the discussion around \eqref{eq:twopntnormal}.}.
 Another important remark is that, in this and the next subsections, we choose the normalization of the operators so that their two-point functions are unit-normalized. This allows us to read off the normalized structure constant directly from the superconformal block expansion.

 In what follows, we analyze the connected contribution $G_{\{p,p\}}$ in the planar limit.
 \paragraph{Perturbative data} In Appendix \ref{ap:4pttreeloop}, we computed $G_{\{p,p\}}$ at tree level and one loop in 't Hooft coupling. For the tree-level computation, we simply used the direct Wick contractions using the PCGG method described in section \ref{sec:PCGG}. For the one-loop computation, we performed the direct Feynman-diagram computation for odd $p$ and then used the lightcone OPE analysis of \cite{Chicherin:2015edu} to generalize the results to even $p$.

 The tree-level result $G^{(0)}_{\{p,p\}}$ is given by
 \beq
 \begin{aligned}
 \frac{G^{(0)}_{\{p,p\}}}{(d_{12})^{N}(d_{34})^{p}}=&-\sum_{s=1}^{\lfloor \frac{p+1}{2}\rfloor}\left(-\frac{z\bar{z}}{\alpha\bar{\alpha}}\right)^{2s-1}\left[\left(\frac{(1-\alpha)(1-\bar{\alpha})}{(1-z)(1-\bar{z})}\right)^{s}+\left(\frac{(1-\alpha)(1-\bar{\alpha})}{(1-z)(1-\bar{z})}\right)^{s-1}\right]\\
 &-2\sum_{s=1}^{\lfloor\frac{p}{2}\rfloor}\left(-\frac{z\bar{z}}{\alpha\bar{\alpha}}\right)^{2s}\left(\frac{(1-\alpha)(1-\bar{\alpha})}{(1-z)(1-\bar{z})}\right)^{s}\period
 \end{aligned}
 \eeq
 As explained in Appendix \ref{ap:4pttreeloop}, each term in the sum describes a different Wick contraction.
 Here $z,\bar{z}, \alpha$ and $\bar{\alpha}$ are the conformal and the R-symmetry cross ratios defined by
 \beq
\begin{aligned}
z\bar{z}&=\frac{x_{12}^2x_{34}^2}{x_{13}^2x_{24}^2}\comma\quad (1-z)(1-\bar{z})=\frac{x_{14}^2x_{23}^2}{x_{13}^2x_{24}^2}\comma\quad \alpha\bar{\alpha}=\frac{y_{12}^2y_{34}^2}{y_{13}^2y_{24}^2}\comma\quad (1-\alpha)(1-\bar{\alpha})=\frac{y_{14}^2y_{23}^2}{y_{13}^2y_{24}^2}\comma
\end{aligned}
\eeq
 with $y_{ij}^2\equiv Y_i\cdot Y_j$.

 The result at one loop $G^{(1)}_{\{p,p\}}$ is given by
 \beq\label{eq:resultatoneloopfourpt}
 \begin{aligned}
 \frac{G^{(1)}_{\{p,p\}}}{(d_{12})^{N}(d_{34})^{p}}=&-2g^2\, r_{1234}\, F^{(1)}(z,\bar{z})\sum_{s=0}^{\lfloor \frac{p}{2}\rfloor-1}\left(\frac{(z\bar{z})^2(1-\alpha)(1-\bar{\alpha})}{(\alpha\bar{\alpha})^2(1-z)(1-\bar{z})}\right)^{s}
 \end{aligned}
 \eeq
 where $r_{1234}$ is a prefactor which follows from the superconformal Ward identity\fn{Note that our definition of the universal prefactor is related to the universal prefactor $R_{1234}$ used in \cite{Chicherin:2015edu} by an overall factor by $R_{1234}=(x_{13}^2x_{24}^2d_{12}^2d_{34}^2)r_{1234}$.}
 \beq
 r_{1234}\equiv \frac{z\bar{z}(1-z)(1-\bar{z})(z-\alpha)(z-\bar{\alpha})(\bar{z}-\alpha)(\bar{z}-\bar{\alpha})}{(\alpha\bar{\alpha})^2 (1-z)(1-\bar{z})}\comma
 \eeq
 and $F^{(1)}$ is the one-loop conformal integral given in \eqref{eq:1loopconformaldef}.

 \paragraph{Superconformal block expansion}To test the predictions from integrability, we perform the superconformal block expansion of the four-point functions in the $12\to 34$ channel.

 At tree level, the four-point function is given by a sum of three contributions, BPS single-trace operators and non-BPS single-trace operators, and double-trace operators:
 \beq\label{eq:treelevelfourpoint}
 \begin{aligned}
 G_{\{p,p\}}^{(0)}=&\sum_{\mathcal{O}^{(L)}_{\circ}}({\sf d}_{\mathcal{O}^{(L)}_{\circ}}{\sf c}_{pp\mathcal{O}^{(L)}_{\circ}})\times \mathcal{F}_{L}+\sum_{\mathcal{O}}\left.\left({\sf d}_{\mathcal{O}}{\sf c}_{pp\mathcal{O}}\right)\right|_{O(g^0)}\times \mathcal{F}_{\Delta^0,s,n,m}\\
 &+\text{(double traces with $\Delta-S\geq 2p$)}\period
 \end{aligned}
 \eeq
Here $\mathcal{F}_{L}$ is the $1/2$-BPS superconformal block for operators with dimension $L$ while $\mathcal{F}_{\Delta,s,n,m}$ are the non-BPS superconformal block for operators with SO(6) Dynkin label $[n,m]$, dimension $\Delta$ and spin $S$.
The superscript in $\Delta^0$ signifies the fact that we are using the tree-level conformal dimension $\mathcal{O}$ to evaluate $\mathcal{F}$.
The explicit expressions for these blocks can be found in Appendix A of \cite{Basso:2017khq} (see also \cite{Dolan:2006ec,Bissi:2015qoa}). ${\sf d}_{\mathcal{O}}(=\sqrt{L}\mathfrak{D}_{\mathcal{O}})$ is the length-stripped structure constant between determinant operators and a single-trace operator while ${\sf c}_{pp\mathcal{O}}$ is the length-stripped structure constant for three single-trace operators defined by
\beq
{\sf c}_{pp\mathcal{O}}\equiv\frac{ N}{\sqrt{p\times p\times L}}\times  C_{\mathcal{O}^{(p)}_{\circ}\mathcal{O}^{(p)}_{\circ}\mathcal{O}}\comma
\eeq
where $L$ is the length of the operator $\mathcal{O}$, and the symbol $C_{\mathcal{O}_1\mathcal{O}_2\mathcal{O}_3}$ denotes the structure constant of operators $\mathcal{O}_{1,2,3}$.
The contribution of the double-trace operators always have $\Delta-S\geq 2p$ since the relevant double-trace operators are made out of the two external single-trace operators. Therefore, as long as we consider the conformal data with $\Delta-S<2p$, we can ignore the contribution from the double trace operators.

At one loop, all the non-BPS OPE data, namely the structure constants and the dimensions, receive corrections. Therefore we expect that the one-loop four-point function can be expanded in the following way:
\beq\label{eq:oneloopfourpoint}
\begin{aligned}
G_{\{p,p\}}^{(1)}=&\sum_{\mathcal{O}}\left[\left.\left({\sf d}_{\mathcal{O}}{\sf c}_{pp\mathcal{O}}\right)\right|_{O(g^2)}\times \mathcal{F}_{\Delta^0,s,n,m} +\left.\left({\sf d}_{\mathcal{O}}{\sf c}_{pp\mathcal{O}}\right)\right|_{O(g^0)}\left.\delta\Delta\right|_{O(g^2)}\times \del_{\Delta^0}\mathcal{F}_{\Delta^0,s,n,m}\right]\\
&+\text{(double traces)}\period
\end{aligned}
\eeq
The first term represents the correction to the structure constant while the second term comes from the correction to the dimension $\delta \Delta$. The conformal data for the BPS operators are tree level exact and therefore do not show up at one loop. An important point worth emphasizing is that the expansion of the one-loop four-point function involves not only the superconformal blocks but also their derivatives $\del_{\Delta}\mathcal{F}_{\Delta,s,n,m}$. By looking at coefficients in front of such derivatives, one can also extract the dimension of the operator $\delta \Delta$.

Now, assuming these structures, one can extract the relevant OPE data from the four-point functions that we computed. However, owing to the degeneracy of the spectrum at weak coupling, what we can extract is not the OPE data of individual operators, but a sum of contributions from several operators with identical tree-level quantum numbers. A convenient way to package the resulting sum is to consider the ``generating function'' introduced in \cite{Basso:2017khq}:
\beq
\mathbb{P}_{\tau,s}^{[n,m]}\equiv \sum_{\mathcal{O} : [\Delta,s,n,m]}\left({\sf d}_{\mathcal{O}}{\sf c}_{pp\mathcal{O}}\right) e^{\blue{y}\delta \Delta} \period
\eeq
Here $\blue{y}$ is the expansion parameter of the generating function that we introduced and the sum is over all the single-trace operators with the tree-level quantum numbers $[\Delta,s,n,m]$: $\tau$ is the (tree-level) twist of the operator defined by $\tau=\Delta-s$ and $n$ and $m$ are the SO(6) Dynkin labels
\beq
[q_1,p,q_2]=[n-m,2m,n-m]\period
\eeq
The expansion of $\mathbb{P}$ at weak coupling contains the perturbative OPE data which appear in \eqref{eq:treelevelfourpoint} and \eqref{eq:oneloopfourpoint}:
\beq
\mathbb{P} =\left.\left({\sf d}_{\mathcal{O}}{\sf c}_{pp\mathcal{O}}\right)\right|_{O(g^0)} +\left.\left({\sf d}_{\mathcal{O}}{\sf c}_{pp\mathcal{O}}\right)\right|_{O(g^2)}+\left.\left({\sf d}_{\mathcal{O}}{\sf c}_{pp\mathcal{O}}\right)\right|_{O(g^0)}\times \left.\delta\Delta\right|_{O(g^2)}\times \blue{y}+\cdots\period
\eeq
\paragraph{Integrability prediction}To compare the perturbative result with the result from integrability, we need to compute ${\sf d}_{\mathcal{O}}{\sf c}_{pp\mathcal{O}}$, or more precisely the generating function $\mathbb{P}_{\tau,s}^{[n,m]}$, from integrability. For ${\sf d}_{\mathcal{O}}$, we simply use our asymptotic formula \eqref{eq:asymptoticformula}. On the other hand, the single-trace structure constant ${\sf c}_{pp\mathcal{O}}$ can be computed from the hexagon approach \cite{Basso:2017khq}. Applying the formula in \cite{Basso:2017khq} to our case, we obtain
\beq\label{eq:hexwithselec}
\left({\sf c}_{pp\mathcal{O}_{{\bf u},\red{\bf v},\blue{\bf w}}}\right)^2=\frac{\langle\red{\bf v}| \red{\bf v}\rangle^2\prod_{k=1}^{M}\mu (u_k)}{\langle {\bf u}|{\bf u}\rangle\prod_{i<j}S_0(u_i,u_j)}\mathcal{A}^2\qquad \qquad \text{with $\red{\bf v}=\blue{\bf w}$}\comma
\eeq
with
\beq
\mathcal{A}=\prod_{i<j}h(u_i,u_j)\sum_{\alpha \cup \bar{\alpha}={\bf u}}(-1)^{|\bar{\alpha}|}\prod_{j\in \bar{\alpha}}T(u_j)e^{ip (u_j)L/2}\prod_{u_i\in \alpha,u_j\in \bar{\alpha}}\frac{1}{h(u_i,u_j)}\period
\eeq
Here $h(u,v)$ is the hexagon form factor, $\mu$ is the measure factor while $T(u)$ is the fundamental $\mathfrak{su}(2|2)$ transfer matrix\fn{In \cite{Basso:2017khq}, $T(u)$ is denoted by $f(u)$. However, as was already noted there, $f(u)$ is nothing but the fundamental $\mathfrak{su}(2|2)$ transfer matrix.}. For more detailed definitions, see \cite{Basso:2017khq}.

As indicated in \eqref{eq:hexwithselec}, ${\sf c}_{pp\mathcal{O}_{{\bf u},\red{\bf v},\blue{\bf w}}}$ vanishes unless the selection rule $\red{\bf v}=\blue{\bf w}$ is satisfied. On the other hand, ${\sf d}_{\mathcal{O}_{{\bf u},\red{\bf v},\blue{\bf w}}}$ is conjectured to vanish unless $\red{\bf v}=\overline{\blue{\bf w}}(\equiv -\blue{\bf w})$. Combining the two selection rules, we reach the conclusion that a given operator $\mathcal{O}_{{\bf u},\red{\bf v},\blue{\bf w}}$ can appear in the OPE expansion only when all the sets of roots are parity-symmetric by themselves and the roots on the left and the right wings coincide:
\beq
{\bf u}=\overline{\bf u}\comma \qquad \red{\bf v}=\overline{\red{\bf v}}=\blue{\bf w}=\overline{\blue{\bf w}}\period
\eeq
This strong selection rule on the OPE is a clear spacetime implication of the worldsheet integrability, which is beyond the scope of the standard representation theory of superconformal algebra. This also suggests that the four-point function of two determinants and two BPS single-trace operators is a nice and simple object which deserves a further study---perhaps even simpler than the four-point functions of four BPS single-trace operators.
\paragraph{Result} Using the methods outlined above, we computed the generating function $\mathbb{P}$ both from the OPE expansion of the perturbative data and from integrability. To generate the prediction from integrability, we first used the $Q$-system code in \cite{Marboe:2017dmb} to obtain the tree-level Bethe roots (see Appendix \ref{apsubsec:BetheRoots} for details of the procedure), computed the one-loop corrections to them, and then plugged them in our asymptotic formulae. When doing so, there are two small points that one has to take into account:
\begin{enumerate}
\item The OPE data $\mathbb{P}^{[n,m]}_{s}$ is given in terms of the quantum number of the superconformal primary states. This is not the same as the representative in the SL(2) grading discussed in section \ref{sec:asymptotic}. Because of this, we need to modify the relation between the quantum numbers and the number of Bethe roots slightly in the following way:
\beq
s=M-K_{\I}^{\bf v}-K_{\III}^{\bf v}-\red{2}\comma\quad n=\frac{L}{2}+K_{\III}^{\bf v}-K_{\II}^{\bf v}-\red{1}\comma\quad m=\frac{L}{2}+K_{\II}^{\bf v}-K_{\I}^{\bf v}-\red{1}\period
\eeq
Here the numbers in red are the required shifts.
\item Our integrability formulae for ${\sf d}_{\mathcal{O}}$ and ${\sf c}_{pp\mathcal{O}}$ contain square roots and therefore can produce sign ambiguities. To resolve this, we first rewrote a product ${\sf d}_{\mathcal{O}}{\sf c}_{pp\mathcal{O}}$ slightly and eliminated the square roots (see Appendix \ref{apsubsec:simpleOPEint} for an explicit expression), and then evaluate the resulting expression. In this way, the answer is unambiguous.
\end{enumerate}

For the actual comparison, we considered the expansion of $G_{\{4,4\}}$ and extracted the single-trace OPE data up to twist 6. The result is summarized in Table \ref{tab:higherrank}. As shown there, in all the cases we tested, the integrability predictions were in agreement with the OPE data\fn{Precisely speaking, we found some sign mismatches for the rows of $\mathbb{P}^{[0,0]}_{4,s}$ and $\mathbb{P}^{[1,1]}_{6,s}$ even after resolving the square roots. They are just overall factors and the results are in perfect agreement once we fix the signs for the tree-level data. There can be several sources of this sign mismatch: First, there might be some subtlety that we missed when we resolved the square root ambiguity in the formulae. Second the overall signs might come from the integrability prediction for ${\sf c}_{pp\mathcal{O}}$ rather than ${\sf d}_{\mathcal{O}}$ since the overall signs for the hexagon approach were never tested extensively. (See for instance \cite{Caetano:2016keh} for a potential sign issue in the hexagon formalism.) Third, fixing the overall sign of the structure constant is rather nontrivial even on the field-theory side since changing the signs of the operators does not modify the two-point function but modifies the three-point function. It would definitely be desirable to understand this point further, but we will postpone it to future investigations.}. Note that we did not test all the numbers in the table since the computation on the integrability side---in particular solving the Bethe equation and finding all the solutions---is computationally costly even with the help of the $Q$-system method. It would be certainly interesting to push the integrability computation further and check more data. Nevertheless we should emphasize the current result already provides extensive tests of our formula in higher-rank sectors and provide strong support for our conjecture.

One interesting outcome of our analysis is that the OPE data for $\mathbb{P}^{[1,0]}_{4,s}$, $\mathbb{P}^{[1,0]}_{6,s}$, and $\mathbb{P}^{[2,1]}_{6,s}$ are all zero although they are allowed from the representation theory of the superconformal symmetry. On the integrability side, this is due to the absence of Bethe roots satisfying the selection rule. This is another manifestation of the hidden simplicity of the four-point functions of two determinants and two single-traces.
\begingroup
\begin{landscape}
\begin{table}
\centering
\caption{The result for the generating function $\mathbb{P}_{\tau,s}^{[n,m]}$ up to one loop, extracted from $G_{\{4,4\}}$. The quantum numbers $\tau$, $s$, $n$, and $m$ are for the top components of the superconformal multiplet. To read off the quantum number for the representatives in the SL$(2)$ sector, one has to shift the spin by 2; $s\to s+2$. We colored the numbers checked against integrability and the generating functions in red correspond to the SL$(2)$ sector. We found that $\mathbb{P}^{[1,0]}_{4,s}$, $\mathbb{P}^{[1,0]}_{6,s}$, and $\mathbb{P}^{[2,1]}_{6,s}$ are all zero both from the OPE and integrability.}
\renewcommand{\arraystretch}{1.8}
\begin{tabular}{l|llll}
\multicolumn{1}{l}{\ \ $s$} & \multicolumn{1}{c}{$0$} & \multicolumn{1}{c}{$2$} & \multicolumn{1}{c}{$4$} & \multicolumn{1}{c}{$6$}\\
\toprule
$\red{\mathbb{P}^{[0,0]}_{2,s}}$ & \textcolor{blue}{$\frac{1}{3}-4g^2+4g^2y$} & \textcolor{blue}{$\frac{1}{35}-\frac{205}{441}g^2+\frac{10}{21}g^2y$} & \textcolor{blue}{$\frac{1}{462}-\frac{1106}{27225}g^2+\frac{7}{165}g^2y$} & \blue{$\frac{1}{6435}-\frac{14380057}{4509004500}g^2+\frac{
761}{225225}g^2y$}\\

$\mathbb{P}^{[0,0]}_{4,s}$ & \textcolor{blue}{$-\frac{1}{30}+\frac{7}{25}g^2-\frac{2}{5} g^2y$} & \textcolor{blue}{$-\frac{1}{378}+\frac{127}{7938}g^2-\frac{2}{63} g^2y$} & \textcolor{blue}{$-\frac{1}{5148}+\frac{1865}{2208492}g^2-\frac{1}{429} g^2y$} & $-\frac{1}{72930}+\frac{24471}{590976100}g^2-\frac{2}{12155} g^2y$\\

$\red{\mathbb{P}^{[1,1]}_{4,s}}$ & \textcolor{blue}{$-\frac{1}{5}+0\times g^2+0\times g^2y$} & \textcolor{blue}{$-\frac{1}{63}+0\times g^2+0\times  g^2y$} & \textcolor{blue}{$-\frac{1}{858}+0\times g^2+0\times  g^2y$} & \blue{$-\frac{1}{12155}+0\times g^2+0\times  g^2y$}\\

$\mathbb{P}^{[0,0]}_{6,s}$ & \textcolor{blue}{$\frac{1}{105}-\frac{218}{2205}g^2+\frac{4}{35} g^2y$} & $\frac{1}{660}-\frac{30607}{1633500}g^2+\frac{101}{4950} g^2y$ & $\frac{2}{10725}-\frac{176412889}{67635067500}g^2+\frac{9419}{3378375} g^2y$ & $\frac{1}{50388}-\frac{1701615497}{5598385949520}g^2+\frac{8549}{26453700} g^2y$\\

$\mathbb{P}^{[1,1]}_{6,s}$ & \textcolor{blue}{$\frac{2}{35}-\frac{2}{5}g^2+\frac{2}{5} g^2y$} & $\frac{1}{110}-\frac{521}{5445}g^2+\frac{16}{165} g^2y$ & $\frac{4}{3575}-\frac{10909}{760500}g^2+\frac{43}{2925} g^2y$ & $\frac{1}{8398}-\frac{4415079}{2556060500}g^2+\frac{101}{56525} g^2y$\\

$\mathbb{P}^{[2,0]}_{6,s}$ & \textcolor{blue}{$-\frac{1}{15}+\frac{2}{3}g^2-\frac{2}{3} g^2y$} & \textcolor{blue}{$-\frac{1}{77}+\frac{521}{3267}g^2-\frac{16}{99} g^2y$} & $-\frac{1}{585}+\frac{10909}{456300}g^2-\frac{43}{1755} g^2y$ & $-\frac{2}{10659}+\frac{1471693}{511212100}g^2-\frac{101}{33915} g^2y$\\

$\red{\mathbb{P}^{[2,2]}_{6,s}}$ & \textcolor{blue}{$\frac{4}{7}-4g^2+4 g^2y$} & \textcolor{blue}{$\frac{1}{11}-\frac{1042}{1089}g^2+\frac{32}{33} g^2y$} & \textcolor{blue}{$\frac{8}{715}-\frac{10909}{76050}g^2+\frac{86}{585} g^2y$} & \blue{$\frac{5}{4199}-\frac{4415079}{25560605}g^2+\frac{202}{11305}g^2y$}
%
%
%
%
%
\end{tabular} \label{tab:higherrank}
\end{table}
\end{landscape}
\endgroup

\subsection{Two-loop four-point functions: dressing phase and large spin\label{subsec:2loop4pt}}
We now perform a nontrivial test of the boundary dressing phase $\sigma_B$ computed in section \ref{subsec:bootstrapphase}. Since its weak-coupling expansion starts at two loops, we need two-loop perturbative data. Computing the two-loop four-point function directly from perturbation is quite a laborious task. Fortunately, for the length-2 BPS single-trace operators---also known as ${\bf 20}^{\prime}$ operators---we succeeded in computing the four-point functions using the idea of {\it bootstrap}, namely by imposing the consistency under the OPE expansion and determining the final answer without direct computations.

\paragraph{Bootstrap method}The method employs a combination of several different techniques, details of which are explained in Appendix \ref{ap:2loop4ptbootstrap}. First, we used the Lagrangian insertion approach \cite{Chicherin:2015edu} and the superconformal Ward identity to write down an allowed form of the {\it integrand} at two loops, up to two unfixed constants $c_{1,2}$. As a result, the four-point function can be expressed as
\beq
\begin{aligned}
G^{(2)}_{\{2,2\}}=&\tilde{R}_{1234}(d_{12})^{N-2}\left[(c_1 z\bar{z}+c_2 (1-z)(1-\bar{z})+c_2)(F^{(1)})^2\right.\\
&\left.+4 c_2 F_z^{(2)}+2 (c_1+c_2)F_{1-z}^{(2)}+4c_2 F_{\frac{z}{z-1}}^{(2)}\right]\period
\end{aligned}
\eeq
Here $F^{(2)}$'s are two-loop conformal integrals and $\tilde{R}_{1234}$ is a prefactor dictated by the superconformal Ward identity (see Appendix \ref{ap:2loop4ptbootstrap} for explicit expressions).

We then analyze the resulting four-point function in the $12\to34$ channel. As a natural extension of \eqref{eq:oneloopfourpoint}, we expect that the two-loop four-point function can be expanded as\fn{Here we omitted writing the double-trace contributions.}
\beq
\begin{aligned}
G_{\{2,2\}}^{(2)}=&\sum_{\mathcal{O}}\left[\left.\left({\sf d}_{\mathcal{O}}{\sf c}_{22\mathcal{O}}\right)\right|_{O(g^4)}\times \mathcal{F}_{\Delta^0,s,n,m} +\left.\left({\sf d}_{\mathcal{O}}{\sf c}_{22\mathcal{O}}\right)\right|_{O(g^2)}\left.\delta\Delta\right|_{O(g^2)}\times \del_{\Delta^0}\mathcal{F}_{\Delta^0,s,n,m}\right.\\
&\left.+\left.\left({\sf d}_{\mathcal{O}}{\sf c}_{22\mathcal{O}}\right)\right|_{O(g^0)}\left.\delta\Delta\right|_{O(g^4)}\,\del_{\Delta^0}\mathcal{F}_{\Delta^0,s,n,m}+\left.\left({\sf d}_{\mathcal{O}}{\sf c}_{22\mathcal{O}}\right)\right|_{O(g^0)}\frac{\left(\left.\delta\Delta\right|_{O(g^2)}\right)^2}{2}\, \del^2_{\Delta^0}\mathcal{F}_{\Delta^0,s,n,m}\right]\period\nn
\end{aligned}
\eeq
All but the first term in the sum are given by the lower loop data or the anomalous dimensions which we already know. Therefore, by imposing that these terms are correctly reproduced, we can constrain the structure of the four-point function. As a result we get $c_2=1$.

We still need to determine $c_1$. This can be done by analyzing the OPE in a different channel $23\to 14$. In this channel, the exchanged operators correspond to open spin chains attached to a determinant operator\fn{More precisely, the ones that are relevant for this analysis are the operators attached to the $Z=0$ Giant Graviton.} (to be denoted by $\mathcal{O}_{\rm open}$), which were studied in \cite{Hofman:2007xp}. Also in this channel, the four-point function can be expanded in a similar manner and we imposed that the second derivative term of the operator with the lowest dimension is correctly reproduced:
\beq
G_{\{2,2\}}^{(2)}\supset \left.\left({\sf c}_{{\bf 20}^{\prime}\mathcal{D}\mathcal{O}_{\rm open}}\right)^2\right|_{O(g^0)} \frac{\left(\delta \Delta_{\mathcal{O}_{\rm open}}|_{O(g^2)}\right)^2}{2}\del^2_{\Delta^0}\mathcal{F}_{\Delta^0,s,n,m}\period
\eeq
To impose this condition, we need to know the anomalous dimension of the open spin-chain operator $\delta \Delta_{\mathcal{O}_{\rm open}}|_{O(g^2)}$, and its structure constant with the ${\bf 20}^{\prime}$ operator and the determinant operator, ${\sf c}_{{\bf 20}^{\prime}\mathcal{D}\mathcal{O}_{\rm open}}$. As for the anomalous dimension, we used the result in \cite{Hofman:2007xp} while the structure constant was read off by performing the same OPE expansion for the one-loop four-point function $G_{\{2,2\}}^{(1)}$. As a result we could determine $c_1$ to be $c_1=-1$.

\paragraph{OPE expansion and comparison} We then performed the OPE expansion of the resulting four-point function and extracted the two-loop OPE coefficients $\left.\left({\sf d}_{\mathcal{O}}{\sf c}_{22\mathcal{O}}\right)\right|_{O(g^4)}$. We in particular focused on the structure constants of the length-2 operators in SL(2) sector, also known as the twist-2 operators, since other OPE data are contaminated by the double-trace contributions. One benefit of studying the twist-2 operators is the absence of the degeneracy; there is only one supermultiplet for a given spin $S$. Thus the expansion gives a single structure constant not a sum of many.
The result of the analysis is summarized in Table \ref{tab:productsOPE}.
\begingroup
\begin{table}[t]
\caption{Products of structure constants ${\sf d}_{\mathcal{O}}{\sf c}_{22\mathcal{O}}$ for spin $S$ twist-$2$ operators $\mathcal{O}_{2,S}$. As shown below, the results are completely rational in contrast to the OPE data for the single-trace four-point functions given in Table 1 of \cite{Basso:2015eqa}. (See also \cite{Eden:2015ija}).}
\centering
\renewcommand{\arraystretch}{1.8}
\begin{tabular}{l|l}
$S$&\multicolumn{1}{c}{${\sf d}_{\mathcal{O}_{2,S}}{\sf c}_{22\mathcal{O}_{2,S}}$}\\\toprule

$2$&$ \frac{1}{3}-4g^2+56g^4$\\

$4$&$\frac{1}{35}-\frac{205 g^2}{441}+\frac{73306 g^4}{9261}$\\

$6$&$\frac{1}{462}-\frac{1106 g^2}{27225}+\frac{826643623 g^4}{1078110000}$\\

$8$&$\frac{1}{6435}-\frac{14380057 g^2}{4509004500}+\frac{2748342985341731 g^4}{42652702617525000}$\\

$10$&$\frac{1}{92378}-\frac{3313402433 g^2}{13995964873800}+\frac{156422034186391633909 g^4}{31100584702491617040000}$
\end{tabular}\label{tab:productsOPE}
\end{table}
\endgroup

We then computed the same combination $\left.\left({\sf d}_{\mathcal{O}}{\sf c}_{22\mathcal{O}}\right)\right|_{O(g^4)}$ from our asymptotic formula and the hexagon approach. As compared to the tree-level and the one-loop computations discussed above, there are two new ingredients; at two loops, the boundary dressing phase $\sigma_B$ starts to give a nontrivial contribution proportional to a zeta function, $\zeta_3$. In addition, the result from the hexagon approach receives a correction from the so-called bottom wrapping, which gives a sum of a rational number and a term proportional to $\zeta_3$ (see \cite{Basso:2015eqa} for further explanations). Remarkably, the terms proportional to $\zeta_3$ completely cancel out making the final result in complete agreement with the numbers in Table \ref{tab:productsOPE}, which are purely rational.
\paragraph{Curious numerology}The absence of terms proportional to $\zeta_3$ is utterly unexpected since the four-point function of ${\bf 20}^{\prime}$'s, which was studied in \cite{Basso:2015eqa}, does contain such terms in contrast to the one discussed above. This is another evidence for the hidden simplicity of the four-point functions with two determinant operators.

As it turns out, a further inspection reveals a more interesting pattern of the OPE data. As mentioned above, the two-loop structure constants of two ${\bf 20}^{\prime}$ and a twist-2 spin $S$ operator ${\sf c}_{22\mathcal{O}_{2,S}}$ are given by a sum of the {\it asymptotic} hexagon formula \eqref{eq:hexwithselec} and the bottom wrapping correction,
\beq
{\sf c}_{22\mathcal{O}_{2,S}}={\rm H}_{\rm asympt}+{\rm H}_{\rm bottom}\period
\eeq
Comparing this with the structure constants with the determinant operators ${\sf d}_{\mathcal{O}_{2,S}}$, we observed an interesting relation (see also Table \ref{tab:squareOPE})
\beq\label{eq:curiousnumerology}
{\sf d}_{\mathcal{O}_{2,S}}=2({\rm H}_{\rm asympt}-{\rm H}_{\rm bottom})\period
\eeq
Namely ${\sf d}_{\mathcal{O}_{2,S}}$ is twice the asymptotic hexagon {\it minus} the bottom wrapping! This is basically the origin of the absence of the $\zeta_3$ terms in the OPE data ${\sf d}_{\mathcal{O}_{2,S}}{\sf c}_{22\mathcal{O}_{2,S}}$.

The relation \eqref{eq:curiousnumerology} is surprising from the integrability point of view. The left hand side ${\sf d}_{\mathcal{O}_{2,S}}$ is given purely by the asymptotic formula without any contributions from the mirror particles. By contrast, the right hand side contains a contribution from mirror particles and is given in terms of integrals. If this relation persists at finite coupling, this would suggest that there should be a simple way to resum (at least a part of) wrapping corrections to the hexagon approach. In view of this, it is quite interesting to compute the structure constant at three loops and see if such a relation still holds\fn{We should however warn the readers that such a relation does not seem to exist for operators with higher twists. (This can be seen already at tree level and one loop.) Thus, pessimistically, one might think that this is merely a coincidence. However, one could also be optimistic and hope that the twist-2 operators are somewhat special and the relation extends to finite coupling.}.
\begingroup
\begin{table}[t]
\caption{Length-stripped structure constants $\left({\sf d}_{\mathcal{O}_{2,S}}/2\right)^{2}$ computed from integrability. The numbers colored in black coincide with the asymptotic hexagon formula while the numbers colored in red coincide with the bottom wrapping corrections.}
\centering
\renewcommand{\arraystretch}{1.8}
\begin{tabular}{l|l}
$S$&\multicolumn{1}{c}{$\left({\sf d}_{\mathcal{O}_{2,S}}/2\right)^2$}\\\toprule

$2$&$\frac{1}{6}-2 g^2+28 g^4-\textcolor[rgb]{1,0,0}{g^4 12\zeta_3}$\\

$4$&$\frac{1}{70}-\frac{205 g^2}{882}+\frac{36653 g^4}{9261}-\textcolor[rgb]{1,0,0}{g^4\left(\frac{10}{7}\zeta_3+\frac{1}{6}\right)}$\\

$6$&$\frac{1}{924}-\frac{553
   g^2}{27225}+\frac{826643623 g^4}{2156220000} -\textcolor[rgb]{1,0,0}{g^4\left(\frac{7}{55}\zeta_3+\frac{199}{7920}\right)}$\\

$8$&$\frac{1}{12870}-\frac{14380057
   g^2}{9018009000}+\frac{2748342985341731 g^4}{85305405235050000}-\textcolor[rgb]{1,0,0}{g^4 \left(\frac{761 }{75075}\zeta _3+\frac{1721}{655200}\right)}$\\

$10$&$\frac{1}{184756} \!-\!\frac{3313402433
   g^2}{27991929747600}\!+\!\frac{156422034186391633909
   g^4}{62201169404983234080000}-\textcolor[rgb]{1,0,0}{g^4 \left(\frac{671 }{881790}\zeta _3+\frac{578887}{2461415040}\right)}$
\end{tabular}\label{tab:squareOPE}
\end{table}
\endgroup
\paragraph{Large spin limit} The relation \eqref{eq:curiousnumerology} also allows us to write ${\sf d}_{\mathcal{O}_{2,S}}$ in terms of harmonic sums, using the harmonic-sum representation of the single-trace structure constants \cite{Eden:2012rr} and the bottom wrapping \cite{Basso:2015zoa}. (The details can be found in Appendix \ref{ap:largespin}.) This provides an expression analytic in spin and makes it straightforward to take the large spin limit. As a result, we found
\beq
\begin{aligned}
\log \left[\frac{{\sf d}_{\mathcal{O}_{2,S}}}{\left.{\sf d}_{\mathcal{O}_{2,S}}\right|_{\rm tree}}\right]=&-2g^2\left[2\log 2 \log S^{\prime}+\zeta_2\right]\\
&+8g^4\left[\left(\zeta_2\log 2+\frac{9}{2}\zeta_3\right)\log S^{\prime}+\frac{4}{5}(\zeta_2)^2+\frac{3}{2}\zeta_3\log 2\right]+O(1/S)\comma
\end{aligned}
\eeq
with $\log S^{\prime}\equiv \log S+\gamma_{E}$.
 Importantly, the result does not contain a term proportional to $(\log S)^2$. This is in stark contrast to the structure constant of a twist $2$ operator and two ${\bf 20}^{\prime}$ operators, which contains a term proportional to $(\log S)^{k}$ at $k$ loops.

Physically, this seems to be related to the absence of the level crossing \cite{Korchemsky:2015cyx}:
In the case of single-trace structure constants \cite{Alday:2013cwa}, the $(\log S)^{k}$ terms get resummed into a Gamma function\fn{$\gamma$ is the anomalous dimension of the twist$-2$ operator which scales as $\log S$ in the large spin limit.}
\beq
{\sf c}_{22\mathcal{O}_{2,S}} \sim \Gamma\left(1-\frac{\gamma}{2}\right)\comma
\eeq
which produces poles of the structure constants at finite coupling. As discussed in \cite{Korchemsky:2015cyx}, the existence of such poles is an indication of the level crossing between the twist $2$ operator and double-trace operators made out of external single-trace operators. On the other hand, we do not expect such a level crossing to happen in our set up since the external operators are determinants and are always much heavier than the single-trace operators. From this perspective, it seems plausible that higher powers of $\log S$ will be absent at higher loops as well. It would be interesting to verify this from integrability.

Also, if what we said is really correct, then it would be quite interesting to determine the $O(\log S)$ and $O(1)$ coefficients at finite 't Hooft coupling using integrability,
\beq
\log\left[\frac{{\sf d}_{\mathcal{O}_{2,S}}}{\left.{\sf d}_{\mathcal{O}_{2,S}}\right|_{\rm tree}}\right] = f (g) \log S^{\prime} +\tilde{f} (g) +O(1/S)\period
\eeq
These coefficients are the analogues of the cusp anomalous dimension \cite{Beisert:2006ez} for the structure constant, and would provide quantities which interpolate the weak-coupling and the strong-coupling physics.

\subsection{Dressing phase at strong coupling\label{subsec:strongdressing}}
The boundary dressing phase can also be tested at strong coupling.

For this purpose, we consider the symmetric configuration of the twisted translated frame, in which the two determinants are at $a_1=+1$ and $a_2=-1$ and the single-trace operator is sitting at the origin. We furthermore assume for now that the single-trace operator is BPS. In the global AdS, which is dual to the radial quantization of $\mathcal{N}=4$ SYM, this corresponds to a point-like string propagating from $\tau=-\infty$ and terminating at $\tau=0$ at the geodesics of Giant Gravitons (see figure \ref{fig:fig30}).

\begin{figure}[t]
\centering
\includegraphics[clip,height=4cm]{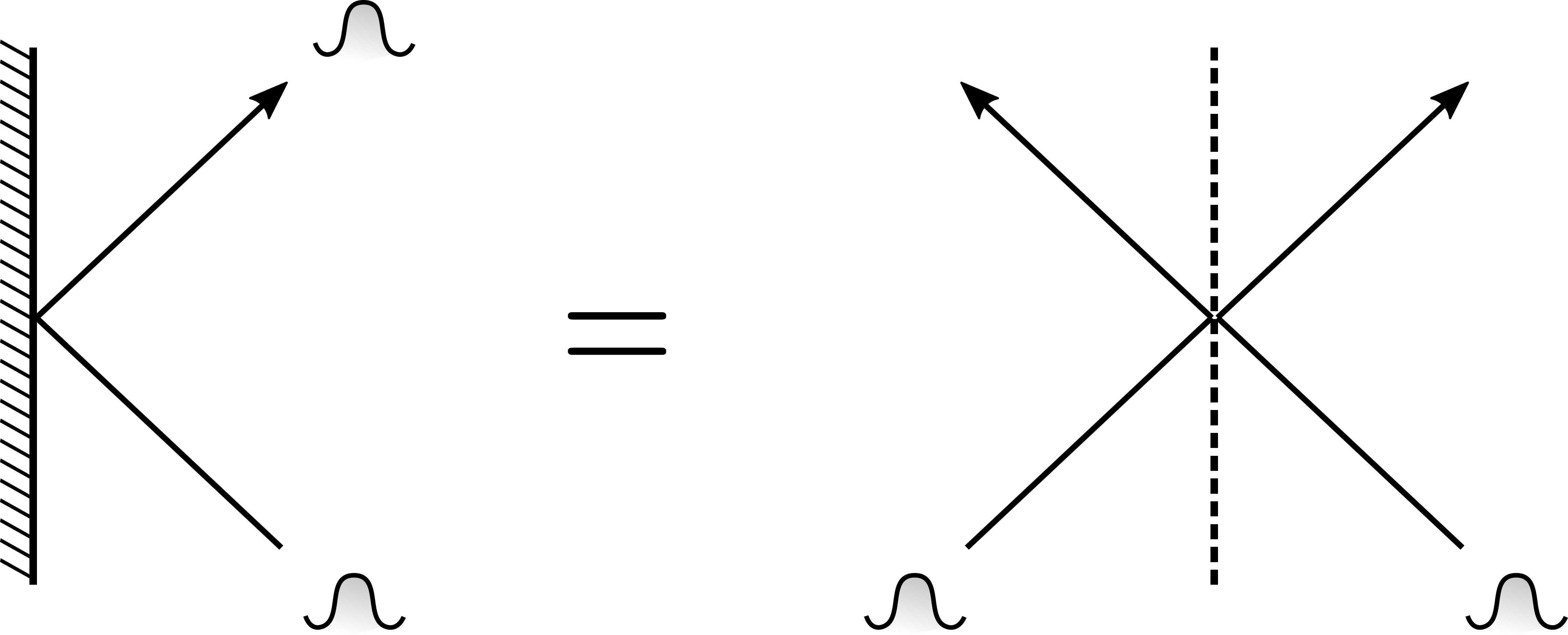}
\caption{The method of images for computing the time-delay at strong coupling.}
\label{fig:fig34}
\end{figure}

\paragraph{Prediction from worldsheet}Now, to read off the boundary dressing phase, we then analyze this classical worldsheet configuration from the mirror channel and study the reflection of a solitonic excitation off the boundary. As in the analysis in \cite{Correa:2012hh}, one can read off the reflection phase from the time delay of the classical solution describing the reflection process. One important difference from the standard analysis is that, since we are considering a reflection process in the mirror channel, we need to consider a {\it mirror giant magnon} instead of the standard giant magnon solutions. While a solution describing a single mirror giant magnon was studied in \cite{Arutyunov:2007tc}, a solution describing a reflection of a mirror giant magnon was never discussed in the literature. Nevertheless, one can predict the relevant phase factor without constructing an explicit solution as we explain below:

To be concrete, let us consider the scattering of the excitation transverse to the D-brane in AdS\fn{In the spin-chain language, it amounts to considering a scattering of a transverse derivative $D^{1\dot{2}}-D^{2\dot{1}}$.}. Being transverse to the D-brane, the excitation obeys the Neumann boundary condition. As was discussed in \cite{Correa:2012hh}, this allows us to use the ``method of images'' to relate a solution describing a reflection process to a solution describing a scattering of two solitons with momenta $-\tilde{p}$ and $\tilde{p}$ (see figure \ref{fig:fig34}). The relation between the time delay $\Delta T$ and the bulk scattering phase $\delta$ reads \cite{Correa:2012hh}
\beq\label{eq:timedelayrelation}
\Delta T=\left.\frac{d\tilde{p}_1}{d\tilde{E}_1}\del_{\tilde{p}_1}\delta (\tilde{p}_1,\tilde{p}_2)\right|_{-\tilde{p}_1=\tilde{p}_2=\tilde{p}}=\left.\frac{d\tilde{p}_2}{d\tilde{E}_2}\del_{\tilde{p}_2}\delta (\tilde{p}_1,\tilde{p}_2)\right|_{-\tilde{p}_1=\tilde{p}_2=\tilde{p}} \period
\eeq
On the other hand, the relation between the boundary scattering phase $\delta_B (\tilde{p})$ and the time delay reads
\beq
\Delta T=\frac{d\tilde{p}}{d\tilde{E}}\del_{\tilde{p}}\delta_B (\tilde{p})\period
\eeq
Comparing the two, we can express the boundary scattering phase in terms of the bulk scattering phase as\fn{A factor of $1/2$ is necessary since the derivative in \eqref{eq:timedelayrelation} can act on both arguments.}
\beq
\delta_B (\tilde{p})=\frac{1}{2}\delta (-\tilde{p},\tilde{p})\period
\eeq
This relation can be translated to the following simple relation between the scalar factor for the reflection matrix $r_0=e^{i\delta_B}$ and the scalar factor for the bulk S-matrix $S_0=e^{i\delta}$:
\beq\label{eq:strongcouplingdressingprediction}
\left(r_0(\bar{u})\right)^2\simeq  S_0 (\bar{u}^{\gamma},u^{\gamma}) \qquad g\to \infty\period
\eeq
Note that the arguments of the bulk S-matrix account for the fact that the scattering occurs in the mirror channel.

\paragraph{Comparison with integrability} Let us now check that the relation \eqref{eq:strongcouplingdressingprediction} is satisfied by our solution. When comparing the two, we need to take into account one subtle point: As discussed in section \ref{subsec:bootstrapphase}, the symmetric configuration corresponds to the minimal boundary dressing phase $\sigma_B^{\rm min} (u)$ rather than the full boundary dressing phase. We thus have
\beq
r_0 (\bar{u})=\frac{u-\frac{i}{2}}{u}\frac{\sigma^{\rm min}_B(u^{\gamma})}{\sigma (\bar{u}^{\gamma},u^{\gamma})}\period
\eeq
On the other hand, $S_{0}(\bar{u}^{\gamma},u^{\gamma})$ reads
\beq
S_0 (\bar{u}^{\gamma},u^{\gamma})=\frac{u-\frac{i}{2}}{u+\frac{i}{2}}\frac{1}{(\sigma (\bar{u}^{\gamma},u^{\gamma}))^2}\period
\eeq

 At strong coupling, all the factors that are rational in the the rapidities are subleading. The only possible contributions come from $\sigma$ and $\sigma^{\rm min}_B$. However, as shown in Appendix \ref{ap:crossing}, $\sigma_B$ is simply unity at strong coupling (even in the mirror kinematics). Therefore we conclude that
 \beq
 (r_0 (\bar{u}))^2 \simeq \frac{1}{(\sigma (\bar{u}^{\gamma},u^{\gamma}))^2}\simeq S_0 (\bar{u}^{\gamma},u^{\gamma})\qquad g\to \infty\comma
 \eeq
 which is in agreement with the prediction \eqref{eq:strongcouplingdressingprediction}.

\section{Conclusion and Future Directions\label{sec:conclusion}}
In this paper, we presented a powerful nonperturbative formalism to compute the three-point function of two determinant operators and one single-trace operator in $\mathcal{N}=4$ SYM. The method is based on integrability and uses the fact that the three-point functions can be interpreted as the overlaps between the integrable boundary state and the states describing the single-trace operator. We determined the boundary state and its reflection matrix from symmetry and integrability, and used them to derive the non-perturbative expressions for the overlaps both for the ground state and the excited states in the SL(2) sector.

We also introduced two new methods to efficiently compute the correlation functions of determinant operators at weak coupling. The first one is based on the large $N$ collective fields, which can be interpreted as the open string field theory on Giant Gravitons, while the other is based on a judicious use of Wick contractions and combinatorics.

The results in this paper are the first fully nonperturbative proposals on correlation functions of operators of finite size in contrast to the previous proposals on the single-trace correlators \cite{Basso:2015zoa,Fleury:2016ykk}, which are so-far proven to be effective only when the operators are large \cite{Coronado:2018ypq,Bargheer:2019kxb}.

However, the quality of the work should be judged not just based on what was achieved, but based also on whether it opens up new directions or raises interesting physical questions. Standing on this viewpoint, below we discuss various potentially interesting questions that could be addressed with the results and the methods developed in this paper.

Let us first list several immediate questions that could be studied with our method. One simple computation would be to make predictions for the structure constants at weak coupling and compare them with the perturbative computations. In particular, the computation at four loops\fn{For single-trace correlators, there have been recent developments in computing the four-point functions perturbatively and extracting the OPE data \cite{Goncalves:2016vir,Georgoudis:2017meq,Chicherin:2018avq,Coronado:2018cxj}. It would be interesting to see if these approaches can be generalized to the correlators with determinants.} is quite interesting since it allows us to see the effect of the wrapping corrections.
It would also be interesting to analyze various limits of our formula. For instance it would be worth studying the structure constant of long operators in the strong coupling limit, and comparing them with the computations based on the classical string solutions\fn{Even if it is difficult to compute a relevant classical solution, one might be able to compute the structure constant directly by using the classical integrability of the worldsheet. See for instance examples in other set ups \cite{Alday:2010vh,Janik:2011bd,Kazama:2013qsa,Kazama:2016cfl,Caetano:2012ac,Toledo:2014koa}.}.

Also interesting would be to perform the numerical computation of the Fredholm determinants and make a plot of the structure constant at finite coupling. For this purpose, it might be helpful to simplify the formula first by rewriting it in terms of more basic quantities such as the $Q$-functions in the quantum spectral curve \cite{Gromov:2013pga,Gromov:2014caa}. At weak coupling, this is in fact possible as we demonstrate in Appendix \ref{ap:Qfunctions}. In any case, the results in this paper provide a concrete reference point, analogous to the TBA for the spectrum \cite{Gromov:2009tv,Gromov:2009bc,Gromov:2009zb,Bombardelli:2009ns,Arutyunov:2009ur}, from which one can develop more sophisticated techniques following the path taken in the study of the spectrum.

It is worth mentioning that the structures similar to the Fredholm determinants were observed in the study of wrapping corrections to the hexagon formalism \cite{Basso:2017muf,Basso:2018cvy}. Since our setup shares several features with the hexagon formalism including the underlying symmetry, it might be interesting to try to draw lessons from our approach.

Using the framework developed in this paper, one can study other observables that also correspond to the worldsheet $g$-functions, for instance the correlation functions on the Coulomb branch \cite{CordovaCoronadoKomatsu} and the correlators of a Wilson loop and a single-trace operator\cite{withAmit}. Other possible setups would be one-point functions in the presence of domain-wall defects discussed in \cite{deLeeuw:2015hxa,Buhl-Mortensen:2015gfd,Buhl-Mortensen:2016pxs,deLeeuw:2016umh,Buhl-Mortensen:2016jqo,deLeeuw:2016ofj,Buhl-Mortensen:2017ind,deLeeuw:2018mkd,Grau:2018keb}.

In the case of defect one-point functions, there are a family of integrable boundary states at weak coupling labelled by the integer $k$, which counts the worldvolume flux. In our setup, the analogue will be to consider a higher-point correlator of a single-trace operator and $k(>2)$ determinant operators. Although it is unlikely that the boundary state remains integrable for generic configurations, there might be some special kinematics of higher-point functions for which integrability is preserved. Finding such integrable higher-point functions would be extremely interesting since this opens up a possibility of analyzing the higher-point functions using the well-established TBA framework.

In addition to these immediate questions, there are also other interesting questions which potentially have deeper physical implications. In what follows, we address them one by one.
\subsection{Four-point functions and spacetime physics\label{subsec:conclusion4pt}}
It is often said that the correlation functions of local operators in CFT are captured entirely by the two- and three-point functions. While it is true that every higher-point function can be decomposed into a product of two- and three-point functions,  interesting physics is often hidden in the higher-point functions in a subtle manner, and cannot be accessed by studying the individual two- and three-point functions. Below we discuss possibilities of applying our framework to study the four-point functions and extract interesting physics.
\subsubsection{Four-point function = finite cylinder}
In the derivation of the $g$-function, we considered a partition function of a cylinder whose ends are capped off by the boundary states. Although this was just a trick to use the thermodynamic Bethe ansatz in this paper, the cylinder partition function actually admits a concrete spacetime interpretation; namely it corresponds to the four-point function of determinant operators.

\begin{figure}[t]
\centering
\includegraphics[clip,height=6cm]{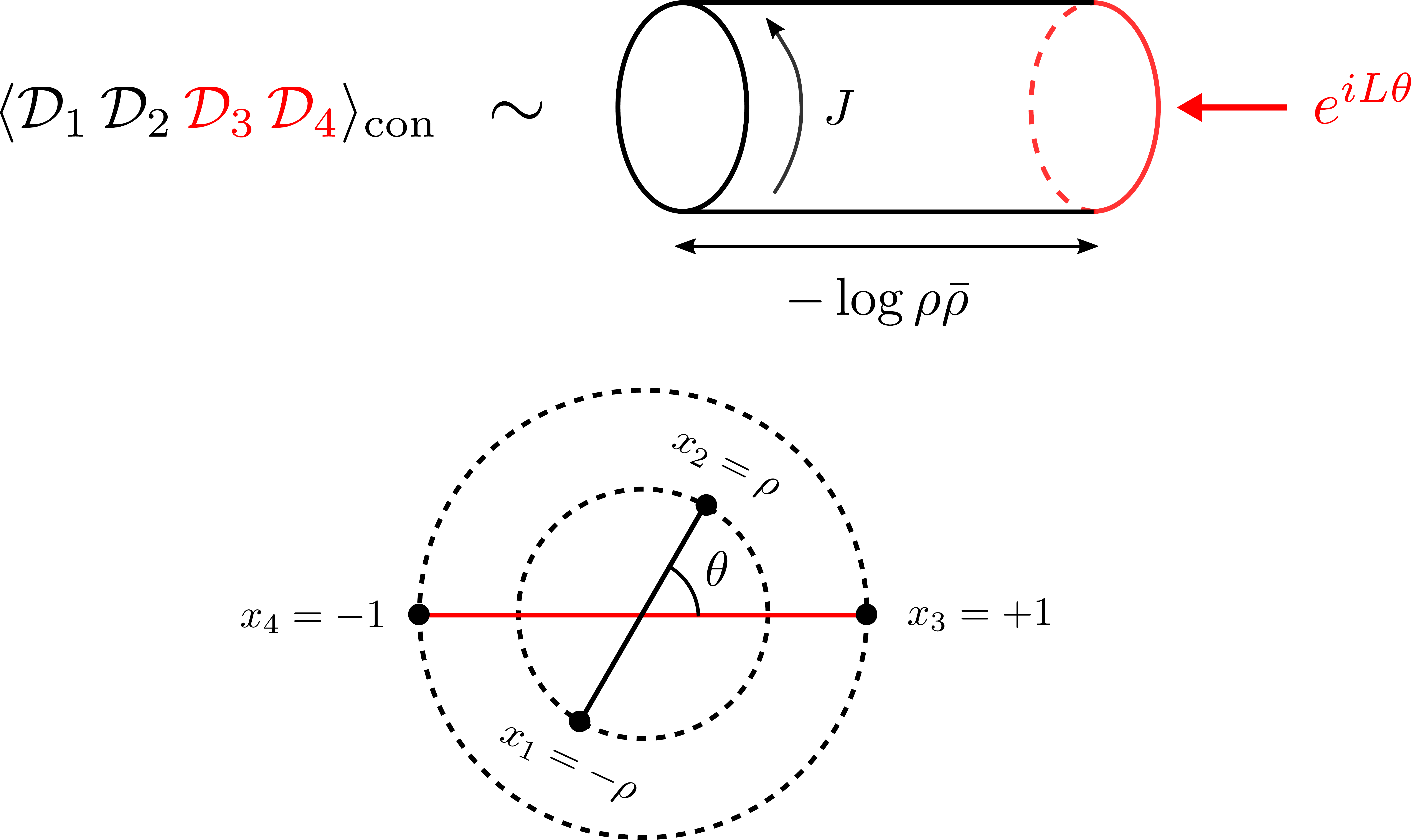}
\caption{The cylinder partition function and the $\rho$-coordinate. The connected four-point function of determinant operators is given by the (exponentiated) cylinder partition function. The length of the cylinder is related to the absolute value of the $\rho$ coordinate, which is defined in the lower figure. The angle of the $\rho$ coordinate is related to the twist of the boundary condition $e^{iL\theta}$, where is the $L$ is the spacetime rotation.}
\label{fig:fig35}
\end{figure}

This basically follows from the fact that a {\it single} boundary state describes a {\it pair} of determinant operators as we saw in this paper. Thus the cylinder with two boundaries would correspond to observables involving {\it four} determinant operators,
\beq
\langle \mathcal{D}_1 \mathcal{D}_2\red{\mathcal{D}_3\mathcal{D}_4}\rangle\comma
\eeq
where the left boundary is created by the operators $\mathcal{D}_{1,2}$ while the right boundary is created by the operators $\red{\mathcal{D}_{3,4}}$.
However, to make this statement more quantitative, we need to understand how to incorporate the dependence on the cross ratios of the four-point functions into the cylinder partition function. For this purpose, it is useful to use the so-called $\rho$-coordinates in CFT \cite{Hogervorst:2013sma}. In terms of the $\rho$-coordinates, the operators $\mathcal{D}_{1,2}$ are on a circle of radius $\sqrt{\rho\bar{\rho}}$ while the other two operators $\mathcal{D}_{3,4}$ are on the unit circle (see figure \ref{fig:fig35}). Recalling the fact that the length of the cylinder $R$ is conjugate to the Hamiltonian in the closed string channel, which in our setup is the spacetime dilatation, it is natural to identify the length with $-\log \rho\bar{\rho}$; the amount of the dilatation one needs to perform in order to map one circle to the other.

On the other hand, the angle of $\rho$, $e^{2i\theta} \equiv \rho/\bar{\rho}$, can be incorporated by ``rotating'' one of the boundary states relative to the other. At the level of the (diagonalized) reflection matrix, this would correspond to multiplying phase factors\fn{One can also perform a similar rotation for the R-symmetry, which also results in the multiplication of some phase factors.} to the right reflection matrix as was shown in an analogous case of the cusped Wilson loop in \cite{Drukker:2012de,Correa:2012hh}.

Thus, to summarize, we expect that the four-point function corresponds to a cylinder partition function of finite length where the circumference of the cylinder determines the $R$-charge of the exchanged operator while the length and the rotation angle encode the conformal cross ratios.

Unfortunately this does not immediately tell us how to compute the full four-point functions with generic conformal cross ratios since the computation of the partition function of a cylinder is tractable only when one of the lengths becomes large. Nevertheless, this does provide a way to analyze various physically interesting limits as we see below.

 \begin{figure}[t]
\centering
\includegraphics[clip,height=3.3cm]{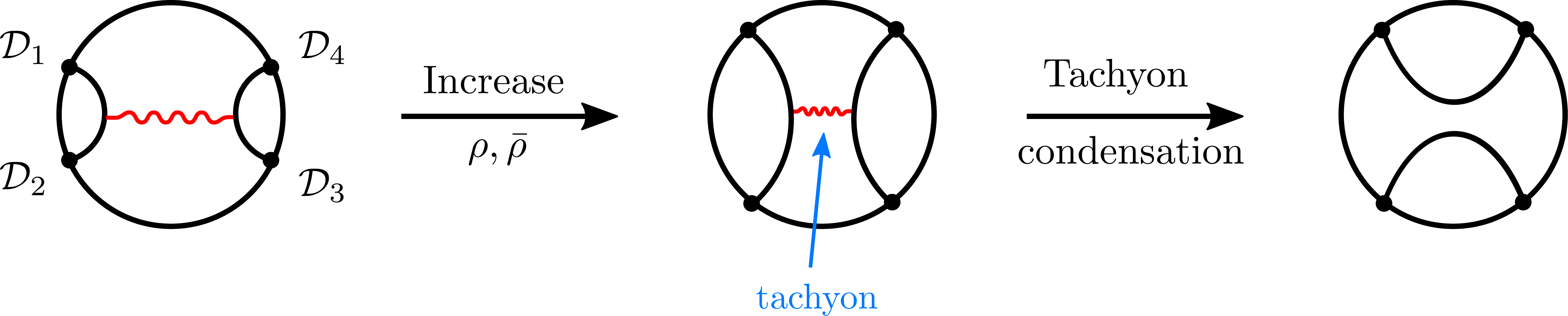}
\caption{The phase transition of the geodesic Witten diagrams and the tachyon condensation. If we increase the cross ratio $\rho\bar{\rho}$, the original configuration of the geodesic Witten diagrams become unstable after some point owing to the open string tachyon. As a result of the tachyon condensation, the D-branes get recombined into different geodesics.}
\label{fig:fig36}
\end{figure}

 \subsubsection{D-brane recombination and Hagedorn behavior}
 One simple limit of the cylinder partition function is the limit in which the length $R=-\log \rho\bar{\rho}$ becomes large, namely $\rho\bar{\rho}\ll 1$. This is precisely the limit discussed in this paper, which is the OPE limit of the four-point function. On the AdS side, this corresponds to a pair geodesics of D-branes, one between $\mathcal{D}_{1,2}$ and the other between $\mathcal{D}_{3,4}$, which are connected by a long thin tube of closed string worldsheet, as shown in figure \ref{fig:fig36}.

 As we increase $\rho\bar{\rho}$, these two geodesics come closer and eventually get recombined into yet another pair of geodesics, which now connect $\mathcal{D}_{2}$ and $\mathcal{D}_{3}$, and $\mathcal{D}_1$ and $\mathcal{D}_4$. From the spacetime point of view, this is basically the phase transition between two geodesic Witten diagrams and is analogous to what is known to happen for the entanglement entropy for disjoint intervals in the large $c$ 2d CFT (see for instance \cite{Hartman:2013mia,Faulkner:2013yia}).

 This phase transition admits an interesting interpretation on the worldsheet. Let us first consider it from the closed string point of view. In the closed string channel, the length of the cylinder $R=-\log \rho\bar{\rho}$ plays the role of the inverse temperature $\beta$, and increasing $\rho\bar{\rho}$ would correspond to increasing the temperature. We then expect that, at some value of $\rho\bar{\rho}$, the partition function would stop converging due to the Hagedorn growth of states as is the case with the standard torus partition function \cite{Atick:1988si}. On the other hand, in the open string channel, this divergence would be associated with the emergence of tachyons in the spectrum of open string. Past this point, the tachyons will get condensed and transform the original D-brane configurations to a new configuration. This is precisely the phase transition\fn{Precisely speaking, the phase transition could also be first order. In that case, the new configuration starts to dominate before the tachyons show up in the original configuration. It would be interesting to figure out which is the case by performing the explicit weak-coupling computations as was done for the transition of the thermal partition function \cite{Aharony:2003sx}.} that we discussed above.

\begin{figure}[t]
\centering
\includegraphics[clip,height=3.5cm]{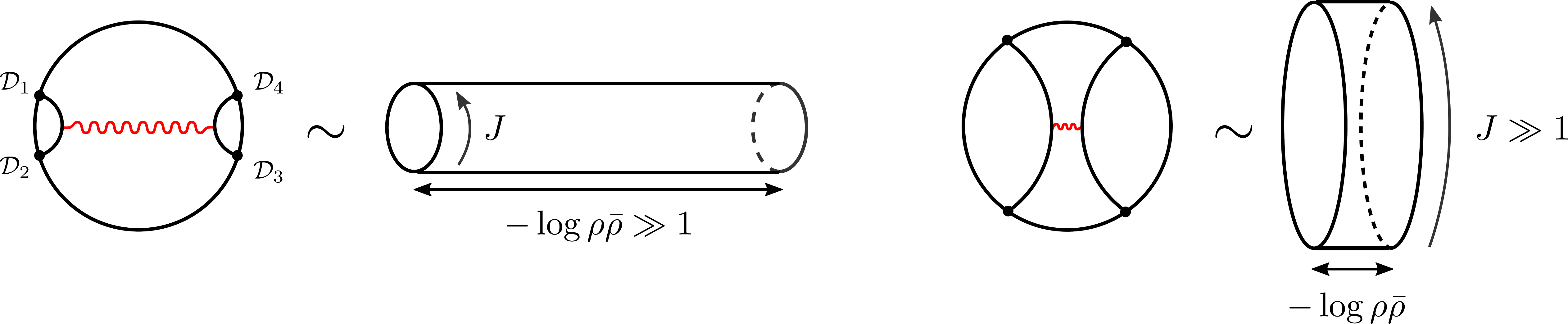}
\caption{The cylinder partition function in two different limits. When $\rho\bar{\rho}$ is small (as in the left figure), we can study the four-point function by applying the TBA in the mirror channel. This is nothing but the OPE limit and is precisely what is done in this paper. On the other hand, when $\rho \bar{\rho}$ is large (as in the right figure), the behavior of the four-point function can be studied by applying the TBA in the physical channel.}
\label{fig:fig37}
\end{figure}

 In order to address these questions using integrability, one needs to study the cylinder partition function in a region where the length $-\log \rho\bar{\rho}$ is small. As mentioned above, computing the partition function at small length is generically hard. However, things will become tractable if we send the other length $L$ to be very large. This was indeed the strategy employed in the recent analysis of the Hagedorn temperature of the torus partition function \cite{Harmark:2017yrv,Harmark:2018red}, and we expect that the same would work also in our setup. Namely the idea is to analyze a cylinder whose circumference $L$ is very large and a length $R$ is small (or finite) by applying the TBA to the {\it physical} channel (see figure \ref{fig:fig37}). This would tell us the spectrum of the (mirror) open string sector thereby allowing us to see the tachyons and analyze the Hagedorn behavior.

In the setup of \cite{Harmark:2017yrv,Harmark:2018red}, the ``configuration'' after the tachyon condensation is the AdS black hole geometry to which the integrability machineries would not be applicable. On the other hand, in our setup, the configuration after the tachyon condensation is still given by Giant Gravitons in AdS. Therefore it might be possible to perform more detailed analysis of the condensation process with the help of integrability. We should also mention that the tachyon spectrum was analyzed in a different context in \cite{Bajnok:2013wsa}, but the advantage in our setup is that we have a better idea about what happens after the tachyon condensation.
\subsubsection{Regge and BFKL limit, and light-ray operators}
\begin{figure}[t]
\centering
\includegraphics[clip,height=3cm]{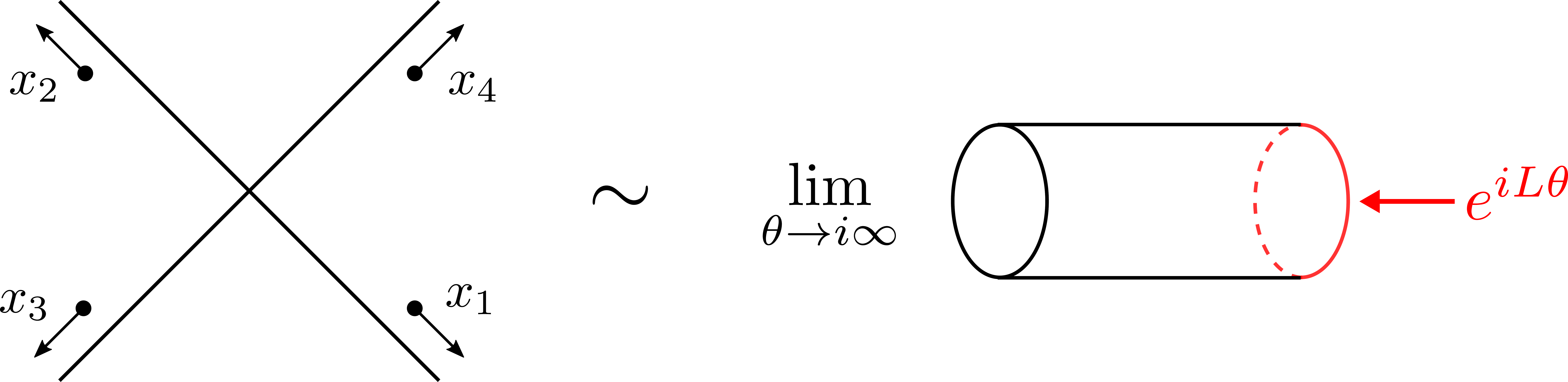}
\caption{The Regge limit of the four-point function and the cylinder partition function. From the worldsheet point of view, the Regge limit corresponds to the limit in which the twist angle goes to $i\infty$.}
\label{fig:fig38}
\end{figure}

\paragraph{Regge and BFKL}
There are also other limits of four-point functions which can potentially be studied within the TBA framework. One such limit is the Regge limit, which corresponds to a large boost limit of the operators $\mathcal{D}_{1,2}$ relative to $\mathcal{D}_{3,4}$. In terms of the cylinder partition function, this simply corresponds to sending the rotation angle $\theta$ to $i\infty$ (see figure \ref{fig:fig38}). Applying our methods to this limit, we might be able to extract the spectrum and the OPE coefficients of pomerons. It would also be possible to apply the TBA in the physical channel to directly explore the behavior of the four-point function in this limit.

Note also that a similar limit was discussed in the context of the cusped Wilson loop \cite{Drukker:2012de,Correa:2012hh}. In that case, the limit corresponds to a cusp of lightlike Wilson loops, whose anomalous dimension is given by the well-known $\Gamma_{\rm cusp}$ \cite{Beisert:2006ez}. It might be interesting to compare it with the setup described above and try to find a possible connection.

From the integrability point of view, the rotation by $\theta$ corresponds to twisting a boundary condition. It is then interesting to consider an analogue of the ``fishnet'' limit \cite{Gurdogan:2015csr,Caetano:2016ydc,Mamroud:2017uyz,Pittelli:2019ceq}; namely the double-scaling limit in which the twist angle goes to $i\infty$ while $\lambda$ goes to zero keeping a product $\lambda e^{-i\theta}$ fixed. In our setup, this would correspond to the famous BFKL limit. Much like the fishnet limit, the BFKL limit is governed by the ladder diagrams and it would be interesting to generalize the computation to include the determinant operators and compare it with the integrability predictions.
\paragraph{Light-ray operators and inversion formula} Another potentially interesting future direction would be to study the light-ray operators \cite{Kravchuk:2018htv}. The light-ray operator is defined by fusing two operators $\mathcal{O}_1$ and $\mathcal{O}_2$ and performing the Lorentzian integral transform. At the level of four-point functions, this amounts to performing a certain integral transformation for the conformal cross ratios, which coincides with the Lorentzian inversion formula \cite{Caron-Huot:2017vep}. It would be interesting to apply it to the cylinder partition function and see if one can directly extract the CFT data for the light-ray operators.

An advantage of using the light-ray operators and the Lorentzian inversion formula is that it allows us to perform the analytic continuation in spin from first principles. As for the spectrum, the analytic continuation in spin was achieved already by the use of the quantum spectral curve \cite{Gromov:2015wca}, but the application of the light-ray operator to the cylinder partition function would put the results on a firmer ground, and also allow us to extract the OPE coefficients as well.
\subsubsection{Bulk-point limit and Loschmidt echo} Yet another interesting limit is the so-called bulk-point limit \cite{Maldacena:2015iua}. In terms of the $\rho$-coordinates, it is defined as a limit
\beq
\rho \sim e^{i\pi}e^{i\theta} \comma\qquad \bar{\rho}\sim e^{i\pi}e^{-i\theta}\period
\eeq
In this limit, there is a point in the bulk of AdS which are lightlike separated from the operator insertion points. Therefore, in the strong coupling limit where we expect to recover the local physics in AdS, this four-point function is expected to exhibit an {\it enhanced} singularity.

\begin{figure}[t]
\centering
\includegraphics[clip,height=5cm]{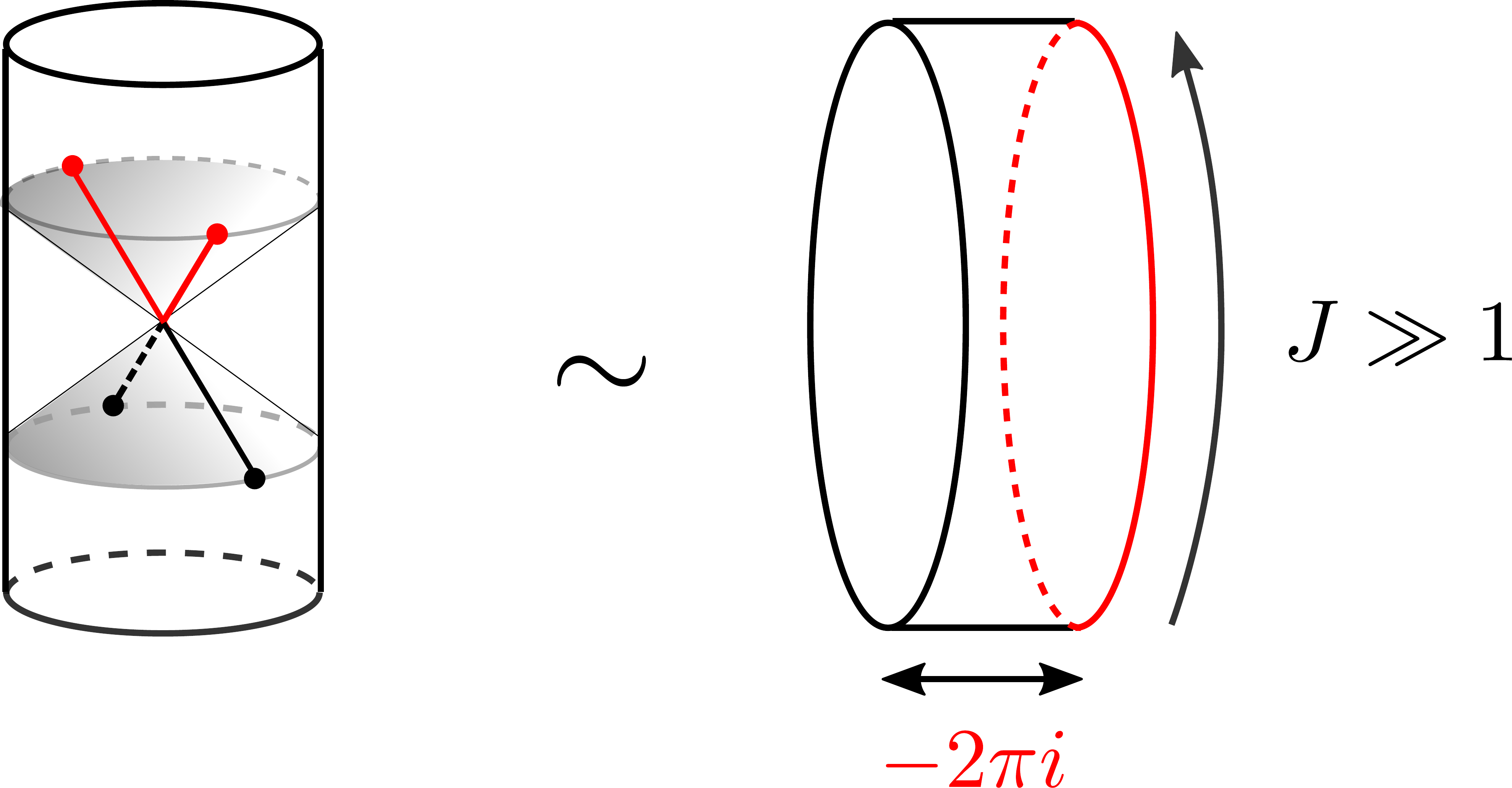}
\caption{The bulk-point limit and the cylinder partition function. The bulk-point limit corresponds to taking the length of the cylinder to be $-2\pi i$. This is similar to the situation studied in the context of the Loschmidt echo \cite{poszsgay:Echo,Piroli:2017mmz,piroli2018non}.}
\label{fig:fig39}
\end{figure}

In terms of the cylinder partition function, this limit corresponds to taking the length of the cylinder to be $-2i\pi$, which is {\it imaginary} and finite. As is the case with the Hagedorn transition, we expect that the singular behavior in this limit can be captured by applying the TBA in the {\it physical} channel, instead of the mirror channel TBA used in this paper (see figure \ref{fig:fig39}).

In the physical channel, setting the length to be $-2i\pi$ would correspond to analytically continuing the temperature to be imaginary. Interestingly, a similar analytic continuation of TBA was analyzed recently in a completely different context: In \cite{piroli2018non}, they analyzed the quantity called the Loschmidt echo in integrable spin chains, which is defined as a return probability of some initial state $|\Psi \rangle$
\beq
G(t)\equiv \frac{\langle \Psi |e^{-iHt}|\Psi\rangle}{\langle \Psi|\Psi\rangle}\period
\eeq
In generic quantum field theories, the Loschmidt echo is expected to provide access to various interesting physical phenomena such as chaos, scrambling and the random matrix behavior. Some of these features will be lost if the theory is integrable, but it still provides a simple nonequilibrium observable.

In particular, if one uses an integrable boundary state as the initial state $|\Psi\rangle$, the computation boils down to evaluating the analytically continued cylinder partition function using the TBA analysis, as was shown in \cite{piroli2018non}. The most interesting outcome of their analysis is a non-analytic behavior in time which is induced by the competitions of two different TBA saddle points. We expect that a similar non-analytic behavior will be observed also in our setup and it can potentially explain various spacetime singularities including the bulk-point singularity mentioned above.

We should also note that the four-point function discussed above is not an ideal setup for probing the bulk-point singularity since, in the same kinematical configuration, there can also be {\it boundary singularities} which come from points on the boundary lightlike separated from all the operators. In order to isolate the bulk singularity, we need to study the six-point function \cite{Maldacena:2015iua}. From this point of view, it is extremely interesting to see if there is an integrable boundary state created by three determinant operators as was mentioned already above. If the answer is positive, we can then study the six-point function using the cylinder partition function and take the bulk-point limit.
\subsection{Application of collective fields\label{subsec:conclusioncollective}}
There are also interesting questions one can ask regarding our collective field approach to the determinant operators.
\subsubsection{Other operators and other theories}
\begin{figure}[t]
\centering
\includegraphics[clip,height=4cm]{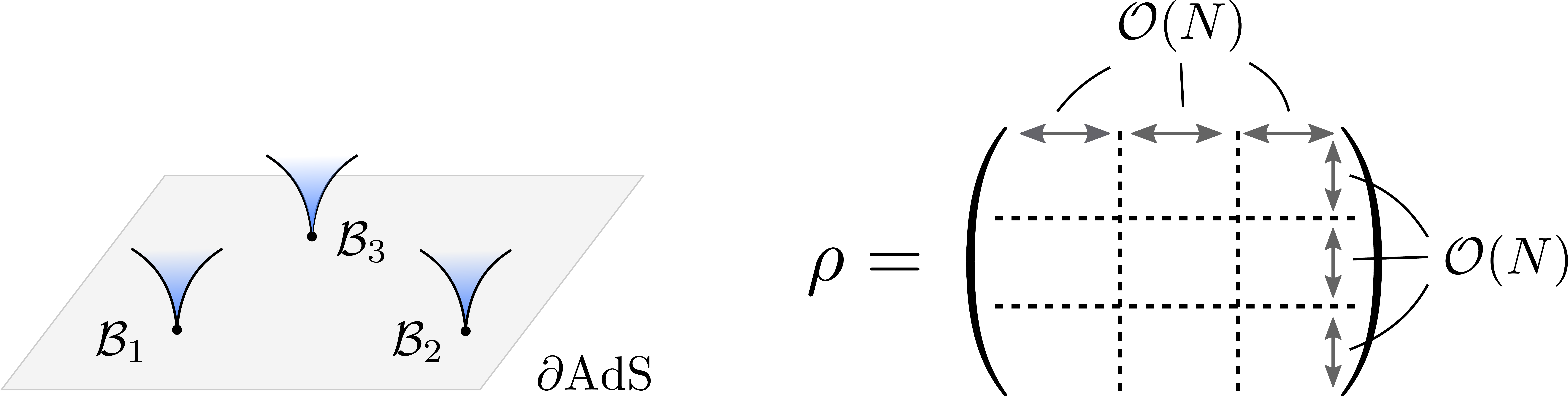}
\caption{The three-point function of the bubbling geometries and the corresponding $\rho$ field. Each operator insertion modifies the geometry near the AdS boundary. The full geometry is expected to be a nontrivial connected geometry connecting three asymptotic regions. The $\rho$ field is now a huge matrix with a block structure where each block is of $O(N)$. We expect that there is a relation among the expectation value of the off-diagonal block, the correlation between two operators and the connectedness of the dual geometry.}
\label{fig:fig40}
\end{figure}

\paragraph{Baryonic operators in other theories}
One interesting possibility is to generalize our idea of collective fields to baryonic operators in other theories, such as the Chern-Simons vector models in three dimensions. The spectrum of the baryonic operators in the large $N$ limit was computed already \cite{Radicevic:2015yla,Aharony:2015mjs}, but it would be interesting to understand the small deformations of the baryonic operators and also their structure constants. It would also be interesting to study baryonic operators in the SYK model \cite{Maldacena:2016hyu,Polchinski:2016xgd} and the tensor models \cite{Witten:2016iux,Klebanov:2016xxf}.
\paragraph{Bubbling geometry} Also interesting would be to analyze the operators of size $O(N^2)$, which correspond to the so-called bubbling geometry \cite{Lin:2004nb}. The simplest way to obtain such an operator would be to take $N$-th product of determinant operators inserted at the same position:
\beq
\mathcal{B}=\left(\det Z\right)^{N}\period
\eeq
Since they are just a product of determinant operators, one can use the same trick to recast the correlators of $\mathcal{B}$'s into an effective theory of $\rho$ variables. The difference from the analysis in this paper is that the size of the $\rho$-matrix now becomes of $O(N)$. This certainly makes the computation---in particular the fluctuation around the saddle point---more complicated but we expect that the main qualitative features discussed in section \ref{sec:effective} are still the same: Namely whenever there are correlations between two operators, the corresponding $\rho$ variable will acquire a nonzero expectation value. See also figure \ref{fig:fig40}. It would be interesting to study such correlators in more details using the $\rho$ variables and try to relate them to the topology of the dual spacetime \cite{Berenstein:2016pcx,Berenstein:2017abm}. It would also be interesting to generalize and apply our method to the so-called fermi liquid operators discussed in \cite{Berkooz:2006wc,Berkooz:2008gc}, which were conjectured to be dual to the $1/16$ BPS black holes.

\subsubsection{Towards a full-fledged open-closed-open duality}
In section \ref{sec:effective}, we discussed the relation between our approach and the open-closed-open duality discussed by Gopakumar. We should however note that the result is still far from being called ``duality'' since they only capture a small subset of observables in $\mathcal{N}=4$ SYM and the computations are mainly done at tree level.

To explain a potential route to make it into a full-fledged duality, let us consider a field theory of a single real scalar in the adjoint of U$(N)$ group,
\beq
Z=\int \mathcal{D}\Phi \exp \left[-\int d^{d}x \frac{1}{2}{\rm tr}\left[\del_{\mu}\Phi\del^{\mu}\Phi\right] +{\rm tr}\left[V(\Phi)\right]\right]\comma
\eeq
where $V(\Phi)$ is some polynomial potential. One way to apply our trick would be to approximate the polynomial potential $V(\Phi)$ by multiple insertions of determinant operators $\det (1-z\Phi)$. Written explicitly, we consider the insertion of the following operators into the {\it Gaussian} path integral of $\Phi$:
\beq
\begin{aligned}
\prod_{x} \prod_{k=1}^{M}\det (1-z_k\Phi(x))&\sim \exp\left[\int d^{d}x\sum_{k}\tr \log (1-z_{k}\Phi(x))\right]\\
&=\exp\left[-\int d^{d}x\sum_{k=1}^{M}\tr\left[ \tilde{V}(\Phi(x))\right]\right]\period
\end{aligned}
\eeq
Here $\prod_x$ denotes the insertion of the operators at every spacetime point and the approximated potential $\tilde{V}$ is given by
\beq
\tilde{V}(\Phi)\equiv  \sum_{n} c_n \Phi^{n}\comma\qquad c_n\equiv \sum_{k=1}^{M}\frac{(z_k)^{n}}{n}\period
\eeq
By increasing $M$, one can in principle approximate any polynomial potential $V$. We can then use our trick to rewrite the path integral into a path integral of $\rho$ fields. As a result of rewriting, the $\rho$ field now becomes a bilocal field of spacetime points $\rho(x,y)$, and it also carries a $M\times M$ matrix structure\fn{Related ideas were discussed in interesting papers \cite{Kazakov:2000ar,Brown:2010af}.}.

Since this rewriting converts the standard field theory into a somewhat peculiar bilocal field theory, it is unclear if this is useful for any practical computation. However, there is one conceptually interesting point in this rewriting. As was the case with the analysis in section \ref{sec:effective}, we expect that $\rho(x,y)$ acquires a nontrivial expectation value whenever there is a nontrivial correlation between the two points. In this sense, this rewriting allows us to convert a standard field theory into a field theory whose fundamental variable is a correlation of the original theory. This may have some interesting implications for holography and the entanglement structure of the field theory \cite{Komatsu}.
\subsection{Closed string from ghost D-brane\label{subsec:conclusionSFT}}
In section \ref{subsec:example1}, we briefly described how to obtain the correlator of single-trace operators from the correlator of determinant operators. The strategy is to either use the replica trick or consider the ratio
\beq\label{eq:dbraneghost}
{\rm tr}\left(\frac{1}{x-Z}\right)=\left.\del_{x^{\prime}} \frac{\det (x^{\prime}-Z)}{\det (x-Z)}\right|_{x^{\prime}=x}\period
\eeq
It would be interesting to see if this trick allows us to relate the integrable structure found in this paper and the hexagon formalism for the single-trace three-point function \cite{Basso:2015zoa}. For this purpose, we need to generalize our analysis to a one-parameter family of determinant operators $\det (x-Z)$. Unfortunately, the preliminary analysis shows that these determinant operators do not correspond to integrable boundary states. This however does not exclude the existence of a more general integrable structure which encompasses both the integrable boundary and the hexagon formalism; we know that the hexagon form factor is not an integrable boundary state but nevertheless has a nice factorized structure.

On the string-theory side, the ratio of determinants in \eqref{eq:dbraneghost} corresponds to a system of a D-brane and a ghost D-brane (to be denoted by the $D/D$-system). From this point of view, the relation \eqref{eq:dbraneghost} can be understood as a mechanism of realizing a perturbative closed string state from the $D/D$-system. An interesting question is whether something similar can be realized in string theory in flat space. More specifically, the relation implies a certain mapping between a space of exactly marginal boundary deformations of the $D/D$-system and the BRST cohomology of closed string states. It would be interesting to prove/disprove this statement. One can also address this question using string field theory since the solution corresponding to a ghost D-brane was constructed in \cite{Masuda:2012kt}.
\subsection{$g$-functions in 2d QFT\label{subsec:gfunctions}}
The analysis of this paper relies on the TBA-approach to the exact $g$-functions in integrable quantum field theories. Although we successfully applied the approach to our problem, there is still room for improvements. For instance, although the Fredholm determinants give a nonperturbative expression for the $g$-function, it is not necessarily easy to compute them in general integrable QFTs. In addition, as was pointed out in \cite{Kostov:2019fvw}, for general integrable theories there is some subtlety\fn{As discussed in section \ref{sec:TBA}, such subtlety seems absent in our problem.} in the application of TBA to $g$-functions, and it would be desirable to better understand the formalism.

One possible way of making progress is to analyze the theories with $\mathcal{N}=(2,2)$ supersymmetry. For such theories, it is likely that $g$-functions for supersymmetric boundary conditions are computable using methods based on supersymmetry\fn{For the computations of hemisphere partition functions (and the disk one-point functions) from localization, see \cite{Honda:2013uca,Hori:2013ika,Sugishita:2013jca,Hashimoto:2015iha,Longhi:2019hdh}.}. For instance it would be nice to find an analogue of the CFIV index \cite{Cecotti:1992qh} which is computable both from integrability and supersymmetry.
\section*{Acknowledgement}
We thank Amit Sever for participation at the initial stage of this project, and collaboration on related topics.
We would also like to thank Jo\~{a}o Caetano, Frank Coronado, Nikolay Gromov, Vladimir Kazakov, Ivan Kostov, Juan Maldacena, Balazs Pozsgay, Michelangelo Preti and Konstantin Zarembo for discussions. SK would like to thank Yuji Okawa for on and off discussions on string field theory in the last couple of years and Rajesh Gopakumar for inspiring conversations on various occasions, from which many of the inspirations of this work came. The research of SK is supported by DOE
grant number DE-SC0009988. The work of EV is funded by the FAPESP grants 2014/18634-9 and 2016/09266-1, and by
the STFC grant ST/P000762/1. He thanks the Erwin Schr{\"o}dinger International Institute
for Mathematics and Physics at the University of Vienna, ETH Zurich and Perimeter
Institute for Theoretical Physics for hospitality.
\appendix
\section{Action, Feynman rules and integrals\label{ap:action}}
In this appendix, we summarize our convention for the color algebra, the action and the Feynman rules. It is the same convention as the one in \cite{Kim:2017sju}, which is equivalent to \cite{Erickson:2000af} with minor modifications. We also summarize relevant integrals which appear in the perturbative computations.
\subsection{Gauge group}
The generators in the fundamental representation of the gauge group U$(N)$ are $N\times N$ matrices that obey
\begingroup \allowdisplaybreaks
\begin{flalign}\label{eq:rel1}
\textrm{tr}(T^{A}T^{B})&=\frac{\delta^{AB}}{2}\comma
\qquad\qquad\qquad\qquad\qquad~~~
[T^{A},T^{B}]=if^{ABC}T^{C}\comma
\\
\label{eq:rel2}
f^{ACD}f^{BCD}&=
\begin{cases}
N\delta^{AB} & \textrm{if } A,B\neq N^{2}\\
0 & \textrm{otherwise}
\end{cases}\comma
\qquad\qquad
(T^{A})^{a}_{~b}(T^{A})^{c}_{~d} =\frac{1}{2}\delta^{a}_{d}\delta^{c}_{b}\comma
\end{flalign}
with $A=1,\dots,N^2$ and $T^{N^{2}}:=\mathbb{I}/\sqrt{2N}$. 
The identities relevant to this paper follow straightforwardly, with the exception of \eqref{eq:rel4} below which needs
\begin{flalign}
\label{eq:rel3}
& f^{AA_{1}A_{2}}f^{AA_{3}A_{4}}(T^{A_{1}})^{a_{1}}_{~b_{1}}(T^{A_{2}})^{a_{2}}_{~b_{2}}(T^{A_{3}})^{a_{3}}_{~b_{3}}(T^{A_{4}})^{a_{4}}_{~b_{4}} \\
=& \frac{1}{8}\left(
-\delta^{a_{1}}_{b_{2}}\delta^{a_{2}}_{b_{3}}\delta^{a_{3}}_{b_{4}}\delta^{a_{4}}_{b_{1}}+\delta^{a_{1}}_{b_{2}}\delta^{a_{2}}_{b_{4}}\delta^{a_{3}}_{b_{1}}\delta^{a_{4}}_{b_{3}}
+\delta^{a_{1}}_{b_{3}}\delta^{a_{2}}_{b_{1}}\delta^{a_{3}}_{b_{4}}\delta^{a_{4}}_{b_{2}}
-\delta^{a_{1}}_{b_{4}}\delta^{a_{2}}_{b_{1}}\delta^{a_{3}}_{b_{2}}\delta^{a_{4}}_{b_{3}}
\right)\period\nonumber
 \end{flalign}
\endgroup
The proof of this uses the second formula in \eqref{eq:rel1} to create pairs of matrices with the same color index
\begin{equation}
f^{AA_{1}A_{2}}f^{AA_{3}A_{4}}(T^{A_{1}})^{a_{1}}_{~b_{1}}(T^{A_{2}})^{a_{2}}_{~b_{2}}(T^{A_{3}})^{a_{3}}_{~b_{3}}(T^{A_{4}})^{a_{4}}_{~b_{4}}
=\left[T^{A_{2}},T^{A}\right]^{a_{1}}_{~b_{1}}(T^{A_{2}})^{a_{2}}_{~b_{2}}\left[T^{A},T^{A_{4}}\right]^{a_{3}}_{~b_{3}}(T^{A_{4}})^{a_{4}}_{~b_{4}}\comma
\end{equation}
and the repeated application of the second formula in \eqref{eq:rel2}.

\subsection{Action and propagators\label{apsubsec:propagator}}
The action of $\mathcal{N}=4$ SYM (in the Euclidean signature)  is given by
\begin{align}
&S=\frac{1}{g_{\rm YM}^2}\int d^{4} x\,\,\mathcal{L}\comma\\
&\mathcal{L}={\rm tr}\left[-\frac{[D_{\mu},D_{\nu}]^2}{2}+(D_{\mu}\Phi^{I})^2+\frac{[\Phi^{I},\Phi^{J}]^2}{2}+i\bar{\Psi}\Gamma^{\mu}D_{\mu}\Psi+\bar{\Psi}\Gamma^{I}[\Phi_I,\Psi]+\del^{\mu}\bar{c}D_{\mu}c+(\del_{\mu}A^{\mu})^2\right]\comma\nonumber
\end{align}
where $D_{\mu}\equiv \del_{\mu}-i[A_{\mu}, \bullet]$ while $c$ and $\bar{c}$ are the BRST ghosts. $\Gamma^{A}=(\Gamma^{\mu},\Gamma^{I})$ are the Dirac matrices in ten dimensions which are normalized as follows:
\beq
{\rm Tr}\left(\Gamma^{A}\Gamma^{B}\right)=16\delta^{AB}\period
\eeq

From this action, one can read off the propagators of the individual fields. For instance, for the gauge field and the scalar field, we have
\beq
\begin{aligned}
\langle (A_{\mu})^{a}{}_{b}(x) (A_{\nu})^{c}{}_{d}(y)\rangle&=\frac{g_{\rm YM}^2\delta^{a}_{d}\delta^{c}_{b}}{8\pi^2}\frac{\delta_{\mu\nu}}{|x-y|^2}\comma\\
\langle (\Phi^{I})^{a}{}_{b}(x)(\Phi^{J})^{c}{}_{d}(y)\rangle&=\frac{g_{\rm YM}^2\delta^{a}_{d}\delta^{c}_{b}}{8\pi^2}\frac{\delta^{IJ}}{|x-y|^2}\period
\end{aligned}
\eeq
\subsection{Toolkit for perturbative computations\label{apsubsec:integrals}}
\paragraph{Conformal integrals} The basic building blocks for the perturbative computation at one loop are the conformal integrals. In particular, the most fundamental is the four-point conformal integral defined by
\beq\label{eq:1loopconformaldef}
 F^{(1)}(z,\bar{z})\equiv \frac{x_{13}^2x_{24}^2}{\pi^2}\int \frac{d^{4}x_5}{x_{15}^2x_{25}^2x_{35}^2x_{45}^2}=\frac{2{\rm Li}_2(z)-{\rm Li}_2 (\bar{z})+\log z\bar{z}\log \frac{1-z}{1-\bar{z}}}{z-\bar{z}}\period
 \eeq
 It satisfies the following relations:
 \beq
 \begin{aligned}
&F^{(1)}(1-z,1-\bar{z})= F^{(1)}(z,\bar{z})\comma\quad F^{(1)}\left(\frac{1}{z},\frac{1}{\bar{z}}\right)= z\bar{z}F^{(1)}(z,\bar{z})\comma\\
&F^{(1)}\left(\frac{z}{z-1},\frac{\bar{z}}{\bar{z}-1}\right)= F^{(1)}\left(\frac{1}{1-z},\frac{1}{1-\bar{z}}\right)=(1-z)(1-\bar{z})F^{(1)}(z,\bar{z})\period
\end{aligned}
 \eeq

\begin{figure}[t]
\centering
\includegraphics[clip,height=5cm]{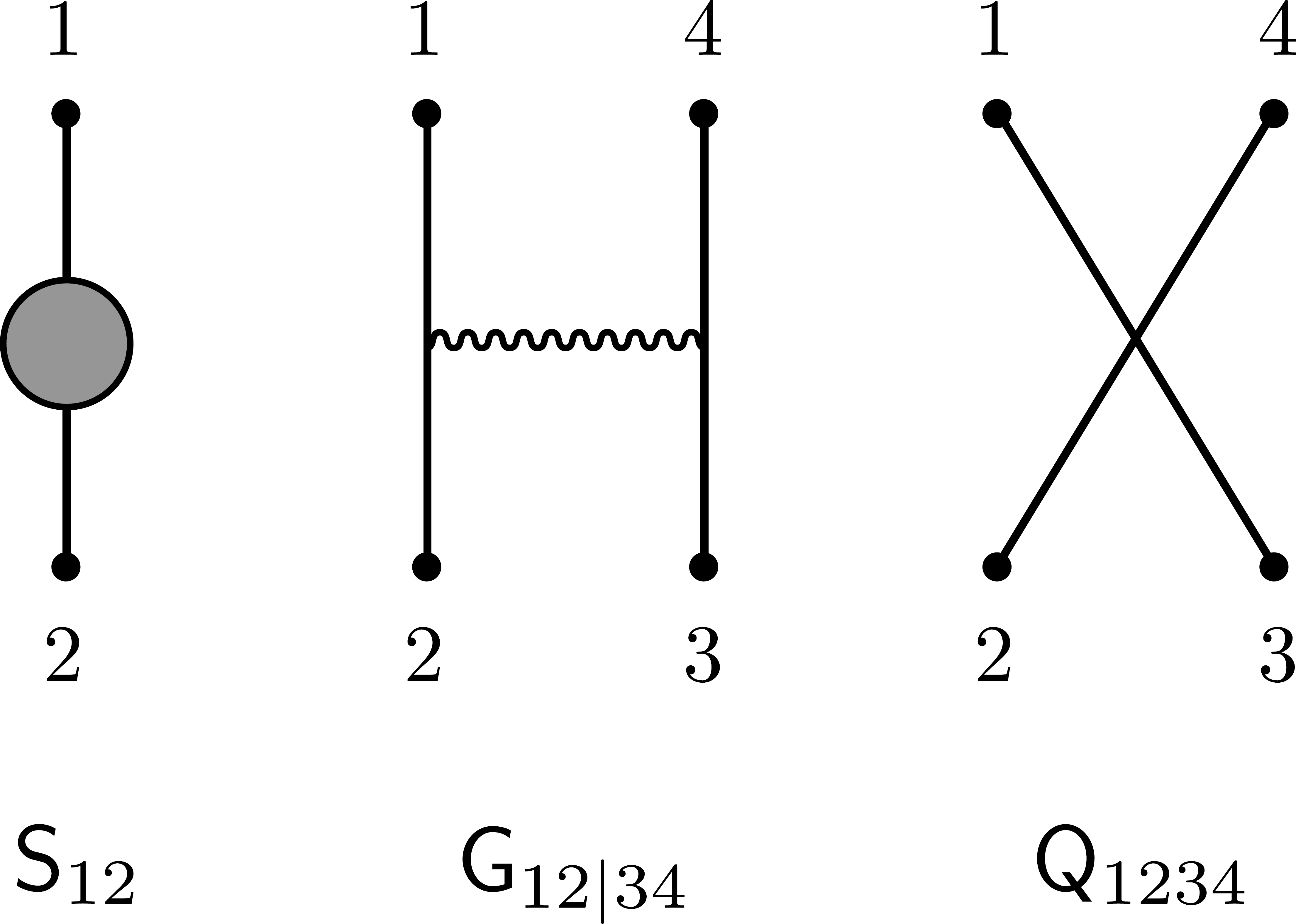}
\caption{The self-energy diagram, the gluon exchange diagram and the scalar quartic diagram.}
\label{apfig1}
\end{figure}

 Another important building block is the three-point integral,
 \beq\label{eq:Y}
 \mathcal{Y}_{123}=\int \frac{d^4 x_5}{x_{15}^2x_{25}^2x_{35}^2}\comma
 \eeq
 which can be obtained by taking the limit of the four-point integral as follows:
 \beq
 \mathcal{Y}_{123}={\rm lim}_{x_4\to \infty}x_4^2\int\frac{d^4 x_5}{x_{15}^2x_{25}^2x_{35}^2x_{45}^2}=\frac{\pi^2F^{(1)}(z^{\prime},\bar{z}^{\prime})}{x_{13}^2}\period
 \eeq
 Here $z^{\prime}$ and $\bar{z}^{\prime}$ are defined by
 \beq
 z^{\prime}\bar{z}^{\prime}=\frac{x_{12}^2}{x_{13}^2}\comma\qquad (1-z^{\prime})(1-\bar{z}^{\prime})=\frac{x_{23}^2}{x_{13}^2}\period
 \eeq
 \paragraph{One-loop diagrams} We now list the planar result for the one-loop diagrams which show up in our computations. At one loop, the relevant diagrams falls into three different classes; the self-energy diagram, the gluon exchange diagram and the scalar quartic diagram (see figure \ref{apfig1}). All these diagrams were computed in the literature and we can simply use the existing results. In what follows, we assume that the end points of the propagtors are contracted with a scalar field $(Y_j\cdot \Phi)(x_j)$. We also assume that the contractions are performed inside a planar diagram and converted the Yang-Mills coupling constant $g_{\rm YM}^2$ into the 't Hooft coupling $g^2$ by multiplying appropriate factors of $N$.

 The contribution from the self-energy diagram ${\sf S}_{ij}$ reads
 \beq\label{eq:1loopS}
 {\sf S}_{ij}=-4g^2 \left(\log \frac{x_{12}}{\epsilon}+1\right)\bar{d}_{ij}\comma
 \eeq
 where $\bar{d}_{ij}\equiv 2g^2 d_{ij}$ is the tree-level propagator. Similarly, the contributions from the gluon exchange diagram ${\sf G}_{ij|kl}$ and the scalar quartic diagram ${\sf Q}_{ijkl}$ are given by
 \beq
 \begin{aligned}\label{eq:1loopGQ}
 {\sf G}_{12|34}&=\frac{g^2}{2}\bar{d}_{12}\bar{d}_{34}F^{(1)}(z,\bar{z})\left(z\bar{z}-z-\bar{z}\right)+{\sf C}_{[12][34]}\comma\\
 {\sf Q}_{1234}&=\frac{g^2}{2}F^{(1)}(z,\bar{z})\left[2\bar{d}_{13}\bar{d}_{24}-(1-z)(1-\bar{z})\bar{d}_{23}\bar{d}_{14}-z\bar{z}\bar{d}_{12}\bar{d}_{34}\right]\period
 \end{aligned}
 \eeq
 Here ${\sf C}_{[12][34]}$ is defined by
 \beq
 \begin{aligned}
 {\sf C}_{[12][34]}=&\frac{g^2}{2}\bar{d}_{12}\bar{d}_{34}\times\\
 &\frac{(x_{13}^2-x_{23}^2)\mathcal{Y}_{123}-(x_{14}^2-x_{24}^2)\mathcal{Y}_{124}+(x_{13}^2-x_{14}^2)\mathcal{Y}_{134}-(x_{23}^2-x_{24}^2)\mathcal{Y}_{234}}{\pi^2}\period
 \end{aligned}
 \eeq

  \begin{figure}[t]
\centering
\includegraphics[clip,height=5cm]{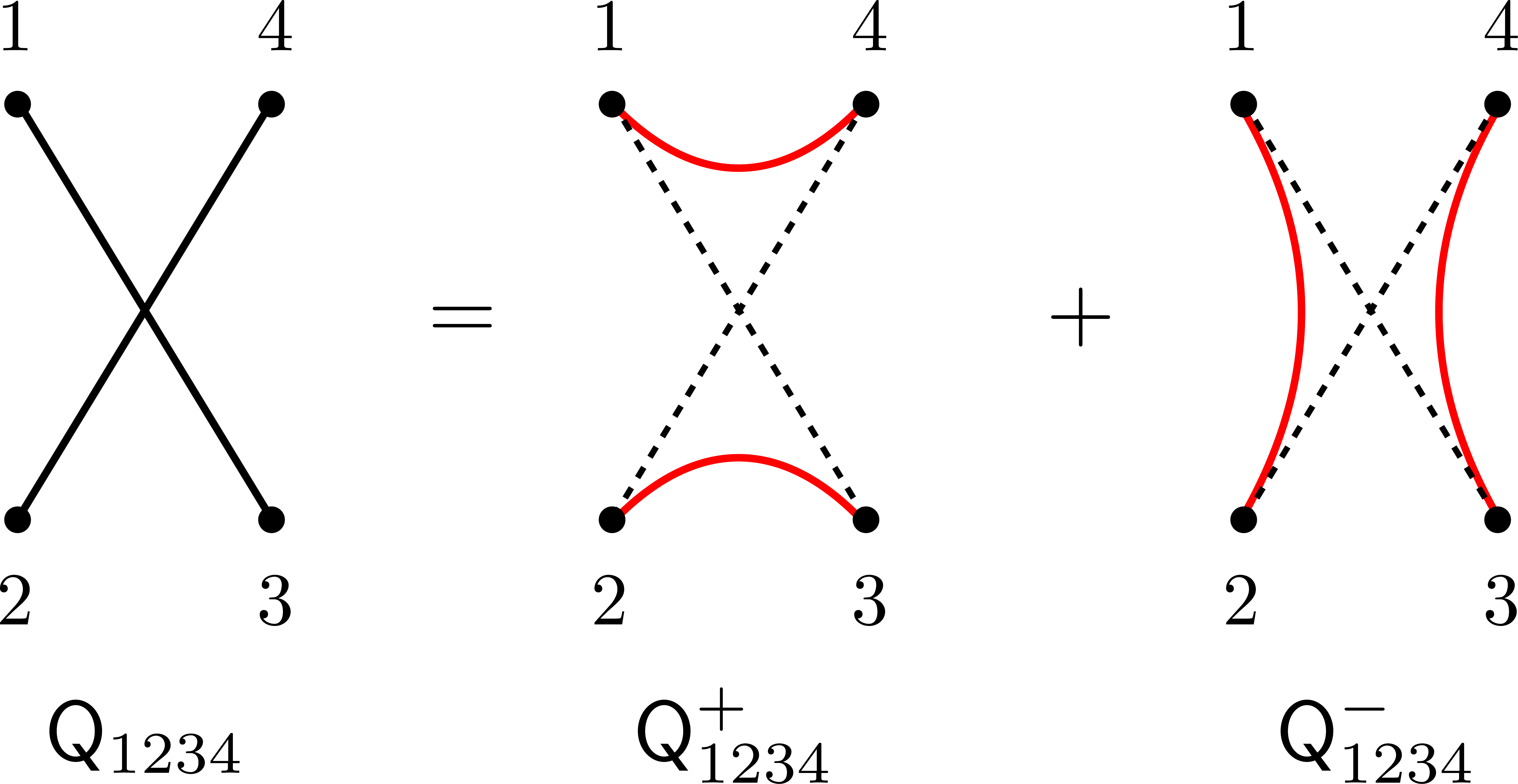}
\caption{The decomposition of the scalar quartic diagram. One can decompose the scalar quartic diagram into two different Wick contraction patterns.}
\label{apfig2}
\end{figure}

 To organize the results of the perturbative computations, it is often useful to split the contribution from the scalar quartic diagram in the following way,
 \beq\label{eq:Qdecompose}
 {\sf Q}_{1234}={\sf Q}^{-}_{1234}+{\sf Q}^{+}_{1234}\comma
 \eeq
 with
 \beq
 \begin{aligned}
 {\sf Q}_{1234}^{-}&=\frac{g^2}{2}F^{(1)}(z,\bar{z})z\bar{z}\left(\frac{1}{\alpha}+\frac{1}{\bar{\alpha}}-1\right)\bar{d}_{12}\bar{d}_{34}\comma\\
 {\sf Q}_{1234}^{+}&=\frac{g^2}{2}F^{(1)}(z,\bar{z})(1-z)(1-\bar{z})\left(\frac{1}{1-\alpha}+\frac{1}{1-\bar{\alpha}}-1\right)\bar{d}_{23}\bar{d}_{14}\period
 \end{aligned}
 \eeq
 This allows us to decompose the quartic interaction into two parts representing different Wick contractions. See figure \ref{apfig2} for a pictorial explanation.
\paragraph{Corner contribution} Other diagrams can be obtained by taking the limit of these basic diagrams. For instance, the ``corner'' contribution defined in \cite{Drukker:2008pi},
 \beq
 {\sf c}_{123}\bar{d}_{12}\bar{d}_{23}=\frac{1}{4}\left({\sf S}_{12}\bar{d}_{23}+\bar{d}_{12}{\sf S}_{23}\right)+{\sf G}_{12|23}+{\sf Q}_{1223}\comma
 \eeq
 can be expressed as
 \beq
 {\sf c}_{123}=\frac{g^2}{2}\frac{x_{13}^2+x_{23}^2-2x_{13}^2}{\pi^2}\mathcal{Y}_{123}\period
 \eeq
 Using the definition of $\mathcal{Y}_{123}$, one can verify that the corner contribution satisfies the following identities:
 \beq\label{eq:cidentities}
 {\sf c}_{123}={\sf c}_{321}\comma\qquad {\sf c}_{123}+{\sf c}_{231}+{\sf c}_{312}=0\period
 \eeq
 In terms of the corner contributions, ${\sf C}_{[12][34]}$ can be rewritten as
 \beq
 {\sf C}_{[12][34]}=\frac{\bar{d}_{12}\bar{d}_{34}}{3}\left[({\sf c}_{312}-{\sf c_{123}})-({\sf c}_{412}-{\sf c_{124}})+({\sf c}_{134}-{\sf c_{341}})-({\sf c}_{234}-{\sf c_{243}})\right]\period
 \eeq
 In the coincident limit, the corner contributions give
 \beq
 {\sf c}_{jjk}=-g^2\left(1+\log \frac{|x_{jk}|}{\epsilon}\right)\comma\qquad {\sf c}_{jkj}=2g^2\left(1+\log \frac{|x_{jk}|}{\epsilon}\right)\period
 \eeq
 \paragraph{Useful identities} Using the results above, one can show various useful relation between the integrals, which we list below:
\begingroup \allowdisplaybreaks
 \begin{align}
 {\sf G}_{12|21}+{\sf Q}_{1221}=&-{\sf S}_{12}\bar{d}_{12}\comma\\
 2{\sf G}_{12|12}+{\sf Q}_{1212}=&2{\sf S}_{12}\bar{d}_{12}\comma\label{eq:apiden1}\\
 -{\sf G}_{21|23}-\frac{1}{2}{\sf Q}_{2123}+\frac{1}{4}\left({\sf S}_{12}\bar{d}_{23}+{\sf S}_{23}\bar{d}_{12}\right)=&{\sf c}_{123}\bar{d}_{12}\bar{d}_{23}\comma\label{eq:apiden2}\\
 {\sf G}_{13|\hat{3}2}+{\sf Q}_{13\hat{3}2}+\frac{1}{4}\left({\sf S}_{13}\bar{d}_{2\hat{3}}+{\sf S}_{2\hat{3}}\bar{d}_{13}\right)=&{\sf c}_{132}\bar{d}_{13}\bar{d}_{2\hat{3}}+\underbrace{g^2 \left[y_{12}^2 y_{3\hat{3}}^2 +2y_{13}^2y_{2\hat{3}}^2 -2y_{1\hat{3}}^2y_{23}^2\right]}_{\displaystyle{=H_{3\hat{3}}}}\nn\\
 \times &\left(\log \frac{\epsilon x_{12}}{x_{13}x_{23}}-1\right)\frac{1}{x_{13}^2x_{23}^2}\period\label{eq:apiden3}
 \end{align}
 \endgroup
In the last inequality, $3$ and $\hat{3}$ denotes the same spacetime points but carry different R-symmetry polarizations. Note also that the notation $y_{ij}^2$ denotes a dot product of the polarization vectors, $y_{ij}^2\equiv Y_i\cdot Y_j$. As indicated, the prefactor on the right hand side of the last equality coincides with the action of the Hamiltonian density in the SO(6) sector acting on the polarizations $3$ and $\hat{3}$. Upon setting $\hat{3}=3$, the last equality reduces to
\beq\label{eq:apiden4}
{\sf G}_{13|32}+{\sf Q}_{1332}+\frac{1}{4}\left({\sf S}_{13}\bar{d}_{23}+{\sf S}_{23}\bar{d}_{13}\right)={\sf c}_{132}\bar{d}_{13}\bar{d}_{23}
\eeq
\section{PCGG in terms of multi-trace operators\label{ap:PCGG}}
In this Appendix we write explicit expressions at finite $N$ for the tree-level and one-loop part of the PCGG in \eqref{eq:PCGGtree1}.
\subsection{Tree level}
We derive the expression for the PCGG in terms of multi-trace operators \eqref{eq:PCGGfinal} by evaluating
\beq\label{eq:apPCGGev}
\begin{aligned}
&\mathcal{G}^{(0)}_{\ell}(x_1,x_2) \equiv \frac{1}{(N!)^2}\left(\begin{array}{c}N\\N-\ell\end{array}\right)^{2}\epsilon_{\red{a_1\cdots a_{N-\ell}}c_1\cdots c_{\ell}}\epsilon^{\red{b_1\cdots b_{N-\ell}}d_1\ldots d_{\ell}}\epsilon_{\red{\bar{a}_1\cdots \bar{a}_{N-\ell}}\bar{c}_1\cdots \bar{c}_{\ell}}\epsilon^{\red{\bar{b}_1\cdots \bar{b}_{N-\ell}}\bar{d}_1\ldots \bar{d}_{\ell}}\\
&\qquad \times\frac{\langle \Phi^{\red{a_1}}{}_{\red{b_1}}\cdots \Phi^{\red{a_{N-\ell}}}{}_{\red{b_{N-\ell}}}\bar{\Phi}^{\red{\bar{a}_1}}{}_{\red{\bar{b}_1}}\cdots \bar{\Phi}^{\red{\bar{a}_{N-\ell}}}{}_{\red{\bar{b}_{N-\ell}}}\rangle_{0}}{\left(g_{\rm YM}^2d_{12}/(8\pi^2)\right)^{N-\ell}}\,\,\Phi^{c_1}{}_{d_1}\cdots \Phi^{c_{\ell}}{}_{d_{\ell}}\bar{\Phi}^{\bar{c}_1}{}_{\bar{d}_1}\cdots \bar{\Phi}^{\bar{c}_{\ell}}{}_{\bar{d}_{\ell}}\period
\end{aligned}
\eeq

First by performing the tree-level Wick contraction, we get
\beq\label{eq:simplifyPCGG}
\begin{aligned}
\frac{\langle \Phi^{\red{a_1}}{}_{\red{b_1}}\cdots \Phi^{\red{a_{N-\ell}}}{}_{\red{b_{N-\ell}}}\bar{\Phi}^{\red{\bar{a}_1}}{}_{\red{\bar{b}_1}}\cdots \bar{\Phi}^{\red{\bar{a}_{N-\ell}}}{}_{\red{\bar{b}_{N-\ell}}}\rangle_{0}}{\left(g_{\rm YM}^2d_{12}/(8\pi^2)\right)^{N-\ell}}&=\sum_{\sigma\in S_{N-\ell}}\delta^{a_1}_{\bar{b}_{\sigma_1}}\cdots \delta^{a_{N-\ell}}_{\bar{b}_{\sigma_{N-\ell}}}\delta^{\bar{a}_1}_{b_{\sigma_1}}\cdots \delta^{\bar{a}_{N-\ell}}_{b_{\sigma_{N-\ell}}}\comma
\end{aligned}
\eeq
where the sum is over all possible permutations of $N-\ell$ elements. Plugging this expression into \eqref{eq:apPCGGev}, one finds that different permutations give the same result owing to the anti-symmetry of the epsilon tensors and therefore we obtain the following expression:
\beq\label{eq:simplifyPCGG2}
\begin{aligned}
\mathcal{G}^{(0)}_{\ell}(x_1,x_2)=&\frac{(N-\ell)!}{(N!)^2}\left(\begin{array}{c}N\\\ell\end{array}\right)^{2} \epsilon_{\red{a_1\cdots a_{N-\ell}}c_1\cdots c_{\ell}}\epsilon^{\blue{b_1\cdots b_{N-\ell}}d_1\ldots d_{\ell}}\epsilon_{\blue{b_1\cdots b_{N-\ell}}\bar{c}_1\cdots \bar{c}_{\ell}}\epsilon^{\red{a_1\cdots a_{N-\ell}}\bar{d}_1\ldots \bar{d}_{\ell}}\\
&\times \Phi^{c_1}{}_{d_1}\cdots \Phi^{c_{\ell}}{}_{d_{\ell}}\bar{\Phi}^{\bar{c}_1}{}_{\bar{d}_1}\cdots \bar{\Phi}^{\bar{c}_{\ell}}{}_{\bar{d}_{\ell}}\period
\end{aligned}
\eeq
The contracted epsilon tensors that appear in \eqref{eq:simplifyPCGG2} can be replaced with anti-symmetric Kronecker delta's introduced in \cite{Berenstein:2003ah},
\beq\label{eq:antisymmetricdelta}
\epsilon_{\red{a_1\cdots a_{N-\ell}}c_1\cdots c_{\ell}}\epsilon^{\red{a_1\cdots a_{N-\ell}}\bar{d}_1\ldots \bar{d}_{\ell}}=(N-\ell)!\,\,\delta^{[\bar{d}_1 \cdots \bar{d}_{\ell}]}_{[c_1\cdots c_{\ell}]}\comma
\eeq
with
\beq\label{eq:antisymmetricdelta2}
\delta^{[\bar{d}_1 \cdots \bar{d}_{\ell}]}_{[c_1\cdots c_{\ell}]}\equiv \sum_{\sigma\in S_{\ell}}(-1)^{|\sigma|}\delta^{\bar{d}_1}_{c_{\sigma_1}}\cdots \delta^{\bar{d}_{\ell}}_{c_{\sigma_{\ell}}}\period
\eeq
The identity \eqref{eq:antisymmetricdelta} can be shown rather straightforwardly by noticing that both sides are totally antisymmetric and nonzero if and only if $\{c_1,\cdots, c_{\ell}\}$ and $\{\bar{d}_1,\cdots, \bar{d}_{\ell}\}$ coincide as a set. The overall coefficient $(N-\ell)!$ can be read off by specifying the indices to particular values.

Substituting these expressions to \eqref{eq:simplifyPCGG2}, we arrive at
\beq\label{eq:simplifyPCGG3}
\begin{aligned}
\mathcal{G}^{(0)}_{\ell}(x_1,x_2)=\frac{(N-\ell)!}{(\ell!)^2}\sum_{\sigma,\rho\in S_{\ell}}(-1)^{|\sigma|+|\rho|}\delta^{\red{\bar{d}_1}}_{\red{c_{\sigma_1}}}\cdots \delta^{\red{\bar{d}_{\ell}}}_{\red{c_{\sigma_{\ell}}}}\delta^{\blue{d_1}}_{\blue{\bar{c}_{\rho_1}}}\cdots \delta^{\blue{d_{\ell}}}_{\blue{\bar{c}_{\rho_{\ell}}}}\Phi^{\red{c_1}}{}_{\blue{d_1}}\cdots \Phi^{\red{c_{\ell}}}{}_{\blue{d_{\ell}}}\bar{\Phi}^{\blue{\bar{c}_1}}{}_{\red{\bar{d}_1}}\cdots \bar{\Phi}^{\blue{\bar{c}_{\ell}}}{}_{\red{\bar{d}_{\ell}}}
\end{aligned}
\eeq
To proceed, we relabel the barred indices as
\beq
\begin{aligned}
\bar{c}_{\rho_k}^{\rm old}=\bar{c}_{k}^{\rm new}\comma\quad \bar{d}_{\rho_k}^{\rm old}=\bar{d}_{k}^{\rm new}\qquad \iff \qquad \bar{c}_{k}^{\rm old}=\bar{c}_{\rho^{-1}_k}^{\rm new}\comma\quad \bar{d}_{k}^{\rm old}=\bar{d}_{\rho^{-1}_k}^{\rm new}\comma
\end{aligned}
\eeq
and rewrite \eqref{eq:simplifyPCGG3} as
\begin{align}
\mathcal{G}^{(0)}_{\ell}(x_1,x_2)=&\frac{(N-\ell)!}{(\ell!)^2}\sum_{\sigma,\rho\in S_{\ell}}(-1)^{|\sigma|+|\rho|}\delta^{\red{\bar{d}_{\rho^{-1}_1}}}_{\red{c_{\sigma_1}}}\cdots \delta^{\red{\bar{d}_{\rho^{-1}_{\ell}}}}_{\red{c_{\sigma_{\ell}}}}\delta^{\blue{d_1}}_{\blue{\bar{c}_{1}}}\cdots \delta^{\blue{d_{\ell}}}_{\blue{\bar{c}_{\ell}}}\Phi^{\red{c_1}}{}_{\blue{d_1}}\cdots \Phi^{\red{c_{\ell}}}{}_{\blue{d_{\ell}}}\bar{\Phi}^{\blue{\bar{c}_1}}{}_{\red{\bar{d}_1}}\cdots \bar{\Phi}^{\blue{\bar{c}_{\ell}}}{}_{\red{\bar{d}_{\ell}}}\nn\\
=&\frac{(N-\ell)!}{\ell!}\underbrace{\sum_{\sigma^{\prime}\in S_{\ell}}(-1)^{|\sigma^{\prime}|}\delta^{\red{\bar{d}_1}}_{\red{c_{\sigma^{\prime}_1}}}\cdots \delta^{\red{\bar{d}_{\ell}}}_{\red{c_{\sigma^{\prime}_{\ell}}}}}_{=\,\,\delta^{[\bar{d}_1\cdots \bar{d}_{\ell}]}_{[c_1\cdots c_{\ell}]}}\left(\Phi\bar{\Phi}\right)^{\red{c_1}}{}_{\red{\bar{d}_1}}\cdots \left(\Phi\bar{\Phi}\right)^{\red{c_{\ell}}}{}_{\red{\bar{d}_{\ell}}}\period
\end{align}
In passing from the first line to the second line, we introduced a new permutation $\sigma^{\prime}\equiv \sigma \circ \rho$ and summed over $\rho$. As indicated, the summation over $\sigma^{\prime}$ reconstructs the antisymmetric Kronecker delta and we see that the result is proportional to a subdeterminant of $(\Phi\bar{\Phi})$ (see \eqref{eq:defsubdet}).
We then use the generating function of subdeterminants to get\fn{Here ${\bf 1}$ is the identity matrix.}
\beq
\begin{aligned}
\mathcal{G}^{(0)}_{\ell}(x_1,x_2)=&(N-\ell)!(-1)^{\ell}\oint \frac{dx}{2\pi i} \frac{1}{x^{N-\ell+1}}\det (x-\Phi\bar{\Phi})\\
=&(N-\ell)!(-1)^{\ell} \oint \frac{dx}{2\pi i} \frac{1}{x^{1-\ell}} \exp \left[{\rm tr}\log ({\bf 1}-x^{-1}\Phi\bar{\Phi})\right]\period
\end{aligned}
\eeq
By expanding ${\rm tr}\log$ in powers of $x^{-1}$, we arrive at the formula in the main text:
\beq
\mathcal{G}^{(0)}_{\ell}(x_1,x_2)=(N-\ell)! (-1)^{\ell}\sum_{\substack{k_1,\ldots,k_{\ell}\\\sum_{s}s k_s=\ell}}\prod_{m=1}^{\ell}\frac{\left(-{\rm tr}\left[(\Phi\bar{\Phi})^m\right]\right)^{k_m}}{m^{k_m}k_m!}\period
\eeq
\subsection{One loop}
In analogy to \eqref{eq:defofPCGG}, the one-loop partially contracted Giant Graviton of length $2\ell$ is
\beq
\begin{aligned}
&\mathcal{G}^{(1)}_{\ell}(x_1,x_2) \equiv \frac{1}{(N!)^2}\left(\begin{array}{c}N\\N-\ell\end{array}\right)^{2}\epsilon_{\red{a_1\cdots a_{N-\ell}}c_1\cdots c_{\ell}}\epsilon^{\red{b_1\cdots b_{N-\ell}}d_1\ldots d_{\ell}}\epsilon_{\red{\bar{a}_1\cdots \bar{a}_{N-\ell}}\bar{c}_1\cdots \bar{c}_{\ell}}\epsilon^{\red{\bar{b}_1\cdots \bar{b}_{N-\ell}}\bar{d}_1\ldots \bar{d}_{\ell}}\\
&\qquad \times\frac{\langle \Phi^{\red{a_1}}{}_{\red{b_1}}\cdots \Phi^{\red{a_{N-\ell}}}{}_{\red{b_{N-\ell}}}\bar{\Phi}^{\red{\bar{a}_1}}{}_{\red{\bar{b}_1}}\cdots \bar{\Phi}^{\red{\bar{a}_{N-\ell}}}{}_{\red{\bar{b}_{N-\ell}}}\rangle_{1}}{\left(g_{\rm YM}^2d_{12}/(8\pi^2)\right)^{N-\ell}}\,\,\Phi^{c_1}{}_{d_1}\cdots \Phi^{c_{\ell}}{}_{d_{\ell}}\bar{\Phi}^{\bar{c}_1}{}_{\bar{d}_1}\cdots \bar{\Phi}^{\bar{c}_{\ell}}{}_{\bar{d}_{\ell}}\comma
\end{aligned}
\eeq
where $\langle \qquad \rangle_{1}$ denotes the one-loop part of the Wick contraction. We rewrite it using \eqref{eq:antisymmetricdelta} and define the tensor $M$ from the contraction of two generalized Kronecker deltas \eqref{eq:antisymmetricdelta2}
\begin{gather}\label{eq:M}
\delta_{[\bar{c}_{1}\dots \bar{c}_{\ell}\bar{c}]}^{[d_{1}\dots d_{\ell}c]}\delta^{[\bar{d}_{1}\dots \bar{d}_{\ell}\bar{c}]}_{[{c}_{1}\dots {c}_{\ell}c]}\equiv N\delta_{[\bar{c}_{1}\dots \bar{c}_{\ell}]}^{[d_{1}\dots d_{\ell}]}\delta^{[\bar{d}_{1}\dots \bar{d}_{\ell}]}_{[{c}_{1}\dots {c}_{\ell}]}+M_{\bar{c}_{1}\dots \bar{c}_{\ell};{c}_{1}\dots {c}_{\ell}}^{d_{1}\dots d_{\ell};\bar{d}_{1}\dots \bar{d}_{\ell}}\period
\end{gather}
There are two ways of dressing the tree-level diagrams with the vertices in figure \ref{apfig1}.

 \begin{figure}[t]
\centering
\includegraphics[clip,height=3cm]{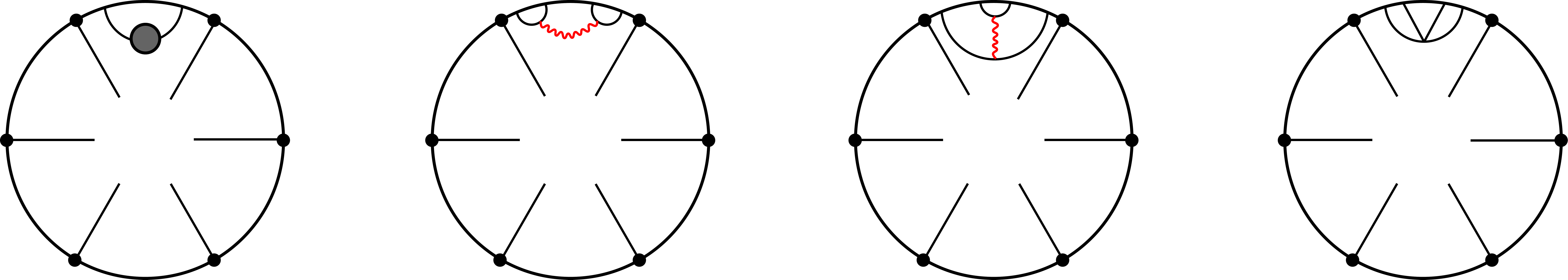}
\caption{One-loop corrections to the partially contracted Giant Graviton.}
\label{apfig8}
\end{figure}

One possibility is to construct a tree-level PCGG of length $2\ell+2$ and insert here one scalar self-energy, which takes two scalars out and leaves the remaining $2\ell$ uncontracted, see first diagram in figure \ref{apfig8}. This contributes with
\begingroup \allowdisplaybreaks
\begin{flalign}\label{eq:PCGG1loopO}
&\frac{\left(N-\ell-1\right)!}{\left(\left(\ell+1\right)!\right)^{2}}\left({\ell}+1\right)^{2}\left[-\frac{g_{\textrm{YM}}^{2}N}{4\pi^{2}}\left(\log\frac{x_{12}}{\epsilon}+1\right)\right]\\
&\times\left\{ \left[\left(N-{\ell}\right)^{2}-1\right]\delta_{[\bar{c}_{1}\dots \bar{c}_{\ell}]}^{[d_{1}\dots d_{\ell}]}\delta^{[\bar{d}_{1}\dots \bar{d}_{\ell}]}_{[{c}_{1}\dots {c}_{\ell}]}-\frac{1}{N}M_{\bar{c}_{1}\dots \bar{c}_{\ell};{c}_{1}\dots {c}_{\ell}}^{d_{1}\dots d_{\ell};\bar{d}_{1}\dots \bar{d}_{\ell}}\right\}
\Phi^{c_1}{}_{d_1}\cdots \Phi^{c_{\ell}}{}_{d_{\ell}}\bar{\Phi}^{\bar{c}_1}{}_{\bar{d}_1}\cdots \bar{\Phi}^{\bar{c}_{\ell}}{}_{\bar{d}_{\ell}}
\period\nonumber
\end{flalign}
\endgroup
The factor $(\ell+1)^2$ is the number of different pairs $\Phi \bar \Phi$ that can be chosen out of the $\ell+1$ pairs in $\mathcal{G}_{\ell+1}^{(0)}$. The tensor in the second line stems from the contraction of two tensors
\begingroup \allowdisplaybreaks
\begin{flalign}
\begin{split}
 & \delta_{[\bar{c}_{1}\dots \bar{c}_{\ell}\bar{c}]}^{[d_{1}\dots d_{\ell}d]}\delta^{[\bar{d}_{1}\dots \bar{d}_{\ell}\bar{d}]}_{[{c}_{1}\dots {c}_{\ell}c]}\left(\delta^c_{\bar{d}}\delta_d^{\bar{c}}-\frac{1}{N}\delta^{c\bar{c}}\delta_{d\bar{d}}\right)
 =
 \left[\left(N-\ell\right)^{2}-1\right]\delta_{[\bar{c}_{1}\dots \bar{c}_{\ell}]}^{[d_{1}\dots d_{\ell}]}\delta^{[\bar{d}_{1}\dots \bar{d}_{\ell}]}_{[{c}_{1}\dots {c}_{\ell}]}-\frac{1}{N}M_{\bar{c}_{1}\dots \bar{c}_{\ell};{c}_{1}\dots {c}_{\ell}}^{d_{1}\dots d_{\ell};\bar{d}_{1}\dots \bar{d}_{\ell}}\comma
\end{split}
\end{flalign}
\endgroup
where one is carried by the tree-level PCGG and the other by the one-loop part of the self-energy \cite{Erickson:2000af,Beisert:2002bb} \footnote{The subscript $S$ selects the first diagram in figure \ref{apfig1}. We cannot use \eqref{eq:1loopS} because we seek an expression exact in $N$.}
\begin{flalign}\label{eq:vertexO}
\left\langle \left(\Phi^{I}\left(x_{1}\right)\right)^{A}\left(\Phi^{J}\left(x_{2}\right)\right)^{B}\right\rangle _{S}&=-g^{4}N^{3}\delta^{IJ}\left(\delta^{AB}-\delta^{AN^{2}}\delta^{BN^{2}}\right)\frac{1}{16\pi^{4}x_{12}^2}\left(\log\frac{x_{12}}{\epsilon}+1\right)
\end{flalign}
dressed up with color matrices.

The other contribution arises from a tree-level PCGG of length $2\ell+4$ with the insertion of the four-scalar interactions, see last three diagrams in figure \ref{apfig8}. It evaluates to
\begingroup \allowdisplaybreaks
\begin{flalign}\label{eq:PCGG1loopXH}
&\frac{\left(N-\ell-2\right)!}{\left(\left(\ell+2\right)!\right)^{2}}\frac{1}{4}\left({\ell}+2\right)^{2}\left({\ell}+1\right)^{2}\left[\frac{g_{\textrm{YM}}^{2}N^{3}}{4\pi^{2}}\left(\log\frac{x_{12}}{\epsilon}+1\right)\right]\\
&\times\left\{ 4\left[\left(N-\ell\right)^{2}+N\right]\left(N-\ell-1\right)^{2}\delta_{[\bar{c}_{1}\dots \bar{c}_{\ell}]}^{[d_{1}\dots d_{\ell}]}\delta^{[\bar{d}_{1}\dots \bar{d}_{\ell}]}_{[{c}_{1}\dots {c}_{\ell}]}+4\left(N-\ell-1\right)^{2}M_{\bar{c}_{1}\dots \bar{c}_{\ell};{c}_{1}\dots {c}_{\ell}}^{d_{1}\dots d_{\ell};\bar{d}_{1}\dots \bar{d}_{\ell}}\right\} \nonumber\\
&\times \Phi^{c_1}{}_{d_1}\cdots \Phi^{c_{\ell}}{}_{d_{\ell}}\bar{\Phi}^{\bar{c}_1}{}_{\bar{d}_1}\cdots \bar{\Phi}^{\bar{c}_{\ell}}{}_{\bar{d}_{\ell}}\period\nonumber
\end{flalign}
\endgroup
The number $\left(\ell+2\right)^{2}\left(\ell+1\right)^{2}/4$ counts the way of choosing two $\Phi$'s and two $\bar \Phi$'s out of those in $\mathcal{G}_{\ell+2}^{(0)}$. The second line is the result of the contraction of two tensors
\begingroup \allowdisplaybreaks
\begin{flalign}\label{eq:rel4}
 & \delta_{[\bar{c}_{1}\dots \bar{c}_{\ell}\bar{c}\bar{c}']}^{[d_{1}\dots d_{\ell}dd']}\delta_{[c_{1}\dots c_{\ell}cc']}^{[\bar{d}_{1}\dots \bar{d}_{\ell}\bar{d}\bar{d}']}\left(-2\delta^d_{\bar{c}}\delta^{d'}_{\bar{c}'}\delta^{\bar{d}}_{c'}\delta^{\bar{d}'}_{c}-2\delta^d_{\bar{c}'}\delta^{d'}_{\bar{c}}\delta^{\bar{d}}_{c}\delta^{\bar{d}'}_{c'}+\delta^{d\bar{d}'}\delta^{d'}_{\bar{c}}\delta^{\bar{d}}_{c'}\delta_{c\bar{c}'}\right.
 \nonumber\\
 &
~~\left. +\delta^d_{\bar{c}}\delta^{d'\bar{d}}\delta^{\bar{d}'}_{c'}\delta_{c\bar{c}'}+\delta^{d\bar{d}'}\delta^{d'}_{\bar{c}'}\delta^{\bar{d}}_{c}\delta_{c'\bar{c}}+\delta^d_{\bar{c}'}\delta^{d'\bar{d}}\delta^{\bar{d}'}_{c}\delta_{c'\bar{c}}\right)\\
= & \,4\left[\left(N-\ell\right)^{2}+N\right]\left(N-\ell-1\right)^{2}\delta_{[\bar{c}_{1}\dots \bar{c}_{\ell}]}^{[d_{1}\dots d_{\ell}]}\delta_{[c_{1}\dots c_{\ell}]}^{[\bar{d}_{1}\dots \bar{d}_{\ell}]}+4\left(N-\ell-1\right)^{2}M_{\bar{c}_{1}\dots \bar{c}_{\ell};c_{1}\dots c_{\ell}}^{d_{1}\dots d_{\ell};\bar{d}_{1}\dots \bar{d}_{\ell}} \comma\nonumber
\end{flalign}
\endgroup
where one is carried by tree-level PCGG and the other by the one-loop part of the connected four-point function\footnote{See \cite{Drukker:2008pi} for the definition of the relevant regularized integrals. Notice that the last formula in (21) therein should read $F_{12,12}=-\left(\log({\epsilon^{2}}/{x_{12}^{2}})-3\right)/(8\pi^{2})$.
}, which is the sum of the gluon exchange\footnote{The subscripts $Q$ and $G$ select the second and third diagram in figure \ref{apfig1} respectively, but do not use the large-$N$ expressions \eqref{eq:1loopGQ}.}
\begingroup \allowdisplaybreaks
\begin{flalign}\label{eq:vertexG}
& \left\langle \left(\Phi^{I}\left(x_{1}\right)\right)^{A}\left(\Phi^{J}\left(x_{2}\right)\right)^{B}\left(\Phi^{K}\left(x_{3}\right)\right)^{C}\left(\Phi^{L}\left(x_{4}\right)\right)^{D}\right\rangle _{G}
\\
=&\frac{g_{\textrm{YM}}^{6}}{\left(2\pi\right)^{4}}\left(f^{ABE}f^{CDE}\delta^{IJ}\delta^{KL}\frac{F_{12,34}}{x_{12}^{2}x_{34}^{2}}
+f^{ACE}f^{BDE}\delta^{IK}\delta^{JL}\frac{F_{13,24}}{x_{13}^{2}x_{24}^{2}}+f^{ILE}f^{BCE}\delta^{IL}\delta^{JK}\frac{F_{14,23}}{x_{14}^{2}x_{23}^{2}}\right)\nonumber
\end{flalign}
\endgroup
and the four-scalar vertex
\begingroup \allowdisplaybreaks
\begin{flalign}\label{eq:vertexQ}
&\left\langle \left(\Phi^{I}\left(x_{1}\right)\right)^{A}\left(\Phi^{J}\left(x_{2}\right)\right)^{B}\left(\Phi^{K}\left(x_{3}\right)\right)^{C}\left(\Phi^{L}\left(x_{4}\right)\right)^{D}\right\rangle _{Q}=g_{\textrm{YM}}^{6}\left[-\left(f^{ACE}f^{BDE}+f^{ADE}f^{BCE}\right)\delta^{IJ}\delta^{KL}\right.
\nonumber\\
&~~\left.+\left(f^{ADE}f^{BCE}-f^{ABE}f^{CDE}\right)\delta^{IK}\delta^{JL}+\left(f^{ABE}f^{CDE}+f^{ACE}f^{BDE}\right)\delta^{IL}\delta^{JK}\right]X_{1234}
 \period
\end{flalign}
\endgroup
One needs to use \eqref{eq:rel3} when \eqref{eq:vertexG} and \eqref{eq:vertexQ} multiply the color matrices. We defined \cite{Beisert:2002bb}
\begingroup \allowdisplaybreaks
\begin{flalign}
X_{1234}&\equiv\frac{1}{\left(2\pi\right)^{8}}\int\frac{d^{4}x_{5}}{x_{15}^{2}x_{25}^{2}x_{35}^{2}x_{45}^{2}}\comma
\qquad\qquad
G_{1,23}\equiv\frac{x_{13}^{2}-x_{12}^{2}}{\left(2\pi\right)^{4}}\mathcal{Y}_{123}\comma
\\
F_{12,34}&\equiv\left(2\pi\right)^{4}\left(x_{13}^{2}x_{24}^{2}-x_{14}^{2}x_{23}^{2}\right)X_{1234}+G_{1,34}-G_{2,34}+G_{3,12}-G_{4,12}
 \nonumber
\end{flalign}
\endgroup
and $\mathcal{Y}$ is \eqref{eq:Y}. The integrals are evaluated in appendix \ref{apsubsec:integrals}.

The sum of \eqref{eq:PCGG1loopO} and \eqref{eq:PCGG1loopXH} yields
\begingroup \allowdisplaybreaks
\begin{align}
\label{eq:PCGG1loop}
\begin{split}
\mathcal{G}_{\ell}^{(1)}
&=-g_{\textrm{YM}}^{2}\frac{\ell(N-\ell-1)}{4\pi^{2}N^{2}}\left(\log\frac{x_{12}}{\epsilon}+1\right)\mathcal{G}_{\ell}^{(0)}
\\
&+g_{\textrm{YM}}^{2}\frac{\left(N-\ell\right)!}{8\pi^{2}\left((\ell)!\right)^{2}}\left(\log\frac{x_{12}}{\epsilon}+1\right)M_{\bar{c}_{1}\dots\bar{c}_{\ell};c_{1}\dots c_{\ell}}^{d_{1}\dots d_{\ell};\bar{d}_{1}\dots\bar{d}_{\ell}}\,\Phi^{c_{1}}{}_{d_{1}}\cdots\Phi^{c_{\ell}}{}_{d_{\ell}}\bar{\Phi}^{\bar{c}_{1}}{}_{\bar{d}_{1}}\cdots\bar{\Phi}^{\bar{c}_{\ell}}{}_{\bar{d}_{\ell}}\period\\
\end{split}
\end{align}
 \endgroup
The first addend can be written as a multiplicative correction to the tree-level PCGG, whereas the other displays the new tensor $M$ \eqref{eq:M}.

We are ready to expand \eqref{eq:PCGG1loop} in terms of multi-trace operators. While the expansion \eqref{eq:PCGGfinal} takes care of the first line, we need to set up an algorithm, in the form of \eqref{eq:Mmultitrace} below, for the scalars contracted with $M$.

The starting point is the identity
\begingroup \allowdisplaybreaks
\begin{flalign}\label{eq:rel16bis}
&\delta_{[\bar{c}_{1}\dots \bar{c}_{\ell+1}]}^{[d_{1}\dots d_{\ell+1}]}\delta^{[\bar{d}_{1}\dots \bar{d}_{\ell+1}]}_{[{c}_{1}\dots {c}_{\ell+1}]}
 \Phi^{c_1}{}_{d_1}\cdots \Phi^{c_{\ell+1}}{}_{d_{\ell+1}}\bar{\Phi}^{\bar{c}_1}{}_{\bar{d}_1}\cdots \bar{\Phi}^{\bar{c}_{\ell+1}}{}_{\bar{d}_{\ell+1}}
\\
=&-\sum_{\substack{k_{1},k_{2},\dots k_{\ell+1}=0\\\sum_{s}s k_s=\ell+1}}\frac{(-)^{\ell}\left(\left({\ell}+1\right)!\right)^{2}}{k_{1}!\dots k_{\ell+1}!}\left(-\textrm{tr}\left(\Phi\bar{\Phi}\right)\right)^{k_{1}}\dots \left(-\frac{\textrm{tr}\left(\Phi\bar{\Phi}\right)^{\ell+1}}{\ell+1}\right)^{k_{\ell+1}}\period\nonumber
\end{flalign}
\endgroup
Next, one defines a replacement rule that turns a collection of $\Phi$'s and $\bar \Phi$'s into the sum over all ways of substituting a pair of scalars $\Phi^{c}_{~d}\bar{\Phi}^{\bar{c}}_{~\bar{d}}$ in it with $\delta^c_{d}\delta^{\bar{c}}_{\bar{d}}$; for example
\begin{flalign}
\label{eq:rule}
\left.
\Phi^{c_1}_{ ~d_1} \Phi^{c_2}_{ ~d_2} \bar{\Phi}^{\bar{c}_1}_{ ~\bar{d}_1} \bar{\Phi}^{\bar{c}_2}_{ ~\bar{d}_2}
\right|_{\textrm{rule}}
&=
\delta^{c_1}_{ d_1} \delta^{\bar{c}_1}_{ \bar{d}_1} \Phi^{c_2}_{ ~d_2} \bar{\Phi}^{\bar{c}_2}_{ ~\bar{d}_2}
+\delta^{c_1}_{ d_1} \delta^{\bar{c}_2}_{ \bar{d}_2} \Phi^{c_2}_{ ~d_2} \bar{\Phi}^{\bar{c}_1}_{ ~\bar{d}_1}
\\&
+ \delta^{c_2}_{ d_2} \delta^{\bar{c}_1}_{ \bar{d}_1} \Phi^{c_1}_{ ~d_1} \bar{\Phi}^{\bar{c}_2}_{ ~\bar{d}_2}
+\delta^{c_2}_{ d_2} \delta^{\bar{c}_2}_{ \bar{d}_2} \Phi^{c_1}_{ ~d_1} \bar{\Phi}^{\bar{c}_1}_{ ~\bar{d}_1}\period\nonumber
\end{flalign}
Applying this operation on \eqref{eq:rel16bis} delivers a new identity. On its rhs, terms linear in $N$ are produced when the rule sends $\textrm{tr}(\Phi\bar \Phi)$ into $\textrm{tr}(\mathbb{I})=N$. On the lhs, the $(\ell+1)^2$ ways of picking one $\Phi$ and one $\bar \Phi$ are equivalent to select the last $\Phi$ and $\bar \Phi$ thanks to the antisymmetry property of the generalized Kronecker delta, hence they all display an index contraction like \eqref{eq:M}:
\begingroup \allowdisplaybreaks
\begin{flalign}
&
\left({\ell}+1\right)^{2}
\left(N\delta_{[\bar{c}_{1}\dots \bar{c}_{\ell}]}^{[d_{1}\dots d_{\ell}]}\delta^{[\bar{d}_{1}\dots \bar{d}_{\ell}]}_{[{c}_{1}\dots {c}_{\ell}]}
+
M_{\bar{c}_{1}\dots \bar{c}_{\ell};{c}_{1}\dots {c}_{\ell}}^{d_{1}\dots d_{\ell};\bar{d}_{1}\dots \bar{d}_{\ell}}\right)
\Phi^{c_1}{}_{d_1}\cdots \Phi^{c_{\ell}}{}_{d_{\ell}}\bar{\Phi}^{\bar{c}_1}{}_{\bar{d}_1}\cdots \bar{\Phi}^{\bar{c}_{\ell}}{}_{\bar{d}_{\ell}}\period
\end{flalign}
\endgroup
After identifying the terms proportional to $N^0$ and $N^1$ on the two sides, we can discard the latter in order to read off the expansion of the tensor $M$. In practice, this is realized by a modified rule that takes the rhs of \eqref{eq:rel16bis} and returns the sum over all ways of substituting one $\Phi^c_{~d}$ and one $\bar{\Phi}^{\bar{c}}_{~\bar{d}}$ \emph{that have ${c}\neq \bar{d}$ and $d\neq \bar{c}$} with $\delta^c_d\delta^{\bar{c}}_{\bar{d}}$. This change has the effect of ignoring length-2 traces $\textrm{tr}(\Phi\bar \Phi)$ that would produce $N$ upon the action of the original rule stated above \eqref{eq:rule}.

In conclusion, the multitrace expansion of the second line in \eqref{eq:PCGG1loop} is
\begingroup \allowdisplaybreaks
\begin{flalign}\label{eq:Mmultitrace}
 &
M_{\bar{c}_{1}\dots \bar{c}_{\ell};{c}_{1}\dots {c}_{\ell}}^{d_{1}\dots d_{\ell};\bar{d}_{1}\dots \bar{d}_{\ell}}
 \Phi^{c_1}{}_{d_1}\cdots \Phi^{c_{\ell}}{}_{d_{\ell}}\bar{\Phi}^{\bar{c}_1}{}_{\bar{d}_1}\cdots \bar{\Phi}^{\bar{c}_{\ell}}{}_{\bar{d}_{\ell}}
 =-\left({\ell}+1\right)^{-2}\\
&\times \left.\sum_{\substack{k_{1},k_{2},\dots k_{\ell+1}=0\\\sum_{s}s k_s=\ell+1}}\frac{(-)^{\ell}\left(\left({\ell}+1\right)!\right)^{2}}{k_{1}!\dots k_{\ell+1}!}\left(-\textrm{tr}\left(\Phi\bar{\Phi}\right)\right)^{k_{1}}\dots \left(-\frac{\textrm{tr}\left(\Phi\bar{\Phi}\right)^{\ell+1}}{\ell+1}\right)^{k_{\ell+1}}\right|_{
\substack{\textrm{modified}\\\hspace{-0.6cm}\textrm{rule}}
}\period\nonumber
\end{flalign}
\endgroup
Since the modified rule removes two scalars at most in a trace and each trace displays the alternating pattern $\textrm{tr}(\Phi\bar{\Phi}\Phi\bar{\Phi}\dots)$, it is easy to realize that \eqref{eq:Mmultitrace} contains non-alternating sequences. The expansion of the one-loop PCGG \eqref{eq:PCGG1loop} in multi-trace operators inherits this feature. This contrasts with \eqref{eq:PCGGfinal} at tree level, where only the alternating pattern appears.

The expressions found for the PCGG can be used in the large-$N$ limit with a word of caution because expansion in multi-trace operators and Wick contractions do not generally commute. For example, taking the naive limit of \eqref{eq:PCGG1loop} and plugging the surviving terms into a correlator overlooks the fact that Wick contractions of the dropped terms may produce powers of $N$ that overcome the suppressions in $N^{-1}$. The correct procedure consists in plugging the finite-$N$ expression \eqref{eq:PCGG1loop} and taking the limit of the expectation value.

\section{Gaudin Norms and Factorization\label{ap:norm}}
In this appendix, we provide explicit expressions for the Gaudin norms in the one-loop SU(2), SL(2) and SO(6) sectors. In particular, we show that the Gaudin norm for the SO(6) sector can be factorized into two determinants if the state obeys the selection rule discussed in the main text. We then show a similar factorization for the SU$(2|2)^2$ chain at finite coupling.
\subsection{SU(2) sector\label{apsubsec:su2norm}}
The norm of the Bethe state in the SU(2) sector is given by\fn{See for instance \cite{Gromov:2012uv}.}
\beq
\begin{aligned}
\langle {\bf u}|{\bf u}\rangle =&\left(\prod_{1\leq i<j\leq M}S_{\rm SU(2)}(u_i,u_j)\right)\left(\prod_{1\leq s\leq M}\frac{1}{\del_u p(u_s)}\right)\\
&\times\det_{1\leq i,j\leq M} \left[\del_{u_i}\left(Lp(u_j)+\frac{1}{i}\sum_{k\neq j}\log S_{\rm SU(2)}(u_j,u_k)\right)\right]\,.
\end{aligned}
\eeq
A point worth mentioning is that in this paper the norm is defined through the two-point function in the twisted-translated frame \eqref{eq:twopntnormal} while it is normally defined in the literature through the two-point functions of complex conjugate operators $\langle \bar{\mathcal{O}}\mathcal{O}\rangle$. This difference results in the difference between the prefactor $\prod S_{\rm SU(2)}(u_i,u_j)$ and the prefactor of (17) in \cite{Gromov:2012uv}. Our normalization is more suited for the computation of the structure constant since it eliminates a phase from the structure constant.

When the rapidities are parity-symmetric, the norm can be further simplified in the following way\fn{Since the (coordinate) Bethe state depends on the order of rapidities, it is important that we order them as $\{u_1,-u_1,u_2,-u_2,\ldots\}$ in order to obtain the expression \eqref{eq:paritynormsu2}.} as was shown in \cite{deLeeuw:2015hxa}:
\beq\label{eq:paritynormsu2}
\langle {\bf u}|{\bf u}\rangle =\left(\prod_{1\leq s\leq \frac{M}{2}}\frac{u_s-i/2}{u_s+i/2}\frac{1}{(\del_u p(u_s))^2}\right)\det G^{\rm SU(2)}_{+}\det G^{\rm SU(2)}_{-}\period
\eeq
\subsection{SL(2) sector\label{apsubsec:sl2norm}}
The results for the SL(2) sector is almost identical to the ones for the SU(2) sector; one simply needs to replace the S-matrix by the SL(2) S-matrix.

For instance the norm of a general Bethe state is given by
\beq
\begin{aligned}
\langle {\bf u}|{\bf u}\rangle =&\left(\prod_{1\leq i<j\leq M}S_{\rm SL(2)}(u_i,u_j)\right)\left(\prod_{1\leq s\leq M}\frac{1}{\del_u p(u_s)}\right)\\
&\times\det_{1\leq i,j\leq M} \left[\del_{u_i}\left(Lp(u_j)+\frac{1}{i}\sum_{k\neq j}\log S_{\rm SL(2)}(u_j,u_k)\right)\right]\comma
\end{aligned}
\eeq
while the norm for a parity-symmetric state reads
\beq\label{eq:paritynormsl2}
\langle {\bf u}|{\bf u}\rangle =\left(\prod_{1\leq s\leq \frac{M}{2}}\frac{u_s+i/2}{u_s-i/2}\frac{1}{(\del_u p(u_s))^2}\right)\det G^{\rm SL(2)}_{+}\det G^{\rm SL(2)}_{-}\period
\eeq
The proof of factorization is the same as the one for the SU(2) sector.
\subsection{SO(6) sector\label{apsubsec:so6norm}}
\paragraph{Gaudin Norm}The norm of the Bethe state in the SO(6) sector (in the coordinate Bethe ansatz normalization) reads
\beq
\begin{aligned}
\langle {\bf u}\,,\red{\bf v}\,,\blue{\bf w}|{\bf u}\,,\red{\bf v}\,,\blue{\bf w}\rangle=&(-1)^{M+\red{K_{\bf v}}+\blue{K_{\bf w}}}\prod_{i<j}^{M}\frac{u_i-u_j-i}{u_i-u_j+i}\prod_{i<j}^{\red{K_{\bf v}}}\frac{v_i-v_j-i}{v_i-v_j+i}\prod_{i<j}^{\blue{K_{\bf w}}}\frac{w_i-w_j-i}{w_i-w_j+i}\\
&\times \det G\comma
\end{aligned}
\eeq
with
\beq
G\equiv\pmatrix{ccc}{\del_{\red{v_i}}\phi_{\red{v_j}}&\del_{\red{v_i}}\phi_{u_{j}}&0\\\del_{u_i}\phi_{\red{v_{j}}}&\del_{u_{i}}\phi_{u_j}&\del_{u_i}\phi_{\blue{w_{j}}}\\0&\del_{\blue{w_i}}\phi_{u_{j}}&\del_{\blue{w_i}}\phi_{\blue{w_j}}}\comma
\eeq
where $\phi$'s are the phase factors of the Bethe equation \eqref{eq:so6betheeq}.

\paragraph{Factorization} We now show that for the Bethe roots which satisfy the selection rule,
\beq
{\bf u}=\{u_1,-u_1,u_2,-u_2,\ldots, u_{\frac{M}{2}},-u_{\frac{M}{2}}\}\comma\quad \red{\bf v}=\{v_1,\ldots, v_{K}\}\comma\quad \blue{\bf w}=\{-v_1,\ldots,-v_{K} \} \comma
\eeq
the Gaudin determinant $\det G$ factorizes into a product of determinants. For this purpose, we first reorder rows and columns of the matrix $G$ to write
\beq
\det G=\det \pmatrix{cc|cc}{V_{++}&B_{++}&V_{+-}&B_{+-}\\\bar{B}_{++}&U_{++}&\bar{B}_{+-}&U_{+-}\\ \hline V_{-+}&B_{-+}&V_{--}&B_{--}\\\bar{B}_{-+}&U_{-+}&\bar{B}_{--}&U_{--}}\comma
\eeq
with
\beq
\begin{aligned}
&\left[V_{++}\right]_{ij}=\del_{\red{v_i}}\phi_{\red{v_j}}\comma\quad\left[V_{--}\right]_{ij}=\del_{\blue{w_i}}\phi_{\blue{w_j}}\comma\quad \left[U_{++}\right]_{ij} =\del_{u_{i_+}}\phi_{u_{j_+}}\comma\quad \left[U_{--}\right]_{ij} =\del_{u_{i_-}}\phi_{u_{j_-}} \comma\\
&\left[B_{++}\right]_{ij}=\del_{\red{v_i}}\phi_{u_{j_{+}}}\comma\quad \left[B_{+-}\right]_{ij}=\del_{\red{v_i}}\phi_{u_{j_{-}}}\comma\quad \left[B_{-+}\right]_{ij}=\del_{\blue{w_i}}\phi_{u_{j_{+}}}\comma\quad \left[B_{--}\right]_{ij}=\del_{\blue{w_i}}\phi_{u_{j_{-}}}\comma\\
&\left[\bar{B}_{++}\right]_{ij}=\del_{u_{j_{+}}}\phi_{\red{v_i}}\comma\quad \left[\bar{B}_{+-}\right]_{ij}=\del_{u_{j_{+}}}\phi_{\blue{w_i}}\comma\quad \left[\bar{B}_{-+}\right]_{ij}=\del_{u_{j_{+}}}\phi_{\red{v_i}}\comma\quad \left[\bar{B}_{--}\right]_{ij}=\del_{u_{j_{-}}}\phi_{\blue{w_i}}\comma\\
&\left[V_{+-}\right]_{ij}=\del_{\red{v_i}}\phi_{\blue{w_j}}\comma\quad \left[V_{-+}\right]_{ij}=\del_{\blue{w_i}}\phi_{\red{v_j}}\period
\end{aligned}
\eeq
Here the indices  $i_+$ and $j_+$ run from $1$ to $\frac{M}{2}$ while $i_-$ and $j_-$ run from $\frac{M}{2}+1$ to $M$. When $\blue{w_j}=-\red{v_j}$, $u_{j_{+}}=u_j$ and $u_{j_{-}}=-u_j$, one can show that the different matrix elements are related by
\beq
U_{-\mp}=U_{+\pm}\comma\quad V_{-\mp}=V_{+\pm}\comma\quad B_{-\mp}=B_{+\pm}\comma\quad \bar{B}_{-\mp}=\bar{B}_{+\pm}\period
\eeq
We can now rewrite the determinant by adding and subtracting the rows and the columns as
\beq
\begin{aligned}
\det G=&\det \pmatrix{cc|cc}{V_{++}&B_{++}&V_{+-}&B_{+-}\\\bar{B}_{++}&U_{++}&\bar{B}_{+-}&U_{+-}\\ \hline V_{+-}&B_{+-}&V_{++}&B_{++}\\\bar{B}_{+-}&U_{+-}&\bar{B}_{++}&U_{++}}\\
&=\det \pmatrix{cc|cc}{V_{++}+V_{+-}&B_{++}+B_{+-}&V_{+-}+V_{++}&B_{+-}+B_{++}\\\bar{B}_{++}+\bar{B}_{+-}&U_{++}+U_{+-}&\bar{B}_{++}+\bar{B}_{+-}&U_{+-}+U_{++}\\ \hline V_{+-}&B_{+-}&V_{++}&B_{++}\\\bar{B}_{+-}&U_{+-}&\bar{B}_{++}&U_{++}}\\
&=\det \pmatrix{cc|cc}{V_{++}+V_{+-}&B_{++}+B_{+-}&0&0\\\bar{B}_{++}+\bar{B}_{+-}&U_{++}+U_{+-}&0&0\\ \hline V_{+-}&B_{+-}&V_{++}-V_{+-}&B_{++}-B_{+-}\\\bar{B}_{+-}&U_{+-}&\bar{B}_{++}-\bar{B}_{+-}&U_{++}-U_{+-}}\period\end{aligned}
\eeq
Then we see that the determinant factorizes into a product determinants of diagonal blocks, $\det G^{\rm SO(6)}_{+}$ and $\det G_{-}^{\rm SO(6)}$ given in \eqref{eq:GpmSO6}. As a result, we obtain the following expression for the norm  of the parity symmetric states:
\beq
\left.\langle {\bf u}\,,\red{\bf v}\,,\blue{\bf w}|{\bf u}\,,\red{\bf v}\,,\blue{\bf w}\rangle\right|_{{\bf u}=-{\bf u}\comma\blue{\bf w}=-\red{\bf v}}=\left(\prod_{j=1}^{\frac{M}{2}}\frac{u_j-\frac{i}{2}}{u_j+\frac{i}{2}}\right)\det G_{+}^{\rm SO(6)}\det G_{-}^{\rm SO(6)}
\eeq
\subsection{SU(2$|$2)$^2$ chain at finite coupling\label{apsubsec:su22norm}}
The factorization of the Gaudin determinant discussed above is applicable also to the full SU$(2|2)^2$ spin chain at finite coupling. To see this, we simply need to note that the Gaudin norm of the SU$(2|2)^2$ spin chain takes the following form
\beq
\langle {\bf u}\comma \red{\bf v}\comma\blue{\bf w}|{\bf u}\comma \red{\bf v}\comma\blue{\bf w}\rangle=\det G \comma
\eeq
with
\beq
G\equiv\pmatrix{ccc}{\del_{\red{V_i}}\phi_{\red{V_j}}&\del_{\red{V_i}}\phi_{u_{j}}&0\\\del_{u_i}\phi_{\red{V_{j}}}&\del_{u_{i}}\phi_{u_j}&\del_{u_i}\phi_{\blue{W_{j}}}\\0&\del_{\blue{W_i}}\phi_{u_{j}}&\del_{\blue{W_i}}\phi_{\blue{W_j}}}\comma
\eeq
where the notations $\red{V_i}$ and $\blue{W_j}$ denote collectively all the nested Bethe roots on the right and the left wings. Written more explicitly, they are defined by
\beq
\begin{aligned}
\red{\bf V}&=\{\red{v_{\I,1}},\ldots,\red{v_{\I,K^{\bf v}_{\I}}}, \red{v_{\II,1}},\ldots,\red{v_{\II,K^{\bf v}_{\II}}},\red{v_{\III,1}},\ldots,\red{v_{\III,K^{\bf v}_{\III}}}\}\comma\\
\blue{\bf W}&=\{\blue{w_{\I,1}},\ldots,\blue{w_{\I,K^{\bf w}_{\I}}}, \blue{w_{\II,1}},\ldots,\blue{w_{\II,K^{\bf w}_{\II}}},\blue{w_{\III,1}},\ldots,\blue{w_{\III,K^{\bf w}_{\III}}}\}\period
\end{aligned}
\eeq
Since the structure of the matrix coincides with that of the SO(6) sector at weak coupling, one can simply follow the same argument and show that the Gaudin norm for the parity-symmetric state can be factorized into a product of two determinants,
\beq
\left.\langle {\bf u}\,,\red{\bf v}\,,\blue{\bf w}|{\bf u}\,,\red{\bf v}\,,\blue{\bf w}\rangle\right|_{{\bf u}=\bar{\bf u}\comma\blue{\bf w}=\bar{\red{\bf v}}}=\det G_{+}\det G_{-}\comma
\eeq
where $G_{\pm}$ are given by \eqref{eq:defGpmAsympt}.
\section{More Rigorous Derivation of TBA and $g$-Functions\label{ap:gfunctionderivation}}
In this appendix, we present a derivation which is in a sense a hybrid of the methods in \cite{Pozsgay:2010tv}, \cite{Woynarovich:2010wt} and \cite{Kostov:2018dmi}. It in particular follows closely the arguments in \cite{Woynarovich:2010wt} but shortcut some of the combinatorial arguments, by introducing an auxiliary field variable $\eta$. It has an advantage that it leads to a simple TBA action with the $Y$-function being a fundamental field variable as is the case with the heuristic argument presented in the main text.

As discussed in the main text, we can label the states in the large $R$ limit using the ``mode numbers'' $n_j$'s which appear in the logarithm of the Bethe equations:
\beq\label{eq:logformbetheap}
2\pi n_j =\frac{1}{i}\log \mathcal{R}(u_j) +\frac{1}{i}\sum_{k} \log \mathcal{S}(u_j,u_k)  \comma\qquad (j=1,\ldots, M)\period
\eeq
Using these mode number basis, we can express the thermal partition function as
\beq
\begin{aligned}
&Z_{ab}=\sum_{M=0}^{\infty}Z_{M}\comma\qquad Z_{M}=\sum_{0<n_1<n_2<\cdots<n_M} e^{-L\tilde{E}(n_1,n_2,\cdots,n_M)}\period
\end{aligned}
\eeq
Here the constraints on the sum $0<n_1<\cdots<n_M$ come from the physical requirements that no two particles can occupy the same mode (see the exclusion property in \eqref{eq:propertiesexclusion}), and that particles with mode numbers $n$ and $-n$ must be identified\fn{This follows from the fact that the momenta of the particles get flipped after the reflection.} in the presence of boundaries. We remind the readers that the energy $\tilde{E}$ is given by a sum of the energies of individual rapidities
\beq\label{eq:energydef}
\tilde{E}(n_1,\ldots,n_M)=\sum_{k=1}^{M}\tilde{E}(u_k)\comma
\eeq
which in turn are determined by the mode numbers through the Bethe equation \eqref{eq:logformbetheap}.

The second step is to convert the constrained sum into an unconstrained sum as in \cite{Woynarovich:2010wt}. This can be done by first computing the sum without any constraints and then subtracting unnecessary terms involving coincident mode numbers. This leads to the following identity derived in \cite{Woynarovich:2010wt},
\beq
\sum_{n_1<\ldots<n_M}f(n_1,\ldots,n_M)=\sum_{p\in P_M}c(p)\sum_{n_1,\ldots, n_{|p|}} f^{(p)}(n_1,\ldots, n_{|p|})  \comma
\eeq
where $\sum_{p\in P_M}$ is a sum over the partitions $p$ of the integer $N$. For a given partition $p=(r_1,r_2,\cdots, r_{|p|})$ with $r_1+\cdots +r_{|p|}=M$, the function $f^{(p)}$ is defined by
\beq
f^{(p)}(n_1,\cdots, n_{|p|})=f(\overbrace{\underbrace{n_1,\ldots, n_1}_{r_{1}},\underbrace{n_2,\ldots, n_2}_{r_{2}},\ldots, \underbrace{n_{|P|},\ldots, n_{|P|}}_{r_{|P|}}}^{M})\period
\eeq
The combinatorial coefficient $c(p)$ can be computed explicitly\fn{The explicit expression can be found in (A.12) of \cite{Woynarovich:2010wt}.}, but all we need is the fact that the constrained sum for a factorized quantity $g(n_1,\ldots,n_M)=\prod_{k=1}^{M}g(n_k)$ is given by
\beq\label{eq:allweneed}
\sum_M\sum_{n_1<\ldots <n_M}g(n_1,\ldots, n_M)=\sum_{M}\sum_{p\in P_{M}}c(p)\sum_{n_1,\ldots, n_{|p|}} g^{(p)}(n_1,\ldots, n_{|p|})=\prod_{n}(1+g(n))\period
\eeq
This can be shown either by using an explicit form of $c(P)$ or by interpreting the sum as a partition function of free fermion. To apply this formula to our problem, we need to clarify what we mean by the energy $\tilde{E}(n_1,\ldots,n_M)$ when mode numbers are coincident. This is simply defined by taking the limit; we solve for the rapidities in terms of mode numbers, take the limit in which several mode numbers coincide and then plug them into the expression \eqref{eq:energydef}. As discussed in \cite{Woynarovich:2010wt}, this in practice amounts to solving the following modified Bethe equation
\beq\label{eq:relationmoderap}
2\pi n_j =\frac{1}{i}\log \mathcal{R}(u_j) +\frac{1}{i}\sum_{k} r_k\log \mathcal{S}(u_j,u_k)  \comma\qquad (j=1,\ldots, |p|)
\eeq
and plug them into the following expression
\beq
\tilde{E}=\sum_{k=1}^{|p|}r_k \tilde{E}(u_k)\period
\eeq

The third step is to approximate the sum over the mode numbers by integrals\fn{The approximation error can be estimated by using the Euler-Maclaurin formula and can be shown to be exponentially small.},
\beq\label{eq:firststepsumtoint}
\sum_{0<n_1,\ldots,n_{|p|}}\qquad \mapsto \qquad \int_{0}^{\infty}(1-\delta (n_1))dn_1\cdots \int_{0}^{\infty}(1-\delta (n_{|p|}))dn_{|p|} \period
\eeq
We included the factor $1-\delta (n_k)$ for each integration variable in order to eliminate the contribution from $n_k=0$ modes. Now we further convert this into the integrals over the rapidities. This can be done by first expanding $\prod_{k}(1-\delta (n_k))$ factor and then using the relation \eqref{eq:relationmoderap}, as was explained in \cite{Kostov:2018dmi}:
\beq
\begin{aligned}\nn
\sum_{\alpha \subset \{1,2,\cdots, |p|\}}\prod_{k=1}^{|p|}\int_{0}^{\infty}dn_k(-1)^{|\alpha|}\prod_{k\in\alpha}\delta (n_k)&=\sum_{\alpha}\prod_{k=1}^{|p|}\int_{0}^{\infty}du_k (-1)^{|\bar{\alpha}|}\det{}_{\bar{\alpha}}\left[\frac{\del n_j}{\del u_k}\right]\prod_{k\in\alpha}\delta (u_k)\\
&=\prod_{k=1}^{|p|}\int_{0}^{\infty}du_k\det \left[\frac{\del n_j}{\del u_k} -\delta (u_j)\delta_{jk}\right]
\end{aligned}
\eeq
Here $\det{}_{\bar{\alpha}}$ is a determinant of a submatrix in which the rows and columns corresponding to the set $\alpha$ are omitted. On the second line, we used the fact that the first line can be regarded as an expansion of a determinant. To proceed, we factorize the Jacobian as follows:
\beq
\det \left[\frac{\del n_j}{\del u_k} -\delta (u_j)\delta_{jk}\right]=\mathcal{J}^{(p)}(u_1,\ldots,u_{|p|})\prod_{k=1}^{|p|}\mu(u_k)\comma
\eeq
with
\beq
\begin{aligned}
\mu(u_j)&\equiv \frac{1}{2\pi i}\del_{u_j}\log \mathcal{R}(u_j)-\delta(u_j)+\frac{1}{2\pi i}\sum_{k} r_k\del_{u_j}\log \mathcal{S}(u_j,u_k)\comma\\
\mathcal{J}^{(p)}(u_1,\ldots,u_{|p|})&\equiv \det \left[\delta_{jk}-\frac{i r_k\del_{u_k}\log \mathcal{S}(u_j,u_k)}{2\pi\mu(u_k)}\right]\period
\end{aligned}
\eeq
As a result we obtain the following expression for the thermal partition function
\beq\label{eq:multipleintegraltoderive}
Z_{ab}=\sum_{M}\sum_{p\in P_{M}}c(p)\left(\prod_{k=1}^{|p|}\int_0^{\infty} \mu (u_{k})du_{k}\right) \,\,\mathcal{J}^{(p)}\,\, \exp \left(-L\sum_{k=1}^{|p|}r_k\tilde{E}(u_k)\right)\period
\eeq

Now comes the crucial step in which we deviate from the discussions in \cite{Woynarovich:2010wt} and \cite{Kostov:2018dmi}. We now insert the following path integral inside the integral over $u_k$'s:
\beq\label{eq:resolutionidentity}
1=\int\mathcal{D}\rho\mathcal{D}\eta\, \exp \left[-i\int_0^{\infty}du\,\, R\eta (u)\left(\rho (u)-\frac{1}{R}\sum_{k=1}^{|p|}r_k \delta(u-u_k)\right) \right]\period
\eeq
If we integrate out $\eta (u)$, we get
\beq
\rho(u)=\frac{1}{R}\sum_{k=1}^{|p|} r_k \delta (u-u_k)\period
\eeq
Therefore, the variable $\rho (u)$ can be interpreted as a {\it generalized density function}. After inserting \eqref{eq:resolutionidentity} to the multiple integrals \eqref{eq:multipleintegraltoderive}, one can then rewrite the integrand in terms of $\rho (u)$ in the following way:
\beq
\begin{aligned}
&\exp \left(-L\sum_{k=1}^{|p|}r_k \tilde{E}(u_k)\right)\mapsto \exp \left(-R\int_0^{\infty}du L\tilde{E}(u)\rho(u)\right)\comma\\
&\mathcal{J}^{(p)}\mapsto \det \left[1-\hat{G}_{-}\right]\comma\qquad \mu (u)\mapsto \frac{\log \mathcal{R}(u)}{2\pi i}+\frac{R}{2\pi i}\int dv \del_u \log \mathcal{S} (u,v)\rho (v)
\end{aligned}
\eeq
Here $\det \left[1-\hat{G}_{-}\right]$ is a Fredholm determinant with a kernel
\beq
\hat{G}_{-}\cdot f(u)\equiv R\int_{0}^{\infty}dv\frac{i \rho (v)\del_{u}\log \mathcal{S}(v,u)}{\mu(v)} f(v)\period
\eeq
This conversion from a finite-dimensional determinant to a Fredholm determinant is slightly nontrivial, but one can check the equivalence explicitly by taking the logarithm of the two expressions and expand both using the $\log \det ={\rm Tr}\log$ relation.

Now, after this rewriting the partition function reads
\beq
\begin{aligned}
Z_{ab}=&\sum_{M}\sum_{p\in P_{M}}c(p)\int \mathcal{D}\rho\mathcal{D\eta}\left(\prod_{k=1}^{|p|}\int_0^{\infty} \mu (u_{k})du_{k} e^{i r_k\eta (u_k)}\right)\\
&\times \det \left[1-\hat{G}_{-}\right]\,\, \exp \left(-L\int_0^{\infty}du \tilde{E}(u)\rho(u)\right)
\end{aligned}
\eeq
In this expression, the dependence on $u_k$ and the partition $p$ only shows up in a factorized form inside the bracket on the first line. We can therefore use the identity \eqref{eq:allweneed}---more precisely, the continuum limit of the identity\fn{Basically, we can discretize the $u$-integral so that $\int \mu (u)du \to \sum_{n}$, apply the identity \eqref{eq:allweneed} and then take the continuum limit again.}---to rewrite it as
\beq
Z_{ab}=\int \mathcal{D}\rho\mathcal{D}\eta \,\,\det \left[1-\hat{G}_{-}\right]\times e^{-R S[\rho,\eta]}\comma
\eeq
with\fn{For rewriting we used
\beq
\sigma (u) =\frac{1}{2\pi i R}\del_{u}\log \mathcal{R}(u)-\frac{\delta (u)}{R}\comma\quad \mathcal{K}_{+}(u,v)=\frac{1}{i}\del_u \mathcal{S} (u,v)\period
\eeq}
\beq
S[\rho,\eta]=\int_0^{\infty}du \left[L\tilde{E}(u)\rho(u)+i\eta (u)\rho (u)-\underbrace{(\sigma (u)+\mathcal{K}_{+}\ast\rho (u))}_{=\,\mu (u)/R}\log (1+e^{i\eta (u)})\right]\period
\eeq
We can also rewrite the action of the kernel $\hat{G}_{-}$ using the relations
\beq
\begin{aligned}
\del_u \log \mathcal{S}(v,u)&=\del_u \left[\log \tilde{S}(v,u)+\log \tilde{S}(v,-u)\right]=-\del_u \left[\log \tilde{S}(u,v)-\log \tilde{S}(u,-v)\right]\comma\\
R\frac{\rho (v)}{\mu (v)}&=\left.\frac{1}{1+1/Y(v)}\right|_{\text{saddle point}}\comma
\end{aligned}
\eeq
where the right hand side of the second equation is evaluated using the saddle-point value of $Y(v)$ \eqref{eq:oursimpleTBA}. As a result, we obtain the formula used in the main text:
\beq
\hat{G}_{-}\cdot f(u)\equiv \int_{0}^{\infty} \frac{dv}{2\pi}\frac{\mathcal{K}_{-}(u,v)}{1+1/Y(v)}f(v) \comma\qquad \mathcal{K}_{-}(u,v) \equiv \frac{1}{i}\del_{u}\log \frac{\tilde{S} (u,v)}{\tilde{S}(u,-v)}\period
\eeq
The rest of the discussions follows the one in the main text.
\section{SU(2$|$2)$^2$ symmetry and S-matrix\label{ap:Smatrix}}
In this appendix, we review the centrally-extended SU(2$|$2)$^2$ symmetry and the S-matrix of the spin chain for $\mathcal{N}=4$ SYM.

The main purpose is to set the convention used in this paper: Throughout this paper, we use the so-called string frame \cite{Arutyunov:2007tc} to describe magnon excitations. The string frame is a natural frame for the worldsheet in the uniform lightcone gauge and has the following two advantages:
\begin{enumerate}
\item The S-matrix does not produce the $\mathcal{Z}$ markers.
\item The length of the spin chain is given by the $R$-charge $J$, which is well-defined nonperturbatively\fn{By contrast, the length in the spin-chain frame is given by the number of fields inside the operators, which is not a well-defined quantum number.}.
\end{enumerate}
Different frames are related by each other by a redefinition of the magnon excitations on the spin chain.

In addition to the choice of the frames, there is also a choice of notations to keep track of the Hopf-algebraic structure of the PSU$(2|2)$ symmetry. In this paper, we use two different (but physically equivalent) notations, both of which will be reviewed in this appendix: The first one is the {\it twisted notation} in \cite{Beisert:2006qh} and makes use of the $\mathcal{Z}$-markers. This notation is used in the main text to determine the matrix structure of the form factor. The second one is the {\it cumulative notation} in \cite{Beisert:2006qh}. This is the notation used often with the string frame in the literature. We will use this notation to discuss the details of the unfolding structure of the reflection matrix in Appendix \ref{ap:reflection}.
\subsection{Symmetry and representation\label{apsubsec:rep}}
The centrally-extended SU$(2|2)^2$ symmetry consists of the following two sets generators, Lorentz, R-symmetry, supersymmetry and superconformal,
\beq
\begin{aligned}
\{L^{\alpha}{}_{\beta}\comma\quad R^{a}{}_{b}\comma\quad Q^{\alpha}{}_{a}\comma\quad S^{a}{}_{\alpha}\}\comma\qquad
\{\dot{L}^{\dot{\alpha}}{}_{\dot{\beta}}\comma\quad \dot{R}^{\dot{a}}{}_{\dot{b}}\comma\quad \dot{Q}^{\dot{\alpha}}{}_{\dot{a}}\comma\quad S^{\dot{a}}{}_{\dot{\alpha}}\}\comma
 \end{aligned}
\eeq
and three central charges
\beq
P\comma\qquad K\comma\qquad C =\frac{D-J}{2}\period
\eeq
Here all the indices run from $1$ to $2$, and $P$ and $K$ originate from the gauge transformation of $\mathcal{N}=4$ SYM.
The nontrivial part of the (anti)commutation relation reads
\beq
\begin{aligned}
&\{Q^{\alpha}{}_{a}\,,\,Q^{\beta}{}_{b} \}=\{\dot{Q}^{\alpha}{}_{a}\,,\,\dot{Q}^{\beta}{}_{b} \}=\epsilon^{\alpha\beta}\epsilon_{ab}P\comma\qquad \{S^{a}{}_{\alpha}\,,\,S^{b}{}_{\beta} \}=\{\dot{S}^{a}{}_{\alpha}\,,\,\dot{S}^{b}{}_{\beta} \}=\epsilon^{ab}\epsilon_{\alpha\beta}K\comma\\
&\{Q^{\alpha}{}_{a}\,,\,S^{b}{}_{\beta}\}=\{\dot{Q}^{\alpha}{}_{a}\,,\,\dot{S}^{b}{}_{\beta}\}=\delta^{b}_{a}L^{\alpha}{}_{\beta}+\delta^{\alpha}_{\beta}R^{b}{}_{a}+\delta^{b}_{a}\delta^{\alpha}_{\beta}C\period
\end{aligned}
\eeq
\paragraph{Twisted notation}
As discussed in section \ref{subsec:formVSreflection}, magnons in the $\mathcal{N}=4$ SYM spin chain transform as a bifundamental representation under this symmetry algebra. One important feature of this symmetry algebra is that the action on the multi-particle states is not simply given by a sum of the action on single-particle states: Rather it has a mild nonlocality described by a coproduct of the Hopf algebra. One way to keep track of such nonlocality is to introduce the so-called $\mathcal{Z}$-markers. In this notation, the action of the symmetry generators reads\fn{Here we only wrote the action for the left PSU$(2|2)$, since the right PSU$(2|2)$ acts in the same way.}
\beq
\begin{aligned}
&R^{a}_{b} \ket{\phi^{c}}= \delta_{b}^{c}\ket{\phi^{a}}-\frac{1}{2}\delta^{a}_{b}\ket{\phi^{c}}\comma\quad
&&L^{\alpha}_{\beta}\ket{\psi^{\gamma}}=\delta^{\gamma}_{\beta} \ket{\psi^{\alpha}}-\frac{1}{2}\delta^{\alpha}_{\beta}\ket{\psi^{\gamma}}\comma\\
&Q^{\alpha}_{a}\ket{\phi^{b}}={\bf a}   \delta^{b}_{a}\ket{\mathcal{Z}^{+1/2}\psi^{\alpha}}\comma\quad &&Q^{\alpha}_{a}\ket{\psi^{\beta}}= {\bf b}  \epsilon^{\alpha\beta}\epsilon_{ab}\ket{\mathcal{Z}^{+1/2}\phi^{b}}\comma\\
&S^{a}_{\alpha}\ket{\phi^{b}}= {\bf c}  \epsilon^{ab}\epsilon_{\alpha\beta}\ket{\mathcal{Z}^{-1/2}\psi^{\beta}}\comma\quad &&S^{a}_{\alpha}\ket{\psi^{\beta}}={\bf d}  \delta^{\beta}_{\alpha}\ket{\mathcal{Z}^{-1/2}\phi^{a}}\period
\end{aligned}
\eeq
\beq
\begin{aligned}
&C\ket{\mathcal{X}}= \frac{1}{2}({\bf a}{\bf d}+{\bf b}{\bf c})\ket{\mathcal{X}}\comma
&&P\ket{\mathcal{X}}={\bf a}{\bf b}
\ket{\mathcal{Z}\mathcal{X}}\comma\\ &K\ket{\mathcal{X}}={\bf c}{\bf d}
\ket{\mathcal{Z}^{-}\mathcal{X}}\comma
\end{aligned}
\eeq
where ${\bf a}$-${\bf d}$ are given by
\beq
{\bf a}(u)=\sqrt{g}\eta\comma \quad {\bf b}(u)= \frac{\sqrt{g}}{\eta}\left( 1-\frac{x^{+}}{x^{-}}\right)\comma \quad {\bf c}(u)=\frac{i\sqrt{g} \eta}{ x^{+}}\comma \quad {\bf d}(u) =\frac{ \sqrt{g}x^{+}}{i\eta}\left( 1- \frac{x^{-}}{x^{+}}\right)
\eeq
 and
 \beq
 \eta=\left(\frac{x^{+}}{x^{-}}\right)^{1/4}\sqrt{i (x^{-}-x^{+})}\comma\qquad U=\left(\frac{x^{+}}{x^{-}}\right)^{1/2}\period
 \eeq
 The positions of the $\mathcal{Z}$-markers can be shifted according to the following rule:
 \beq
 \begin{aligned}
 &&\ket{\mathcal{X}\mathcal{Z}^{\pm}}=\left( \frac{x^{+}}{x^{-}}\right)^{\pm1}\!\!\!\!\ket{\mathcal{Z}^{\pm}\mathcal{X}}\comma
\end{aligned}
 \eeq
 which effectively reproduces the nonlocality of the action of the symmetry generators.
\paragraph{Cumulative notation} There is another way to deal with the coproduct structure of the symmetry algebra, which is called {\it cumulative notation}. In this notation, we do not use the $\mathcal{Z}$ markers. Instead we introduce an extra parameter $\zeta$ and write the action of the symmetry generators as
\beq
\begin{aligned}
&Q^{\alpha}_{a}\ket{\phi^{b}}={\bf a}   \delta^{b}_{a}\ket{\psi^{\alpha}}\comma\quad &&Q^{\alpha}_{a}\ket{\psi^{\beta}}= {\bf b}  \epsilon^{\alpha\beta}\epsilon_{ab}\ket{\phi^{b}}\comma\\
&S^{a}_{\alpha}\ket{\phi^{b}}= {\bf c}  \epsilon^{ab}\epsilon_{\alpha\beta}\ket{\psi^{\beta}}\comma\quad &&S^{a}_{\alpha}\ket{\psi^{\beta}}={\bf d}  \delta^{\beta}_{\alpha}\ket{\phi^{a}}\comma
\end{aligned}
\eeq
where ${\bf a}$-${\bf d}$ are now given by
\beq
\begin{aligned}
{\bf a}(u,\zeta)&=\sqrt{g\zeta}\eta\comma \quad &&{\bf b}(u,\zeta)= \frac{\sqrt{g\zeta}}{\eta}\left( 1-\frac{x^{+}}{x^{-}}\right)\comma \\
 {\bf c}(u,\zeta)&=\sqrt{\frac{g}{\zeta}}\frac{i \eta}{ x^{+}}\comma \quad &&{\bf d}(u,\zeta) =\sqrt{\frac{g}{\zeta}}\frac{ x^{+}}{i\eta}\left( 1- \frac{x^{-}}{x^{+}}\right)\period
\end{aligned}
\eeq
The newly introduced phase factor $\zeta$ takes into account the nonlocality of the action of the symmetry generators: More precisely, for a multiparticle state $|\chi_1(u_1)\chi_2(u_2)\cdots \rangle$, $\zeta$'s for the $j$-th particle and $(j+1)$-th particle are related by
\beq
\zeta_{j+1}=\zeta_je^{ip_j}\period
\eeq
\subsection{S-matrix\label{apsubsec:Smatrix}}
The full SU$(2|2)^2$ S-matrix consists of a product of two matrix parts and an overall factor
\beq
\mathbb{S}(u_1,u_2)=S_0 (u_1,u_2)\mathcal{S}_{12} \otimes \dot{\mathcal{S}}_{12}\period
\eeq
The overall factor is given by
\beq
S_0 (u_1,u_2)=\frac{x_1^{+}-x_2^{-}}{x_1^{-}-x_2^{+}}\frac{1-1/x_1^{-}x_2^{+}}{1-1/x_1^{+}x_2^{-}}\frac{1}{(\sigma (u_1,u_2))^2}\period
\eeq
with $\sigma$ being the bulk dressing phase.

The action of the matrix part of the S-matrix in the string frame reads
\beq
\begin{aligned}
\mathcal{S}_{12}\ket{\phi^{a}_1\phi^{b}_2}&=A_{12}\ket{\phi_2^{\{a}\phi_1^{b\}}}
+B_{12}\ket{\phi_2^{[a}\phi_1^{b]}}+\frac{1}{2}C_{12}\epsilon^{ab}\epsilon_{\alpha\beta}\ket{\psi^{\alpha}_2\psi^{\beta}_1}\comma\\
\mathcal{S}_{12}\ket{\psi^{\alpha}_1\psi^{\beta}_2}&=D_{12}\ket{\psi_2^{\{\alpha}\psi_1^{\beta\}}}
+E_{12}\ket{\psi_2^{[\alpha}\psi_1^{\beta]}}+\frac{1}{2}F_{12}\epsilon^{\alpha\beta}\epsilon_{ab}\ket{\phi_2^{a}\phi_1^{b}}\comma\\
\mathcal{S}_{12}\ket{\phi^{a}_1\psi^{\beta}_2}&=G_{12}\ket{\psi^{\beta}_2\phi_1^{a}}
+H_{12}\ket{\phi_2^{a}\psi^{\beta}_1}\comma\\
\mathcal{S}_{12}\ket{\psi^{\alpha}_1\phi^{b}_2}&=K_{12}\ket{\psi_2^{\alpha}\phi_1^{b}}
+L_{12}\ket{\phi_2^{b}\psi_1^{\alpha}}\comma
\end{aligned}
\eeq
with
\begingroup \allowdisplaybreaks
\begin{flalign}
A_{12}&=U_1 U_2^{-1}\frac{x_2^{+}-x_1^{-}}{x_{2}^{-}-x_1^{+}}\comma\nonumber\\
B_{12}&=U_1U_2^{-1}\frac{x_2^{+}-x_1^{-}}{x_{2}^{-}-x_1^{+}}\left(1-2\frac{1-1/(x_2^{-}x_1^{+})}{1-1/(x_2^{+}x_1^{+})}\frac{x_2^{-}-x_1^{-}}{x_2^{+}-x_1^{-}} \right)\comma\nonumber\\
C_{12}&=U_1\frac{2\eta_1 \eta_2}{ x_1^{+}x_2^{+}}\frac{1}{1-1/(x_1^{+}x_2^{+})}\frac{x_2^{-}-x_1^{-}}{x_2^{-}-x_1^{+}}\comma\nonumber\\
D_{12}&=-1\comma\nonumber\\
E_{12}&=-\left( 1-2\frac{1-1/(x_2^{+}x_1^{-})}{1-1/(x_2^{-}x_1^{-})}\frac{x_2^{+}-x_1^{+}}{x_2^{-}-x_1^{+}}\right)\comma\\
F_{12}&=-U_2^{-1}\frac{2(x_1^{+}-x_1^{-})(x_2^{+}-x_2^{-})}{\eta_1\eta_2 x_1^{-}x_2^{-}}\frac{1}{1-1/(x_1^{-}x_2^{-})}\frac{x_2^{+}-x_1^{+}}{x_2^{-}-x_1^{+}}\comma\nonumber\\
G_{12}&=U_2^{-1}\frac{x_2^{+}-x_1^{+}}{x_2^{-}-x_1^{+}}\comma\nonumber\\
H_{12}&=\frac{\eta_1}{\eta_2}\frac{x_2^{+}-x_2^{-}}{x_2^{-}-x_1^{+}}\comma\nonumber\\
K_{12}&=U_1 U_2^{-1}\frac{\eta_2}{\eta_1}\frac{x_1^{+}-x_1^{-}}{x_2^{-}-x_1^{+}}\comma\nonumber\\
L_{12}&=U_1\frac{x_2^{-}-x_1^{-}}{x_2^{-}-x_1^{+}}\comma\nonumber
\end{flalign}
\endgroup
where $x^{\pm}_{1,2}$ are the Zhukovsky variables of the first and the second particles. Although not obvious at first sight, one can verify that the S-matrix elements satisfy $H_{12}=K_{12}$ and $C_{12}=F_{12}$.

\section{Reflection Matrix\label{ap:reflection}}
In this appendix, we explicitly write down the action of the reflection matrix in the mirror channel. We then compare it with the bulk S-matrix and show that they are related by the ``unfolding'' relation.
\subsection{Action of the reflection matrix\label{apsubsec:reflectionaction}}
The reflection matrix on the right boundary can be computed from the form factor by using the relation \eqref{eq:relationformreflection}. The result consists of two parts, the overall scalar phase and the matrix part which we denote by $\mathcal{R}$:
\beq
R_L(u)=g_0 (u^{\gamma})\times \hat{\mathcal{R}}(u)\period
\eeq
 Using the bifundamental notation, the action of $\mathcal{R}$ reads
\beq
\begin{aligned}
\hat{\mathcal{R}}|\phi^{a}\phi^{\dot{a}}(u)\rangle&=\frac{\tilde{A}_{\mathcal{G}}+\tilde{B}_{\mathcal{G}}}{2}|\phi^{\{a}\phi^{\dot{a}\}}(\bar{u})\rangle+\frac{-3\tilde{A}_{\mathcal{G}}+\tilde{B}_{\mathcal{G}}}{2}|\phi^{[a}\phi^{\dot{a}]}(\bar{u})\rangle-\tilde{L}_{\mathcal{G}}\epsilon^{a\dot{a}}\epsilon_{\alpha\dot{\alpha}}|\psi^{\alpha}\psi^{\dot{\alpha}}(\bar{u})\rangle\\
\hat{\mathcal{R}}|\psi^{\alpha}\psi^{\dot{\alpha}}(u)\rangle&=\frac{\tilde{D}_{\mathcal{G}}+\tilde{E}_{\mathcal{G}}}{2}|\psi^{\{\alpha}\psi^{\dot{\alpha}\}}(\bar{u})\rangle+\frac{-3\tilde{D}_{\mathcal{G}}+\tilde{E}_{\mathcal{G}}}{2}|\psi^{[\alpha}\psi^{\dot{\alpha}]}(\bar{u})\rangle+\tilde{G}_{\mathcal{G}}\epsilon^{\alpha\dot{\alpha}}\epsilon_{a\dot{a}}|\phi^{a}\phi^{\dot{a}}(\bar{u})\rangle\\
\hat{\mathcal{R}}|\phi^{a}\psi^{\dot{\alpha}}(u)\rangle&=-i\tilde{K}_{\mathcal{G}}|\psi^{\dot{\alpha}}\phi^{a}(\bar{u})\rangle+i\tilde{C}_{\mathcal{G}}|\phi^{a}\psi^{\dot{\alpha}}(\bar{u})\rangle\\
\hat{\mathcal{R}}|\psi^{\alpha}\phi^{\dot{a}}(u)\rangle&=i\tilde{H}_{\mathcal{G}}|\phi^{\dot{a}}\psi^{\alpha}(\bar{u})\rangle+i\tilde{F}_{\mathcal{G}}|\psi^{\alpha}\phi^{\dot{a}}(\bar{u})\rangle\comma
\end{aligned}
\eeq
where $\tilde{A}_{\mathcal{G}}$-$\tilde{H}_{\mathcal{G}}$ are defined simply by $\tilde{\mathcal{A}}_{\mathcal{G}}(u)\equiv \mathcal{A}_{\mathcal{G}}(u^{\gamma})$ (see \eqref{eq:finalwithz}). Written more explicitly they read
with
\beq
\begin{aligned}
&\tilde{A}_{\mathcal{G}}=1\comma\quad
\tilde{B}_{\mathcal{G}}=\frac{1+x^{-}(x^{+})^3}{x^{+}(x^{+}+x^{-})} \comma\quad \tilde{E}_{\mathcal{G}}=-\frac{1+x^{+}(x^{-})^3}{x^{-}(x^{+}+x^{-})}\comma\\
 &\tilde{C}_{\mathcal{G}}=\tilde{F}_{\mathcal{G}}=i\frac{1-(x^{+}x^{-})^2}{2\sqrt{x^{+}x^{-}}x^{+}(x^{+}+x^{-})}\comma\quad \tilde{D}_{\mathcal{G}}=-1\comma\\
&\tilde{G}_{\mathcal{G}}=\tilde{L}_{\mathcal{G}}=\mp i \frac{x^{+}x^{-}-1}{2x^{+} \sqrt{x^{+}x^{-}}}\comma\quad \tilde{H}_{\mathcal{G}}=\tilde{K}_{\mathcal{G}}=\mp i\frac{x^{+}x^{-}+1}{2 x^{+}\sqrt{x^{+}x^{-}}}\period
\end{aligned}
\eeq
Here the upper and lower signs for $\tilde{G}_{\mathcal{G}}$-$\tilde{K}_{\mathcal{G}}$ correspond to $z=+i$ and $z=-i$ respectively.
\subsection{Unfolding interpretation\label{apsubsec:unfolding}}
We now want to ``unfold'' the reflection matrix and relate it to the bulk S-matrix. To understand this, it is useful to analyze how each magnon transforms under the diagonal PSU$(2|2)$ symmetry.

Let us first go back to a form factor description, in which all magnons are in the ``ket'',
\beq
\begin{aligned}
F_{A\dot{A},B\dot{B}}(u)&=\langle \mathcal{G}|\mathcal{X}_{A\dot{A}}(u)\mathcal{X}_{B\dot{B}}(\bar{u})\rangle\\
&=\langle \mathcal{G}|\chi_{A}(u)\dot{\chi}_{\dot{A}}(u)\chi_{B}(\bar{u})\dot{\chi}_{\dot{B}}(\bar{u})\rangle\period
\end{aligned}
\eeq
In the second line, we expressed the excitations using the bifundamental notation, anticipating the expected unfolding structure.

In the next step, we act the diagonal PSU$(2|2)$ generators to this state. In particular, we focus on the action of the fermionic charges
\beq
\mathcal{Q}^{\alpha}{}_{a}=Q^{\alpha}{}_{a}+i\kappa \epsilon^{\alpha\dot{\beta}}\epsilon_{a\dot{b}}\dot{S}^{\dot{b}}{}_{\dot{\beta}}\qquad \mathcal{S}^{a}{}_{\alpha}={S}^{a}{}_{\alpha}+\frac{i}{\kappa}\epsilon^{a\dot{b}}\epsilon_{\alpha\dot{\beta}}\dot{Q}^{\dot{\beta}}{}_{\dot{b}}\period
\eeq
On the undotted part of the first particle $\chi_{A}(u)$, the diagonal PSU$(2|2)$ act in the same way as the standard PSU$(2|2)$ generators. Using the cumulative notation, in which the representation of each fundamental magnon is specified with the rapidity $u$ and the phase $\zeta$, we conclude that $\chi_{A}(u)$ belongs to the following module:
\beq
\chi_{A}(u): \qquad \mathcal{V}(u,\zeta)\period
\eeq

On the other hand, the action on the dotted parts is nonstandard since the labelling of the generators get reshuffled. For instance, the action on $\chi_{\dot{A}}(u)$ reads
\beq\label{eq:actionotherparts}
\begin{aligned}
\chi_{\dot{A}}(u):\qquad &\mathcal{Q}^{\alpha}_{a}\ket{\phi^{b}}=i\kappa{\bf c}   \delta^{b}_{a}\ket{\psi^{\alpha}}\comma\quad &&\mathcal{Q}^{\alpha}_{a}\ket{\psi^{\beta}}=i\kappa {\bf d}  \epsilon^{\alpha\beta}\epsilon_{ab}\ket{\phi^{b}}\comma\\
&\mathcal{S}^{a}_{\alpha}\ket{\phi^{b}}= \frac{i{\bf a}}{\kappa}  \epsilon^{ab}\epsilon_{\alpha\beta}\ket{\psi^{\beta}}\comma\quad &&\mathcal{S}^{a}_{\alpha}\ket{\psi^{\beta}}=\frac{i{\bf b}}{\kappa}  \delta^{\beta}_{\alpha}\ket{\phi^{a}}\comma
\end{aligned}
\eeq
with ${\bf a}\equiv {\bf a}(u,\zeta)$ etc.
 Now, the crucial observation is that reshuffling of the coefficients ${\bf a}$-${\bf d}$ can be ``undone'' if we use the following relation
\beq
{\bf c}=-\frac{{\bf a}(\bar{u}^{2\gamma},\zeta^{\prime})}{\zeta}\comma\qquad  {\bf d}=\frac{{\bf b}(\bar{u}^{2\gamma},\zeta^{\prime})}{\zeta}\comma\qquad {\bf a}=-\zeta{\bf c}(\bar{u}^{2\gamma},\zeta^{\prime})\comma\qquad {\bf b}=\zeta{\bf d}(\bar{u}^{2\gamma},\zeta^{\prime})\comma
\eeq
with
\beq
 \zeta^{\prime}=\zeta e^{-ip(\bar{u}^{2\gamma})}\period
 \eeq
 One can further eliminate the extra factors of $i$, $\kappa$ and $\zeta$ by noticing that the overall phase factor $\zeta$ is related to an insertion of the $\mathcal{Z}$ marker in the twisted notation, and therefore it is natural to set it to be
 \beq
 \zeta =i\kappa z\qquad (z=\pm i)\comma
 \eeq
 in view\fn{We must admit that this part of the argument is heuristic. It however leads to a nice and consistent unfolding description as we see below.} of the rule of pulling out the $\mathcal{Z}$-markers in \eqref{eq:pullingoutZ}. By doing so, we can rewrite \eqref{eq:actionotherparts} as
 \beq
\begin{aligned}
\chi_{\dot{A}}(u):\qquad &\mathcal{Q}^{\alpha}_{a}\ket{\phi^{b}}={\bf a}(\bar{u}^{2\gamma},\zeta^{\prime})   \delta^{b}_{a}\ket{\hat{\psi}^{\alpha}}\comma\quad &&\mathcal{Q}^{\alpha}_{a}\ket{\hat{\psi}^{\beta}}= {\bf b}(\bar{u}^{2\gamma},\zeta^{\prime})  \epsilon^{\alpha\beta}\epsilon_{ab}\ket{\phi^{b}}\comma\\
&\mathcal{S}^{a}_{\alpha}\ket{\phi^{b}}= {\bf c}(\bar{u}^{2\gamma},\zeta^{\prime})  \epsilon^{ab}\epsilon_{\alpha\beta}\ket{\hat{\psi}^{\beta}}\comma\quad &&\mathcal{S}^{a}_{\alpha}\ket{\hat{\psi}^{\beta}}={\bf d}(\bar{u}^{2\gamma},\zeta^{\prime})  \delta^{\beta}_{\alpha}\ket{\phi^{a}}\comma
\end{aligned}
 \eeq
 where we redefined the fermionic basis as
 \beq
 |\hat{\psi}\rangle \equiv z|\psi\rangle\period
 \eeq

 We thus conclude that a single-particle state $\chi_{A}\chi_{\dot{A}}(u)$ transforms under the diagonal PSU$(2|2)$ as follows:
 \beq
 \chi_{A}\chi_{\dot{A}}(u):\qquad \underbrace{\mathcal{V}(\bar{u}^{2\gamma}, \zeta^{\prime})}_{\dot{A}}\otimes \underbrace{\mathcal{V}(u,\zeta^{\prime}e^{ip(\bar{u}^{2\gamma})})}_{A}\period
 \eeq
 Note that we brought the dotted excitation to the left in order to conform with the standard rule for the phase factor $\zeta_{j+1}=\zeta_je^{ip_j}$. This gives an extra sign $(-1)^{|A||\dot{A}|}$ in the final result \eqref{eq:finalunfoldingdetail}.

 Performing a similar analysis to the second particle $\chi_B\chi_{\dot{B}}(\bar{u})$, we arrive at the conclusion that the two-particle state $|\chi_{A}(u)\dot{\chi}_{\dot{A}}(u)\chi_{B}(\bar{u})\dot{\chi}_{\dot{B}}(\bar{u})\rangle$ transforms as
 \beq
 \underbrace{\mathcal{V}(\bar{u}^{2\gamma}, \zeta^{\prime})}_{\dot{A}}\otimes \underbrace{\mathcal{V}(u,\zeta^{\prime}e^{ip(\bar{u}^{2\gamma})})}_{A}\otimes \underbrace{\mathcal{V}(\bar{u}, \zeta^{\prime}e^{ip(\bar{u}^{2\gamma})+ip(u)})}_{B}\otimes \underbrace{\mathcal{V}(u^{-2\gamma},\zeta^{\prime}e^{ip(\bar{u}^{2\gamma})+ip(u)+ip(\bar{u})})}_{\dot{B}}\comma
 \eeq
 where the fermionic basis of the dotted indices are redefined as follows:
 \beq
 \chi_{\dot{A}}:\quad |\hat{\psi}\rangle \equiv z|\psi\rangle\comma\qquad \chi_{\dot{B}}:\quad |\hat{\psi}\rangle \equiv -z|\psi\rangle\period
 \eeq

Having understood how magnons transform under the diagonal PSU$(2|2)$, we can now perform the mirror transformation $u\to u^{\gamma}$ to get the reflection matrix. This brings the first two excitations to the left edge while the last two excitations to the right edge. As a result, we conclude that the reflection matrix can be understood as a map between the following representations of the diagonal PSU$(2|2)$:
\beq
\hat{\mathcal{R}}: \quad  \underbrace{\mathcal{V}(\bar{u}^{\gamma}, \zeta^{\prime})}_{\dot{A}}\otimes \underbrace{\mathcal{V}(u^{\gamma},\zeta^{\prime}e^{ip(\bar{u}^{\gamma})})}_{A}\mapsto \underbrace{\mathcal{V}(u^{\gamma},\zeta^{\prime})}_{\dot{B}}\otimes  \underbrace{\mathcal{V}(\bar{u}^{\gamma},\zeta^{\prime}e^{ip(u^{\gamma})})}_{B}\period
\eeq
This is precisely the same map as the one induced by the PSU$(2|2)$-invariant S-matrix determined by Beisert. Taking into account prefactors which arise from swapping the indices $A$ and $\dot{A}$ and the redefinition of fermionic basis, we expect the following relation to hold:
\beq\label{eq:relationbetweenRandS}
\hat{\mathcal{R}}_{A\dot{A}}^{B\dot{B}}(u)\propto z^{|\dot{A}|+|\dot{B}|}(-1)^{|A||\dot{A}|}\mathcal{S}_{\dot{A}A}^{B\dot{B}}(\bar{u}^{\gamma},u^{\gamma})\period
\eeq

We can then check explicitly if the relation \eqref{eq:relationbetweenRandS} is satisfied or not. For $z=-i$, we found that the relation is indeed satisfied. On the other hand, the relation is not satisfied as it is for $z=i$. However, it turns out that, once we redo the analysis exchanging  the roles of dotted indices and undotted indices, we obtain a slightly modified relation which is satisfied for $z=i$. In summary, we found that the relation between the S-matrix and the reflection matrix is given as follows:
\beq\label{eq:finalunfoldingdetail}
\begin{aligned}
\hat{\mathcal{R}}_{A\dot{A}}^{B\dot{B}}=\frac{(x^{-}+1/x^{-})(1+x^{+}x^{-})}{2 (x^{+}+x^{-})}\times \begin{cases}z^{|\dot{A}|+|\dot{B}|}(-1)^{|B||\dot{B}|}\mathcal{S}_{A\dot{A}}^{\dot{B}B}&\qquad (z=+i)\\z^{|\dot{A}|+|\dot{B}|}(-1)^{|A||\dot{A}|}\mathcal{S}_{\dot{A}A}^{B\dot{B}}&\qquad (z=-i)\end{cases}\period
\end{aligned}
\eeq
We thus arrive at the unfolding representation used in the main text,
\beq
R_{L}= r_0 (\bar{u}) \times \mathcal{S} \comma
\eeq
with
\beq
r_0 (\bar{u})\equiv g_0(u^{\gamma})\times \frac{(x^{-}+1/x^{-})(1+x^{+}x^{-})}{2(x^{+}+x^{-})}\period
\eeq
\section{Crossing Equation\label{ap:crossing}}
In this appendix, we give details of the derivation of the crossing equation and its solution, together with the expansions at weak and strong couplings.
\subsection{Details of the derivation\label{apsubsec:crossingderivation}}
The decoupling condition that we want to solve is
\beq
1=\langle \mathcal{G}| \psi^{1}\dot{\psi}^{1}(u)\quad \psi^2 \dot{\psi}^2(\bar{u})\quad \psi^1\dot{\psi}^1(\bar{u}^{2\gamma})\quad \psi^2 \dot{\psi}^2(u^{-2\gamma})\rangle\period
\eeq
Using the explicit form of the two-particle form factor \eqref{eq:ansatzfullwithg} and \eqref{eq:finalwithz}, one can rewrite the right hand side as
\beq
1=\underbrace{g_0 (u)g_0(\bar{u}^{2\gamma})}_{{\tt phase}}\underbrace{\frac{D_{\mathcal{G}}(u)+E_{\mathcal{G}}(u)}{2}\frac{D_{\mathcal{G}}(\bar{u}^{2\gamma})+E_{\mathcal{G}}(\bar{u}^{2\gamma})}{2}}_{{\tt matrix}}\period
\eeq
Using
\beq
-\frac{D_{\mathcal{G}}(u)+E_{\mathcal{G}}(u)}{2}=\frac{\left(x^{-}+\frac{1}{x^{-}}\right)(x^{+}+x^{-})}{2(1+x^{+}x^{-})}\comma
\eeq
the matrix part can be computed as follows:
\beq
{\tt matrix}=\frac{(x^{+}+\frac{1}{x^{+}})(x^{-}+\frac{1}{x^{-}})}{4 (x^{+}x^{-})^2}\left(\frac{x^{+}+x^{-}}{1+1/x^{+}x^{-}}\right)^2\period
\eeq
On the other hand, the phase factor is given by
\beq
{\tt phase}=g_0 (u)g_0 (\bar{u}^{2\gamma})=\left(\frac{x^{+}}{x^{-}}\right)^2\frac{u+i/2}{u}\frac{u-i/2}{u}\frac{\sigma_B (u)\sigma_B(\bar{u}^{2\gamma})}{\sigma (u,\bar{u})\sigma (\bar{u}^{2\gamma},u^{-2\gamma})}\period
\eeq
The product of the bulk dressing phase can be evaluated using the crossing equation
\beq
\sigma (u_1,u_2)\sigma (u_1^{2\gamma},u_2)=\frac{(1-1/x^{+}_1x^{+}_2)(1-x_1^{-}/x_2^{+})}{(1-1/x^{+}_1x^{-}_2)(1-x_1^{-}/x_2^{-})}\comma
\eeq
as follows
\beq
\begin{aligned}
\sigma (u,\bar{u})\sigma (\bar{u}^{2\gamma},u^{-2\gamma})&=\left(\sigma (u,\bar{u})\sigma (u^{-2\gamma},\bar{u})\right)\times\left(\sigma (\bar{u},u^{-2\gamma})\sigma (\bar{u}^{2\gamma},u^{-2\gamma})\right)\\
&=\frac{(x^{+}+\frac{1}{x^{+}})(x^{-}+\frac{1}{x^{-}})}{4 (x^{-})^4}\left(\frac{x^{+}+x^{-}}{1+1/x^{+}x^{-}}\right)^2\period
\end{aligned}
\eeq

Putting things together and using the Watson's equation $\sigma_B(\bar{u})=\sigma_B(u)$, we obtain
\beq\label{eq:apcrosstosolve}
\sigma_B(u)\sigma_B(u^{2\gamma})=\frac{u}{u+i/2}\frac{u}{u-i/2}\period
\eeq
\subsection{Solving the crossing equation\label{apsubsec:crossingsolution}}
We now solve the equation \eqref{eq:apcrosstosolve}. For this, we make an ansatz for the minimal solution, $\sigma_B^{\rm min}(u)=G (x^{+})/G(x^{-})$. We can then rewrite the crossing equation as
\beq
\left(G(x)G(1/x)\right)^{D-D^{-1}}=u^{2-D-D^{-1}}\comma
\eeq
with $D\equiv e^{\frac{i}{2}\del_u}$. Rewriting this equation, we obtain
\beq
G(x)G(1/x)=u^{F(D)}\comma
\eeq
where $F(D)$ is given formally as
\beq
F(D)=\frac{2-D-D^{-1}}{D-D^{-1}}\period
\eeq
This expression however needs to be interpreted with care. Depending on how we expand $F(D)$ as the Laurent series of $D$, the resulting answer, in particular its analytic properties, will be quite different. The choice which gives the correct analytic properties is to expand $F$ as
\beq
F(D)=-\frac{D}{1+D}+\frac{D^{-1}}{1+D^{-1}}=\sum_{k=1}^{\infty}(-1)^{k}(D^{k}-D^{-k})\period
\eeq
We then get
\beq
\log G(x)G(1/x)=\sum_{k=1}^{\infty}(-1)^{k}\log \left[\frac{u+\frac{ik}{2}}{u-\frac{ik}{2}}\right]\period
\eeq
The right hand side is not convergent as it is. One can make it convergent is by first taking the {\it second derivative}, perform the sum and then integrate twice. As a result, we get
\beq\label{eq:Gcrossing}
G(x)G(1/x)=\mathfrak{G}(u)\comma
\eeq
with
\beq
\mathfrak{G}(u)\equiv \frac{\Gamma (\frac{1}{2}-iu)\Gamma (1+iu)}{\Gamma (\frac{1}{2}+iu)\Gamma (1-iu)}\period
\eeq

Equation \eqref{eq:Gcrossing} can be solved by the standard Wiener-Hopf method and the solution reads\fn{An important property of the kernel in \eqref{eq:tentativeG} is that it is invariant under the simultaneous transformations $x\to1/x$ and $z\to 1/z$ up to an overall sign:
\beq
\left(\frac{1}{x-z}+\frac{1}{2z}\right)dz \mapsto - \left(\frac{1}{x-z}+\frac{1}{2z}\right)dz\period
\eeq
This is necessary to guarantee that \eqref{eq:tentativeG} is a solution to \eqref{eq:Gcrossing}.
}
\beq\label{eq:tentativeG}
\tilde{G}(x)=\exp \oint \frac{dz}{2\pi i}\left(\frac{1}{x-z}+\frac{1}{2z}\right)\log \mathfrak{G}(g(z+1/z))\qquad |x|>1\period
\eeq
In the actual computation, one can drop $1/2z$ term in the kernel since the dressing phase is given by a ratio $G(x^{+})/G(x^{-})$, and the contributions from $1/2z$ cancel in the final answer. Thus, in what follows we define $G(x)$ by
\beq
G(x)=\exp \oint \frac{dz}{2\pi i}\frac{1}{x-z}\log \mathfrak{G}(g(z+1/z))\qquad |x|>1\period
\eeq

From this integral expression, we can readily verify $G(-x)=G(x)^{-1}$. This guarantees that our ansatz satisfies the Watson's equation:
\beq
\sigma^{\rm min}_B(\bar{u})=\frac{G(-x^{-})}{G(-x^{+})}=\frac{G(x^{+})}{G(x^{-})}=\sigma^{\rm min}_B (u)\period
\eeq
\subsection{Weak and strong coupling expansions\label{apsubsec:expansions}}
Let us now expand the result at weak and strong couplings. For this purpose, we expand the logarithm of $G(x)$ as
\beq
\log G(x)=\sum_{n=1}^{\infty}\frac{c_{n}(g)}{x^{n}}\comma
\eeq
with
\beq\label{eq:expanddressing1}
c_{n}(g)=\oint\frac{dz}{2\pi i}z^{n-1}\log \mathfrak{G}(g(z+z^{-1}))\comma
\eeq
and expand $c_n(g)$. Note that this expansion is valid only for $|x|>1$. When $|x|<1$, we need to use the relation \eqref{eq:Gcrossing} to rewrite it as $G(x)=\mathfrak{G}(u)/G(1/x)$ and expand $G(1/x)$.
\paragraph{Weak coupling}
To  compute the weak-coupling expansion of $c_n(g)$, we use the following integral representation of $\log \Gamma(z)$:
\beq
\log \Gamma(z)=\int_{0}^{\infty}dt\frac{e^{-t}}{t}\left(z-1-\frac{1-e^{-t(z-1)}}{1-e^{-t}}\right)\period
\eeq
We then get
\beq\label{eq:expanddressing2}
\log \mathfrak{G}(g(z+z^{-1}))=\int_0^{\infty}dt \frac{e^{igt (z+1/z)}-e^{-igt (z+1/z)}}{t(1+e^{t/2})}\period
\eeq
Now, using the generating function of Bessel functions, we get
\beq
e^{igt (z+z^{-1})}-e^{-igt (z+z^{-1})}=2i\sum_{n=-\infty}^{\infty}(-1)^{n}J_{2n+1}(2gt)z^{-(2n+1)}\period
\eeq
Plugging this expression into \eqref{eq:expanddressing1} and \eqref{eq:expanddressing2}, we arrive at the following integral representation:
\beq
c_{2r+1}(g)=2i(-1)^{r}\int_0^{\infty}dt \frac{J_{2r+1}(2gt)}{t(1+e^{t/2})}\period
\eeq

Expanding the integrand in powers of $g$, we obtain
\beq
c_{2r+1}(g)=\sum_{n=0}^{\infty}c_{2r+1}^{(n)}g^{1+2(n+r)}\comma
\eeq
with
\beq
c_{2r+1}^{(n)}=\frac{4i(-1)^{n+r}(4^{n+r}-1)\Gamma (1+2r+n)}{\Gamma (n+1)\Gamma (2+2r+n)}\zeta (1+2r+2n)\period
\eeq
This gives odd-zeta values except for $n=r=0$, in which case we obtain
\beq
c_1^{(0)}=4i\log 2\period
\eeq

Using these results, one can write down the weak-coupling expansion of the minimal solution $\sigma_B(u)$ as
\beq
\sigma_B^{\rm min} (u)=\frac{2^{4E(u)}}{4}\left[1-g^{4}\frac{12\zeta(3)}{u^2+\frac{1}{4}}+g^6\left(\frac{4(1-12u^3)}{(u^2+\frac{1}{4})^3}\zeta(3)+\frac{120}{u^2+\frac{1}{4}}\zeta (5)\right)+\cdots\right]\period
\eeq
Here we separated the prefactor $2^{4E(u)}/4$, which comes from $c_1^{(0)}$. Combining this with the CDD factor \eqref{eq:ourCDD}, we obtain the result in the main text \eqref{eq:ourdressexpand}.
\paragraph{Strong coupling}
The strong-coupling expansion can be obtained directly from \eqref{eq:expanddressing1} using the asymptotic expansion of the Gamma functions. We then get
\beq\label{eq:cnexpandedstrong}
c_{n}(g)=\frac{i^{n}(-1+(-1)^{n})}{8g}\left[1+\frac{1-n^2}{192g^2}+\cdots\right]\period
\eeq
We can then resum \eqref{eq:cnexpandedstrong} to get
\beq
\log G(x)=-\frac{ix}{4g(1+x^2)}\left[1+\frac{x^2}{24g^2 (1+x^2)^2}+\cdots\right]\qquad |x|>1\period
\eeq
The important point of this result is the absence of the term linear in $g$. Therefore, at the leading strong coupling limit, $G(x)$ can simply be approximated by unity.

To compare with the phase shift of the mirror giant magnon, we also need an expansion of $G(x)$ for $|x|<1$. For this purpose, we use the crossing equation to reexpress $G(x)$ in terms of $G(x)$ and expand. However, we should keep in mind that it is $\tilde{G}$ rather than $G$ which satisfy the crossing equation \eqref{eq:Gcrossing}. Taking into account this subtle point, we obtain the following expansion
\beq
\log G(x)=O(g^{-5})\qquad |x|<1\period
\eeq
We again see that $G(x)$ is unity at the leading order at strong coupling.

\section{$g$-Functions for Nested Bethe Ansatz System\label{ap:gfunctionNested}}
In this Appendix, we derive the $g$-function for theories described by the nested Bethe ansatz generalizing the argument in \cite{Pozsgay:2010tv}. To explain the derivation in a simple setup relevant for the analysis in the main text, we consider the TBA associated with the mirror ABA with fundamental excitations only\fn{For $\mathcal{N}=4$ SYM, this is not a consistent truncation since the S-matrix contains bound-state poles which require us to include the bound states when deriving the TBA. However the discussion below does not rely on the detailed form of the S-matrix and can be applicable to any theories with the same symmetry structure.}.

\subsection{Derivation}
Our starting point is the nested Bethe ansatz equation (nBAE) derived in section \ref{subsec:ABA}, which can be rewritten as follows:

\beq
\begin{aligned}
1=e^{ i\phi^{\zero}_j}=&\,\mathcal{R}(v_{\zero,j})\prod_{k=1}^{M}\mathcal{S}_{\zero \zero}(v_{\zero,j},v_{\zero,k})\prod_{l=1}^{K_{\I}}\mathcal{S}_{\zero \I}(v_{\zero,j},v_{\I,l}),\\
1=e^{ i\phi^{\I}_j}=&\,\prod_{l=1}^{K_{\zero}}\mathcal{S}_{\I\zero}(v_{\I,j},v_{\zero,l})\prod_{m=1}^{K_{\II}}\mathcal{S}_{\I\II}(v_{\I,j},v_{\II,m}),\\
1=e^{ i\phi^{\II}_j}=&\,-\prod_{l=1}^{K_{\I}}\mathcal{S}_{\II\I}(v_{\II,j},v_{\I,l})\prod_{k=1}^{K_{\II}}\mathcal{S}_{\II\II}(v_{\II,j},v_{\II,k}).
\end{aligned}
\eeq
Here we denoted the middle node rapidities by $v_{\zero,j}$ to simplify the notations and $\mathcal{R}$ and $\mathcal{S}$ are given by
\beq
\begin{aligned}
\mathcal{S}_{\zero\zero}(u,v)&\equiv \tilde{S}_0 (u,v)\tilde{S}_0(u,\bar{v})\comma\qquad&&\mathcal{S}_{\zero\I}(u,v)\equiv \tilde{S}^{\zero,\I}(u,v)\tilde{S}^{\zero,\I}(u,\bar{v}) \comma\\
\mathcal{S}_{\I\zero}(u,v)&\equiv\tilde{S}^{\I,\zero}(u,v)\tilde{S}^{\I,\zero}(u,\bar{v})\comma\qquad &&\mathcal{S}_{\I\II}(u,v)=\left(\mathcal{S}_{\II\I}(v,u)\right)^{-1}\equiv\tilde{S}^{\I,\II}(u,v)\comma\\
\mathcal{S}_{\II\II}&\equiv \tilde{S}^{\II,\II}(u,v)\comma\qquad &&\mathcal{R}(u)\equiv \frac{e^{2i\tilde{p}(u)R} r_0 (u)^2}{\mathcal{S}_{\zero\zero}(u,u)}\period
\end{aligned}
\eeq

As usual, we introduce densities of rapidities for both the main roots and the auxiliary roots to analyze the thermodynamic limit. The nBAE in this limit gives the number of available energy levels $\rho_{\rm tot}$,
\beq
\begin{aligned}
\label{eq:nBAEcontinous}
\rho_{\rm tot}^{\zero}(u)&\equiv \sigma(u) +\mathcal{K}_{+}^{\zero\zero}\ast \rho^{\zero}(u)+\mathcal{K}_{+}^{\zero\I}\ast\rho^{\I}(u)\comma\\
\rho_{\rm tot}^{\I}(u)&\equiv \mathcal{K}_{+}^{\I\zero}\ast \rho^{\zero}(u)+K^{\I\II}\star\rho^{\II}(u)\comma\\
\rho_{\rm tot}^{\II}(u)&\equiv K^{\II\I}\star \rho^{\I}(u)+K^{\II\II}\star\rho^{\II}(u)\comma
\end{aligned}
\eeq
where $\star$ is the convolution along the full real axis while $\ast$ is the convolution along the positive real axis, and
\beq
\begin{aligned}
\sigma(u)&\equiv\frac{1}{2\pi iR}\partial_u\log\mathcal{R}(u)-\frac{\delta(u)}{R}\comma\qquad K_{AB}(u,v)\equiv \frac{1}{i}\del_u \log \tilde{S}^{A,B}(u,v)\comma\\
 \mathcal{K}_{+}^{A\zero}(u,v)&\equiv\frac{1}{i}\partial_u\log\mathcal{S}_{A\zero}(u,v)=K^{A\zero}(u,v)+K^{A\zero}(u,-v)\comma\\
 \mathcal{K}_{+}^{\zero A}(u,v)&\equiv\frac{1}{i}\partial_u\log\mathcal{S}_{\zero A}(u,v)=K^{\zero A}(u,v)+K^{\zero A}(u,-v)\period
\end{aligned}
\eeq
\paragraph{Effective action}  Proceeding as before, we write the path integral for the degeneracies for the three sets of rapidities
\begin{align}
\Omega[\rho^{\zero}]=&\,\int\mathcal{D}\eta^{\zero}(u)\,\exp\left[R\int_0^{\infty}du\left\{\rho_{\rm tot}^{\zero}(u)\log(1+e^{i\eta^{\zero}(u)})-i\eta^{\zero}(u)\rho^{\zero}(u)\right\}\right],\\\nonumber
\Omega[\rho^a]=&\,\int\mathcal{D}\eta^a(u)\,\exp\left[R\int_{-\infty}^{\infty}du\left\{\rho_{\rm tot}^a(u)\log(1+e^{i\eta^a(u)})-i\eta^a(u)\rho^a(u)\right\}\right],\quad a=\II,\III.
\end{align}
Note that the integration ranges for the middle-node roots and others are different. We can combine these path integral for the degeneracy and the thermal sum to write down the partition function as a path integral
\begin{align}
\label{eq:pathint}
Z_{ab}=\sum_{\psi}e^{-LE_{\psi}}=\mathcal{N}\times\int\prod_{A=\zero}^{\III}\mathcal{D}\eta^A\mathcal{D}\rho^A\, e^{-RS_{\text{eff}}[\rho^A,\eta^A]}.
\end{align}
where the effective action $S_{\text{eff}}[\eta^A,\rho^A]$ is given by
\begin{align}
\label{eq:Seff}
S_{\text{eff}}[\eta^a,\rho^a]=&\,\int_0^{\infty}du\left[LE(u)\rho^{\zero}(u)+i\eta^{\zero}(u)\rho^{\zero}(u)-\rho_{\rm tot}^{\zero}(u)\log(1+e^{i\eta^{\zero}(u)}) \right]\\\nonumber
&\,+\sum_{a=\I}^{\II}\int_{-\infty}^{\infty}du \left[i\eta^a(u)\rho^a(u)-\rho_{\rm tot}^a(u)\log(1+e^{i\eta^a(u)}) \right]
\end{align}

\paragraph{Saddle point} In the limit $R\gg1$, the path integral (\ref{eq:pathint}) is dominated by the saddle point and the leading quadratic fluctuations. The saddle point equations for $\eta^B$
\begin{align}
\frac{\delta}{\delta\eta^B}S_{\text{eff}}[\eta^A,\rho^A]=0
\end{align}
relate $\eta^A$ with the corresponding $Y$-functions
\begin{align}
\eta^A=-i\log Y_A,\qquad Y_A=\frac{\rho^A}{\rho_{\rm tot}^{A}-\rho^A}.
\end{align}
The saddle point equation for $\rho^A$ then leads to the TBA equations,
\begin{align}
&\log Y_{\zero}=-LE+\log\left(1+{Y_{\zero}}\right)\ast \mathcal{K}_{+}^{\zero \zero}+\log\left(1+{Y_{\I}}\right)\star \mathcal{K}_{+}^{\I\zero},\\\nonumber
&\log Y_{\I}=\log\left(1+{Y_\zero}\right)\ast \mathcal{K}_{+}^{\zero\I}+\log\left(1+{Y_{\II}}\right)\star K^{\II\I},\\\nonumber
&\log Y_{\II}=\log\left(1+{Y_\I}\right)\star K^{\I\II}+\log\left(1+{Y_\II}\right)\star K^{\II\II}.
\end{align}
Dropping the terms that are linear in $\rho^A$, we obtain the saddle point of the partition function
\begin{align}
\left.S_{\text{eff}}[\eta^A,\rho^A]\right|_{\text{saddle}}=-\int_0^{\infty}du\,\sigma(u)\log(1+Y_\zero(u))
\end{align}

\paragraph{Quadratic fluctuation} To obtain the quadratic fluctuation, we compute the Hessian of the effective action as before. It is straightforward to find that
\begin{align}
\texttt{fluctuation}=(\det\mathcal{G}_{\text{eff}})^{-1}
\end{align}
where\footnote{The ``$i$'' factor has the same origin as in the toy model discussed before.}
\begin{align}
i\mathcal{G}_{\text{eff}}=
\pmatrix{ccc}{{\displaystyle \frac{\delta^2 S_{\text{eff}}}{\delta \eta^{\zero} \delta \rho^\zero}}&{\displaystyle \frac{\delta^2 S_{\text{eff}}}{\delta \eta^\zero \delta \rho^\I}}&{\displaystyle{0}}\vspace{10pt}\\{\displaystyle \frac{\delta^2 S_{\text{eff}}}{\delta \eta^\I \delta \rho^\zero}}&{\displaystyle \frac{\delta^2 S_{\text{eff}}}{\delta \eta^\I \delta \rho^\I}}&{\displaystyle \frac{\delta^2 S_{\text{eff}}}{\delta \eta^\I \delta \rho^\II}}\vspace{10pt}\\{\displaystyle{0}}&{\displaystyle \frac{\delta^2 S_{\text{eff}}}{\delta \eta^\II \delta \rho^\I}}&{\displaystyle \frac{\delta^2 S_{\text{eff}}}{\delta \eta^\II \delta \rho^\II}}}
\end{align}
The matrix elements of $\mathcal{G}_{\text{eff}}$ can be computed using (\ref{eq:Seff})
\begin{align}
\left(\mathcal{G}_{\text{eff}}\right)_{A,B}=\frac{1}{i}\frac{\delta^2 S_{\text{eff}}}{\delta\rho^A\delta\eta^B}=\delta_{A,B}-\hat{G}_+^{A,B}
\end{align}
The operator $\hat{G}_+^{A,B}$ is defined as
\beq
\begin{aligned}
\hat{G}_+^{A,\zero}\cdot f(u)&=\int_0^{\infty}\frac{dv}{2\pi}\frac{\mathcal{K}_{+}^{A,\zero}(u,v)}{1+1/Y_{\zero}(v)}f(v)dv\comma\\
\hat{G}_+^{\zero,b}\cdot f(u)&=\int_{-\infty}^{\infty}\frac{dv}{2\pi}\frac{\mathcal{K}_{+}^{\zero,b}(u,v)}{1+1/Y_{b}(v)}f(v)dv\comma\\
\hat{G}_+^{b,c}\cdot f(u)&=\int_{-\infty}^{\infty}\frac{dv}{2\pi}\frac{K^{b,c}(u,v)}{1+1/Y_c(v)}f(v)dv\comma
\end{aligned}
\eeq
with $A=\zero,\I,\II$ and $b,c=\I,\II$. Therefore the fluctuation is again given by a Fredholm determinant
\begin{align}
\texttt{fluctuation}=\left(\det\big[1-\hat{G}_+\big] \right)^{-1}.
\end{align}

\paragraph{Normalization factor}
Finally we compute the normalization factor $\mathcal{N}$ in (\ref{eq:pathint}). This factor basically comes from the Jacobian of the change of variables from mode numbers $\{\phi^A_i\}$ to rapidities $\{v_{B,j}\}$. It is therefore proportional to the Gaudin determinant of the nBAE,
\begin{align}
\mathcal{N}\propto \det\left(\frac{\partial\phi^A_i}{\partial v_{B,j}}\right)\period
\end{align}
We compute $\partial\phi^\zero_i/\partial v_{\zero, j}$ as an example,
\begin{align}
\nn
\frac{\partial\phi^\zero_i}{\partial v_{\zero, j}}=\delta_{ij}\left[2\pi R\sigma(v_{\zero,i})+\sum_{k=1}^{M}\mathcal{K}_{+}^{\zero\zero}(v_{\zero,i},v_{\zero,k})+\sum_{k=1}^{K_{\I}} \mathcal{K}_{+}^{\zero\II}(v_{\zero,i},v_{\I,k})\right]+\frac{1}{i}\frac{\partial}{\partial v_{\zero,j}}\log\mathcal{S}^{\zero \zero}(v_{\zero, i},v_{\zero, j})
\end{align}
The quantity in the bracket on the right hand side coincides with $2\pi R\rho_{\rm tot}^\zero(v_{\zero,j})$ in the thermodynamic limit. On the other hand, the last term can be rewritten as
\begin{align}
\frac{1}{i}\partial_u\log\mathcal{S}^{\zero \zero}(v,u)= \frac{1}{i}\partial_u\left(\underbrace{\log \tilde{S}_0(v,u)}_{=-\log \tilde{S}_0(u,v)}+\underbrace{\log \tilde{S}_0(v,-u)}_{=\log \tilde{S}_0(u,-v)}\right)= -\mathcal{K}^{\zero\zero}_{-}(u,v)\comma
\end{align}
with $\mathcal{K}_{-}^{AB}(u,v)\equiv K^{AB}(u,v)- K^{AB}(u,-v)$. Similarly, we can compute other elements and show that the determinant can be factorized as
\begin{align}
\det\left(\frac{\partial\phi^A_i}{\partial v_{B,j}}\right)=\varrho_{\rm tot}\times\det\left[1-\hat{G}_-\right]
\end{align}
where
\begin{align}
\varrho_{\rm tot}=\prod_{k=1}^{M}2\pi R\rho_{\rm tot}^{\zero}(v_{\zero,k})\times \prod_{k=1}^{K_{\I}}2\pi R\rho_{\rm tot}^{\I}(v_{\I,k})\times \prod_{k=1}^{K_{\II}}2\pi R\rho^{\II}_{\rm tot}(v_{\II,k})\period
\end{align}
As discussed in \cite{Pozsgay:2010tv}, the prefactor $\varrho_{\rm tot}$ can be absorbed into the definition of the measure when writing the partition function in the path integral form in the thermodynamic limit. Alternatively we can apply the argument in Appendix \ref{ap:gfunctionderivation} to derive the path integral and see explicitly that this factor drops out from the integrand.
In the thermodynamic limit, the determinant part $\det\left[ 1-\hat{G}_-\right]$ becomes a Fredholm determinant
whose action is given by
\beq
\begin{aligned}
\hat{G}_-^{A,\zero}\cdot f(u)&=\int_0^{\infty}\frac{dv}{2\pi}\frac{\mathcal{K}_{-}^{A,\zero}(u,v)}{1+1/Y_{\zero}(v)}f(v)\comma\\
\hat{G}_-^{\zero,b}\cdot f(u)&=\int_{-\infty}^{\infty}\frac{dv}{2\pi}\frac{\mathcal{K}_{-}^{\zero,b}(u,v)}{1+1/Y_{b}(v)}f(v)\comma\\
\hat{G}_-^{b,c}\cdot f(u)&=\int_{-\infty}^{\infty}\frac{dv}{2\pi}\frac{K^{b,c}(u,v)}{1+1/Y_c(v)}f(v)\comma
\end{aligned}
\eeq
with $A=\zero,\I,\II$ and $b,c=\I,\II$.

Combing what we obtained and reading off the O(1) piece, we obtain the following expression for the $g$-function,
\beq
g=\exp\left[\int_0^{\infty}\frac{du}{2\pi}\Theta (u) \log (1+Y_{\zero}(u))\right]\sqrt{\frac{\det \left[1-\hat{G}_{-}\right]}{\det \left[1-\hat{G}_{+}\right]}}\comma
\eeq
with
\beq
\Theta (u)=\frac{1}{i}\del_u r_0 (u)-\pi \delta (u)-\frac{1}{2i}\del_u \log \tilde{S}^{\zero\zero} (u,\bar{u})\period
\eeq
\subsection{Rewriting}
We now show that one can rewrite the ratio of Fredholm determinants as in section \ref{subsec:TBAgeneral}. Namely we want to show
\beq\label{eq:nicerewritingfornested}
\frac{\det\left[1-\hat{G}_{-}\right]}{\det\left[1-\hat{G}_{+}\right]}=\frac{\det\left[1-\hat{G}\right]}{\left(\det\left[1-\hat{G}_{+}\right]\right)^2}\period
\eeq
Here $\hat{G}$ is a convolution kernel associated with the TBA of the periodic system which contains both the left and the right wings. Written explicitly, its action is defined by
\beq
\begin{aligned}
\hat{G}^{A,B}\cdot f(u)&=\int_{-\infty}^{\infty}\frac{dv}{2\pi}\frac{K^{A,B}(u,v)}{1+1/Y_{\zero}(v)}f(v)\comma
\end{aligned}
\eeq
where the indices $A$ and $B$ take $\zero,\pm \I , \pm \II$, and the $Y$-functions are extended by
\beq
Y_{-a}(u)\equiv Y_{a}(\bar{u})\comma\quad Y_{\zero}(\bar{u})\equiv Y_{\zero}(u)\period
\eeq

The derivation of \eqref{eq:nicerewritingfornested} closely follows the one in section \ref{subsec:TBAgeneral}. We first decompose the action of $\hat{G}^{A,\zero}$ as
\beq
\begin{aligned}
\hat{G}^{A,\zero}\cdot f_{+}(u)&=\int^{\infty}_{0}\frac{dv}{2\pi}\frac{K^{A,\zero}(u,v)}{1+1/Y_{\zero}(v)}f_{+}(v)+\int^{\infty}_{0}\frac{dv}{2\pi}\frac{K^{A,\zero}(u,-v)}{1+1/Y_{\zero}(v)}f_{-}(v)\comma\\
\hat{G}^{A,\zero}\cdot f_{-}(u)&=\int^{\infty}_{0}\frac{dv}{2\pi}\frac{K^{A,\zero}(-u,v)}{1+1/Y_{\zero}(v)}f_{+}(v)+\int^{\infty}_{0}\frac{dv}{2\pi}\frac{K^{A,\zero}(-u,-v)}{1+1/Y_{\zero}(v)}f_{-}(v)\comma
\end{aligned}
\eeq
where both $u$ and $v$ are positive real and $f_{\pm }(v)\equiv f (\pm v)$ with $v>0$. We can then express the full action of $\hat{G}$ in the following matrix form,
\begin{align}
&\hat{G}\cdot \pmatrix{c}{f_{\II}(u)\\f_{\I}(u)\\f_{+}(u)\\f_{-\II}(u)\\f_{-\I}(u)\\f_{-}(u)}=\\
&\int \frac{dv}{2\pi}\pmatrix{cccccc}{\frac{K^{\II,\II}(u,v)}{1+1/Y_{\II}(v)}&\frac{K^{\II,\I}(u,v)}{1+1/Y_{\II}(v)}&0&0&0&0\\
\frac{K^{\I,\II}(u,v)}{1+1/Y_{\II}(v)}&0&\left[\frac{K^{\I,\zero}(u,v)}{1+1/Y_{\zero}(v)}\right]&0&0&\left[\frac{K^{\I,\zero}(u,-v)}{1+1/Y_{\zero}(v)}\right]\\
0&\frac{K^{\zero,\I}(u,v)}{1+1/Y_{\I}(v)}&\left[\frac{K^{\zero,\zero}(u,v)}{1+1/Y_{\zero}(v)}\right]&0&\frac{K^{\zero,\I}(u,-v)}{1+1/Y_{\I}(v)}&\left[\frac{K^{\zero,\zero}(u,-v)}{1+1/Y_{\zero}(v)}\right]\\
 0&0&0&\frac{K^{\II,\II}(u,v)}{1+1/Y_{\II}(v)}&\frac{K^{\II,\I}(u,v)}{1+1/Y_{\II}(v)}&0\\
0&0&\left[\frac{K^{\I,\zero}(u,-v)}{1+1/Y_{\zero}(v)}\right]&\frac{K^{\I,\II}(u,v)}{1+1/Y_{\II}(v)}&0&\left[\frac{K^{\I,\zero}(u,v)}{1+1/Y_{\zero}(v)}\right]\\
0&\frac{K^{\zero,\I}(u,-v)}{1+1/Y_{\I}(v)}&\left[\frac{K^{\zero,\zero}(u,-v)}{1+1/Y_{\zero}(v)}\right]&0&\frac{K^{\zero,\I}(u,v)}{1+1/Y_{\I}(v)}&\left[\frac{K^{\zero,\zero}(u,v)}{1+1/Y_{\zero}(v)}\right]
}
\pmatrix{c}{f_{\II}(v)\\f_{\I}(v)\\f_{+}(v)\\f_{-\II}(v)\\f_{-\I}(v)\\f_{-}(v)}\comma\nn
\end{align}
where the integration range for the terms inside the brackets are from $0$ to $\infty$ while others are from $-\infty$ to $\infty$.

Now the crucial observation is that the structure of the matrix is the same as the Gaudin norm in the SO(6) spin chain discussed in Appendix \ref{ap:norm}. Namely it can be decomposed into four $3\times 3$ blocks and the two off-diagonal blocks and the two diagonal blocks are identical. This structure will not be modified even if we consider the full operator $1-\hat{G}$ since ``$1$'' only adds a piece proportional to the identity matrix. We then add and subtract rows and columns as in Appendix \ref{ap:norm} without modifying the determinant to get the following matrix structure:
\begin{align}
&\pmatrix{cccccc}{\frac{K^{\II,\II}(u,v)}{1+1/Y_{\II}(v)}&\frac{K^{\II,\I}(u,v)}{1+1/Y_{\II}(v)}&0&0&0&0\\
\frac{K^{\I,\II}(u,v)}{1+1/Y_{\II}(v)}&0&\left[\frac{\mathcal{K}_{+}^{\I,\zero}(u,v)}{1+1/Y_{\zero}(v)}\right]&0&0&0\\
0&\frac{\mathcal{K}_{+}^{\zero,\I}(u,v)}{1+1/Y_{\I}(v)}&\left[\frac{\mathcal{K}_{+}^{\zero,\zero}(u,v)}{1+1/Y_{\zero}(v)}\right]&0&0&0\\
 \ast&\ast&\ast&\frac{K^{\II,\II}(u,v)}{1+1/Y_{\II}(v)}&\frac{K^{\II,\I}(u,v)}{1+1/Y_{\II}(v)}&0\\
\ast&\ast&\ast&\frac{K^{\I,\II}(u,v)}{1+1/Y_{\II}(v)}&0&\left[\frac{\mathcal{K}_{-}^{\I,\zero}(u,v)}{1+1/Y_{\zero}(v)}\right]\\
\ast&\ast&\ast&0&\frac{\mathcal{K}_{-}^{\zero,\I}(u,v)}{1+1/Y_{\I}(v)}&\left[\frac{\mathcal{K}_{-}^{\zero,\zero}(u,v)}{1+1/Y_{\zero}(v)}\right]
}\period
\end{align}
It is then straightforward to see that $1-\hat{G}$ factorizes into $(1-\hat{G}_{+})(1-\hat{G}_{-})$ proving the statement \eqref{eq:nicerewritingfornested}.

\section{Asymptotic limit for Nested Bethe Ansatz System}
\label{ap:asymptotic}
In this appendix, we take the asymptotic limit of the exact $g$-function for models that can be solved by nested Bethe ansatz. We focus on the simplified version of the TBA system discussed in Appendix~\ref{ap:gfunctionNested}, which only contains the fundamental excitations. Nevertheless the argument can be readliy generalized to other systems. In particular, we comment on how it can be applied to the giant graviton OPE coefficient in the SL(2) sector.

As the first step, we consider the ratio of the Fredholm determinants \eqref{eq:nicerewritingfornested} in the exact formula and perform the analytic continuation of the momentum carrying Y-function, $Y_{\zero}(u)$. We then get
\beq
\frac{\det\big[1-\hat{G}_-^{\bullet (A,B)}\big]}{\det\big[1-\hat{G}_+^{\bullet(A,B)}\big]}\comma
\eeq
where $G^{\bullet}$'s are the analytically continued kernels
\begin{align}
\begin{aligned}
\label{eq:analyticContin}
\hat{G}^{\bullet (A,\mathbf{0})}_{\pm}\cdot f(u)=&\,\sum_{k=1}^{M/2}\frac{i \mathcal{K}^{A\mathbf{0}}_{\pm}(u,\tilde{u}_k)}{\partial_u\log Y_\mathbf{0}(\tilde{u}_k)}f(\tilde{u}_k)
+\int_{0}^{\infty}\frac{dv}{2\pi}\frac{\mathcal{K}_{\pm}^{A\mathbf{0}}(u,v)}{1+1/Y_\mathbf{0}(v)}f(v),\\
\hat{G}^{\bullet (\mathbf{0},b)}_{\pm}\cdot f(u)=&\,\int_{-\infty}^{\infty}\frac{dv}{2\pi}\frac{\mathcal{K}^{\mathbf{0}b}_{\pm}(u,v)}{1+1/Y_b(v)}f(v),\\
\hat{G}^{\bullet (a,b)}_{\pm}\cdot f(u)=&\,\hat{G}^{(a,b)}\cdot f(u)=\int\frac{dv}{2\pi}\frac{{K}^{ab}(u,v)}{1+1/Y_b(v)}f(v)\comma
\end{aligned}
\end{align}
with $A=\zero,\I,\II$ and $a,b=\I,\II$.

Our derivation follows closely the one for the toy model in the main text. We first write
\begin{align}
\hat{G}_{\pm}^{\bullet (A,B)}\cdot f(u)=\mathsf{S}_{\pm}^{(A,B)}\cdot f(u)+\mathsf{I}_{\pm}^{(A,B)}\cdot f(u)
\end{align}
where $\mathsf{S}_{\pm}$ and $\mathsf{I}_{\pm}$ denote the convolution kernel for the discrete sum and the continuous integral. From the structure of the TBA kernels (\ref{eq:analyticContin}), we have
\begin{align}\label{eq:propertykerneliab}
\mathsf{S}_{\pm}^{(A,b)}=0,\qquad \mathsf{I}_{+}^{(a,b)}=\mathsf{I}_-^{(a,b)}=\mathsf{I}^{(a,b)}.
\end{align}
We then rewrite the ratio of determinant as
\begin{align}
\label{eq:generalizedFac}
\frac{\det\big[1-\hat{G}^{\bullet(A,B)}_{-}\big]}{\det\big[1-\hat{G}^{\bullet(A,B)}_+\big]}=
\frac{\det\big[1-\mathsf{I}_-^{(A,B)}\big]}{\det\big[1-\mathsf{I}_+^{(A,B)} \big]}
\times \frac{\det\big[1-\widehat{\mathsf{S}}_-^{(A,B)}\big]}{\det\big[1-\widehat{\mathsf{S}}_+^{(A,B)} \big]}
\end{align}
where the dressed summation kernel is given by
\begin{align}
\widehat{\mathsf{S}}_{\pm}^{(A,B)}(u,v)=\sum_C\mathsf{S}_{\pm}^{(A,C)}\ast\big(1+\mathsf{I}_{\pm}+\mathsf{I}_{\pm}^2+\cdots\big)^{(C,B)}(u,v).
\end{align}
The next step is to re-express the second term on the right hand side of (\ref{eq:generalizedFac}) in terms of the exact Gaudin determinants.
The exact quantization condition for the physical parity symmetric rapidities are given by
\begin{align}
\phi(u_j)=2\pi n_j,\qquad n_j\in\mathbb{Z}
\end{align}
where the explicit form of $\phi(u)$ is
\begin{align}
\phi(u)=L p(u)+\frac{1}{i}\sum_{l=1}^{M/2}\log\big[S^{\mathbf{00}}(u,u_l)S^{\mathbf{00}}(u,-u_l)\big]+\frac{1}{i}\sum_A\log(1+Y_A)\ast \mathcal{K}_+^{A\mathbf{0}}(u).
\end{align}
We can first rewrite
\begin{align}
\label{eq:KKanalytic}
\frac{\det\big(1-\widehat{\mathsf{S}}_-^{(A,B)}\big)}{\det\big(1-\widehat{\mathsf{S}}_+^{(A,B)}\big)}=
\frac{\det\left[\partial_u\phi(u_i)\delta_{ij}-\widehat{\mathcal{K}}^{\mathbf{00}}_{+}(u_i,u_j)\right]}
{\det\left[\partial_u\phi(u_i)\delta_{ij}-\widehat{\mathcal{K}}^{\mathbf{00}}_{-}(u_i,u_j)\right]}
\end{align}
where the dressed TBA kernels are given by
\begin{align}
\widehat{{K}}^{AB}_{\pm}(u,v)={K}^{AC}_{\pm}\ast\big(1+\mathsf{I}_\pm+\mathsf{I}_\pm^2+\cdots\big)^{(C,B)}(u,v).
\end{align}
In (\ref{eq:KKanalytic}) we have used the relation
\begin{align}
\widehat{\mathcal{K}}^{\mathbf{00}}_{\pm}(\tilde{u}_j,\tilde{u}_k)=\widehat{\mathcal{K}}^{\mathbf{00}}_{\mp}(u_j,u_k).
\end{align}
We can prove that
\begin{align}
\partial_u\phi(u_i)\delta_{ij}-\widehat{\mathcal{K}}^{\mathbf{00}}_-(u_i,u_j)=\frac{\partial\phi(u_i)}{\partial u_j}
\end{align}
following the same steps as the toy model and using the analytically continued TBA equation
\begin{align}
\label{eq:mirrorTBAgen}
\log Y_A(u)=-L\tilde{E}(u)\delta_{\mathbf{0}A}+\sum_B\log(1+Y_B)\ast \mathcal{K}_+^{BA}(u)
-\sum_{l=1}^{M/2}\log\big[\tilde{S}^{\mathbf{0}A}(\tilde{u}_l,u)\tilde{S}^{\mathbf{0}A}(\tilde{u}_l,-u) \big]
\end{align}
Similarly, we can define the quantity $\tilde{\phi}$
\begin{align}
\tilde{\phi}(u)=L\,p(u)+\frac{1}{i}\sum_{l=1}^M\log S^{\mathbf{00}}(u,u_l)+\frac{1}{i}\sum_A\log(1+Y_A)\ast {K}^{A\mathbf{0}}(u).
\end{align}
Together with the corresponding excited state TBA equation
\begin{align}
\log Y_A(u)=-L\tilde{E}(u)\delta_{\mathbf{0}A}+\sum_B\log(1+Y_B)\ast K^{BA}(u)-\sum_{l=1}^M \log\tilde{S}^{\mathbf{0}\,A}(\tilde{u}_l,u)
\end{align}
we can show that
\begin{align}
\det\left[\partial_u\phi(u_i)\delta_{ij}-\widehat{\mathcal{K}}_+^{\mathbf{00}}(u_i,u_j)\right]
\det\left[\partial_u\phi(u_i)\delta_{ij}-\widehat{\mathcal{K}}_-^{\mathbf{00}}(u_i,u_j)\right]=\det_{i,j}\left(\frac{\partial\tilde{\phi}(u_j)}{\partial u_j} \right)
\end{align}
for the parity symmetric $Y$-functions. Therefore we arrived at the same decomposition
\begin{align}
\label{eq:decomposeNest}
\frac{\det\big[1-\hat{G}^{\bullet(A,B)}_-\big]}{\det\big[1-\hat{G}_+^{\bullet(A,B)}\big]}
=\frac{\det\big[1-\mathsf{I}_-^{(A,B)}\big]}{\det\big[1-\mathsf{I}_+^{(A,B)}\big]}
\times\frac{\det\big[\partial_{u_j}\tilde{\phi}(u_i)\big]}{\big(\det\big[\partial_{u_j}\phi(u_i)\big]\big)^2}
\end{align}
Using this decomposition, it is straightforward to take the asymptotic limit. As in the toy model, the exact quantization condition reduces to the asymptotic Bethe equation in the $L\to \infty$ limit. We thus have
\begin{align}
\lim_{L\to\infty}\frac{\det\big[\partial_{u_j}\tilde{\phi}(u_i)\big]}{\big(\det\big[\partial_{u_j}\phi(u_i)\big]\big)^2}
=\frac{\det\big[\partial_{u_j}\tilde{\phi}^{\text{asym}}(u_i)\big]}{\big(\det\big[\partial_{u_j}\phi^{\text{asym}}(u_i)\big]\big)^2}
=\frac{\det\big(1-G_+^{{(A,B)}}\big)}{\det\big(1-G_-^{{(A,B)}}\big)}
\end{align}
Now, to show that the first ratio in (\ref{eq:decomposeNest}) becomes trivial, we use the fact that in the asymptotic limit $Y_\mathbf{0}$ is exponentially suppressed. Thus the kernels $\mathsf{I}^{(A,\mathbf{0})}$ all vanish in the asymptotic limit and we have
\begin{align}
\begin{aligned}
\tr\big[\left(\mathsf{I}_{\pm}^{(A,B)}\right)^{n}\big]=&\,\tr\big[\mathsf{I}_{\pm}^{(A_1,A_2)}\mathsf{I}_{\pm}^{(A_2,A_3)}\ast\cdots\ast\mathsf{I}_{\pm}^{(A_n,A_1)} \big]\\
\mapsto&\,\tr\big[\mathsf{I}_{\pm}^{(a_1,a_2)}\mathsf{I}_{\pm}^{(a_2,a_3)}\ast\cdots\ast\mathsf{I}_{\pm}^{(a_n,a_1)} \big]=\,\tr\big[\mathsf{I}^{(a_1,a_2)}\mathsf{I}^{(a_2,a_3)}\ast\cdots\ast\mathsf{I}^{(a_n,a_1)} \big].
\end{aligned}
\end{align}
(Here we used \eqref{eq:propertykerneliab} in the second line).
We then get
\begin{align}
\begin{aligned}
\log\left(\frac{\det\big[1-\mathsf{I}^{(A,B)}_-\big]}{\det\big[1-\mathsf{I}^{(A,B)}_+\big]}\right)=&\,\log\left(\det\big[1-\mathsf{I}^{(A,B)}_-\big] \right)-\log\left(\det\big[1-\mathsf{I}^{(A,B)}_+\big] \right)\\
=&\,-\sum_{n=1}^{\infty}\frac{1}{n}\left[(\mathsf{I}_-^{(A,B)})^n-(\mathsf{I}_+^{(A,B)})^n \right]\mapsto 0\comma
\end{aligned}
\end{align}
which shows that the ratio is indeed trivial in the asymptotic limit. Combining the two ratios, we recovers the asymptotic formula in the limit $L\to \infty$,
\begin{align}
\lim_{L\to\infty}\frac{\det\big[1-\hat{G}^{\bullet(A,B)}_-\big]}{\det\big[1-\hat{G}_+^{\bullet(A,B)}\big]}
=\frac{\det\big(1-G_+^{{(A,B)}}\big)}{\det\big(1-G_-^{{(A,B)}}\big)}\period
\end{align}

\paragraph{Giant Graviton OPE coefficient in SL(2) sector}
Let us make brief comments on  the asymptotic limit of the Giant Graviton OPE coefficient in the SL(2) sector. To apply our argument, we denote the $Y$-functions as $Y_{\mathbf{A}}$ and split the indices as $\mathbf{A}=(\mathbf{0},\mathbf{a})$ where $\mathbf{0}$ corresponds to momentum carrying nodes $\mathbf{0}\mapsto(a,0)$ while $\mathbf{a}$  corresponds to all the rest, $\mathbf{a}\mapsto(a,s)$ with $s\neq 0$. 
As compared to the simplified version discussed above, there are two main complications in this case; First there are infinitely many momentum carrying nodes, each of which corresponds to different bound states. Second among those infinitely many momentum carrying nodes, we only perform the analytic continuation of $Y_{(1,0)}$, which corresponds to the $Y$-function for the fundamental excitation \cite{Gromov:2009zb}. This is because we are interested in the operators in the SL(2) sector\fn{For operators outside of the SL(2) sector, the analytic continuation one needs to perform is more complicated and is not well-understood in general \cite{Arutyunov:2009ax}. This is why we focus on the SL(2) sector.}. Nevertheless, it is still true that all the $Y$-functions for the momentum carrying nodes vanish in the asymptotic limit $Y_{\mathbf{0}}\to 0$, and $\mathsf{I}_{+}^{(\mathbf{a},\mathbf{b})}=\mathsf{I}_-^{(\mathbf{a},\mathbf{b})}$ for the other nodes. These two properties are enough to apply our argument and derive the asymptoptic formula \eqref{eq:asymptoticformula}.

\section{Thermodynamic Bethe Ansatz for $\mathcal{N}=4$ SYM\label{ap:TBAfull}}
In this appendix, we display the standard TBA equations for the periodic system, from which one can derive the TBA equation for the Giant Gravitons through the folding procedures discussed in section \ref{sec:TBA}.

By going through the standard procedures, we find that the ``raw'' TBA's are given by
\beq
\begin{aligned}
\log Y_{a,0}=\varphi_{a,0}\equiv &\,-L\tilde{E}_a+\sum_{\blue{\pm}}\Big[\log(1+Y_{b,0})\star K^{\bullet\bullet}_{b,a}
+\log(1+Y_{m,\blue{\pm}1})\star K^{\triangleright\bullet}_{m-1,a}\\
&\,+\log(1+1/Y_{2,\blue{\pm}2})\star K^{y\bullet }_{+a}+\log(1+Y_{1,\blue{\pm}1})\star K^{y\bullet }_{-a}\Big]\,,\\
\log Y_{a,\blue{\pm} 1}=\varphi_{a,\blue{\pm} 1}\equiv &\,\log(1+Y_{b,0})\star K^{\bullet\triangleright}_{b,a-1}
+\log(1+Y_{b,\blue{\pm} 1})\star K^{\triangleright\triangleright}_{b-1,a-1}\\
&\,+\log\frac{1+Y_{1,\blue{\pm} 1}}{1+1/Y_{2,\blue{\pm} 2}}\,\widehat{\star}\, K^{y\triangleright }_{-,a-1},\\
\log Y_{1,\blue{\pm} s}=\varphi_{1,\blue{\pm} s}\equiv &\,\log(1+1/Y_{1,\blue{\pm} t})\star K^{\circ\circ}_{t-1,s-1}+\log\frac{1+Y_{1,\blue{\pm} 1}}{1+1/Y_{2,\blue{\pm} 2}}\,\widehat{\star}\, K^{y\circ }_{-,s-1},\\
\log Y_{2,\blue{\pm} 2}=\varphi_{2,\blue{\pm} 2}\equiv&\,-\log(1+Y_{a,0})\star K^{\bullet y}_{a,+}-\log\frac{1+1/Y_{1,m}}{1+Y_{m,1}}\star K^{\circ y}_{m-1,+}+\pi i,\\
\log Y_{1,\blue{\pm} 1}=\varphi_{1,\blue{\pm} 1}\equiv &\,\log(1+Y_{a,0})\star K^{\bullet y}_{a,-}+\log\frac{1+1/Y_{1,\blue{\pm} m}}{1+Y_{m,\blue{\pm} 1}}\star K^{\circ y}_{m-1,-}+\pi i\comma
\end{aligned}
\eeq
Here $\blue{\pm}$ denote the contributions from the left and the right wings, and the $Y$-functions here are related to the $Y$-functions in the review article \cite{Bajnok:2010ke} in the following way:
\beq
\begin{aligned}
Y_Q^{\bullet}\mapsto Y_{Q,0}\comma\quad &Y_M^{\triangleright}\mapsto Y_{M+1,\pm 1}\comma\quad Y_N^{\circ}\mapsto Y_{1,\pm (N+1)}\comma \quad Y_+^{y}\mapsto Y_{2,\pm 2}\comma\quad Y_-^{y}\mapsto Y_{1,\pm 1}\period
\end{aligned}
\eeq
\section{Eliminating Nested Levels from Fredholm Determinants\label{ap:Eliminating}}
In this appendix, we show that one can (at least formally) eliminate the dependence on the $Y$-functions at the nested level generalizing the idea in section \ref{subsec:eliminating}.

The idea is based on the fact that the Fredholm determinants in \eqref{eq:gfiniteGandGplus} can both be obtained from the functional variations of the TBA equations. This procedure of computing the Fredholm determinants can be regarded as the functional analogue of computing the Gaudin-like determinants from the derivatives of the Bethe equations. We can therefore argue, at least formally, that the Fredholm determinants $\det [1-\hat{G}]$ and $\det [1-\hat{G}_{+}]$ can be factorized as follows:
\beq
\begin{aligned}
\det \left[1-\hat{G}\right]=&\det \left[1-\hat{G}_{\red{\bf v}}\right]\det \left[1-\hat{G}_{\blue{\bf w}}\right]\det \left[1-\hat{\tilde{G}}_{\bf u}\right]\comma\\
\det \left[1-\hat{G}\right]=&\det \left[1-\hat{G}_{\red{\bf v}}\right]\det \left[1-\hat{\tilde{G}}_{+}\right]\period
\end{aligned}
\eeq
Here the kernels $\hat{G}_{\red{\bf v}}$ and $\hat{G}_{\blue{\bf w}}$ are given by
\beq
\begin{aligned}
\left[\hat{G}_{\red{\bf v}}\right]_{(a,s),(b,t)}(u,v)&\equiv \left[\frac{\delta \varphi_{a,s}(u)}{\delta(\log Y_{b,t}(v))}\right]^{T}\qquad \text{with $s,t>0$}\comma\\
 \left[\hat{G}_{\blue{\bf w}}\right]_{(a,s),(b,t)}(u,v)&\equiv \left[\frac{\delta \varphi_{a,s}(u)}{\delta(\log Y_{b,t}(v))}\right]^{T}\qquad \text{with $s,t<0$}\period
\end{aligned}
\eeq
On the other hand $\hat{\tilde{G}}_{\bf u}$ and $\hat{\tilde{G}}_{+}$ are {\it induced Fredholm operators} defined by
\beq
\begin{aligned}
\left[\hat{\tilde{G}}_{{\bf u}}\right]_{a,b}(u,v)&\equiv \left[\frac{\bar{\delta} \varphi_{a,0}(u)}{\bar{\delta}(\log Y_{b,0}(v))}\right]^{T}\comma\\
 \left[\hat{\tilde{G}}_{+}\right]_{a,b}(u,v)&\equiv \left[\frac{\bar{\delta} \varphi_{a,0}(u)}{\bar{\delta}(\log Y_{b,0}(v))}\right]^{T}\comma
\end{aligned}
\eeq
where the variation $\bar{\delta}$ means that we first express all the $Y$-functions at the nested levels ($Y_{a,s}$ with $s\neq 0$) in terms of the middle-node $Y$-functions $Y_{a,0}$ and then perform the variation. In practice, this amounts to the following manipulations: First we take the variations of $\varphi_{a,0}$ and $\varphi_{a,0}^{p}$ assuming that all the $Y$-functions at the nested level are also functionals of $Y_{a,0}$. This produces terms like
\beq
\frac{\delta \log Y_{b,t}}{\delta \log Y_{a,0}}\comma
\eeq
after the variation. We then rewrite these terms in terms of the middle-node $Y$-functions using the TBA equations at the nested levels.

For the parity-symmetric states, it is easy to see that the Fredholm operators for the two wings are identical, $\hat{G}_{\red{\bf v}}=\hat{G}_{\blue{\bf w}}$. Thus, we can express the final result simply in terms of the induced Fredholm operators as
\beq
g=\exp\left[\sum_{a=1}^{\infty}\int_0^{\infty}\frac{du}{2\pi}\Theta_a (u) \log (1+Y_{a,0}(u))\right] \times \frac{\sqrt{\det\left[1-\hat{\tilde{G}}_{\bf u}\right]}}{\det \left[1-\hat{\tilde{G}}_{+}\right]}\period
\eeq
Of course, the procedures explained here are rather formal and not mathematically rigorous. Nevertheless it has an advantage that the integral kernels for the induced Fredholm operators only contain $Y_{a,0}$ and are suppressed in the asymptotic limit. However, it also has a disadvantage that the kernels one gets from this are rather complicated and it is not so obvious how to perform the analytic continuation to get the excited states. It would be interesting to see if one could simplify the kernels for the induced Fredholm operators and put to a more practical use such as the numerical computations.
\section{One-Loop Three-Point Functions in SO(6) Sector\label{ap:1loop3pt}}
In this appendix, we compute the one-loop diagrams relevant for the three-point function in the SO(6) sector. We perform the computation using the method of the partially-contracted Giant Graviton (PCGG).

As explained in section \ref{sec:PCGG}, the basic idea of the PCGG approach is to first perform partial free-field Wick contractions between two determinants and then later consider the contractions with a single-trace operator or with interaction vertices. As a result of the partial contractions, we obtain a sum of non-local multi-trace operators,
\beq\label{eq:appcggreminder}
\begin{aligned}
&\left.\mathcal{D}_1 (x_1)\mathcal{D}_2 (x_2)\right|_{\text{partial contractions}}\overset{N\to \infty}{=}\\&\sum_{\ell=0}^{N}(N-\ell)!\left(\frac{g_{\rm YM}^2 d_{12}}{4\pi^2}\right)^{N-\ell}\left[ (-1)^{\ell}\sum_{\substack{k_1,\ldots,k_{\ell}\\\sum_{s}s k_s=\ell}}\prod_{m=1}^{\ell}\frac{\left(-{\rm tr}\left[(\Phi\bar{\Phi})^m\right]\right)^{k_m}}{m^{k_m}k_m!}\right]\period
\end{aligned}
\eeq
As in section \ref{sec:PCGG}, $\Phi$ and $\bar{\Phi}$ denote fields in the first and the second determinant operators ($\mathcal{D}_1$ and $\mathcal{D}_2$) respectively.
Except for special multi-trace terns that we will discuss later, only the single-trace terms in \eqref{eq:appcggreminder} survive in the large $N$ limit. Since the contractions with the interaction vertices can change the length of the operators, the length of the relevant single-trace terms can be $L$, $L+2$ or $L+4$ (with $L$ being the length of the single-trace operator $\mathcal{O}$).

 \begin{figure}[t]
\centering
\includegraphics[clip,height=11.5cm]{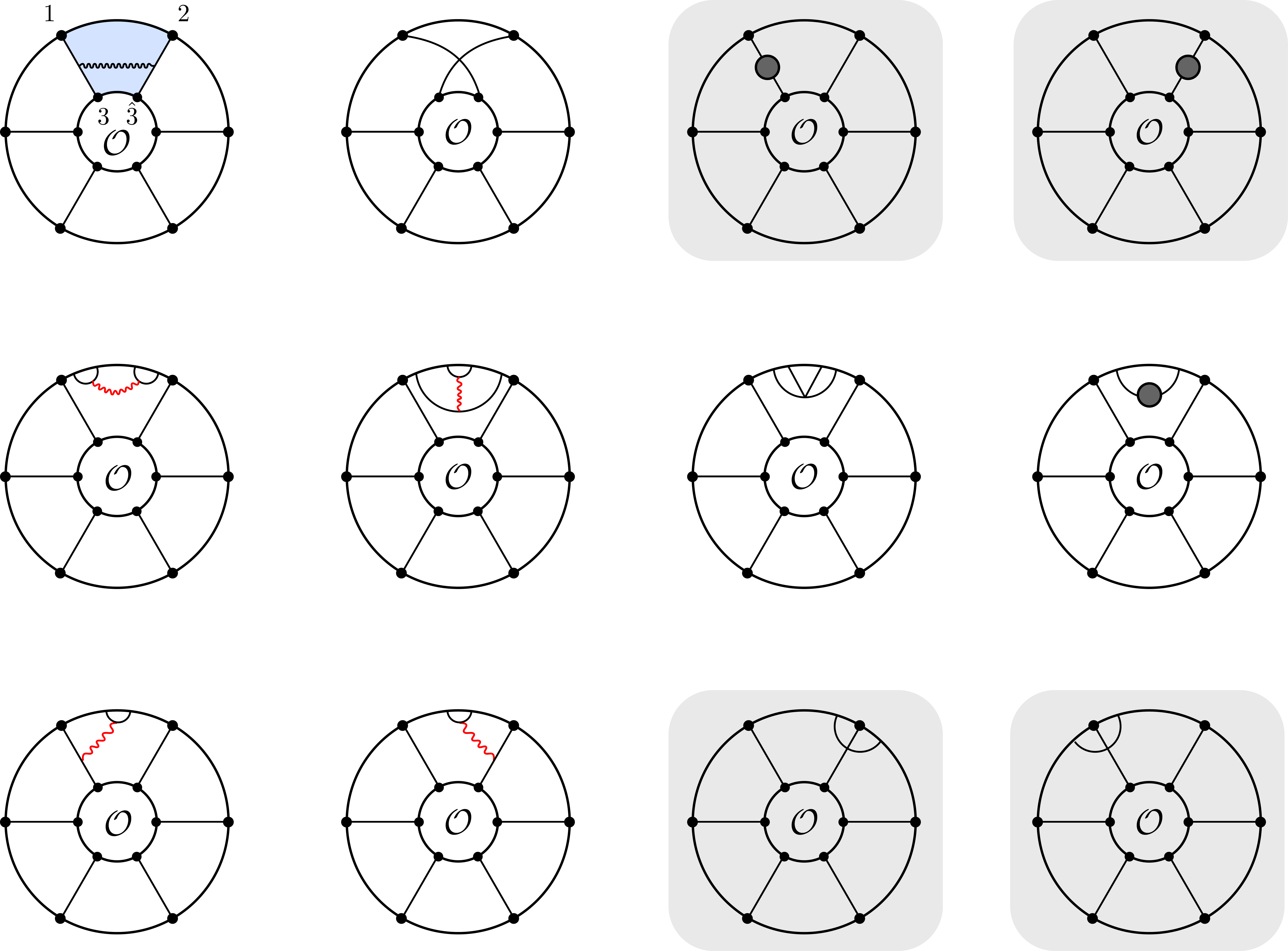}
\caption{One-loop corrections to the three-point function. Here we depicted corrections associated with the face colored in blue in the first diagram. The diagrams drawn on a gray background come with a factor of $1/2$.}
\label{apfig3}
\end{figure}

To efficiently compute the sum of one-loop corrections, we first draw the tree-level contraction between the single-trace operator $\mathcal{O}$ and the PCGG (see figure \ref{apfig3}). As shown in the figure, the propagators split the region between the PCGG and $\mathcal{O}$ into $L$ faces. At one loop, each face gets dressed by the interaction diagram. For instance, the face colored in figure \ref{apfig3} gets dressed by the diagrams listed there. Among the diagrams listed in the figure, the ones drawn on a gray background are associated with edges rather than faces. Since each edge is shared between two adjacent faces, we need to include a factor of $1/2$ when we sum such terms. We also need to keep in mind that. when the length of the PCGG is $L+2$, there will be an additional relative sign\fn{There is also the extra $1/m$ factor in \eqref{eq:appcggreminder}. This factor is cancelled if we sum over all possible physically equivalent contractions which are related to each other by the cyclic permutation of fields inside the trace of the PCGG.} coming from $(-1)^{\ell}$ in \eqref{eq:appcggreminder}.

Taking these points into account, we get the following result as the corrections to the face in figure \ref{apfig3}:
\beq
\begin{aligned}\label{eq:apface3pt}
\frac{{\tt face}|_{\text{single}}}{x_{13}^2x_{23}^2}=&\underbrace{{\sf G}_{13|\hat{3}2}+{\sf Q}_{13\hat{3}2}}_{\text{first line; \eqref{eq:apiden3}}}+\underbrace{\left[2{\sf G}_{12|12}+{\sf Q}_{1212}-{\sf S}_{12}\bar{d}_{12}\right]\frac{\bar{d}_{13}\bar{d}_{2\hat{3}}}{(\bar{d}_{12})^2}}_{\text{second line; \eqref{eq:apiden1}}}+\underbrace{\frac{1}{2}\left[{\sf S}_{13}\bar{d}_{2\hat{3}}+{\sf S}_{2\hat{3}}\bar{d}_{13}\right]}_{\text{third line}}\\
&\underbrace{-\frac{1}{\bar{d}_{12}}\left[{\sf G}_{12|13}\bar{d}_{2\hat{3}}+\frac{1}{2}{\sf Q}_{1213}\bar{d}_{2\hat{3}}+{\sf G}_{21|23}\bar{d}_{13}+\frac{1}{2}{\sf Q}_{2123}\bar{d}_{13}\right]}_{\text{last line; \eqref{eq:apiden2}}}\period
\end{aligned}
\eeq
Here we stripped off the factor proportional to the tree-level propagator $1/(x_{13}^2x_{23}^2)$ from the definition of ${\tt face}|_{\rm single}$.
This can be evaluated explicitly using the identities \eqref{eq:apiden1}-\eqref{eq:apiden3} as indicated below each term in \eqref{eq:apface3pt}.

\begin{figure}[t]
\centering
\includegraphics[clip,height=3cm]{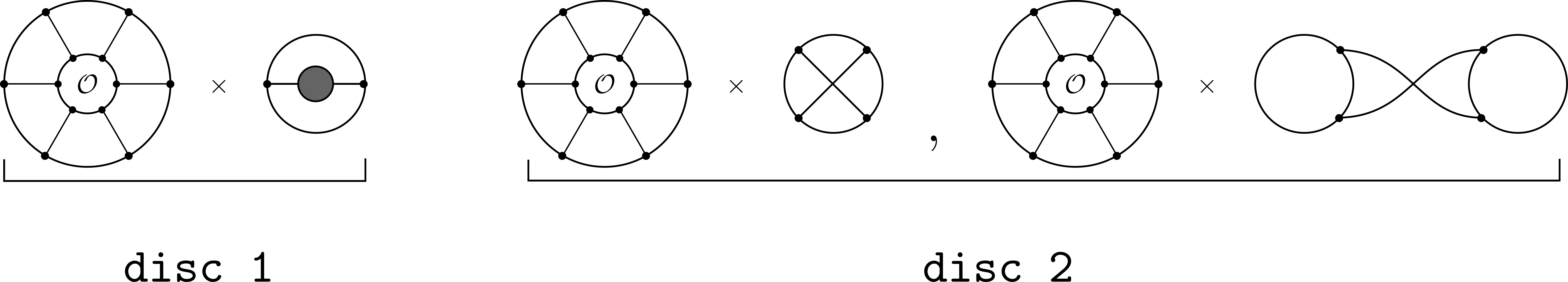}
\caption{The corrections coming from the multi-trace terms.}
\label{apfig4}
\end{figure}

In addition to these contributions, we need to consider the contributions from the multi-trace terms in \eqref{eq:appcggreminder} where the interaction vertices get contracted with extra traces: For instance, we need to consider the contribution in which the self-energy diagram gets contracted with ${\rm tr}[\bar{\Phi}\Phi]$ in the double-trace term (see figure \ref{apfig4})
\beq
\left(N-\frac{L}{2}-1\right)!\times \left(\frac{g_{\rm YM}^2d_{12}}{4\pi^2}\right)^{N-\frac{L}{2}-1}\times (-1)^{\frac{L}{2}+1}\frac{{\rm tr}\left[(\Phi\bar{\Phi})^{L/2}\right]}{L/2}\times {\rm tr}\left[\Phi\bar{\Phi}\right]\period
\eeq
As a result of the contraction, this gives
\beq
\begin{aligned}
{\tt disc1}=&\left\{\left(N-\frac{L}{2}\right)!\times \left(\frac{g_{\rm YM}^2d_{12}}{4\pi^2}\right)^{N-\frac{L}{2}}(-1)^{\frac{L}{2}+1}\left<\frac{{\rm tr}\left[(\Phi\bar{\Phi})^{L/2}\right]}{L/2}\mathcal{O}\right>_0 \right\}\\
&\times \left(N+\frac{L}{2}+O(1/N)\right)\frac{{\sf S}_{12}}{\bar{d}_{12}}\period
\end{aligned}
\eeq
Here the terms on the first line coincides with the tree-level contribution. Similarly, the gluon exchange diagrams and the scalar quartic diagram can get contracted with
\beq
\left(N-\frac{L}{2}-2\right)!\times \left(\frac{g_{\rm YM}^2d_{12}}{4\pi^2}\right)^{N+\frac{L}{2}-2}\times (-1)^{\frac{L}{2}+2}\frac{{\rm tr}\left[(\Phi\bar{\Phi})^{L/2}\right]}{L/2}\times \frac{{\rm tr}\left[(\Phi\bar{\Phi})^2\right]}{2}\comma
\eeq
or
\beq
\left(N-\frac{L}{2}-2\right)!\times \left(\frac{g_{\rm YM}^2d_{12}}{4\pi^2}\right)^{N+\frac{L}{2}-2}\times (-1)^{\frac{L}{2}+1}\frac{{\rm tr}\left[(\Phi\bar{\Phi})^{L/2}\right]}{L/2}\times \frac{\left({\rm tr}\left[\Phi\bar{\Phi}\right]\right)^2}{2}\period
\eeq
After the contraction, this gives
\beq
\begin{aligned}
{\tt disc2}=&\left\{\left(N-\frac{L}{2}\right)!\times \left(\frac{g_{\rm YM}^2d_{12}}{4\pi^2}\right)^{N-\frac{L}{2}}(-1)^{\frac{L}{2}+1}\left<\frac{{\rm tr}\left[(\Phi\bar{\Phi})^{L/2}\right]}{L/2}\mathcal{O}\right>_0 \right\}\\
&\times \left(-\frac{N}{2}-\frac{L}{2}+O(1/N)\right)\left(\frac{2{\sf G}_{12|12}+{\sf Q}_{1212}}{(\bar{d}_{12})^2}\right)\period
\end{aligned}
\eeq
Summing the two contributions ${\tt disc1}$ and ${\tt disc2}$ using \eqref{eq:apiden1}, we find that the leading $O(N)$ terms cancel out while the subleading $O(1)$ term is proportional to $L$ and can be understood as a correction to each face
\beq
\frac{{\tt face}|_{\rm double}}{x_{13}^2x_{23}^2}=-\frac{1}{2}\left[2{\sf G}_{12|12}+{\sf Q}_{1212}-{\sf S}_{12}\bar{d}_{12}\right]\frac{\bar{d}_{13}\bar{d}_{2\hat{3}}}{(\bar{d}_{12})^2}\period
\eeq

Then, after eliminating the term proportional to ${\sf c}_{123}+{\sf c}_{231}+{\sf c}_{312}$ using \eqref{eq:cidentities}, we find that the full result for the face in figure \ref{apfig3} is given simply by
\beq
{\tt face}|_{\rm single}+{\tt face}|_{\rm double}=H_{3\hat{3}}\left(\log \frac{\epsilon x_{12}}{x_{13}x_{23}}-1\right)\period
\eeq
The first term in the bracket correctly reproduces the expected space-time behavior coming from the anomalous dimension
\beq
\sim \delta \Delta \log \frac{x_{12}}{x_{13}x_{23}}\comma
\eeq
while the second term corresponds to the correction to the structure constant. However one has to keep in mind that what we computed is un-normalized three-point function and one needs to subtract off the correction to (the square root of) the normalization in order to read off the scheme-independent structure constant. The correction to the normalization can be read off from the results in the literature \cite{Alday:2005nd} and it boils down to multiplying the following term for each face
\beq
{\tt norm}=-H_{3\hat{3}}\period
\eeq
As a result, we conclude that the scheme-independent structure constant can be computed by inserting
\beq\label{eq:finalHthree}
-H_{3\hat{3}}-\frac{1}{2}{\tt norm}=-\frac{1}{2}H_{3\hat{3}}\comma
\eeq
to each face. When summed up, it reconstructs a $-1/2$ times the full Hamiltonian in the SO(6) sector $\mathbb{H}$. We thus conclude that the correction coming from the one-loop diagram is given by the following substitution
\beq
\langle \text{N\'{e}el}_0| \mapsto \langle \text{N\'{e}el}_0|\mathbb{H}\period
\eeq

\section{Four-Point Functions at Tree Level and One Loop\label{ap:4pttreeloop}}
In this appendix, we compute the connected four-point function of two determinant operators and two length-$p$ BPS single-trace operators
\beq\label{eq:4pt}
\langle \mathcal{D}_1 (x_1,Y_1)\mathcal{D}_2 (x_2,Y_2)\mathcal{O}_{\circ}^{(p)}(x_3,Y_3)\mathcal{O}_{\circ}^{(p)}(x_4,Y_4)\rangle_{\rm connected}\equiv \frac{p}{N}G_{\{p,p\}}
\eeq
 at tree level and one loop. To perform the computation, we use the PCGG approach. For the three-point function in the SO(6) sector discussed in Appendix \ref{ap:1loop3pt}, we only needed special multi-trace terms in the PCGG expansion \eqref{eq:appcggreminder} in the large $N$ limit. For the four-point functions, there is yet another contribution which we need to take into account when $p$ is even: For even $p$, the following double-trace term in the PCGG produces a term of $O(1)$,
 \beq
 \begin{aligned}
& \langle{\rm tr}\left[(\Phi\bar{\Phi})^{p/2}\right]{\rm tr}\left[(\Phi\bar{\Phi})^{p/2}\right]  \mathcal{O}_{\circ}^{(p)}(x_3,Y_3)\mathcal{O}_{\circ}^{(p)}(x_4,Y_4)\rangle =\\
 &\langle{\rm tr}\left[(\Phi\bar{\Phi})^{p/2}\right] \mathcal{O}_{\circ}^{(p)}(x_3,Y_3)\rangle\,\times \,\langle{\rm tr}\left[(\Phi\bar{\Phi})^{p/2}\right] \mathcal{O}_{\circ}^{(p)}(x_4,Y_4)\rangle \sim O(1)\period
 \end{aligned}
 \eeq
 with $\Phi=Y_1\cdot \Phi (x_1)$ and $\bar{\Phi}=Y_2\cdot \Phi (x_2)$.
 At tree level, one can easily compute such contributions and confirms that they only affect the OPE data with twist $\tau\geq 2p$, which is not discussed in the main text. Therefore as long as the tree-level analysis is concerned, we can simply discard such contributions. At one loop, to avoid any possible subtleties coming from such contributions, we first perform the computation assuming $p$ to be odd. We then extend the final result to even $p$ using the lightcone OPE analysis in \cite{Chicherin:2015edu}.
\subsection{Tree level}
\begin{figure}[t]
\centering
\includegraphics[clip,height=2.5cm]{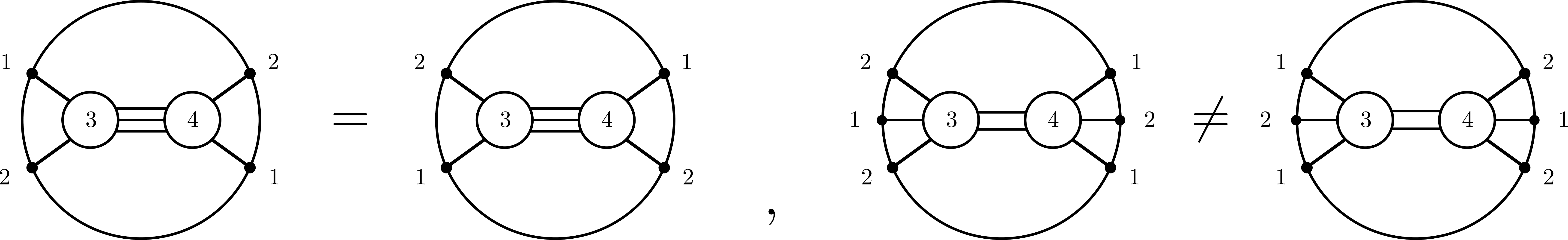}
\caption{For even $\ell$, the diagrams that are related by the permutation of $1$ and $2$ give the same contribution while for odd $\ell$ they are different.}
\label{apfig5}
\end{figure}

Using the PCGG approach, the connected four-point function large $N$ can be computed by evaluating the following sum of correlators
\beq\label{eq:aptocompute4pt}
({\tt 4pt})_{0}=\sum_{\ell=1}^{p}(2g^2d_{12})^{N-\ell}\frac{(-1)^{\ell+1}}{\ell}\left< {\rm tr}\left[\left(\Phi\bar{\Phi}\right)^{\ell}\right] \mathcal{O}_{\circ}^{(p)}(x_3,Y_3) \,\mathcal{O}_{\circ}^{(p)}(x_4,Y_4)\right>_{(0)}\period
\eeq
The subscript $(0)$ reminds that Wick contractions are at tree level. Since each summand is a planar three-point function, one can simply compute them by performing the Wick contractions. One should however keep in mind that there is a small difference between even $\ell$ and odd $\ell$: For even $\ell$, different contractions which are related by the cyclic permutation of fields inside ${\rm tr}\left[\left(\Phi\bar{\Phi}\right)^{\ell}\right]$ give always the same contribution while for odd $\ell$ there are two distinct contributions which arise from the cyclic permutation. See figure \ref{apfig5}.

Taking this into account, we perform the computation and get the following expression for the normalized four-point function
\beq
 \begin{aligned}
 \frac{G^{(0)}_{\{p,p\}}}{(d_{12})^{N}(d_{34})^{p}}=&-\sum_{s=1}^{\lfloor \frac{p+1}{2}\rfloor}\left(-\frac{z\bar{z}}{\alpha\bar{\alpha}}\right)^{2s-1}\left[\left(\frac{(1-\alpha)(1-\bar{\alpha})}{(1-z)(1-\bar{z})}\right)^{s}+\left(\frac{(1-\alpha)(1-\bar{\alpha})}{(1-z)(1-\bar{z})}\right)^{s-1}\right]\\
 &-2\sum_{s=1}^{\lfloor\frac{p}{2}\rfloor}\left(-\frac{z\bar{z}}{\alpha\bar{\alpha}}\right)^{2s}\left(\frac{(1-\alpha)(1-\bar{\alpha})}{(1-z)(1-\bar{z})}\right)^{s}\comma
 \end{aligned}
 \eeq
 where the first line is from odd $\ell$ while the second line is from even $\ell$.
\subsection{One loop\label{apsubsec:4ptPCGG1loop}}
We plug \eqref{eq:PCGGexpansion} with the expansion \eqref{eq:PCGGtree1} into the four-point function. The one loop part \beq
 \begin{aligned}
({\tt 4pt})_{1}=
&
\sum_{\ell=1}^{p}\left(\frac{g_{\rm YM}^2d_{12}}{8\pi^2}\right)^{N-\ell}
\left< \mathcal{G}^{(0)}_{\ell}(x_1,x_2) \mathcal{O}_{\circ}^{(p)}(x_3,Y_3) \,\mathcal{O}_{\circ}^{(p)}(x_4,Y_4)\right>_{(1)}
\\
+&\sum_{\ell=1}^{p}\left(\frac{g_{\rm YM}^2d_{12}}{8\pi^2}\right)^{N-\ell}
\left< \mathcal{G}^{(1)}_{\ell}(x_1,x_2) \mathcal{O}_{\circ}^{(p)}(x_3,Y_3) \,\mathcal{O}_{\circ}^{(p)}(x_4,Y_4)\right>_{(0)}
 \end{aligned}
 \eeq
takes contributions from the tree-level PCGG $\mathcal{G}^{(0)}_{\ell}$, in the correlator evaluated at subleading order (with subscript $1$), and the one-level PCGG $\mathcal{G}^{(1)}_{\ell}$, in the correlator evaluated at leading order ($0$). Keeping the relevant traces in \eqref{eq:PCGGfinal} and \eqref{eq:PCGG1loop}, the correlators read explicitly
 \begin{flalign}
 \label{eq:aptocompute4ptat1loop}
&({\tt 4pt})_{1}=
\sum_{\ell=1}^{p}(2g^2d_{12})^{N-\ell}\frac{(-1)^{\ell+1}}{\ell}\left< {\rm tr}\left[\left(\Phi\bar{\Phi}\right)^{\ell}\right] \mathcal{O}_{\circ}^{(p)}(x_3,Y_3) \,\mathcal{O}_{\circ}^{(p)}(x_4,Y_4)\right>_{(1)}
\nonumber
\\
&
+4g^{2}\left(\log\frac{x_{12}}{\epsilon}+1\right)\sum_{\ell=1}^{p}\left(2g^{2}d_{12}\right)^{N-\ell}\left(-\right)^{\ell}\left\langle \textrm{tr}\left[\left(\Phi\bar{\Phi}\right)^{\ell}\right]\,\mathcal{O}_{\circ}^{(p)}(x_{3},Y_{3})\,\mathcal{O}_{\circ}^{(p)}(x_{4},Y_{4})\right\rangle _{\textrm{(0)}}
\\
&
+\frac{4g^{2}}{N}\left(\log\frac{x_{12}}{\epsilon}+1\right)\left(2g^{2}d_{12}\right)^{N-p}\left\langle \textrm{tr}\left[\Phi(\Phi\bar{\Phi})^{\frac{p-1}{2}}\right]\,\textrm{tr}\left[\bar{\Phi}(\Phi\bar{\Phi})^{\frac{p-1}{2}}\right]\mathcal{O}_{\circ}^{(p)}(x_{3},Y_{3})\,\mathcal{O}_{\circ}^{(p)}(x_{4},Y_{4})\right\rangle _{(0)}
\period\nonumber
 \end{flalign}
In particular, the last line comes from the part in \eqref{eq:PCGG1loop} carrying the tensor $M$ and comment below \eqref{eq:Mmultitrace} becomes relevant: such term survives the planar limit because contractions generate a power of $N$ that overcomes the suppression $1/N$ of the prefactor.

\begin{figure}
\centering
\includegraphics[clip,height=18cm]{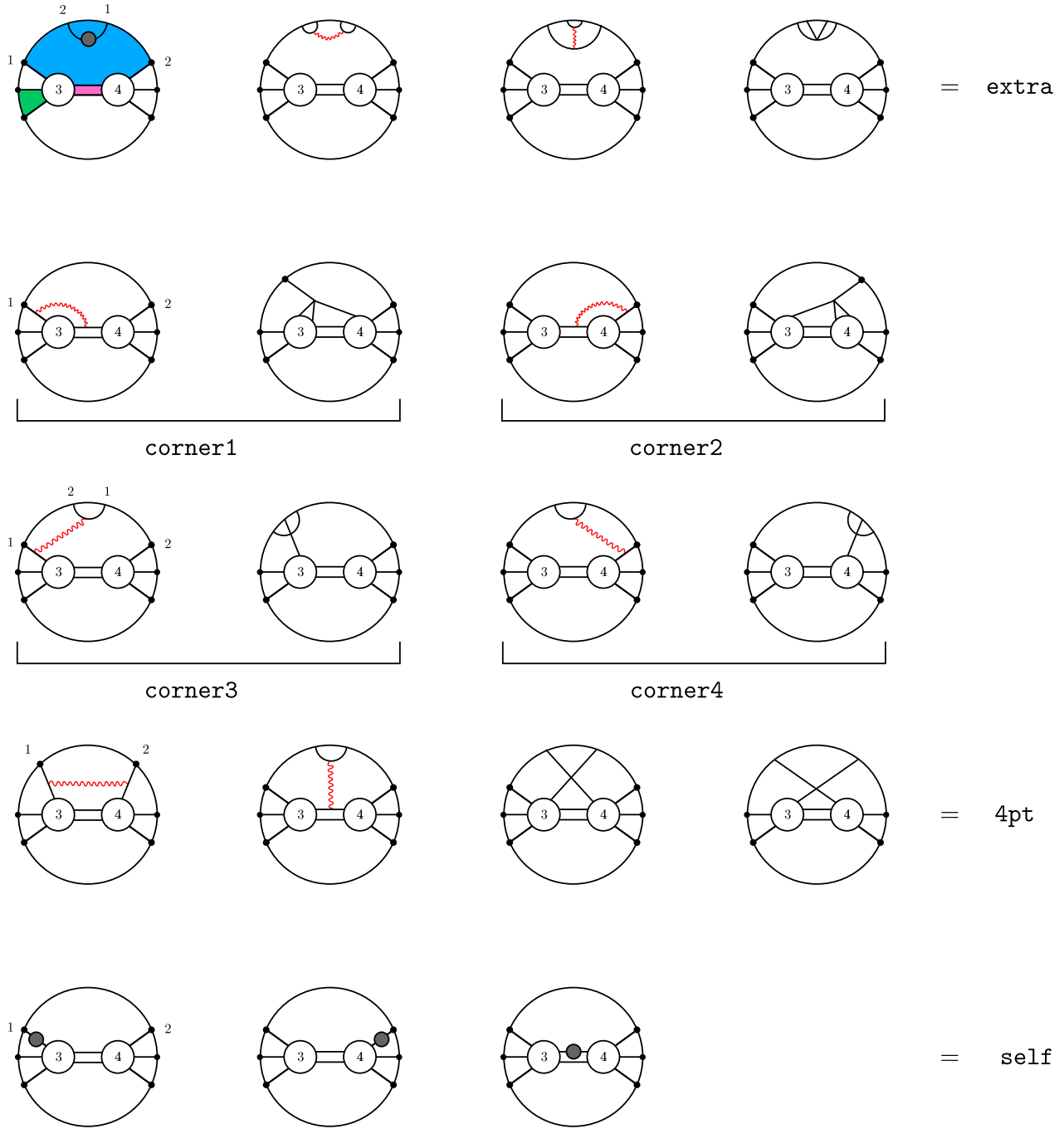}
\caption{One-loop corrections to the four-point function. The diagrams ${\tt extra}$ correspond to the second line in \eqref{eq:aptocompute4ptat1loop} and the others to the first line therein. For the four-point functions, there are three different kinds of faces; the two-point faces, the three-point faces and the four-point faces (colored in different colors in the first diagram). Here we listed corrections associated with the four-point face in the first diagram.}
\label{apfig6}
\end{figure}

The third line in \eqref{eq:aptocompute4ptat1loop} splits into the product of two-point functions at tree level:
 \beq
 \begin{aligned}
{\tt partM}
&=
-\frac{2p^{2}}{Nd_{12}}\left(2g^{2}d_{12}\right)^{N+p+1}\left(\log\frac{x_{12}}{\epsilon}+1\right)\left(\frac{d_{13}d_{14}d_{23}d_{24}}{\left(d_{12}\right)^{4}}\right)^{p/2}
\\
&\times
\left[\left(\frac{d_{13}d_{24}}{d_{23}d_{14}}\right)^{1/2}+\left(\frac{d_{23}d_{14}}{d_{13}d_{24}}\right)^{1/2}\right]\period
 \end{aligned}
 \eeq
 As for the first two lines, the basic strategy of the computation is similar to the one in Appendix \ref{ap:1loop3pt}: We first consider planar tree-level diagrams and dress each face of the diagrams by the interaction vertices. As shown in figure \ref{apfig6}, there are three different kinds of faces for the four-point functions; faces which contain two spatial points (to be called {\it two-point faces}), faces which contain three spatial points ({\it three-point faces}) and faces which contain all the four points ({\it four-point faces}). The corrections to the three-point faces are basically the same as the one computed in Appendix \ref{ap:1loop3pt}, whose result is given by \eqref{eq:finalHthree}. Since here we are considering the BPS operators, which are annihilated by the one-loop Hamiltonian, the corrections of this type simply vanish. The same is true also for the corrections to the two-point faces. We can therefore focus on the corrections to the four-point faces.

\newpage
\newpage

A list of diagrams associated with a single four-point face is given in figure \ref{apfig6}. A point worth mentioning is that, unlike the gluon exchange diagrams, the scalar quartic diagrams do not really correspond to a particular tree-level diagram. To associate them with a definite tree-level diagram, we performed the decomposition of the scalar quartic diagram given in \eqref{eq:Qdecompose}.

As shown in the figure, the diagrams can be grouped into several different contributions. The first one is ${\tt extra}$, which combines the diagrams that give rise to the second line of \eqref{eq:aptocompute4ptat1loop}. This reads\fn{The results below are for the ratios between one-loop diagrams and tree-level diagrams.}
\beq
{\tt extra}= -\frac{{\sf S}_{12}}{\bar{d}_{12}}+  \frac{{\sf Q}_{1212}+  2{\sf G}_{12|12}}{(\bar{d}_{12})^2}=\frac{{\sf S}_{12}}{\bar{d}_{12}}\comma
\eeq
where we used \eqref{eq:apiden1}. All diagrams below contribute to the first line of \eqref{eq:aptocompute4ptat1loop}. The next four contributions are the ``corner contributions'' which read
\beq
\begin{aligned}
{\tt corner 1}&=\frac{{\sf G}_{13|34}+{\sf Q}_{1334}}{\bar{d}_{13}\bar{d}_{34}}={\sf c}_{134}-\frac{1}{4}\left(\frac{{\sf S}_{13}}{\bar{d}_{13}}+\frac{{\sf S}_{34}}{\bar{d}_{34}}\right)\comma\\
{\tt corner 2}&=\frac{{\sf G}_{34|42}+{\sf Q}_{3442}}{\bar{d}_{34}\bar{d}_{42}}={\sf c}_{342}-\frac{1}{4}\left(\frac{{\sf S}_{34}}{\bar{d}_{34}}+\frac{{\sf S}_{24}}{\bar{d}_{24}}\right)\comma\\
{\tt corner 3}&=-\frac{{\sf G}_{12|13}+\frac{1}{2}{\sf Q}_{1213}}{\bar{d}_{12}\bar{d}_{13}}={\sf c}_{213}-\frac{1}{4}\left(\frac{{\sf S}_{12}}{\bar{d}_{12}}+\frac{{\sf S}_{13}}{\bar{d}_{13}}\right)\comma\\
{\tt corner 4}&=-\frac{{\sf G}_{21|24}+\frac{1}{2}{\sf Q}_{2124}}{\bar{d}_{12}\bar{d}_{24}}={\sf c}_{421}-\frac{1}{4}\left(\frac{{\sf S}_{12}}{\bar{d}_{12}}+\frac{{\sf S}_{24}}{\bar{d}_{24}}\right)\comma
\end{aligned}
\eeq
where we used \eqref{eq:apiden4} and \eqref{eq:apiden2}. We also have the ${\tt 4pt}$ contribution which reads
\beq
\begin{aligned}
{\tt 4pt}=&\frac{{\sf G}_{13|42}}{\bar{d}_{13}\bar{d}_{42}}-\frac{{\sf G}_{12|34}}{\bar{d}_{12}\bar{d}_{34}}-\frac{{\sf Q}_{1234}^{-}}{\bar{d}_{12}\bar{d}_{34}}-\frac{{\sf Q}_{2134}^{+}}{\bar{d}_{13}\bar{d}_{24}}\\
=&-({\sf c}_{134}+{\sf c}_{342}+{\sf c}_{213}+{\sf c}_{421})+g^2F^{(1)}(z,\bar{z})\left[z+\bar{z}-1+\frac{1}{2}(\alpha+\bar{\alpha})\left(1-\frac{z\bar{z}}{\alpha\bar{\alpha}}\right)\right]\period
\end{aligned}
\eeq
In addition, there are contributions from the self-energy diagrams and the multi-trace terms as discussed in Appendix \ref{ap:1loop3pt}, which give
\beq
\begin{aligned}
{\tt self}&=\frac{1}{2}\left[\frac{{\sf S}_{13}}{\bar{d}_{13}}+\frac{{\sf S}_{34}}{\bar{d}_{34}}+\frac{{\sf S}_{24}}{\bar{d}_{24}}\right]\comma\\
{\tt disconnected}&=-\frac{1}{2}\left[-\frac{{\sf S}_{12}}{\bar{d}_{12}}+  \frac{{\sf Q}_{1212}+  2{\sf G}_{12|12}}{(\bar{d}_{12})^2}\right]=-\frac{1}{2}\frac{{\sf S}_{12}}{\bar{d}_{12}}\period
\end{aligned}
\eeq

Summing up all these contributions, we find that ${\sf c}_{ijk}$ and ${\sf S}_{ij}$ all cancel, giving the following result for the correction to the face in figure \ref{apfig6}:
\beq
{\tt f}_{z,\alpha}\equiv g^2F^{(1)}(z,\bar{z})\left[z+\bar{z}-1+\frac{1}{2}(\alpha+\bar{\alpha})\left(1-\frac{z\bar{z}}{\alpha\bar{\alpha}}\right)\right]\period
\eeq
Since there are two four-point faces for each tree-level diagram, the result of dressing faces with the interaction vertices is given by
\beq\label{eq:apbulk}
\begin{aligned}
 {\tt bulk}=&-2\sum_{s=1}^{ \frac{p-1}{2}}\left(-\frac{z\bar{z}}{\alpha\bar{\alpha}}\right)^{2s-1}\left[{\tt f}_{\frac{z}{z-1},\frac{\alpha}{\alpha-1}}\left(\frac{(1-\alpha)(1-\bar{\alpha})}{(1-z)(1-\bar{z})}\right)^{s}+{\tt f}_{z,\alpha}\left(\frac{(1-\alpha)(1-\bar{\alpha})}{(1-z)(1-\bar{z})}\right)^{s-1}\right]\\
 &-4\sum_{s=1}^{\frac{p-1}{2}}\left(-\frac{z\bar{z}}{\alpha\bar{\alpha}}\right)^{2s}\left({\tt f}_{\frac{z}{z-1},\frac{\alpha}{\alpha-1}}+{\tt f}_{z,\alpha}\right)\left(\frac{(1-\alpha)(1-\bar{\alpha})}{(1-z)(1-\bar{z})}\right)^{s}
 \end{aligned}
\eeq
Here we assumed that $p$ is odd in order to avoid the subtlety mentioned in the beginning of this appendix. Note also that some of the arguments of ${\tt f}$ get transformed since, depending on the tree-level diagrams that we dress, the roles of $\Phi$ and $\bar{\Phi}$ are swapped.

 \begin{figure}[t]
\centering
\includegraphics[clip,height=11cm]{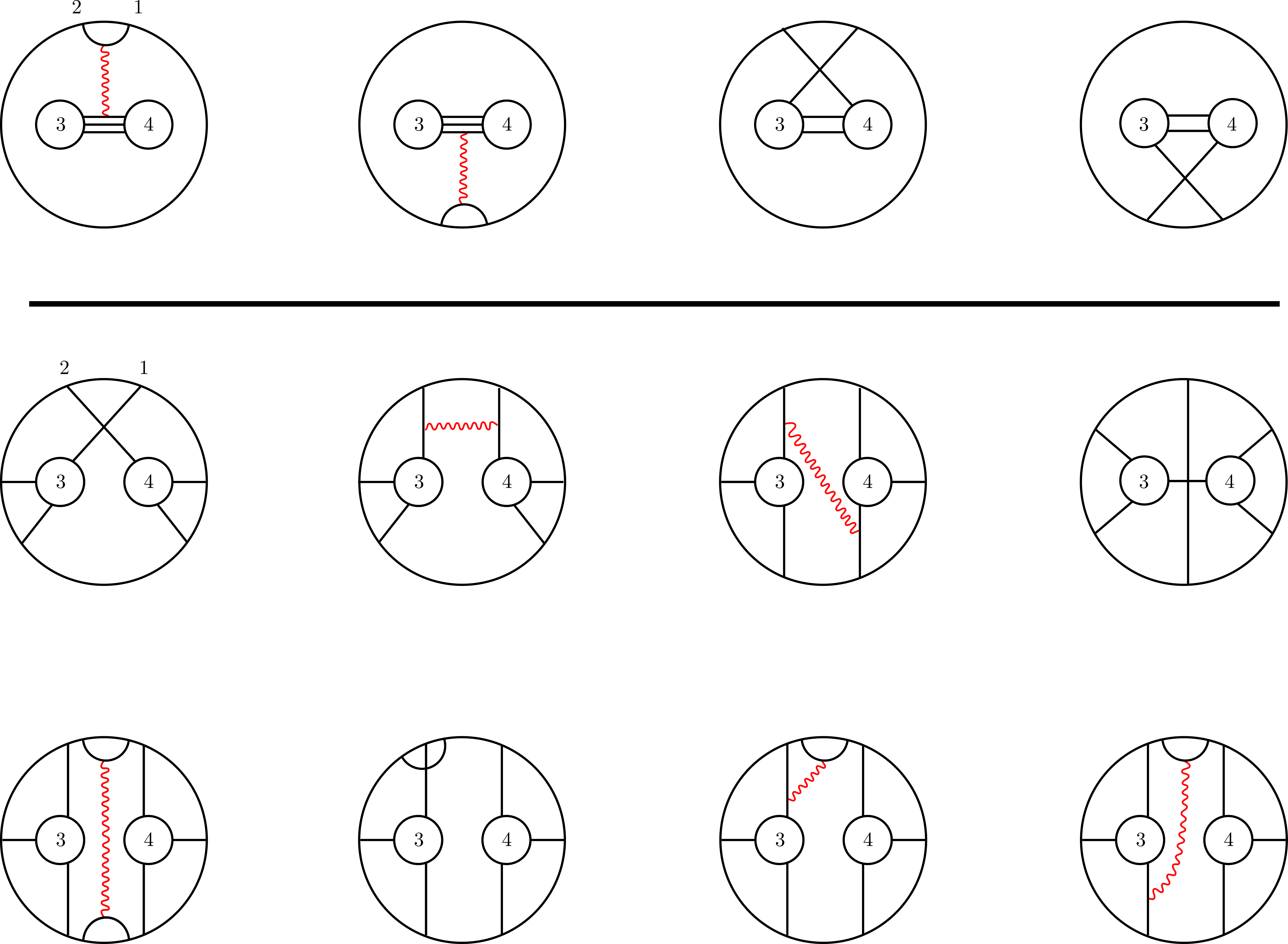}
\caption{The diagrams that contribute for $L=0$ and $2p$. The first four diagrams are for $L=0$ while the remaining ones are for $L=2p$. For $L=2p$, we omitted drawing diagrams with the same topology; we have only drawn one representative for each.}
\label{apfig7}
\end{figure}

The result \eqref{eq:apbulk} only captures when the length of the PCGG is $2\leq L \leq 2p-2$. For $L=0$ and $L=2p$, some of the diagrams in figure \ref{apfig7} will be missing while there will be some new diagrams. See figure \ref{apfig7} for the diagrams that contribute for $L=0$ and $2p$. The result for $L=0$ reads
\beq
\begin{aligned}
\left.{\tt boundary}\right|_{L=0}&=2\frac{{\sf G}_{12|34}+{\sf G}_{21|34}+{\sf Q}^{-}_{1234}+{\sf Q}^{-}_{2134}}{\bar{d}_{12}\bar{d}_{34}}\period
\end{aligned}
\eeq
One can also compute the correction for $L=2p$ ($\left.{\tt boundary}\right|_{L=2p}$) explicitly. However, since it contains various diagrams and the expression is lengthy, we will not present them here\fn{In addition, this contribution does not affect the analysis in the main text since it only produces the OPE data with twist $\tau \geq 2p$.}.

Summing up all the contributions ${\tt bulk}$, ${\tt boundary}|_{L=0}$, ${\tt boundary}|_{L=2p}$ and (after normalization) ${\tt partM}$, we find that the final result drastically simplifies and is given by
\beq
 \begin{aligned}
 \frac{G^{(1)}_{\{p,p\}}}{(d_{12})^{N}(d_{34})^{p}}=&-2g^2\, r_{1234}\, F^{(1)}(z,\bar{z})\sum_{s=0}^{ \frac{p-1}{2}}\left(\frac{(z\bar{z})^2(1-\alpha)(1-\bar{\alpha})}{(\alpha\bar{\alpha})^2(1-z)(1-\bar{z})}\right)^{s}\period
 \end{aligned}
 \eeq
 with
 \beq
  r_{1234}\equiv \frac{z\bar{z}(1-z)(1-\bar{z})(z-\alpha)(z-\bar{\alpha})(\bar{z}-\alpha)(\bar{z}-\bar{\alpha})}{(\alpha\bar{\alpha})^2 (1-z)(1-\bar{z})}\period
 \eeq

 The result so far is valid only for odd $p$. In order to extend this result to even $p$, we impose the lightcone OPE constraint used first in \cite{Chicherin:2015edu}.  The constraints requires the correlator to satisfy the following property:
 \beq
\lim_{\substack{x_{34}^2\to 0\\Y_3\to Y_4 \\d_{12}:{\rm fixed}}}\left[ G^{(1)}_{\{p+1,p+1\}}-d_{12}G^{(1)}_{\{p,p\}}\right]=O(d_{12})\period
 \eeq
 For a derivation, see \cite{Chicherin:2015edu}. Using this result, we can extend the result to even $p$ and obtain the result shown in the main text \eqref{eq:resultatoneloopfourpt}.
\section{Details on Comparison between OPE and Integrability}
\label{ap:OPEvsInt}
\subsection{Bethe Roots from $Q$-system}\label{apsubsec:BetheRoots}
An important ingredient for computing the integrability result is the Bethe roots, or solutions of the Beisert-Staudacher equations up to $O(g^2)$ order. We first find the solutions at tree level and then find loop corrections by perturbation. The tree level Bethe roots can be found by the $Q$-system method of Marboe-Volin \cite{Marboe:2017dmb}, which has been coded up in a \texttt{Mathematica} file. In this appendix, we give some details for the computation of Bethe roots at the leading order using $Q$-system method.\par

\paragraph{Quantum numbers} The functions which finds all the $Q$-functions on a given lattice are $\texttt{DQ}[\texttt{ns}]$ and $\texttt{cDQ}[\texttt{ns}]$ in the \texttt{Mathematica} file. The difference between these two functions is that $\texttt{cDQ}[\texttt{ns}]$ imposes additional constraints on the solutions which selects the ones that satisfy the zero momentum condition. Since we consider solutions corresponding to single trace operators in $\mathcal{N}=4$ SYM, we shall use the function $\texttt{cDQ}[\texttt{ns}]$ to find the solutions. The input of both functions $\texttt{ns}$ is a set of 8 oscillator numbers
\begin{align}
[n_{\mathbf{b}_1},n_{\mathbf{b}_2}|n_{\mathbf{f}_1},n_{\mathbf{f}_2},n_{\mathbf{f}_3},n_{\mathbf{f}_4}|n_{\mathbf{a}_1},n_{\mathbf{a}_2}]
\end{align}
The oscillator numbers are related to the Dynkin labels of $\mathfrak{so}(4)$ (denoted by $[s_1,s_2]$) and $\mathfrak{so}(6)$ (denoted by $[q_1,p,q_2]$) by
\begin{align}
s_1=n_{\mathbf{b}_2}-n_{\mathbf{b}_1},\quad s_2=n_{\mathbf{a}_1}-n_{\mathbf{a}_2},
\quad q_1=n_{\mathbf{f}_1}-n_{\mathbf{f}_2},\quad p=n_{\mathbf{f}_2}-n_{\mathbf{f}_3},\quad q_2=n_{\mathbf{f}_3}-n_{\mathbf{f}_4}
\end{align}
In addition, we have
\begin{align}
\Delta_0=\frac{n_{\mathbf{f}}}{2}+n_{\mathbf{a}},\qquad L=\frac{n_{\mathbf{f}}}{2}+\frac{n_{\mathbf{a}}}{2}-\frac{n_{\mathbf{b}}}{2}
\end{align}
which allows us to fix the oscillator numbers (we need to impose some obvious condition such as $n_{\mathbf{a}}$, $n_{\mathbf{f}}$, $n_{\mathbf{b}}$ are non-negative). From (\ref{eq:quantumNum}), we can find how it is related to the number of Bethe roots.\par

To apply the $Q$-system method to our case, we notice that there is one subtlety related to the grading of the Dynkin diagram. The Beisert-Staudacher equations in (\ref{eq:BSequation}) corresponds to the so-called SL(2) grading or \emph{non-compact ABA} (ncABA) grading according to Marboe-Volin.  However, the default grading for the input of the function $\texttt{cDQ}[\texttt{ns}]$ is the \emph{compact beauty} (cb) grading. Therefore, we  first convert the quantum numbers to the compact beauty grading and then plug in $\texttt{DQ}[\texttt{ns}]$. This can be achieved by a function in that notebook called
\begin{align}
\texttt{nTOn[ns,ncABA,cb]}.
\end{align}

The output of $\texttt{cDQ}[\texttt{ns}]$ is a list of all possible solutions of $Q$-functions $\mathbb{Q}_{a,s}$ with $a,s=0,1,\cdots,4$. To select the $Q$-functions that are related to the solutions of Beisert-Staudacher equations (\ref{eq:BSequation}), we need to read the Dynkin diagram using the non-compact ABA grading. In more detail, the Bethe roots of different nodes are related to the $Q$-functions as
\begin{align}
&\prod_{j=1}^{K_{\III}^{\bf v}}(u-v_{\III,j})=\textcolor{blue}{u^{-L}}\,\mathbb{Q}_{0,1}(u),\quad
\prod_{j=1}^{K_{\II}^{\bf v}}(u-v_{\II,j})=\mathbb{Q}_{1,1}(u),\quad
\prod_{j=1}^{K_{\I}^{\bf v}}(u-v_{\I,j})=\textcolor{blue}{u^L}\,\mathbb{Q}_{2,1}(u),\\\nonumber
&\prod_{j=1}^{K_{\I}^{\bf w}}(u-w_{\I,j})=\mathbb{Q}_{2,3}(u),\quad
\prod_{j=1}^{K^{\bf v}_{\II}}(u-w_{\II,j})=\mathbb{Q}_{3,3}(u),\quad
\prod_{j=1}^{K^{\bf v}_{\III}}(u-w_{\III,j})=\mathbb{Q}_{4,3}(u)
\end{align}
where the factors of $u^{\pm L}$ originate from the asymptotic behaviors of $Q$-functions and the main roots are given by
\begin{align}
\prod_{j=1}^{M}(u-u_j)=\mathbb{Q}_{2,2}(u).
\end{align}
Having the $Q$-functions, it is straightforward to find their zeros numerically. This gives the solutions of Beisert-Staudacher equation at the leading order. We can further impose the selection rules and choose the ones that give non-zero OPE coefficients.
\subsection{Asymptotic OPE Data from Integrability\label{apsubsec:simpleOPEint}}
The integrability formulae for ${\sf d}_{\mathcal{O}}$ and ${\sf c}_{pp\mathcal{O}}$ separately contain the square roots. In this appendix, we show that the product can be rewritten in a form free of square roots.

For this purpose, we first express ${\sf d}_{\mathcal{O}}$ as
\beq
{\sf d}_{\mathcal{O}}=-(i^{J}+(-i)^{J})\sqrt{\left(\prod_{1\leq s\leq \frac{M}{2}}\frac{u_s^2+\frac{1}{4}}{u_s^2}\sigma_B^2(u_s)\right)\frac{\langle {\bf u}|{\bf u}\rangle}{(\det G_{-})^2}}\comma
\eeq
where $\langle {\bf u}|{\bf u}\rangle$ is the full Gaudin norm $\langle {\bf u}|{\bf u}\rangle=\det G_{+}\det G_{-}$. On the other hand, the single-trace structure constant can be expressed as
\beq
{\sf c}_{pp\mathcal{O}}=\langle \red{\bf v}|\red{\bf v}\rangle\sqrt{\frac{\prod_{k=1}^{M}\mu (u_k)\prod_{i<j}h(u_i,u_j)h(u_j,u_i)}{\langle {\bf u}|{\bf u}\rangle}}\tilde{\mathcal{A}}\comma
\eeq
with
\beq
\tilde{\mathcal{A}}=\sum_{\alpha \cup \bar{\alpha}={\bf u}}(-1)^{|\bar{\alpha}|}\prod_{j\in \bar{\alpha}}T(u_j)e^{ip (u_j)L/2}\prod_{u_i\in \alpha,u_j\in \bar{\alpha}}\frac{1}{h(u_i,u_j)}\period
\eeq

Multiplying the two expressions and using the fact that the rapidities are parity symmetric, we can simplify the expression for the product as follows:
\beq
{\sf d}_{\mathcal{O}}{\sf c}_{pp\mathcal{O}}=-(i^{J}+(-i)^{J})\frac{\langle \red{\bf v}|\red{\bf v}\rangle}{\det G_{-}}\left(\prod_{k=1}^{\frac{M}{2}}\tilde{\mu}(u_k)\right)\left(\prod_{1\leq i<j\leq \frac{M}{2}}H(u_i,u_j)\right)\tilde{A}\comma
\eeq
with
\beq
\begin{aligned}
\tilde{\mu}(u)&\equiv \sigma_B(u)\mu(u)\sqrt{h(u,-u)h(-u,u)\frac{u^2+\frac{1}{4}}{u^2}}\\
&=\sigma_B(u)\frac{\left(1+\frac{1}{(x^{+})^2}\right)\left(1+\frac{1}{(x^{-})^2}\right)}{\left(1-\frac{1}{(x^{+})^2}\right)\left(1-\frac{1}{(x^{-})^2}\right)}\left(\frac{1-\frac{1}{x^{+}x^{-}}}{1+\frac{1}{x^{+}x^{-}}}\right)^2\comma
\end{aligned}
\eeq
and
\begin{align}
H(u,v)&\equiv \sqrt{h(u,v)h(v,u)h(\bar{u},\bar{v})h(\bar{v},\bar{u})h(\bar{u},v)h(v,\bar{u})h(u,\bar{v})h(\bar{v},u)}\\
&=\frac{(u^2-v^2)^2}{\left((u-v)^2+1\right)\left((u+v)^2+1\right)}\left[\frac{\left(1+\frac{1}{x^{+}y^{+}}\right)\left(1+\frac{1}{x^{-}y^{-}}\right)}{\left(1-\frac{1}{x^{+}y^{+}}\right)\left(1-\frac{1}{x^{-}y^{-}}\right)}\frac{\left(1-\frac{1}{x^{+}y^{-}}\right)\left(1-\frac{1}{x^{-}y^{+}}\right)}{\left(1+\frac{1}{x^{+}y^{-}}\right)\left(1+\frac{1}{x^{-}y^{+}}\right)}\right]^2\comma\nn
\end{align}
with $x^{\pm}\equiv x^{\pm}(u)$ and $y^{\pm}\equiv x^{\pm}(v)$.
\section{Two-Particle States in SO(6) and SO(4,2) Sectors\label{ap:2particle}}
In this appendix, we directly construct two-particle Bethe wave functions in the SO(6) and SO(4,2) sectors at tree level. The results are used in the main text to test the matrix structure of the form factor.
\subsection{SO(6) sector\label{apsubsec:2particleSO6}}
The Hamiltonian of the SO(6) spin chain is
\beq
\mathbb{H}=2g^2\sum_{n}\frac{1}{2}\mathbb{K}_{n,n+1}+(\mathbb{I}_{n,n+1}-\mathbb{P}_{n,n+1})\comma
\eeq
where the subscripts denote on which site the operators act, and $\mathbb{I}$, $\mathbb{P}$ and $\mathbb{K}$ are defined by
\beq
\begin{aligned}
\mathbb{I}|\ldots, \underset{n}{I},\underset{n+1}{J},\ldots\rangle&=|\ldots, \underset{n}{I},\underset{n+1}{J},\ldots\rangle\comma\qquad
\mathbb{P}|\ldots, \underset{n}{I},\underset{n+1}{J},\ldots\rangle=|\ldots, \underset{n}{J},\underset{n+1}{I},\ldots\rangle\comma\\
\mathbb{K}|\ldots, \underset{n}{I},\underset{n+1}{J},\ldots\rangle&=\delta_{I,J}\sum_{K=1}^{6}|\ldots, \underset{n}{K},\underset{n+1}{K},\ldots\rangle\period
\end{aligned}
\eeq

To construct a two-particle eigenstate, we first make the following plane wave ansatz,
\beq
| \Psi_{c\dot{c}|d\dot{d}} \rangle =\sum_{n}\chi_{c\dot{c}|d\dot{d}} (n)|\ldots , \underset{n}{\bar{Z}},\ldots \rangle +\sum_{n<m}\psi_{c\dot{c}|d\dot{d}}^{a\dot{a}|b\dot{b}}(n,m)|\cdots, \underset{n}{\Phi}^{a\dot{a}},\cdots,\underset{m}{\Phi}^{b\dot{b}},\cdots\rangle\comma
\eeq
with
\beq
\begin{aligned}
\psi^{a\dot{a}|b\dot{b}}_{c\dot{c}|d\dot{d}}(n,m) &= \delta^{a}_{c}\delta^{b}_{d}\delta^{\dot{a}}_{\dot{c}}\delta^{\dot{b}}_{\dot{d}} e^{ip n +iqm}+ S^{ab}_{cd}\dot{S}^{\dot{a}\dot{b}}_{\dot{c}\dot{d}} e^{ipm+iqn}\comma \\
 \chi_{c\dot{c}|d\dot{d}}(n)&=\tilde{\chi}_{c\dot{c}|d\dot{d}}e^{i (p+q)n}\period
 \end{aligned}
\eeq
Here $\ldots$ denote a sea of $Z$ fields and we used the SU(2)$^2$ notation to express magnons:
\beq
X =\Phi^{1\dot{1}}\comma\qquad \bar{X}\to -\Phi^{2\dot{2}}\comma\qquad Y\to \Phi^{1\dot{2}}\comma\qquad \bar{Y}\to \Phi^{2\dot{1}}\period
\eeq
Note that the indices $a\dot{a}$ and $b\dot{b}$ are summed while the indices $c\dot{c}$ and $d\dot{d}$ are labels for different wave functions and not summed.

To determine the unfixed constants $\tilde{\chi}$, $S$ and $\dot{S}$, we impose the eigenvalue equation
\beq\label{eq:apeigenvalue}
\mathbb{H}| \Psi_{cd|\dot{c}\dot{d}} \rangle=E | \Psi_{cd|\dot{c}\dot{d}} \rangle \comma
\eeq
with
\beq
E=4g^2(1-\cos p)+4 g^2(1-\cos q)\period
\eeq
This is a straightforward yet tedious exercise. As a result we get
\beq\label{eq:chiSso6exp}
\begin{aligned}
\tilde{\chi}_{c\dot{c}|d\dot{d}}&=\frac{u-v}{u-v-i}\epsilon_{cd}\epsilon_{\dot{c}\dot{d}}\comma\\
S_{cd}^{ab}\dot{S}_{\dot{c}\dot{d}}^{\dot{a}\dot{b}}&=\frac{1}{(u-v)^2+1}\Big((u-v)\delta^{a}_{c}\delta^{b}_{d}-i \delta^{b}_{c}\delta^{a}_{d}\Big)\left((u-v)\delta^{\dot{a}}_{\dot{c}}\delta^{\dot{b}}_{\dot{d}}-i \delta^{\dot{b}}_{\dot{c}}\delta^{\dot{a}}_{\dot{d}}\right)\comma
\end{aligned}
\eeq
where the rapidities are defined by $e^{ip}=(u+i/2)/(u-i/2)$ and $e^{iq}=(v+i/2)/(v-i/2)$.
\subsection{SO(4,2) sector\label{apsubsec:2particlSO42}}
The Hamiltonian for the full PSU$(2,2|4)$ spin chain at one loop was determined in \cite{Beisert:2003jj} by Beisert. To write it down, we need to express fields in $\mathcal{N}=4$ SYM in terms of oscillators,
\beq
\begin{aligned}
&[a_{\alpha},\bar{a}^{\beta}]=\delta_{\alpha}^{\beta}\comma\qquad
&&[b_{\dot{\alpha}},\bar{b}^{\dot{\beta}}]=\delta_{\dot{\alpha}}^{\dot{\beta}}\comma\\
&\{c_{a},\bar{c}^{b}\}=\delta_{a}^{b}\comma\qquad &&\{d_{\dot{a}},\bar{d}^{\dot{b}}\}=\delta_{\dot{a}}^{\dot{b}}\period
\end{aligned}
\eeq
Different states in the oscillator correspond to different letters in $\mathcal{N}=4$ SYM as follows\fn{Note that the oscillator Hilbert space is actually larger than the space of fundamental fields. To restrict it to the space of fundamental fields, we need to impose that the total numbers of dotted and undotted oscillators are the same.}:
\beq
\begin{aligned}
&|0\rangle \mapsto Z\comma\quad \bar{a}^{\alpha}\bar{b}^{\dot{\beta}}|0\rangle\mapsto D^{\alpha\dot{\beta}}Z\comma\period
\end{aligned}
\eeq

In terms of these oscillators, the Hamiltonian density acting on the two neighboring sites is given by
\beq
\begin{aligned}
H_{12}|A\rangle_{1}\otimes |B\rangle_{2}=\sum_{A^{\prime},B^{\prime}}c_{n_{\rm tot},n_{12},n_{21}}|A^{\prime}\rangle_1 \otimes |B^{\prime}\rangle_{2}\comma
\end{aligned}
\eeq
where $n_{\rm tot}$ is the total number\fn{The total number of oscillators is conserved upon the action of the Hamiltonian.} of oscillators of states $|A\rangle$ and $|B\rangle$, and $n_{12}$ and $n_{21}$ are the number of oscillators hopping from the site $1$ to the site $2$ and vice versa.
$c_{n_{\rm tot},n_{12},n_{21}}$ is given by
\beq
c_{n_{\rm tot},n_{12},n_{21}}=(-1)^{1+n_{12}n_{21}}\frac{\Gamma \left(\frac{1}{2}(n_{12}+n_{21})\right)\Gamma \left(1+\frac{1}{2}(n_{\rm tot}-n_{12}-n_{21})\right)}{\Gamma (1+\frac{1}{2}n_{\rm tot})}\comma
\eeq
if $n_{12}\neq 0$ or $n_{21}\neq 0$. When both $n_{12}$ and $n_{21}$ are zero, we instead have
\beq
c_{n_{\rm tot},0,0}=h(n_{\rm tot}/2)\comma
\eeq
with $h$ being the harmonic number.

Let us now consider the action of this Hamiltonian on the two-particle states in the SO(4,2) sector. The relevant terms for the computation are\fn{These actions are written in the oscillator normalization of states.}
\begingroup \allowdisplaybreaks
\begin{flalign*}
Z\otimes Z\mapsto& 0\comma\\
D^{\alpha\dot{\alpha}}Z\otimes Z\mapsto&  D^{\alpha\dot{\alpha}}Z\otimes Z -Z\otimes D^{\alpha\dot{\alpha}}Z\comma\\
Z\otimes D^{\alpha\dot{\alpha}}Z\mapsto&  Z\otimes D^{\alpha\dot{\alpha}}Z -D^{\alpha\dot{\alpha}}Z\otimes Z\comma\\
D^{\alpha\dot{\alpha}}Z\otimes D^{\beta\dot{\beta}}Z\mapsto& \frac{3}{2}D^{\alpha\dot{\alpha}}Z\otimes D^{\beta\dot{\beta}}Z-\frac{1}{2}\left(D^{\alpha\dot{\alpha}}D^{\beta\dot{\beta}}Z\otimes Z+Z\otimes D^{\alpha\dot{\alpha}}D^{\beta\dot{\beta}}Z\right)\\
&+\frac{1}{2}\left(D^{\alpha\dot{\beta}}Z\otimes D^{\beta\dot{\alpha}}Z+D^{\beta\dot{\alpha}}Z\otimes D^{\alpha\dot{\beta}}Z\right)-\frac{1}{2}D^{\beta\dot{\beta}}Z\otimes D^{\alpha\dot{\alpha}}Z\comma\\
D^{\alpha\dot{\alpha}}D^{\beta\dot{\beta}}Z\otimes Z\mapsto &\frac{3}{2}D^{\alpha\dot{\alpha}}D^{\beta\dot{\beta}}Z\otimes Z-\frac{1}{2}Z\otimes D^{\alpha\dot{\alpha}}D^{\beta\dot{\beta}}Z \\
&-\frac{1}{2}\left(D^{\alpha\dot{\alpha}}Z\otimes D^{\beta\dot{\beta}}Z+D^{\beta\dot{\beta}}Z\otimes D^{\alpha\dot{\alpha}}Z+D^{\alpha\dot{\beta}}Z\otimes D^{\beta\dot{\alpha}}Z+D^{\beta\dot{\alpha}}Z\otimes D^{\alpha\dot{\beta}}Z\right)\comma\\
Z\otimes D^{\alpha\dot{\alpha}}D^{\beta\dot{\beta}}Z\mapsto &\frac{3}{2}Z\otimes D^{\alpha\dot{\alpha}}D^{\beta\dot{\beta}}Z-\frac{1}{2}D^{\alpha\dot{\alpha}}D^{\beta\dot{\beta}}Z\otimes Z \\
&-\frac{1}{2}\left(D^{\alpha\dot{\alpha}}Z\otimes D^{\beta\dot{\beta}}Z+D^{\beta\dot{\beta}}Z\otimes D^{\alpha\dot{\alpha}}Z+D^{\alpha\dot{\beta}}Z\otimes D^{\beta\dot{\alpha}}Z+D^{\beta\dot{\alpha}}Z\otimes D^{\alpha\dot{\beta}}Z\right)
\end{flalign*}
\endgroup
The wave function for the two-particle states can be written in general as
\beq
\begin{aligned}
\ket{\Psi_{\gamma\dot{\gamma}|\delta\dot{\delta}}}=& \sum_n \hat{\chi}^{\alpha\dot{\alpha}|\beta\dot{\beta}}_{\gamma\dot{\gamma}|\delta\dot{\delta}}(n) \ket{\ldots, \underbrace{D^{\alpha\dot{\alpha}}D^{\beta\dot{\beta}}Z}_{n},\ldots}_{\rm osc}\\&+\sum_{n<m}\hat{\psi}_{\gamma\dot{\gamma}| \delta\dot{\delta}}^{\alpha\dot{\alpha}|\beta\dot{\beta}}(n,m)\ket{\ldots,\underbrace{D^{\alpha\dot{\alpha}}Z}_{n},\ldots,\underbrace{D^{\beta\dot{\beta}}Z}_{m},\ldots}_{\rm osc}\period
 \end{aligned}
 \eeq
 Let us make a cautionary remark: The states defined above do not correspond directly to the fields with derivatives owing to the difference of the normalization factors. This is similar to what we encountered in section \ref{sec:weak} for the SL(2) sector. To convert them to the fields with derivatives, we need to use redefined states,
\beq
\begin{aligned}
\ket{\Psi_{\gamma\dot{\gamma}|\delta\dot{\delta}}}=& \sum_n \chi^{\alpha\dot{\alpha}|\beta\dot{\beta}}_{\gamma\dot{\gamma}|\delta\dot{\delta}}(n) \ket{\ldots, \underbrace{D^{\alpha\dot{\alpha}}D^{\beta\dot{\beta}}Z}_{n},\ldots}\\&+\sum_{n<m}\psi_{\gamma\dot{\gamma}| \delta\dot{\delta}}^{\alpha\dot{\alpha}|\beta\dot{\beta}}(n,m)\ket{\ldots,\underbrace{D^{\alpha\dot{\alpha}}Z}_{n},\ldots,\underbrace{D^{\beta\dot{\beta}}Z}_{m},\ldots}\comma
 \end{aligned}
\eeq
with
\beq
\ket{D^{\alpha\dot{\alpha}}D^{\beta\dot{\beta}}Z}=\begin{cases}4\ket{D^{\alpha\dot{\alpha}}D^{\beta\dot{\beta}}Z}_{\rm osc}\quad &\alpha\neq\beta\comma\dot{\alpha}\neq \dot{\beta}\\2\ket{D^{\alpha\dot{\alpha}}D^{\alpha\dot{\beta}}Z}_{\rm osc}\quad &\dot{\alpha}\neq \dot{\beta}\\
2\ket{D^{\alpha\dot{\alpha}}D^{\beta\dot{\alpha}}Z}_{\rm osc}\quad &\alpha\neq\beta\\
\ket{D^{\alpha\dot{\alpha}}D^{\alpha\dot{\alpha}}Z}_{\rm osc}\quad &\end{cases}\period
\eeq
The constants of proportionality can be determined by comparing the two-point functions and the norms of the oscillator states.

The following steps are similar to the ones for the SO(6) sector. We first make a plane-wave ansatz
\beq
\begin{aligned}
\psi^{\alpha\dot{\alpha}|\beta\dot{\beta}}_{\gamma\dot{\gamma}|\delta\dot{\delta}}(n,m) &= \delta^{\alpha}_{\gamma}\delta^{\beta}_{\delta}\delta^{\dot{\alpha}}_{\dot{\gamma}}\delta^{\dot{\beta}}_{\dot{\delta}} e^{ip n +iqm}+ S^{\alpha\beta}_{\gamma\delta}\dot{S}^{\dot{\alpha}\dot{\beta}}_{\dot{\gamma}\dot{\delta}} e^{ipm+iqn}\comma \\
 \chi^{\alpha\dot{\alpha}|\beta\dot{\beta}}_{\gamma\dot{\gamma}|\delta\dot{\delta}}(n)&=\tilde{\chi}^{\alpha\dot{\alpha}|\beta\dot{\beta}}_{\gamma\dot{\gamma}|\delta\dot{\delta}}e^{i (p+q)n}\comma
 \end{aligned}
\eeq
and impose the eigenvalue equation \eqref{eq:apeigenvalue}. As a result we get
\beq
\begin{aligned}
\tilde{\chi}^{\alpha\dot{\alpha}|\beta\dot{\beta}}_{\gamma\dot{\gamma}|\delta\dot{\delta}}&=\frac{(u-v)}{4(u-v-i)}\left(\delta^{\alpha}_{\gamma}\delta^{\beta}_{\delta}+\delta^{\alpha}_{\delta}\delta^{\beta}_{\gamma}\right)\left(\delta^{\dot{\alpha}}_{\dot{\gamma}}\delta^{\dot{\beta}}_{\dot{\delta}}+\delta^{\dot{\alpha}}_{\dot{\delta}}\delta^{\dot{\beta}}_{\dot{\gamma}}\right)\comma
\\
S^{\alpha\beta}_{\gamma\delta}\dot{S}^{\dot{\alpha}\dot{\beta}}_{\dot{\gamma}\dot{\delta}}&=\frac{1}{(u-v)^2+1}\left(i\delta^{\alpha}_{\gamma}\delta^{\beta}_{\delta}+(u-v)\delta^{\alpha}_{\delta}\delta^{\beta}_{\gamma}\right)\left(i\delta^{\dot{\alpha}}_{\dot{\gamma}}\delta^{\dot{\beta}}_{\dot{\delta}}+(u-v)\delta^{\dot{\alpha}}_{\dot{\delta}}\delta^{\dot{\beta}}_{\dot{\gamma}}\right)\period
\end{aligned}
\eeq
\section{Bootstrapping Two-Loop Four-Point Functions\label{ap:2loop4ptbootstrap}}
In this appendix, we provide details on how we determined the two-loop correlator of two determinant operators and two ${\bf 20}^{\prime}$ operators.
\paragraph{Lagrangian insertion and superconformal Ward identity}
To determine the two-loop correlator, we first use the {\it Lagrangian insertion formula} and express $G^{(k)}_{\{2,2\}}$ as an integral of a $(4+k)$-point function \cite{Chicherin:2015edu},
\beq
G^{(k)}_{\{2,2\}}\sim \int d^{4}x_5\cdots  d^{4}x_{4+k}\langle \mathcal{D}_1\mathcal{D}_2 \mathcal{O}_3\mathcal{O}_4 \mathcal{L}(x_5)\cdots \mathcal{L}(x_{4+k})\rangle_{\rm Born}\comma
\eeq
where $\mathcal{O}_{3,4}$ are ${\bf 20}^{\prime}$ operators, $\mathcal{L}$ is the chiral Lagrangian density and the correlator on the right hand side is evaluated at the Born level. The basic strategy of the computation is to determine an {\it integrand} of $G^{(k)}_{\{2,2\}}$ (to be denoted by $\mathcal{G}^{(k)}_{\{2,2\}}$) by imposing various consistency conditions:
\beq
G^{(k)}_{\{2,2\}}=\int d^{4}x_5 \cdots d^{4}x_{4+k} \,\,\mathcal{G}_{\{2,2\}}^{(k)} (x_1,\ldots, x_4;x_5,\ldots, x_{4+k})\period
\eeq

The first constraint comes from the supersymmetry.  Owing to the superconformal Ward identity, the loop correction to the four-point function is proportional to a universal factor
\beq
R_{1234}=\frac{(z-\alpha)(z-\bar{\alpha})(\bar{z}-\alpha)(\bar{z}-\bar{\alpha})}{z\bar{z}(1-z)(1-\bar{z})}d_{13}^2 d_{24}^2x_{13}^2 x_{24}^2\comma
\eeq
What is important in the subsequent analysis is that the factor $R_{1234}$ defined here is {\it purely quadratic}\fn{This can be mostly easily seen by writing $R_{1234}$ as
\beq
\begin{aligned}
R_{1234}=&d_{12}^2 d_{34}^2 x_{12}^2 x_{34}^2 +d_{13}^2 d_{24}^2 x_{13}^2 x_{24}^2+d_{14}^2d_{23}^2x_{14}^2x_{23}^2\\
&+d_{12}d_{23}d_{34}d_{14}(x_{13}^2 x_{24}^2 -x_{12}^2 x_{34}^2 -x_{14}^2 x_{23}^2)\\
&+d_{12}d_{13}d_{24}d_{34}(x_{14}^2 x_{23}^2 -x_{12}^2 x_{34}^2 -x_{13}^2 x_{24}^2)\\
&+d_{13}d_{14}d_{23}d_{24}(x_{12}^2 x_{34}^2 -x_{14}^2 x_{23}^2 -x_{13}^2 x_{24}^2)\period
\end{aligned}
\eeq}
 in the harmonic variables $Y_i$'s.

Let us now express the integrand $\mathcal{G}^{(k)}_{\{2,2\}}$ ($k\geq 1$) as\fn{Note that here the factor $(d_{12})^{N-2}$ is the only possible propagator factor which has the correct harmonic weights and is a polynomial in $Y_k$'s. This is however not true if the lengths of the single-trace are longer than $2$. In such a case, the result is given by a sum of several different structures.}
\beq
\mathcal{G}^{(k)}_{\{2,2\}}= R_{1234} (d_{12})^{N-2}\frac{P^{(k)}}{\displaystyle{\prod_{\substack{1\leq p\leq 4\\5\leq q\leq 4+k}}x_{pq}^2\prod_{5\leq p<q\leq 4+k}x_{pq}^2}}\period
\eeq
By counting the conformal and harmonic weights, one can readily see that $P^{(k)}$ carries harmonic weight $0$ and conformal weight $2(1-k)$ at every point, $x_1,\ldots x_{4+k}$. Since the final result must be a polynomial in $Y_k$'s, this means that $P^{(k)}$ must be independent of the harmonic variables $Y_k$'s. This argument still leaves a possibility that $P^{(k)}$ is a rational function of the distances $x_{ij}^2$. However, by analyzing the singularity structure of the integrand, one can show that $P^{(k)}$ must be a polynomial of $x_{ij}^2$. See sections 3.1 and 3.3.3 in \cite{Chicherin:2015edu}.

This observation drastically simplifies the analysis at low loop orders. In particular, it immediately follows that the one-loop polynomial $P^{(1)}$ must be constant and the computation of the four-point function boils down to computing that constant. To determine the two-loop answer, we impose yet another symmetry constraint on $P^{(2)}$: Namely, the polynomial must be completely symmetric under the permutation of the positions of ${\bf 20}^{\prime}$ and the positions of the Lagrangian insertions $\mathcal{L}$; namely the permutation of $x_3,\ldots, x_6$. Imposing this constraint, we find that there are two allowed polynomials with conformal weight $-2$:
\beq
\begin{aligned}
p_1=&x_{12}^2(x_{34}^2 x_{56}^2+x_{35}^2 x_{46}^2+x_{36}^2 x_{45}^2)\comma\\
p_2=&x_{16}^2x_{25}^2x_{34}^2+x_{15}^2x_{26}^2x_{34}^2+x_{16}^2x_{24}^2x_{35}^2+x_{15}^2x_{24}^2x_{36}^2+x_{14}^2x_{26}^2x_{35}^2+x_{14}^2x_{25}^2x_{36}^2\\
&+x_{16}^2x_{23}^2x_{45}^2+x_{15}^2x_{23}^2x_{46}^2+x_{13}^2x_{26}^2x_{45}^2+x_{13}^2x_{25}^2x_{46}^2+x_{14}^2x_{23}^2x_{56}^2+x_{13}^2x_{24}^2x_{56}^2\period
\end{aligned}
\eeq
Therefore, the computation of the four-point function boils down to the computation of two constants multiplying $p_1$ and $p_2$\fn{
Here we factored out $1/\pi^4$ for convenience.}:
\beq
P^{(2)}=\frac{c_1 p_1+c_2 p_2}{\pi^4}\period
\eeq
Performing the integration, this leads to the following expression for the four-point function:
\beq
\begin{aligned}
G_{\{2,2\}}^{(2)}=&\tilde{R}_{1234}(d_{12})^{N-2}\left[\left(c_1 z\bar{z}+c_2 (1-z)(1-\bar{z})+c_2\right)\left(F^{(1)}(z,\bar{z})\right)^2\right.\\
&\left.+4c_2F_{z}^{(2)}+2(c_1+c_2)F_{1-z}^{(2)}+4c_2F_{\frac{z}{z-1}}^{(2)}\right]\comma
\end{aligned}
\eeq
where $F^{(1)}$ is the one-loop conformal integral \eqref{eq:1loopconformaldef} and $\tilde{R}_{1234}=R_{1234}/(x_{13}^2x_{24}^2)$. We used the shorthand notations for the two-loop conformal integrals,
\beq\label{eq:2loopconformaldef}
F_{z}^{(2)}=F^{(2)}(z,\bar{z}),\quad F_{1-z}^{(2)}=F^{(2)}(1-z,1-\bar{z}),\quad F_{\frac{z}{z-1}}^{(2)}=\frac{F^{(2)}\left(\frac{z}{z-1},\frac{\bar{z}}{\bar{z}-1}\right)}{(1-z)(1-\bar{z})}\comma
\eeq
with
\begin{align}
F^{(2)}(z,\bar{z})&=\frac{x_{13}^2x_{24}^2x_{14}^2}{\pi^4}\int \frac{d^{4}x_5 d^{4}x_{6}}{x_{15}^2x_{25}^2x_{45}^2x_{56}^2 x_{16}^2x_{36}^2 x_{46}^2}\\
&=\frac{1}{z-\bar{z}}\left[\frac{\log (z\bar{z})^2}{2}({\rm Li}_2(z)-{\rm Li}_2 (\bar{z}))-3 \log (z\bar{z})({\rm Li}_3 (z)-{\rm Li}_3 (\bar{z}))+6 ({\rm Li}_4 (z)-{\rm Li}_4 (\bar{z}))\right]\period\nn
\end{align}
\paragraph{OPE constraint in the s-channel}
Now, to determine the constants $c_{1,2}$, we next analyze the four-point function in the s-channel, namely in the $12\to34$ channel. Expanding the resulting four-point function in the OPE limit $0<z,\bar{z}\ll 1$, we obtain the following leading contribution:
\beq
G^{(2)}_{\{2,2\}}=z\bar{z}\left[56c_2 +12 (c_1+c_2)\zeta(3)-32c_2\log z\bar{z}+6 c_2 \left(\log z\bar{z}\right)^2\right]+\cdots\period
\eeq
This can be identified with a contribution from the Konishi operator $\sum_{I}{\rm tr}\left(\Phi^{I}\Phi^{I}\right)$ which has $\left.\Delta\right|_{O(g^0)}=2$ and $S=0$, and is singlet\fn{This identification follows from the fact that 1.~the operator with dimension $\Delta$ and spin $S$ contributes to the OPE as $z^{(\Delta+S)/2}\bar{z}^{(\Delta-S)/2}$ and 2.~the absence of the R-symmetry cross ratios $\alpha$ and $\bar{\alpha}$ indicates that the operator is singlet.} in SO(6). In particular, one should identify the $(\log z\bar{z})^2$ term with the expansion of the anomalous dimension $\delta \Delta$,
\beq
\left.\left({\sf d}_{\mathcal{O}}{\sf c}_{22\mathcal{O}}\right)\right|_{O(g^0)}(z\bar{z})^{\frac{2+\delta \Delta}{2}}=\left.\left({\sf d}_{\mathcal{O}}{\sf c}_{22\mathcal{O}}\right)\right|_{O(g^0)}z\bar{z}\left[1+\frac{\delta \Delta}{2}\log z\bar{z}+\frac{(\delta \Delta)^2}{8}(\log z\bar{z})^2+\cdots\right]\period
\eeq
Comparing the two expressions and
using the anomalous dimension of the Konishi operator and the tree-level structure constant $\left.\left({\sf d}_{\mathcal{O}}{\sf c}_{22\mathcal{O}}\right)\right|_{O(g^0)}$, which can be read off from the lower-loop four-point functions,
\beq
\delta \Delta =12g^2+O(g^4)\comma\qquad \left.\left({\sf d}_{\mathcal{O}}{\sf c}_{22\mathcal{O}}\right)\right|_{O(g^0)}=\frac{1}{3}\comma
\eeq
we conclude that $c_2$ is given by
\beq\label{eq:c2=1}
c_2 =1\period
\eeq
\paragraph{OPE constraint in the t-channel} To determine the remaining constant $c_1$, we need to consider the OPE expansion in the t-channel, or equivalently $14\to 23$ channel. This amounts to expanding the four-point function around $z=\bar{z}=1$. Performing the expansion for the one-loop and two-loop four-point functions, we get
\beq\label{eq:expandedinwrong}
\begin{aligned}
G_{\{2,2\}}^{(1)}=&\frac{(1-\alpha)^2(1-\bar{\alpha})^2}{(1-z)(1-\bar{z})\alpha^2\bar{\alpha}^2}\left(2\log \left[(1-z)(1-\bar{z})\right]-4\right)+\cdots\comma\\
G_{\{2,2\}}^{(2)}=&\frac{(1-\alpha)^2(1-\bar{\alpha})^2}{(1-z)(1-\bar{z})\alpha^2\bar{\alpha}^2}\left(-2(c_1+2c_2)\Big(\log \left[(1-z)(1-\bar{z})\right]\Big)^2\right.\\
&\left.-2(5c_1+11c_2)\log \left[(1-z)(1-\bar{z})\right]+\cdots\right)+\cdots\period
\end{aligned}
\eeq
In this channel, the operator with dimension $\Delta =N+2-\alpha$ comes with a factor $1/|1-z|^{\alpha}$. Therefore the terms in \eqref{eq:expandedinwrong} can be identified with a contribution from an operator with tree-level dimension $N$. It turns out that there is only a single candidate of such a supermultiplet; the multiplet which contains the ground state of a length 1 open spin chain attached to the so-called $Z=0$ brane:
\beq
\mathcal{O}_{\rm open} \sim \epsilon^{j_1\ldots j_{N-1} a}_{i_1\ldots i_{N-1}b}Z^{i_1}_{j_1}\cdots Z^{i_{N-1}}_{j_{N-1}}Y_{a}^{b}\period
\eeq
Matching with the expected structure of the OPE expansion, we can express the numbers in \eqref{eq:expandedinwrong} in terms of the following OPE data:
\begin{align}
2=&\left.\left({\sf c}_{{\bf 20}^{\prime}\mathcal{D}\mathcal{O}_{\rm open}}\right)^2\right|_{O(g^0)}\frac{\left.\delta \Delta_{\mathcal{O}_{\rm open}}\right|_{O(g^2)}}{2}\comma\qquad
-4=\left.\left({\sf c}_{{\bf 20}^{\prime}\mathcal{D}\mathcal{O}_{\rm open}}\right)^2\right|_{O(g^2)}\comma\nn\\
2(c_1+2c_2)=&\left.\left({\sf c}_{{\bf 20}^{\prime}\mathcal{D}\mathcal{O}_{\rm open}}\right)^2\right|_{O(g^0)}\frac{\left(\left.\delta \Delta_{\mathcal{O}_{\rm open}}\right|_{O(g^2)}\right)^2}{8}\comma\label{eq:variousreltwoloopf}\\
-2(5c_1+11c_2)=&\left.\left({\sf c}_{{\bf 20}^{\prime}\mathcal{D}\mathcal{O}_{\rm open}}\right)^2\right|_{O(g^0)}\frac{\left.\delta \Delta_{\mathcal{O}_{\rm open}}\right|_{O(g^4)}}{2}+\left.\left({\sf c}_{{\bf 20}^{\prime}\mathcal{D}\mathcal{O}_{\rm open}}\right)^2\right|_{O(g^2)}\frac{\left.\delta \Delta_{\mathcal{O}_{\rm open}}\right|_{O(g^2)}}{2}\period\nn
\end{align}
The dimension of this operator can be computed by a sum of the energies of two ``boundary magnons'' in \cite{Hofman:2007xp},
\beq
\delta \Delta_{\mathcal{O}_{\rm open}} =4g^2 -4g^4\period
\eeq
Using this result and the first two equations in \eqref{eq:variousreltwoloopf}, we can determine the structure constants as follows:
\beq
\left({\sf c}_{{\bf 20}^{\prime}\mathcal{D}\mathcal{O}_{\rm open}}\right)^2=1-4g^2\period
\eeq
We can then plug this result into the latter two equations in \eqref{eq:variousreltwoloopf} to obtain
\beq
2=2(c_1+2c_2)\comma\qquad -2 (5c_1+11c_2)=-12\period
\eeq
Solving these equations we get
\beq
c_1=-1\comma\qquad c_2=1\period
\eeq
The fact that this solution is consistent with \eqref{eq:c2=1} provides additional support for the validity of our analysis.

\section{Harmonic Sums and Large Spin Limit\label{ap:largespin}}
\paragraph{Harmonic Sum}In this appendix, we present an analytic expression for the two-loop structure constants of determinant operators and a twist-$2$ operator.
 Owing to the relation \eqref{eq:curiousnumerology}, this becomes almost a trivial task since such expressions for ${\sf c}_{22\mathcal{O}_{2,j}}$ and the bottom wrappings are already known. Below we simply summarize the results emphasizing the difference from ${\sf c}_{22\mathcal{O}_{2,j}}$. To express the results, we use the nested harmonic sums defined by
\beq
\begin{aligned}
S_{a}(j)\equiv \begin{cases}\sum_{n=1}^{j}\frac{1}{n^{|a|}}&a\geq 0\\
\sum_{n=1}^{j}\frac{(-1)^{n}}{n^{|a|}}&a< 0\end{cases}\comma \qquad S_{a,b,\ldots}(j)\equiv \begin{cases}\sum_{n=1}^{j}\frac{1}{n^{|a|}}S_{b,\ldots}(n)&a\geq 0\\
\sum_{n=1}^{j}\frac{(-1)^{n}}{n^{|a|}}S_{b,\ldots}(n)&a< 0\end{cases}\period
\end{aligned}
\eeq
In what follows, we omit writing the argument of the harmonic sum, but it is always spin $j$. Note also that in this appendix we use $j$ to denote spin in order to avoid the clash of notations.

Up to two loops, the result is given by
\beq\label{eq:dOsquaredharmonic}
\left(\frac{{\sf d}_{\mathcal{O}_{2,j}}}{\left.{\sf d}_{\mathcal{O}_{2,j}}\right|_{\rm tree}}\right)^2= ({\tt prefactor}) \times \Big[1-4g^2 S_2 +8g^4(r_2 -6\zeta_3S_{1})\Big]\comma
\eeq
with
\beq
\begin{aligned}
r_2 \equiv \,&5S_{-4}+8S_{-3}S_{1}+4S_{-2}(S_{1})^2+2S_{-2}S_{2}+2(S_{2})^2 +8S_{1}S_{3}\\&+7S_{4}-8S_{-3,1}-8S_{1}S_{-2,1}-6S_{-2,2}-4S_{1,3}-4S_{3,1}+8S_{-2,1,1}\period
\end{aligned}
\eeq
The prefactor is given by\fn{$\zeta_2$ is just $\pi^2/6$, but we chose to use it to have a compact formula.}
\beq
\begin{aligned}
({\tt prefactor})=&\frac{\Gamma (j+1+\frac{\gamma}{2})^2}{\Gamma (j+1)^2}\frac{\Gamma (2j+1)}{\Gamma (2j+1+\gamma)}+\frac{1}{4}\zeta_2\gamma^2\comma
\end{aligned}
\eeq
where $\gamma$ is the anomalous dimension,
\beq
\gamma =8g^2S_1+16g^4\left(-S_{-1}-2S_{-2}S_{1}-2S_{1}S_{2}-S_{3}+2S_{-2,1}\right)\period
\eeq

On the other hand, the result for the single-trace structure constant reads
\beq
\begin{aligned}
\left(\frac{{\sf c}_{22\mathcal{O}_{2,j}}}{\left.{\sf c}_{22\mathcal{O}_{2,j}}\right|_{\rm tree}}\right)^2= ({\tt prefactor}) \times \Big[1-4g^2 S_2 +8g^4(\red{\tilde{r}_2}\,\, \red{+}\,\,6\zeta_3S_{1})\Big]\comma
\end{aligned}
\eeq
with
\beq\label{eq:harmonicsumtilr}
\begin{aligned}
\tilde{r}_2 \equiv \,&5S_{-4}+2(S_{-2})^2+4S_{-3}S_{1}+2S_{-2}S_{2}+2(S_{2})^2\\
&+4S_{1}S_{3}+5S_{4}-4S_{-3,1}-2S_{-2,2}-4S_{-1,3}\period
\end{aligned}
\eeq
As highlighted in red, the difference from \eqref{eq:dOsquaredharmonic} is a rational term \eqref{eq:harmonicsumtilr} and the sign in front of $\zeta_3$.

Just as a reference, here we also show the tree level structure constants:
\beq
\left.\left({\sf d}_{\mathcal{O}_{2,j}}/2\right)^2\right|_{\rm tree}=\left.\left({\sf c}_{22\mathcal{O}_{2,j}}\right)^2\right|_{\rm tree}=\frac{\Gamma (2j+1)}{\Gamma (j+1)^2}\period
\eeq
\paragraph{Large spin limit} From the harmonic sum representation, we can compute the large spin behavior of the three-point function. It simply follows from the asymptotic behavior of the harmonic sum given in \cite{Alday:2013cwa}. Referring the details of the analysis to \cite{Alday:2013cwa}, here we simply present the final answer:
\beq
\begin{aligned}
\log \left[\left(\frac{{\sf d}_{\mathcal{O}_{2,j}}}{\left.{\sf d}_{\mathcal{O}_{2,j}}\right|_{\rm tree}}\right)^2\right]=&-4g^2\left[2\log 2 \log j^{\prime}+\zeta_2\right]\\
&+16g^4\left[\left(\zeta_2\log 2+\frac{9}{2}\zeta_3\right)\log j^{\prime}+\frac{4}{5}(\zeta_2)^2+\frac{3}{2}\zeta_3\log 2\right]\period
\end{aligned}
\eeq
Here $\log j^{\prime}\equiv \log j +\gamma_{E}$. As emphasized in the main text, an important feature of this formula is the absence of a term proportional to $(\log j)^2$. This makes the large-spin behavior drastically different from the single-trace three-point functions. It would be interesting to understand its physical origin.
\section{Q-Functions and Overlap with N\'{e}el State\label{ap:Qfunctions}}
In this appendix, we show that the overlap between the generalized N\'{e}el state and a Bethe state in the SU(2) sector can be expressed as multiple integrals of $Q$-functions.

\paragraph{Bethe states in algebraic Bethe ansatz}
Throughout the main text, we have been using the Bethe states in the coordinate Bethe ansatz, which depend explicitly on the order of the rapidities ${\bf u}=\{u_1,u_2,\cdots \}$. However, for the purpose of deriving the integral expressions, it turned out to be more convenient to use the Bethe states in the algebraic Bethe ansatz, which do not depend on the order. The relation between the two is well-known (see for instance \cite{Escobedo:2010xs}) and reads
\beq
|{\bf u}\rangle_{\rm algebraic}=\left(\prod_{i=1}^{M}\frac{i(u-i/2)^{L}}{u+i/2}\right)\left(\prod_{i<j}\frac{u_i-u_j+i}{u_i-u_j}\right)|{\bf u}\rangle_{\rm coordinate}\comma
\eeq
with $L$ being the length of the spin chain and $M$ being the number of magnons, ${\bf u}=\{u_1,\ldots, u_M\}$. Since we exclusively use the algebraic Bethe ansatz normalization in this appendix, we will omit writing the subscript ${\rm algebraic}$ in what follows. This change of normalization affects the value of the overlap $\langle \text{N\'{e}el}_0|{\bf u}\rangle$, but the normalized overlap $\langle \text{N\'{e}el}_0|{\bf u}\rangle/\sqrt{\langle{\bf u} |{\bf u}\rangle}$ is invariant under the change of the normalization.

\paragraph{Rewriting for $\bm{M=L/2}$}
To rewrite the overlap into integrals, we use the relation between the overlap and a certain partition function in the six-vertex model discussed in \cite{Foda:2015nfk}.. As shown there, the overlap of interest can be expressed as a sum of two partition functions $Z_{{\bf u},{\bm{\theta}}}$ as follows\fn{The result in \cite{Foda:2015nfk} contains yet another parameter $\lambda$. Here we set it to be $-2i$ throughout this appendix.}:
\beq
\langle \text{N\'{e}el}_0| {\bf u}\rangle =(-1)^{M/2}\left.\left(\left. Z_{{\bf u},\bm{\theta}}\right|_{\xi=\frac{i}{2}}+\left. Z_{{\bf u},\bm{\theta}}\right|_{\xi=-\frac{i}{2}}\right)\right|_{\theta_i\to 0}\period
\eeq
Here $\langle\text{N\'{e}el}_0|$ is the weighted N\'{e}el state defined in \eqref{eq:weightedneel}, which is a sum of the generalized N\'{e}el state $\langle\text{N\'{e}el}^{\text{\cite{deLeeuw:2015hxa}}}_{M}|$ introduced in \cite{deLeeuw:2015hxa}:
\beq\label{eq:generalizedandweighted}
\langle \text{N\'{e}el}_0|=\sum_{M=0}^{L/2}\langle\text{N\'{e}el}^{\text{\cite{deLeeuw:2015hxa}}}_{M}|\period
\eeq
and $\bm{\theta}$ is a set of inhomogeneities of the spin chain which are parity symmetric, $\bm{\theta}=\{\theta_1,-\theta_1,\ldots, \theta_{\frac{L}{2}},-\theta_{\frac{L}{2}}\}$. $\xi$ is an extra parameter whose definition can be found in \cite{Foda:2015nfk}. In what follows, we assume that both the length $L$ and the number of magnons $M$ to be even.

The partition function $Z_{{\bf u},\bm{\theta}}$ admits a simple $\frac{L}{2}\times \frac{L}{2}$ determinant expression when $M=\frac{L}{2}$, as was shown in \cite{Foda:2015nfk},
\beq\label{eq:izergintype}
\begin{aligned}
Z_{{\bf u},\bm{\theta}}=&\, i^{\frac{L^2}{4}+2L}\prod_{j=1}^{L/2}(u_j+\xi) \prod_{a=1}^{L/2}(2\theta_a +i)\times\frac{\prod_{j,a}\left[(u_j-\theta_a)^2+\frac{1}{4}\right]\left[(u_j+\theta_a)^2+\frac{1}{4}\right]}{\prod_{j<k}(u_j^2-u_k^2)\prod_{a<b}(\theta_a^2-\theta_b^2)}\\
&\times \det \mathcal{M}\comma
\end{aligned}
\eeq
with
\beq
\mathcal{M}_{ij}\equiv \frac{1}{\left[(\theta_i-u_j)^2+\frac{1}{4}\right]\left[(\theta_i+u_j)^2+\frac{1}{4}\right]}\period
\eeq
Note that the determinants that appeared in the main text are of the form of the Gaudin norm. By contrast, the determinant \eqref{eq:izergintype} is similar to the so-called Izergin determinant for the overlaps of the Bethe states.

In what follows, we rewrite this determinant expression into a sum over partitions. We then later show that the same sum over partitions arise from an integral representation, thereby establishing the equivalence between the determinant formula and the integral representation. The derivation closely follows the one in Appendix B of \cite{Kazama:2013rya}\fn{See also \cite{Gromov:2016itr}.}.

The first step of rewriting is to factorize the matrix $\mathcal{M}$ and convert it to  a sum of two matrices,
\beq
\mathcal{M}=\Theta\cdot (M^{+}-M^{-})\comma
\eeq
with
\beq
\begin{aligned}
M^{\pm}_{ij} \equiv \frac{1}{(\theta_j \pm\frac{i}{2})^2-u_j^2}\comma\qquad \Theta_{ij}=\frac{i\delta_{ij}}{2 \theta_i }\period
\end{aligned}
\eeq
We then have
\beq
\det \mathcal{M}=\left(\frac{i}{2}\right)^{L/2}\prod_{k=1}^{L/2}\frac{1}{\theta_k}\det \left(M^{+}-M^{-}\right)\period
\eeq
The determinant of a sum of two matrices can be decomposed (simply using the definition of the determinant) as
\beq\label{eq:expanddetap}
\begin{aligned}
\det  (M^{+}-M^{-})=&\sum_{\sigma \in S_{L/2}}(-1)^{|\sigma|}(M_{1\sigma_1}^{+}-M_{1\sigma_1}^{-})\cdots (M_{\frac{L}{2}\,\sigma_{L/2}}^{+}-M_{\frac{L}{2}\,\sigma_{L/2}}^{-})\\
=&\sum_{\epsilon_k=\pm}(-1)^{n_{-}}\sum_{\sigma \in S_{L/2}}(-1)^{|\sigma|}M_{1\sigma_1}^{\epsilon_1}\cdots M_{\frac{L}{2}\,\sigma_{L/2}}^{\epsilon_{L/2}}\comma
\end{aligned}
\eeq
where $n_{-}$ is the total number of $-$'s in the set $\{\epsilon_k\}$. Since the second line in \eqref{eq:expanddetap} by itself takes a form of an expansion of a determinant, we can reexpress it as
\beq
\det (M^{+}-M^{-})=\sum_{\epsilon_{k}=\pm}(-1)^{n_{-}}\det M_{ij}^{\epsilon_i}\period
\eeq
At this point, one can use the Cauchy determinant formula to express the determinant as
\beq
\det M_{ij}^{\epsilon_i}=\det \left(\frac{1}{(\theta_i+ \frac{i\epsilon_i}{2})^2-u_j^2}\right)=\frac{\prod_{j<k}(u_j^2-u_k^2)((\theta_j^{\epsilon_j})^2-(\theta_k^{\epsilon_k})^2 )}{\prod_{j,k}((\theta_j^{\epsilon_j})^2-u_k^2)}\comma
\eeq
where we introduced the notation
\beq
\theta_j ^{\epsilon_j}=\theta_j+\frac{i\epsilon_j}{2}\period
\eeq
Substituting this expression into \eqref{eq:izergintype}, we get
\beq
\begin{aligned}\label{eq:tocompareSoV1}
\left.  Z_{{\bf u},\bm{\theta}} \right|_{M=L/2}=&\, i^{\frac{L^2}{4}+\frac{5L}{2}}\prod_{j=1}^{L/2}(u_j+\xi)\prod_{a=1}^{L/2}\left(1 +\frac{i}{2\theta_a}\right)\\
&\times \sum_{\epsilon_k=\pm}(-1)^{n_{-}}\prod_{j,k}^{L/2}\left[(\theta_j^{-\epsilon_j})^2-u_k^2\right]\prod_{j<k}^{L/2}\frac{(\theta_j^{\epsilon_j})^2-(\theta_k^{\epsilon_k})^2}{\theta_j^2-\theta_k^2}\period
\end{aligned}
\eeq

\paragraph{Integral representation} Let us now consider the following integral representation:
\beq
\mathcal{I}=\prod_{k=1}^{L/2}\left(\oint_{\theta_k\pm i/2}\frac{x_kdx_k}{2\pi i}\right)\prod_{j<k}^{L/2}(x_j^2-x_k^2)\prod_{k=1}^{L/2}\frac{Q_{\bf u}(x_k)Q_{\bf u}(-x_k)}{Q_{\bm{\theta}}^{+}(x_k)Q_{\bm{\theta}}^{-}(x_k)}\comma
\eeq
with
\beq
Q_{\bf u}(x)\equiv \prod_{k=1}^{M}(x-u_k)\comma\qquad Q_{\bm{\theta}}(x)=\prod_{k=1}^{L/2}(x^2-\theta_k^2)\period
\eeq
Here the integration contour of $x_k$ goes around the two points $\theta_k\pm i/2$ counterclockwise.
The function $Q_{\bf u}(x)$ is often called the Baxter function or $Q$-function in the literature.

The integral $\mathcal{I}$ can be computed explicitly by taking the residues. The result reads
\beq
\begin{aligned}
\mathcal{I}=&(-1)^{\frac{L^2}{4}}\prod_{k=1}^{L/2}\frac{1}{2\theta_k}\sum_{\epsilon_k=\pm}\prod_k \frac{1}{2i\epsilon_k}\prod_{j,k}^{L/2}\left[(\theta_j^{\epsilon_j})^2-u_k^2\right]\\
&\times\prod_{j<k}^{L/2}\frac{(\theta_j^{\epsilon_j})^2-(\theta_k^{\epsilon_k})^2}{\left[\theta_j^2-\theta_k^2\right]^2\left[(\theta_j^{2\epsilon_j})^2-\theta_k^2\right]\left[\theta_j^2-(\theta_k^{2\epsilon_k})^2\right]}
\end{aligned}
\eeq
Now, using the following identity which can be verified for every pair of $\epsilon_{k,j}=\pm$,
\beq
\frac{(\theta_j^{\epsilon_j})^2-(\theta_k^{\epsilon_k})^2}{\left[(\theta_j^{2\epsilon_j})^2-\theta_k^2\right]\left[\theta_j^2-(\theta_k^{2\epsilon_k})^2\right]}=\frac{(\theta_j^{-\epsilon_j})^2-(\theta_k^{-\epsilon_k})^2}{\left[(\theta_j-\theta_k)^2+1\right]\left[(\theta_j+\theta_k)^2+1\right]}\comma
\eeq
we can further rewrite it as\fn{To derive the right hand side, we flipped the signs of $\epsilon_k$'s and used the identity $\prod_{k}\epsilon_k =(-1)^{n_{-}}$.}
\beq
\begin{aligned}\label{eq:tocompareSoV2}
\mathcal{I}
=&i^{\frac{L^2}{2}+\frac{L}{2}}\prod_{k=1}^{L/2}\frac{1}{4\theta_k}\prod_{j<k}^{L/2}\frac{1}{\left[(\theta_j-\theta_k)^2+1\right]\left[(\theta_j+\theta_k)^2+1\right]}\\
&\times\sum_{\epsilon_k=\pm} \prod_k (-1)^{n_{-}}\prod_{j,k}^{L/2}\left[(\theta_j^{-\epsilon_j})^2-u_k^2\right]\prod_{j<k}\frac{(\theta_j^{\epsilon_j})^2-(\theta_k^{\epsilon_k})^2}{(\theta_j^2-\theta_k^2)^2}\period
\end{aligned}
\eeq

Comparing \eqref{eq:tocompareSoV1} and \eqref{eq:tocompareSoV2}, we arrive at the following formula:
\beq
\begin{aligned}\label{eq:SoVzero}
\left.Z_{{\bf u},\bm{\theta}}\right|_{M=L/2}=&\frac{i^{-\frac{L^2}{4}}\prod_k^{L/2} (2\theta_k +i)(2u_k+2\xi)}{\prod_{j<k}^{L/2}\left[(\theta_j-\theta_k)^2+1\right]\left[(\theta_j+\theta_k)^2+1\right]}\prod_{j<k}^{L/2}(\theta_j^2-\theta_k^2)\\
&\times\prod_{k=1}^{L/2}\left(\oint_{\theta_k \pm i/2} \frac{x_kdx_k}{2\pi i} \right) \prod_{j<k}^{L/2} (x_j^2-x_k^2) \prod_k \frac{Q_{\bf u}(x_k)Q_{\bf u}(-x_k)}{Q_{\bm{\theta}}^{+}(x_k)Q_{\bm{\theta}}^{-}(x_k)}\period
\end{aligned}
\eeq
Although the formula establishes the relation between the partition function and the integral, it is unfortunately singular in the homogenous limit $\theta_k\to 0$ since the integration contours get pinched by the poles of the integrand. To make it nonsingular, we perform a trick developed in \cite{Kazama:2013rya};  we add an extra Vandermonde-like factor,
\beq
\prod_{j<k}^{L/2}\sinh [\pi (x_j-x_k)]\sinh [\pi (x_j+x_k)]\comma
\eeq
extend the contours so that they encircle all $\theta_k\pm i/2$, and divide by the overall factor\fn{For more detailed discussion, see \cite{Kazama:2013rya}.}. This however turned out to be insufficient in the present case since the integrand contains poles also at $-\theta_k\pm i/2$. To resolve this problem, we perform the change of variables $x_k\to -x_k$ to each $x_k$ separately. This gives $2^{L}$ different integral expressions and we average over all those expressions. Effectively, this amounts to extending the contour so that they include both $\theta_k\pm i/2$ and $-\theta_k\pm i/2$. We thus finally get
\beq
\begin{aligned}\label{eq:SoVtwo}
\left.Z_{{\bf u},\bm{\theta}}\right|_{M=L/2}=&\frac{i^{-L/2}\prod_k^{L/2} (2\theta_k +i)(u_k+\xi)}{\left(\frac{L}{2}\right)!\prod_{j<k}^{L/2}((\theta_j-\theta_k)^2+1)((\theta_j+\theta_k)^2+1)}\prod_{j<k}^{L/2}\frac{\pi^2(\theta_j^2-\theta_k^2)}{\sinh [\pi(\theta_j-\theta_k)]\sinh [\pi(\theta_j+\theta_k)]}\\
&\times\oint_{\mathcal{C}}\tilde{\Delta}(x_1,\ldots, x_{\frac{L}{2}})\times \prod_{k=1}^{L/2} \frac{x_kdx_k }{2\pi i} \frac{Q_{{\bf u}}(x_k)Q_{\bf u}(-x_k)}{Q_{\bm{\theta}}^{+}(x_k)Q_{\bm{\theta}}^{-}(x_k)}\comma
\end{aligned}
\eeq
with
\beq
\begin{aligned}
\tilde{\Delta}(x_1,\ldots, x_\frac{L}{2})&=\prod_{j<k}^{L/2}(x_j^2-x_k^2)\frac{\sinh [\pi (x_j-x_k)]\sinh [\pi (x_j+x_k)]}{\pi^2}\\
&=\prod_{j<k}^{L/2}(x_j^2-x_k^2)\frac{\sinh^2 (\pi x_j)-\sinh^2(\pi x_k)}{\pi^2}\period
\end{aligned}
\eeq
The contour $\mathcal{C}$ now encircles all the poles of the integrand.
It is now trivial to take the homogenous limit $\theta_k\to 0$, and the result reads
\beq
\begin{aligned}\label{eq:SoVthree}
\left.Z_{{\bf u},{\bf 0}}\right|_{M=L/2}=&\frac{\prod_k (u_k+\xi)}{\left(\frac{L}{2}\right)!}\oint_{\mathcal{C}}\tilde{\Delta}(x_1,\ldots, x_{L/2})\times \prod_{k=1}^{L/2} \frac{x_kdx_k }{2\pi i} \frac{Q_{\bf u}(x_k)Q_{\bf u}(-x_k)}{Q_{\bm{0}}^{+}(x_k)Q_{\bm{0}}^{-}(x_k)}\period
\end{aligned}
\eeq

\paragraph{Generalization to $\bm{M<L/2}$} We now extend the result to a state with a less number of magnons. This can be done by sending some of the rapidities to infinite. Since the creation operator in the algebraic Bethe ansatz behaves as
\beq
B(u)\sim i u^{L-1}S^{-}+\cdots\comma
\eeq
the limit converts the overlap in the following way:
\beq
\lim_{u_{L/2}\to \infty}\frac{\langle \text{N\'{e}el}_0|u_1,\ldots, u_{L/2}\rangle}{i u^{L-1}_{L/2}}=\langle \text{N\'{e}el}_0|S^{-}|u_1,\ldots, u_{L/2-1}\rangle \period
\eeq
Repeating this procedure, we get
\beq\label{eq:howtotakelimit}
\lim_{u_{M+1},\ldots u_{\frac{L}{2}}\to \infty}\frac{Z_{{\bf u},\bm{0}}}{\left(\frac{L}{2}-M\right)!\times i^{\frac{L}{2}-M}\prod_{k=M+1}^{L/2}u_k^{L-1}}\period
\eeq
The extra factor $\left(\frac{L}{2}-M\right)!$ comes from the following equality
\beq
\langle \text{N\'{e}el}_0|(S^{-})^{k}|{\bf u}\rangle=k!\langle \text{N\'{e}el}_0|{\bf u}\rangle\comma
\eeq
which can be verified by expanding the weighted N\'{e}el state as a sum of generalized N\'{e}el states and using the property
\beq
\langle\text{N\'{e}el}^{\text{\cite{deLeeuw:2015hxa}}}_{L/2}|(S^{-})^k=k!\langle\text{N\'{e}el}^{\text{\cite{deLeeuw:2015hxa}}}_{L/2-k}|\period
\eeq
Taking the limit \eqref{eq:howtotakelimit} carefully, we obtain the following expression for the partition function
\beq
\begin{aligned}\label{eq:SoVfour}
Z_{{\bf u},{\bf 0}}=&\frac{i^{\frac{L}{2}-M}\prod_{k=1}^{M} (u_k+\xi)}{\left(\frac{L}{2}\right)!\left(\frac{L}{2}-M\right)!}\oint_{\mathcal{C}}\tilde{\Delta}(x_1,\ldots, x_{L/2})\left(\sum_{k}^{L/2}x_{k}^2\right)^{\frac{L}{2}-M} \prod_{k=1}^{\frac{L}{2}} \frac{x_kdx_k }{2\pi i} \frac{Q_{\bf u}(x_k)Q_{\bf u}(-x_k)}{Q_{\bm{0}}^{+}(x_k)Q_{\bm{0}}^{-}(x_k)}\period
\end{aligned}
\eeq

 The expression \eqref{eq:SoVfour} does not take quite a simple form owing to the factor $\left(\sum_{k}^{L/2}x_{k}^2\right)^{\frac{L}{2}-M}$. This however can be remedied by rewriting the integral by introducing the ``twist'' parameter $\phi$ as
\beq
\begin{aligned}\label{eq:SoVfive}
Z_{{\bf u},{\bf 0}}=&\frac{\prod_{k=1}^{M} (u_k+\xi)}{\left(\frac{L}{2}\right)!}\int_{0}^{2\pi}\frac{d\phi}{2\pi}e^{i\phi \left(\frac{L}{2}-M\right)}\oint_{\mathcal{C}}\tilde{\Delta}\times  \prod_{k=1}^{\frac{L}{2}} \frac{x_kdx_k }{2\pi i} \frac{Q_{\bf u}(x_k)Q_{\bf u}(-x_k)e^{-i\phi x_{k}^2}}{Q_{\bm{0}}^{+}(x_k)Q_{\bm{0}}^{-}(x_k)}\period
\end{aligned}
\eeq

\paragraph{Final result} So far we did not use the fact that ${\bf u}$'s is parity symmetric. When ${\bf u}$ is parity symmetric we have $Z_{{\bf u},{\bf 0}}|_{\xi=\frac{i}{2}}=Z_{{\bf u},{\bf 0}}|_{\xi=-\frac{i}{2}}$, and therefore the overlap can be expressed simply as
\beq\label{eq:finaloverlapintegralSoV}
\begin{aligned}
&\langle \text{N\'{e}el}_0| {\bf u}\rangle =\\
&2\frac{\prod_{k=1}^{M/2} (u_k^2+\frac{1}{4})}{\left(\frac{L}{2}\right)!}\int_{0}^{2\pi}\frac{d\phi}{2\pi}e^{i\phi \left(\frac{L}{2}-M\right)}\oint_{\mathcal{C}}\tilde{\Delta}\times  \prod_{k=1}^{\frac{L}{2}} \frac{x_kdx_k }{2\pi i} \frac{Q_{\bf u}(x_k)Q_{\bf u}(-x_k)e^{-i\phi x_{k}^2}}{Q_{\bm{0}}^{+}(x_k)Q_{\bm{0}}^{-}(x_k)}\comma
\end{aligned}
\eeq
where $Q_{\bf u}$ is now given by
\beq
Q_{\bf u}(x)=\prod_{k=1}^{M/2}(x^2-u_k^2)\period
\eeq
This is our final expression for the overlap. To write down the normalized overlap $\langle \text{N\'{e}el}_0|{\bf u}\rangle/\sqrt{\langle{\bf u} |{\bf u}\rangle}$, we combine this result with the integral expression for the norm\fn{Here we slightly rewrote the integrand so that the expression looks similar to \eqref{eq:finaloverlapintegralSoV}.} obtained in \cite{Kazama:2013rya}:
\beq
\begin{aligned}
&\langle {\bf u}| {\bf u}\rangle =\frac{1}{L!}\int_{0}^{2\pi}\frac{d\phi}{2\pi}e^{i\phi \left(L-2M\right)}\oint_{\mathcal{C}}\Delta\times  \prod_{k=1}^{L} \frac{dx_k }{2\pi i} \frac{Q_{\bf u}(x_k)Q_{\bf u}(x_k)e^{-i\phi x_{k}}}{Q_{\bm{0}}^{+}(x_k)Q_{\bm{0}}^{-}(x_k)}\comma
\end{aligned}
\eeq
with
\beq
\Delta(x_1,\ldots,x_L)=\prod_{i<j}\frac{(x_i-x_j)\sinh [\pi (x_i-x_j)]}{\pi}\period
\eeq
It is intriguing that the results for the norm and the overlap take remarkably similar forms. It would be interesting to develop a deeper understanding of this similarity.

The results we obtained resemble the integral expressions in the so-called Sklyanin's separation of variable (SoV) approach. In recent years, the SoV approach has been applied successfully to several interesting physical observables; the norms of Bethe states \cite{Gromov:2016itr}, the expectation value of the null polygonal Wilson loop at weak coupling \cite{Belitsky:2014rba,Belitsky:2016fce}, the three-point function at weak coupling \cite{Kazama:2013rya,Jiang:2015lda}, certain fishnet type diagrams in two dimensions \cite{Derkachov:2018rot}, and the lightlike limit of correlation functions at weak coupling \cite{Belitsky:2019ygi}. More recently, similar integral expressions showed up\fn{These results were computed not by integrability but by some other means, such as the supersymmetric localization and the resummation of diagrams. Nevertheless, the result turned out to take a remarkably simple form when expressed in terms of Q-functions.} also in the computation of correlation functions on the BPS Wilson loop \cite{Giombi:2018qox,Giombi:2018hsx} and the Wilson loop in the ladder limit \cite{Cavaglia:2018lxi}. Recurrent appearance of such integral expressions strongly suggest that the SoV method might be a useful framework to study the correlation functions at finite coupling.
\bibliographystyle{utphys}
\bibliography{GGdraftref2}
\end{document}